\documentclass[a4paper,12pt,numbered,print,index]{Classes/PhDThesisPSnPDF}
\input{Preamble/preamble}

\title{Weather-based forecasting of energy generation, consumption and price for microgrids management}

\author{Jonathan Dumas}

\dept{Faculty of Applied Sciences \\Department of Computer Science and Electrical Engineering}

\university{Li\`ege University}
\crest{\includegraphics[width=0.45\textwidth]{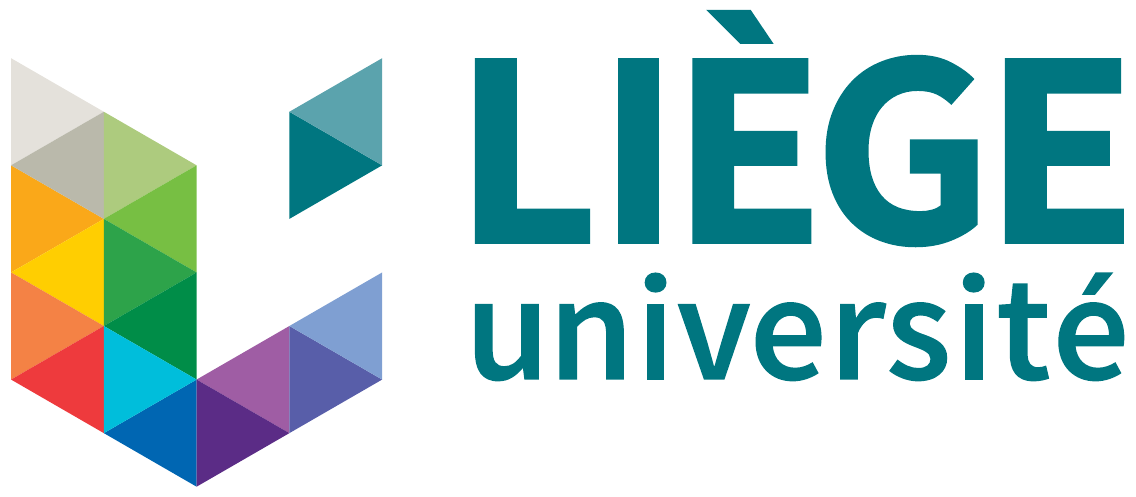}}

\supervisor{Prof. Bertrand Corn\'elusse}

\supervisorlinewidth{0.5\textwidth}

\renewcommand{\submissiontext}{This version of the manuscript is pending the approval of the jury. \\ This dissertation is submitted for the degree of}

\degreetitle{Doctor of Philosophy}

\college{Montefiore research unit}

\subject{LaTeX} \keywords{{LaTeX} {PhD Thesis} {Engineering} {University of
Li\`ege}}

\ifdefineAbstract
\fi

\ifdefineChapter
\fi

\begin{document}

\frontmatter

\maketitle

\begin{jury}

\noindent \textbf{\textcolor{RoyalBlue}{Prof. Quentin Louveaux (President)}} University of Li\`ege, Belgium.\\
\noindent \textbf{\textcolor{RoyalBlue}{Prof. Bertrand Corn\'elusse (Supervisor)}} University of Li\`ege, Belgium.\\
\noindent \textbf{\textcolor{RoyalBlue}{Prof. Gilles Louppe}} University of Li\`ege, Belgium.\\
\noindent \textbf{\textcolor{RoyalBlue}{Prof. Pierre Pinson}} Technical University of Denmark, Denmark.\\
\noindent \textbf{\textcolor{RoyalBlue}{Prof. Ricardo Bessa}} INESC TEC, Portugal.\\
\noindent \textbf{\textcolor{RoyalBlue}{Prof. Simone Paoletti}} University of Siena, Italy.\\
\end{jury}

\begin{dedication} 

\LARGE `` Nothing is impossible, just do it and let's see.  \LARGE '' \\
--- \textbf{A true coconut-shaker} \\ [5mm]

\LARGE ``There are admirable potentialities in every human being. Believe in your strength and your youth. Learn to repeat endlessly to yourself: It all depends on me. \LARGE '' \\
--- \textbf{Andr\'e Gide} \\ [5mm]

\LARGE ``In vino veritas.  \LARGE '' \\
--- \textbf{Pliny the Elder, 23–79 A.D.} \\

\end{dedication}

\begin{declaration}

I hereby declare that except where specific reference is made to the work of others, the contents of this dissertation are original and have not been submitted in whole or in part for consideration for any other degree or qualification in this or any other university. This dissertation is my own 
work and contains nothing which is the outcome of work done in collaboration with others, except as specified in the text and Acknowledgements.


\end{declaration}

\begin{acknowledgements}      

Throughout the writing of this dissertation, I have received a great deal of support and assistance. \\

\noindent \textbf{\textcolor{RoyalBlue}{Family}} \\
I would like to acknowledge my family, and especially Perrine and my daughter Lily. I could not have achieved this thesis without their emotional and material support. They always have been there for me. Perrine saved me much time taking care of our beloved daughter, \textbf{she did, in fact, the most challenging part of the job}. In my opinion, it is simple to work, especially in a research environment. It is much harder to raise a child. In addition, I could not have been so implied, without her support, in non-profit organizations to help make energy transition to a carbon-free economy: the Shifters Belgium\footnote{\url{https://theshiftproject.org/equipe/\#benevoles}}, the working group on the sustainability of the ReD\footnote{\url{https://www.facebook.com/reduliege/}} (an association aiming to gather the Ph.D. candidates from the University of Li\`ege), the climate\footnote{\url{https://climatefresk.org/}} and digital collages\footnote{\url{https://digitalcollage.org/}}, and the ClimActes 2021 summer school\footnote{\url{http://climactes.org/}}. \\

\noindent \textbf{\textcolor{RoyalBlue}{Supervisor}} \\
I express my gratitude to my supervisor Prof. Bertrand Corn\'elusse, whose expertise was invaluable in formulating the research questions and methodology. Bertrand is significantly implicated in designing exciting lectures for bachelor and master students. He is also committed to supervising his Ph.D. students despite its busy agenda. Professors are supposed to be supermen or superwomen: teach, teach and teach, then, being at the front end of the research, and of course, implicated in the University and Faculty challenges. However, they have as everybody 24 hours per day :). He made it possible to work with the Sienna group and encouraged me throughout the past three years. We have had helpful discussions on technical topics like optimization and machine learning and friendly talks on many other questions. Your insightful feedback pushed me to sharpen my thinking and brought my work to a higher level. We had the opportunity to spend an excellent stay at the first microgrids summer school at Belfort in 2019. He also provided the funds for three summer schools:  microgrids summer school at Belfort [2019], sustainable ICT summer school at Catholic University of Louvain [2020], and ClimActes at Li\`ege University [2021]. He provided the funds for attending several conferences: 16th European Energy Market Conference (EEM 2019), XXI Power Systems Computation Conference (PSCC 2020), 16th International Conference on Probabilistic Methods Applied to Power Systems (PMAPS 2020), 14th IEEE PowerTech (PowerTech 2021). Finally, the two stays at Sienna University in 2019 and 2020 (just before the lockdown of the covid-19 crisis). \\

\noindent \textbf{\textcolor{RoyalBlue}{Thesis committee}} \\
I am grateful to Prof. Simone Paoletti, Prof. Quentin Louveaux, and Xavier Fettweis, research associate FNRS, for being part of the thesis committee, their support, and valuable guidance throughout this thesis and the discussions about optimization and weather forecasting. In addition, Xavier Fettweis taught how to use the MAR regional climate model. I want to thank all three of you for your regular support and the opportunities to further my research. \\

\noindent \textbf{\textcolor{RoyalBlue}{Thesis jury}} \\
I am deeply honored by this jury. Each one of the members is profoundly remarkable in his field and inspired my research during this thesis. I could not have dreamt of a better jury and hope to be at the level expected. In addition, I sincerely thank the jury members for devoting themselves to reading this manuscript and all the related administrative duties. \\

\noindent \textbf{\textcolor{RoyalBlue}{Li\`ege University Applied Science Faculty colleagues}} \\
I thank my colleagues and friends of the Applied Science Faculty for their support and for the exciting discussions we had together during the last three years: especially Selmane Dakir, Antonio Sutera, Antoine Wehenkel, Laurine Duchesne, Pascal Leroy, Adrien Bolland, Raphael Fonteneau, Sylvain Quoilin, Gaulthier Gain, Ioannis Noukas, Miguel Manuel de Villena, and all the others that I did not mention. I also thank Gilles Louppe, associate professor at Li\`ege University, for the quality of his course "INFO8010 - Deep Learning". Gilles Louppe is very involved in designing exciting courses. I enjoyed the lectures that I followed in Spring 2021 at the end of this thesis. I wish I had attended them at the beginning of my thesis to understand better the deep learning topic that I learned from scratch. \\

\noindent \textbf{\textcolor{RoyalBlue}{Coconut-shakers group and ReD}} \\
I thank my beloved friends of the \textit{coconut-shakers group}: Antoine Dubois, Fran\c cois Rigo, Thibaut Theate, Gilles Corman, and Adrien Corman. We have worked together intensively to propose actions and ideas at the Applied Science Faculty to make tiny changes about the climate change issue. There is still a massive gap between rhetoric and actions. Nevertheless, we keep on believing that one-day things will change!
I also thank the active members of the ReD: Kathleen Jacquerie, Chloe Stevenne, Coline Gr\'egoire, and all the others that I did not mention. \\

\noindent \textbf{\textcolor{RoyalBlue}{Sienna University colleagues}} \\
I appreciated working with Prof. Simone Paoletti, Prof. Antonello Giannitrapani, and Prof. Antonio Vicino from Sienna University. It was terrific to work with them at Sienna University, and we have had enjoyable moments. They are devoted to their work, and they know how to make a good working relationship. They helped to understand stochastic programming better and to develop exciting research directions. \\

\noindent \textbf{\textcolor{RoyalBlue}{Master students}} \\
I thank all the master students who did incredible work on various topics such as forecasting, optimization during their master thesis, especially during the covid-19 crisis. It was a great pleasure to work with them: Audrey Lempereur [2020-Applied Science Faculty], Clement Liu [2020-Ecole Polytechnique], Jean Cyrus de Gourcuff [2021-Ecole Polytechnique], Colin Cointe [2021-Mines ParisTech], Damien Lanaspeze [2021-Mines ParisTech], Benjamin Delvoye [2021-Applied Science Faculty], and Gabriel Guerra Martin [2021-Polytechnic University of Catalonia]. \\

\noindent \textbf{\textcolor{RoyalBlue}{Li\`ege University Applied Science Faculty staff}} \\
I sincerely thank the shadow workers of the Applied Science Faculty: the cleaning workers (thank you Patricia!), the administrative staff (Eric Vangenechten), and all the others that I did not mention. 
I also thank the academic staff of the Applied Science Faculty: Eric Delhez (for the discussions on sustainability even if I think can do better!), Ang\'elique Leonard, Dominique Toye, and all the others that I did not mention. \\

\noindent \textbf{\textcolor{RoyalBlue}{Li\`ege University human resource staff}} \\
I thank all the human resource staff. We are so fortunate at the University of Li\`ege to have free and very high-quality training courses on various topics such as self-development, negotiation, teamwork. In particular, I appreciate the training courses of Jean Yves Girin. He is incredible, and he helped me improve my human skills (but there is still much work to do). I can only encourage all the academics (including professors), administrative staff, Ph.D. candidates, and researchers to attend these training courses. I was the only man and the only one from the Applied Science Faculty most of the time. Does that mean that the 'men' do not need these training courses about human skills? \\

\noindent \textbf{\textcolor{RoyalBlue}{Li\`ege University IFRES}} \\
I thank the IFRES for the pedagogical training that we attend as teaching assistants. We are lucky that Li\`ege University provides these training courses. \\

\noindent \textbf{\textcolor{RoyalBlue}{P. Pinson}} \\
I thank professor Prof. Pierre Pinson from the Technical University of Denmark. His work has been a great inspiration during this thesis. In addition, he is open to discussion and sharing all material from his courses. In particular, I used a lot of the course materials "Renewables in Electricity Markets" to design the forecasting lessons that I taught in the course "ELEN0445 Microgrids" at the University of Li\`ege during the years 2018, 2019, 2020, and 2021. \\

\noindent \textbf{\textcolor{RoyalBlue}{Summer schools}} \\
\textit{Microgrids Belfort 2019 summer school.} I thank all the members of the microgrids Belfort 2019 summer school, and especially Robin Roche. It was the first summer school on microgrids, and we learned a lot in an enjoyable working environment. The Typhoon Hil team did a great job presenting its tools and organize the challenge. \\

\noindent \textit{Sustainable ICT 2020 summer school.} I also thank the members of the Sustainable ICT 2020 summer school that took place at the Catholic University of Louvain (UCL). It was an excellent summer school with very skilled researchers and Ph.D. candidates. There is at UCL a group of Ph.D. candidates (Thibault Pirson, Gregoire Lebrun, and many others) and professors (Jean-Pierre Raskin, Herv\'e Jeanmart, David Bol) very motivated to address the sustainability challenges. It is a great source of inspiration.  \\

\noindent \textit{ClimACTES 2021 summer school.} I am grateful to all the members of the ClimACTES 2021 summer school that took place at Li\`ege University. This event was outstanding where we have faced the challenges of climate change and the urgent need to reduce the emission of greenhouse gases drastically. 
The first week was dedicated to lessons from professors from several universities about these challenges. Then, we worked in the second week to develop projects to achieve ambitious targets to help to decarbonize the economy. I was part of the education team, and I learned from students and people from various backgrounds. It was one of the richest experiences in my life. Philippe Gilson, that founded ClimACTES, is a great source of inspiration. The work of all the ClimACTES volunteers was incredible. They dedicated two entire years to build this summer school. I can only say congratulations!\\

\noindent \textbf{\textcolor{RoyalBlue}{Reviewers}} \\
I would like to thank all the reviewers of the papers submitted to conferences and journals. They helped me to improve the work proposed and to get out of my comfort zone.  \\

\noindent \textbf{\textcolor{RoyalBlue}{Li\`ege University}} \\
Finally, I thank the University of Li\`ege. Indeed, they financed me as a teaching assistant during these two last years and provided me with a nice working environment. 

\end{acknowledgements}

\begin{abstract}

\epi{Do. Or do not. There is no try.}{Yoda}

\noindent \textbf{\textcolor{RoyalBlue}{Climate Change}} \\
The decade 2009-2019 was particularly intense in rhetoric about efforts to tackle the climate crisis, such as the 2015 United Nations Climate Change Conference, COP 21. However, the carbon dioxide emissions at the world scale increased constantly from 29.7 (GtCO\textsubscript{2}) in 2009 to 34.2 in 2019. \textit{The current gap between rhetoric and reality on emissions was and is still huge}. The Intergovernmental Panel on Climate Change proposed different mitigation strategies to achieve the net emissions reductions that would be required to follow a pathway that \textbf{limits global warming to 1.5°C} with no or limited overshoot. 
There are still pathways to reach net-zero by 2050. Several reports propose detailed scenarios and strategies to achieve these targets. They remain narrow and highly challenging, requiring all stakeholders, governments, businesses, investors, and citizens to take action this year and every year after so that the goal does not slip out of reach. In most of these trajectories, \textit{electrification} and an increased share of \textit{renewables} are some of the key pillars. The transition towards a carbon-free society goes through \textbf{an inevitable increase in the share of renewable generation in the energy mix} and a drastic decrease in the total consumption of fossil fuels. \\

\noindent \textbf{\textcolor{RoyalBlue}{Thesis topic}} \\
In contrast to conventional power plants, renewable energy is subject to uncertainty. Most of the generation technologies based on renewable sources are non-dispatchable, and their production is stochastic and complex to predict in advance. A high share of renewables is challenging for power systems that have been designed and sized for dispatchable units. Therefore, this thesis studies the \textit{integration of renewables in power systems} by investigating forecasting and decision-making tools. 

Since variable generation and electricity demand both fluctuate, they must be forecast ahead of time to inform real-time electricity scheduling and longer-term system planning. Better short-term forecasts enable system operators to reduce reliance on polluting standby plants and proactively manage increasing amounts of variable sources. Better long-term forecasts help system operators and investors to decide where and when to build variable plants. In this context, \textit{probabilistic} forecasts, which aim at modeling the distribution of all possible future realizations, have become a vital tool to equip decision-makers, hopefully leading to better decisions in energy applications. 

When balancing electricity systems, system operators use scheduling and dispatch to determine how much power every controllable generator should produce. This process is slow and complex, governed by NP-hard optimization problems such as unit commitment and optimal power flow. 
Scheduling becomes even more complex as electricity systems include more storage, variable generators, and flexible demand. Thus, scheduling must improve significantly, allowing operators to rely on variable sources to manage systems. Therefore, stochastic or robust optimization strategies have been developed along with decomposition techniques to make the optimization problems tractable and efficient.\\

\noindent \textbf{\textcolor{RoyalBlue}{Thesis content}} \\
These two challenges raise two central research questions studied in this thesis: (1) How to produce reliable probabilistic renewable generation forecasts, consumption, and electricity prices? (2) How to make decisions with uncertainty using probabilistic forecasts to improve scheduling? The thesis perimeter is the energy management of "small" systems such as \textit{microgrids} at a residential scale on a day-ahead basis. The manuscript is divided into two main parts to propose directions to address both research questions: (1) a forecasting part; (2) a planning and control part. \\

\noindent \textbf{\textcolor{RoyalBlue}{Thesis first part}} \\
The forecasting part presents several techniques and strategies to produce and evaluate probabilistic forecasts. We provide the forecasting basics by introducing the different types of forecasts to characterize the behavior of stochastic variables, such as renewable generation, and the tools to assess the different types of forecasts. An example of forecast quality evaluation is given by assessing PV and electrical consumption point forecasts computed by standard deep-learning models such as recurrent neural networks. Then, the following Chapters investigate the quantile forecasts, scenarios, and density forecasts on several case studies. First, more advanced deep-learning models such as the encoder-decoder architecture produce PV quantile forecasts. Second, a density forecast-based approach computes probabilistic forecasts of imbalance prices on the Belgian case. Finally, a recent class of deep generative models, normalizing flows, generates renewable production and electrical consumption scenarios. Using an energy retailer case study, this technique is extensively compared to state-of-the-art generative models, the variational autoencoders and generative adversarial networks. \\

\noindent \textbf{\textcolor{RoyalBlue}{Thesis second part}} \\
The planning and control part proposes approaches and methodologies based on optimization for decision-making under uncertainty using probabilistic forecasts on several case studies. We introduce the basics of decision-making under uncertainty using optimization strategies: stochastic programming and robust optimization. Then, we investigate these strategies in several case studies in the following Chapters. First, we propose a value function-based approach to propagate information from operational planning to real-time optimization in a deterministic framework. Second, three Chapters focus on the energy management of a grid-connected renewable generation plant coupled with a battery energy storage device in the capacity firming market. This framework promotes renewable power generation facilities in small non-interconnected grids. The day-ahead planning of the system uses either a stochastic or a robust approach. Furthermore, a sizing methodology of the system is proposed. Finally, we consider the day-ahead market scheduling of an energy retailer to evaluate the forecast value of the deep learning generative models introduced in the forecasting part. \\

\noindent \textbf{\textcolor{RoyalBlue}{Perspectives}} \\
We propose four primary research future directions. 
(1) Forecasting techniques of the future. The development of new machine learning models that take advantage of the underlying physical process opens a new way of research. For instance, new forecasting techniques that take advantage of the power system characteristics, such as the graphical normalizing flows capable of learning the power network structure, could be applied to hierarchical forecasting. 
(2) Machine learning for optimization. Models that simplify optimization planning problems by learning a sub-optimal space. For instance, a deep learning model can partially learn the sizing space to provide a fast and efficient tool. A neural network can also emulate the behavior of a physics solver that solves electricity differential equations to compute electricity flow in power grids. Furthermore, such proxies could evaluate if a given operation planning decision would lead to acceptable trajectories where the reliability criterion is met in real-time. 
(3) Modelling and simulation of energy systems. New flexible and open-source optimization modeling tools are required to capture the growing complexity of such future energy systems. To this end, in the past few years, several open-source models for the strategic energy planning of urban and regional energy systems have been developed. EnergyScope TD and E4CLIM are two of them where we think it may be relevant to implement and test the forecasting techniques and scheduling strategies developed in this thesis.
(4) Psychology and machine learning. Achieving sustainability goals requires as much the use of relevant technology as psychology. Therefore, one of the main challenges is not designing relevant technological tools but changing how we consume and behave in our society. Thus, machine learning and psychology could help to identify appropriate behaviors to reduce carbon footprint. Then, inform individuals, and provide constructive opportunities by modeling individual behavior.
\end{abstract}


\tableofcontents




\mainmatter


\chapter*{Foreword}  
\addcontentsline{toc}{chapter}{Foreword}

\epi{The greatest glory in living lies not in never falling, but in rising every time we fall.}{Nelson Mandela}
\epi{Adults keep saying we owe it to the young people, to give them hope, but I don’t want your hope. I don’t want you to be hopeful. I want you to panic. I want you to feel the fear I feel every day. I want you to act. I want you to act as you would in a crisis. I want you to act as if the house is on fire, because it is.}{Greta Thunberg}

\begin{figure}[htbp]
	\centering
	\includegraphics[width=1.\linewidth]{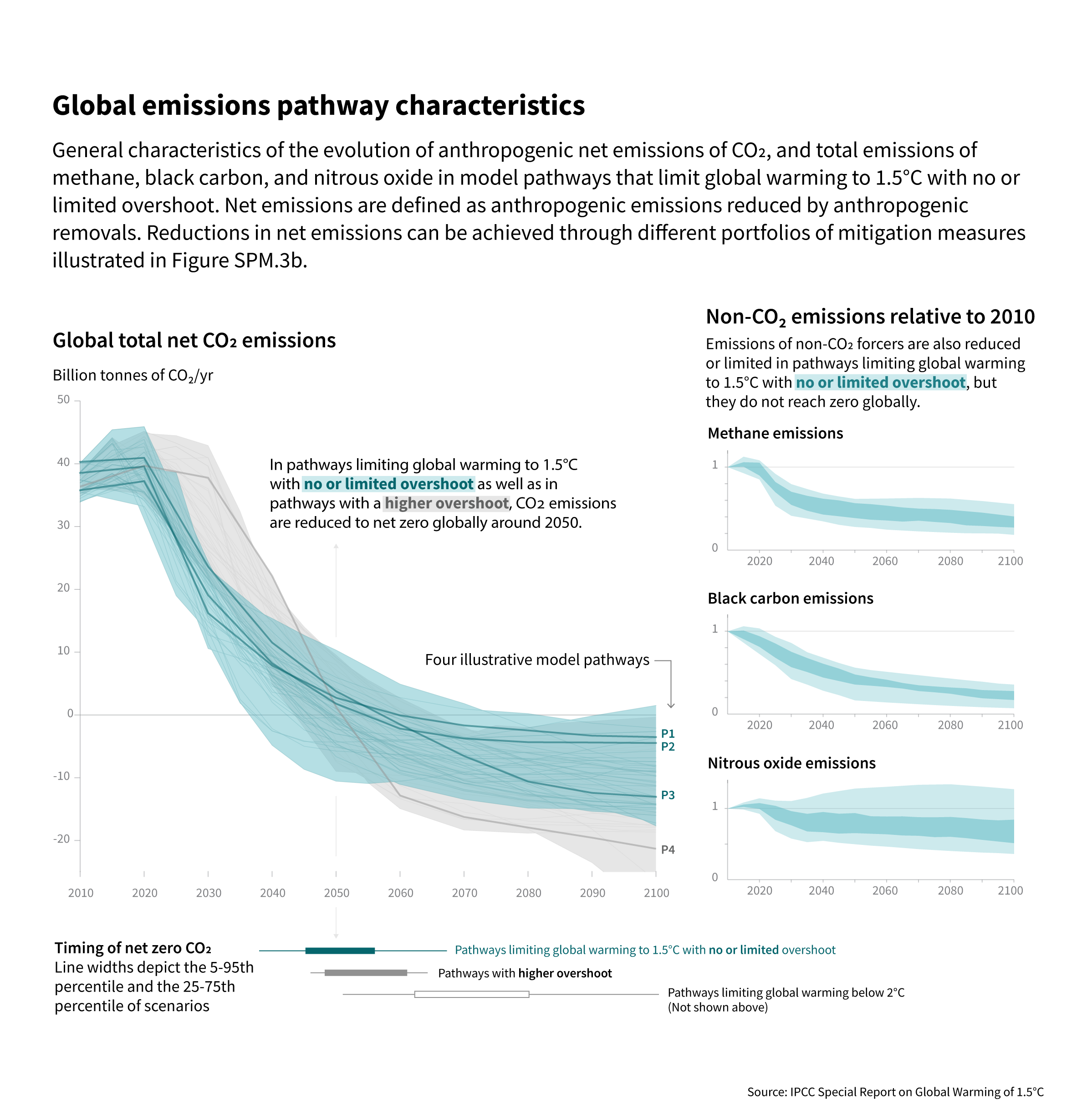}
	\caption{IPCC global net anthropogenic CO\textsubscript{2} emission pathways and targets. Credits: \citep[Figure SPM.3a]{ipcc2018sr15}.}
	\label{fig:ipcc}
\end{figure}

\noindent \textbf{\textcolor{RoyalBlue}{Climate Change}} \\
The Intergovernmental Panel on Climate Change (IPCC) set the global net anthropogenic CO\textsubscript{2} emission trajectories and targets, depicted in Figure \ref{fig:ipcc}, to limit climate change \citep{ipcc2018sr15}. It is a summary for policymakers (SPM) that presents the key findings of the special report published in 2018\footnote{The IPCC released in August 2021 the AR6 SPM \citep{ipcc2021ar6spm}. It presents key findings of the Working Group I contribution to the IPCC’s Sixth Assessment Report (AR6) on the physical science basis of climate change. It is a must to read for every decision-maker, professor, researcher, or person that wants to understand the challenges at stake.}, based on the assessment of the available scientific, technical and socio-economic literature relevant to Global Warming of 1.5°C and for the comparison between Global Warming of 1.5°C and 2°C above pre-industrial levels. The SPM presents the emission scenarios consistent with 1.5°C Global Warming:
\begin{quote}
'In model pathways with no or limited overshoot of 1.5°C, global net anthropogenic CO\textsubscript{2} emissions decline by about 45\% from 2010 levels by 2030 (40–60\% interquartile range), reaching net-zero around 2050 (2045–2055 interquartile range).' \citep{ipcc2018sr15}[C.1]\footnote{\label{note1}\url{https://www.ipcc.ch/sr15/chapter/spm/}}
\end{quote}
The IPCC proposes different mitigation strategies to achieve the net emissions reductions required to follow a pathway to limit Global Warming to 1.5°C with no or limited overshoot. These strategies require a tremendous amount of effort from all countries.
\begin{quote}
'Pathways limiting Global Warming to 1.5°C with no or limited overshoot would require rapid and far-reaching transitions in energy, land, urban and infrastructure (including transport and buildings), and industrial systems (high confidence). These systems transitions are unprecedented in terms of scale, but not necessarily in terms of speed, and imply deep emissions reductions in all sectors, a wide portfolio of mitigation options and a significant upscaling of investments in those options (medium confidence).' \citep{ipcc2018sr15}[C.2]\footref{note1}
\end{quote} 

\noindent \textbf{\textcolor{RoyalBlue}{Trajectories to achieve IPCC targets}} \\
Several reports propose pathways and strategies to achieve these targets. The International Energy Agency (IEA\footnote{\url{https://www.iea.org/}}) presents a comprehensive study of how to transition to a net-zero energy system by 2050 in the special report 'Net-zero by 2050: A roadmap for the global energy system' \citep{iea2021}. It is consistent with limiting the global temperature rise to 1.5 °C without a temperature overshoot (with a 50 \%  probability). The Net‐Zero Emissions by 2050 Scenario (NZE) shows what is needed for the global energy sector to achieve net‐zero CO\textsubscript{2} emissions by 2050:
\begin{quote}
'In the NZE, global energy‐related and industrial process CO\textsubscript{2} emissions fall by nearly 40\% between 2020 and 2030 and to net-zero in 2050. Universal access to sustainable energy is achieved by 2030. There is a 75\% reduction in methane emissions from fossil fuel use by 2030. These changes take place while the global economy  more  than doubles through to 2050 and the global population increases by 2 billion.' \citep[Chapter 2: Summary]{iea2021}
\end{quote}

The key pillars of decarbonization of the global energy system proposed are (1) energy efficiency, (2) behavioral changes, (3) \textbf{electrification}, (4) \textbf{renewables}, (5) hydrogen and hydrogen‐based fuels, (6) bioenergy, and (7) carbon capture, utilization, and storage. 

First, concerning electrification, the direct use of low‐emissions electricity in place of fossil fuels is one of the most significant drivers of emissions reductions in the NZE, accounting for around 20\% of the total reduction achieved by 2050. The share of the electricity in the final consumption increases from 20\% in 2020 to 49\% in 2050. It concerns the sectors of the industry, transport, and buildings. Second, concerning renewables:
\begin{quote}
'At a global level, renewable energy technologies are the key to reducing emissions from electricity supply. Hydropower has been a leading low‐emission source for many decades, but it is mainly the expansion of wind and solar that triples renewables generation by 2030 and increases it more than eightfold by 2050 in the NZE. The share of renewables in total electricity generation globally increases from 29\% in 2020 to over 60\% in 2030 and to nearly 90\% in 2050. To achieve this, annual  capacity  additions  of  wind  and  solar between 2020 and 2050 are five‐times higher than the average over the last three years. Dispatchable renewables are critical to maintain electricity  security, together with other low‐carbon generation, energy storage and robust electricity networks. In the NZE, the main dispatchable renewables globally in 2050 are hydropower (12\% of generation), bioenergy (5\%), concentrating solar power (2\%) and geothermal (1\%).' \citep[Section 2.4.5]{iea2021}
\end{quote}

The IEA report is not the ground truth but has the merit to propose guidelines and directions. There are many other reports and organizations that present strategies and scenarios to achieve the IPCC targets. For instance, The Shift Project (TSP) is a European think tank\footnote{\url{https://theshiftproject.org/en/home/}} advocating the shift to a post-carbon economy. It proposes guidelines and information on energy transition in Europe. The take-home message is that \textbf{there are still pathways to reach net-zero by 2050}. They remain narrow and challenging, requiring all stakeholders, governments, businesses, investors, and citizens to take action this year and every year after so that the goal does not slip out of reach. \\

\begin{figure}[htbp]
	\centering
	\includegraphics[width=0.4\linewidth]{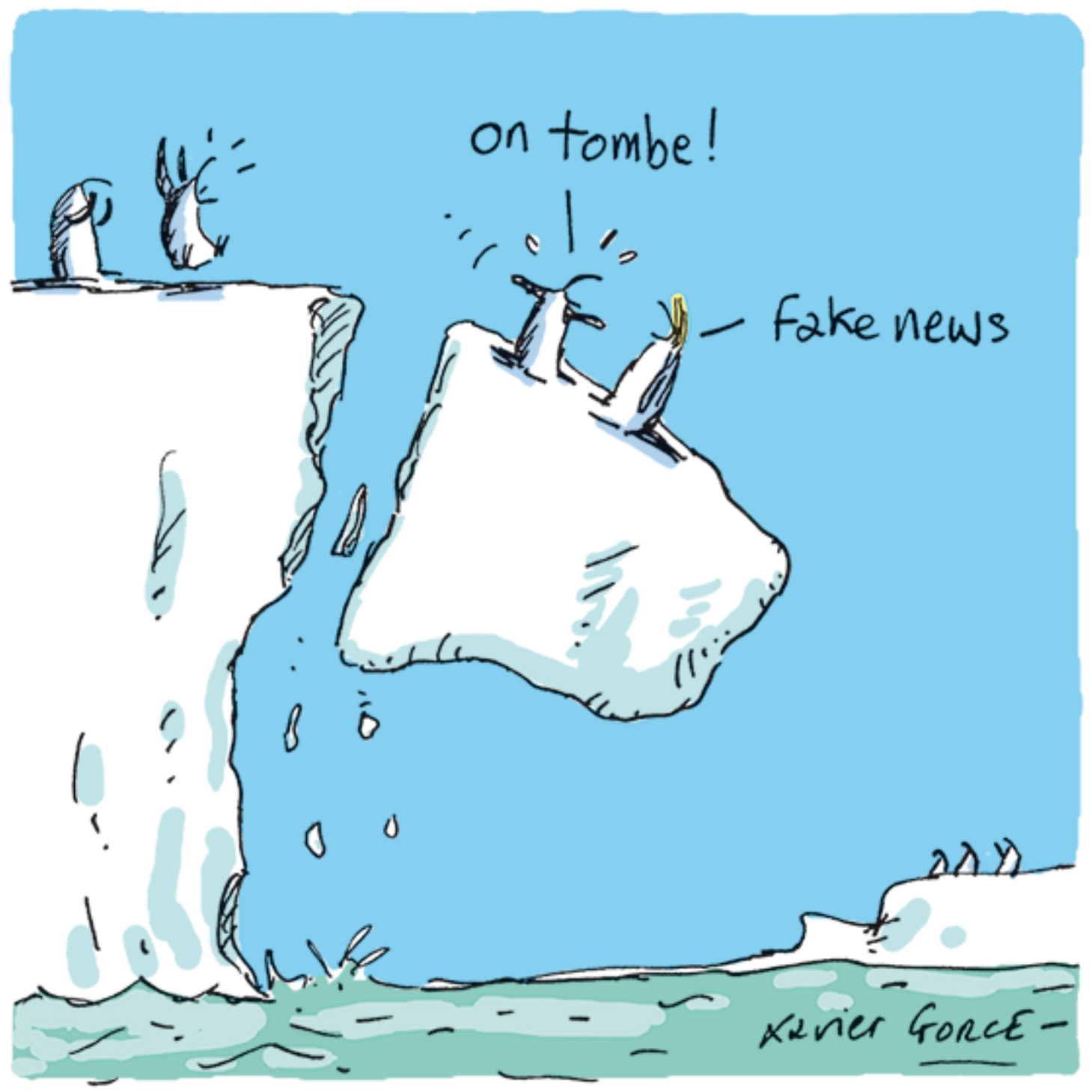}
	\caption{Climate change fake news. Credits: Xavier Gorce.}
	\label{fig:climate_change_fake_news}
\end{figure}

\begin{figure}[htbp]
	\centering
	\includegraphics[width=0.8\linewidth]{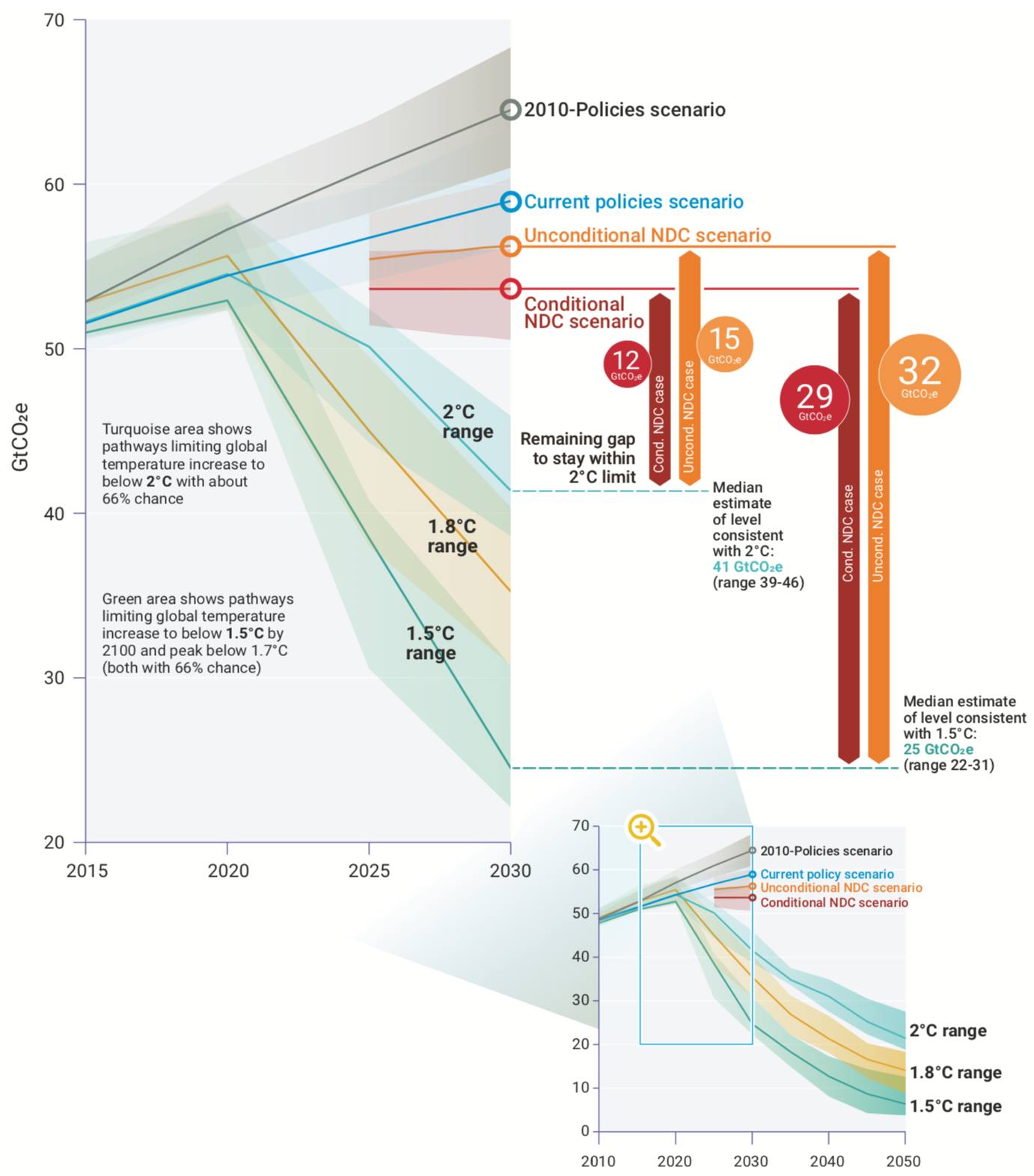}
	\caption{Global GHG emissions under different scenarios and the emissions gap in 2030 (median and 10-th to 90-th percentile range; based on the pre-COVID-19 current policies scenario). Credits: \citep[Figure ES.5]{olhoff2020emissions}.}
	\label{fig:un-gap-report}
\end{figure}

\begin{figure}[htbp]
	\centering
	\includegraphics[width=1.\linewidth]{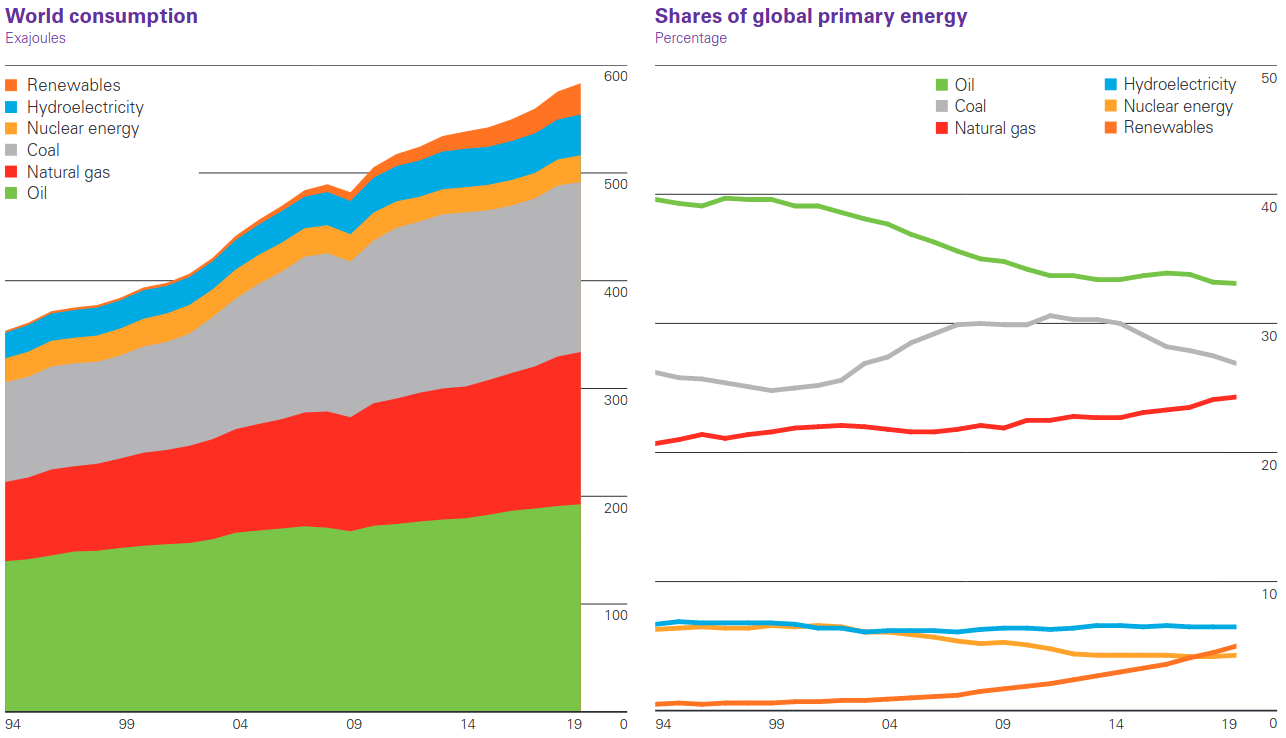}
	\caption{World consumption of primary energy (left) and shares of global primary energy (right) from 1994 to 2020. Credits: BP’s Statistical Review of World Energy 2020 \citep{looney2020statistical}.}
	\label{fig:bp2020}
\end{figure}

\noindent \textbf{\textcolor{RoyalBlue}{Gap between rhetoric and reality}} \\
However, the \textbf{current gap between rhetoric and reality on emissions is still huge}:
\begin{quote}
'We are approaching a decisive moment for international efforts to tackle the climate crisis – a great challenge of our times. The number of countries that have pledged to reach net‐zero emissions by mid‐century or soon after continues to grow, but so do global greenhouse gas emissions. This gap between rhetoric and action needs to close if we are to have a fighting chance of reaching net-zero by 2050 and limiting the rise in global temperatures to 1.5 °C.' \citep[Foreword]{iea2021}
\end{quote}
Figure \ref{fig:climate_change_fake_news} illustrates humorously this gap.
The UNEP Emissions Gap Report provides every year a review of the difference between the greenhouse emissions forecast in 2030 and where they should be to avoid the worst impacts of climate change. Figure \ref{fig:un-gap-report} depicts the global GHG emissions under different scenarios and the emissions gap in 2030.
%
2009-2019 was particularly intense in rhetoric tackling the climate crisis like the 2015 United Nations Climate Change Conference, COP 21. However, the carbon dioxide emissions at the world scale constantly rose from 29.7 (GtCO\textsubscript{2}) in 2009 to 34.2 in 2019 \citep{looney2020statistical}\footnote{\url{https://www.bp.com/content/dam/bp/business-sites/en/global/corporate/pdfs/energy-economics/statistical-review/bp-stats-review-2020-full-report.pdf}}. 
In the meantime, the primary energy consumption increased from 134 000 TWh to 162 000. In 2019, the primary energy consumption by fuel was composed of oil 53 600 (33.1\%), coal 39 300 (24.2\%), natural gas 43 900 (27.0\%), nuclear energy 6 900 (4.3\%), hydro-electricity 10 500 (6\%), and renewables 8 100 (5\%), as depicted in Figure \ref{fig:bp2020}.
The share of renewables in the energy mix progressed and reached 5\% in 2019 (a record). However, the total consumption of fossil fuels such as oil, coal, and natural gas also rose. Oil continues to hold the largest share of the energy mix, coal is the second-largest fuel, and natural gas grew to a record share of 24.2\%.

The Covid‐19 pandemic has delivered a shock to the world economy. It resulted in an unprecedented 5.8\% decline in CO\textsubscript{2} emissions in 2020. The IPCC targets for 2050 require a reduction of 5\% each year from 2020 to 2050. However, the IEA data shows that global energy‐related CO\textsubscript{2} emissions have started to climb again since December 2020.
Nevertheless, there is still hope, and every action to decrease the CO\textsubscript{2} emissions to gain a reduction of 0.1°C is a winning!

\clearpage

\chapter{General introduction}  
\begin{infobox}{Overview}
This chapter introduces the context, motivations, content, and contributions of the thesis. Two main parts compose the manuscript: (1) forecasting; (2) planning and control. Finally, it lists the publications.
\end{infobox}
\epi{Life has no meaning a priori... It is up to you to give it a meaning, and value is nothing but the meaning that you choose.}{Jean-Paul Sartre}

\section{Context and motivations}

\begin{assumption}
Let suppose a utopian world where the current gap between rhetoric and reality on GHG emissions has been drastically decreasing to limit Climate Change and achieve the ambitious targets prescribed by the IPCC.
\end{assumption}
Therefore, according to the IPCC targets, the transition to a carbon-free society goes through an inevitable increase in the renewable generation's share in the energy mix and a drastic decrease in the total consumption of fossil fuels.
\begin{assumption}
This thesis does not debate or study whether and where renewables should be implemented. 
\end{assumption}
Renewables have pros and cons and are not carbon-free. We take the scenarios proposed by organizations such as the IPCC or IEA that evaluate the relevance of a particular type of renewable energy in the energy transition pathways. In addition, the development of variable renewable energies systems results from and produces many interactions between several actors such as citizens, electricity networks, markets, and ecosystems \citep{labussiere2018energy}. This thesis does not discuss energy transformations' political and social aspects and focuses on integrating renewable energies into an existing interconnected system.

\begin{assumption}
This thesis studies the integration of renewables in power systems by investigating forecasting and decision-making tools based on machine learning.
\end{assumption}
In contrast to conventional power plants, renewable energy is subject to uncertainty. Generation technologies based on renewable sources, with the notable exception of hydro and biomass, are \textit{non-dispatchable}, \textit{i.e.}, their output cannot or can only partly be controlled at the will of the producer. Their production is \textit{stochastic} \citep{morales2013integrating} and therefore, hard to forecast. 
A high share of renewables is challenging for power systems that have been designed and sized for dispatchable units \citep{ueckerdt2015analyzing}. Variable renewable energies depend on meteorological conditions and thus pose challenges to the electricity system’s adequacy when conventional capacities are reduced. Therefore, it is necessary to redefine the flexible power system features \citep{impram2020challenges}.

Machine learning can contribute on all fronts by informing the research, deployment, and operation of electricity system technologies. High leverage contributions in power systems include \citep{rolnick2019tackling}: accelerating the development of clean energy technologies, \textit{improving demand and clean energy forecasts}, \textit{improving electricity system optimization and management}, and enhancing system monitoring.
This thesis focuses on two leverages: (1) the supply and demand forecast; (2) the electricity system optimization and management.

Since variable generation and electricity demand both fluctuate, they must be forecast ahead of time to inform real-time electricity scheduling and longer-term system planning. 
Better short-term forecasts enable system operators to reduce reliance on polluting standby plants and proactively manage increasing amounts of variable sources. Better long-term forecasts help system operators and investors to decide where and when to build variable plants. Forecasts need to become more accurate, span multiple horizons in time and space, and better quantify uncertainty to support these use cases. 
In this context, \textit{probabilistic} forecasts \citep{gneiting2014probabilistic}, which aim at modeling the distribution of all possible future realizations, have become an important tool to equip decision-makers, hopefully leading to better decisions in energy applications \citep{morales2013integrating,hong2016probabilistic,hong2020energy}.

When balancing electricity systems, system operators use scheduling and dispatch to determine how much power every controllable generator should produce. This process is slow and complex, governed by NP-hard optimization problems \citep{rolnick2019tackling} such as unit commitment and optimal power flow that must be coordinated across multiple time scales, from sub-second to days ahead. 
Scheduling becomes even more complex as electricity systems include more storage, variable generators, and flexible demand. Indeed, operators manage even more system components while simultaneously solving scheduling problems more quickly to account for real-time variations in electricity production. Thus, scheduling must improve significantly, allowing operators to rely on variable sources to manage systems.

Therefore, the two main research questions are:
\begin{enumerate}
    \item How to produce reliable probabilistic forecasts of renewable generation, consumption, and electricity prices?
    \item How to make decisions with uncertainty using probabilistic forecasts to improve scheduling?
\end{enumerate}
Modeling tools for energy systems differ in terms of their temporal and spatial resolutions, level of technical details, simulation horizons. In particular, two classes of models can be distinguished depending on their focus: (1) system operation and (2) system design \citep{collins2017integrating}. \citep[Chapter 1]{limpens2021generating} depicts both of these classes.

In the first class, the operational models of energy systems generally optimize a single sector's operation, such as the electricity sector, and do not consider investment costs. They describe the constraints of the studied system with a high level of accuracy and can model rapid variations in renewable energy production, forecast errors, or reserve markets. The energy system must be controlled at any time, which requires an excellent technical resolution of the components and accounts for the risks of failure. The system must decide how the load is shared between the different units below the hour at a short time horizon. Solving this problem involves accurately representing production units, such as the power ramps up or load constraints. Therefore, models will decide which units must be committed to optimally dispatching the load or generating shortly (hours - days). 

In the second class, the long-term planning models generate scenarios for an energy system's long-term evolution. They include investments, optimize the system design over multiple years, and have a lower technical resolution of operational constraints. With a horizon of one year, new units can be built in existing sites or former units modernized. With a longer horizon, the overall system can be changed. When the horizon is long enough to neglect the existing system, the models can then optimize the future design of the energy system from scratch and assess the impact of different long-term policies on the design of the system.

\begin{figure}[htbp]
	\centering
	\includegraphics[width=0.6\linewidth]{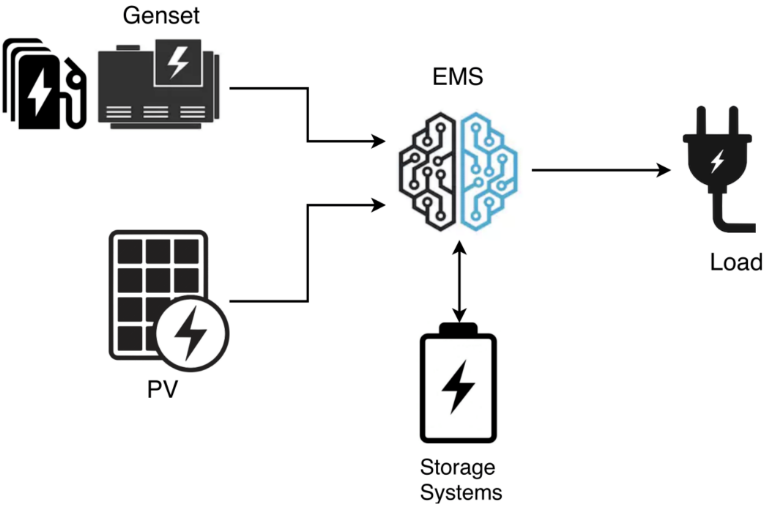}
	\caption{Microgrid scheme. Credits: ELEN0445 Microgrids course \url{https://github.com/bcornelusse/ELEN0445-microgrids}, Li\`ege University.}
	\label{fig:microgrid}
\end{figure}

Each model class is essential and answers different needs to tackle the energy transition and help integrate renewable energies. This thesis addresses the first class of models. An example of a thesis investigating the second one is \citet{limpens2021generating}.
\begin{assumption}
This thesis considers the energy management of "small" systems such as microgrids at a residential scale on a day-ahead basis. 
\end{assumption}
Indeed, the development of microgrids provides an effective way to integrate renewable energy sources and exploit the available flexibility in a decentralized manner. Microgrids are small electrical networks composed of decentralized energy resources and loads controlled locally. They can be operated either interconnected or in islanded mode. Figure \ref{fig:microgrid} depicts a microgrid composed of PV generation, diesel generator (Genset), storage systems, load, and an energy management system (EMS). Energy storage is a crucial component for the stable and safe operation of a microgrid. Storage devices can compensate for the variability of the renewable energy sources and the load to balance the system. The reader is referred to \citet{zia2018microgrids} that proposes a comparative and critical analysis on decision-making strategies and their solution methods for microgrid energy management systems.

\begin{assumption}
This thesis considers the energy management of grid-connected microgrids on a day-ahead basis.
\end{assumption}
Therefore, we are interested in producing reliable day-ahead probabilistic forecasts of renewable generation, consumption, and electricity prices\footnote{This thesis considers only the imbalance prices in the Belgian case study.} for a microgrid composed of PV or wind generation and electrical consumption.
However, in some specific cases, this perimeter is not strictly respected. For instance, the sizing of a grid-connected PV plant with a battery energy storage system is studied in the specific framework of capacity firming or the day-ahead planning of an energy retailer. 

\section{Classification of forecasting studies}  

One of the first works of this thesis was to conduct a literature review of forecasting studies. The forecasting literature is vast and composed of thousands of papers, even when selecting a particular field such as load or PV forecasting. Therefore, a \textit{classification} into \textit{two dimensions} of load forecasting studies is proposed by \citet{dumas2018classification} to decide which forecasting tools to use in which case. The approach can be extended to electricity prices, PV, or wind power forecasting.
This classification aims to provide a synthetic view of the relevant forecasting techniques and methodologies by forecasting problem. This methodology is illustrated by reviewing several papers and summarizing the leading techniques and methodologies' fundamental principles. 

The classification process relies on two parameters that define a forecasting problem: a temporal couple with the forecasting horizon and the resolution and a load couple with the system size and the load resolution. Each article is classified with key information about the dataset used and the forecasting tools implemented: the forecasting techniques (probabilistic or deterministic) and methodologies, the data cleansing techniques, and the error metrics. 
%
The process to select the articles reviewed was conducted into two steps. First, a set of load forecasting studies was built based on relevant load forecasting reviews and forecasting competitions. The second step consisted of selecting the most relevant studies of this set based on the following criteria: the quality of the description of the forecasting techniques and methodologies implemented, the description of the results and the contributions.
%
For the sake of clarity, this manuscript does not detail this study. It can be read in two passes. 
\begin{enumerate}
    \item The first one identifies the forecasting problem of interest to select the corresponding class into one of the four classification tables. Each one references all the articles classified across a forecasting horizon. They provide a synthetic view of the forecasting tools used by articles addressing similar forecasting problems. Then, a second level composed of four Tables summarizes key information about the forecasting tools and the results of these studies.
    \item  The second pass consists of reading the key principles of the main techniques and methodologies of interest and the reviews of the articles.
\end{enumerate}

\section{Content and contributions}  

\begin{figure}[htbp]
	\centering
	\includegraphics[width=1\linewidth]{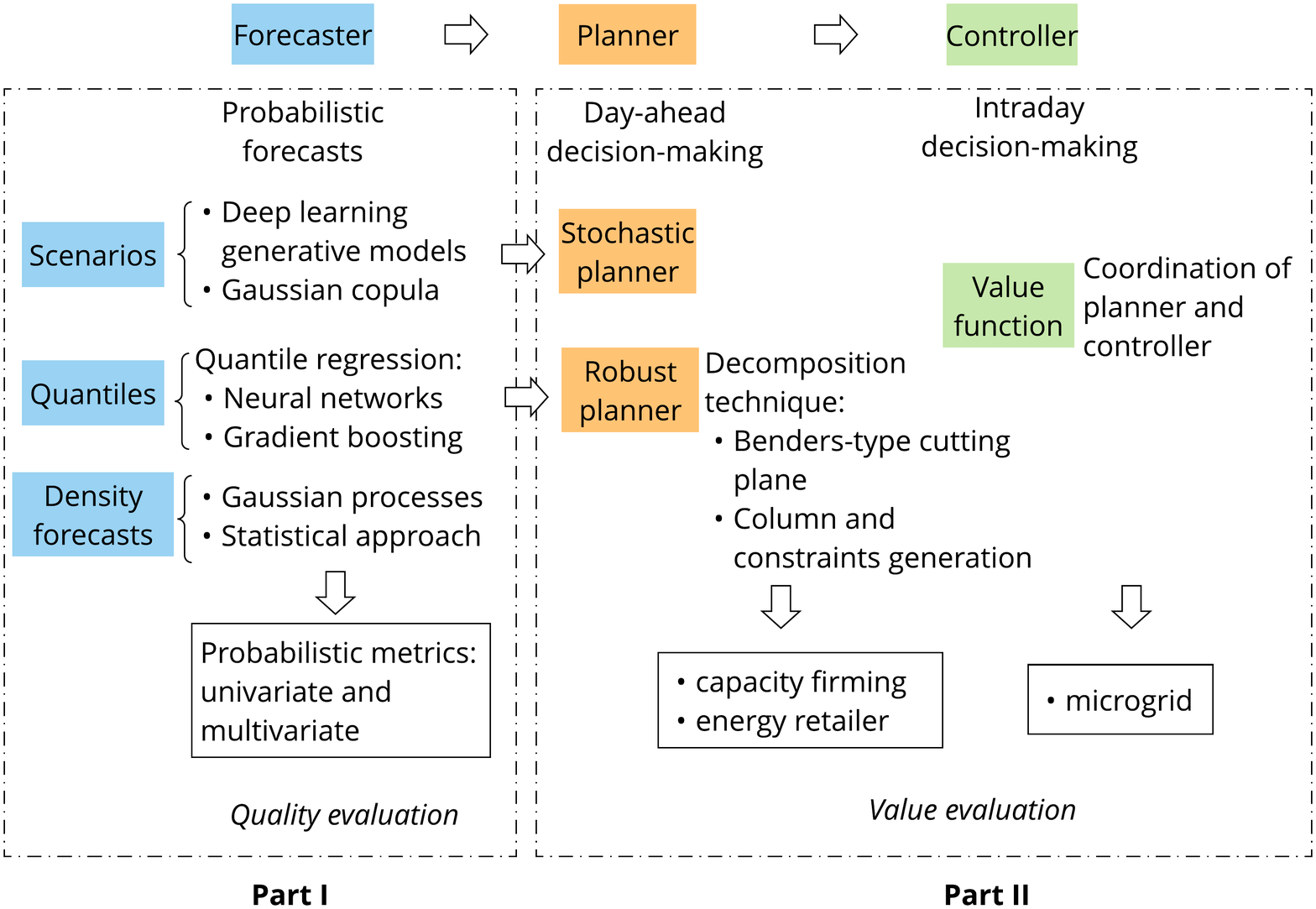}
	\caption{Thesis skeleton.}
	\label{fig:thesis_skeleton}
\end{figure}

The manuscript is divided into two main parts: Part \ref{part:forecasting} forecasting; Part \ref{part:optimization} planning and control. Figure \ref{fig:thesis_skeleton} depicts the thesis skeleton.
Part \ref{part:forecasting} provides the forecasting tools and metrics required to produce and evaluate reliable point and probabilistic forecasts to be used as input of decision-making models in Part \ref{part:optimization}. The latter proposes approaches and methodologies based on optimization for decision-making under uncertainty using probabilistic forecasts on several case studies.

\clearpage
\subsection*{Part \ref{part:forecasting} content and contributions}  

The content and main contributions of the forecasting part are:
\begin{itemize}
    \item Chapter \ref{chap:forecasting-general_background} introduces different types of forecasts to characterize the behavior of stochastic variables, such as renewable generation, electrical consumption, and electricity prices.
    
    \item Chapter \ref{chap:forecast_evaluation} provides the tools to assess the different types of forecasts. For predictions in any form, one must differentiate between their quality and their value. This Chapter focus on forecast quality. Part \ref{part:optimization} addresses the forecast value. An example of forecast quality assessment is conducted on PV and electrical consumption point forecasts. They are computed using common deep-learning models such as recurrent neural networks and used as inputs of day-ahead planning in Chapter \ref{chap:coordination-planner-controller}.
    \\[2mm]
    \textbf{\textcolor{RoyalBlue}{References:}} The point forecast quality evaluation is an extract of \bibentry{dumas2021coordination}.
    
    
    
    \item Chapter \ref{chap:quantile-forecasting} investigates PV quantiles forecasts using deep-learning models such as the encoder-decoder architecture. Then, Chapter \ref{chap:capacity-firming-robust} uses the PV intraday point and quantiles forecasts as inputs of a robust optimization planner. \\[2mm] \textbf{\textcolor{RoyalBlue}{References:}} This chapter is an adapted version of \bibentry{dumas2020deep}.
    
    \item Chapter \ref{chap:density-forecasting} proposes imbalance prices density forecasts with a particular focus on the Belgian case. A two-step density forecast-based approach computes the net regulation volume state transition probabilities to infer the imbalance prices. \\  [2mm]
    \textbf{\textcolor{RoyalBlue}{References:}} This chapter is an adapted version of \bibentry{dumas2019probabilistic}.
    
    \item Chapter \ref{chap:scenarios-forecasting} studies the generation of scenarios for renewable production and electrical consumption by implementing a recent class of deep generative models, normalizing flows. It provides a fair comparison of the quality and value of this technique with the state-of-the-art deep learning generative models, Variational AutoEncoders, and Generative Adversarial Networks. Chapter \ref{chap:energy-retailer} assesses the forecast value. \\  [2mm]
    \textbf{\textcolor{RoyalBlue}{References:}} This chapter is an adapted version of \bibentry{dumas2021nf}.
    
    \item Finally, Chapter \ref{chap:forecasting-conclusions} draws the general conclusions and perspectives of Part \ref{part:forecasting}.
\end{itemize}

\subsection*{Part \ref{part:optimization} content and contributions}  

The content and main contributions of the planning and control part are:
\begin{itemize}
    \item Chapter \ref{chap:optimization-general_background} introduces different types of optimization strategies for decision-making under uncertainty: stochastic and robust optimization. Then, it presents succinctly two decomposition methods to address the two-stage robust non-linear optimization problem: the Benders dual-cutting plane and the column and constraints generation algorithms.
    
    \item Chapter \ref{chap:coordination-planner-controller} presents a value function-based approach as a way to propagate information from operational planning to real-time optimization. \\ [2mm]
    \textbf{\textcolor{RoyalBlue}{References:}} This chapter is an adapted version of \bibentry{dumas2021coordination}.
    
    \item Chapter \ref{chap:capacity-firming-stochastic} addresses the energy management, using a stochastic approach, of a grid-connected renewable generation plant coupled with a battery energy storage device in the capacity firming market. This framework has been designed to promote renewable power generation facilities in small non-interconnected grids. \\ [2mm]
    \textbf{\textcolor{RoyalBlue}{References:}} This chapter is an adapted version of \bibentry{dumas2020stochastic}.
    
    \item Chapter \ref{chap:capacity-firming-sizing} extends Chapter \ref{chap:capacity-firming-stochastic} and proposes a sizing methodology of the system. \\ [2mm]
    \textbf{\textcolor{RoyalBlue}{References:}} This chapter is an adapted version of \bibentry{dumas2020probabilistic}. 
    
    \item Chapter \ref{chap:capacity-firming-robust} extends Chapter \ref{chap:capacity-firming-stochastic} and investigates the day-ahead planning using robust optimization. \\
    \textbf{\textcolor{RoyalBlue}{Python code:}} \url{https://github.com/jonathandumas/capacity-firming-ro} \\ [2mm]
    \textbf{\textcolor{RoyalBlue}{References:}} This chapter is an adapted version of \bibentry{dumas2021probabilistic}.
    
    \item Chapter \ref{chap:energy-retailer}, is the extension of Chapter \ref{chap:scenarios-forecasting}, and presents the forecast value evaluation of the deep learning generative models by considering the day-ahead market scheduling of electricity aggregators, such as energy retailers or generation companies. \\
    \textbf{\textcolor{RoyalBlue}{Python code:}} \url{https://github.com/jonathandumas/generative-models} \\[2mm]
    \textbf{\textcolor{RoyalBlue}{References:}} This chapter is an adapted version of \bibentry{dumas2021nf}. 
    
    \item Finally, Chapter \ref{chap:optimization-conclusions} draws the general conclusions and perspectives of Part \ref{part:optimization}.
\end{itemize}
\clearpage
\subsection*{Publications}  

The thesis is mainly based on the following studies, all available in open-access on arXiv, listed in chronological order:
\begin{itemize}
    \item \bibentry{dumas2018classification}
    \item \bibentry{dumas2019probabilistic}
    \item \bibentry{dumas2021coordination}
    \item \bibentry{dumas2020stochastic}
    \item \bibentry{dumas2020deep} 
    \item \bibentry{dumas2020probabilistic}
    \item \bibentry{dumas2021probabilistic}
    \item \bibentry{dumas2021nf}.
\end{itemize}

\part{Forecasting}\label{part:forecasting}

\begin{infobox}{Overview}
Part \ref{part:forecasting} presents the forecasting techniques and metrics required to
produce and evaluate reliable point and probabilistic forecasts to be used as input of
decision-making models in Part \ref{part:optimization}. Then, it investigates the various types of forecasts in several case studies: point forecasts, quantile forecasts, prediction intervals, density forecasts, and scenarios.
\end{infobox}
\epi{I never think of the future — it comes soon enough.}{Albert Einstein}
\epi{We have two classes of forecasters: Those who don’t know — and those who don’t know they don’t know.}{John Kenneth Galbraith}

\begin{figure}[htbp]
	\centering
	\includegraphics[width=1\linewidth]{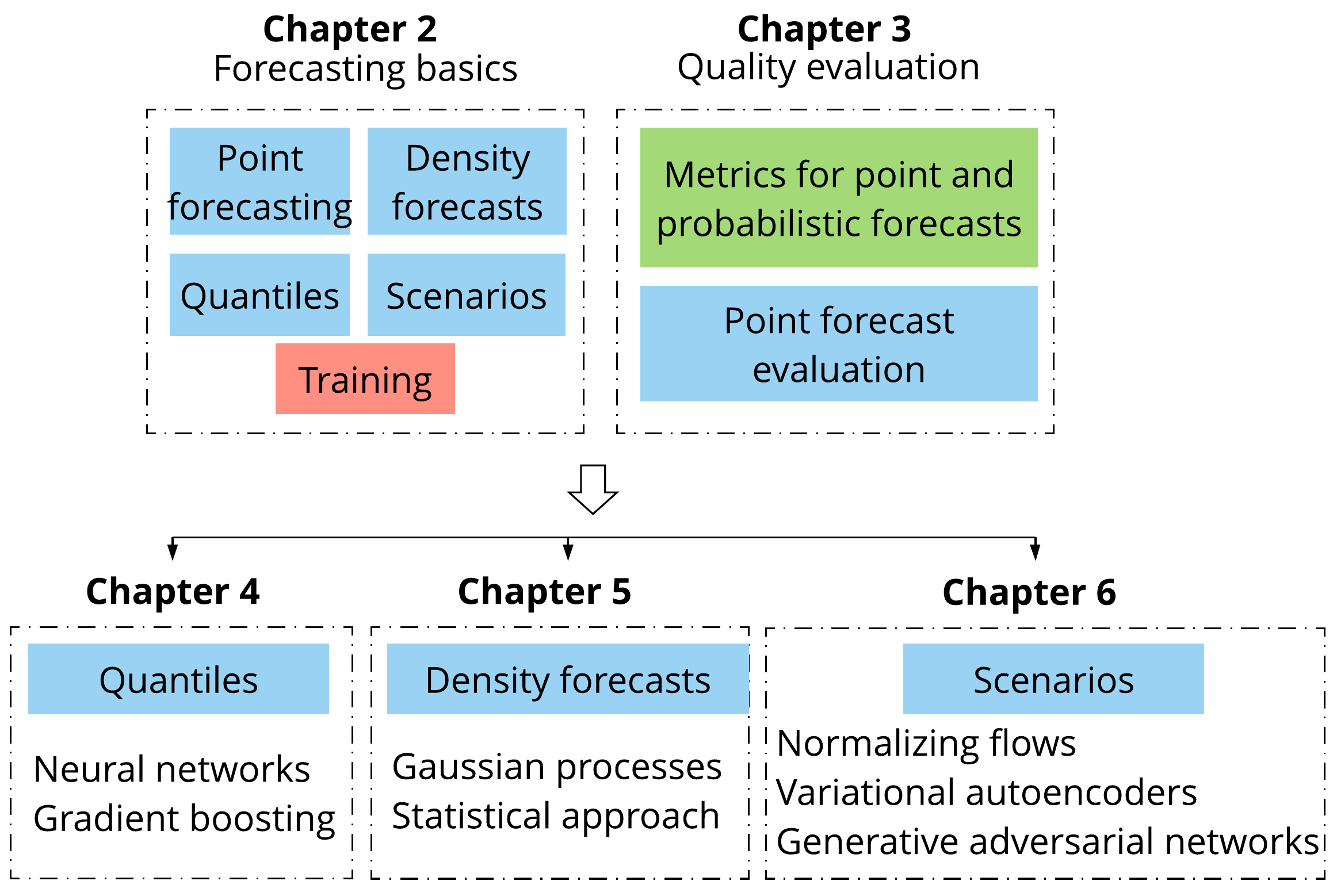}
	\caption{Part \ref{part:forecasting} skeleton.}
	\label{fig:forecasting_part}
\end{figure}

Figure \ref{fig:forecasting_part} illustrates the organization of Part \ref{part:forecasting}, which can be read in two passes depending on the forecasting knowledge.
First, a forecasting practitioner may identify the forecasting type of interest and select the corresponding Chapter. For instance, Chapter \ref{chap:scenarios-forecasting} studies the scenarios of renewable generation and electrical consumption.
Second, a forecasting entrant should be interested in reading Chapters \ref{chap:forecasting-general_background} and \ref{chap:forecast_evaluation} to acquire the forecasting basics. Then, the following Chapters are the application of each type of forecast on case studies.

\chapter*{Part \ref{part:forecasting} general nomenclature}
\addcontentsline{toc}{chapter}{Part \ref{part:forecasting} general nomenclature}

\subsection*{Acronyms}
\begin{supertabular}{l p{0.8\columnwidth}}
	Name & Description \\
	\hline
	PDF & Probability Density Function. \\
	CDF & Cumulative Distribution Function. \\
    (N)MAE & (Normalized) mean absolute error. \\
	(N)RMSE & (Normalized) root mean square error. \\
	CRPS & Continuous ranked probability score. \\
	IS & Interval score. \\
	ES & Energy score. \\
	VS & Variogram score. \\
	DM & Diebold-Mariano. \\
	RNN & Recurrent Neural Network. \\
	LSTM & Long Short-Term Memory.\\
	BLSTM & Bidirectional LSTM. \\
	GBR & Gradient boosting regression. \\
	MLP & Multi-layer perceptron. \\
	ED & Encoder-decoder.\\
	NFs & Normalizing Flows.\\
	GANs & Generative adversarial Networks.\\
	VAEs & Variational AutoEncoders.\\
\end{supertabular}

\subsection*{Variables and parameters}
\begin{supertabular}{l l p{0.8\columnwidth}}
	Name & Range & Description \\
	\hline
	$X$ & & Continuous random variable. \\
	$\mathbf{X}$ & & Continuous multivariate random variable. \\
	$x$ & $\in \mathbb{R}$ & Realization of the random variable $X$. \\
	$\mathbf{x}$ &  $\in \mathbb{R}^T$ & Realization of the multivariate random variable $\mathbf{X}$. \\
	$\hat{x}$ & $\in \mathbb{R}$ & Point forecast of $x$. \\
	$\mathbf{\hat{x}}$ & $\in \mathbb{R}^T$ & Multi-output point forecast of $\mathbf{x}$. \\
	$\hat{x}^{(q)}$ & $\in \mathbb{R}$ & Quantile forecast of $x$. \\
	$\mathbf{\hat{x}}^{(q)}$ & $\in \mathbb{R}^T$ & Multi-output quantile forecast of $\mathbf{x}$. \\
	$\hat{I}^{(\alpha)}$ & $\in \mathbb{R}^2$ & Prediction interval with a coverage rate of $(1-\alpha)$. \\
	$\hat{\mathbf{I}}^{(\alpha)}$ & $\in \mathbb{R}^{2T}$ & Multi-output prediction interval with a coverage rate of $(1-\alpha)$. \\
	$\hat{x}^i$ & $\in \mathbb{R}$ & Scenario $i$ of $x$. \\
	$\mathbf{\hat{x}}^i$ & $\in \mathbb{R}^T$ & Scenario $i$ of $\mathbf{x}$. \\
	$\epsilon$ & $\in \mathbb{R}$ & Point forecast error. \\
	$\xi$ & $\in \{0,1\}$ & Indicator variable. \\
	$\tilde{\xi}$ & $\in \mathbb{R}^+$& Empirical level. \\
\end{supertabular}

\subsection*{Symbols}
\begin{supertabular}{l p{0.8\columnwidth}}
	Name  & Description \\
	\hline
	$g_\theta$ & Forecasting model with parameters $\theta$. \\
	$\mathop{\mathbb{E}}$  & Expectation. \\
	$f$ & Probability Density Function. \\
	$F$ & Cumulative Distribution Function. \\
	$\hat{f}$ & Density forecast of the pdf. \\
	$\hat{F}$ & Density forecast of the cdf. \\
	$\rho$ & Pinball loss function. \\
\end{supertabular}

\subsection*{Sets and indices}
\begin{supertabular}{l p{0.8\columnwidth}}
	Name & Description \\
	\hline
	$t$ & Time index. \\
	$k$ & Forecasting lead time index. \\
	$T$ & Forecasting horizon. \\
	$q$ & Quantile index. \\
	$\omega$ & Scenario index. \\
	$\#\Omega$ & Number of scenarios. \\
	$\mathcal{T}$ & Set of time periods, $\mathcal{T}= \{1,2, \ldots, T\}$. \\
	$\Omega$ & Set of scenarios, $\Omega= \{1,2, \ldots, \#\Omega\}$. \\
	$\mathcal{D}$ & Information set. \\

\end{supertabular}

\chapter{Forecasting background}\label{chap:forecasting-general_background}

\begin{infobox}{Overview}
This chapter introduces the main concepts in forecasting as a background for the work developed in the following chapters of this manuscript. It presents the various types of forecasts: point forecasts, quantile forecasts, prediction intervals, density forecasts, and scenarios. In addition, it provides some knowledge on how to train a forecasting model. \\
General textbooks \citep{morales2013integrating,hastie2009elements,zhang2020dive,Goodfellow-et-al-2016} provide further information for the interested reader. Two courses on this topic also provide interesting material: (1) "Renewables in Electricity Markets"\footnote{\url{http://pierrepinson.com/index.php/teaching/}} given by professor Pierre Pinson at the Technical University of Denmark. It covers some basics of electricity markets, the impact of renewables on markets, participation of renewable energy producers in electricity markets, and renewable energy analytics (mainly forecasting); (2) "INFO8010 - Deep Learning"\footnote{\url{https://github.com/glouppe/info8010-deep-learning}}, ULi\`ege, Spring 2021, given by associate professor Gilles Louppe at Li\`ege University. It covers the foundations and the landscape of deep learning.
\end{infobox}
\epi{If life were predictable it would cease to be life, and be without flavor. }{Eleanor Roosevelt}
\begin{figure}[htbp]
	\centering
	\includegraphics[width=1\linewidth]{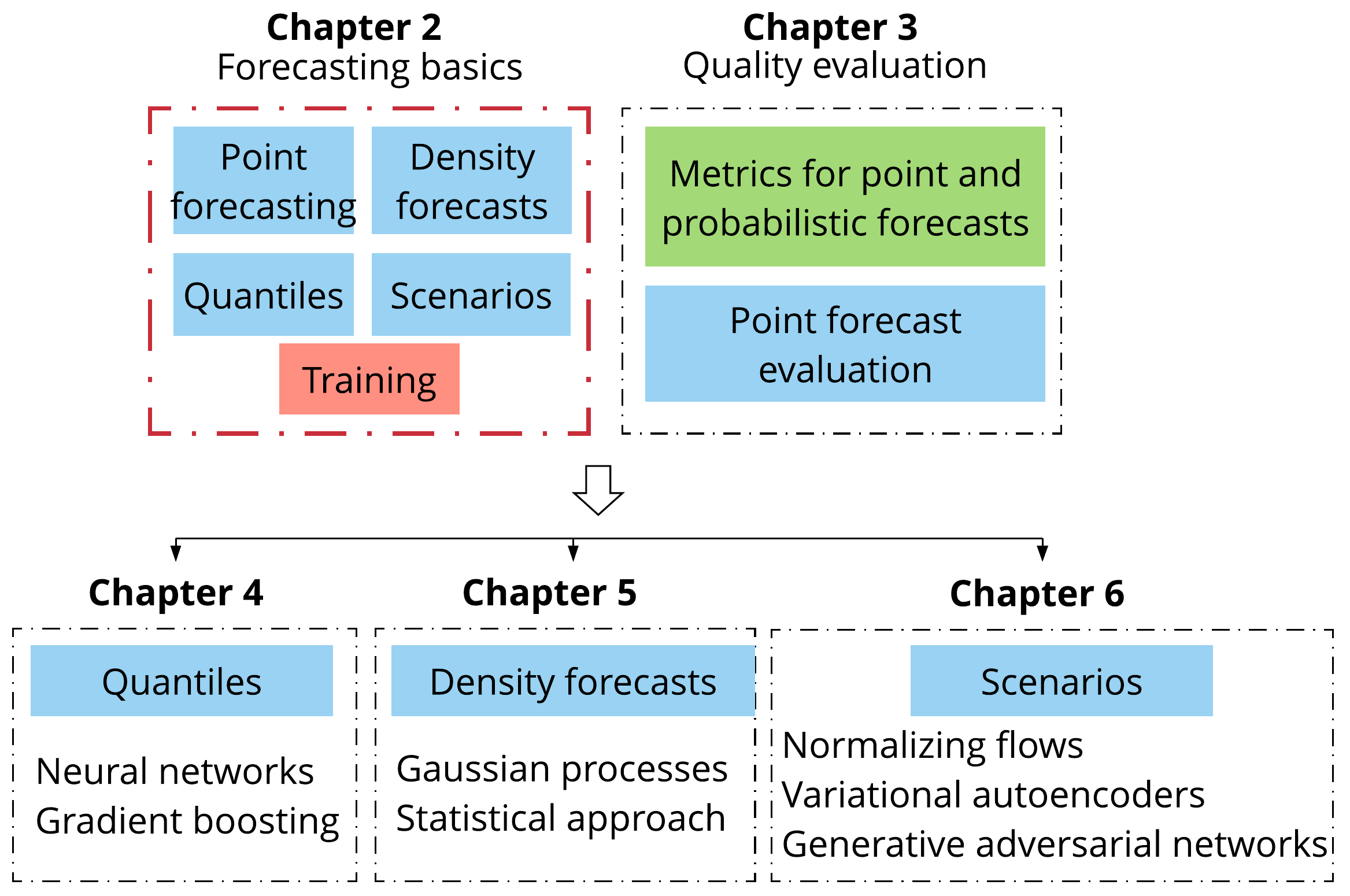}
	\caption{Chapter \ref{chap:forecasting-general_background} position in Part \ref{part:forecasting}.}
\end{figure}
\clearpage

Following \citet{morales2013integrating} power generation from renewable energy sources, such as wind and solar, are referred to as \textit{stochastic power generation} in this thesis. Electrical consumption and electricity prices are also modeled as stochastic variables.
Predictions of renewable energy generation, consumption, and electricity prices can be obtained and presented differently. The choice of a particular kind of forecast depends on the process characteristics of interest to the decision-maker and the type of operational problem. The various types of forecasts and their presentation are introduced in the following, starting from the most common \textit{point forecasts} and building up towards the more advanced products that are \textit{probabilistic forecasts} and scenarios. 

This Chapter is organized as follows. Section \ref{sec:forecasting-background-point} introduces the point forecasts. Section \ref{sec:forecasting-background-probabilistic} presents the various types of probabilistic forecasts. Section \ref{sec:forecasting-background-model} proposes an abstract formulation of a model-based forecaster. Section \ref{sec:forecasting-model-training} provides some knowledge on how to train a forecasting model. Finally, conclusions are drawn in Section \ref{sec:forecasting-background-conclusions}.

\section{Point forecast}\label{sec:forecasting-background-point}

Let $x_t \in \mathbb{R}$ be the variable of interest, \textit{e.g.}, renewable energy generation, consumption, or electricity prices, measured at time $t$, which corresponds to a realization of the random variable $X_t$. 
\begin{definition}[A model-based forecast]\citep[Chapter 2]{morales2013integrating}
A (model-based) forecast $ \hat{x}_{t+k|t} \in \mathbb{R}$ is an estimate of some of the characteristics of the stochastic process $X_{t +k}$ issued at time $t$ for time $t+k$ given a model $g_\theta$, with parameters $\theta$ and the information set $\mathcal{D}$ gathering all data and knowledge about the processes of interest up to time $t$, such as weather forecasts, historical observations, calendars variables, \textit{etc}.
\end{definition}

In the above definition, $k$ is the \textit{lead time}, sometimes also referred to as \textit{forecast horizon}. The 'hat' symbol expresses that $ \hat{x}_{t+k|t}$ is an estimate only. It models the presence of uncertainty both in our knowledge of the process and inherent to the process itself. The forecast for time $t + k$ is conditional on our knowledge of stochastic process up to time $t$, including the data used as input to the forecasting process and the models identified and parameters estimated. Therefore, a forecaster somewhat makes the crucial assumption that the future will be like the past.

Forecasts are series of consecutive values $ \hat{x}_{t+k|t}, \ k=k_1, \ldots, k_T$, that is, for regularly spaced lead times up to the \textit{forecast length} $T$. That regular spacing $\Delta t$ is called the \textit{temporal resolution} of the forecasts. For instance, when one talks of day-ahead forecasts with an hourly resolution, forecasts consist of a series gathering predicted power values for each of the following 24 hours of the next day.

\begin{definition}[Point forecast]\citep[Chapter 2]{morales2013integrating}
A \textit{point forecast} $ \hat{x}_{t+k|t} \in \mathbb{R}$ is a single-valued issued at time $t$ for $t + k$, and corresponds to the conditional expectation of $X_{t+k}$  
\begin{equation}\label{eq:point_forecast_def}
\hat{x}_{t+k|t} := \mathop{\mathbb{E}}[X_{t+k|t} | g_\theta, \mathcal{D} ] ,
\end{equation}
given $g_\theta$, and the information set $\mathcal{D}$.
\end{definition}

A forecast in the form of a conditional expectation translates to acknowledging the presence of uncertainty, even though it is not quantified and communicated.

\begin{definition}[Multi-output point forecast]
A \textit{multi-output point forecast} computed at $t$ for $t+k_1$ to $t+k_T$ is the vector 
\begin{align}
	\label{eq:multi_output_point_forecast}	
	\mathbf{\hat{x}}_t := &  [\hat{x}_{t+k_1|t}, \ldots,\hat{x}_{t+k_T|t}]^\intercal	\in \mathbb{R}^T.
\end{align}
\end{definition}
Depending on the problem formulation, it can be computed directly as a vector or as an aggregate of single output forecasts.

\section{Probabilistic forecasts}\label{sec:forecasting-background-probabilistic}

In contrast to point predictions, probabilistic forecasts aim at providing decision-makers with the full information about potential future outcomes. Let $f_t$ and $F_t$ be the \textit{probability density function} (PDF) and related \textit{cumulative distribution function} (CDF) of $X_t$, respectively. 
\begin{definition}[Probabilistic forecast]\citep[Chapter 2]{morales2013integrating}
A probabilistic forecast issued at time $t$ for time $t + k$ consists of a prediction of the PDF (or equivalently, the CDF) of $X_{t +k}$, or of some summary features.
\end{definition}
Various types of probabilistic forecasts have been developed: \textit{quantile}, \textit{prediction intervals}, \textit{scenarios}, and \textit{density} forecasts.

\subsection{Quantiles}

\begin{definition}[Quantile forecast]\citep[Chapter 2]{morales2013integrating}
A \textit{quantile forecast} $\hat{x}^{(q)}_{t+k|t} \in \mathbb{R}$ with nominal level $q$ is an estimate, issued at time $t$ for time $t+k$ of the quantile $x^{(q)}_{t+k|t}$ for the random variable $X_{t+k|t}$ 
\begin{equation}\label{eq:quantile_forecast_def}
P[X_{t+k|t} \leq \hat{x}^{(q)}_{t+k|t} | g_\theta, \mathcal{D}]  = q,
\end{equation}
given $g_\theta$, and the information set $\mathcal{D}$. Or equivalently $\hat{x}^{(q)}_{t+k|t}  = \hat{F}^{-1}_{t+k|t}(q)$, with $\hat{F}$ the estimated cumulative distribution function of the continuous random variable $X$.
\end{definition}

By issuing a quantile forecast $ \hat{x}^{(q)}_{t+k|t}$, the forecaster tells at time $t$ that there is a probability $q$ that $x_{t+k}$ will be less than $ \hat{x}^{(q)}_{t+k|t}$ at time $t + k$. Quantile forecasts are of interest for several operational problems. For instance, the optimal day-ahead bidding of wind or PV generation uses quantile forecasts whose nominal level is a simple function of day-ahead and balancing market prices \citep{bitar2012bringing}. Furthermore, quantile forecasts also define prediction intervals that can be used for robust optimization.

\begin{definition}[Multi-output quantile forecast]
A \textit{multi-output quantile forecast} of length $T$ with nominal level $q$ computed at $t$ for $t+k_1$ to $t+k_T$ is the vector 
\begin{align}
	\label{eq:multi_output_quantile_forecast_def}	
	\mathbf{\hat{x}}^{(q)}_t := &  [\hat{x}_{t+k_1|t}^{(q)}, \ldots,\hat{x}_{t+k_T|t}^{(q)}]^\intercal \in \mathbb{R}^T.
\end{align}
\end{definition}

\subsection{Prediction intervals}

Quantile forecasts give probabilistic information in the form of a threshold level associated with a probability. Even though they may be of direct use for several operational problems, they cannot provide forecast users with a feeling about the level of forecast uncertainty for the coming period. For that purpose, prediction intervals define the range of values within which the observation is expected to be with a certain probability, \textit{i.e.}, its nominal coverage rate \citep{pinson2007non}. 

\begin{definition}[Prediction interval]\citep[Chapter 2]{morales2013integrating}
A \textit{prediction interval} $\hat{I}^{(\alpha)}_{t+k|t}\in \mathbb{R}^2$ issued at $t$ for $t + k$, defines a range of potential values for $X_{t+k}$, for a certain level of probability $(1-\alpha)$, $\alpha \in [0,1]$. Its nominal coverage rate is
\begin{equation}\label{eq:PI_def}
P[X_{t+k} \in \hat{I}^{(\alpha)}_{t+k|t}| g_\theta, \mathcal{D}]  = 1 - \alpha.
\end{equation}
\end{definition}

\begin{definition}[Central prediction interval]\citep[Chapter 2]{morales2013integrating}
A \textit{central prediction interval} consists of centering the prediction interval on the median where there is the same probability of risk below and above the median. A central prediction interval with a coverage rate of $(1-\alpha)$ is estimated by using the quantiles $q=(\alpha/2)$ and $q=(1-\alpha/2)$. Its nominal coverage rate is
\begin{equation}\label{eq:PI_centered_def}
\hat{I}^{(\alpha)}_{t+k|t}= [\hat{x}^{(q=\alpha/2)}_{t+k|t}, \hat{x}^{(q=1-\alpha/2)}_{t+k|t}].
\end{equation}
\end{definition}
For instance, central prediction interval with a nominal coverage rate of 90\%, \textit{i.e.}, $(1 - \alpha) = 0.9$, are defined by quantile forecasts with nominal levels of 5 and 95\%.

\begin{definition}[Multi-output central prediction interval]
A \textit{multi-output central prediction interval} with a coverage rate of $(1-\alpha)$ computed at $t$ for $t+k_1$ to $t+k_T$ is the vector 
\begin{equation}\label{eq:PI_multi_def}
\hat{\mathbf{I}}^{(\alpha)}_{t+k|t}= [\hat{I}^{(\alpha)}_{t+k_1|t}, \ldots ,\hat{I}^{(\alpha)}_{t+k_T|t}]^\intercal \in \mathbb{R}^{2T}.
\end{equation}
\end{definition}

\subsection{Scenarios}

Let us introduce the \textit{multivariate} random variable
\begin{equation}\label{eq:rv_multi_def}
\mathbf{X}_t := \{X_{t+k} \}, \quad k=k_1, \ldots, k_T,
\end{equation}
which gathers the random variables characterizing the stochastic power generation process for the $T$ following lead times. Hence, it covers their marginal densities as well as their interdependence structure.

\begin{definition}[Scenarios]
\textit{Scenarios} issued at time $t$ and for a set of $T$ successive lead times, \textit{i.e.}, with $k=k_1, \ldots, k_T$ consist of a set of $M$ time trajectories
\begin{equation}\label{eq:scenarios_def}
\mathbf{\hat{x}}^i_t := [\hat{x}_{t+k_1|t}^i, \ldots,\hat{x}_{t+k_T|t}^i ]^\intercal \in \mathbb{R}^T \quad i=1, \ldots, M.
\end{equation}
\end{definition}

The resulting time trajectories comprise scenarios like those commonly used in stochastic programming.

\subsection{Density forecasts}

All the various types of predictions presented in the above, \textit{i.e.}, point, quantile, and interval forecasts, are only partly describing the complete information about the future of $X$ at every lead time. Density forecasts would give this whole information for each point of time in the future.
\begin{definition}[Density forecast]\citep[Chapter 2]{morales2013integrating}
A \textit{density forecasts} $\hat{f}_{t+k|t}$ ($\hat{F}_{t+k|t}$) issued at time $t$ for $t+k$, is a complete description of the pdf (or cdf) of $X_{t+k}$ conditional on a given model $g_\theta$, and the information set $\mathcal{D}$.
\end{definition}

\section{Model-based formulation}\label{sec:forecasting-background-model}

Let assume the information set $\mathcal{D} := \{ \mathbf{x}_t, \mathbf{c}_t \}_{t=1}^N$ is composed of $N$ independent and identically distributed samples from the joint distribution $p(\mathbf{x},\mathbf{c})$ of two continuous variables $X$ and $C$. $X$ is the variable of interest, \textit{e.g.}, renewable energy generation, consumption, or electricity prices, and $C$ is the context, \textit{e.g.}, the weather forecasts, calendar variables, or exogenous variables.
Generically, any prediction of a random variable $X$ issued at time $t$, being point or probabilistic forecast is a linear or nonlinear function of $\mathcal{D}$. The goal of Part \ref{part:forecasting} is to generate multi-output context-based forecasts $\mathbf{\hat{x}}$ that are distributed under $p(\mathbf{x}|\mathbf{c})$.

\begin{definition}[Multi-output model-based forecasts]
A \textit{multi-output model-based forecasts} of length $T$ computed by $g_\theta$ at $t$ for $t+k_1$ to $t+k_T$ is the vector 
\begin{align}
	\label{eq:multi_output_model_based}	
	\mathbf{\hat{x}}_t \sim &  g_\theta (\mathbf{x}_{<t}, \mathbf{c}_t) \in \mathbb{R}^T,
\end{align}
given the context $\mathbf{c}_t$, and observations $\mathbf{x}_{<t} = [x_1, \ldots, x_{t-1}]^\intercal$ up to time $t$.
\end{definition}
Its purpose is to generate synthetic but realistic data $\mathbf{\hat{x}}_t$ whose distribution is as close as possible to the unknown data distribution $p(\mathbf{x}|\mathbf{c})$. When considering point forecasts, quantile forecasts, and scenarios, $\mathbf{\hat{x}}_t$ is defined by (\ref{eq:multi_output_point_forecast}), (\ref{eq:multi_output_quantile_forecast_def}), and (\ref{eq:scenarios_def}), respectively. This abstract formulation is used in Section \ref{sec:point-forecasting} and Chapters \ref{chap:quantile-forecasting} and \ref{chap:scenarios-forecasting} where the forecasting models and the related inputs including the context are specified.

\section{Model training}\label{sec:forecasting-model-training}

This section provides the basics of supervised learning that is used in Part \ref{part:forecasting} to train the forecasting models $g_\theta$. It relies mainly on Lecture 1 of INFO8010 - Deep Learning\footnote{\url{https://github.com/glouppe/info8010-deep-learning}}, ULi\`ege, Spring 2021, from associate professor Gilles Louppe \citep{louppe2014understanding} at Li\`ege University. 
The interested reader may found interesting materials in \citet[Chapter 3]{duchesne2021machine} and \citet[Chapter 2]{sutera2021importance}. They introduce the different types of machine learning problems, describe their characteristics, and the procedure to apply supervised learning.

\subsection{Regression with supervised learning}

Consider the unknown joint probability distribution $p(\mathbf{x},\mathbf{c})$ of two continuous variables $X$, \textit{e.g.}, renewable energy generation, and $C$, \textit{e.g.}, the weather forecasts, introduced in the previous Section. Let assume some training data $\mathcal{D} = \{ \mathbf{x}_t, \mathbf{c}_t \}_{t=1}^N$ composed of $N$ independent and identically distributed samples.

\textit{Supervised learning} is usually concerned with the two following \textit{inference} problems: \textit{classification} and \textit{regression}. Classification consists of identifying a decision boundary between objects of distinct classes. Regression aims at estimating relationships among (usually continuous) variables. In this thesis, we focus on regression.
\begin{definition}[Regression]
Given $\mathbf{x}_t, \mathbf{c}_t \in \mathcal{D}$, for $t=1, \ldots, N$ we would like to estimate for any new $\mathbf{c}$
\begin{align}
	\label{eq:regression}	
    \mathop{\mathbb{E}} [ X=\mathbf{x} | C=\mathbf{c}].
\end{align}
\end{definition}

For instance, let assume $g_\theta$ is parameterized with a neural network where the last layer does not contain any final activation. If we make the assumption that $p(\mathbf{x},\mathbf{c}) \sim \mathcal{N}(\mathbf{x}; \mu = g_\theta(\mathbf{x}|\mathbf{c}); \sigma^2 = 1)$, we can perform maximum likelihood estimation to estimate the model's parameters $\theta$ that is equivalent to minimize the negative log likelihood
\begin{subequations}
\label{eq:background-likelihhod}	
\begin{align}
	\arg \max_\theta p(\mathcal{D}|\theta) & \sim \arg \min_\theta - \log \prod_{\mathbf{x}_t, \mathbf{c}_t \in \mathcal{D}} p(\mathbf{x} = \mathbf{x}_t|\mathbf{c}_t, \theta) \\
    & = \arg \min_\theta - \sum_{\mathbf{x}_t, \mathbf{c}_t \in \mathcal{D}}  \log p(\mathbf{x} = \mathbf{x}_t|\mathbf{c}_t, \theta) \\
    & = \arg \min_\theta - \sum_{\mathbf{x}_t, \mathbf{c}_t \in \mathcal{D}}  \log \bigg(\frac{1}{\sqrt{2\pi}} \exp  \big(-\frac{1}{2} \lVert \mathbf{x}_t - g_\theta(\mathbf{x}_t|\mathbf{c}_t) \rVert^2  \big)\bigg) \\
    & \sim \arg \min_\theta \sum_{\mathbf{x}_t, \mathbf{c}_t \in \mathcal{D}}\lVert \mathbf{x}_t - g_\theta(\mathbf{x}_t|\mathbf{c}_t) \rVert^2,
\end{align}
\end{subequations}
which recovers the common squared error loss $L(\mathbf{x}, \mathbf{\hat{x}} \sim  g_\theta(\mathbf{x} | \mathbf{c}) ) = \lVert \mathbf{x} - \mathbf{\hat{x}} \rVert^2 $ that will be used in Section \ref{sec:point-forecasting} for point forecasting.

\subsection{Empirical risk minimization}

Consider a function $g_\theta$ produced by a learning algorithm. The predictions of this function can be evaluated through a loss $L:\mathbb{R}^T \times \mathbb{R}^T \rightarrow \mathbb{R}$, such that $L(\mathbf{x}, 	\mathbf{\hat{x}} \sim  g_\theta(\mathbf{x} | \mathbf{c}) ) \geq 0 $ measures how close the prediction $\mathbf{\hat{x}}$ from $\mathbf{x}$ is. For instance, in point forecasting $L$ is the mean squared error or the pinball loss for quantile forecasting.

The key idea of the model training relies on the \textit{empirical risk minimization}. Let $\mathcal{G}$ denote the hypothesis space, \textit{i.e}. the set of all functions $g_\theta$ than can be produced by the chosen learning algorithm.
\begin{definition}[Empirical risk minimization]\citep{vapnik1992principles}
We are looking for a function $g_\theta$ with a small expected risk
\begin{align}
	\label{eq:expected-risk}	
    R(g_\theta) = \mathop{\mathbb{E}}_{\mathbf{x},\mathbf{c} \sim p(\mathbf{x},\mathbf{c})}[L(\mathbf{x},	\mathbf{\hat{x}} \sim  g_\theta(\mathbf{x} | \mathbf{c})],
\end{align}
also called the generalization error.
\end{definition}

Therefore, for a given data generating distribution $p(\mathbf{x},\mathbf{c})$ and for a given hypothesis space $\mathcal{G}$, the optimal model is
\begin{align}
\label{eq:optimal-model}	
 g_\theta^\star = \arg \min_{g_\theta \in \mathcal{G}} R(g_\theta).
\end{align}
However, since $p(\mathbf{x},\mathbf{c})$ is unknown, the expected risk cannot be evaluated and the optimal model cannot be determined.
Nevertheless, if we have some training data $\mathcal{D}= \{ \mathbf{x}_t, \mathbf{c}_t \}_{t=1}^N$ composed of $N$ independent and identically distributed samples, we can compute an estimate that is the empirical risk \citep{vapnik1992principles} (or training error)
\begin{align}
\label{eq:empirical-expected-risk}	
 \hat{R}(g_\theta, \mathcal{D}) = \frac{1}{N} \sum_{\mathbf{x}_t,  \mathbf{c}_t \in \mathcal{D}} L(\mathbf{x}_t, \mathbf{\hat{x}}_t \sim  g_\theta(\mathbf{x}_t | \mathbf{c}_t).
\end{align}
This estimator is unbiased and can be used for finding a good enough approximation of $g_\theta^\star$, resulting in the empirical risk minimization principle
\begin{align}
\label{eq:optimal-model-emprirical}	
g_{\theta, \mathcal{D}}^\star = \arg \min_{g_\theta \in \mathcal{G}} \hat{R}(g_\theta, \mathcal{D}).
\end{align}
Note: most machine learning algorithms, including neural networks, implement empirical risk minimization. Under regularity assumptions, empirical risk minimizers converge: $\lim_{N \rightarrow \infty} g_{\theta, \mathcal{D}}^\star \rightarrow g_\theta^\star$.

\begin{figure}[htbp]
	\centering
	\includegraphics[width=0.4\linewidth]{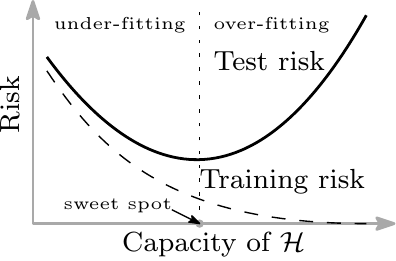}
	\caption{The classical U-shaped risk curve arising from the bias-variance trade-off. Curves for training risk (dashed line) and test risk (solid line). Note: $\mathcal{H}$ on the Figure is $\mathcal{G}$. Credits: \citep{belkin2019reconciling}.}
	\label{fig:u-shaped}
\end{figure}

The capacity of a hypothesis space induced by a learning algorithm intuitively represents the ability to find a suitable model $g_\theta \in \mathcal{G}$ for any function, regardless of its complexity. In practice, capacity can be controlled through hyper-parameters $\theta$ of the learning algorithm. Then, the goal is to adjust the capacity of the hypothesis space $\mathcal{G}$ such that the expected risk of the empirical risk minimizer gets as low as possible. To this end, it is essential to understand the concept of the bias-variance trade-off \citep{geman1992neural}. First, reducing the capacity makes $ g_{\theta, \mathcal{D}}^\star$ fit the data less on average, which increases the bias term (under-fitting). Second, increasing the capacity makes $ g_{\theta, \mathcal{D}}^\star$ vary a lot with the training data, which increases the variance term (over-fitting). Therefore, the bias-variance trade-off implies that a model should balance under-fitting and over-fitting: rich enough to express underlying structure in data, simple enough to avoid fitting spurious patterns. It is summarized in the classical U-shaped risk curve \citep{belkin2019reconciling}, shown in Figure \ref{fig:u-shaped}.

\subsection{The "double descent" curve}

\begin{figure}[htbp]
	\centering
	\includegraphics[width=0.7\linewidth]{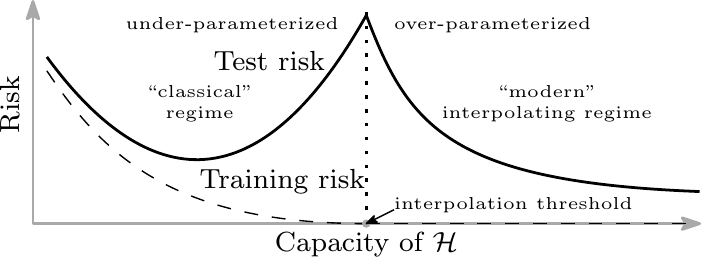}
	\caption{The double descent risk curve, which incorporates the U-shaped risk curve (\textit{i.e.}, the "classical" regime) together with the observed behavior from using high capacity function classes (\textit{i.e.}, the "modern" interpolating regime), separated by the interpolation threshold. Credits: \citep{belkin2019reconciling}.}
	\label{fig:doubledescent}
\end{figure}

In the past few years, several studies have raised questions about the mathematical foundations of machine learning and their relevance to practitioners. Indeed, very complex machine learning models such as neural networks are trained to exactly fit the data. Classically, following the concept of bias-variance trade-off, these models should be considered over-fit. However, they often obtain high accuracy on test data.
%
%
It is illustrated by Figure \ref{fig:doubledescent} from \citet{belkin2019reconciling} that discuss empirical evidence for the double descent curve.
This study demonstrated that the capacity of the function class does not necessarily reflect how well the predictor matches the \emph{inductive bias} appropriate for the problem at hand. By training models on a range of real-world datasets and synthetic data, the inductive bias that seems appropriate is the regularity or smoothness of a function as measured by a specific function space norm. Then, by considering larger function classes, which include more candidate predictors compatible with the data, the training allowed to determine interpolating functions with smaller norms and are thus "simpler".  Therefore, increasing function class capacity improves the performance of classifiers.
This connection investigated by \citet{belkin2019reconciling} between the performance and the structure of machine learning models delineates the limits of classical analyses. It has implications for both the theory and practice of machine learning.


\subsection{Training methodology}

In practice, one has a finite dataset of input-output pairs $\mathcal{D}$ and no further information about the joint probability distribution $p(\mathbf{x},\mathbf{c})$. Then, the model is selected by minimizing the empirical risk (\ref{eq:empirical-expected-risk}) over the dataset. Therefore, it is typically strongly biased in an optimistic way and is a bad estimate of the generalization error, also called the testing error. A "good" model should predict well data independent from the dataset used for training but drawn from the same distribution.

\subsection{Learning, validation, and testing sets}

\begin{figure}[htbp]
	\centering
	\includegraphics[width=0.7\linewidth]{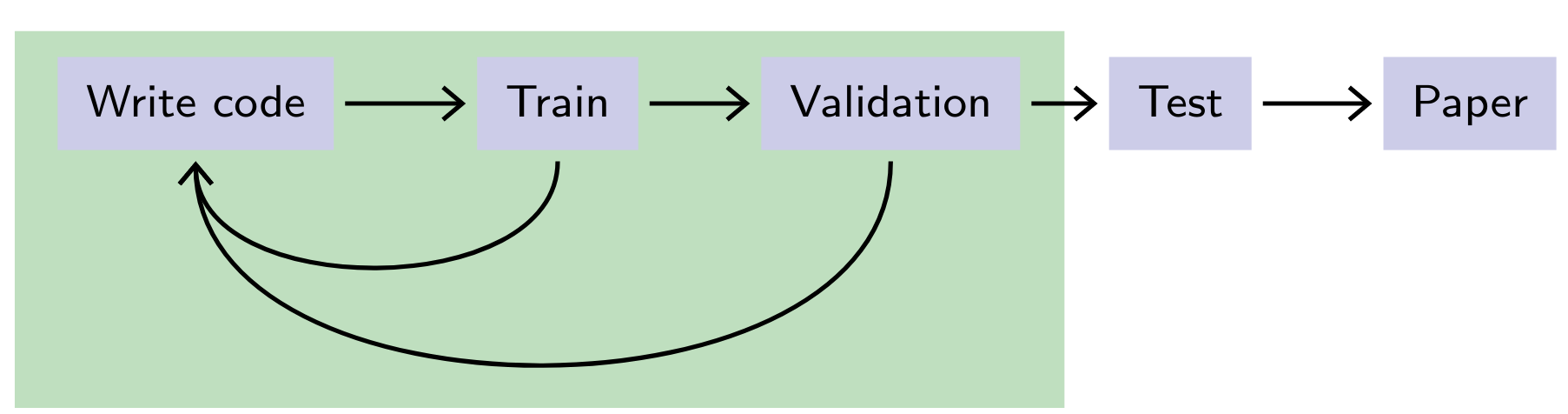}
	\caption{Proper evaluation protocols. Credits: Francois Fleuret, EE559 Deep Learning, EPFL \url{https://fleuret.org/dlc/}.}
	\label{fig:eval_prococol}
\end{figure}

A \textit{good practice} in machine learning is to use a dedicated part of the dataset as an independent testing set that is not used to train the models but only to estimate the generalization error. Ideally, the dataset $\mathcal{D}$ is divided randomly into three parts, as depicted in Figure \ref{fig:eval_prococol}.
\begin{enumerate}
    \item The models are trained on the learning set. 
    \item The validation set is used: (1) to evaluate the generalization error of the trained models to select among several learning algorithms the one more suited to the studied problem; (2) to optimize some algorithm’s hyper-parameters $\theta$ and to avoid over-fitting. 
    \item Finally, the testing set is kept until the end of the process. It allows assessing the performance of the selected model on independent data. 
\end{enumerate}
A common rule is to build the training set as large as possible to obtain good predictors while keeping enough samples in the validation and testing sets to correctly conduct the hyper-parameters selection and estimate the generalization error properly. 

\subsection{k-fold cross-validation}

When the dataset is composed of only a few months of data, dividing it into three parts can lead to a too "small" learning set to learn good predictors regarding both the empirical risk and the generalization error. In this case, the $k$-fold cross-validation methodology allows evaluating the generalization error of the algorithm correctly \citep{hastie2009elements}.

\begin{figure}[htbp]
	\centering
	\includegraphics[width=0.8\linewidth]{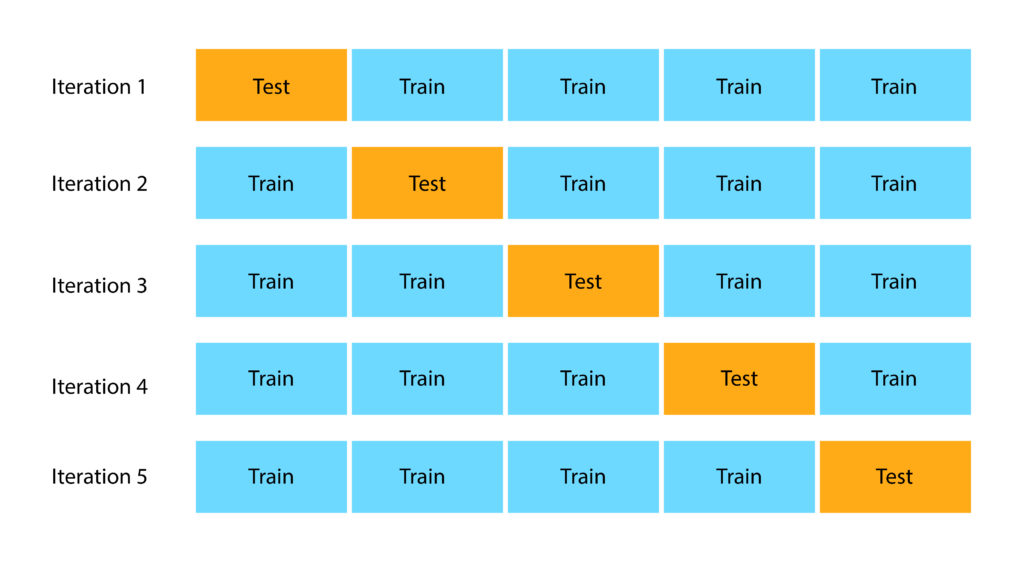}
	\caption{$k$-fold cross-validation.}
	\label{fig:crossvalidation}
\end{figure}

It consists of dividing the dataset into $k$-folds. Then, the model is trained on $k-1$ folds, and one fold is left out to evaluate the testing error. In total, the model is trained $k$ times with $k$ pairs of learning and testing sets. The generalization error is estimated by averaging the $k$ errors computed on $k$ testing sets. This procedure is adopted in Chapters \ref{chap:quantile-forecasting} and \ref{chap:density-forecasting} where the dataset is composed of only a few months. Figure \ref{fig:crossvalidation} depicts a 5-folds cross-validation.

However, $k$-fold cross-validation may suffer from high variability, which can be responsible for bad choices in model selection and erratic behavior in the estimated expected prediction error. A study conducted by \citet{bengio2004no} demonstrated there is no unbiased estimator of the variance of $k$-fold cross-validation. Therefore, the assessment of the significance of observed differences in cross-validation scores should be treated with caution.

\section{Conclusions}\label{sec:forecasting-background-conclusions}

This Chapter introduces the forecasting basics with the various types of forecasts, starting from the most common point forecasts and building up towards the more advanced probabilistic products: quantiles, prediction intervals, scenarios, and density forecasts. It also provides some basics on how to train and evaluate a forecasting model properly.
The next Chapter presents the methodologies to evaluate the quality of these various types of forecasts.

\chapter{Forecast evaluation}\label{chap:forecast_evaluation}

\begin{infobox}{Overview}
This chapter introduces the main concepts in forecasting evaluation as a background for the work developed in the following chapters of this manuscript. \\
General textbooks such as \citep{morales2013integrating} provide further information to the interested reader. In addition, the course "Renewables in Electricity Markets"\footnote{\url{http://pierrepinson.com/index.php/teaching/}} given by professor Pierre Pinson at the Technical University of Denmark proposes valuable materials on this topic.
\end{infobox}
\epi{Everything that can be counted does not necessarily count; everything that counts cannot necessarily be counted.}{Albert Einstein}
\begin{figure}[htbp]
	\centering
	\includegraphics[width=1\linewidth]{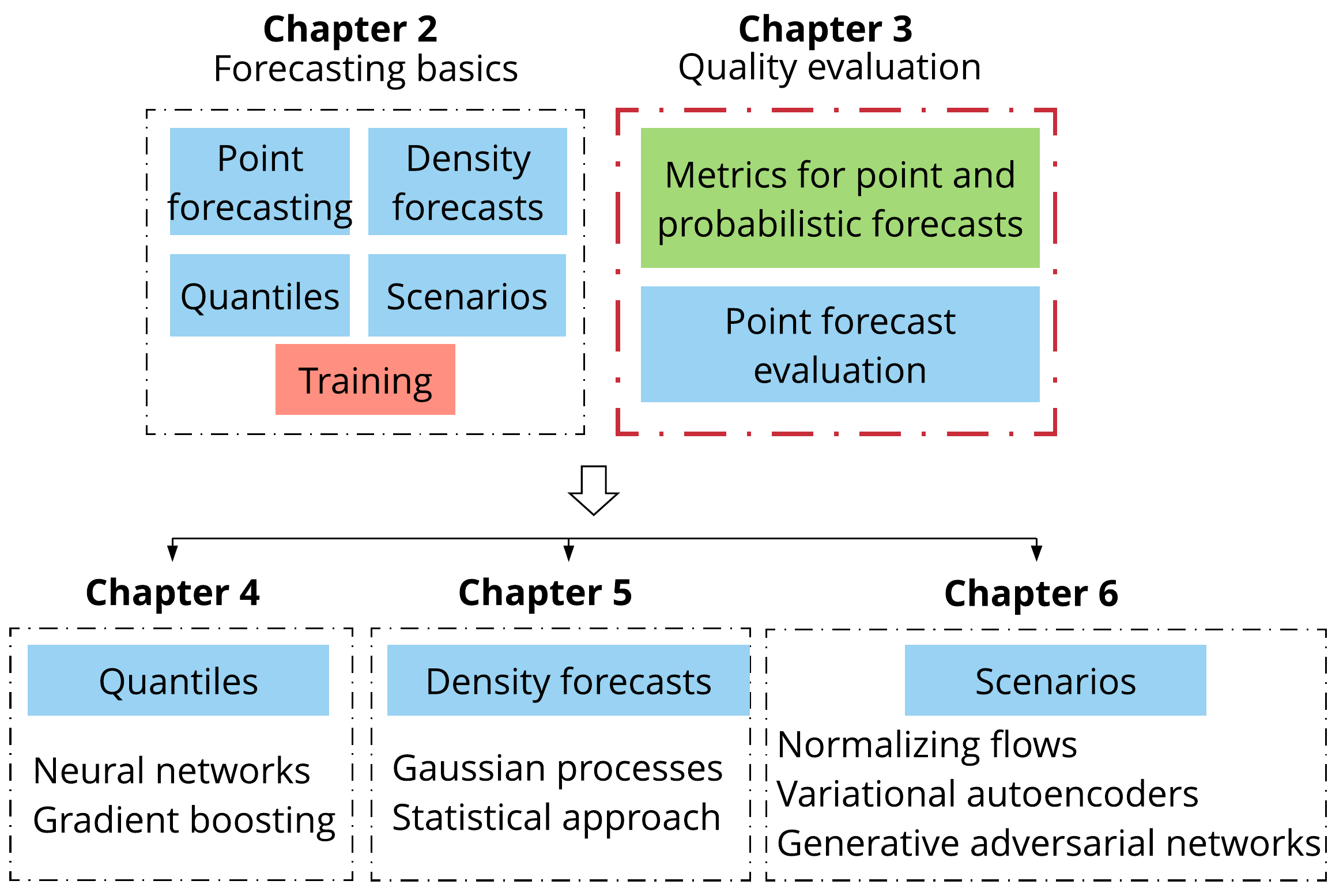}
	\caption{Chapter \ref{chap:forecast_evaluation} position in Part \ref{part:forecasting}.}
\end{figure}
\clearpage

For predictions in any form, one must differentiate between their \textit{quality} and their \textit{value} \citep{morales2013integrating}. The forecast quality corresponds to the ability of the forecasts to genuinely inform of future events by mimicking the characteristics of the processes involved. The forecast value, studied in Part \ref{part:optimization}, relates to the benefits of using forecasts in decision-making, such as participation in the electricity market. Consequently, forecast quality is independent of the current operational problem while not forecast value. Intuitively, it is relevant to evaluate the forecast quality before using the predictions as input to operational problems. Even if a good quality does not necessarily imply a good forecast value, the quality assessment provides interesting information about the model at hand.

This Chapter adopts the general framework of \citet{morales2013integrating} to provide the basics required to appraise the quality of predictions.
It is organized as follows. Sections \ref{sec:forecasting-quality-point} and \ref{sec:forecasting-quality-probabilistic} introduce the metrics for point and probabilistic forecasting, respectively. Finally, conclusions are drawn in Section \ref{sec:forecasting-quality-conclusions}.

\section{Metrics for point forecasts}\label{sec:forecasting-quality-point}

The base quantity for evaluating point forecasts of a continuous random variable is the forecast error. %
\begin{definition}[Forecast error]\citep[Chapter 2]{morales2013integrating}
The \textit{forecast error} $\epsilon_{t+k|t}$ is the difference between observed and predicted values
\begin{equation}\label{eq:forecast_error_def}
\epsilon_{t+k|t} := x_{t+k|t} - \hat{x}_{t+k|t}.
\end{equation}
\end{definition}
Note that the forecast error may be normalized so that verification results can be comparable for different time series. If so, normalization is most commonly performed by the nominal capacity of the site of interest. The first error criterion that may be computed is the bias of the forecasts, which corresponds to the systematic part of the forecast error. It may be corrected in a straightforward manner using simple statistical models.
\begin{definition}[Bias]\citep[Chapter 2]{morales2013integrating}
The \textit{bias} is the mean of all errors over the evaluation period of length $N$, considered indifferently. For lead time $k$, it is
\begin{equation}\label{eq:bias_def}
\text{bias}(k) := \frac{1}{N} \sum_{t =1}^{N} \epsilon_{t+k|t} .
\end{equation}
\end{definition}
This summary measure does not tell much about the quality of point forecasts but only about a systematic error that should be corrected. Therefore, for a better appraisal of the forecasts, it is advised to use scores such as the mean absolute and root mean square errors.

\begin{definition}[Mean absolute error]\citep[Chapter 2]{morales2013integrating}
The \textit{mean absolute error} (MAE) is defined as the average of the absolute forecast errors over an evaluation set of length $N$
\begin{equation}\label{eq:MAE_def}
\text{MAE}(k) := \frac{1}{N} \sum_{t =1}^{N} |\epsilon_{t+k|t}| .
\end{equation}
\end{definition}

\begin{definition}[Root mean square error]\citep[Chapter 2]{morales2013integrating}
The \textit{root mean square error} (RMSE) is defined as the square root of the sum of squared errors over an evaluation of length $N$
\begin{equation}\label{eq:RMSE_def}
\text{RMSE}(k) := \sqrt{\frac{1}{N} \sum_{t =1}^{N} (\epsilon_{t+k|t})^2} .
\end{equation}
\end{definition}

All the above error criteria are independent of the length of the evaluation set. The nominal capacity of the renewable energy site of interest can be used to normalize them, and then they are referred to Nbias, NRMSE, and NMAE.

\section{Metrics for probabilistic forecasts}\label{sec:forecasting-quality-probabilistic}

\subsection{Calibration}

The first requirement of probabilistic forecasts is to consistently inform about the probability of events. It leads to the concept of \textit{probabilistic calibration}, also referred to as \textit{reliability}. The assessment of probabilistic calibration only informs about a form of bias of probabilistic forecasts. 
A frequentist approach, based on an evaluation set of sufficient length, can be performed by assessing the reliability of each of the defining quantile forecasts using the indicator variable.
\begin{definition}[Indicator variable]\citep[Chapter 2]{morales2013integrating}
The \textit{indicator variable} $\xi^q_{t,k}$, for a given quantile forecast $\hat{x}^{(q)}_{t+k|t}$ and corresponding realization $x_{t+k}$ is
\begin{equation}\label{eq:indicator_variable_def}
\xi^q_{t,k} := 1_{\{x_{t+k} < \hat{x}^{(q)}_{t+k|t}\}} = \begin{cases} 1 &    x_{t+k} < \hat{x}^{(q)}_{t+k|t} \\
0  &  x_{t+k} \geq \hat{x}^{(q)}_{t+k|t} \end{cases}.
\end{equation}
\end{definition}
$\xi^q_{t,k}$ is a binary variable indicating if the quantile forecasts cover, or not, the measurements. The empirical level of quantile forecasts can be defined, estimated, and eventually compared with their nominal one by using this indicator variable.
\begin{definition}[Empirical level]\citep[Chapter 2]{morales2013integrating}
The \textit{empirical level} $\tilde{\xi}^q_{t,k}$, for a nominal level $q$ and lead time $k$, is obtained by calculating the mean of the $\{\xi^q_{t,k}\}_{t=1}^N$ time-series over an evaluation set of length $N$
\begin{equation}\label{eq:empirical_level_def}
\tilde{\xi}^q_{t,k} := \frac{1}{N} \sum_{t=1}^N \xi^q_{t,k}.
\end{equation}
\end{definition}

The difference between nominal and empirical levels of quantile forecasts is to be seen as a probabilistic bias. Then, probabilistic calibration may be appraised visually by using \textit{reliability diagrams} plotting empirical \textit{vs.} nominal levels of the quantiles defining density forecasts.

\subsection{Univariate skill scores}

Perfectly calibrated probabilistic forecasts do not guarantee that the forecasts are "good". For instance, they may not discriminate among situations with various uncertainty levels, while these aspects are crucial in decision-making. The overall quality of probabilistic forecasts may be assessed based on skill scores. 
First, we consider the \textit{univariate} skill score that can only assess the quality of the forecasts with respect to their marginals (time period). Second, we focus on the \textit{multivariate} skill score that can directly assess multivariate scenarios.

\subsubsection{Continuous ranked probability score}

The continuous ranked probability score (CRPS) \citep{gneiting2007strictly} is a univariate scoring rule that penalizes the lack of resolution of the predictive distributions as well as biased forecasts. It is negatively oriented, \textit{i.e.}, the lower, the better, and for deterministic forecasts, it turns out to be the mean absolute error (MAE). Thus, it can be directly compared to the MAE criterion used for point forecasts since the CRPS is its generalization in a probabilistic forecasting framework. The CRPS is used to compare the skill of predictive marginals for each component of the random variable of interest. In our case, for the twenty-four time periods of the day.
\begin{definition}[CRPS-integral]\citep{gneiting2007strictly}
The \textit{CRPS} for predictive densities $\hat{F}_{t+k|t}$ and corresponding measurement $x_{t+k|t}$, is calculated over an evaluation set of length $N$
\begin{equation}\label{eq:CRPS_def}
\text{CRPS}(k) := \frac{1}{N} \sum_{t=1}^N \int_{x'} [\hat{F}_{t+k|t}(x') - 1_{\{ x' >  x_{t+k|t} \}}]^2 dx'.
\end{equation}
\end{definition}
%
However, it is not easy to estimate the integral form of the CRPS defined by (\ref{eq:CRPS_def}).
\begin{definition}[CRPS-energy]\citep{gneiting2007strictly}
\citet{gneiting2007strictly} propose a formulation called the \textit{energy form} of the CRPS since it is just the one-dimensional case of the energy score, defined in negative orientation as follows
\begin{align}\label{eq:CRPS_NRG_def}
\text{CRPS}(P, x_{t+k|t}) = & \mathbb{E}_P [|X-x_{t+k|t}|] - \frac{1}{2} \mathbb{E}_P [|X-X'|],
\end{align}
where X and X' are independent random variables with distribution P and finite first moment, and $\mathbb{E}_P$ is the expectation according to the probabilistic distribution P. 
\end{definition}

The CRPS can be computed for quantile forecasts and scenarios.
%
Let $ \{\hat{x}^i_{t+k|t}  \}_{i=1}^M$ be the set of $M$ scenarios generated at time $t$ for lead time $k$. The estimator of (\ref{eq:CRPS_NRG_def}) proposed by \citet{zamo2018estimation}, over an evaluation set of length $N$ for lead time $k$, is 
\begin{align}\label{eq:CRPS_eNRG_scenarios}
\text{CRPS}(k) & = \frac{1}{N} \bigg[\sum_{t=1}^N \frac{1}{M}\sum_{i = 1}^M|\hat{x}^i_{t+k|t}-x_{t+k}|  - \frac{1}{2 M^2} \sum_{i,j = 1}^M |\hat{x}^i_{t+k|t}- \hat{x}^j_{t+k|t}| \bigg].
\end{align}
Let $ \{\hat{x}^{(q)}_{t+k|t} \}_{q=1}^{Q}$ be the set of $Q$ quantiles generated at time $t$ for lead time $k$. (\ref{eq:CRPS_NRG_def}) is over an evaluation set of length $N$ for lead time $k$
\begin{align}\label{eq:CRPS_eNRG_quantiles}
\text{CRPS}(k) & = \frac{1}{N} \bigg[\sum_{t=1}^N \frac{1}{Q}\sum_{q = 1}^Q|\hat{x}^{(q)}_{t+k|t}-x_{t+k}|  - \frac{1}{2 Q^2} \sum_{q,q' = 1}^Q |\hat{x}^{(q)}_{t+k|t}- \hat{x}^{(q')}_{t+k|t}| \bigg].
\end{align}

\subsubsection{Quantile score}
The quantile score (QS), also known as the \textit{pinball loss}, is complementary to the CRPS. It permits obtaining detailed information about the forecast quality at specific probability levels, \textit{i.e.}, over-forecasting or under-forecasting, and particularly those related to the tails of the predictive distribution \citep{lauret2019verification}. It is negatively oriented and assigns asymmetric weights to negative and positive errors for each quantile. 

\begin{definition}[Pinball loss]
The pinball loss function for a given quantile $q$ is
\begin{equation}\label{eq:pinball_loss}
\rho_q(\hat{x}, x) :=\max \big\{(1-q) (\hat{x} - x), q (x - \hat{x}) \big\}.
\end{equation}
\end{definition}

\begin{definition}[Quantile score]
The \textit{quantile score}, over an evaluation set of length $N$ for lead time $k$, is
\begin{equation}\label{eq:quantile_score_def}
\text{QS} (k) := \frac{1}{N} \sum_{t=1}^N \frac{1}{Q} \sum_{q=1}^Q \rho_q(\hat{x}^{(q)}_{t+k|t}, x_{t+k}).
\end{equation}
The quantile score, over an evaluation set of length $N$ for a given quantile $q$ over all lead times $k$, is
\begin{equation}\label{eq:quantile_score_per_quantile_def}
\text{QS} (q) := \frac{1}{N} \sum_{t=1}^N \frac{1}{T}\sum_{k=k_1}^{k_T} \rho_q(\hat{x}^{(q)}_{t+k|t}, x_{t+k}).
\end{equation}
\end{definition}

\subsubsection{Interval score}
The interval score (IS) \citep{gneiting2007strictly} assesses the quality of central prediction interval forecasts specifically. It rewards narrow prediction intervals. It penalizes the forecasts where the observation is outside the interval thanks to the penalty term that depends on $\alpha$
\begin{definition}[IS]
The \textit{interval score}, for a central prediction interval with a coverage rate of $(1-\alpha)$, over an evaluation set of length $N$ and for lead time $k$ is
\begin{align}\label{eq:IS_def}
\text{IS} (k) & = \frac{1}{N} \sum_{t=1}^{N} (\hat{x}^{(1-\alpha/2)}_{t+k|t} - \hat{x}^{(\alpha/2)}_{t+k|t})  + \frac{2}{\alpha} (\hat{x}^{(\alpha/2)}_{t+k|t} - x_{t+k}) 1_{\{x_{t+k} \leq \hat{x}^{(\alpha/2)}_{t+k|t} \}}  \notag \\
& + \frac{2}{\alpha} ( x_{t+k} - \hat{x}^{(1-\alpha/2)}_{t+k|t})1_{\{  x_{t+k} \geq \hat{x}^{(1-\alpha/2)}_{t+k|t}\} }.
\end{align}
\end{definition}

Note: the skill scores can be normalized by the total installed capacity for a renewable generation plant.

\subsection{Multivariate skill scores}

\subsubsection{Energy score}

The energy score (ES) is the most commonly used scoring rule when a finite number of trajectories represents distributions. It is a multivariate generalization of the CRPS and has been formulated and introduced by \citet{gneiting2007strictly}. The ES is proper and negatively oriented, \textit{i.e.}, a lower score represents a better forecast. 
The ES is used as a multivariate scoring rule by \citet{golestaneh2016generation} to investigate and analyze the spatio-temporal dependency of PV generations. They emphasize the ES pros and cons. It is capable of evaluating forecasts relying on marginals with correct variances but biased means. Unfortunately, its ability to detect incorrectly specified correlations between the components of the multivariate quantity is somewhat limited.

\begin{definition}[Energy score]
\citet{gneiting2007strictly} introduced a generalization of the continuous ranked probability score defined in negative orientation as follows
\begin{align}\label{eq:ES-def}
\text{ES}(P, \mathbf{x}) = & \mathbb{E}_P  \Vert X-\mathbf{x} \Vert - \frac{1}{2} \mathbb{E}_P \Vert X-X'\Vert,
\end{align}
where and X and X' are independent random variables with distribution P and finite first moment, $\mathbb{E}_P$ is the expectation according to the probabilistic distribution P, and $ \Vert \cdot \Vert $ the Euclidean norm. 
\end{definition}

The ES is computed following \citet{gneiting2008assessing} over an evaluation set of length $N$ as follows
\begin{align}\label{eq:ES-estimator}
\text{ES} & = \frac{1}{N} \sum_{t = 1}^M\bigg[\frac{1}{M}\sum_{i = 1}^M \Vert \mathbf{\hat{x}}_t^i-\mathbf{x}_t \Vert - \frac{1}{2 M^2} \sum_{i,j = 1}^M \Vert\mathbf{\hat{x}}_t^i - \mathbf{\hat{x}}_t^j \Vert \bigg].
\end{align}
Note: when we consider the marginals of $\mathbf{x}$, it is easy to recognize that (\ref{eq:ES-estimator}) is the CRPS.

\subsubsection{Variogram score}

An alternative class of proper scoring rules based on the geostatistical concept of variograms is proposed by \citet{scheuerer2015variogram}. They study the sensitivity of these variogram-based scoring rules to inaccurate predicted means, variances, and correlations. The results indicate that these scores are distinctly more discriminative concerning the correlation structure. Thus, the Variogram score (VS) captures correlations between multivariate components in contrast to the Energy score.

\begin{definition}[Variogram score]
For a given day $t$ of the testing set and a $T$-variate observation $\mathbf{x}_t \in \mathbb{R}^T$, the Variogram score metric of order $\gamma$ is formally defined as
\begin{align}\label{eq:VS-def}
\text{VS}_t & = \sum_{k,k'}^T w_{kk'} \bigg ( |x_{t,k} -x_{t,k'} |^\gamma - \mathbb{E}_P |\hat{x}_{t,k} -\hat{x}_{t,k'} |^\gamma \bigg)^2,
\end{align}
where $\hat{x}_{t,k}$ and $\hat{x}_{t,k'}$ are the $k$-th and $k'$-th components of the random vector $\mathbf{\hat{x}}_t$ distributed according to P for which the $\gamma$-th absolute moment exists, and $w_{kk'}$ are non-negative weights. Given a set of $M$ scenarios $\{  \mathbf{\hat{x}}_t^i \}_{i=1}^M$ for this given day $t$, the forecast variogram $\mathbb{E}_P |\hat{x}_{t,k} -\hat{x}_{t,k'} |^\gamma$ can be approximated $\forall k,k' =1, \cdots, T$ by 
\begin{align}\label{eq:VS-approximation}
\mathbb{E}_P |\hat{x}_{t,k} -\hat{x}_{t,k'} |^\gamma & \approx \frac{1}{M} \sum_{i=1}^M |\hat{x}_{t,k}^i -\hat{x}_{t,k'}^i |^\gamma.
\end{align}
\end{definition}

\noindent Then, the VS is averaged over the testing set of length $N$
\begin{align}\label{eq:VS-TEST}
\text{VS} & = \frac{1}{N} \sum_{t=1}^N \text{VS}_t.
\end{align}
In this thesis, we evaluate the Variogram score with equal weights across all hours of the day $w_{kk'}=1$ and using a $\gamma$ of 0.5, which for most cases provides a good discriminating ability as reported in \citet{scheuerer2015variogram}. 

\section{Point forecasting evaluation example}\label{sec:point-forecasting}

This section proposes an example of point forecasting quality evaluation. It is based on an extract of \citet{dumas2021coordination}.
In addition, it introduces multi-output weather-based point forecasting by defining the problem formulation and implementation on a real-case study to compute PV and electrical consumption forecasts. They are used by a day-ahead deterministic planner presented in Chapter \ref{chap:coordination-planner-controller}. 
Note: PV intraday point forecast will also be considered in this thesis using an encoder-decoder architecture developed in \citet{dumas2020deep}. Chapter \ref{chap:quantile-forecasting} details the approach and the results.


\subsection{Formulation}\label{sec:forecasting-point-formulation}

Point forecasting corresponds to the conditional expectation of the stochastic process for every lead time (\ref{eq:point_forecast_def}). This definition is linked to the so-called \textit{loss function} $L(\hat{x}, x)$. Loss functions assign a penalty to forecast errors as a proxy of the cost of these errors for those making decisions based on such forecasts. There exist various types of loss functions, such as the mean squared error or the pinball loss. A special relevant case of a loss function to compute point forecasts is the mean squared error
\begin{equation}
\label{eq:loss_function_quadratic}
L_2(\hat{x}, x) = (\hat{x}- x)^2.
\end{equation}
At the time $t$ a point forecast $\hat{x}_{t+k|t}$ for time $t + k$ is the value of the process such that it minimizes the expected loss for the forecast user for all potential realizations of the process, given our state of knowledge at that time. Therefore, when considering multi-output point forecasts of length $T$ the estimation of the model parameters is performed by solving 
\begin{subequations}
\begin{align}
\label{eq:point_forecast_estimation2}
\theta^\star = \arg \min_\theta &   \frac{1}{N} \sum_{t=1}^N {\lVert  \mathbf{\hat{x}}_t - \mathbf{x}_t \rVert}^2_2, \\
\mathbf{\hat{x}}_t \sim & g_\theta(\mathbf{x}_{<t}, \mathbf{c}_t),
\end{align}
\end{subequations}
given the information set $\mathcal{D}$ of length $N$, and ${\lVert \cdot \rVert}_2$ the Euclidean norm.

\subsection{Forecasting models}\label{sec:forecasting-point-models}

Two standard forecasting techniques are implemented to forecast the PV production and the consumption, with a dedicated model per variable of interest. The first model uses a Recurrent Neural Network (RNN) of the Keras Python library \citep{chollet2015keras}. The RNN is a Long Short Term Memory (LSTM) with one hidden layer composed of $2 \times n+1$ neurons with $n$ the number of input features. The second model is a Gradient Boosting Regressor (GBR) of the Scikit-learn Python library \citep{scikit-learn}. 
They both use past values of the PV production and consumption, and the weather forecasts provided by the Laboratory of Climatology of the Li\`ege University, based on the MAR regional climate model \citep{fettweis2017reconstructions}. It is an atmosphere model designed for meteorological and climatic research, used for a wide range of applications, from km-scale process studies to continental-scale multi-decade simulations.
The study \citet{dumas2021coordination} focuses on the real-time control of microgrids based on planning that requires a forecast horizon of a few hours up to a few days.
Both models are trained by solving (\ref{eq:point_forecast_estimation2}) that becomes
\begin{subequations}
\begin{align}
\mathbf{\hat{x}}_t \sim & g_\theta(\mathbf{x}_{<t}, \mathbf{c}_t), \\
\mathbf{x}_{<t} = & [x_t, ..., x_{t-k_4}] , \\
\mathbf{c}_t = & [\hat{c}_{t+k_1|t}^i, ..., \hat{c}_{t+k_T|t}^i]  ,
\end{align}
\end{subequations}
with $\mathbf{\hat{x}}_t$ the variable to forecast, \textit{i.e}, PV, consumption, and $\hat{c}_i$ the forecast of the $i^\text{th}$ weather variable, \textit{e.g.}, direct solar irradiance, wind speed, air temperature. 
In the case study considered, the point forecasts are computed each quarter for the subsequent $T=96$ periods with a resolution $\Delta t = 15$ minutes. The forecasting process is implemented by using a rolling forecast methodology where the learning set is updated every six hours, with a fixed size limited to the week preceding the forecasts, to maintain a reasonable computation time.

\subsection{Results}\label{sec:forecasting-point-results}

\begin{figure}[tb]
	\centering
	\includegraphics[width=0.7\linewidth]{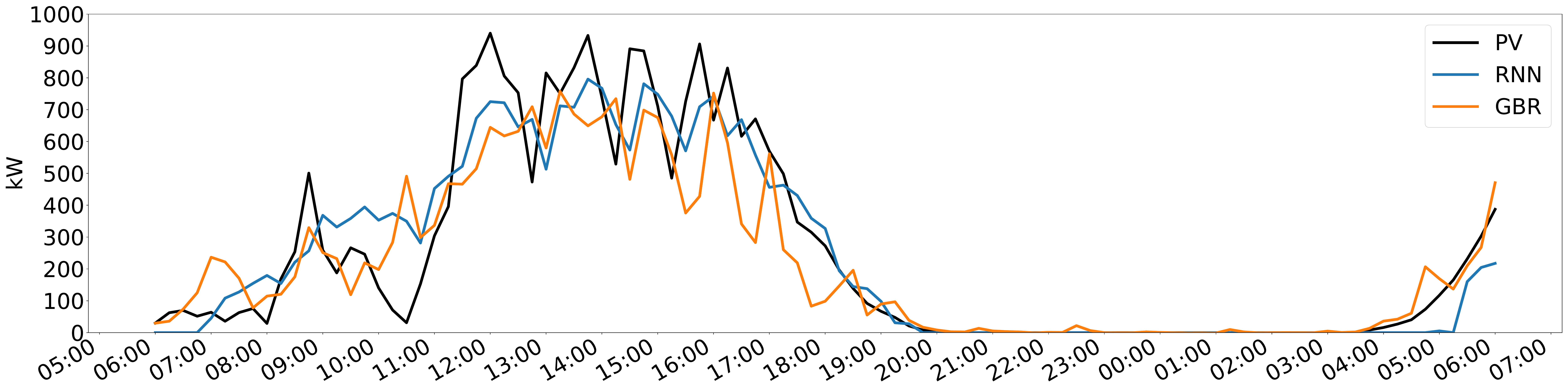}  %
	\captionsetup{justification=centering}
	\caption{PV point forecasts on $\pscctwelveofjune$ computed at 06h00 UTC. The PV observations are in black.}
	\label{fig:case3_pv_forecast_12062019}
\end{figure}

Figure~\ref{fig:case3_pv_forecast_12062019} illustrates the PV point forecasts of the RNN and GBR models compared to the observations (black) on a particular day of the testing set. It is interesting to point out the sudden drop in the PV production around 10 a.m. that the models do not accurately forecast.

\begin{figure}[tb]
	\centering
	\includegraphics[width=0.7\linewidth]{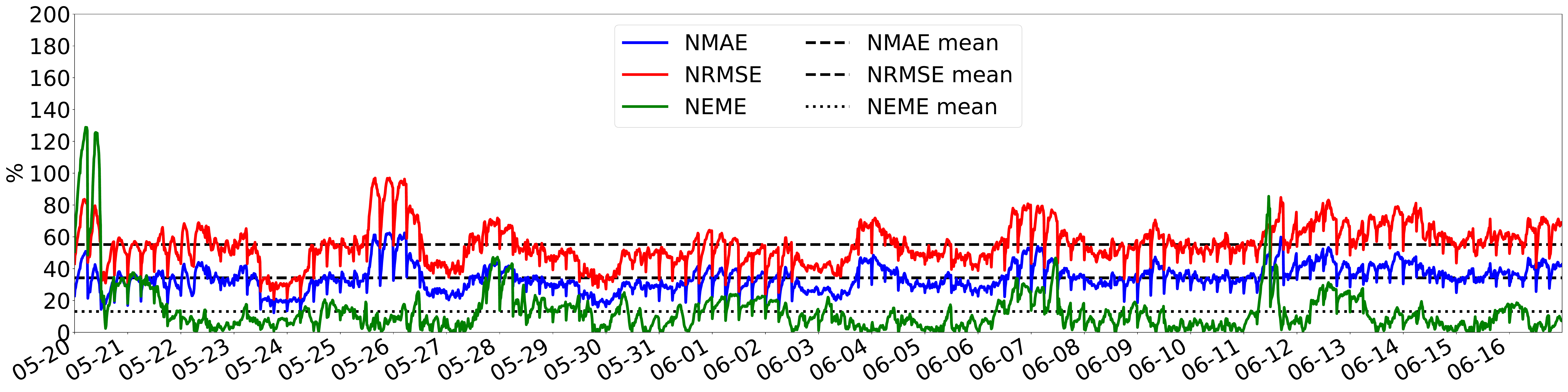}
	\includegraphics[width=0.7\linewidth]{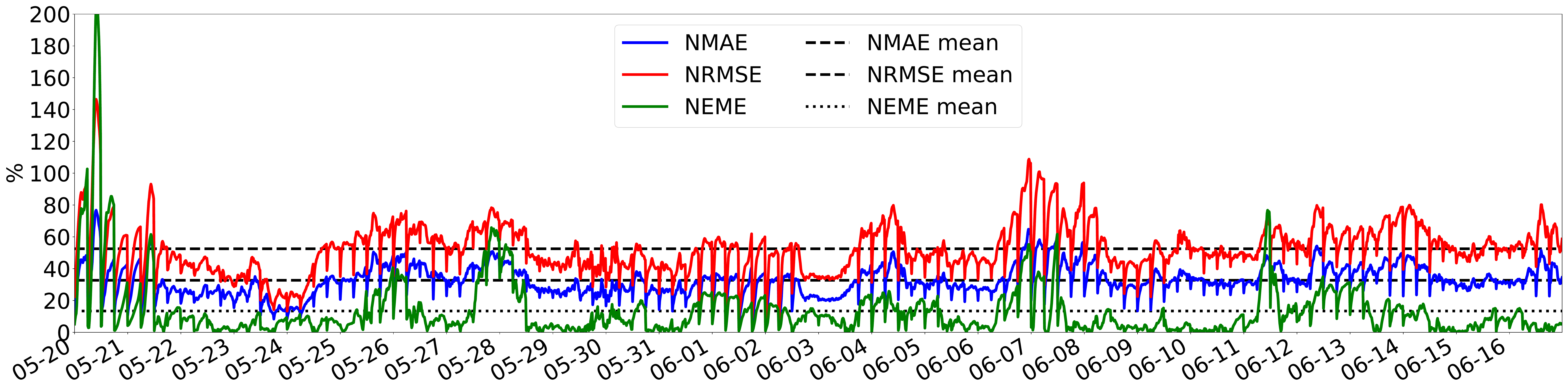} 
	\captionsetup{justification=centering}
	\caption{PV forecast scores, GBR (top) and LSTM (bottom).}
	\label{fig:pvforecast_scores}
	\centering
	\includegraphics[width=0.7\linewidth]{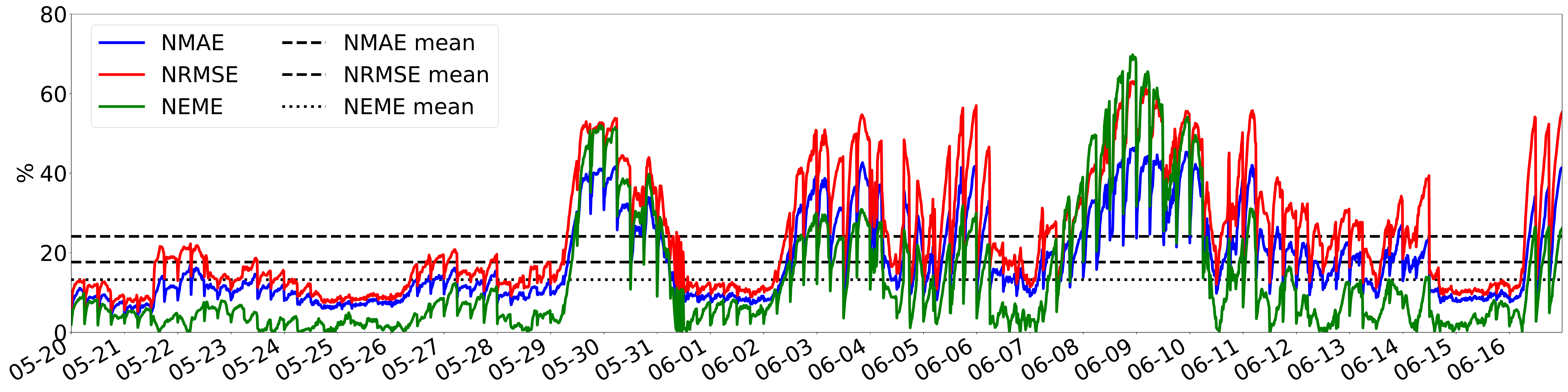} 
	\includegraphics[width=0.7\linewidth]{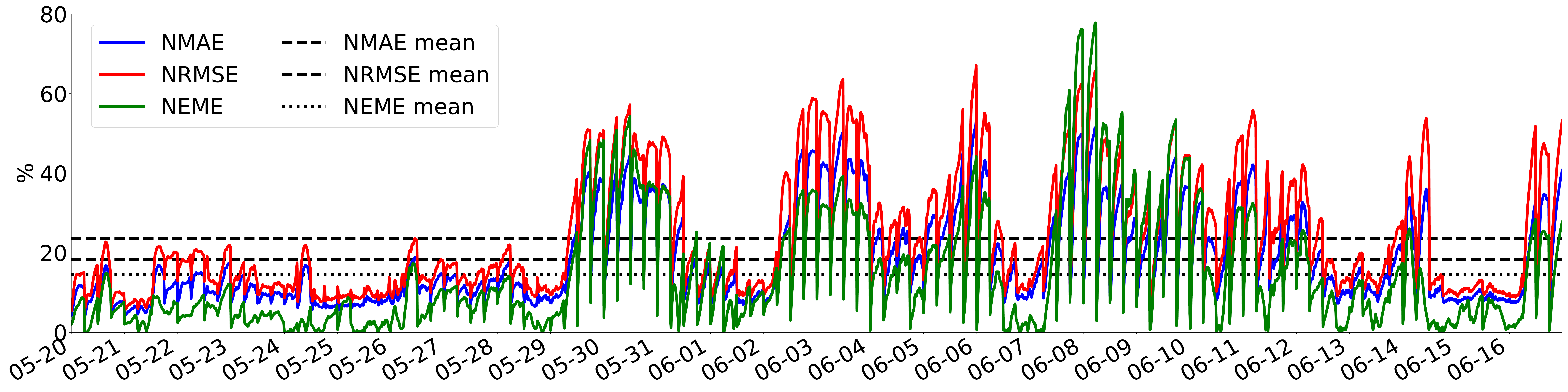} 
	\captionsetup{justification=centering}
	\caption{Consumption forecast scores, GBR (top) and LSTM (bottom).}
	\label{fig:consoforecast_scores}
\end{figure}

The point forecasts are evaluated each quarter over the entire forecasting horizon, the next 96 periods, using the Normalized Mean Absolute Error (NMAE), the Normalized Root Mean Squared Error (NRMSE), and the Normalized Energy Measurement Error (NEME). The normalizing coefficients for both the NMAE and NRMSE are the mean of the absolute values of the PV and consumption over all the simulation data set. The NEME is the NMAE of the energy summed over the entire forecasting horizon. Figures \ref{fig:pvforecast_scores} and \ref{fig:consoforecast_scores} provide the scores for both GBR and LSTM models for each quarter, plain lines, and the average over the entire simulation dataset, dashed lines. On average, the LSTM model yields slightly smaller average NMAE, NRMSE, and NEME values. However, the differences are not significant, and the forecast value (see Chapter \ref{chap:coordination-planner-controller}) needs to be evaluated to decide which model is the best. 
\clearpage

\section{Conclusions}\label{sec:forecasting-quality-conclusions}

This Chapter provides the basics of forecast verification to acquire background knowledge on forecast quality that corresponds to the ability of the forecasts to genuinely inform of future events by mimicking the characteristics of the processes involved.
In contrast, the forecast value, investigated in Part \ref{part:optimization},  relates to the interests of using forecasts in decision-making, such as participation in the electricity market. However, it is intuitively expected that higher-quality forecasts will yield better policies and decisions. Therefore, quality evaluation is complementary to value evaluation and provides an insight into forecasting model skills. 
The Chapter illustrates the evaluation methodology with a multi-output point forecasting model and an implementation of a real-case study using two standard techniques. The weather-based forecasting models are trained to compute quarterly day-ahead PV and electrical consumption point forecasts. Then, the quality is assessed by computing point forecasts metrics.
In the following Chapters, the quality metrics are used to evaluate the various probabilistic forecasts on several case studies.

\chapter{Quantile forecasting}\label{chap:quantile-forecasting}

\begin{infobox}{Overview}
Overall, the Chapter contributions can be summarized as follows.
\begin{enumerate}
    \item A deep learning-based multi-output quantile architecture computes prediction intervals of PV generation on a day-ahead and intraday basis. The goal is to implement an improved probabilistic intraday forecaster, the encoder-decoder, to benefit from the last PV generation observations. This architecture is compared to a feed-forward neural network.
    \item The weather forecasts are used to directly take into account the impact of the weather forecast updates generated every six hours.
    \item A proper assessment of the quantile forecasts is conducted by using a $k$-fold cross-validation methodology and probabilistic metrics. It allows computing average scores over several testing sets and mitigating the dependency of the results to specific days of the dataset. 
    \item Finally, a comparison of deep learning quantile regression models is conducted with quantiles derived from deep learning generative models.
\end{enumerate}

\textbf{\textcolor{RoyalBlue}{References:}} This chapter  is an adapted version of the following publications: \\[2mm]\bibentry{dumas2020deep}. \\[2mm] \bibentry{dumas2021nf}.
%
\end{infobox}
\epi{Prediction is very difficult, especially if it's about the future.}{Niels Bohr}
\begin{figure}[htbp]
	\centering
	\includegraphics[width=1\linewidth]{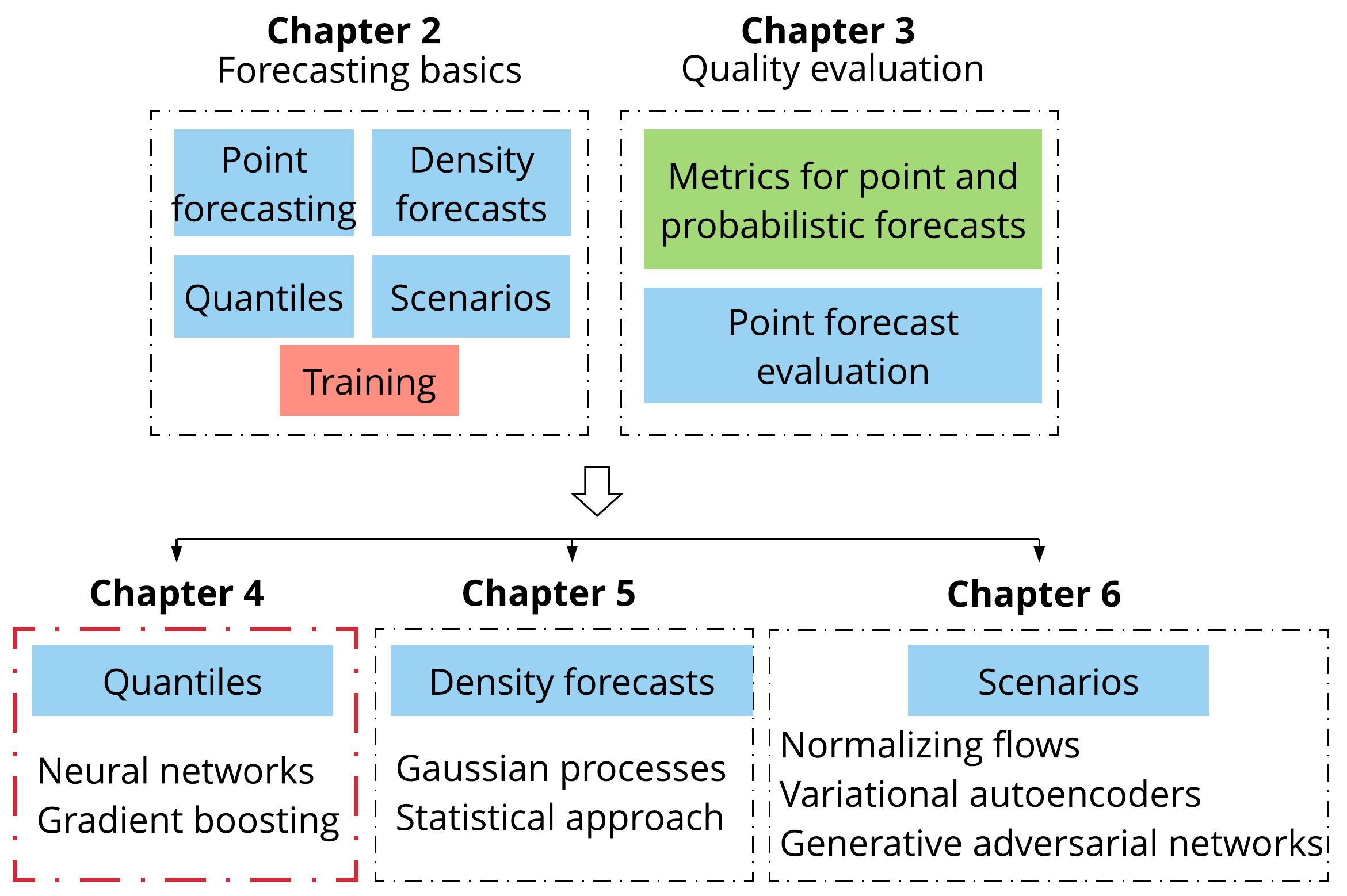}
	\caption{Chapter \ref{chap:quantile-forecasting} position in Part \ref{part:forecasting}.}
\end{figure}
\clearpage

This Chapter focuses on \textit{quantile} forecasts that provide probabilistic information about future renewable power generation, in the form of a threshold level associated with a probability. 
This approach is investigated in \citet{dumas2020deep} with probabilistic PV forecasters that exploit recent breakthroughs in the field of data science by using advanced deep learning structures, such as the \textit{encoder-decoder} architecture \citep{bottieau2019very}. 
It is implemented to compute day-ahead and intraday multi-output PV quantiles forecasts, to efficiently capture the correlation between time periods, and to be used as input of a robust optimization model. For instance, to address the energy management system of a grid-connected renewable generation plant coupled with a battery energy storage device detailed in \citet{dumas2021probabilistic} and Chapter \ref{chap:capacity-firming-robust}.
The case study is composed of PV production monitored on-site at the University of Li\`ege (ULi\`ege), Belgium. The weather forecasts from the MAR climate regional model \citep{fettweis2017reconstructions} provided by the Laboratory of Climatology are used as inputs of the deep learning models.

The remainder of this Chapter is organized as follows. Section \ref{sec:forecasting-quantile-formulation} presents the non-parametric quantile forecasting framework. Section \ref{sec:forecasting-quantile-related} provides the related work. Section \ref{sec:forecasting-quantile-models} details the forecasting techniques. Section \ref{sec:forecasting-quantile-case} describes the case study and presents the results. Section \ref{sec:forecasting-quantile-comparison} proposes a comparison of the quantile regression models with quantiles derived from deep learning generative models. Finally, Section \ref{sec:forecasting-quantile-conclusions} summarizes the main findings and highlights ideas for further work. 

\section{Formulation}\label{sec:forecasting-quantile-formulation}

\textit{Quantile regression }\citep{koenker1978regression} is one of the most famous non-parametric approach. It does not assume the shape of the predictive distributions. It can be implemented with various types of forecasting techniques, \textit{e.g.}, neural networks, linear regression, gradient boosting, or any other regression techniques.
The estimation of the model parameters $\theta$ to compute multi-output quantile forecasts of length $T$ is performed by minimizing the pinball loss (\ref{eq:pinball_loss}) over all lead times
\begin{align}
\label{eq:quantile_forecast_estimation}
\theta^\star = \arg \min_\theta  &  \frac{1}{N} \sum_{t=1}^N \frac{1}{Q} \sum_{q=1}^Q \sum_{k=k_1}^{k_T} \rho_q(\hat{x}_{t+k}^{(q)}, x_{t+k}), \\
\mathbf{\hat{x}}_t^{(q)} \sim & g_\theta(\mathbf{x}_{<t}, \mathbf{c}_t),
\end{align}
given the information set $\mathcal{D}$ of length $N$, and a set of $Q$ quantiles. 

\section{Related work}\label{sec:forecasting-quantile-related}

The literature on quantile forecasting is broad. We present a few papers that have gained our attention in quantile PV probabilistic forecasting.
At the Global Energy Forecasting Competition 2014 \citep{hong2016probabilistic} solar forecasts are expressed in the form of 99 quantiles with various nominal proportions between zero and one. The models are evaluated by using the pinball loss function. This study summarizes the recent research progress on probabilistic energy forecasting at that time, and this competition made it possible to develop innovative techniques.
A systematic framework for generating PV probabilistic forecasts is developed by \citet{golestaneh2016very}. A non-parametric density forecasting method based on Extreme Learning Machine is adopted to avoid restrictive assumptions on the shape of the forecast densities. 
A combination of bidirectional Recurrent Neural Networks (RNNs) with Long Short-Term Memory (LSTM), called Bidirectional LSTM (BLSTM), is proposed by \citet{toubeau2018deep}. It has the benefits of both long-range memory and bidirectional processing. The BLSTM is trained by minimizing the quantile loss to compute quantile forecasts of aggregated load, wind and PV generation, and electricity prices on a day-ahead basis. Finally, an innovative architecture, referred to as encoder-decoder (ED), is developed by \citet{bottieau2019very} to generate reliable predictions of the future system imbalance used for robust optimization.

\section{Forecasting models}\label{sec:forecasting-quantile-models}

This Section presents the forecasting techniques implemented to compute multi-output point and quantile forecasts of PV generation.

\subsection{Gradient boosting regression (GBR)}

\textit{Gradient boosting} builds an additive model in a forward stage-wise fashion \citep{hastie2009elements}. It allows for the optimization of arbitrary differentiable loss functions. In each stage, a regression tree is fit on the negative gradient of the given loss function.
The gradient boosting regressor (GBR) from the Scikit-learn \citep{scikit-learn} Python library is trained by minimizing the quantile loss. The learning rate is set to $10^{-2}$, the max depth to 5, and the number of estimators to 500. There is a GBR model trained per quantile as this library does not handle a single model for several quantiles.

\subsection{Multi-layer perceptron (MLP)}

A description of the most widely used "vanilla" neural network, the \textit{multi-layer perceptron} (MLP), is provided by \citep{hastie2009elements}. A MLP with a single hidden layer is considered for the day-ahead forecasts and as the benchmark for the intraday forecasts. Note, MLPs with two and three hidden layers did not provide any significant improvement on the case study considered. The activation function is the Rectified Linear Unit (ReLU). The number of neurons of the hidden layer is $n_\text{input} + (n_\text{output} - n_\text{input}) / 2$, with $n_\text{input}$ and $n_\text{output}$ the number of neurons of the input and output layers, respectively. The learning rate is set to $10^{-2}$ and the number of epoch to 500 with a batch size of 8. It is implemented using the PyTorch Python library \citep{paszke2017automatic}. 

\subsection{Encoder-decoder (ED)}

Several technical information about recent advances in neural networks is provided by \citet{toubeau2018deep,bottieau2019very}. In particular, recurrent neural networks have shown a high potential in processing and predicting complex time series with multi-scale dynamics. However, RNNs are known to struggle in accessing time dependencies more than a few time steps long due to the vanishing gradient problem. Indeed, back-propagated errors during the training stage either fades or blows up over time. Long Short-Term Memory and Gated Recurrent Units networks tackle this problem by using internal memory cells \citep{bottieau2019very}. A neural network composed of a LSTM and feed-forward layers, referred to as LSTM in the rest of the Chapter, is implemented for the day-ahead and intraday forecasts. The number of LSTM units is $n_\text{input} + (n_\text{output} - n_\text{input}) / 3$, and the number of neurons of the feed-forward layer $n_\text{input} + 2 \times (n_\text{output} - n_\text{input}) / 3$. 

\begin{figure}[htbp]
	\centering
    \includegraphics[width=0.3\linewidth]{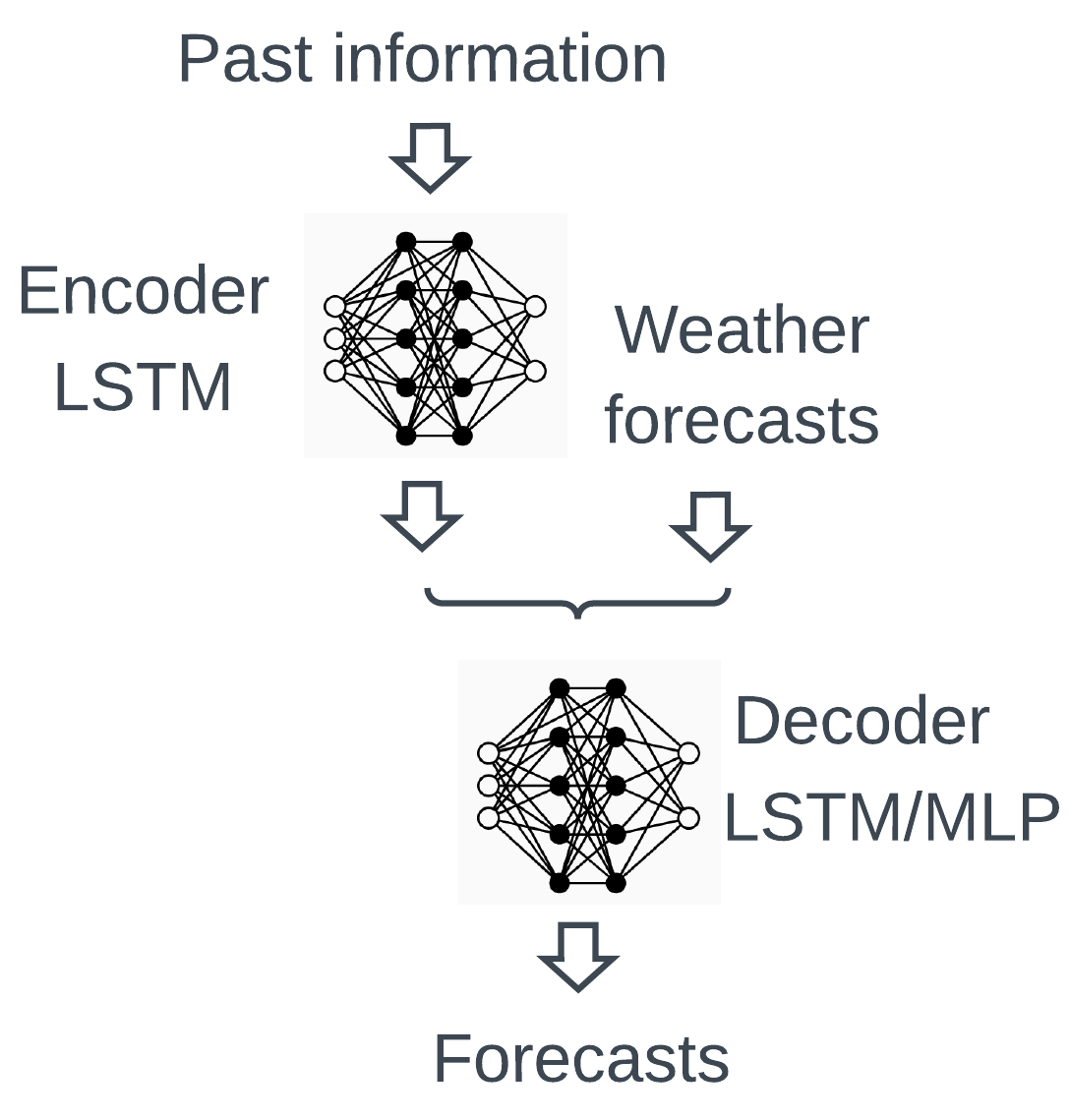}
	\caption{Encoder-decoder architecture.}
	\label{fig:encoder-decoder}
\end{figure}

An innovative architecture referred to as encoder-decoder \citep{bottieau2019very}, is composed of two different networks and has recently shown promising results for translation tasks and speech recognition applications and imbalance price forecasting. The encoder-decoder, depicted in Figure \ref{fig:encoder-decoder}, processes features from the past, such as past PV observations, to extract the relevant historical information that is contained into a reduced vector of fixed dimensions, based on the last hidden state. Then, the decoder processes this representation along with the known future information such as weather forecasts.
A version of the encoder-decoder architecture (ED-1) is implemented with a LSTM as the encoder and a MLP as the decoder. In a second version (ED-2), the decoder is a LSTM followed by an additional feed-forward layer. Both versions of the encoder-decoder are used as intraday forecasters. In ED-1, the encoder has $2 \times n_\text{input}$ units with $n_\text{input}$ the number of neurons of the encoder input layer, features from the past. Then, the encoder output is merged with the weather forecasts becoming the decoder input layer that has $n_\text{output} /2$ neurons. In ED-2, the decoder has the same number of cells as the encoder, and the feed-forward layer is composed of $n_\text{output} /2$ neurons. The LSTM, ED-1, and ED-2 models are implemented using the TensorFlow Python library \citep{tensorflow2015-whitepaper}. The activation functions are the ReLU. The learning rate is set to $10^{-3}$, the number of epoch to 500 with a batch size of 64 for the three models.

A sensitivity analysis has been conducted to select the hyperparameters: number of hidden layers, neurons, epochs, and learning rate. Overall, increasing the number of hidden layers and neurons increases the model complexity. It can enhance the accuracy, but only up to a limited number of layers and neurons due to overfitting issues. In addition, the hyperparameter solution is closely related to the size of the historical database \citep{toubeau2018deep}. A deep learning model with more significant hidden layers and neurons requires a large amount of data to estimate the parameters accurately. In the case study considered, there are only 157 days of data with a 15 minutes resolution. Thus, we decided to restrict the number of layers and neurons to select a smaller model that performs better with the available information.

\section{The ULi\`ege case study}\label{sec:forecasting-quantile-case}

\subsection{Case study description}

The ULi\`ege case study is composed of a PV generation plant with an installed capacity of 466.4 kW. The PV generation has been monitored on a minute basis for 157 days. The data is resampled to 15 minutes. The set of quantiles is $ \{q= 10\%, \ldots, 90\%\}$ for both the day-ahead and intraday forecasts. 
Numerical experiments are performed on an Intel Core i7-8700 3.20 GHz based computer with 12 physical CPU cores and 32 GB of RAM running on Ubuntu 18.04 LTS. 

\subsection{Numerical settings}

The MAR regional climate model \citep{fettweis2017reconstructions} provided by the Laboratory of Climatology of the Li\`ege University is forced by GFS (Global Forecast System) to compute weather forecasts on a six hours basis, four-time gates per day at 00:00, 06:00, 12:00, and 18:00 with a 10-day horizon and a 15 minutes resolution. The solar irradiance and air temperature at 2 meters are normalized by a standard scaler and used as inputs to the forecasting models.

A $k$-fold cross-validation strategy computes average scores over several testing sets to mitigate the dependency of the results to specific days of the dataset. The dataset is divided into $k$ parts of equal length, and there are $k$ possible testing sets $1 \leq i \leq k$. For a given testing set $i$, the models are trained over the $k-1$ parts of the dataset. Eleven pairs of fixed lengths of 142 and 15 days are built. One pair is used to conduct the hyperparameters sensitivity analysis, and the ten others for testing where the scores are averaged.
The Mean Absolute Error and Root Mean Squared Error are introduced to evaluate the point forecasts. The MAE, RMSE, CRPS, and IS are normalized by the PV total installed capacity with NMAE and NRMSE the normalized MAE and RMSE.

The day-ahead models, MLP, LSTM, and GBR compute forecasts at noon for the next day.
Four intraday time gates are considered at 00:00, 06:00, 12:00, and 18:00. The intraday forecasts of time gate 00:00 are computed by the day-ahead models using only the weather forecasts. Then, the following three intraday forecasts are computed by intraday models where the MLP, ED-1, and ED-2 models use the weather forecasts and the last three hours of PV generation.

The day-ahead and the first intraday predictions are delivered for the 96 quarters of the next day from 00:00 to 23:45 indexed by time steps $0 \leq k \leq 95$. The prediction horizons span from 12 to 36 hours, for the day-ahead gate 12:00, and 0 to 24 hours, for the intraday gate 00:00.
The prediction horizon is cropped to $11 \leq k \leq 80$ because the PV generation is always 0 for time steps $0 \leq k \leq 10$ and $81 \leq k \leq 95$ on the ULi\`ege case study.
The next three intraday predictions are performed for the 72, 48, and 24 next quarters of the day corresponding to the gates 06:00, 12:00, and 18:00. Therefore, the prediction horizons span from 0 to 18 hours, 0 to 12 hours, and 0 to 6 hours. The intraday forecasting time periods are $24 \leq k \leq 80$, $48 \leq k \leq 80$, and $72 \leq k \leq 80$.
Table \ref{tab:tcpu_comparison} compares the mean and the standard deviation of the computation times over the ten learning sets to train the point and quantile forecast models\footnote{The day-ahead and intraday LSTM training times are identicals for both point and quantile forecasts as they only take the weather forecasts as inputs.}.
\begin{table}[htbp]
\renewcommand{\arraystretch}{1.25}
	\begin{center}
		\begin{tabular}{lrrr}
			\hline  \hline
			day-ahead &  MLP & LSTM & GBR \\ \hline
			point    & 5.3  (0.1)  & 23.7  (0.3) & 3.4 (0.1)  \\
			quantile & 7.6  (0.2)  & 69.0  (0.6) & 44.6 (0.4)  \\ \hline 
			
			intraday  &  MLP & ED-1  & ED-2 \\ \hline
			point     & 5.0  (0.1)   & 5.2  (0.1) & 17.2 (0.2) \\
			quantile  & 17.9  (0.2)  & 6.4  (0.2) & 18.0 (0.3) \\ \hline \hline
		\end{tabular}
		\caption{Training computation time (s).}
		\label{tab:tcpu_comparison}
	\end{center}
\end{table}

\subsection{Day-ahead results}

\begin{figure}[htbp]
	\centering
	\begin{subfigure}{.45\textwidth}
		\centering
		\includegraphics[width=\linewidth]{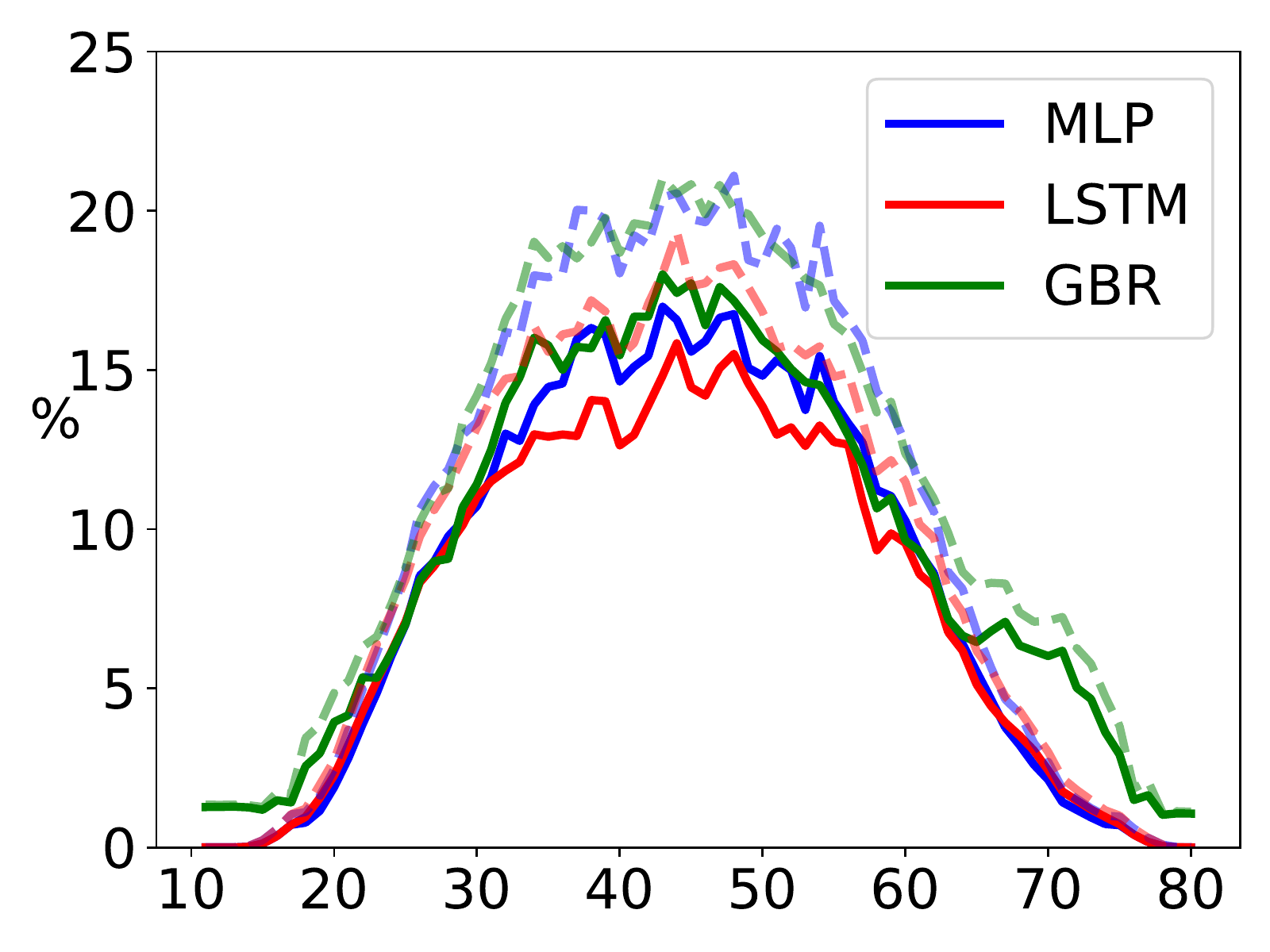}
		\caption{NMAE (plain) and NRMSE (dashed).}
		\label{fig:dad_point_comparison}
	\end{subfigure}%
	\begin{subfigure}{.45\textwidth}
		\centering
		\includegraphics[width=\linewidth]{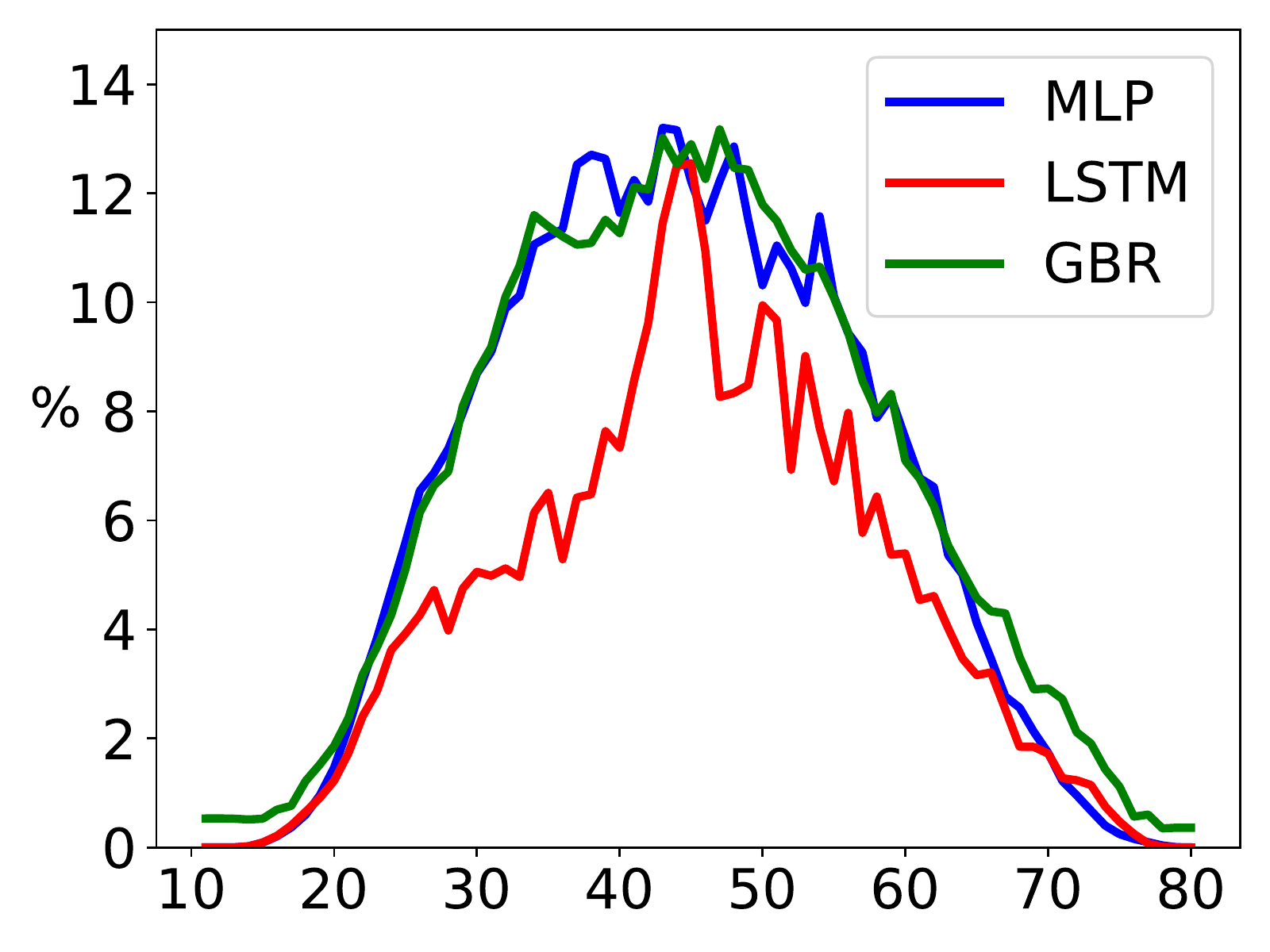}
		\caption{CRPS.}	
		\label{fig:dad_quantile_comparison}
	\end{subfigure}
	\caption{NMAE, NRMSE, and CRPS of the day-ahead models per lead time $k$.}
	\label{fig:dad_comparison}
\end{figure}

Figure \ref{fig:dad_point_comparison} compares the NMAE (plain lines), NRMSE (dashed lines), and Figure \ref{fig:dad_quantile_comparison} the CRPS per lead time $k$ of the day-ahead models of gate 12:00. Table \ref{tab:dad_comparison} provides the mean and standard deviation of the NMAE, NRMSE, and CRPS. The LSTM achieved the best results for both point and quantile forecasts. Figures \ref{fig:forecasts_plot_MLP_dad}, \ref{fig:forecasts_plot_LSTM_dad}, and \ref{fig:forecasts_plot_GBR_dad} compare the MLP, LSTM, and GBR day-ahead quantile and point forecasts (black line named dad 12) of gate 12:00 on $\quantiledate$ with the observation in red. One can see that the predicted intervals of the LSTM model better encompass the actual realizations of uncertainties than the MLP and GBR.
\begin{table}[htbp]
\renewcommand{\arraystretch}{1.25}
	\begin{center}
		\begin{tabular}{llrrr}
			\hline  \hline
Score & Gate &  MLP & LSTM & GBR\\
\hline
\multirow{2}{*}{NMAE} & 12 & 8.2 (1.2) & \textbf{7.6}  (1.5) &  9.2 (0.9)\\
& 24 & 7.9 (1.2) & \textbf{7.7}  (1.6) &  9.0 (0.8)\\ \hline  %
\multirow{2}{*}{NRMSE} & 12 & 10.2  (1.4) & \textbf{9.2}  (1.6) & 11.2 (0.9) \\
& 24 & 9.7   (1.2) & \textbf{9.4}  (1.8) & 10.9 (0.8) \\ \hline 
\multirow{2}{*}{CRPS} & 12 & 6.2  (1.1) & \textbf{4.4}  (0.2) & 6.4 (0.7)  \\
& 24 & 6.2  (1.0) & \textbf{4.4}  (0.2) & 6.3 (0.6)   \\ \hline	 \hline		
		\end{tabular}
\caption{The NMAE, NRMSE, CRPS of day-ahead models are averaged over all periods. The best-performing model for each score and gate is written in bold.}
\label{tab:dad_comparison}
	\end{center}
\end{table}


\subsection{Intraday results}

\begin{figure}[tb]
	\centering
	\begin{subfigure}{.45\textwidth}
		\centering
		\includegraphics[width=\linewidth]{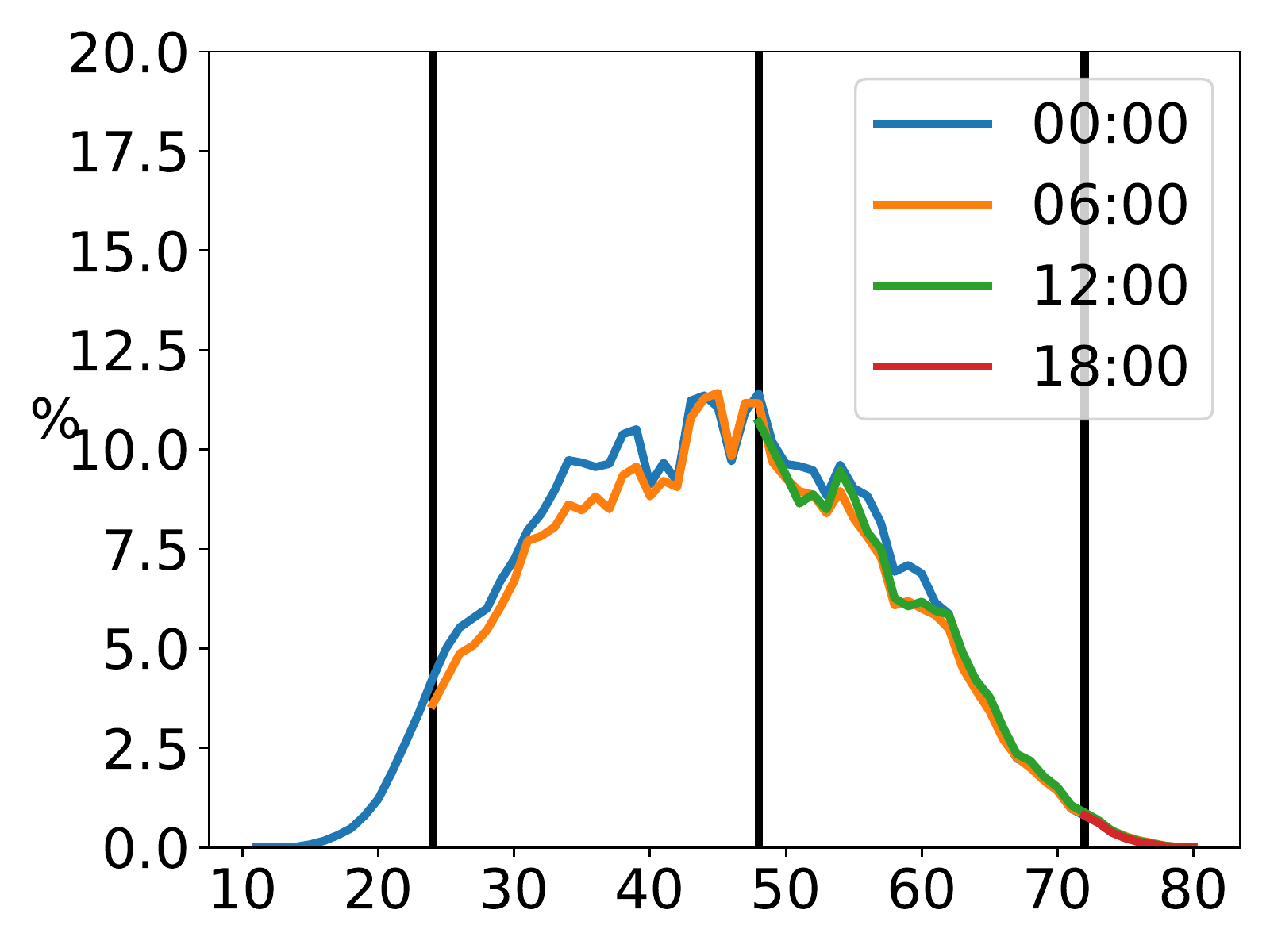}
		\caption{ED-1.}
	\end{subfigure}%
	\begin{subfigure}{.45\textwidth}
		\centering
		\includegraphics[width=\linewidth]{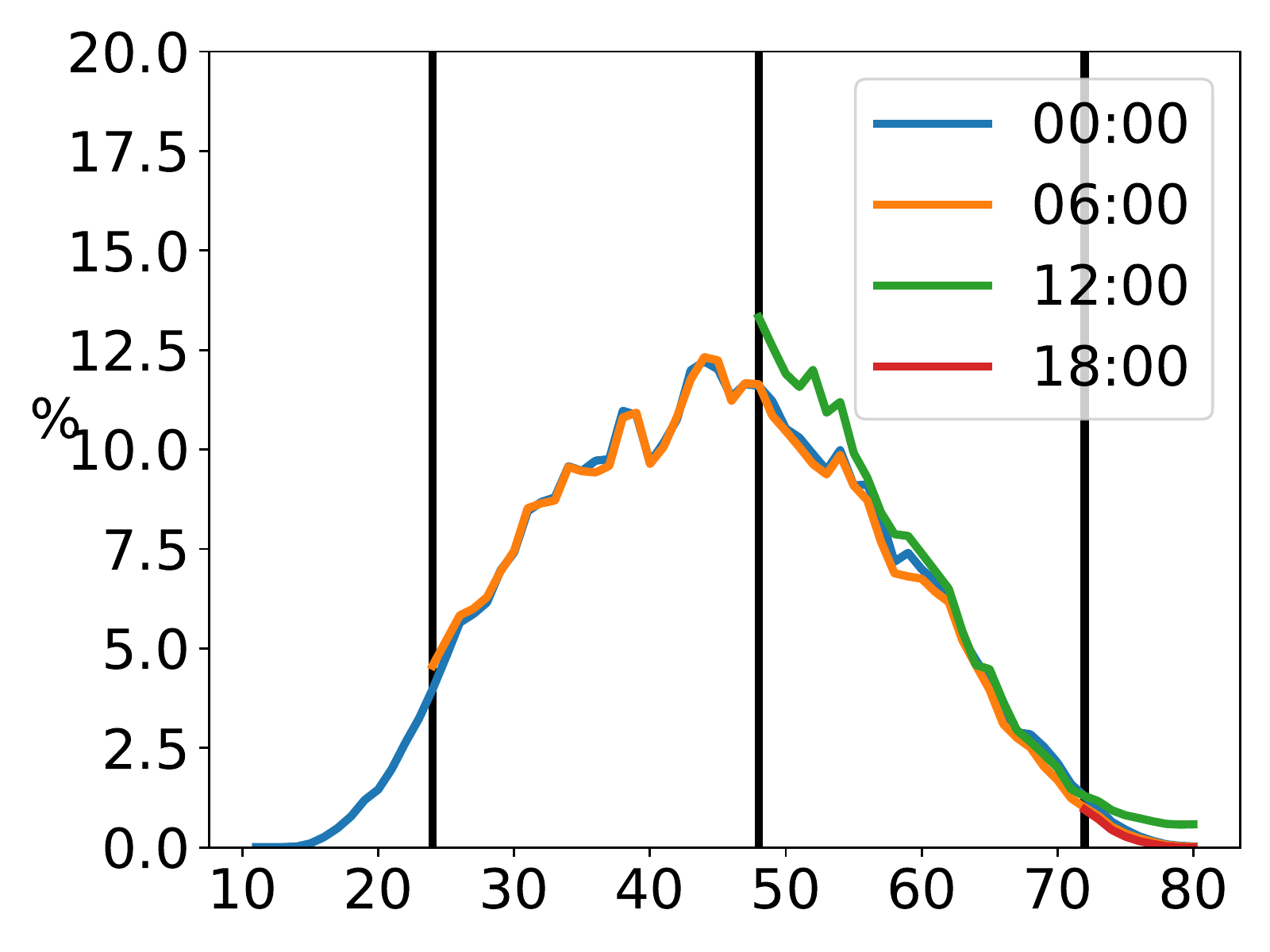}
		\caption{ED-2.}
\end{subfigure}
\begin{subfigure}{.45\textwidth}
	\centering
	\includegraphics[width=\linewidth]{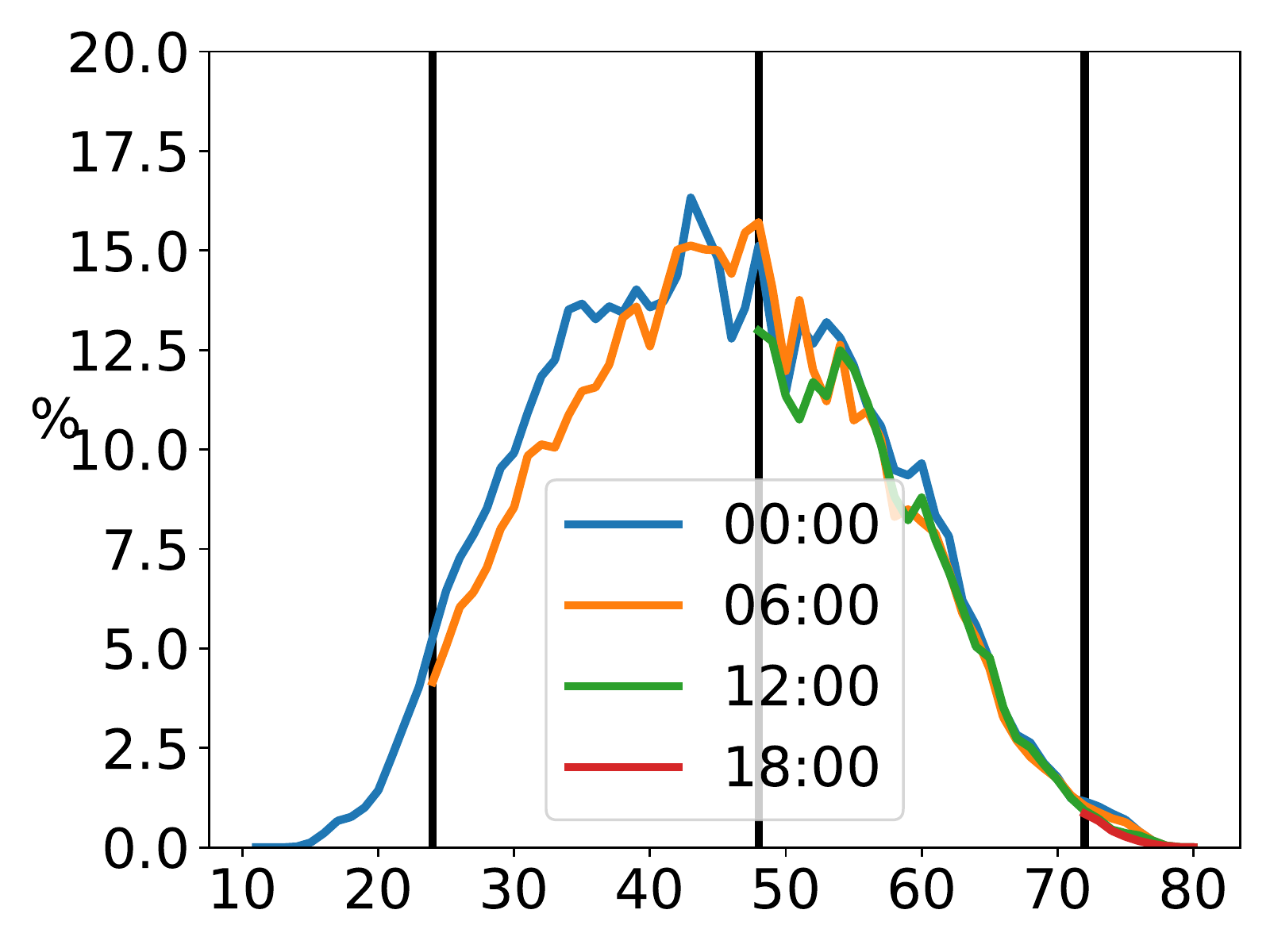}
	\caption{MLP.}
\end{subfigure}%
\begin{subfigure}{.45\textwidth}
	\centering
	\includegraphics[width=\linewidth]{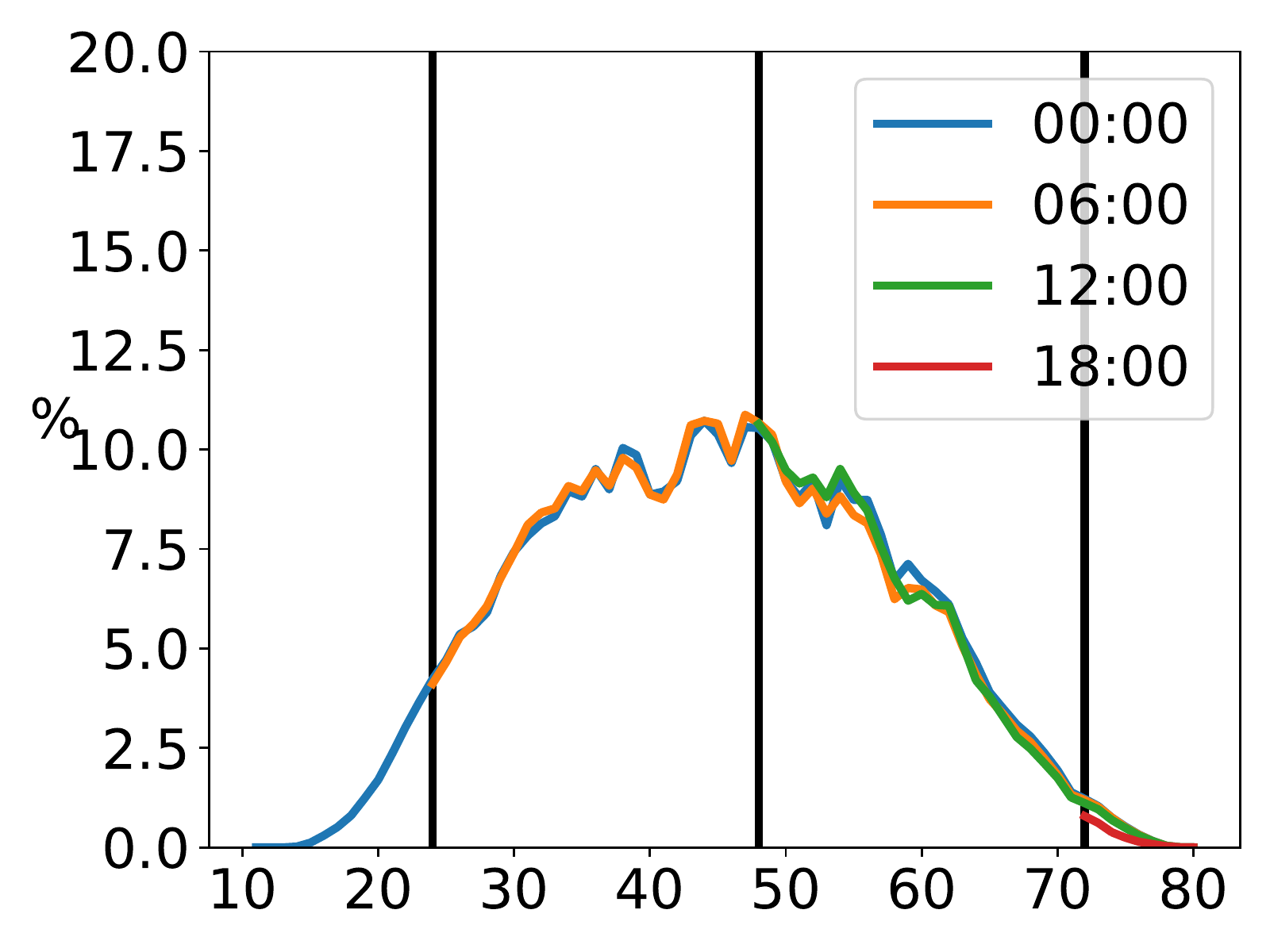}
	\caption{LSTM.}
\end{subfigure}
	\caption{CRPS of intraday models per lead time $k$.}
	\label{fig:intra_CRPS}
\end{figure}

\begin{figure}[htbp]
	\centering
	\begin{subfigure}{.45\textwidth}
		\centering
		\includegraphics[width=\linewidth]{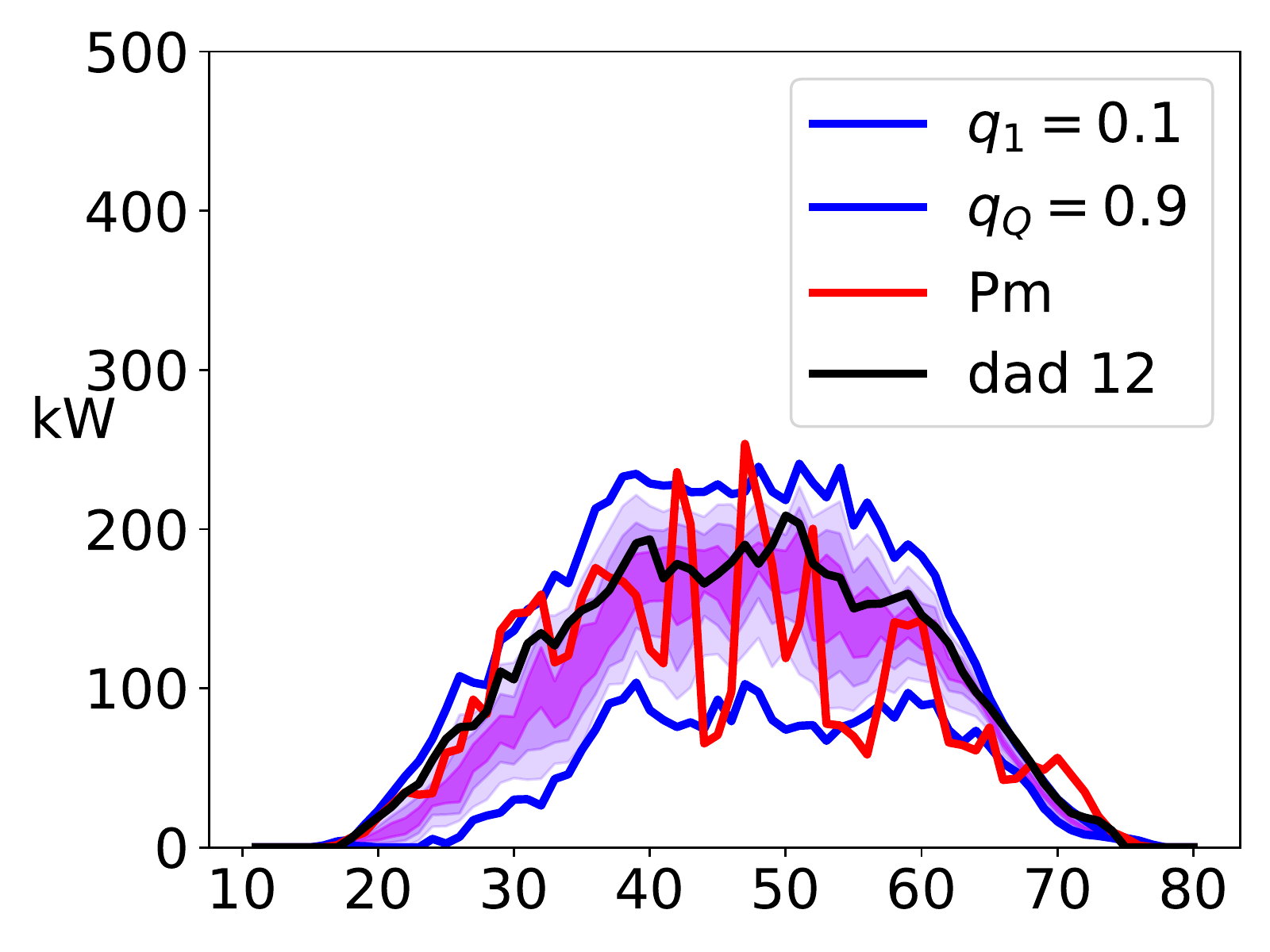}
		\caption{MLP day-ahead.}
		\label{fig:forecasts_plot_MLP_dad}
	\end{subfigure}%
	\begin{subfigure}{.45\textwidth}
	\centering
	\includegraphics[width=\linewidth]{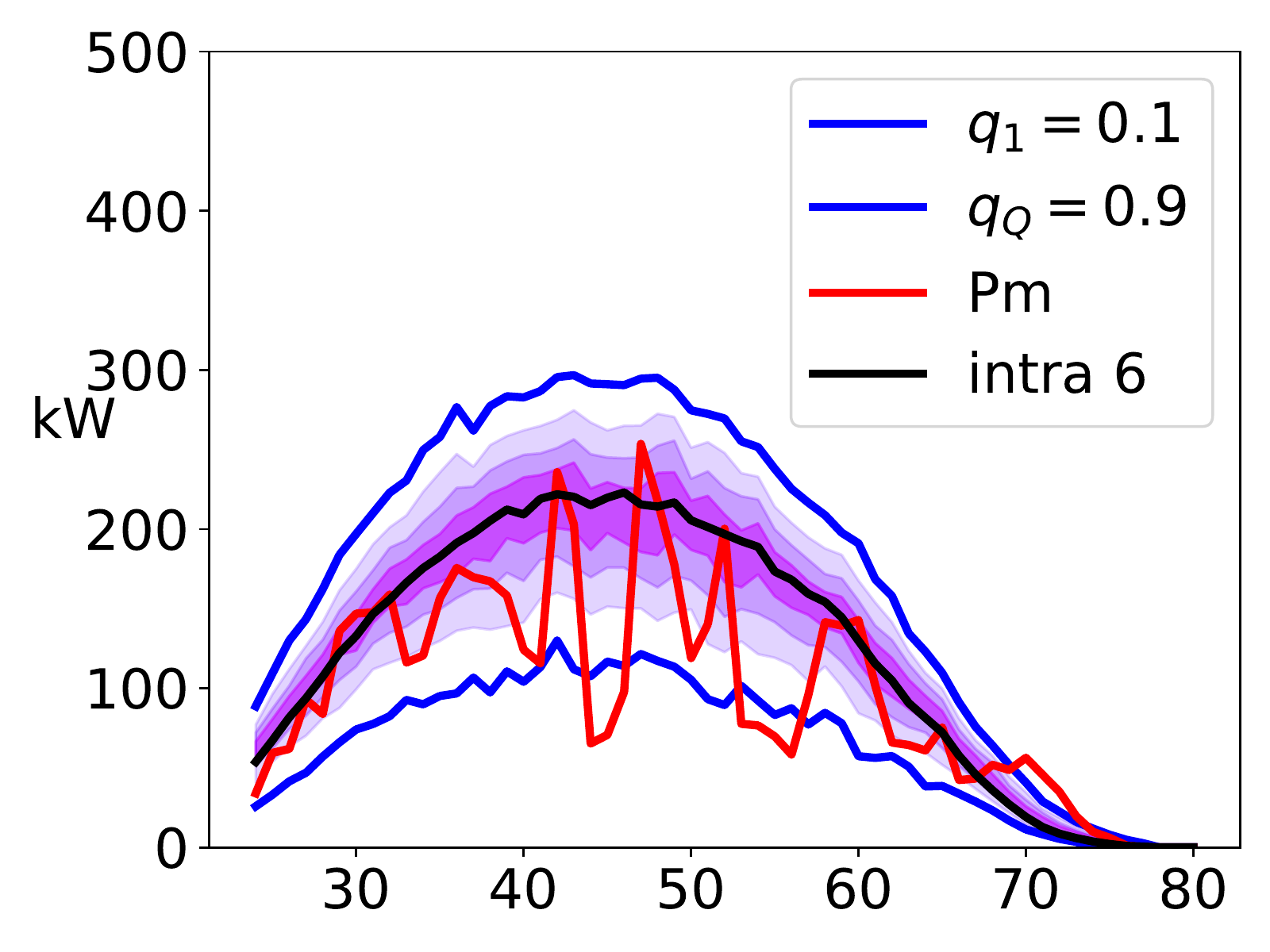}
	\caption{ED-1 intraday.}
	\label{fig:forecasts_plot_ED1_intra}
\end{subfigure}
	\begin{subfigure}{.45\textwidth}
		\centering
		\includegraphics[width=\linewidth]{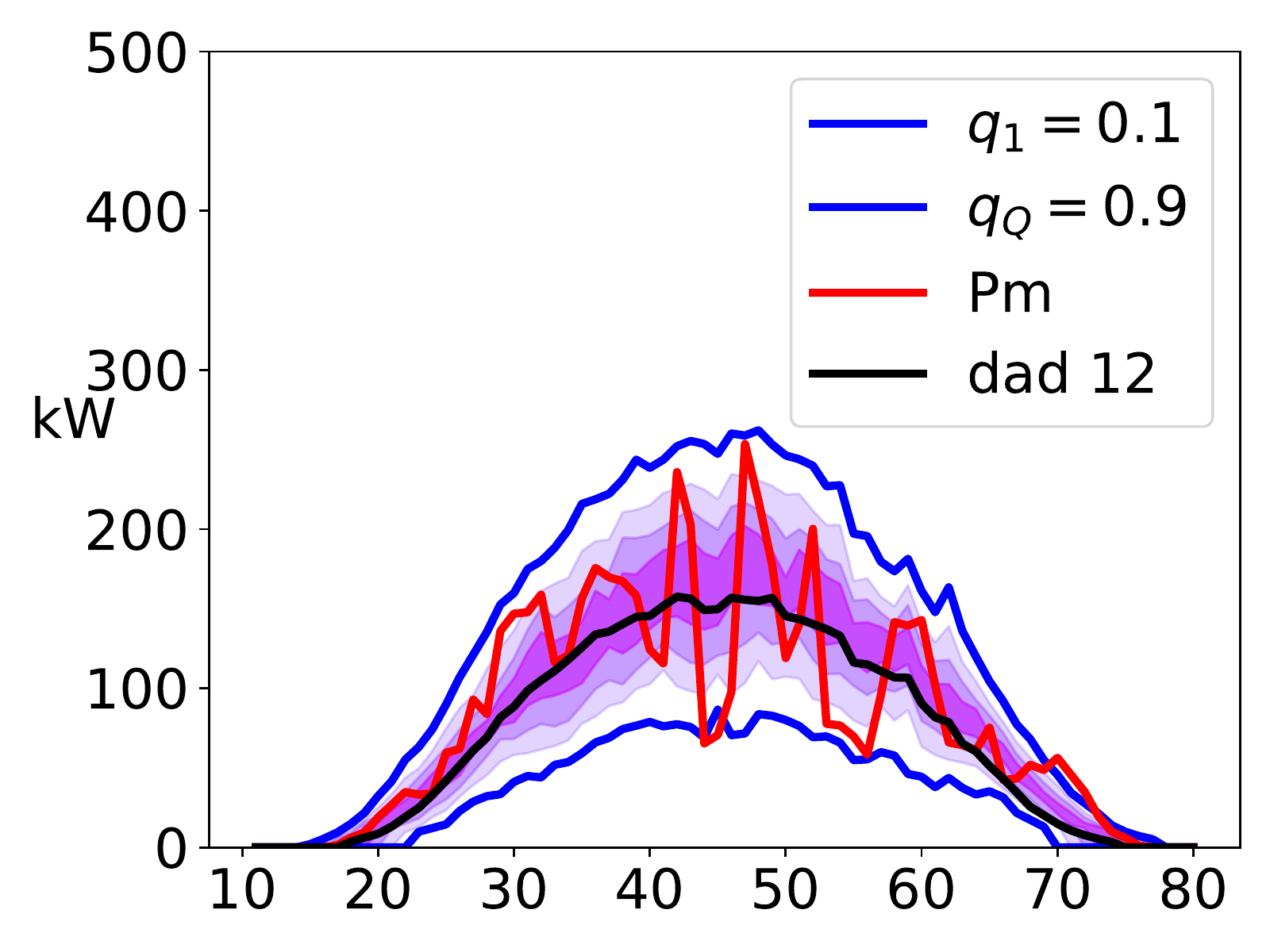}
		\caption{LSTM day-ahead.}
		\label{fig:forecasts_plot_LSTM_dad}
	\end{subfigure}%
	\begin{subfigure}{.45\textwidth}
	\centering
	\includegraphics[width=\linewidth]{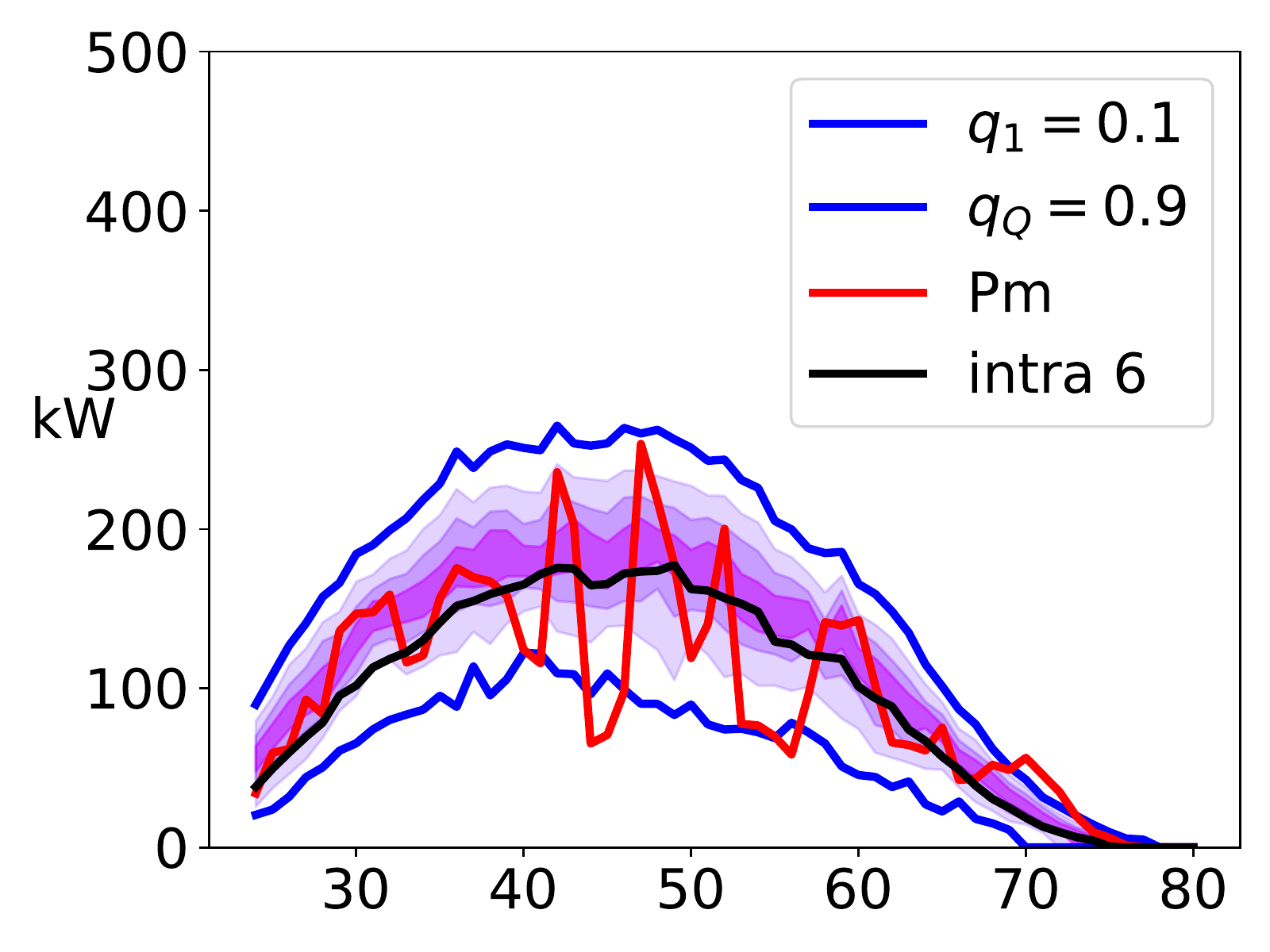}
	\caption{LSTM intraday.}
	\label{fig:forecasts_plot_LSTM_intra}
\end{subfigure}
	\begin{subfigure}{.45\textwidth}
		\centering
		\includegraphics[width=\linewidth]{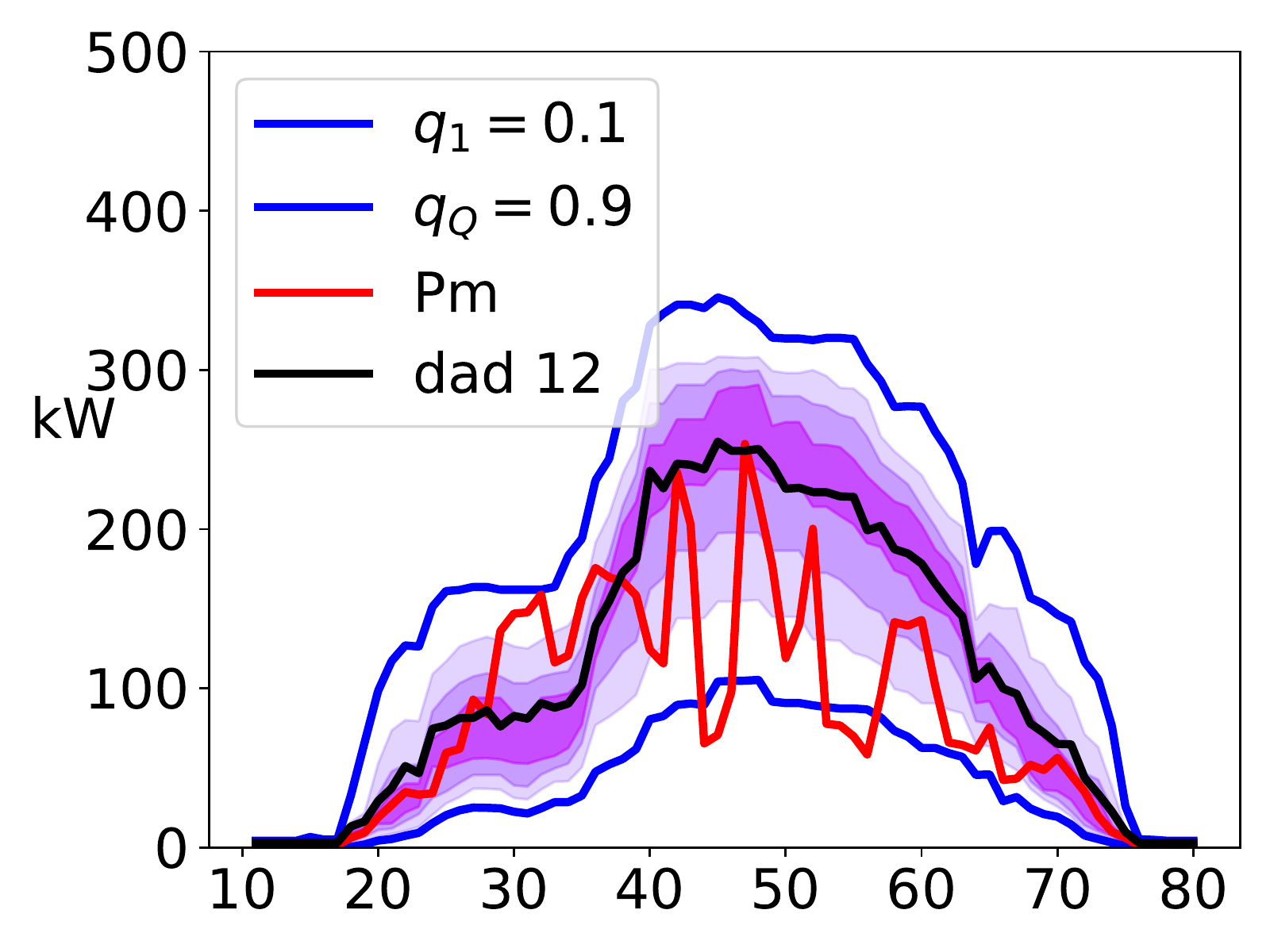}
		\caption{GBR day-ahead.}
		\label{fig:forecasts_plot_GBR_dad}
	\end{subfigure}%
	\begin{subfigure}{.45\textwidth}
		\centering
		\includegraphics[width=\linewidth]{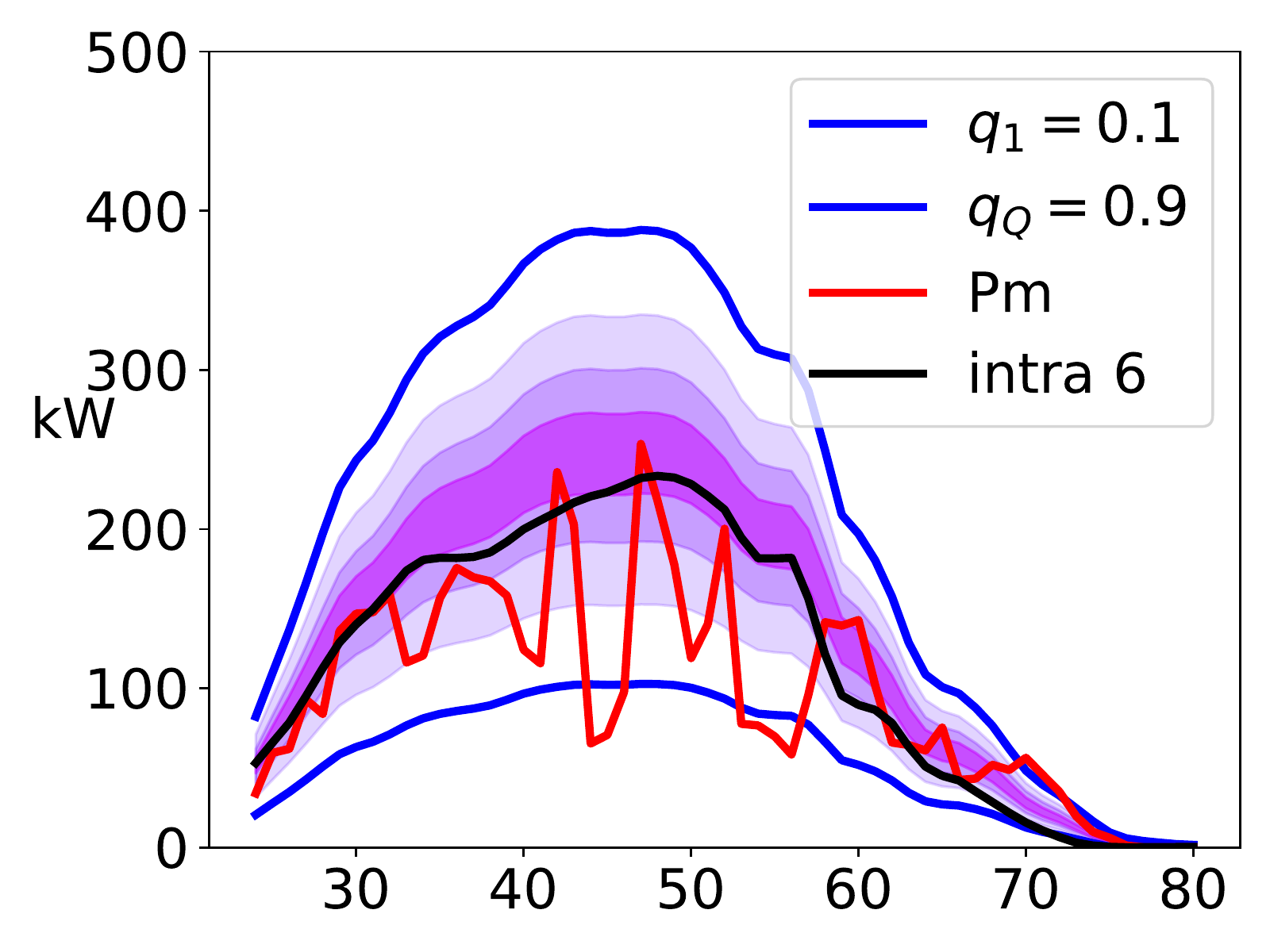}
		\caption{ED-2 intraday.}
		\label{fig:forecasts_plot_ED2_intra}
	\end{subfigure}
	\caption{Quantiles \textit{vs}. point forecasts of day-ahead models of gate 12:00 (left), and intraday models of gate 06:00 (right) on $\quantiledate$, the observations are in red.}
	\label{fig:forecasts_plot}
\end{figure}

Table \ref{tab:intra_scores} provides the averaged NMAE, NRMSE, and CRPS per gate of intraday models. The LSTM achieved the best NMAE and NRMSE for the 06:00 gate, and the ED-1 achieved the best NMAE and NRMSE for the noon gate and the best CRPS for both gates. Figure \ref{fig:intra_CRPS} compares the CRPS per lead time $k$ of the intraday models. The ED-1 benefits from the last PV generation observations. Indeed, some CRPS values for both 06:00 and 12:00 gates are below the ones of 00:00 gate.
Table \ref{tab:intra_WS} provides the interval score of intraday models for 80\%, 60\%, 40\%, and 20\% width of central intervals. The ED-1 model achieved the best results except for the prediction interval width of 80\%, where it is ED-2 and LSTM for the 06:00 and 12:00 gates, respectively. Overall, The LSTM achieved close results to the ED-1. 
Figures \ref{fig:forecasts_plot_ED1_intra}, \ref{fig:forecasts_plot_LSTM_intra}, and \ref{fig:forecasts_plot_ED2_intra} compare the ED-1, LSTM, and ED-2 intraday quantile and point forecasts (black line named intra 6) of 06:00 gate on $\quantiledate$ with the observation in red. Generally, one can see that the predicted intervals of ED-1 and LSTM models better encompass the actual realizations of uncertainties than ED-2.
\begin{table}[!htb]
\renewcommand{\arraystretch}{1.25}
	\begin{center}
		\begin{tabular}{llrrrr}
			\hline  \hline
			Score & Gate &  MLP & ED-1 & ED-2 & LSTM\\
			\hline
\multirow{2}{*}{NMAE} & 6  & 8.9  (1.0)  & 8.5  (1.4)  &9.4  (1.0 ) & \textbf{7.6}  (1.5) \\
& 12 & 6.7  (1.4) & \textbf{6.4}  (1.3 ) & 7.1  (1.1)  & 7.2  (1.1)\\ \hline 

\multirow{2}{*}{NRMSE} &6  & 10.9  (0.9) & 10.3   (1.3)   & 11.3  (1.1)  & \textbf{7.7}  (1.6) \\
&12 & 8.7   (1.3) & \textbf{7.8}    (1.2)   & 8.5   (1.2)  & 9.4  (1.8)\\ \hline 
			
\multirow{2}{*}{CRPS} &6  & 8.1  (0.7) & \textbf{5.9}  (0.9) & 6.6  (0.7) & 6.2  (0.7) \\
&12 & 5.8  (1.2) & \textbf{4.5}  (0.7) & 5.6  (1.8) & 4.7  (0.5) \\ \hline \hline
		\end{tabular}
\caption{The NMAE, NRMSE, CRPS of intraday models are averaged over all periods. The best-performing model for each score and gate is written in bold.}
		\label{tab:intra_scores}
	\end{center}
\end{table}
\begin{table}[!htb]
\renewcommand{\arraystretch}{1.25}
	\begin{center}
		\begin{tabular}{llrrrr}
			\hline  \hline
Width & Gate & MLP & ED-1 & ED-2 & LSTM \\
			\hline
\multirow{2}{*}{80\%} & 6  & 24.4  (2.9) &  14.9  (4.0) & \textbf{13.9}  (4.9) & 19.3 (4.2)\\
& 12 & 17.4  (3.5) &  10.6  (1.8) & 11.6  (10.1) & \textbf{9.6} (2.0)\\ \hline  
\multirow{2}{*}{60\%} & 6  & 37.6  (3.2) &  \textbf{29.9}  (5.0) & 32.2  (4.2) & 30.7 (4.6)\\
& 12 & 27.2  (4.3) &  \textbf{22.4}  (4.2) & 27.5  (10.8)&  22.6 (3.1)\\ \hline  
\multirow{2}{*}{40\%} & 6  & 58.0  (4.5) &  \textbf{50.1}  (6.5) & 57.2  (6.0) & 51.6 (5.8)\\
& 12 & 42.4  (6.9)  & \textbf{37.7}  (5.9) & 48.1  (16.8) & 39.2 (4.9)\\ \hline  
\multirow{2}{*}{20\%} & 6  & 111.8  (8.4) & \textbf{97.1}  (11.7) & 112.1  (10.3)& 99.5 (10.4)\\
& 12 & 81.5  (13.8) &  \textbf{72.7}  (10.0) & 94.8  (32.1) & 76.5 (8.0)\\ \hline  \hline
		\end{tabular}
\caption{The interval score of intraday models is averaged over all periods. The best-performing model for each score and gate is written in bold.}
		\label{tab:intra_WS}
	\end{center}
\end{table}

\clearpage
\subsection{Conclusions}

An encoder-decoder architecture is implemented on the intraday scale to produce accurate forecasts. It efficiently captures the contextual information composed of past PV observations and future weather forecasts while capturing the temporal dependency between forecasting periods over the entire forecasting horizon.  The models are compared by using a $k$-fold cross-validation methodology and quality metrics on a real case study composed of the PV generation of the parking rooftops of the Li\`ege University.
The best day-ahead model for point and quantile forecasts is a neural network composed of a LSTM cell and an additional feed-forward layer. 
Then, the encoder-architecture composed of a LSTM-MLP yields accurate and calibrated forecast distributions learned from the historical dataset compared to the MLP and LSTM-LSTM models for the intraday point and quantile forecasts. However, the LSTM produced similar results. \\

\noindent Several extensions are under investigation. First, using a larger dataset of at least one full year to consider the entire PV seasonality. Second, developing a PV scenario approach based on the encoder-decoder architecture. 

\section{Comparison with generative models}\label{sec:forecasting-quantile-comparison}

This Section proposes a quality evaluation of the normalizing flows (NFs) and LSTM PV quantiles, used in Chapter \ref{chap:capacity-firming-robust} for robust optimization, using the quantile score, the reliability diagram, and the continuous ranked probability score. Indeed, the two-phase engagement control of the capacity firming framework requires day-ahead and intraday \textit{top-quality} forecasts. The more accurate the forecasts, the better the planning and the control. 
To this end, the Normalizing Flows technique computes quantile day-ahead forecasts compared to a common alternative technique using a Long Short-Term Memory neural network. The controller requires intraday point forecasts computed by an encoder-decoder architecture. NFs are investigated in Section \ref{sec:ijf-background} of Chapter \ref{chap:scenarios-forecasting}.
Both NFs and LSTM models use as input the weather forecasts of the MAR climate regional model provided by the Laboratory of Climatology of the Li\`ege University \citep{fettweis2017reconstructions}. The NFs model generates day-ahead scenarios, and the quantiles are derived. The LSTM model computes the quantiles directly and is trained by minimizing the quantile loss. The set of PV quantiles considered for the assessment is $ \{q=10\%, \ldots, 90\%\}$.

In this study, the class of Affine Autoregressive flows is implemented\footnote{\url{https://github.com/AWehenkel/Normalizing-Flows} \citep{wehenkel2019unconstrained}}. A five-step Affine Autoregressive flow is trained by maximum likelihood estimation with 500 epochs and a learning rate of $10^{-4}$. 
The LSTM learning rate is $10^{-3}$, the number of epoch to 500 with a batch size of 64. 
\begin{figure}[htbp]
	\centering
	\begin{subfigure}{.45\textwidth}
		\centering
		\includegraphics[width=\linewidth]{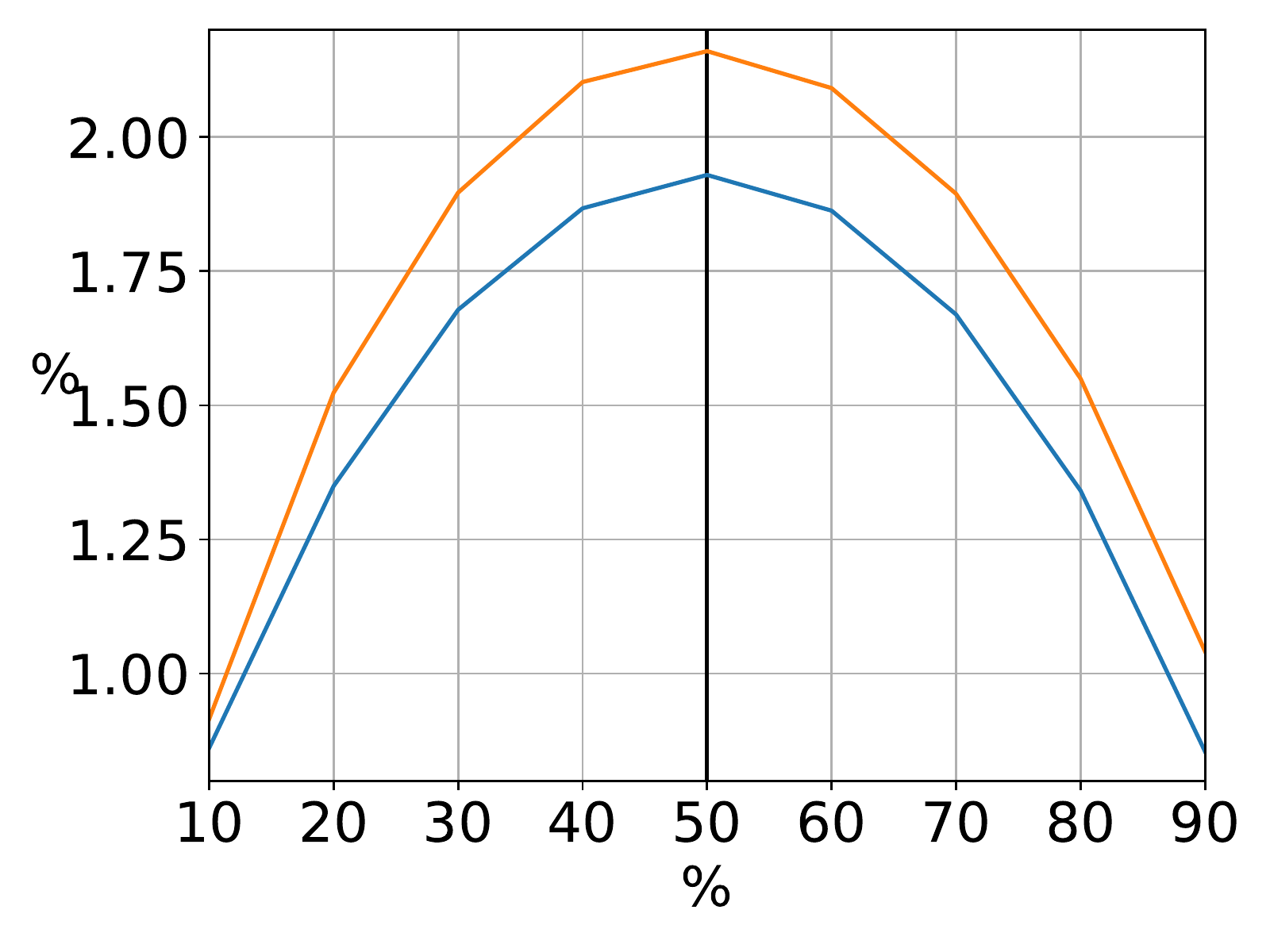}
		\caption{QS$(q)$.}
	\end{subfigure}%
	\begin{subfigure}{.45\textwidth}
		\centering
		\includegraphics[width=\linewidth]{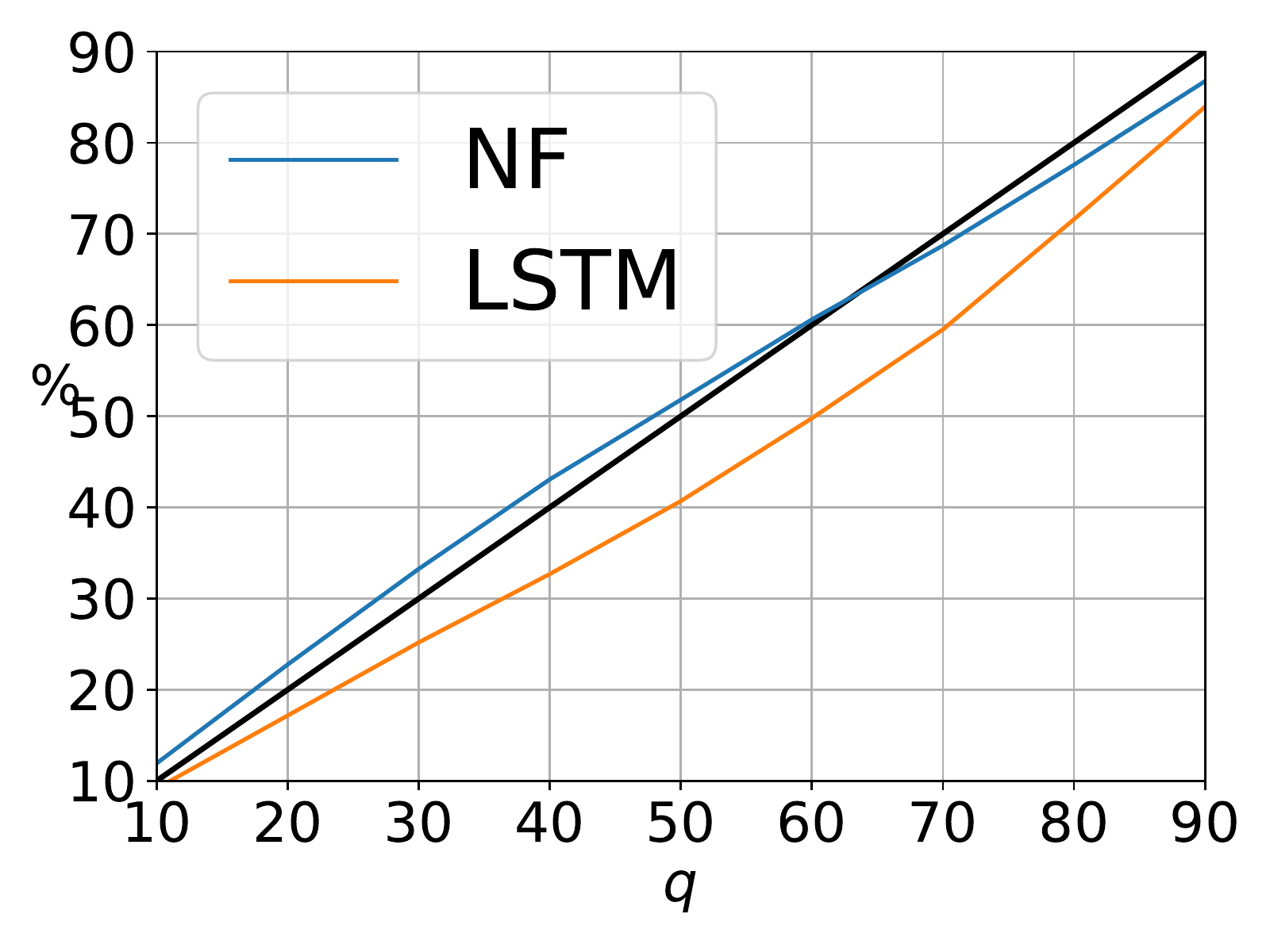}
		\caption{Reliability diagram.}
	\end{subfigure}	
	\begin{subfigure}{.45\textwidth}
		\centering
		\includegraphics[width=\linewidth]{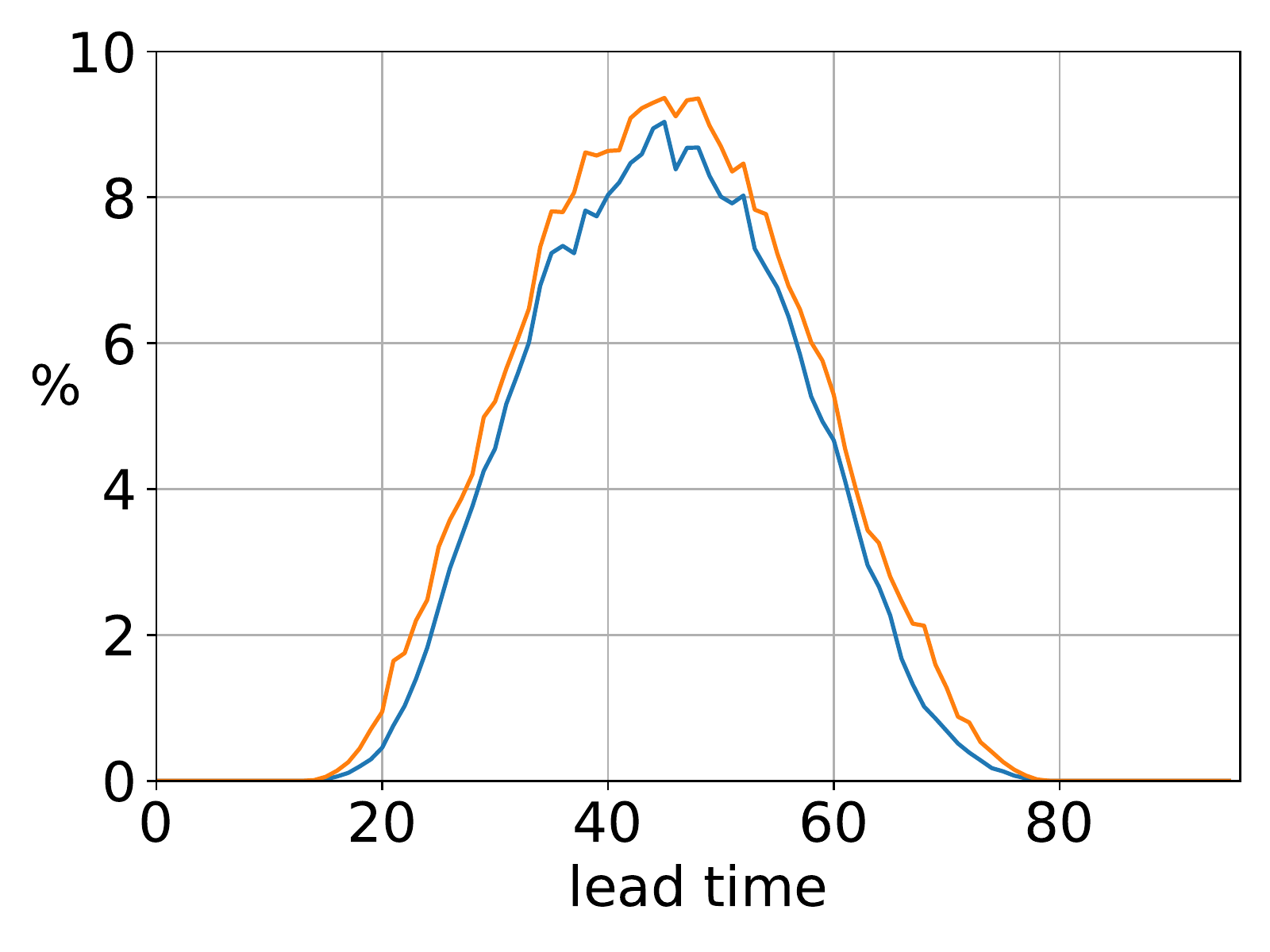}
		\caption{CRPS$(k)$.}
	\end{subfigure}%
	\caption{Quantile forecast quality evaluation of LSTM \textit{v.s} NFs models.}
	\label{fig:forecast_eval}
\end{figure}
Figure \ref{fig:forecast_eval} provides the results for these quality metrics computed over the entire dataset normalized by the total installed capacity. The NFs model outperforms the LSTM model with average values of 1.49\% and 2.80\% \textit{vs.} 1.69\% and 3.15\% for the QS and CRPS, respectively. The NFs quantiles are also more reliable, as indicated by the reliability diagram. These results motivate the use of the NFs as they outperform common deep learning approaches such as LSTM models. 

\clearpage
\section{Conclusions}\label{sec:forecasting-quantile-conclusions}

This Chapter proposes a formulation of the quantile forecasts problem using deep learning models trained by quantile regression to compute multi-output PV quantiles. The forecast quality is evaluated on a real case study composed of the PV generation of the parking rooftops of the Li\`ege University. 
In addition, these quantile regression models are compared to PV quantiles derived from deep learning generative models, which will be investigated in detail in Chapter \ref{chap:scenarios-forecasting}. In terms of forecast quality, the generative models outperform on this case study the quantile regression models. However, it does not mean they are better in terms of forecast value which will be assessed in Chapter \ref{chap:capacity-firming-robust} where a robust planner uses the PV quantiles in the form of prediction intervals.


\chapter{Density forecasting of imbalance prices}\label{chap:density-forecasting}

\begin{infobox}{Overview}
The contributions of this Chapter are two-fold.
\begin{enumerate}
    \item A novel two-step probabilistic approach (TSPA) is proposed for forecasting the Belgium imbalance prices. The TSPA uses a direct forecasting strategy \citep{taieb2012review}. It consists of forecasting an imbalance price for each quarter of the horizon independently from the others, requiring a model per quarter.
    \item It sets a reference for other studies as this subject is rarely addressed.
\end{enumerate}

\textbf{\textcolor{RoyalBlue}{References:}} This chapter  is an adapted version of the following publication: \\[2mm]\bibentry{dumas2019probabilistic}. 
%
\end{infobox}
\epi{If you don't know where you're going any road will do.}{Lewis Carroll}
\begin{figure}[htbp]
	\centering
	\includegraphics[width=1\linewidth]{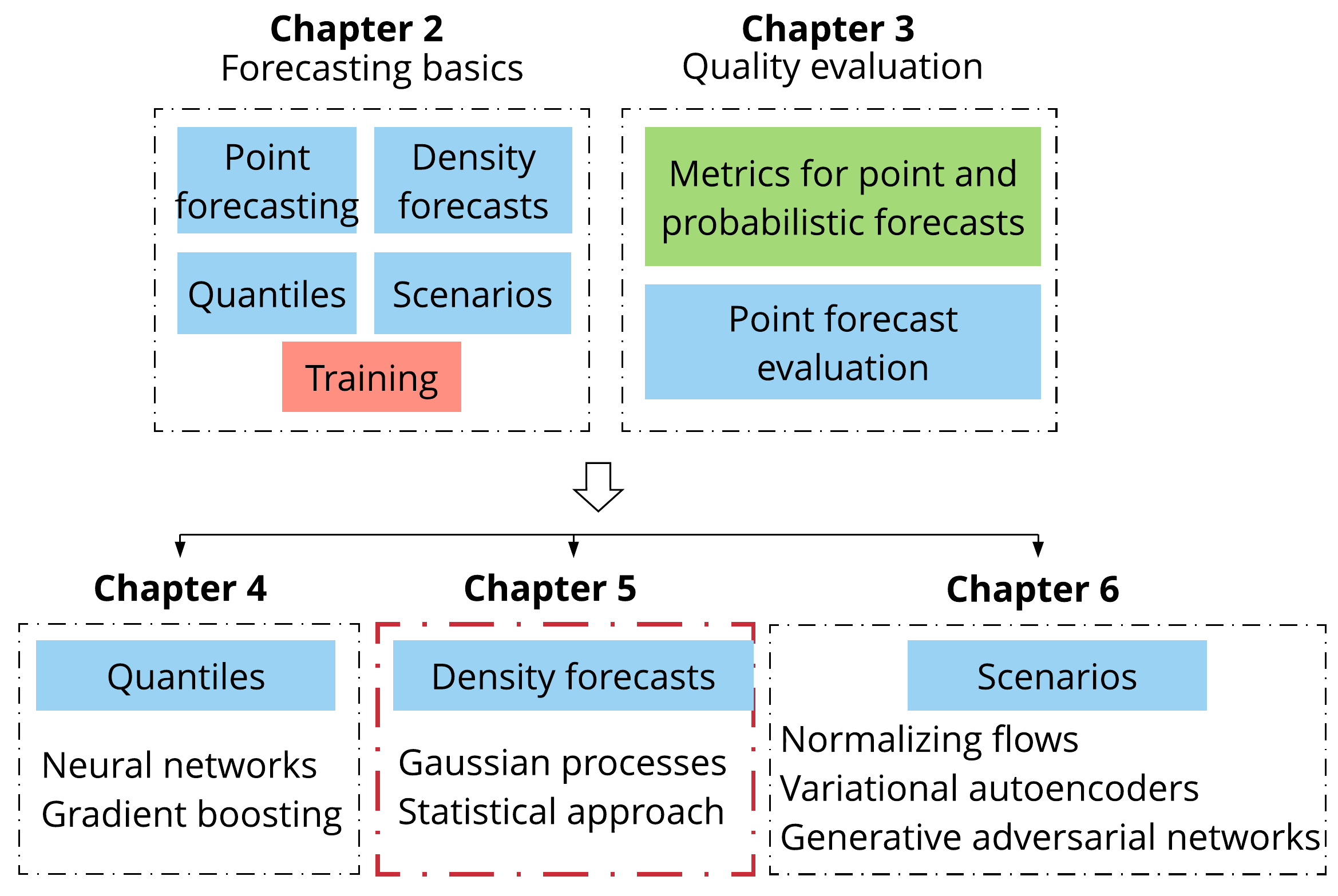}
	\caption{Chapter \ref{chap:density-forecasting} position in Part \ref{part:forecasting}.}
\end{figure}
\clearpage

This Chapter presents the probabilistic forecasting of imbalance prices methodology developed in \citet{dumas2019probabilistic}. A novel \textit{two-step probabilistic approach} is proposed, with a particular focus on the Belgian case. The first step consists of computing the net regulation volume (NRV) \textit{state transition probabilities}. It is modeled as a matrix estimated using historical data. This matrix is then used to infer the imbalance prices. Indeed, the NRV can be related to the level of reserves activated, and the corresponding marginal prices for each activation level are published by the Belgian Transmission System Operator (TSO) one day before electricity delivery. The model is compared to: (1) a multi-layer perceptron, implemented in a deterministic setting; (2) the widely used probabilistic technique, the Gaussian Processes (GP). 

This Chapter is organized as follows. Section \ref{sec:forecasting-eem-related-work} details the related work. Section \ref{sec:forecasting-eem-formulation} introduces the novel two-step probabilistic approach formulation and the assumptions made. Section \ref{sec:forecasting-eem-case-study} describes the numerical tests on the Belgian case. Section \ref{sec:forecasting-eem-results} reports the results. Conclusions are drawn in Section \ref{sec:forecasting-eem-conclusions}.  Appendix \ref{appendix:forecasting-eem-notation} lists the acronyms, parameters, and forecasted or computed variables. Finally, Annex \ref{appendix:forecasting-eem} provides a short reminder of the imbalance market and the Belgian balancing mechanisms.

\section{Related work}\label{sec:forecasting-eem-related-work}

The progressive, large-scale integration of renewable energy sources has altered electricity market behavior and increased the electricity price volatility over the last few years \citep{de2015negative, green2010market, ketterer2014impact}. 
In this context, imbalance price forecasting is an essential tool the strategic participation in short-term energy markets. Several studies take into account the imbalance prices as penalties for deviation from the bids to compute the optimal bidding strategy \citep{giannitrapani2016bidding, pinson2007trading, bitar2012bringing, boomsma2014bidding}. However, these penalties are known only a \textit{posteriori}. A forecast indicating the imbalance prices and the system position, short or long, with an interval to inform about the range of potential outcomes is a powerful tool for decision making. Probabilistic forecasting usually outperforms deterministic models when used with the appropriate bidding strategies \citep{pinson2007trading}.
Whereas the literature on day-ahead electricity forecast models is extensive, studies about balancing market prices forecast have received less attention. We recommend three papers related to this topic. First, a statistical description of imbalance prices for shortage and surplus is presented by \citet{saint2002wind}. Second, a combination of classical and data mining techniques to forecast the system imbalance volume is addressed by \citet{garcia2006forecasting}.  Finally, a review and benchmark of time series-based methods for balancing market price forecasting are proposed by \citet{klaeboe2015benchmarking}. One-hour and one-day-ahead forecasts are considered for state determination, balancing volume, and price forecasting on the Nord Pool price zone NO2 in Norway.

\section{Formulation}\label{sec:forecasting-eem-formulation}

This study focuses on the intraday market time scale that requires a forecasting horizon from a few minutes to a few hours with a resolution $\Delta t$. The day-ahead time scale requires forecasts of the imbalance prices from 12 to 36 hours, which is not realistic at this stage. The input data are the imbalance price history, the NRV, and the marginal prices for activation published by the TSO.
The probabilistic approach consists of forecasting the imbalance prices in two steps: computing the NRV state transition probabilities, then forecasting the imbalance prices, as depicted in Figure \ref{fig:eem:forecasting_process}. It is motivated by the ELIA imbalance price mechanisms detailed in Appendix \ref{appendix:forecasting-eem}.
\begin{figure}[tb]
	\centering
	\includegraphics[width=0.5\linewidth]{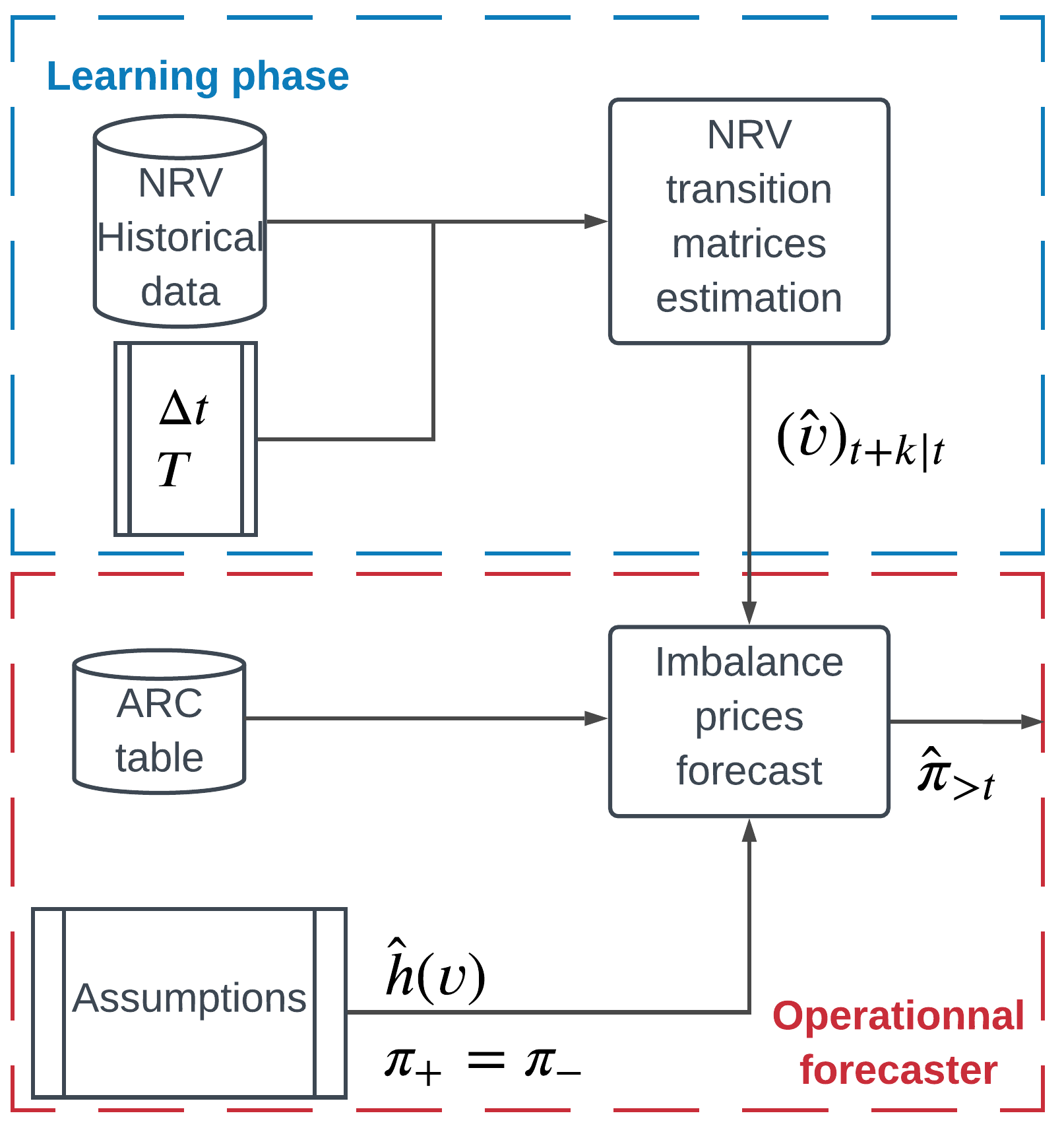}
	\captionsetup{justification=centering}
	\caption{Two-step probabilistic approach (TSPA) imbalance price forecasting process.}
	\label{fig:eem:forecasting_process}
\end{figure}

\subsection{Net regulation volume forecasting}

\begin{figure}[tb]
	\centering
	\includegraphics[width=0.4\linewidth]{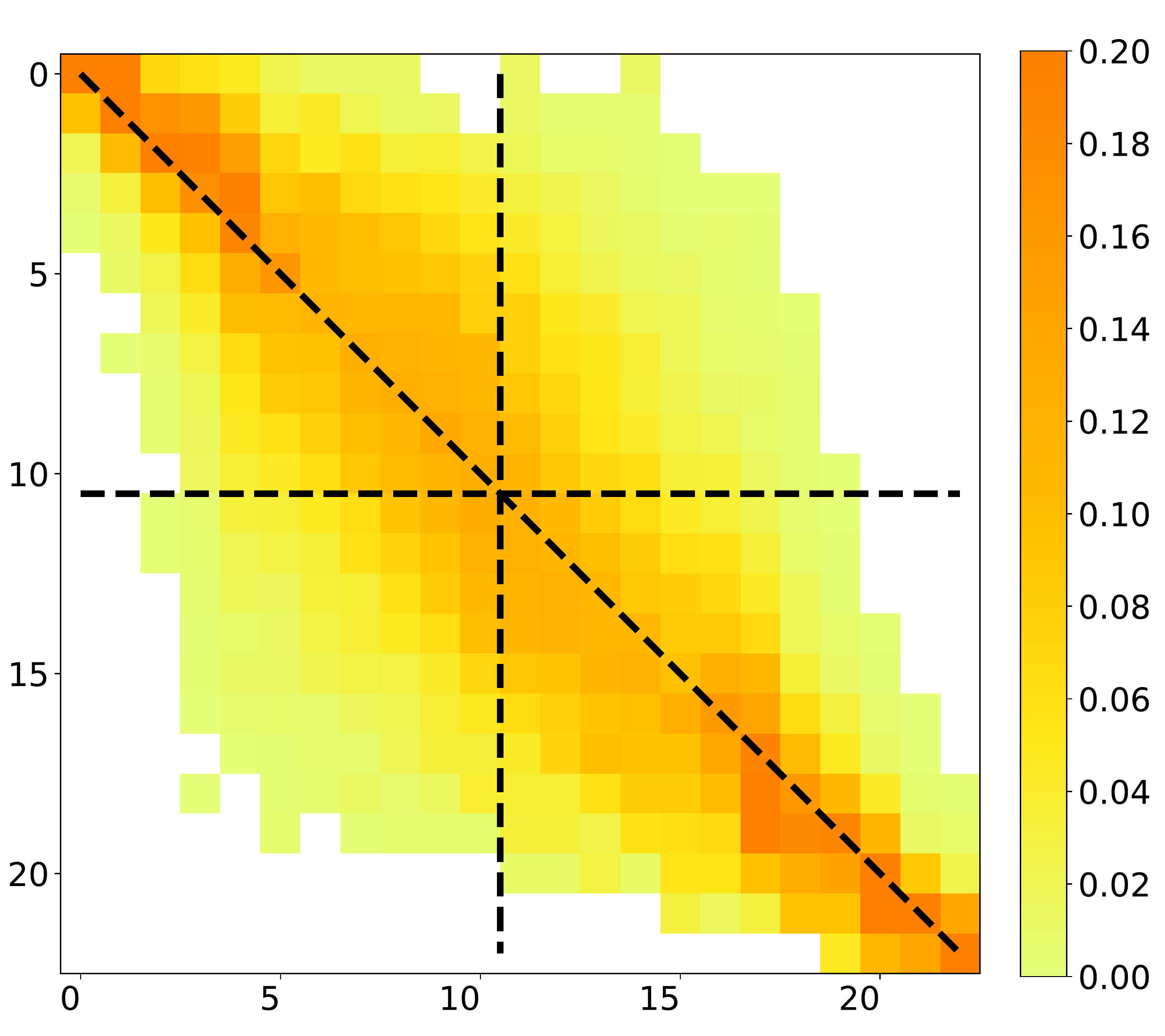}%
	\includegraphics[width=0.4\linewidth]{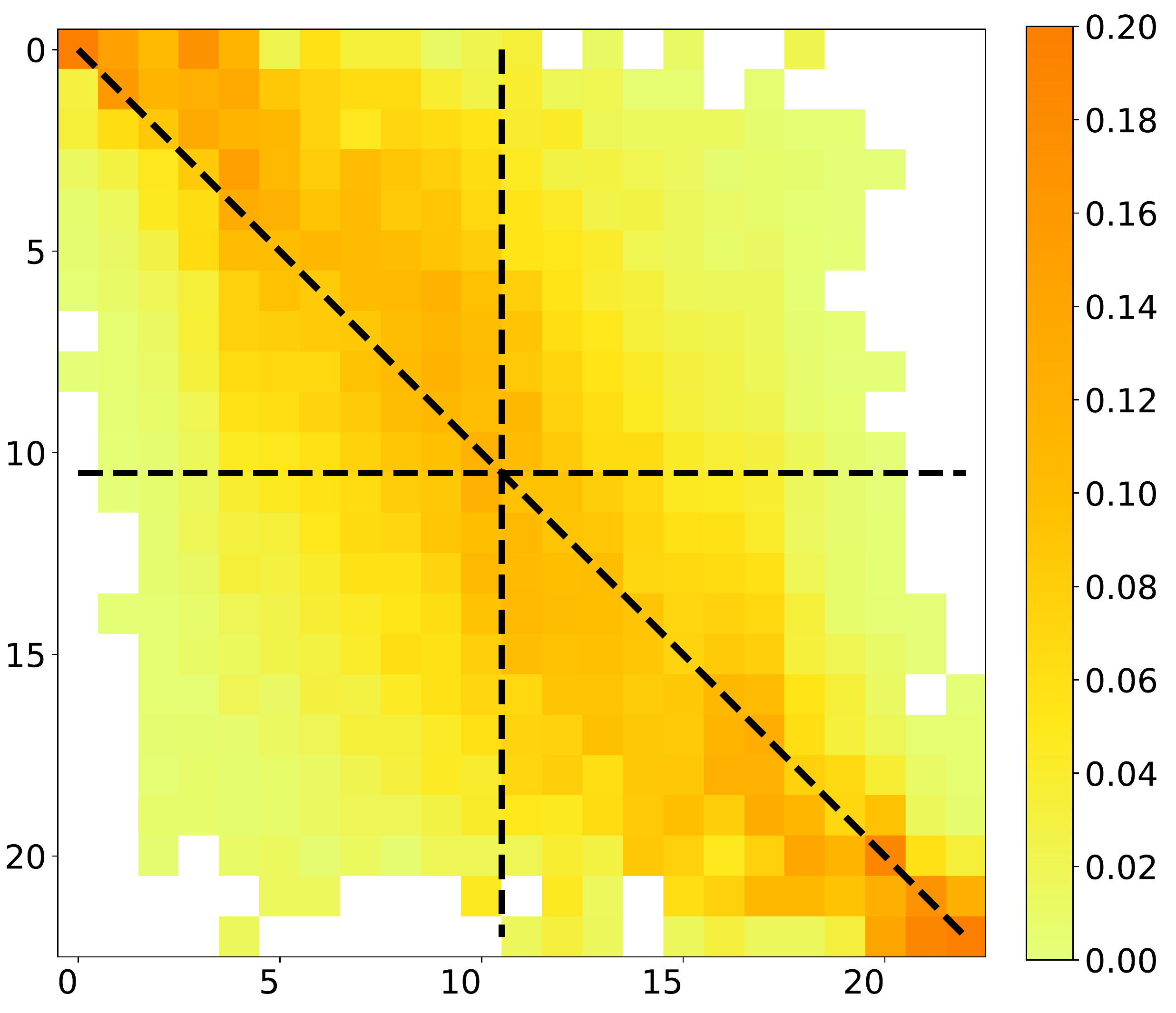}
	\captionsetup{justification=centering}
	\caption{NRV transition matrix from $t$ to $t+15$ min (left) and $t+60$ min (right).}
	\label{fig:imbalance-transition_matrix_2017-12_60_heatmap}
\end{figure}

Let consider the $T$ forecasting horizons $k_1=\Delta t, \ldots, k_T = T \Delta t$ with $\Delta t$ the market period, 15 minutes for Belgium. The NRV historical data is discretized into $N$ bins, $\NRV_i$, centered around $\NRV_{i,1/2}$. Note: this discretization has been determined after a statistical study of the NRV distribution.
The $T$ NRV transition matrices $(\NRV)_{t+k|t}$, of dimensions $N \times N$, from a known state at time $t$ to a future state at time $t+k$ are estimated by using the NRV historical data, and referred to as $(\hat{\NRV})_{t+k|t}$. They are composed of the following conditional probabilities $\forall k=k_1, \ldots, k_T$
\begin{equation}
\label{eq:pij_def} 
p^{ij}_{t+k|t}= \Pr[\NRV(t+k) \in \NRV_j \mid \NRV(t) \in \NRV_i ], \quad  i,j \in {\llbracket 1; N \rrbracket}^2,
\end{equation}
with $\NRV(t)$ the measured NRV at time $t$, and $\sum_{j=1}^N p^{ij}_{t+k|t} = 1$ $\forall i \in  \llbracket 1; N \rrbracket$. The conditional probabilities (\ref{eq:pij_def}) are estimated statistically over the learning set (LS) $\forall k=k_1, \ldots, k_T$
\begin{equation}
\label{eq:pij_estimated_def} 
\hat{p}^{ij}_{t+k|t} = \frac{\sum_{t\in LS } 1_{\{ \NRV(t+k) \in \NRV_j \ | \ \NRV(t) \in \NRV_i\}}}{\sum_{t\in LS } 1_{\{ \NRV(t) \in \NRV_i\}}}, \quad  i,j \in {\llbracket 1; N \rrbracket}^2.
\end{equation}
Figure~\ref{fig:imbalance-transition_matrix_2017-12_60_heatmap} illustrates the matrices $(\hat{\NRV})_{t+k_1|t}$ and $(\hat{\NRV})_{t+k_4|t}$ with 2017 as learning set.
The estimated mean $\hat{\NRV}^m_{t+k|t}$ and standard deviation $\hat{\NRV}^{std}_{t+k|t}$ of the NRV at time $t$ for $t+k$ are calculated as follows 
\begin{equation}
\begin{aligned}
\hat{\NRV}^m_{t+k|t} & = \sum_{j=1}^N \hat{p}^{ij}_{t+k|t}  \NRV_{j,1/2} \\
\hat{\NRV}^{std}_{t+k|t} &= \sqrt{\sum_{j=1}^N \hat{p}^{ij}_{t+k|t} (\NRV_{j,1/2}  - \hat{\NRV}^m_{t+k|t})^2},
\end{aligned}
\end{equation}
with $i$ such as $\NRV(t) \in \NRV_i$.

\subsection{Imbalance price forecasting}

The NRV can be related to the level of reserves activated and the corresponding marginal prices for each activation level, published by the TSO one day before electricity delivery. We thus first forecast the NRV and its spread among the Gross Upward regulation Volume (GUV) and Gross Downward regulation Volume (GDV). Then, we forecast the reserve products activated (contracted or not) to select the most probable MIP and MDP into the ARC table. Finally, the mean and the standard deviation of the imbalance price forecast are derived.

However, the ARC table contains only the contracted reserve products. Most of the time, the first activated reserve products come from the non contracted International Grid Control Cooperation platform (IGCC-/+), the contracted secondary reserve (R2-/+) and the non contracted regulation reserves (Bids-/+)\footnote{More information about the reserve products is available at \url{http://www.elia.be}.}.
For instance, consider a quarter of an hour with an NRV of 150 MW, spread into 170 MW of GUV and 20 MW of GDV. Suppose ELIA activated 80 MW of IGCC+ and 90 MW of R2+. Then, the MIP is given in the marginal activation price of R2+ in the ARC table at the range $[0,100]$ MW. Suppose that ELIA has activated 20 MW of IGCC+, 20 MW of R2 +, and 130 MW of Bids+. Then, the MIP is given in the marginal activation price of Bids+. However, this is not a contracted reserve, and its price is not in the ARC table. Then, it is more complicated to predict the MIP and consequently the imbalance prices. Therefore, we introduce several simplifying assumptions justified by a statistical study on the 2017 ELIA imbalance data.
\begin{assumption} The NRV is entirely spread into the GUV if the NRV is positive or GDV if the NRV is negative.
\end{assumption} 

\noindent The mean and standard deviation of the GUV and GDV are $109 \pm 82$ MW \textit{vs.} $17 \pm 27$ MW when the NRV is positive, while it is $13 \pm 20$ MW \textit{vs.} $110 \pm 73$ MW when the NRV is negative. This assumption enables to select directly in the ARC table the marginal price for activation corresponding to the range of activation equal to the NRV, minus IGCC.
\begin{assumption} We do not consider the Bids reserve product. Thus, we assume the NRV spreads over the IGCC and reserve products of the ARC table.
\end{assumption}

\noindent The percentage of Bids reserve product, positive or negative, activated over each quarter of 2017 is 11.5\%.

\begin{assumption} The level of activated IGCC reserve product is modeled by a function $\hat{h}$ of the NRV. 
\end{assumption}

\noindent $\hat{h}$ assigns for a given value of NRV a range of activation $p$ into the ARC table. $c_t^p$ is the ARC marginal price at $t$ and for the activation range $p$, with $p \in \llbracket 1; P \rrbracket $. If $\hat{h}(\NRV)$ falls into the activation range $p$, then $c_t^p(\hat{h}(\NRV))$ is equal to $c_t^p$. Due to the 2017 statistical distribution of the IGCC versus the NRV, $\hat{h}$ is defined as follows
\begin{equation}
\hat{h}(x) = \left\{\begin{array}{lcr}
x  & \textnormal{if} & |x| \leq 100, \\
x -100 & \textnormal{if} & x > 100, \\
x + 100 & \textnormal{if} & x < 100.
\end{array} %
\right. 
\label{eq:IGCCmodelisation}
\end{equation}
The mean and standard deviation (MW) of the IGCC+ and IGCC- are
\begin{equation} \notag
 \left\{\begin{array}{lcr}
17 \pm 25 \ \& \ 23 \pm 24  & \textnormal{if} & |NRV| \leq 100, \\
50 \pm 48 \ \& \ 5 \pm 15 & \textnormal{if} & NRV > 100, \\
2 \pm 10 \ \& \ 67 \pm 47   & \textnormal{if} & NRV < 100.
\end{array} %
\right. 
\end{equation}
Generally, ELIA first tries to activate the IGCC product to balance the system. However, when the system imbalance is too high other reserve products are required. 

\begin{assumption} The positive imbalance price is equal to the negative one. 
\end{assumption}

The positive and negative imbalance means prices are $42.23$ and $43.04$ \euro/MWh. They are different $30.38$\% of the time, but the NMAE and NRMSE are $0.02$ and $0.06$\%. Indeed, the positive and negative prices differ only by a small correction parameter if the system imbalance is greater than 140 MW, cf. Appendix~\ref{appendix:forecasting-eem-balancing-belgium}.

Under these assumptions, the estimated mean $\hat{\pi}^m_{t+k|t}$ and standard deviation $\hat{\pi}^{std}_{t+k|t}$ of the imbalance prices at time $t$ for $t+k$ are calculated as follows
\begin{equation}
\begin{aligned}
\hat{\pi}^m_{t+k|t} & = \sum_{j=1}^N \hat{p}^{ij}_{t+k|t} c_j^{t+k} (\hat{h}(\NRV_{j,1/2})) \\
\hat{\pi}^{std}_{t+k|t} &= \sqrt{\sum_{j=1}^N \hat{p}^{ij}_{t+k|t} (c_j^{t+k}\left(\hat{h}(\NRV_{j,1/2})) -\hat{\pi}^m_{t+k|t} \right) ^2},
\end{aligned}
\end{equation} 
with $i$ such as $\NRV(t) \in \NRV_i$. Finally, on a quarterly basis a forecast is issued at time $t$ and composed of a set of $T$ couples $\hat{\pi}_{>t} := \Big\{ (\hat{\pi}^m_{t+k|t}, \pi^{std}_{t+k|t})  \Big\}_{k=k_1}^{k_T}$.

\section{Case study}\label{sec:forecasting-eem-case-study}

\begin{figure}[tb]
	\centering
	\includegraphics[width=0.8\linewidth]{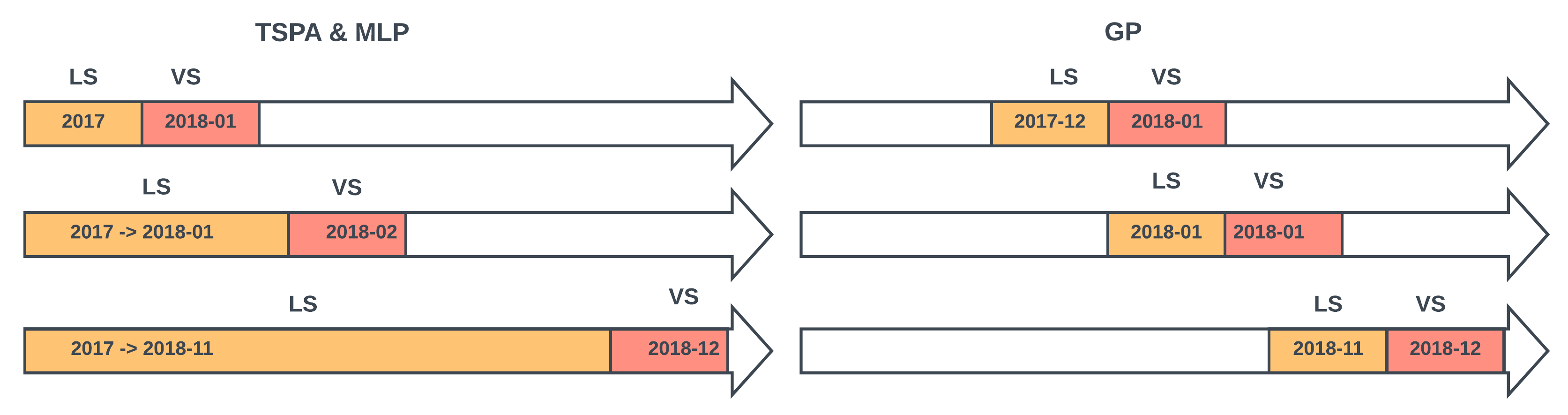}
	\captionsetup{justification=centering}
	\caption{Rolling forecast strategy.}
	\label{fig:eem-rolling-forecast}
\end{figure}

This approach is compared to a widely used probabilistic technique, the Gaussian Processes, and a "classic" deterministic technique, a Multi-Layer Perceptron (MLP). Both techniques are implemented using the Scikit-learn Python library \citep{scikit-learn}. The GP uses Mat\'{e}rn, constant and white noise kernels. The MLP has one hidden layer composed of $2 \times n+1$ neurons with $n$ input features. The dataset comprises the 2017 and 2018 historical Belgium imbalance price and NRV, available on Elia's website. Both the MLP and GP models forecast the imbalance prices based on the previous twenty-four hours of NRV and imbalance prices, representing a total of $2 \times 96$ input features. The MLP uses a Multi-Input Multi-Output (MIMO) strategy, and the GP a Direct strategy \citep{taieb2012review}\footnote{GP regression with multiple outputs is non-trivial and still a field of active research \citep{wang2015gaussian,liu2018remarks}.}. The Direct strategy consists of training a model $\hat{g_k}$ per market period
\begin{equation}
\hat{\pi}_{t+k|t} = \hat{g_k}(\pi_t, \ldots, \pi_{t-k_T}, \NRV_t, \ldots, \NRV_{t-k_T}  ), \quad \forall k=k_1, \ldots, k_T,
\end{equation}
and the forecast is composed of the $T$ predicted values computed by the $T$ models $\hat{g_k}$. In contrast, the MIMO strategy consists of training only one model $\hat{g}$ to directly compute the $T$ values of the variable of interest
\begin{equation}
[\hat{\pi}_{t+k_1|t}, \ldots, \hat{\pi}_{t+k_T|t}]^\intercal = \hat{g} (\pi_t, \ldots, \pi_{t-k_T}, \NRV_t, \ldots, \NRV_{t-k_T}  ).
\end{equation}
For both MIMO and Direct strategies, the forecast is computed quarterly and composed of $T$ values. The forecasting process uses a \textit{rolling forecast strategy} where the training set is updated every month, depicted in Figure \ref{fig:eem-rolling-forecast}. Its size increases by one month for both the MLP and TSPA techniques, with the initial training set being 2017. However, it is limited to the month preceding the forecast for the GP technique to maintain a reasonable computation time. The validation set is 2018, where each month is forecasted by a model trained on a different learning set. 

\section{Results}\label{sec:forecasting-eem-results}

The probabilistic forecasts are evaluated using the Pinball Loss Function (PLF) and the Continuous Ranked Probability Score. They are compared to the deterministic metrics with the Normalized Mean Absolute Error (NMAE), and the Normalized Root Mean Squared Error (NRMSE) of the mean predicted imbalance prices. The scores NMAE$(k)$, NRMSE$(k)$, PLF$(k)$, and CRPS$(k)$ for a lead time $k$ are computed over the entire validation set. The normalizing coefficient for both the NMAE and NRMSE is 55.02 \texteuro $/MWh$, the mean absolute value of the imbalance prices over 2018.

\begin{figure}[tb]
\centering
	\begin{subfigure}{0.5\textwidth}
	\centering
	\includegraphics[width=\linewidth]{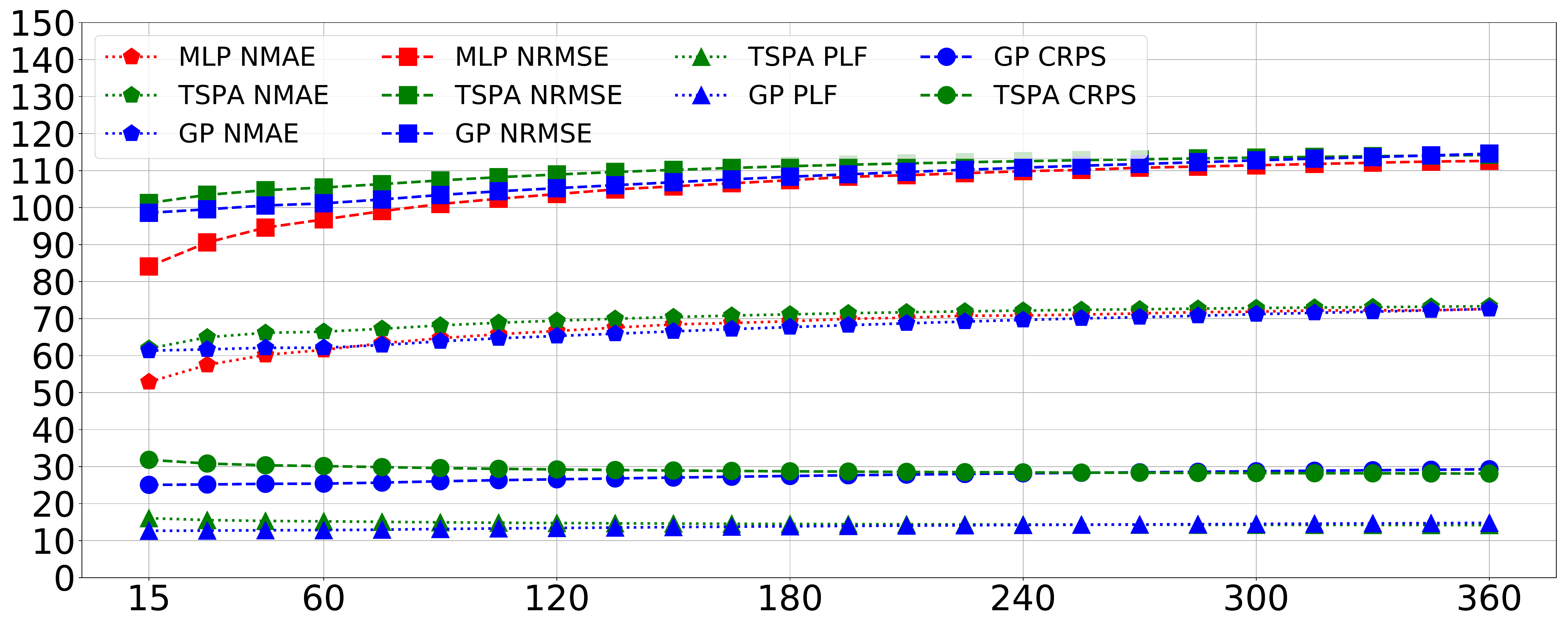}
	\captionsetup{justification=centering}
	\caption{Average scores over all lead times $k$ for each forecasting horizon.}
	\label{fig:scores_per_horizon_M2_GP_MLP_360}
	\end{subfigure}%
	\begin{subfigure}{0.5\textwidth}
	\centering
	\includegraphics[width=\linewidth]{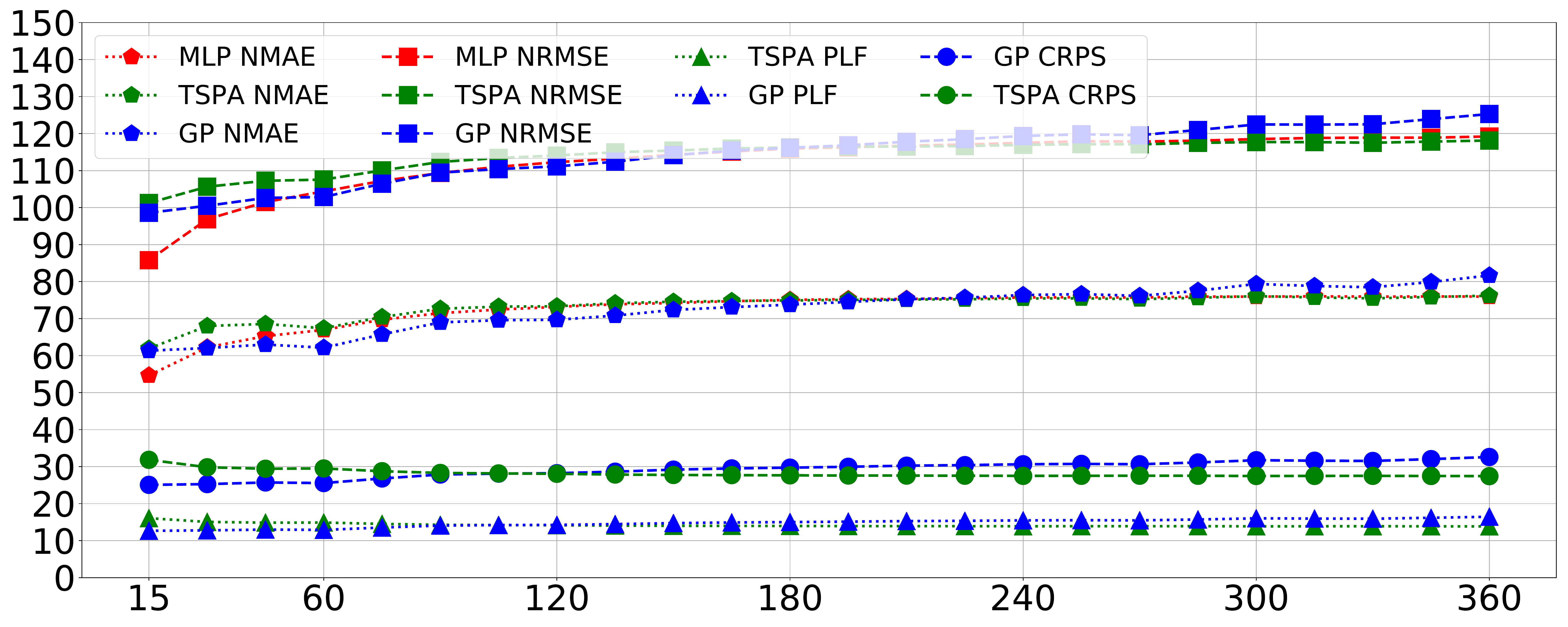}
	\captionsetup{justification=centering}
	\caption{Score per lead time $k$ for the forecasting horizon of 360 minutes.}
	\label{fig:scores_M2_GP_MLP_360}
	\end{subfigure}
	\caption{Scores.}
	\label{fig:EEM_2019_scores}
\end{figure}

The forecaster computes the mean and standard deviation of the imbalance prices per market period. Then, a Gaussian distribution generates samples, and the percentiles $1, \ldots, 99$ are derived. They are used to compute the PLF. The CRPS is computed using the \textit{crps\_gaussian} function of the Python properscoring\footnote{\url{https://pypi.org/project/properscoring/}} library.
Table \ref{tab:meanscore} presents the average scores over all lead times $k$ for the horizons of 15, 60 and 360 minutes. Figure \ref{fig:scores_per_horizon_M2_GP_MLP_360} provides the average scores over all lead times $k$ for each forecasting horizon, and Figure \ref{fig:scores_M2_GP_MLP_360} depicts the score per lead time $k$ for the forecasting horizon of 360 minutes.
\begin{table}[tb]
	\begin{center}
	\renewcommand\arraystretch{1.25}
		\begin{tabular}[b]{l l r r r r}
			\hline \hline
			$k$  & Technique & NMAE & NRMSE & PLF & CRPS  \\ \hline
			\multirow{3}{*}{15 min} & MLP  & \textbf{52.74}  & \textbf{84.37}  & -     & -       \\
			 & GP   & 61.33 & 98.59  & 16.48 & 32.64   \\
			 & TSPA & 61.91 & 101.24 & \textbf{16.07} & \textbf{31.84}   \\
			\hline
			\multirow{3}{*}{60 min} & MLP  & \textbf{61.85} & \textbf{97.26}  & -     & -       \\
			 & GP    & 62.13 & 101.14 & 16.09 & 31.87   \\
			 & TSPA  & 66.47 & 105.43 & \textbf{15.22} & \textbf{30.15}   \\
			\hline
			\multirow{3}{*}{360 min} & MLP  & 72.64 & \textbf{112.90} & -     & -       \\
		     & GP   & \textbf{72.61} & 114.56 & 14.79 & 29.29   \\
			 & TSPA & 73.35 & 114.2 & \textbf{14.2} & \textbf{28.12}   \\ \hline \hline
		\end{tabular}
			\caption{Average scores over all lead times $k$. The best-performing model for each lead time is written in bold.}
		\label{tab:meanscore}
	\end{center}
\end{table}
%
%
\begin{figure}[tb]
	\centering
	\includegraphics[width=0.5\linewidth]{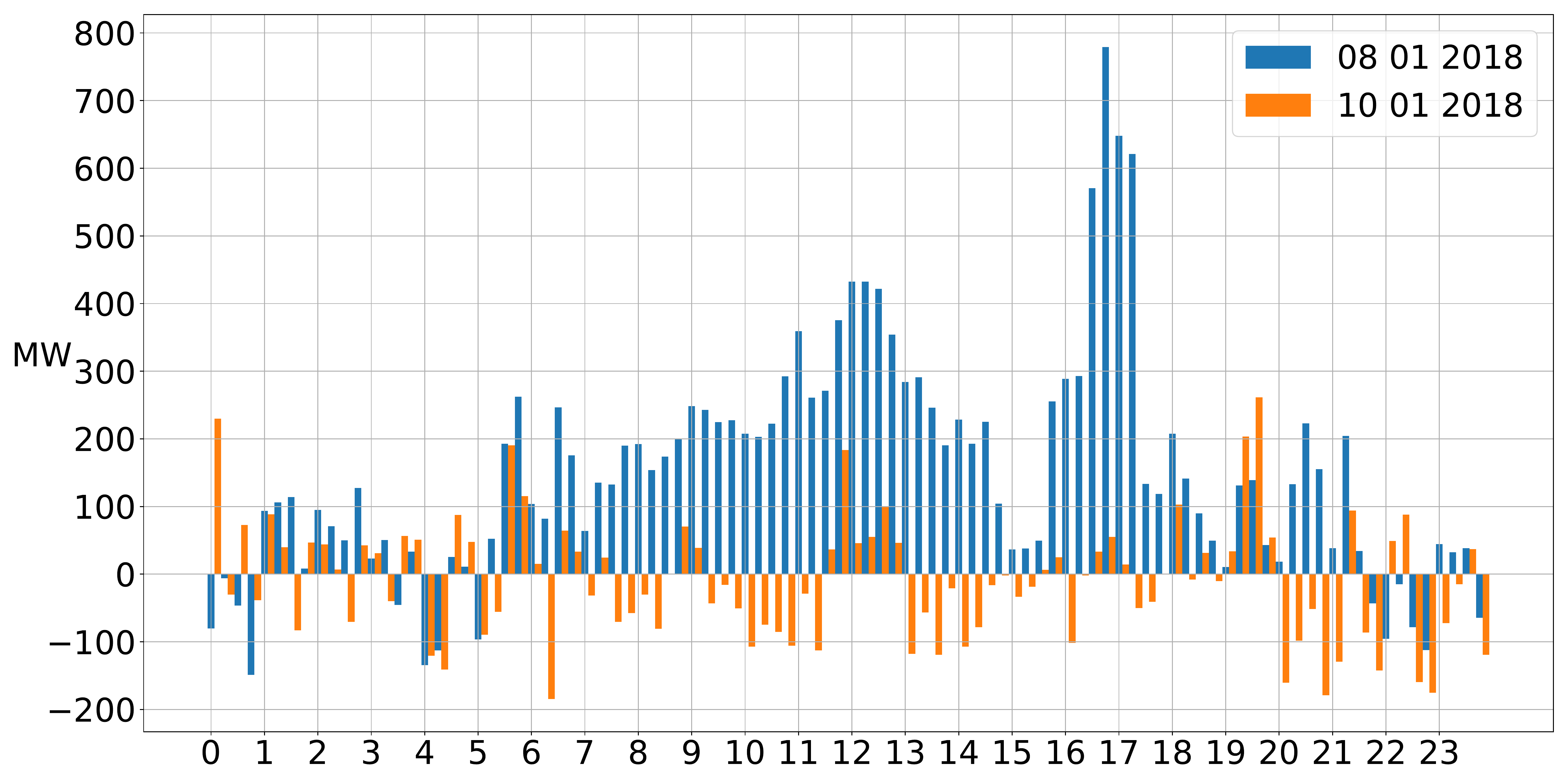}
	\captionsetup{justification=centering}
	\caption{ELIA NRV on $\EEMdateone$ (blue) and $\EEMdatetwo$ (orange).}
	\label{fig:nrv08_and_10_012018}
\end{figure}
\begin{figure}[tb]
	\centering
	\begin{subfigure}{0.5\textwidth}
	\includegraphics[width=\linewidth]{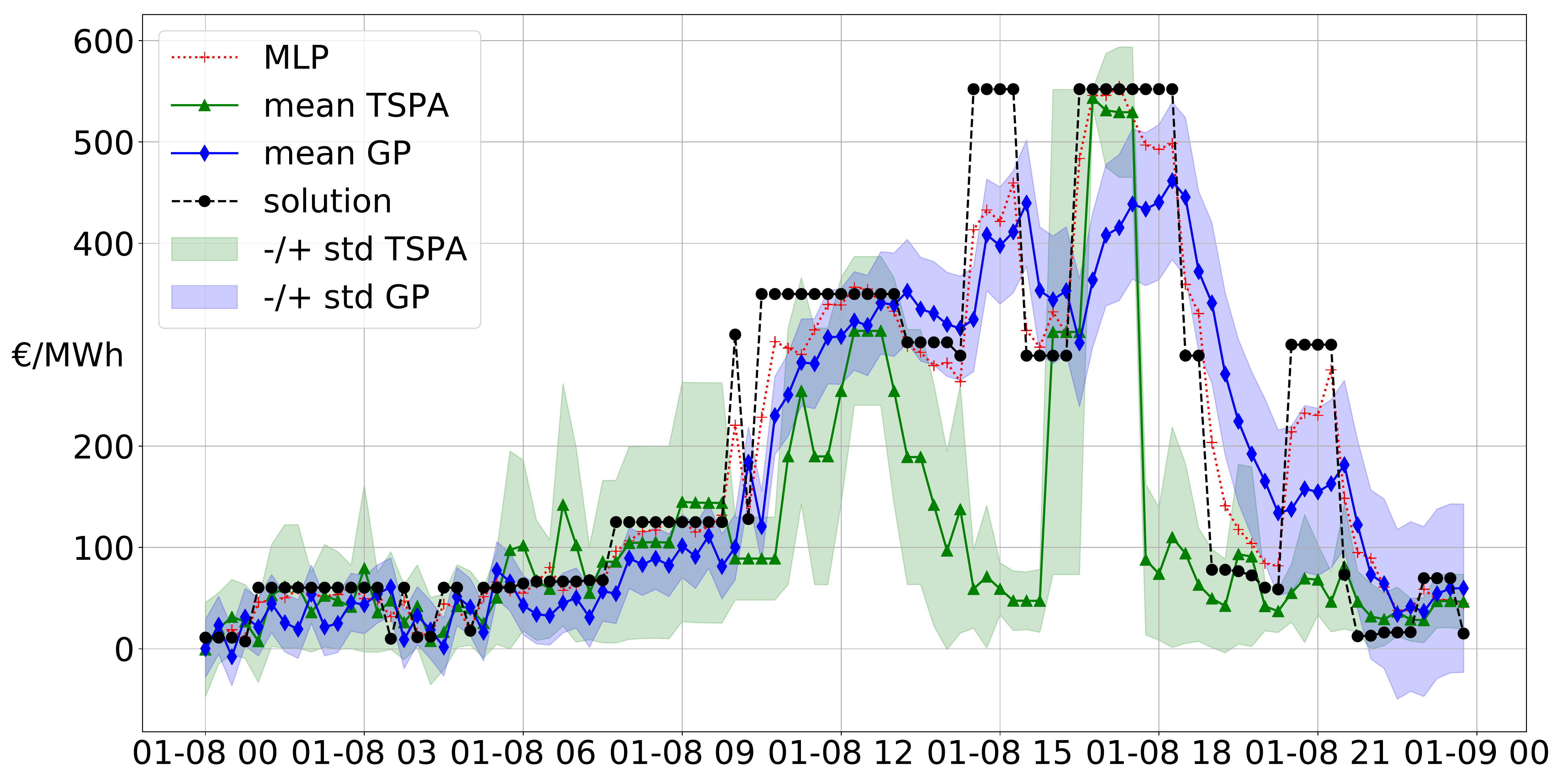}
	\captionsetup{justification=centering}
	\caption{$\EEMdateone$}
	\end{subfigure}%
	\begin{subfigure}{0.5\textwidth}
	\includegraphics[width=\linewidth]{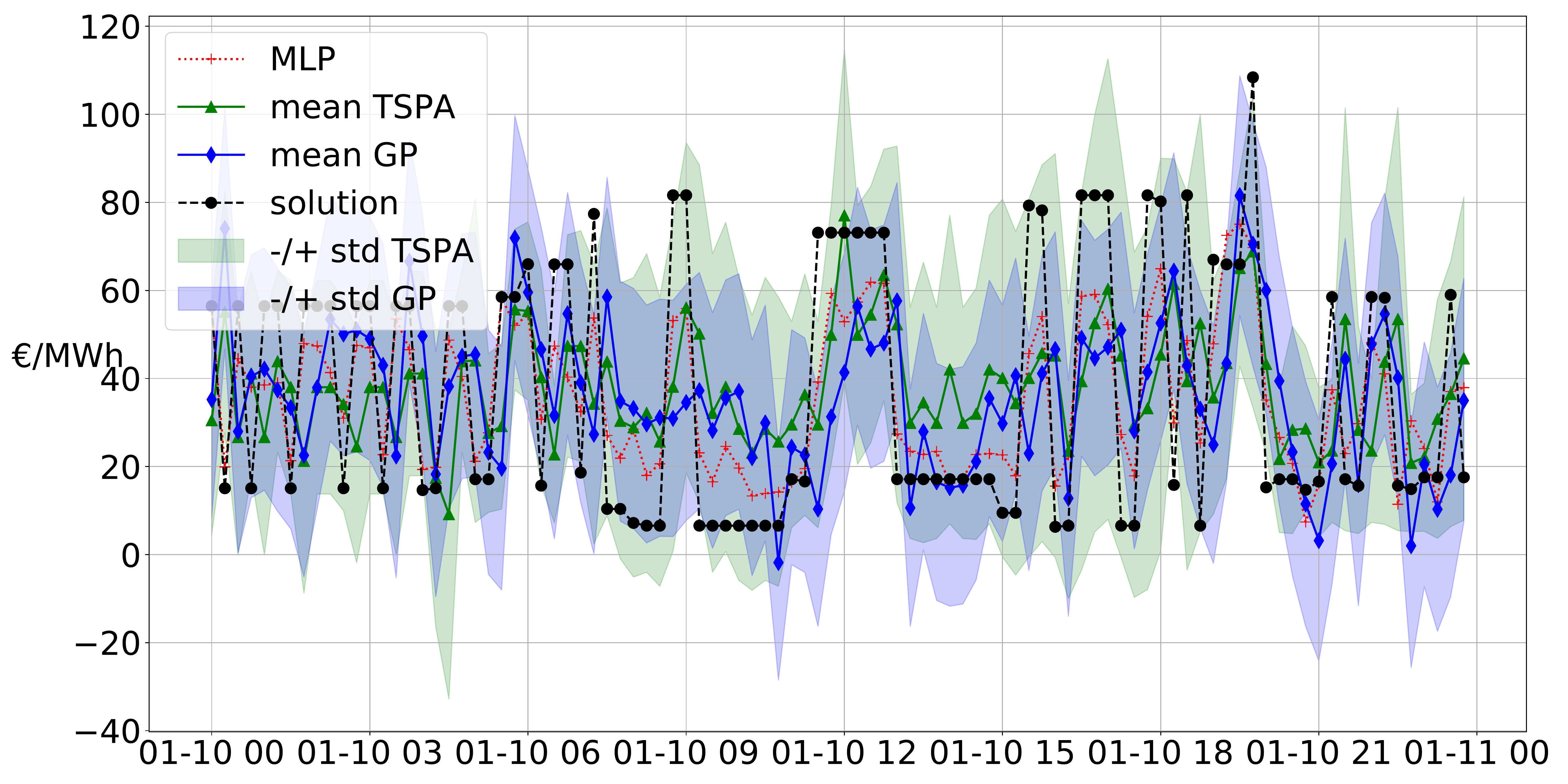}
	\captionsetup{justification=centering}
	\caption{$\EEMdatetwo$}
	\end{subfigure}
	\caption{MLP, GP and TSPA 15 minutes horizon forecasts.}
	\label{fig:M2_GP_MLP_15_08012018}
\end{figure}

Two days, depicted in Figure \ref{fig:nrv08_and_10_012018}, from the validation set are selected to illustrate the results. On $\EEMdateone$, the ELIA system was short on average, leading to a high NRV and imbalance prices. On $\EEMdatetwo$, the ELIA system was alternatively short and long leading to fluctuating NRV and imbalance prices. The 15 minutes horizon forecasts are depicted in Figure~\ref{fig:M2_GP_MLP_15_08012018}, where only the last forecasted value for each quarter is shown. The 60 and 360 minutes horizon forecasts are depicted in Figure~\ref{fig:M2_GP_MLP_60_and_360_08012018_10012018} in Appendix \ref{annex:EEM_figures}. On $\EEMdateone$ the GP provides better results on average as it accurately follows the actual prices. On $\EEMdatetwo$, there is no clear winner. Other figures are reported in Appendix \ref{annex:EEM_figures} for other forecasting horizons.
The MLP provides the best NMAE and NRMSE, except for the horizon of 360 minutes, and the TSPA the best CRPS and PLF scores for the three forecasting horizons considered. However, to select the best forecasting model, it would be necessary to measure the accuracy of the global bidding chain composed of the forecasting and decision-making modules.


\section{Conclusions}\label{sec:forecasting-eem-conclusions}

This study addressed the problem of forecasting the imbalance prices in a probabilistic framework. The novel two-step probabilistic approach consists of computing the net regulation volume state transition probabilities. Then, it infers the imbalance price from the ELIA ARC table and computes a probabilistic forecast. It is compared to the MLP and GP techniques in the Belgium case. The results indicate it outperforms them using probabilistic metrics, but it is less accurate at predicting the precise imbalance prices. 

Learning models could improve this novel probabilistic approach in three directions:
(1) To avoid making our simplifying assumptions.
(2) By adding input features to describe the market situation better.
(3) Extending the approach to implement the whole bidding strategy chain would allow determining which approach is the best.

\clearpage

\section*{Appendix: notation}\label{appendix:forecasting-eem-notation}

\subsection*{\textbf{Acronyms}}

\begin{supertabular}{l p{0.75\columnwidth}}
ARC & Available Regulation Capacity \\
BRP & Balancing Responsible Party \\
GP & Gaussian Processes \\
GDV & Gross Downward regulation Volume\\
GUV & Gross Upward regulation Volume\\
IGCC & International Grid Control Cooperation \\
MDP & Marginal price for Downward Regulation \\
$\text{metric}$ & NMAE, NRMSE, PLF, CRPS \\
MIMO & Multi-Input Multi-Output \\
MIP & Marginal price for Upward Regulation \\
NRV & Net Regulation Volume \\
R2 & Secondary reserve, upwards or downwards \\
TSO & Transmission System Operator \\
TSPA & Two-Step Probabilistic Approach \\
\end{supertabular}

\vspace*{1mm}

\subsection*{\textbf{Parameters}}

\begin{supertabular}{l p{0.6\columnwidth} l}
Symbol & Description & Unit \\
\hline
$\pi_{+}$, $\pi_{-}$ & Positive/Negative imbalance price  & \euro/MWh\\
$ \alpha_1$, $ \alpha_2$ & ELIA parameters for $\pi_{+}$ and $\pi_{-}$ & \euro/MWh \\
$c_t^p$ & ARC marginal price at $t$ and for activation range $p$ & \euro/MWh\\
$\NRV(t)$ & NRV measured at time $t$ & MW\\
$\NRV_i$ & NRV bin $i$ & MW\\
$\NRV_{i,1/2}$ & Center of NRV bin $i$ & MW\\
$(\NRV)_{t+k|t}$ & NRV transition matrix from $t$ to $t+k$ & - \\
$p^{ij}_{t+k|t}$ & NRV conditional probabilities at $t$ for $t+k$  & -\\
\end{supertabular}

\subsection*{\textbf{Forecasted or computed variables}}

\begin{supertabular}{l p{0.6\columnwidth} l}
Symbol & Description & Unit \\
\hline
$\hat{\pi}^m_{t+k|t}$  & Predicted mean imbalance price at $t$ for $t+k$ & \euro/MWh\\
$\hat{\pi}^{std}_{t+k|t}$ & Standard deviation of $\hat{\pi}^m_{t+k|t}$ at $t$ for $t+k$ & \euro/MWh\\
$\hat{\pi}_{>t}$ & Set $\Big\{ (\hat{\pi}^m_{t+k|t}, \hat{\pi}^{std}_{t+k|t})  \Big\}_{k=k_1}^{k_T}$ & \euro/MWh\\
$\hat{\NRV}^m_{t+k|t}$  & Predicted mean NRV at $t$ for $t+k$ & MW\\
$\hat{\NRV}^{std}_{t+k|t}$ & Standard deviation of $\hat{\NRV}^m_{t+k|t}$ at $t$ for $t+k$ & MW\\
$(\hat{\NRV})_{t+k|t}$ & Estimated NRV transition matrix from $t$ to $t+k$ &- \\
$(\hat{\NRV})_{>t}$ & Set $\Big\{ (\hat{\NRV})_{t+k|t}\Big\}_{k=k_1}^{k_T}$ & -\\
$\hat{p}^{ij}_{t+k|t}$ & Estimated NRV conditional probability at $t$ for $t+k$ & -\\
\end{supertabular}

\section{Appendix: balancing mechanisms}\label{appendix:forecasting-eem}

\subsection{Balancing mechanisms}

A balancing mechanism aims to balance a given geographical area and control sudden imbalances between injection and off-take. Generally, this mechanism relies on exchanges with neighboring TSOs, the responsible balance parties, and the usage of reserve capacities.
Each party that desires to inject or off-take to the grid must be managed by a Balancing Responsible Party (BRP). The BRP is responsible for balancing all off-takes and injections within its customer's portfolio. The TSO applies an imbalance tariff when it identifies an imbalance between total physical injections, imports, and purchases on the one hand and total off-takes, exports, and sales on the other.
When the BRPs cannot balance their customer's portfolios, the TSO activates reserves to balance the control area. These reserves are mainly from conventional power plants, which can be quickly activated upward or downward to cover real-time system imbalances. The main types of reserve are the Frequency Containment Reserve (FCR), the Automatic Frequency Restoration Reserve (aFRR), the Manual Frequency Restoration Reserve (mFRR), and the Replacement Reserve (RR)\footnote{\url{https://www.entsoe.eu/}}. 
The activation of these reserves results from a merit order representing the activation cost of reserve capacity. If the system faces a power shortage, the TSO activates upward reserves that result in a positive marginal price on the reserve market. Then, the TSO pays the Balancing Service Provider. The cost of this activation is transferred to the BRPs. BRPs facing short positions are reinforcing the system imbalance. They must pay the marginal price to the TSO. BRPs facing long positions are restoring the system imbalance. They receive the marginal price from the TSO. This mechanism incentives market players to maintain their portfolios in balance, as well as to reduce the net system imbalance.

\subsection{Belgium balancing mechanisms}\label{appendix:forecasting-eem-balancing-belgium}
 
This section describes the ELIA imbalance price mechanisms and the data publication part of the TSPA inputs. On a 15 minutes basis, the NRV is defined as the sum of the GUV and GDV. The Gross Upward Volume (GUV) is the sum of the volumes of all upward regulations. The Gross Downward Volume (GDV) is the sum of the volumes of all downward regulations. If the NRV is positive, the highest price of all upward activated products, the Marginal price for Upward Regulation (MIP), is applied for the imbalance price calculation. If the NRV is negative, the lowest price of all downward activated products, the Marginal price for Downward Regulation (MDP),  is applied. The definitions of the positive $\pi_{+}$ and negative $\pi_{-}$ imbalance prices are provided in Table \ref{tab:elia_imbalance_prices}. The correction parameters $\alpha_1$ and $ \alpha_2$ are zero when the system imbalance is lower than 140 MW and proportional to it when greater than 140 MW.

The MIP and MDP prices are in the third Available Regulation Capacity (ARC) table most of the time. The ARC publication considers the applicable merit order, \textit{i.e.}, the order in which Elia must activate the reserve products. Then, the volumes are ranked by activation price (cheapest first). The marginal price is the highest price for every extra MW upward volume and the lowest for every extra MW downward volume. The ARC table, showing the activation price of the contracted reserves per activation range of 100 MW, displays the estimated activation price considering a certain NRV. For a given quarter-hour $t$ there are $P$ marginal prices for activation $c^p_t$, $p \in \llbracket 1; P \rrbracket $, each one of them related to the activation range $p$. $P$ is equal to 22 with 11 negatives ranges and 11 positives ranges. The first activation range, $p=1$, corresponds to the interval $[-\infty, -1000]$ MW, the second one to $[-1000, -900]$, ..., $[-100, 0]$, $[0, 100]$ up to $[1000, +\infty]$. The data of day $D$ are published on $D-1$ at 6 pm based on the nomination of day-ahead and intraday programs and bids submitted by the concerned parties. The values of each quarter-hour of the day are refreshed quarterly. Therefore, the published values are an estimation. However, they are likely to include the MIP and MDP prices. At the condition to determine the NRV and its spread between the GUV and GDV. The TSPA takes as input the third ARC table to determine the most probable MIP and MDP prices.
\begin{table}[!htb]
	\begin{center}
		\renewcommand\arraystretch{1.25}
		\begin{tabular}[b]{lll}
			\hline \hline
			BRP perimeter & $NRV < 0$ & $NRV > 0$  \\ \hline
			$ > 0$ & $\pi_{+}  = MDP - \alpha_1 $ & $\pi_{+}  = MIP$  \\ 
			$ < 0$ & $\pi_{-}  = MDP$  & $\pi_{-}  = MIP + \alpha_2$   \\	\hline \hline
		\end{tabular}
			\caption{Elia imbalance prices.}
		\label{tab:elia_imbalance_prices}
	\end{center}
\end{table}

\subsection{Additional results}\label{annex:EEM_figures}

%
\begin{figure}[htb]
	\centering
	\begin{subfigure}{0.5\textwidth}
	\includegraphics[width=\linewidth]{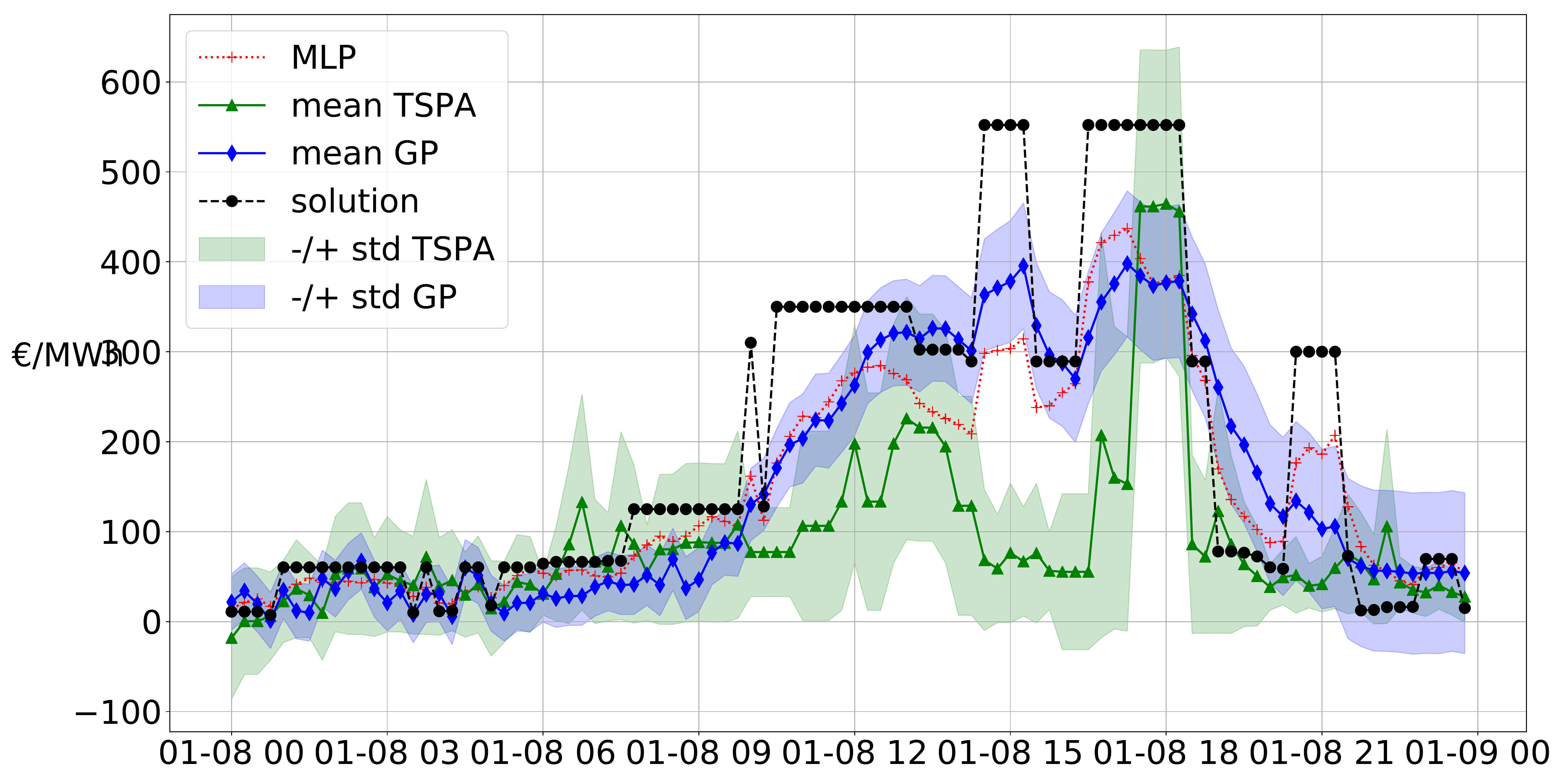}
	\captionsetup{justification=centering}
	\caption{$\EEMdateone$}
	\end{subfigure}%
	\begin{subfigure}{0.5\textwidth}
	\includegraphics[width=\linewidth]{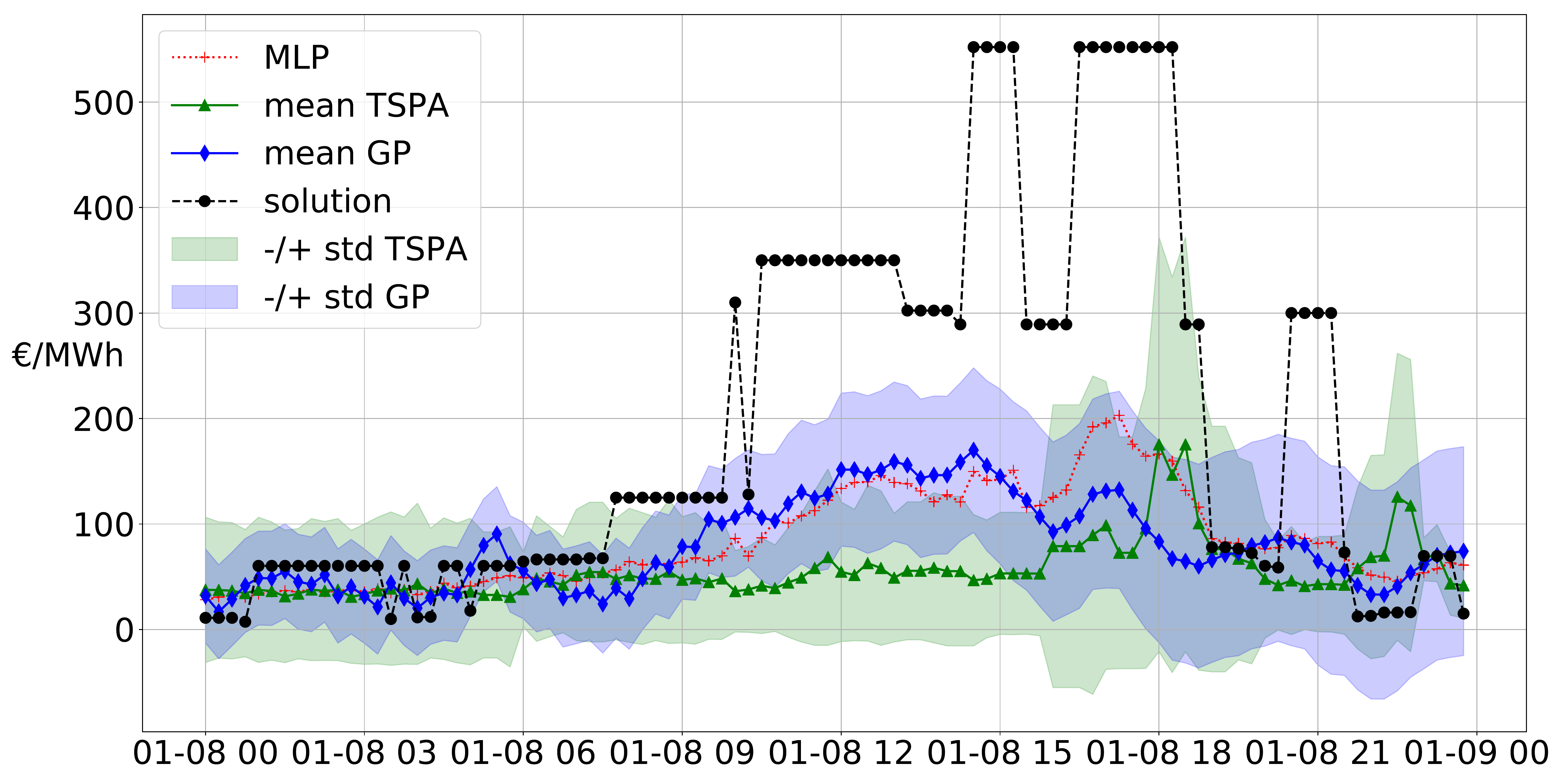}
	\captionsetup{justification=centering}
	\caption{$\EEMdateone$}
	\end{subfigure}
	\begin{subfigure}{0.5\textwidth}
	\includegraphics[width=\linewidth]{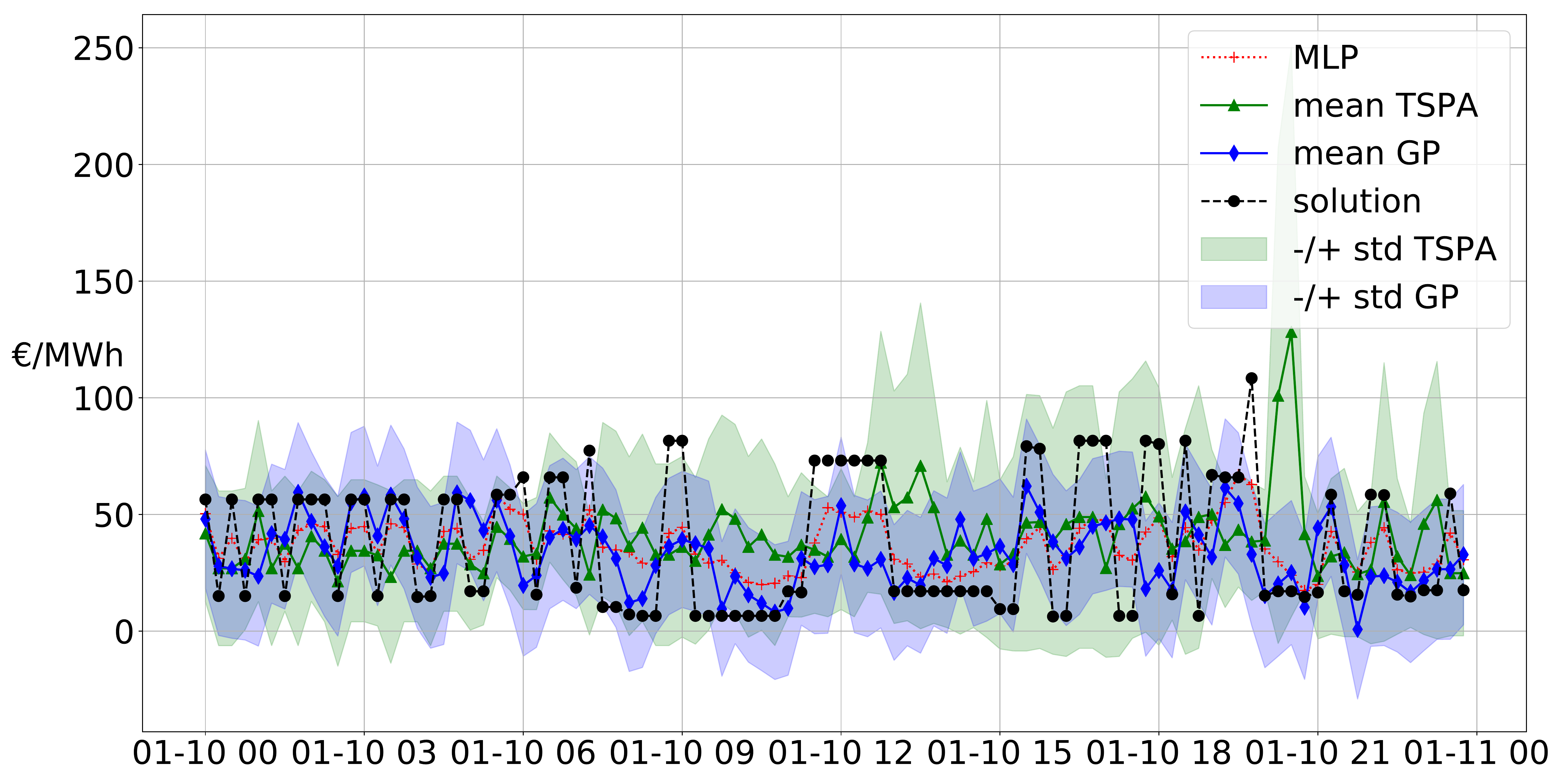}
	\captionsetup{justification=centering}
	\caption{$\EEMdatetwo$}
	\end{subfigure}%
	\begin{subfigure}{0.5\textwidth}
	\includegraphics[width=\linewidth]{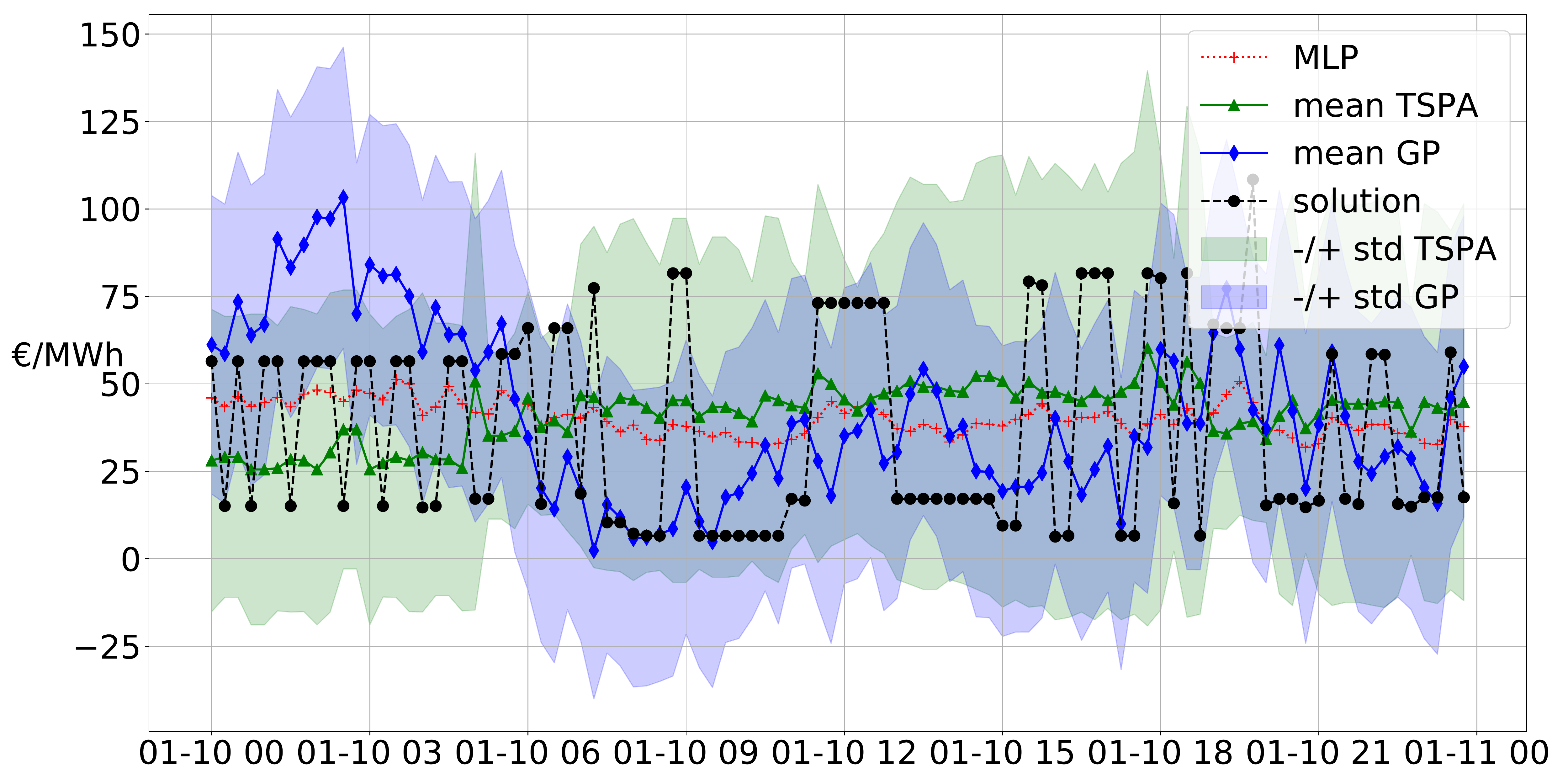}
	\captionsetup{justification=centering}
	\caption{$\EEMdatetwo$}
	\end{subfigure}
	\caption{MLP, GP and TSPA 60 (left) and 360 (right) minutes horizon forecasts.}
	\label{fig:M2_GP_MLP_60_and_360_08012018_10012018}
\end{figure}
%
%
%
\begin{figure}[htb]
	\centering
	\begin{subfigure}{0.5\textwidth}
	\includegraphics[width=\linewidth]{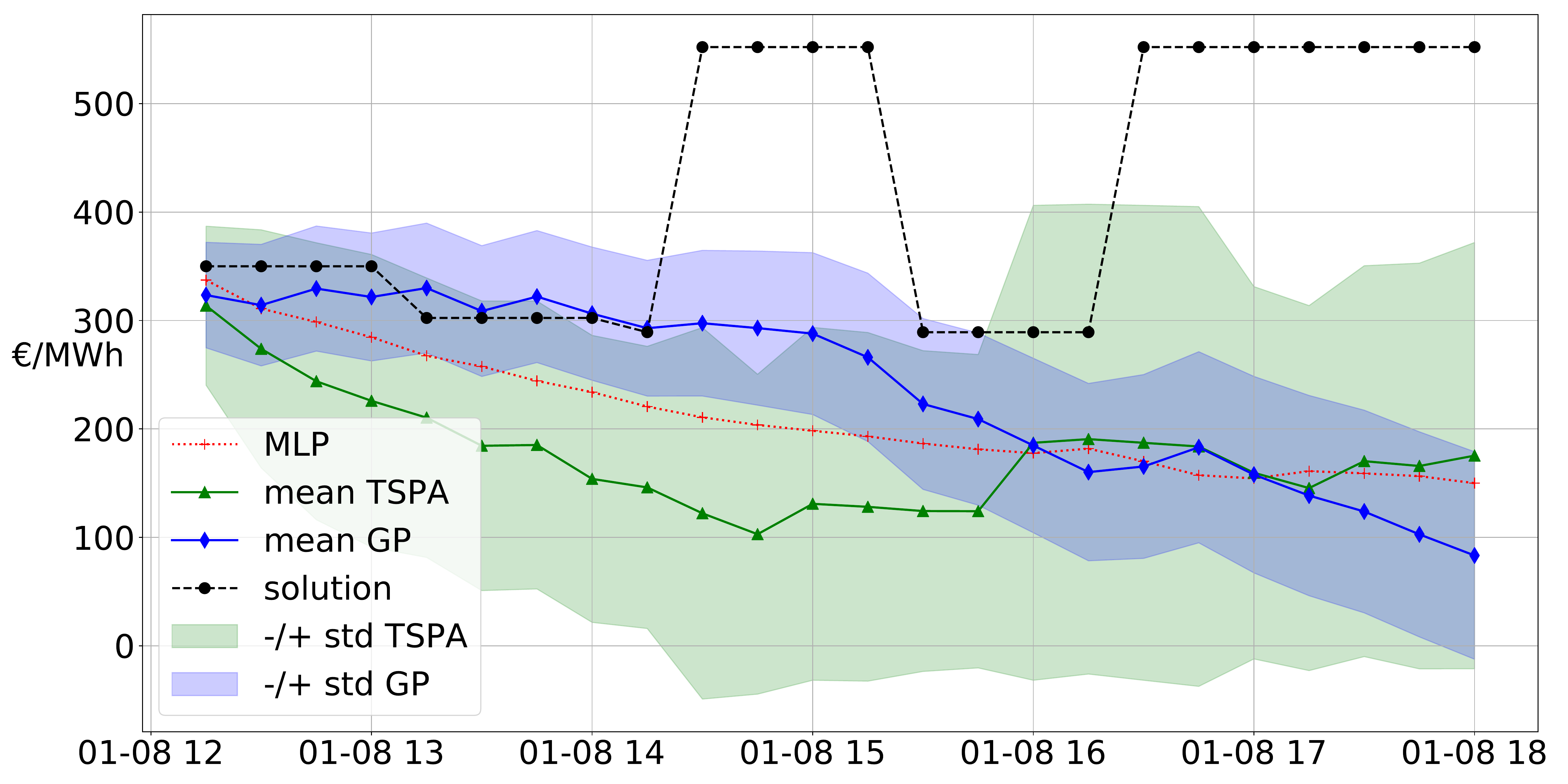}
	\captionsetup{justification=centering}
	\caption{$\EEMdateone$}
	\end{subfigure}%
	\begin{subfigure}{0.5\textwidth}
	\includegraphics[width=\linewidth]{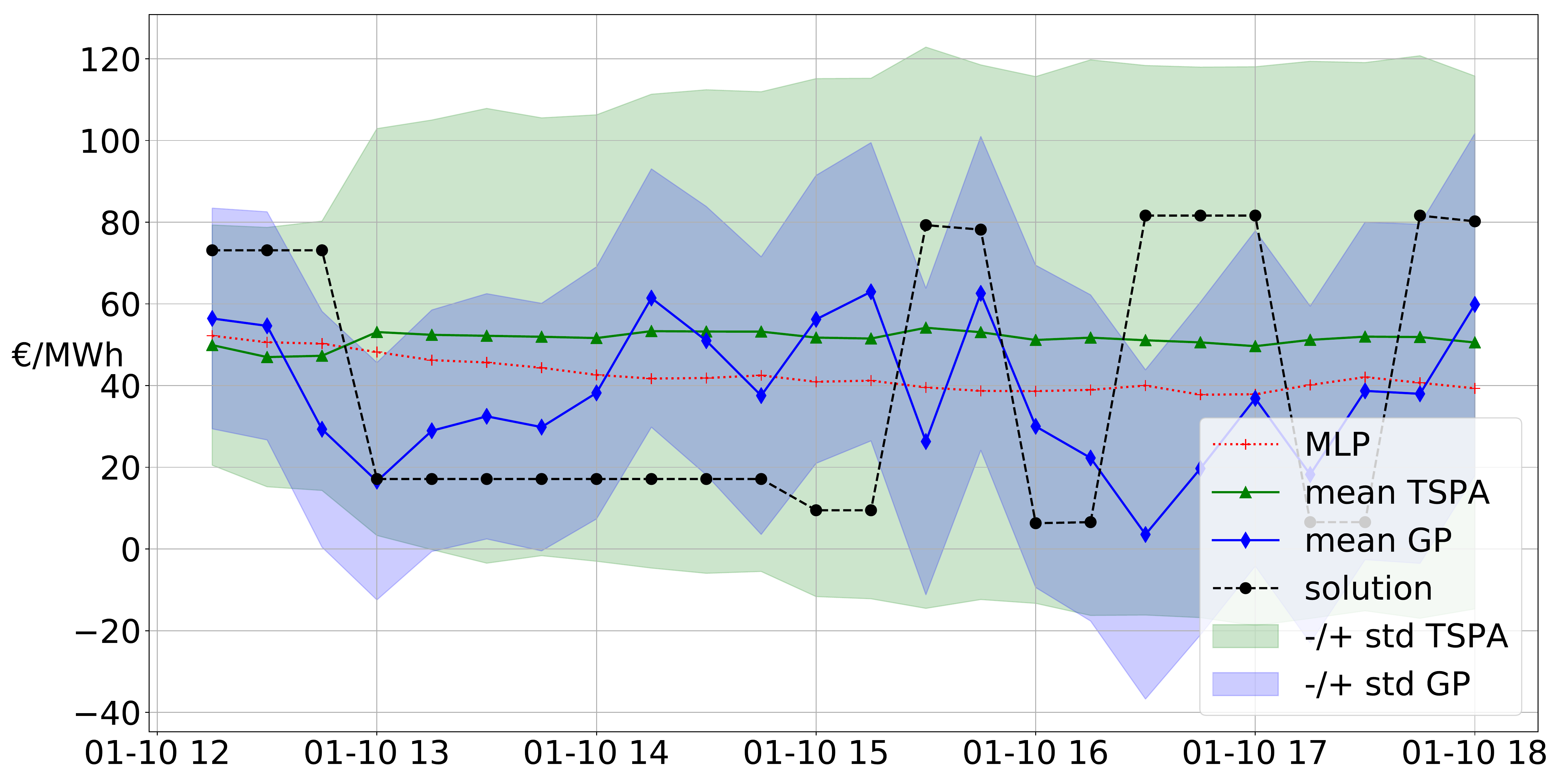}
	\captionsetup{justification=centering}
	\caption{$\EEMdatetwo$}
	\end{subfigure}
	\caption{MLP, GP and TSPA forecasts, 12h00 UTC, with an horizon of 360 minutes.}
	\label{fig:M2_GP_MLP_360_08_and_10_012018_1200}
\end{figure}


\chapter{Scenarios}\label{chap:scenarios-forecasting}

\begin{infobox}{Overview}
The main contributions of this Chapter are three-fold:
\begin{enumerate}
	\item We present to the power systems community a recent class of deep learning generative models: the Normalizing Flows (NFs). Then, we provide a fair comparison of this technique with the state-of-the-art deep learning generative models, Generative Adversarial Networks (GANs) and Variational AutoEncoders (VAEs). It uses the open data of the Global Energy Forecasting Competition 2014 (GEFcom 2014) \citep{hong2016bprobabilistic}. To the best of our knowledge, it is the first study that extensively compares NFs, GANs, and VAEs on several datasets, PV generation, wind generation, and load with a proper assessment of the quality and value based on complementary metrics, and an easily reproducible case study; \item We implement conditional generative models to compute improved weather-based PV, wind power, and load scenarios. In contrast to most of the previous studies that focused mainly on past observations;
	\item Overall, we demonstrate that NFs are more accurate in quality and value, providing further evidence for deep learning practitioners to implement this approach in more advanced power system applications.
\end{enumerate}
This study provides open-access to the \textcolor{RoyalBlue}{Python code}: \url{https://github.com/jonathandumas/generative-models}. 

\textbf{\textcolor{RoyalBlue}{References:}} This chapter  is an adapted version of the following publication: \\[2mm]\bibentry{dumas2021nf}. 
%
\end{infobox}
\epi{I won't say 'See you tomorrow' because that would be like predicting the future, and I'm pretty sure I can't do that.}{Ludwig Wittgenstein}
\begin{figure}[htbp]
	\centering
	\includegraphics[width=1\linewidth]{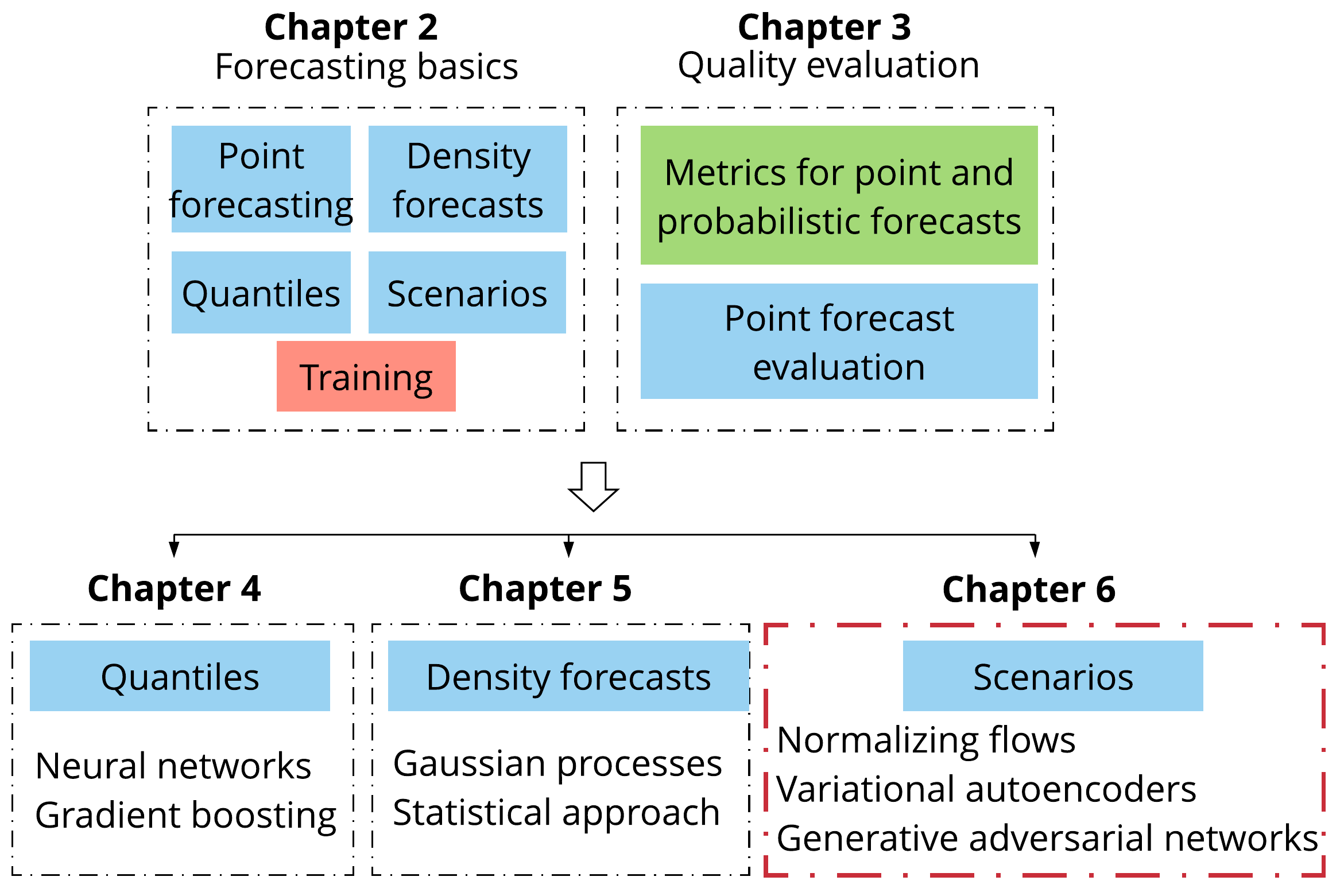}
	\caption{Chapter \ref{chap:scenarios-forecasting} position in Part \ref{part:forecasting}.}
\end{figure}
\clearpage

\section{Introduction}
This Chapter focuses on \textit{scenario generation}, a popular probabilistic forecasting method to capture the uncertainty of load, photovoltaic (PV) generation, and wind generation. It consists of producing sequences of possible load or power generation realizations for one or more locations. 

Forecasting methodologies can typically be classified into two groups: \textit{statistical} and \textit{machine learning} models. On the one hand, statistical approaches are more interpretable than machine learning techniques, sometimes referred to as black-box models. On the other hand, they are generally more robust, user-friendly, and successful in addressing the non-linearity in the data than statistical techniques. We provide in the following a few examples of statistical approaches. More references can be found in \citet{khoshrou2019short} and \citet{mashlakov2021assessing}. 

Multiple linear regression models \citep{wang2016electric} and autoregressive integrated moving average \citep{de200625} are among the most fundamental and widely-used models. The latter generates spatiotemporal scenarios with given power generation profiles at each renewables generation site \citep{morales2010methodology}. These models mostly learn a relationship between several explanatory variables and a dependent target variable. However, they require some expert knowledge to formulate the relevant interaction between different variables. Therefore, the performance of such models is only satisfactory if the dependent variables are well formulated based on explanatory variables.
Another class of statistical approaches consists of using simple parametric distributions, \textit{e.g.}, the Weibull distribution for wind speed \citep{karaki2002probabilistic}, or the beta distribution for solar irradiance \citep{karaki1999probabilistic} to model the density associated with the generative process. In this line, the (Gaussian) copula method has been widely used to model the spatial and temporal characteristics of wind \citep{pinson2009probabilistic} and PV generation \citep{zhang2019coordinated}. For instance, the problem of generating probabilistic forecasts for the aggregated power of a set of renewable power plants harvesting different energy sources is addressed by \citet{camal2019scenario}.

Overall, these approaches usually make statistical assumptions increasing the difficulty to model the underlying stochastic process. The generated scenarios approximate the future uncertainty but cannot correctly describe all the salient features in the power output from renewable energy sources. \textit{Deep learning} is one of the newest trends in artificial intelligence and machine learning to tackle the limitations of statistical methods with promising results across various application domains.

\subsection{Related work}

Recurrent neural networks (RNNs) are among the most famous deep learning techniques adopted in energy forecasting applications. A novel pooling-based deep recurrent neural network is proposed by \citet{shi2017deep} in the field of short-term household load forecasting. It outperforms statistical approaches such as autoregressive integrated moving average and classical RNN. A tailored forecasting tool, named encoder-decoder, is implemented in \citet{dumas2020deep} to compute intraday multi-output PV quantiles forecasts.
Guidelines and best practices are developed by \citet{hewamalage2020recurrent} for forecasting practitioners on an extensive empirical study with an open-source software framework of existing RNN architectures. In the continuity, \citet{toubeau2018deep} implemented a bidirectional long short-term memory (BLSTM) architecture. It is trained using quantile regression and combined with a copula-based approach to generate scenarios. A scenario-based stochastic optimization case study compares this approach to other models regarding forecast quality and value. Finally, \citet{salinas2020deepar} trained an autoregressive recurrent neural network on several real-world datasets. It produces accurate probabilistic forecasts with little or no hyper-parameter tuning.


Deep generative modeling is a class of techniques that trains deep neural networks to model the distribution of the observations. In recent years, there has been a growing interest in this field made possible by the appearance of large open-access datasets and breakthroughs in both general deep learning architectures and generative models. Several approaches exist such as energy-based models, variational autoencoders, generative adversarial networks, autoregressive models, normalizing flows, and numerous hybrid strategies. They all make trade-offs in terms of computation time, diversity, and architectural restrictions. We recommend two papers to get a broader knowledge of this field. (1) The comprehensive overview of generative modeling trends conducted by \citet{bond2021deep}. It presents generative models to forecasting practitioners under a single cohesive statistical framework. (2) The thorough comparison of normalizing flows, variational autoencoders, and generative adversarial networks provided by \citet{ruthotto2021introduction}. It describes the advantages and disadvantages of each approach using numerical experiments in the field of computer vision. In the following, we focus on the applications of generative models in power systems.

In contrast to statistical approaches, deep generative models such as \textit{Variational AutoEncoders} (VAEs) \citep{kingma2013auto} and \textit{Generative Adversarial Networks} (GANs) \citep{goodfellow2014generative} directly learn a generative process of the data. They have demonstrated their effectiveness in many applications to compute accurate probabilistic forecasts, including power system applications. 
They both make probabilistic forecasts in the form of Monte Carlo samples that can be used to compute consistent quantile estimates for all sub-ranges in the prediction horizon. Thus, they cannot suffer from the issue raised by \citet{ordiano2020probabilistic} on the non-differentiable quantile loss function. Note: the generative models such as GANs and VAEs allow directly generating scenarios of the variable of interest. In contrast with methods that first compute weather scenarios to generate probabilistic forecasts such as implemented by \citet{sun2020probabilistic} and \citet{khoshrou2019short}.
A VAE composed of a succession of convolutional and feed-forward layers is proposed by \citet{zhanga2018optimized} to capture the spatial-temporal complementary and fluctuant characteristics of wind and PV power with high model accuracy and low computational complexity. Both single and multi-output PV forecasts using a VAE are compared by \citet{dairi2020short} to several deep learning methods such as LSTM, BLSTM, convolutional LSTM networks, and stacked autoencoders. The VAE consistently outperformed the other methods.
A GAN is used by \citet{chen2018model} to produce a set of wind power scenarios that represent possible future behaviors based only on historical observations and point forecasts. This method has a better performance compared to Gaussian Copula. A Bayesian GAN is introduced by \citet{chen2018bayesian} to generate wind and solar scenarios, and a progressive growing of GANs is designed by \citet{yuan2021multi} to propose a novel scenario forecasting method.
In a different application, a GAN is implemented for building occupancy modeling without prior assumptions \citep{chen2018building}. Finally, a conditional version of the GAN using several labels representing some characteristics of the demand is introduced by \citet{lan2018demand} to output power load data considering demand response programs.

Improved versions of GANs and VAEs have also been studied in the context of energy forecasting. The Wasserstein GAN enforces the Lipschitz continuity through a gradient penalty term (WGAN-GP), as the original GANs are challenging to train and suffer from mode collapse and over-fitting. Several studies applied this improved version in power systems: (1) a method using unsupervised labeling and conditional WGAN-GP models the uncertainties and variation in wind power \citep{zhang2020typical}; (2) a WGAN-GP models both the uncertainties and the variations of the load \citep{wang2020modeling}; (3) \citet{jiang2021day} implemented scenario generation tasks both for a single site and for multiple correlated sites without any changes to the model structure.
Concerning VAEs, they suffer from inherent shortcomings, such as the difficulties of tuning the hyper-parameters or generalizing a specific generative model structure to other databases. An improved VAE is proposed by \citet{qi2020optimal} with the implementation of a $\beta$ hyper-parameter into the VAE objective function to balance the two parts of the loss. This improved VAE is used to generate PV and power scenarios from historical values.

However, most of these studies did not benefit from conditional information such as weather forecasts to generate improved PV power, wind power, and load scenarios. In addition, to the best of our knowledge, only \citet{ge2020modeling} compared NFs to these techniques for the generation of daily load profiles. Nevertheless, the comparison only considers quality metrics, and the models do not incorporate weather forecasts.

\subsection{Research gaps and scientific contributions}

This study investigates the implementation of \textit{Normalizing Flows} \citep[NFs]{rezende2015variational} in power system applications. NFs define a new class of probabilistic generative models. They have gained increasing interest from the deep learning community in recent years. A NF learns a sequence of transformations, a \textit{flow}, from a density known analytically, \textit{e.g.}, a \textit{Normal} distribution, to a complex target distribution. In contrast to other deep generative models, NFs can directly be trained by maximum likelihood estimation. They have proven to be an effective way to model complex data distributions with neural networks in many domains. First, speech synthesis \citep{oord2018parallel}. Second, fundamental physics to increase the speed of gravitational wave inference by several orders of magnitude \citep{green2021complete} or for sampling Boltzmann distributions of lattice field theories \citep{albergo2021introduction}. Finally, in the capacity firming framework by \citet{dumas2021probabilistic}.

This present work goes several steps further than \citet{ge2020modeling} that demonstrated the competitiveness of NFs regarding GANs and VAEs for generating daily load profiles. First, we study the conditional version of these models to demonstrate that they can handle additional contextual information such as weather forecasts or geographical locations. Second, we extensively compare the model's performances both in terms of forecast value and quality. The forecast quality corresponds to the ability of the forecasts to genuinely inform of future events by mimicking the characteristics of the processes involved. The forecast value relates to the benefits of using forecasts in decision-making, such as participation in the electricity market. Third, we consider PV and wind generations in addition to load profiles. Finally, in contrast to the affine NFs used in their work, we rely on monotonic transformations, which are universal density approximators \citep{huang2018neural}.\\

Given that Normalizing Flows are rarely used in the power systems community despite their potential, our main aim is to present this recent deep learning technique and demonstrate its interest and competitiveness with state-of-the-art generative models such as GANs and VAEs on a simple and easily reproducible case study. The research gaps motivating this study are three-fold:
\begin{enumerate}
	\item To the best of our knowledge, only \citet{ge2020modeling} compared NFs to GANs and VAEs for the generation of daily load profiles. Nevertheless, the comparison is only based on quality metrics, and the models do not take into account weather forecasts; 
	\item Most of the studies that propose or compare forecasting techniques only consider the forecast quality such as \citet{ge2020modeling}, \citet{sun2020probabilistic}, and \citet{mashlakov2021assessing};
	\item The conditional versions of the models are not always addressed, such as in \citet{ge2020modeling}. However, weather forecasts are essential for computing accurate probabilistic forecasts.
\end{enumerate}

\noindent With these research gaps in mind, the main contributions of this Chapter are three-fold:
\begin{enumerate}
	\item We provide a fair comparison both in terms of quality and value with the state-of-the-art deep learning generative models, GANs and VAEs, using the open data of the Global Energy Forecasting Competition 2014 (GEFcom 2014) \citep{hong2016bprobabilistic}. To the best of our knowledge, it is the first study that extensively compares the NFs, GANs, and VAEs on several datasets, PV generation, wind generation, and load with a proper assessment of the quality and value based on complementary metrics, and an easily reproducible case study; 
	\item We implement conditional generative models to compute improved weather-based PV, wind power, and load scenarios. In contrast to most of the previous studies that focused mainly on past observations;
	\item Overall, we demonstrate that NFs are more accurate in quality and value, providing further evidence for deep learning practitioners to implement this approach in more advanced power system applications.
\end{enumerate}

\begin{figure}[htbp]
	\centering
	\includegraphics[width=0.8\linewidth]{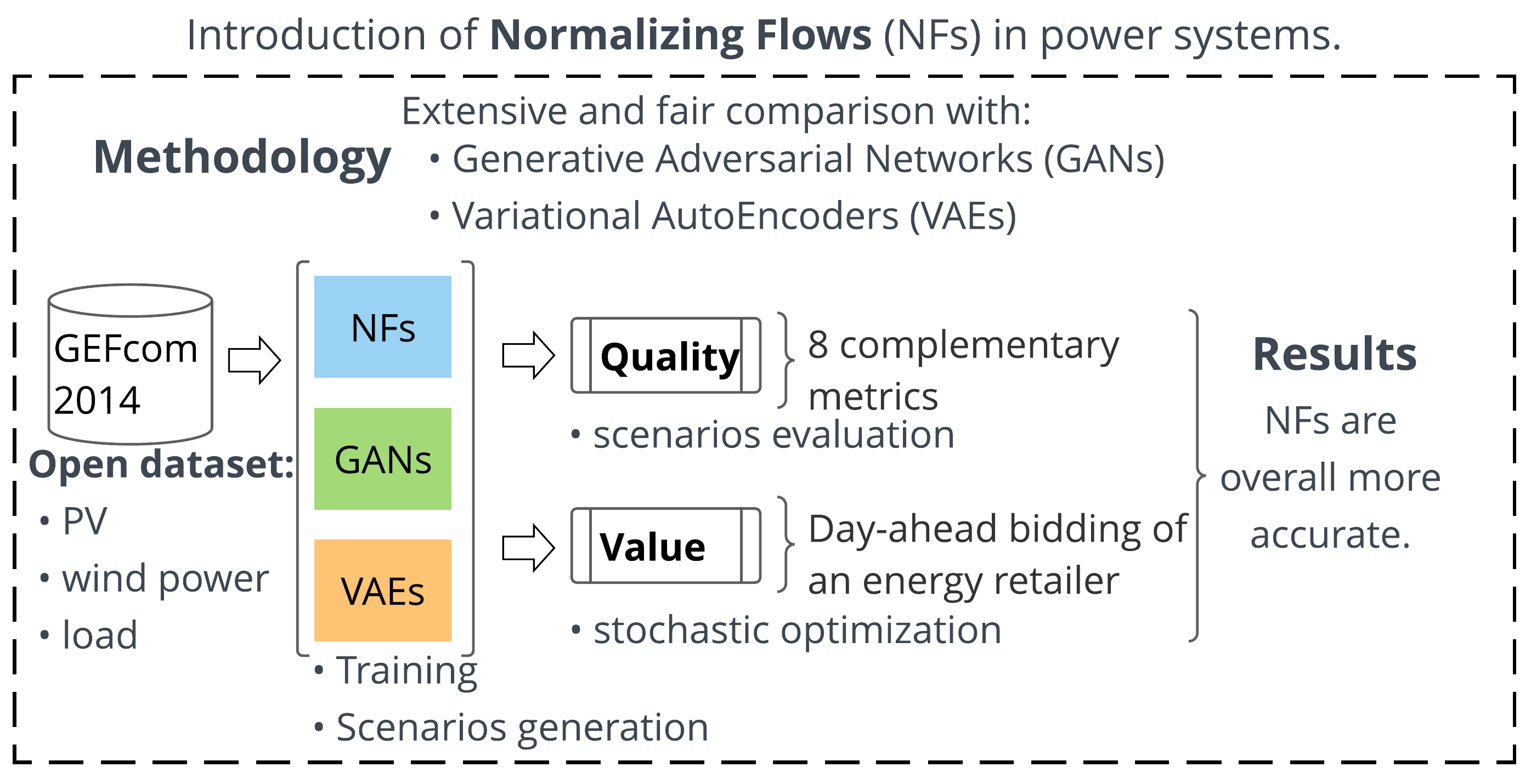}
	\caption{The framework of the study. \\
	The Chapter's primary purpose is to present and demonstrate the potential of NFs in power systems. A fair comparison is conducted both in terms of quality and value with the state-of-the-art deep learning generative models, GANs and VAEs, using the open data of the Global Energy Forecasting Competition 2014 \citep{hong2016bprobabilistic}. The PV, wind power, and load datasets are used to assess the models. The quality evaluation is conducted by using eight complementary metrics, and the value assessment by considering the day-ahead bidding of an energy retailer using stochastic optimization. Overall, NFs tend to be more accurate both in terms of quality and value and are competitive with GANs and VAEs.}
	\label{fig:paper-framework}
\end{figure}

In addition to these contributions, this study also provides open-access to the Python code\footnote{\url{https://github.com/jonathandumas/generative-models}} to help the community to reproduce the experiments.
Figure \ref{fig:paper-framework} provides the framework of the proposed method and Table \ref{tab:AE-contributions} presents a comparison of the present study to three state-of-the-art papers using deep learning generative models to generate scenarios.
\begin{table}[htbp]
\renewcommand{\arraystretch}{1.25}
	\begin{center}
		\begin{tabular}{lcccc}
			\hline \hline
Criteria			     & \citep{wang2020modeling}  &  \citep{qi2020optimal}  & \citep{ge2020modeling} & study \\ \hline
GAN                      & \checkmark & $\times$   & \checkmark & \checkmark \\
VAE                      & $\times$   & \checkmark & \checkmark & \checkmark   \\
NF                       & $\times$   & $\times$   & \checkmark & \checkmark   \\
Number of models         & 4          & 1          &  3         & 3  \\
PV 	                     & $\times$   & \checkmark & $\times$   & \checkmark  \\
Wind power 	             & $\times$   & \checkmark & $\times$   & \checkmark \\
Load 	                 & \checkmark & $\sim$     & \checkmark & \checkmark    \\
Weather-based            & \checkmark & $\times$   & $\times$   & \checkmark    \\
Quality assessment 	     & \checkmark & \checkmark & \checkmark & \checkmark   \\
Quality metrics  	     & 5          & 3          &   5        & 8 \\
Value assessment     	 & $\times$   & \checkmark & $\times$   & \checkmark \\
Open dataset             & $\sim$     & $\times$   & \checkmark & \checkmark \\
Value replicability      & -          & $\sim$     &  -         & \checkmark \\
Open-access code         & $\times$   & $\times$   &  $\times$  & \checkmark \\
\hline \hline
\end{tabular}
\caption{Comparison of the study's contributions to three state-of-the-art studies using deep generative models. \\
\checkmark: criteria fully satisfied, $\sim$: criteria partially satisfied, $\times$: criteria not satisfied, ?: no information, -: not applicable. 
GAN: a GAN model is implemented; VAE: a VAE model is implemented; NF: a NF model is implemented; PV: PV scenarios are generated; Wind power: wind power scenarios are generated; Load: load scenarios are generated; Weather-based: the model generates weather-based scenarios; Quality assessment: a quality evaluation is conducted: Quality metrics: number of quality metrics considered; Value assessment: a value evaluation is considered with a case study; Open dataset: the data used for the quality and value evaluations are in open-access; Value replicability: the case study considered for the value evaluation is easily reproducible; Open-access code: the code used to conduct the experiments is in open-access.
Note: the justifications are provided in Appendix \ref{annex:AE-table1}.}
\label{tab:AE-contributions}
\end{center}
\end{table}

\subsection{Applicability of the generative models}
Probabilistic forecasting of PV, wind generation, electrical consumption, and electricity prices plays a vital role in renewable integration and power system operations. The deep learning generative models presented in this Chapter can be integrated into practical engineering applications. We present a non-exhaustive list of five applications in the following. (1) The forecasting module of an energy management system (EMS) \citep{silva2021optimal}. Indeed, EMSs are used by several energy market players to operate various power systems such as a single renewable plant, a grid-connected or off-grid microgrid composed of several generations, consumption, and storage devices. An EMS is composed of several key modules: monitoring, forecasting, planning, control, \textit{etc}. The forecasting module aims to provide the most accurate forecast of the variable of interest to be used as inputs of the planning and control modules. (2) Stochastic unit commitment models that employ scenarios to model the uncertainty of weather-dependent renewables. For instance, the optimal day-ahead scheduling and dispatch of a system composed of renewable plants, generators, and electrical demand are addressed by \citet{camal2019scenario}. (3) Ancillary services market participation. A virtual power plant aggregating wind, PV, and small hydropower plants is studied by \citet{camal2019scenario} to optimally bid on a day-ahead basis the energy and automatic frequency restoration reserve. (4) More generally, generative models can be used to compute scenarios for any variable of interest, \textit{e.g.}, energy prices, renewable generation, loads, water inflow of hydro reservoirs, as long as data are available. (5) Finally, quantiles can be derived from scenarios and used in robust optimization models such as in the capacity firming framework \citep{dumas2021probabilistic}. 

\subsection{Organization}

The remainder of this Chapter is organized as follows. Section~\ref{sec:ijf-background} presents the generative models implemented: NFs, GANs, and VAEs. Section~\ref{sec:ijf-quality_assessment} provides the quality and assessment methodology.
Section~\ref{sec:ijf-numerical_results} details empirical results on the GEFcom 2014 dataset, and Section~\ref{sec:ijf-conclusion} summarizes the main findings and highlights ideas for further work. 
Chapter \ref{chap:energy-retailer} in Part \ref{part:optimization} provides the forecast value assessment.
Appendix \ref{annex:AE-table1} presents the justifications of Table \ref{tab:AE-contributions}, and Appendix \ref{annex:ijf-background} provides additional information on the generative models.

\section{Background}\label{sec:ijf-background}

This section formally introduces the conditional version of NFs, GANs, and VAEs implemented in this study. We assume the reader is familiar with the neural network's basics. However, for further information, \citet{goodfellow2016deep,zhang2020dive} provide a comprehensive introduction to modern deep learning approaches.

\subsection{Multi-output forecasts using a generative model}

Let us consider some dataset $\mathcal{D} = \{ \mathbf{x}^i, \mathbf{c}^i \}_{i=1}^N$ of $N$ independent and identically distributed samples from the joint distribution $p(\mathbf{x},\mathbf{c})$ of two continuous variables $X$ and $C$. $X$ being the wind generation, PV generation, or load, and $C$ the weather forecasts. They are both composed of $T$ periods per day, with $\mathbf{x}^i := [x_1^i, \ldots , x_T^i]^\intercal \in \mathbb{R}^T$ and $\mathbf{c}^i := [c_1^i, \ldots , c_T^i]^\intercal \in \mathbb{R}^T$. The goal of this work is to generate multi-output weather-based scenarios $\mathbf{\hat{x}} \in \mathbb{R}^T$ that are distributed under $p(\mathbf{x}|\mathbf{c})$.

A generative model is a probabilistic model $p_\theta(\cdot)$, with parameters $\theta$, that can be used as a generator of the data. Its purpose is to generate synthetic but realistic data $\mathbf{\hat{x}} \sim p_\theta(\mathbf{x}|\mathbf{c})$ whose distribution is as close as possible to the unknown data distribution $p(\mathbf{x}|\mathbf{c})$. In our application, it computes on a day-ahead basis a set of $M$ scenarios at day $d-1$ for day $d$
\begin{align}
	\label{eq:multi_output_scenario}	
	\mathbf{\hat{x}}_d^i := &  \big[\hat{x}_{d, 1}^i, \ldots,\hat{x}_{d, T}^i\big]^\intercal	\in \mathbb{R}^T \quad i=1, \ldots, M.
\end{align}
For the sake of clarity, we omit the indexes $d$ and $i$ when referring to a scenario $\mathbf{\hat{x}}$ in the following. 

\subsection{Deep generative models}

\begin{figure}[htbp]
	\centering
	\includegraphics[width=0.6\linewidth]{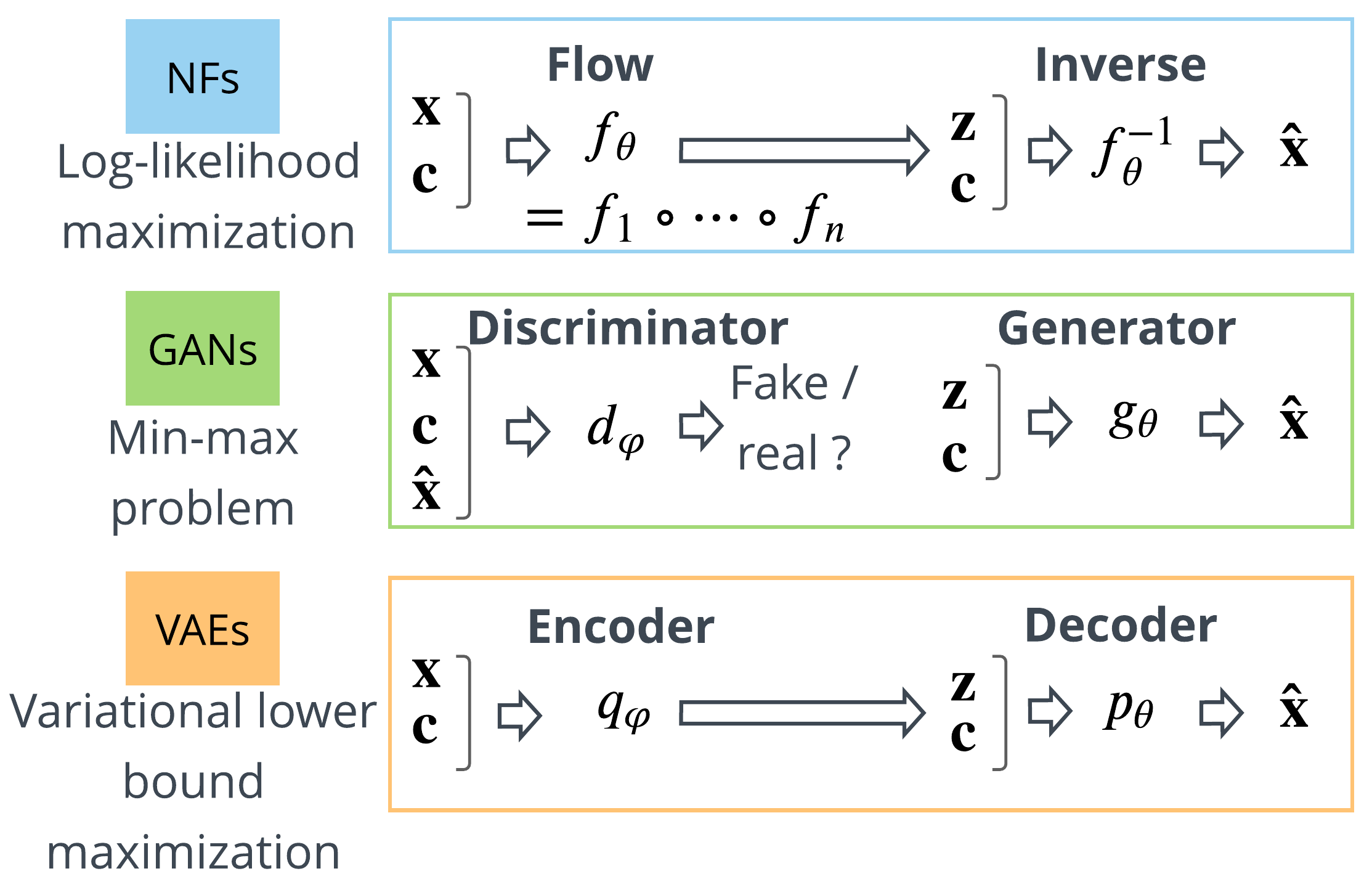}
	\caption{High-level comparison of three categories of generative models considered in this Chapter: normalizing flows, generative adversarial networks, and variational autoencoders. \\
	All models are conditional as they use the weather forecasts $\mathbf{c}$ to generate scenarios $\mathbf{\hat{x}}$ of the distribution of interest $\mathbf{x}$: PV generation, wind power, load.
	Normalizing flows allow exact likelihood calculation. In contrast to generative adversarial networks and variational autoencoders, they explicitly learn the data distribution and provide direct access to the exact likelihood of the model’s parameters. The inverse of the flow is used to generate scenarios.
	The training of generative adversarial networks relies on a min-max problem where the generator and the discriminator parameters are jointly optimized. The generator is used to compute the scenarios.
	Variational autoencoders indirectly optimize the log-likelihood of the data by maximizing the variational lower bound. The decoder computes the scenarios.
	Note: Section \ref{sec:AE-comparison} provides a theoretical comparison of these models. 
	}
	\label{fig:AE-methods-comparison}
\end{figure}

Figure \ref{fig:AE-methods-comparison} provides a high-level comparison of three categories of generative models considered in this Chapter: Normalizing Flows, Generative Adversarial Networks, and Variational AutoEncoders.


\subsubsection{Normalizing flows}

A normalizing flow is defined as a sequence of invertible transformations $f_k : \mathbb{R}^T  \rightarrow \mathbb{R}^T$, $k = 1, \ldots, K$, composed together to create an expressive invertible mapping $f_\theta := f_1 \circ \ldots  \circ f_K : \mathbb{R}^T  \rightarrow \mathbb{R}^T$. This composed function can be used to perform density estimation, using $f_\theta$ to map a sample $\mathbf{x} \in  \mathbb{R}^T $ onto a latent vector $\mathbf{z} \in  \mathbb{R}^T $ equipped with a known and tractable probability density function $p_z$, \textit{e.g.}, a Normal distribution. The transformation $f_\theta$ implicitly defines a density $p_\theta(\mathbf{x})$ that is given by the change of variables
\begin{align}
\label{eq:change_formula}	
p_\theta(\mathbf{x})  & = p_z(f_\theta(\mathbf{x}))|\det J_{f_\theta}(\mathbf{x})| , 
\end{align}
where $ J_{f_\theta}$ is the Jacobian of $f_\theta$ regarding $\mathbf{x}$. The model is trained by maximizing the log-likelihood $\sum_{i=1}^N \log p_\theta(\mathbf{x}^i, \mathbf{c}^i)$ of the model's parameters $\theta$ given the dataset $\mathcal{D}$. For simplicity let us assume a single-step flow $f_\theta$ to drop the index $k$ for the rest of the discussion.

In general, $f_\theta$ can take any form as long as it defines a bijection. However, a common solution to make the Jacobian computation tractable in (\ref{eq:change_formula}) consists of implementing an \textit{autoregressive} transformation \citep{kingma2016improving}, \textit{i.e.}, such that $f_\theta$ can be rewritten as a vector of scalar bijections $f^i$
\begin{subequations}
\label{eq:f_autoregressive}	
\begin{align}
	f_\theta(\mathbf{x}) & := [f^1(x_1; h^1), \ldots, f^T(x_T; h^T)]^\intercal , \\
	h^i & := h^i (\mathbf{x}_{<i} ; \varphi^i) \quad 2\leq i\leq T, \\
	\mathbf{x}_{<i} & := [x_1, \ldots, x_{i-1}]^\intercal  \quad 2\leq i\leq T, \\
	h^1 & \in  \mathbb{R},
\end{align}
\end{subequations}
where $f^i(\cdot; h^i ) : \mathbb{R} \rightarrow \mathbb{R}$ is partially parameterized by an autoregressive conditioner $h^i(\cdot; \varphi^i): \mathbb{R}^{i-1} \rightarrow \mathbb{R}^{|h^i|}$ with parameters $\varphi^i$, and $\theta$ the union of all parameters $\varphi^i$. 

There is a large choice of transformers $f^i$: affine, non-affine, integration-based, \textit{etc}. This work implements an integration-based transformer by using the class of Unconstrained Monotonic Neural Networks (UMNN) \citep{wehenkel2019unconstrained}. It is a universal density approximator of continuous random variables when combined with autoregressive functions.
The UMNN consists of a neural network architecture that enables learning arbitrary monotonic functions. It is achieved by parameterizing the bijection $f^i$ as follows
\begin{align}
\label{eq:UMNN_bijective_mapping}	
	f^i(x_i; h^i ) & =\int_0^{x_i} \tau^i(x_i, h^i) dt + \beta^i(h^i),
\end{align}
where $\tau^i(\cdot; h^i) : \mathbb{R}^{|h^i|+1}  \rightarrow \mathbb{R}^+$ is the integrand neural network with a strictly positive scalar output, $h^i \in  \mathbb{R}^{|h^i|} $ an embedding made by the conditioner, and $ \beta^i(\cdot) : \mathbb{R}^{|h^i|}  \rightarrow \mathbb{R}$ a neural network with a scalar output. The forward evaluation of $f^i$ requires solving the integral (\ref{eq:UMNN_bijective_mapping}) and is efficiently approximated numerically by using the Clenshaw-Curtis quadrature. The pseudo-code of the forward and backward passes is provided by \citet{wehenkel2019unconstrained}.

\begin{figure}[htbp]
	\centering
	\includegraphics[width=0.7\linewidth]{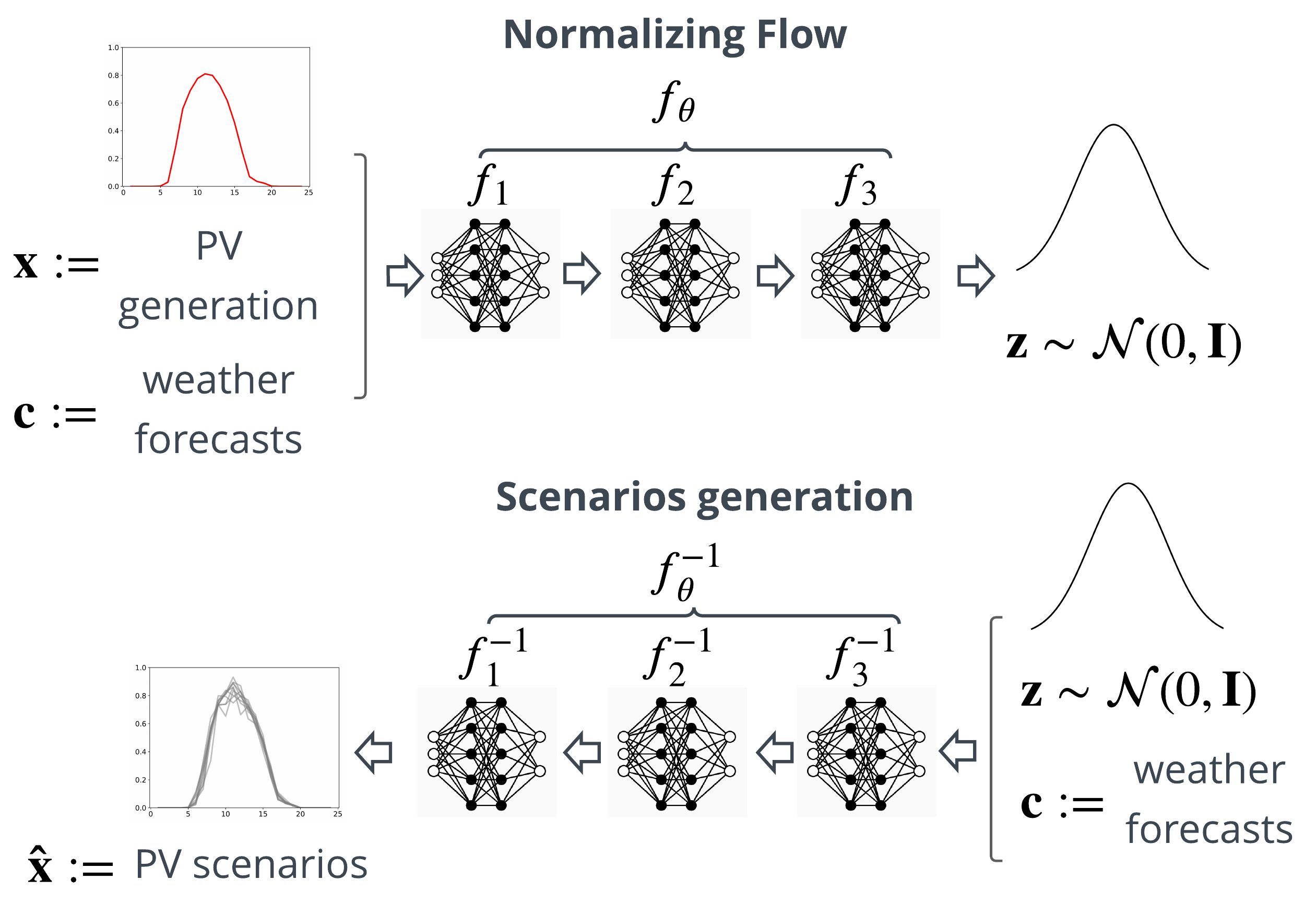}
	\caption{The process of conditional normalizing flows is illustrated with a three-step NF for PV generation. \\
	The model $f_\theta$ is trained by maximizing the log-likelihood of the model's parameters $\theta$ given a dataset composed of PV observations and weather forecasts. Recall $f_\theta$ defines a bijection between the variable of interest $\mathbf{x}$, PV generation, and a Normal distribution $\mathbf{z}$. Then, the PV scenarios $\mathbf{\hat{x}}$ are generated by using the inverse of $f_\theta$ that takes as inputs samples from the Normal distribution $\mathbf{z}$ and the weather forecasts $\mathbf{c}$.}
	\label{fig:UMNN_structure}
\end{figure}

\citet{papamakarios2017masked}'s Masked Autoregressive Network (MAF) is implemented to simultaneously parameterize the $T$ autoregressive embeddings $h^i$ of the flow (\ref{eq:f_autoregressive}). Then, the change of variables formula applied to the UMMN-MAF transformation results in the following log-density when considering weather forecasts
\begin{subequations}
\label{eq:UMNN_likelihood_estimation}
\begin{align}
\log p_\theta(\mathbf{x}, \mathbf{c}) & = \log p_z(f_\theta(\mathbf{x}, \mathbf{c}))|\det J_{f_\theta}(\mathbf{x}, \mathbf{c})| , \\
 & = \log p_z(f_\theta(\mathbf{x}, \mathbf{c})) + \sum_{i=1}^T \log \tau^i\big(x_i, h^i(\mathbf{x}_{<i}), \mathbf{c}\big),
\end{align}
\end{subequations}
that can be computed exactly and efficiently with a single forward pass. The UMNN-MAF approach implemented is referred to as NF in the rest of the Chapter. Figure~\ref{fig:UMNN_structure} depicts the process of conditional normalizing flows with a three-step NF for PV generation. Note: Appendix \ref{annex:ijf-nf} provides additional information on NFs.

\subsubsection{Variational autoencoders}

\begin{figure}[htbp]
	\centering
	\includegraphics[width=0.7\linewidth]{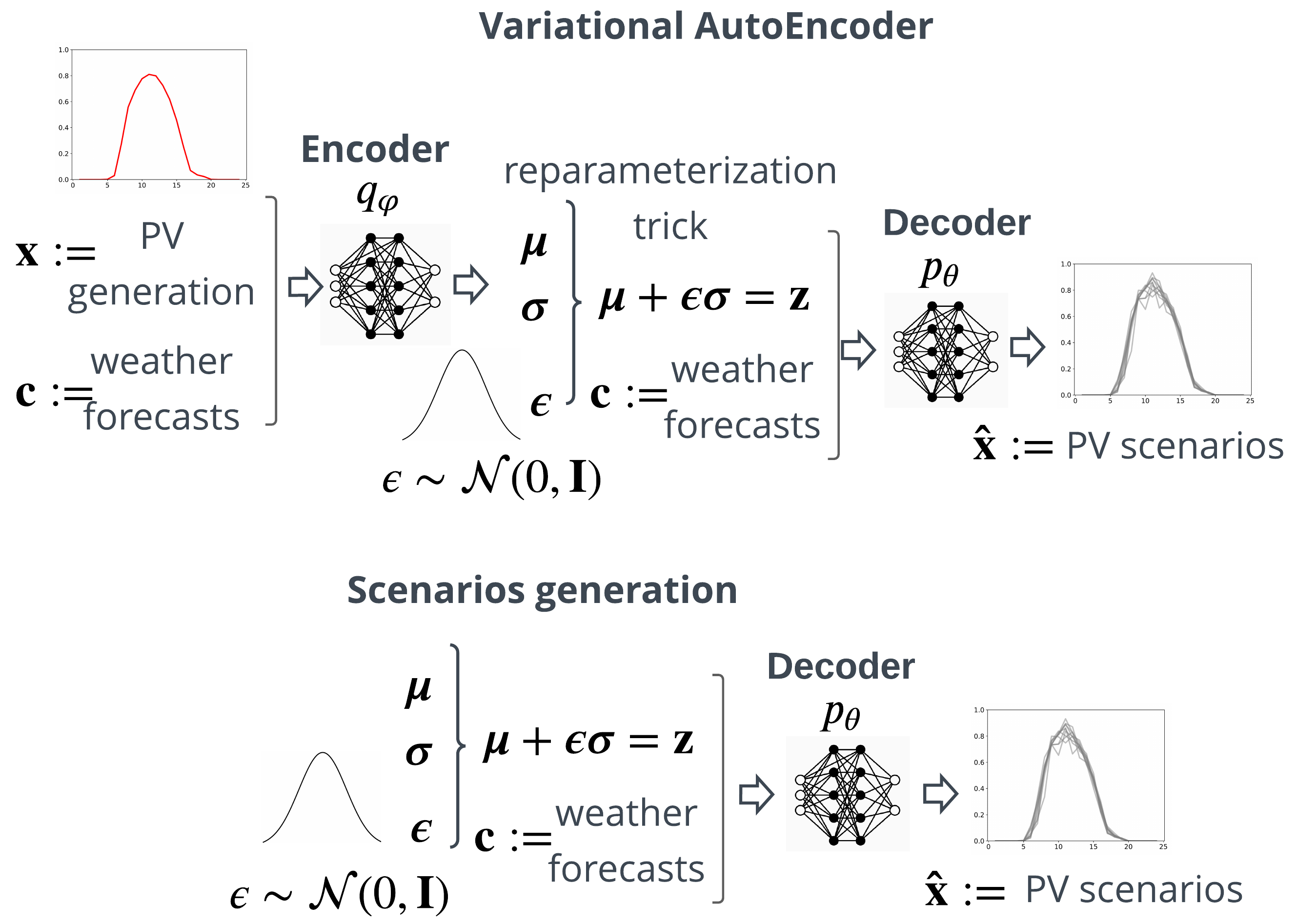}
	\caption{The process of the conditional variational autoencoder is illustrated for PV generation. \\
	The VAE is trained by maximizing the variational lower bound given a dataset composed of PV observations and weather forecasts. The encoder $q_\varphi$ maps the variable of interest $\mathbf{x}$ to a latent space $\mathbf{z}$. The decoder $p_\theta$ generates the PV scenarios $\mathbf{\hat{x}}$ by taking as inputs samples $\mathbf{z}$ from the latent space and the weather forecasts $\mathbf{c}$.}
	\label{fig:VAE_structure}
\end{figure}

A VAE is a deep latent variable model composed of an \textit{encoder} and a \textit{decoder} which are jointly trained to maximize a lower bound on the likelihood. The encoder $q_\varphi(\cdot) : \mathbb{R}^T \times \mathbb{R}^{|\mathbf{c}|}  \rightarrow \mathbb{R}^d$ approximates the intractable posterior $p(\mathbf{z}|\mathbf{x}, \mathbf{c})$, and the decoder $p_\theta(\cdot) : \mathbb{R}^d \times \mathbb{R}^{|\mathbf{c}|} \rightarrow \mathbb{R}^T $ the likelihood $p(\mathbf{x}|\mathbf{z}, \mathbf{c})$ with $\mathbf{z} \in \mathbb{R}^d$.
Maximum likelihood is intractable as it would require marginalizing with respect to all possible realizations of the latent variables $\mathbf{z}$. \citet{kingma2013auto} addressed this issue by maximizing the \textit{variational lower bound} $\mathcal{L}_{\theta, \varphi}(\mathbf{x, c})$ as follows
\begin{subequations}
\label{eq:variational_lower_bound}	
\begin{align}
\log p_\theta(\mathbf{x}|\mathbf{c}) = & KL[ q_\varphi(\mathbf{z}|\mathbf{x}, \mathbf{c}) || p(\mathbf{z}|\mathbf{x},\mathbf{c})] +\mathcal{L}_{\theta, \varphi}(\mathbf{x, c}) , \\
\geq & \mathcal{L}_{\theta, \varphi}(\mathbf{x, c}), \\
\mathcal{L}_{\theta, \varphi}(\mathbf{x, c}) := & \mathbb{E}_{q_\varphi(\mathbf{z}|\mathbf{x}, \mathbf{c})}\left[\log\frac{p(\mathbf{z}) p_\theta(\mathbf{x}|\mathbf{z}, \mathbf{c})}{q_\varphi(\mathbf{z}|\mathbf{x}, \mathbf{c})}\right],
\end{align}
\end{subequations}
as the Kullback-Leibler (KL) divergence \citep{perez2008kullback} is non-negative. Appendix \ref{annex:ijf-vae} details how to compute the gradients of $\mathcal{L}_{\theta, \varphi}(\mathbf{x, c})$, and its exact expression for the implemented VAE composed of fully connected neural networks for both the encoder and decoder. Figure~\ref{fig:VAE_structure} depicts the process of a conditional variational autoencoder for PV generation.

\subsubsection{Generative adversarial networks}

\begin{figure}[htbp]
	\centering
	\includegraphics[width=0.7\linewidth]{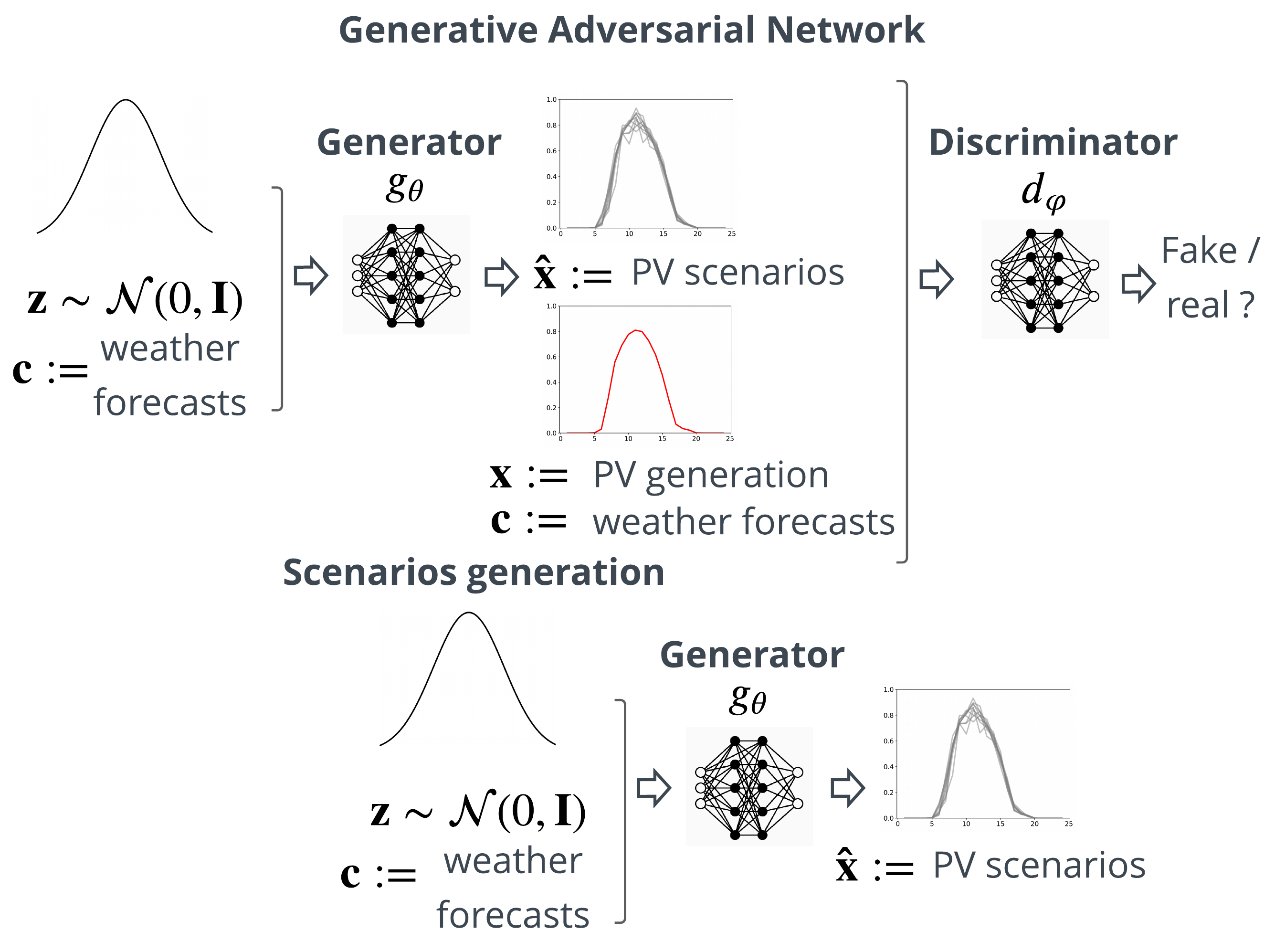}
	\caption{The process of the conditional generative adversarial network is illustrated for PV generation. \\
	The GAN is trained by solving a min-max problem given a dataset composed of PV observations $\mathbf{x}$ and weather forecasts. The generator $g_\theta$ computes PV scenarios $\mathbf{\hat{x}}$ by taking as inputs samples from the Normal distribution $\mathbf{z}$ and the weather forecasts $\mathbf{c}$, and the decoder $d_\phi$ tries to distinguishes true data from scenarios.}
	\label{fig:GAN_structure}
\end{figure}

GANs are a class of deep generative models proposed by \citet{goodfellow2014generative} where the key idea is the adversarial training of two neural networks, the \textit{generator} and the \textit{discriminator}, during which the generator learns iteratively to produce realistic scenarios until they cannot be distinguished anymore by the discriminator from real data. The generator $g_\theta(\cdot) : \mathbb{R}^d \times \mathbb{R}^{|\mathbf{c}|} \rightarrow \mathbb{R}^T $ maps a latent vector $\mathbf{z} \in \mathbb{R}^d $ equipped with a known and tractable prior probability density function $p(\mathbf{z})$, \textit{e.g.}, a Normal distribution, onto a sample $\mathbf{x} \in  \mathbb{R}^T $,  and is trained to fool the discriminator. The discriminator $d_\phi(\cdot) :\mathbb{R}^T\times \mathbb{R}^{|\mathbf{c}|} \rightarrow [0, 1]$ is a classifier trained to distinguish between true samples $\mathbf{x}$ and generated samples $\mathbf{\hat{x}}$. \citet{goodfellow2014generative} demonstrated that solving the following min-max problem
\begin{align}
\label{eq:WGAN_GP}	
 \theta^\star = & \arg \min_\theta \max_\phi V(\phi, \theta),
\end{align}
where $V(\phi, \theta)$ is the value function, recovers the data generating distribution if $g_\theta(\cdot)$ and $d_\phi(\cdot)$ are given enough capacity. The state-of-the-art conditional Wasserstein GAN with gradient penalty (WGAN-GP) proposed by \citet{gulrajani2017improved} is implemented with $V(\phi, \theta)$ defined as
\begin{subequations}
\label{eq:WGAN-GP}	
\begin{align}
V(\phi, \theta) = &   - \bigg(\underset{\mathbf{\hat{x}}}{\mathbb{E}} [ d_\phi (\mathbf{\hat{x}}|\mathbf{c}) ] -\underset{\mathbf{x} }{\mathbb{E}} [ d_\phi (\mathbf{x}|\mathbf{c}) ]  + \lambda \text{GP} \bigg) , \\
\text{GP} = & \underset{\mathbf{\tilde{x}}}{\mathbb{E}}  \bigg[ \bigg( \Vert \nabla_{\mathbf{\tilde{x}}}  d_\phi (\mathbf{\tilde{x}}|\mathbf{c})  \Vert_2 - 1 \bigg)^2 \bigg ],
\end{align}
\end{subequations}
where $\mathbf{\tilde{x}}$ is implicitly defined by sampling convex combinations between the data and the generator distributions $\mathbf{\tilde{x}} = \rho \mathbf{\hat{x}} + (1-\rho) \mathbf{x}$ with $\rho \sim \mathbb{U}(0,1)$. The WGAN-GP constrains the gradient norm of the discriminator's output with respect to its input to enforce the 1-Lipschitz conditions. This strategy differs from the weight clipping of WGAN that sometimes generates only poor samples or fails to converge.
Appendix \ref{annex:ijf-gan} details the successive improvements from the original GAN to the WGAN and the final WGAN-GP implemented, referred to as GAN in the rest of the Chapter.
Figure~\ref{fig:GAN_structure} depicts the process of a conditional generative adversarial network for PV generation.

\clearpage

\subsection{Theoretical comparison}\label{sec:AE-comparison}

Normalizing flows are a generative model that allows exact likelihood calculation. They are efficiently parallelizable and offer a valuable latent space for downstream tasks. In contrast to GANs and VAEs, NFs explicitly learn the data distribution and provide direct access to the exact likelihood of the model’s parameters, hence offering a sound and direct way to optimize the network parameters \citep{wehenkel2020graphical}. However, NFs suffer from drawbacks \citep{bond2021deep}. One disadvantage of requiring transformations to be invertible is that the input dimension must be equal to the output dimension, making the model difficult to train or inefficient. Each transformation must be sufficiently expressive while being easily invertible to compute the Jacobian determinant efficiently. The first issue is also raised by \citet{ruthotto2021introduction} where ensuring sufficient similarity of the distribution of interest and the latent distribution is of high importance to obtain meaningful and relevant samples. However, in our numerical simulations, we did not encounter this problem. Concerning the second issue, the UMNN-MAF transformation provides an expressive and effective way of computing the Jacobian.

VAEs indirectly optimize the log-likelihood of the data by maximizing the variational lower bound. The advantage of VAEs over NFs is their ability to handle non-invertible generators and the arbitrary dimension of the latent space. However, it has been observed that when applied to complex datasets such as natural images, VAEs samples tend to be unrealistic. There is evidence that the limited approximation of the true posterior, with a common choice being a normally distributed prior with diagonal covariance, is the root cause \citep{zhao2017towards}. This statement comes from the field of computer vision. However, it may explain the shape of the scenarios observed in our numerical experiments in Section \ref{sec:ijf-numerical_results}.

The training of GANs relies on a min-max problem where the generator and the discriminator are jointly optimized. Therefore, it does not rely on estimates of the likelihood or latent variable. The adversarial nature of GANs makes them notoriously difficult to train due to the saddle point problem \citep{arjovsky2017towards}. Another drawback is the mode collapsing, where one network stays in bad local minima, and only a small subset of the data distribution is learned. Several improvements have been designed to address these issues, such as the Wasserstein GAN with gradient penalty. Thus, GANs models are widely used in computer vision and power systems. However, most GAN approaches require cumbersome hyperparameter tuning to achieve similar results to VAEs or NFs. In our numerical simulations, the GAN is highly sensitive to hyperparameter variations, which is consistent with \citep{ruthotto2021introduction}.

Each method has its advantages and drawbacks and makes trade-offs in terms of computing time, hyper-parameter tuning, architecture complexity, \textit{etc}. Therefore, the choice of a particular method is dependent on the user criteria and the dataset considered. In addition, the challenges of power systems are different from computer vision. Therefore, the limitations established in the computer vision literature such as \citet{bond2021deep} and \citet{ruthotto2021introduction} must be addressed with caution. Therefore, we encourage the energy forecasting practitioners to test and compare these methods in power systems applications.

\section{Quality assessment}\label{sec:ijf-quality_assessment}

For predictions in any form, one must differentiate between their quality and their value \citep{morales2013integrating}. Forecast quality corresponds to the ability of the forecasts to genuinely inform of future events by mimicking the characteristics of the processes involved. Forecast value relates, instead, to the benefits from using forecasts in a decision-making process, such as participation in the electricity market. The forecast value is assessed in Part \ref{part:optimization} Chapter \ref{chap:energy-retailer} by considering the day-ahead market scheduling of electricity aggregators, such as energy retailers or generation companies.

Evaluating and comparing generative models remains a challenging task. Several measures have been introduced with the emergence of new models, particularly in the field of computer vision. However, there is no consensus or guidelines as to which metric best captures the strengths and limitations of models. Generative models need to be evaluated directly to the application they are intended for \citep{theis2015note}. Indeed, good performance to one criterion does not imply good performance to the other criteria. Several studies propose metrics and make attempts to determine the pros and cons. We selected two that provide helpful information. (1) 24 quantitative and five qualitative measures for evaluating generative models are reviewed and compared by \citet{borji2019pros} with a particular emphasis on GAN-derived models. (2) several representative sample-based evaluation metrics for GANs are investigated by \citet{xu2018empirical} where the kernel Maximum Mean Discrepancy (MMD) and the 1-Nearest-Neighbour (1-NN) two-sample test seem to satisfy most of the desirable properties. The key message is to combine several complementary metrics to assess the generative models. Some of the metrics proposed are related to image generation and cannot directly be transposed to energy forecasting.

Therefore, we used eight complementary quality metrics to conduct a relevant quality analysis inspired by the energy forecasting and computer vision fields.
They can be divided into four groups: (1) the \textit{univariate} metrics composed of the continuous ranked probability score, the quantile score, and the reliability diagram. They can only assess the quality of the scenarios with respect to their marginals; (2) the \textit{multivariate} metrics are composed of the energy and the variogram scores. They can directly assess multivariate scenarios; (3) the \textit{specific} metrics composed of a classifier-based metric and the correlation matrix between scenarios for a given context; (4) the Diebold and Mariano statistical test.
The univariate and multivariate metrics are already defined in Chapter \ref{chap:forecast_evaluation}. Therefore, only the specific metrics and statistical test are introduced in this Section.

\subsection{Classifier-based metric}
%

Modern binary classifiers can be easily turned into robust two-sample tests where the goal is to assess whether two samples are drawn from the same distribution \citep{lehmann2006testing}. In other words, it aims at assessing whether a generated scenario can be distinguished from an observation. 
To this end, the generator is evaluated on a held-out testing set that is split into a testing-train and testing-test subsets. The testing-train set is used to train a classifier, which tries to distinguish generated scenarios from the actual distribution. Then, the final score is computed as the performance of this classifier on the testing-test set.

In principle, any binary classifier can be adopted for computing classifier two-sample tests (C2ST). A variation of this evaluation methodology is proposed by \citet{xu2018empirical} and is known as the 1-Nearest Neighbor (NN) classifier. The advantage of using 1-NN over other classifiers is that it requires no special training and little hyper-parameter tuning. This process is conducted as follows. Given two sets of observations $S_r$ and generated $S_g$ samples with the same size, \textit{i.e.}, $|S_r| = |S_g|$, it is possible to compute the leave-one-out (LOO) accuracy of a 1-NN classifier trained on $S_r$ and $S_g$ with positive labels for $S_r$ and negative labels for $S_g$. The LOO accuracy can vary from 0 \% to 100 \%. The 1-NN classifier should yield a 50 \% LOO accuracy when $|S_r| = |S_g|$ is large. This is achieved when the two distributions match. Indeed, the level 50 \% happens when a label is randomly assigned to a generated scenario. It means the classifier is not capable of discriminating generated scenarios from observations. If the generative model over-fits $S_g$ to $S_r$, \textit{i.e.}, $S_g = S_r$, and the accuracy would be 0 \%. On the contrary, if it generates widely different samples than observations, the performance should be 100 \%. Therefore, the closer the LOO accuracy is to 1, the higher the degree of under-fitting of the model. The closer the LOO accuracy is to 0, the higher the degree of over-fitting of the model. The C2ST approach using LOO with 1-NN is adopted by \citet{qi2020optimal} to assess the PV and wind power scenarios of a $\beta$ VAE.

However, this approach has several limitations. First, it uses the testing set to train the classifier during the LOO. Second, the 1-NN is very sensitive to outliers as it simply chose the closest neighbor based on distance criteria. This behavior is amplified when combined with the LOO where the testing-test set is composed of only one sample. Third, the euclidian distance cannot deal with a context such as weather forecasts. Therefore, we cannot use a conditional version of the 1-NN using weather forecasts to classify weather-based renewable generation and the observations. Fourth, C2ST with LOO cannot provide ROC curve but only accuracy scores. 
An essential point about ROC graphs is that they measure the ability of a classifier to produce good relative instance scores. In our case, we are interested in discriminating the generated scenarios from the observations, and the ROC provides more information than the accuracy metric to achieve this goal. A standard method to reduce ROC performance to a single scalar value representing expected performance is to calculate the area under the ROC curve, abbreviated AUC. The AUC has an important statistical property: it is equivalent to the probability that the classifier will rank a randomly chosen positive instance higher than a randomly chosen negative instance \citep{fawcett2004roc}.

To deal with these issues, we decided to modify this classifier-based evaluation by conducting the C2ST as follows: (1) the scenarios generated on the learning set are used to train the classifier using the C2ST. Therefore, the classifier uses the entire testing set and can compute ROC; (2) the classifier is an Extra-Trees classifier that can deal with context such as weather forecasts.

\begin{figure}[htb]
\centering
\includegraphics[width=0.3\linewidth]{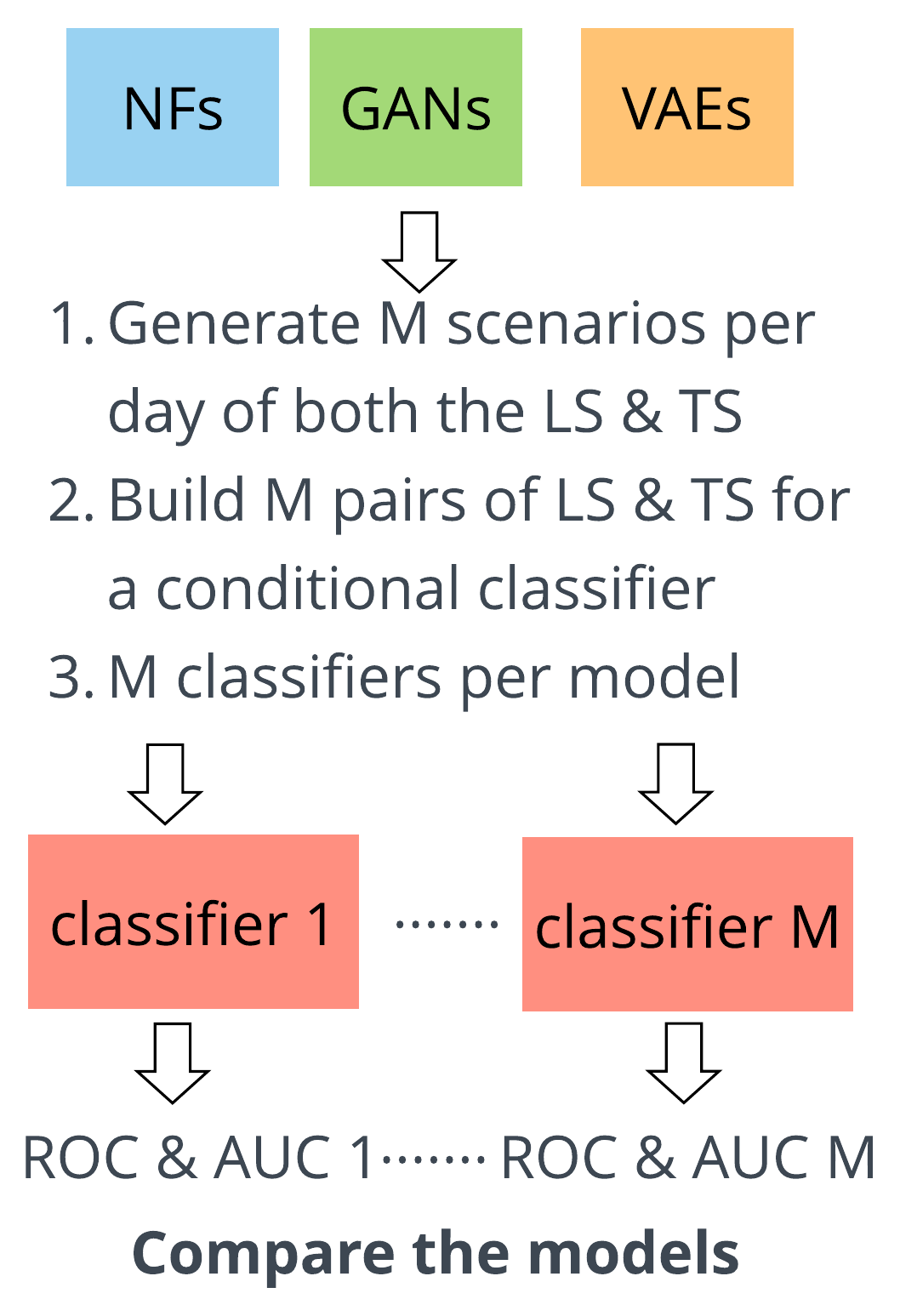}
\caption{Classifier-based metric methodology. \\ 
Each generative model generates $M$ scenarios per day of the learning and testing sets. They are used to build $M$ pairs of learning and testing sets for a conditional classifier by including an equal proportion of observations and weather forecasts. $M$ conditional classifiers, per model, are trained and make predictions. The $M$ ROC and AUC are computed per model, and the results are compared.
}
\label{fig:AE-classifier-based}
\end{figure}

More formally, for a given generative model $g$, the following steps are conducted:
\begin{enumerate}
    \item Initialization step: the generative model $g$ has been trained on the LS and has generated $M$ weather-based scenarios per day of both the LS
     and TS: $\{  \mathbf{\hat{x}}_\text{LS}^i \}_{i=1}^M :=\cup_{d \in \text{LS}} \{  \mathbf{\hat{x}}_d^i \}_{i=1}^M$ and $\{  \mathbf{\hat{x}}_\text{TS}^i \}_{i=1}^M :=\cup_{d \in \text{TS}} \{  \mathbf{\hat{x}}_d^i \}_{i=1}^M$. For the sake of clarity the index $g$ is omitted, but both of these sets are dependent on model $g$.
    \item $M$ pairs of learning and testing sets are built with an equal proportion of generated scenarios and observations: $\mathcal{D}_\text{LS}^i := \bigg\{ \{\mathbf{\hat{x}}_\text{LS}^i, 0\} \cup \{    \{\mathbf{x}_\text{LS}^i, 1\}  \bigg\} $ and $\mathcal{D}_\text{TS}^i = \bigg\{ \{\mathbf{\hat{x}}_\text{TS}^i, 0\} \cup \{    \{\mathbf{x}_\text{TS}^i, 1\}  \bigg\} $. Note: $|\mathcal{D}_\text{LS}^i| = 2|\text{LS}|$ and $|\mathcal{D}_\text{TS}^i| = 2|\text{TS}|$.
    \item For each pair of learning and testing sets $\{\mathcal{D}_\text{LS}^i, \mathcal{D}_\text{TS}^i\}_{i=1}^M$ a classifier $d_g^i$ is trained and makes predictions.
    \item The $\text{ROC}_g^i$ curves and corresponding $\text{AUC}_g^i$ are computed for $i=1, \cdots, M$.
\end{enumerate}
This classifier-based methodology is conducted for all models $g$, and the results are compared. Figure \ref{fig:AE-classifier-based} depicts the overall approach.
The classifiers $d_g^i$ are all Extra-Trees classifier made of $1000$ unconstrained trees with the hyper-parameters "$\text{max\_depth}$" set to "None", and ``$\text{n\_estimators}$" to 1 000.

\subsection{Correlation matrix between scenarios}
The second specific metric consists of computing the correlation matrix between the scenarios generated for given weather forecasts. Formally, let $ \{    \mathbf{\hat{x}}^i  \}_{i=1}^M$ be the set of $M$ scenarios generated for a given day of the testing set. It is a matrix ($M \times 24$) where each row is a scenario. Then, the Pearson’s correlation coefficients are computed into a correlation matrix ($24 \times 24$). This metric indicates the variety of scenario shapes.

\subsection{Diebold-Mariano test}

Using relevant metrics to assess the forecast quality is essential. However, it is also necessary to analyze whether any difference in accuracy is statistically significant. Indeed, when different models have almost identical values in the selected error measures, it is not easy to draw statistically significant conclusions on the outperformance of the forecasts of one model by those of another.
The Diebold-Mariano (DM) test \citep{diebold2002comparing} is probably the most commonly used statistical testing tool to evaluate the significance of differences in forecasting accuracy. It is model-free, \textit{i.e.}, it compares the forecasts of models, and not models themselves.
The DM test is used in this study to assess the CRPS, QS, ES, and VS metrics. The CRPS and QS are univariate scores, and a value of CRPS and QS is computed per marginal (time period of the day). Therefore, the multivariate variant of the DM test is implemented following \citet{ziel2018day}, where only one statistic for each pair of models is computed based on the 24-dimensional vector of errors for each day.

For a given day $d$ of the testing set, let $\epsilon_d \in \mathbb{R}$ be the error computed by an arbitrary forecast loss function of the observation and scenarios. The test consists of computing the difference between the errors of the pair of models $g$ and $h$ over the testing set
\begin{equation}
\label{eq:univariate-loss}
\Delta(g,h)_d  = \epsilon^g_d - \epsilon^h_d, \quad \forall d \in \text{TS},
\end{equation}
and to perform an asymptotic $z$-test for the null hypothesis that the expected forecast error is equal and the mean of differential loss series is zero $ \mathbb{E}[\Delta(g,h)_d] = 0$. It means there is no statistically significant difference in the accuracy of the two competing forecasts.
The statistic of the test is deduced from the asymptotically standard normal distribution as follows
\begin{equation}
\label{eq:DM-value}
\text{DM}(g,h) = \sqrt{\# \text{TS}}\frac{\hat{\mu}}{\hat{\sigma}},
\end{equation}
with $\# \text{TS}$ the number of days of the testing set, $\hat{\mu}$ and $\hat{\sigma}$ the sample mean and the standard deviation of $\Delta(g,h)$.
Under the assumption of covariance stationarity of the loss differential series $\Delta(g,h)_d$, the DM statistic is asymptotically standard normal.
The lower the $p$-value, \textit{i.e.}, the closer it is to zero, the more the observed data is inconsistent with the null hypothesis: $ \mathbb{E}[\Delta(g,h)_d] < 0$ the forecasts of the model $h$ are more accurate than those of model $g$. If the $p$-value is less than the commonly accepted level of 5 \%, the null hypothesis is typically rejected. It means that the forecasts of model $g$ are significantly more accurate than those of model $h$.

When considering the ES or VS scores, there is a value per day of the testing set $\text{ES}_d$ or $\text{VS}_d$. In this case, $\epsilon_d = \text{ES}_d$ or $\epsilon_d = \text{VS}_d$.
However, when considering the CRPS or QS, there is a value per marginal and per day of the testing set $\text{CRPS}_{d,k}$ or $\text{QS}_{d,k}$. 
A solution consists of computing 24 independent tests, one for each hour of the day. Then, to compare the models based on the number of hours for which the predictions of one model are significantly better than those of another. Another way consists of a multivariate variant of the DM-test with the test performed jointly for all hours using the multivariate loss differential series.
In this case, for a given day $d$, $\mathbf{\epsilon}^g_d = [\epsilon^g_{d,1}, \ldots,  \epsilon^g_{d,24}]^\intercal$, $\mathbf{\epsilon}^h_d = [\epsilon^h_{d,1}, \ldots,  \epsilon^h_{d,24}]^\intercal$ are the vectors of errors for a given metric of models $g$ and $h$, respectively. Then the multivariate loss differential series
\begin{equation}
\label{eq:multivariate-loss}
\Delta(g,h)_d  = \Vert \mathbf{\epsilon}^g_d\Vert_1 - \Vert \mathbf{\epsilon}^h_d\Vert_1,
\end{equation}
defines the differences of errors using the $\Vert \cdot \Vert_1$ norm. Then, the $p$-value of two-sided DM tests is computed for each model pair as described above.
The univariate version of the test has the advantage of providing a more profound analysis as it indicates which forecast is significantly better for which hour of the day.
The multivariate version enables a better representation of the results as it summarizes the comparison in a single $p$-value, which can be conveniently visualized using heat maps arranged as chessboards. In this study, we decided to adopt the multivariate DM-test for the CRPS and QS.

\section{Case study}\label{sec:ijf-implementation_details}

\begin{figure}[htbp]
	\centering
	\includegraphics[width=0.6\linewidth]{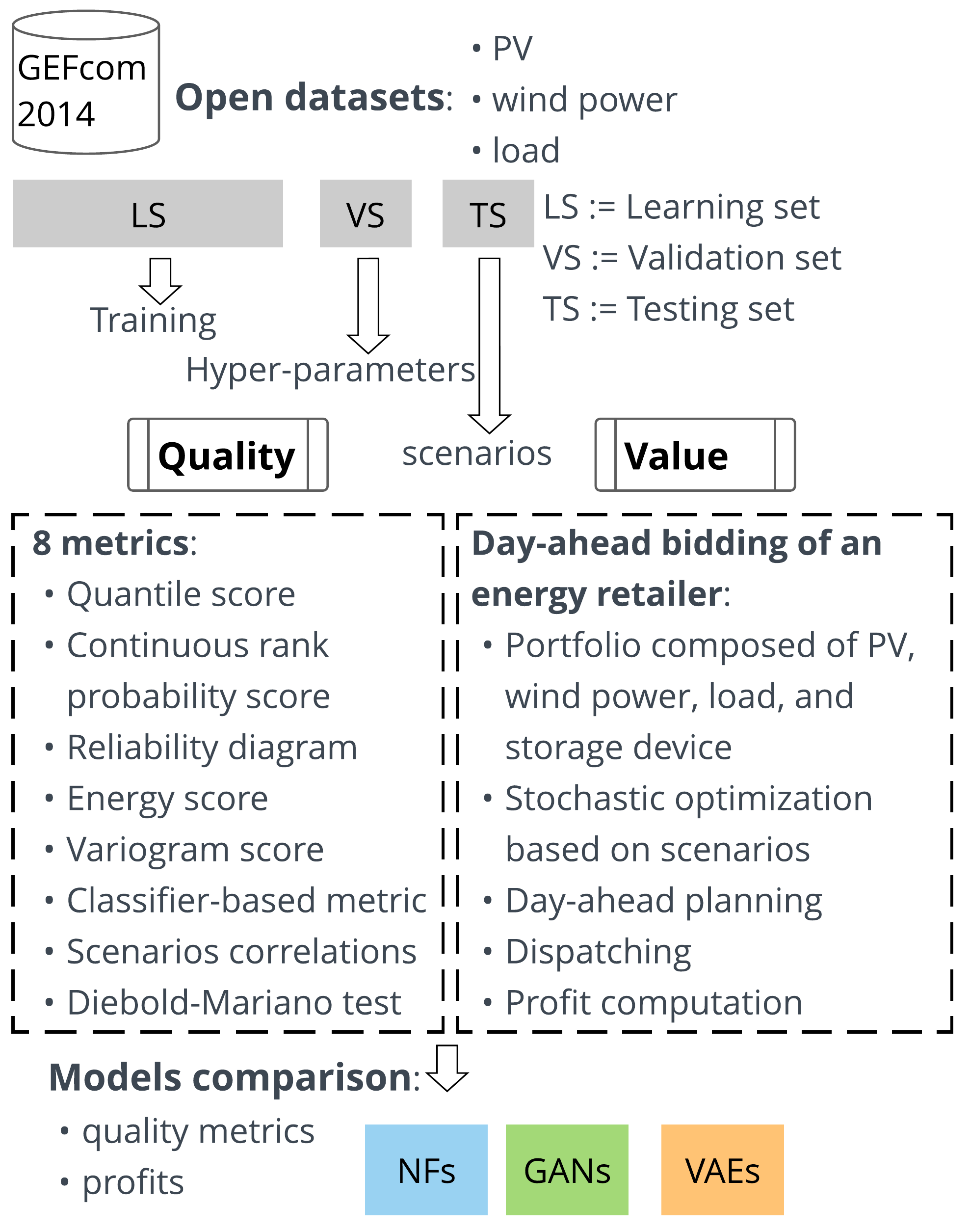}
	\caption{Methodology to assess the quality and value of the GAN, VAE, and NF models implemented in this study. \\
    The PV, wind power, and load datasets of the open-access Global Energy Forecasting Competition 2014 are divided into three parts: learning, validation, and testing sets. The learning set is used to train the models, the validation set to select the optimal hyper-parameters, and the testing set to conduct the numerical experiments. The quality and value of the models are assessed by using the scenarios generated on the testing set. The quality evaluation consists of eight complementary metrics. The value assessment uses the simple and easily reproducible case study of the day-ahead bidding of an energy retailer. The energy retailer portfolio is composed of PV, wind power generation, load, and a storage system device. The retailer bids on the day-ahead market by computing a planning based on stochastic optimization. The dispatch is computed by using the observations of the PV generation, wind power, and load. Then, the profits are evaluated and compared.
	}
	\label{fig:AE-numerical-experiments-methodology}
\end{figure}

The quality and value evaluations of the models are conducted on the load, wind, and PV tracks of the open-access GEFCom 2014 dataset \citep{hong2016bprobabilistic}, composed of one, ten, and three zones, respectively. See Chapter \ref{chap:energy-retailer} for the value evaluation and the energy retailer problem statement. 
Figure \ref{fig:AE-numerical-experiments-methodology} depicts the methodology to assess both the quality and value of the GAN, VAE and NF models implemented in this study.

\subsection{Implementation details}

By appropriate normalization, we standardize the weather forecasts to have a zero mean and unit variance. Table~\ref{tab:features} provides a summary of the implementation details described in what follows. For the sake of proper model training and evaluation, the dataset is divided into three parts per track considered: learning, validation, and testing sets. The learning set (LS) is used to fit the models, the validation set (VS) to select the optimal hyper-parameters, and the testing set (TS) to assess the forecast quality and value. The number of samples~($\#$), expressed in days, of the VS and TS, is $50 \cdot n_z$, with $n_z$ the number of zones of the track considered. The 50 days are selected randomly from the dataset, and the learning set is composed of the remaining part with $D \cdot n_z$ samples, where $D$ is provided for each track in Table~\ref{tab:features}.
The NF, VAE, and GAN use the weather forecasts as inputs to generate on a day-ahead basis $M$ scenarios $\mathbf{\hat{x}} \in \mathbb{R}^T$. 


\subsection*{Wind track}

The zonal $\mathbf{u}^\text{10}$, $\mathbf{u}^\text{100}$ and meridional $\mathbf{v}^\text{10}$, $\mathbf{v}^\text{100}$ wind components at 10 and 100 meters are selected, and six features are derived following the formulas provided by \citet{landry2016probabilistic} to compute the wind speed $\mathbf{ws}^\text{10}$, $\mathbf{ws}^\text{100}$, energy $\mathbf{we}^\text{10}$, $\mathbf{we}^\text{100}$ and direction $\mathbf{wd}^\text{10}$, $\mathbf{wd}^\text{100}$ at 10 and 100 meters
\begin{subequations}
\label{eq:wind_derived_features}	
\begin{align}
\mathbf{ws} = & \sqrt{\mathbf{u} + \mathbf{v}} ,\\
\mathbf{we} = & \frac{1}{2} \mathbf{ws}^3,\\
\mathbf{wd} = & \frac{180}{\pi} \arctan(\mathbf{u}, \mathbf{v}).
\end{align}
\end{subequations}
For each generative model, the wind zone is taken into account with one hot-encoding variable $Z_1, \ldots, Z_{10}$, and the wind feature input vector for a given day $d$ is
\begin{align}
\label{eq:wind_input}	
\mathbf{c}_d^\text{wind} = & [\mathbf{u}^\text{10}_d, \mathbf{u}^\text{100}_d, \mathbf{v}^\text{10}_d, \mathbf{v}^\text{100}_d, \mathbf{ws}^\text{10}_d, \mathbf{ws}^\text{100}_d, \notag \\
& \mathbf{we}^\text{10}_d, \mathbf{we}^\text{100}_d, \mathbf{wd}^\text{10}_d, \mathbf{wd}^\text{100}_d, Z_1, \ldots, Z_{10}],
\end{align}
of dimension $n_f \cdot T + n_z = 10 \cdot 24 + 10$. 

\subsection*{PV track}

The solar irradiation $\mathbf{I}$, the air temperature $\mathbf{T}$, and the relative humidity $\mathbf{rh}$ are selected, and two features are derived by computing $\mathbf{I}^2$ and $\mathbf{IT}$. For each generative model, the PV zone is taken into account with one hot-encoding variable $Z_1, Z_2, Z_3$, and the PV feature input vector for a given day $d$ is
\begin{align}
\label{eq:PV_input}	
\mathbf{c}_d^\text{PV} = & [\mathbf{I}_d, \mathbf{T}_d, \mathbf{rh}_d, \mathbf{I}^2_d, \mathbf{IT}_d, Z_1, Z_2, Z_3],
\end{align}
of dimension $n_f \cdot T + n_z$. For practical reasons, the periods where the PV generation is always 0, across all zones and days, are removed, and the final dimension of the input feature vector is $n_f \cdot T + n_z = 5 \cdot 16 + 3$.

\subsection*{Load track}

The 25 weather station temperature $\mathbf{w}_1, \ldots, \mathbf{w}_{25}$ forecasts are used. There is only one zone, and the load feature input vector for a given day $d$ is
\begin{align}
\label{eq:load_input}	
\mathbf{c}_d^\text{load} = & [\mathbf{w}_1, \ldots, \mathbf{w}_{25}],
\end{align}
of dimension $n_f \cdot T = 25 \cdot 24$.
\begin{table}[htbp]
\renewcommand{\arraystretch}{1.25}
	\begin{center}
		\begin{tabular}{lrrr}
			\hline \hline
			            			     & Wind &  PV   & Load  \\ \hline
		$T$ periods 	                 & 24 &  16     & 24  \\
		$n_z$ zones     			     & 10 &  3      & ---  \\
		$n_f$ features  			     & 10 &  5      & 25  \\
		$\mathbf{c}_d$ dimension           & $n_f \cdot T + n_z$ &  $n_f \cdot T + n_z$   & $n_f \cdot T$  \\
		$\#$ LS (days)   		         & $631\cdot n_z$ &  $720 \cdot n_z$        	  & $1999$  \\
		$\#$ VS/TS (days)   	         & $50\cdot n_z$ &  $50 \cdot n_z$         	  & $50$\\
         \hline \hline
		\end{tabular}
		\caption{Dataset and implementation details. \\
		Each dataset is divided into three parts: learning, validation, and testing sets. The number of samples~($\#$) is expressed in days and is set to 50 days for the validation and testing sets. $T$ is the number of periods per day considered, $n_z$ the number of zones of the dataset, $n_f$ the number of weather variables used, and $\mathbf{c}_d$ is the dimension of the conditional vector for a given day that includes the weather forecasts and the one hot-encoding variables when there are several zones. Note: the days of the learning, validation, and testing sets are selected randomly.}
		\label{tab:features}
	\end{center}
\end{table}

\subsection{Hyper-parameters}\label{annex:hp}

\noindent Table~\ref{tab:model_hp} provides the hyper-parameters of the NF, VAE, and GAN implemented. The Adam optimizer \citep{kingma2014adam} is used to train the generative models with a batch size of 10\% of the learning set.
\begin{table}[htbp]
\renewcommand{\arraystretch}{1.25}
	\begin{center}
		\begin{tabular}{m{0.1cm}lrrr}
			\hline  \hline
			                        & & Wind &  PV  & Load  \\ \hline
			\multirow{5}{*}{(a)} & 
			Embedding Net   	    & $4 \times 300$ & $4 \times 300$    & $4 \times 300$ \\ 
		&	Embedding size          & 40 & 40  & 40\\ 
		&	Integrand Net   	    & $3 \times 40$  & $3 \times 40$	 & $3 \times 40$ \\ 
		&	Weight decay            & 5.10$^{-4}$    & 5.10$^{-4}$       & 5.10$^{-4}$   \\
        &   Learning rate           & 10$^{-4}$      & 5.10$^{-4}$       & 10$^{-4}$  \\ \hline 
			\multirow{4}{*}{(b)} & 
			Latent dimension        & 20 & 40  & 5  \\
		&	E/D Net                 & $1 \times 200$ & $2 \times 200$   & $1 \times 500 $ \\ 
		&	Weight decay            & 10$^{-3.4}$    & 10$^{-3.5}$      & 10$^{-4}$  \\
		&	Learning rate           & 10$^{-3.4}$    & 10$^{-3.3}$      & 10$^{-3.9}$  \\ \hline 
			 \multirow{4}{*}{(c)}& 
			Latent dimension        & 64 & 64 & 256  \\
		&	G/D Net                 & $2 \times 256$ & $3 \times 256$   & $2 \times 1024 $ \\ 
		&	Weight decay            & 10$^{-4}$      & 10$^{-4}$        & 10$^{-4}$  \\
		&	Learning rate           & 2.10$^{-4}$    & 2.10$^{-4}$      & 2.10$^{-4}$  \\ \hline \hline
		\end{tabular}
		\caption{(a) NF, (b) VAE, and (c) GAN hyper-parameters. \\
		The hyper-parameters selection is performed on the validation set using the Python library Weights \& Biases \citep{wandb}. This library is an experiment tracking tool for machine learning, making it easier to track experiments. The GAN model was the most time-consuming during this process, followed by the VAE and NF. Indeed, the GAN is highly sensitive to hyper-parameter modifications making it challenging to identify a relevant set of values. In contrast, the NF achieved satisfactory results, both in terms of scenarios shapes and quality, by testing only a few sets of hyper-parameter values.}
		\label{tab:model_hp}
	\end{center}
\end{table}
%
The NF implemented is a one-step monotonic normalizer using the UMNN-MAF\footnote{\url{https://github.com/AWehenkel/Normalizing-Flows}}. The embedding size $|h^i|$ is set to 40, and the embedding neural network is composed of $l$ layers of $n$ neurons ($l \times n$). The same integrand neural network $\tau^i(\cdot)$ $\forall i=1, \ldots, T$ is used and composed of 3 layers of $ |h^i|$ neurons ($3 \times 40$).
Both the encoder and decoder of the VAE are feed-forward neural networks ($l \times n$), ReLU activation functions for the hidden layers, and no activation function for the output layer.
Both the generator and discriminator of the GAN are feed-forward neural networks ($l \times n$). The activation functions of the hidden layers of the generator (discriminator) are ReLU (Leaky ReLU). The activation function of the discriminator output layer is ReLU, and there is no activation function for the generator output layer. The generator is trained once after the discriminator is trained five times to stabilize the training process, and the gradient penalty coefficient $\lambda$ in (\ref{eq:WGAN_GP}) is set to 10 as suggested by \citet{gulrajani2017improved}.

\begin{figure}[tb]
	\centering
    \includegraphics[width=\linewidth]{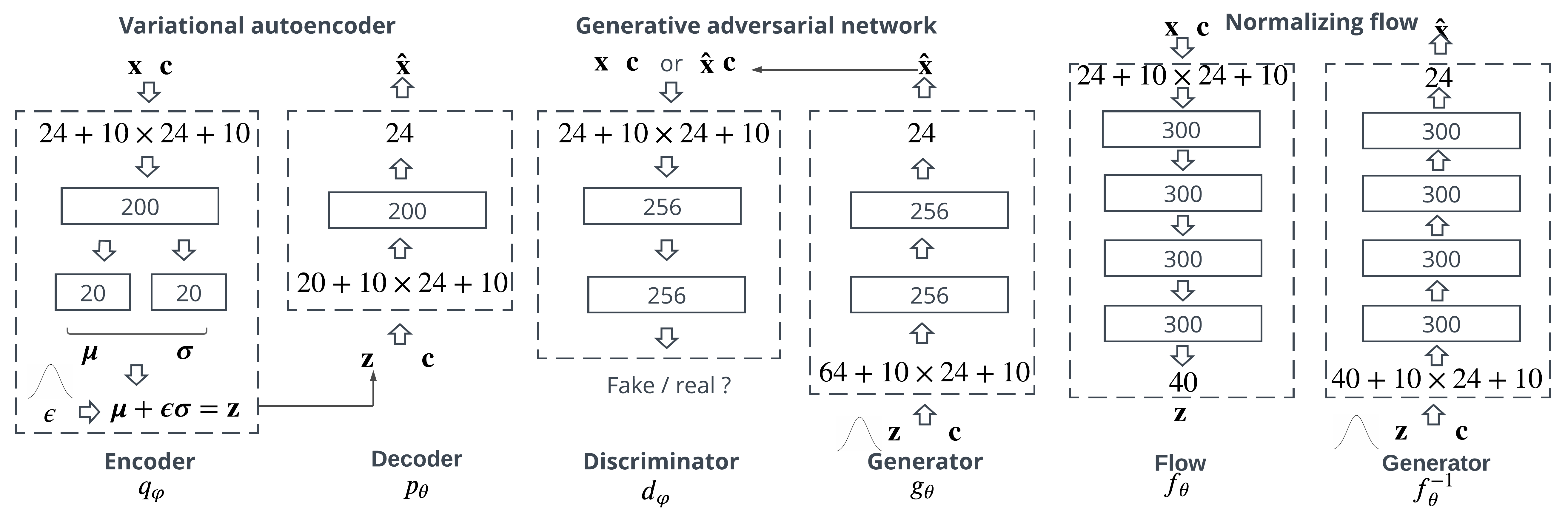}
	\caption{Variational autoencoder, generative adversarial network, and normalizing flow structures implemented for the wind dataset. \\
	VAE: both the encoder and decoder are feed-forward neural networks composed of one hidden layer with 200 neurons. Increasing the number of layers did not improve the results for this dataset. The latent space dimension is 20. \\
	GAN: both the discriminator and generator are feed-forward neural networks composed of two hidden layers with 256 neurons. The latent space dimension is 64. \\
	NF: a single-step monotonic normalizing flow is implemented with a feed-forward neural network composed of four hidden layers with 300 neurons. The latent space dimension is 40. Note: For clarity, the integrand network is not included but is a feed-forward neural network composed of three hidden layers with 40 neurons.
    Increasing the number of steps of the normalizing flow did not improve the results. The monotonic transformation is complex enough to capture the stochasticity of the variable of interest. However, when considering affine autoregressive normalizing flows, the number of steps should be generally higher. Numerical experiments indicated that a five-step autoregressive flow was required to achieve similar results for this dataset. Note: the results are not reported in this study for the sake of clarity}
	\label{fig:gan-vae-nf-details}
\end{figure}

Figure \ref{fig:gan-vae-nf-details} illustrates the VAE, GAN, and NF structures implemented for the wind dataset where the number of weather variables selected and the number of zones is 10, and 10, respectively. Recall, $\mathbf{c}:=$ weather forecasts,  $\mathbf{\hat{x}}:=$ scenarios $\mathbf{x}:=$ wind power observations, $\mathbf{z}:=$ latent space variable, $\mathbf{\epsilon}:=$ Normal variable (only for the VAE).

\clearpage 

\section{Quality results}\label{sec:ijf-numerical_results}

First, a thorough comparison of the models is conducted on the wind track. Second, all tracks are considered for the sake of clarity. Note that the model ranking slightly differs depending on the track.

\subsection{Wind track}

\begin{figure}[htbp]
	\centering
	\begin{subfigure}{.45\textwidth}
		\centering
		\includegraphics[width=\linewidth]{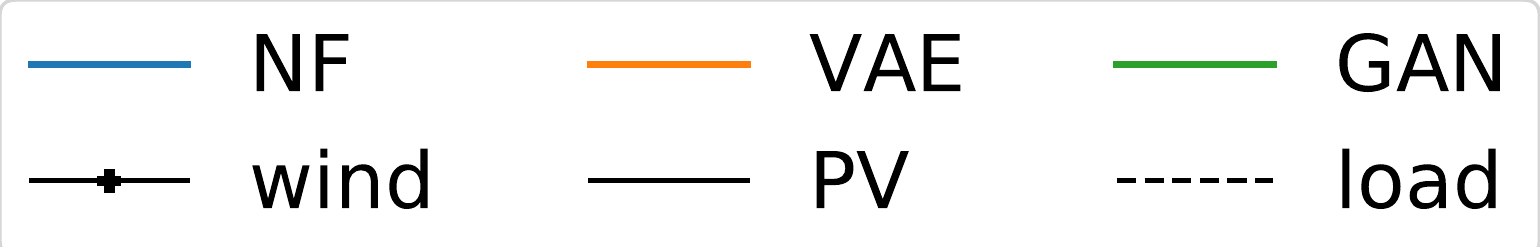}
	\end{subfigure}
	\begin{subfigure}{.33\textwidth}
		\centering
		\includegraphics[width=\linewidth]{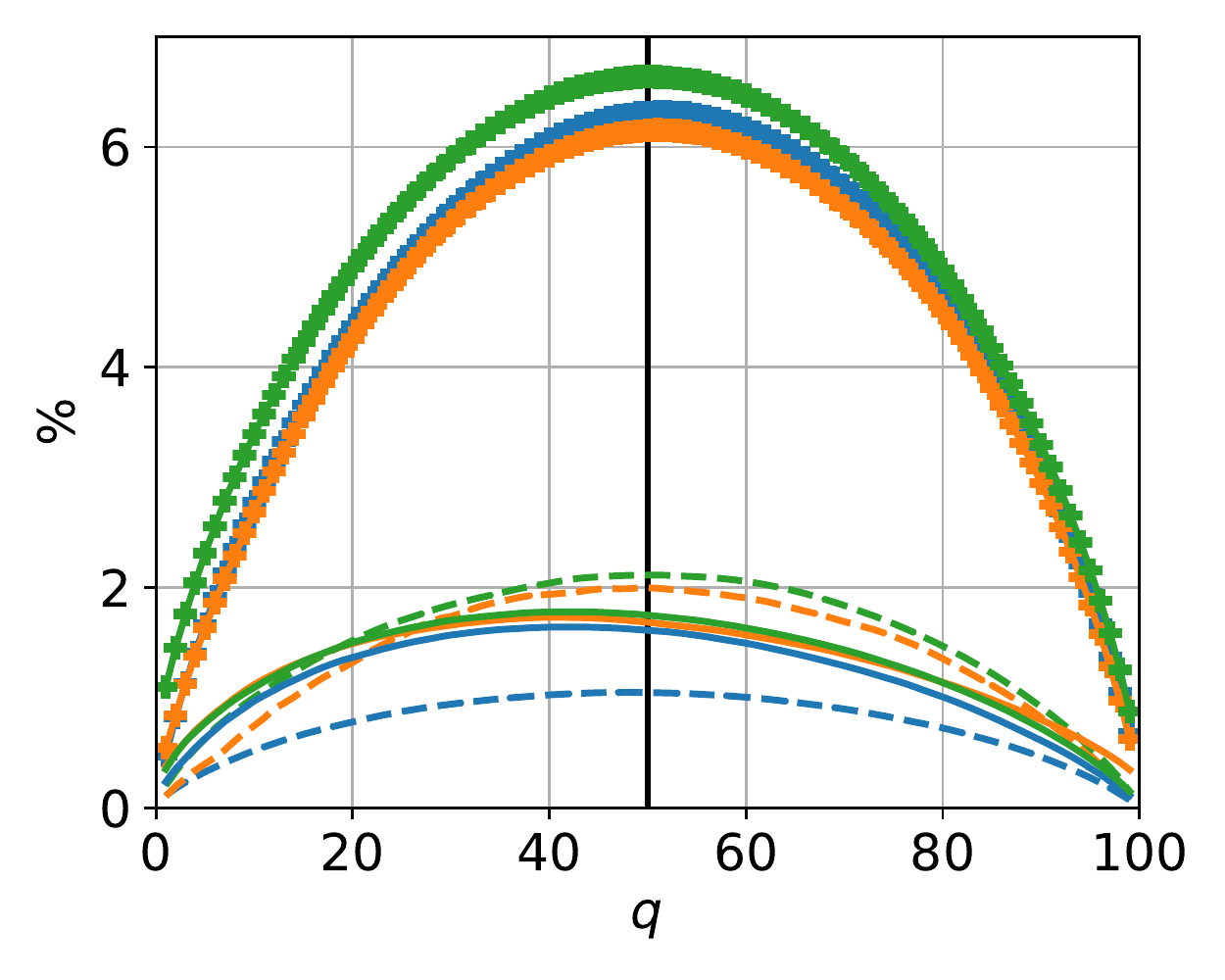}
		\caption{Quantile score.}
	\end{subfigure}%
	\begin{subfigure}{.33\textwidth}
		\centering
		\includegraphics[width=\linewidth]{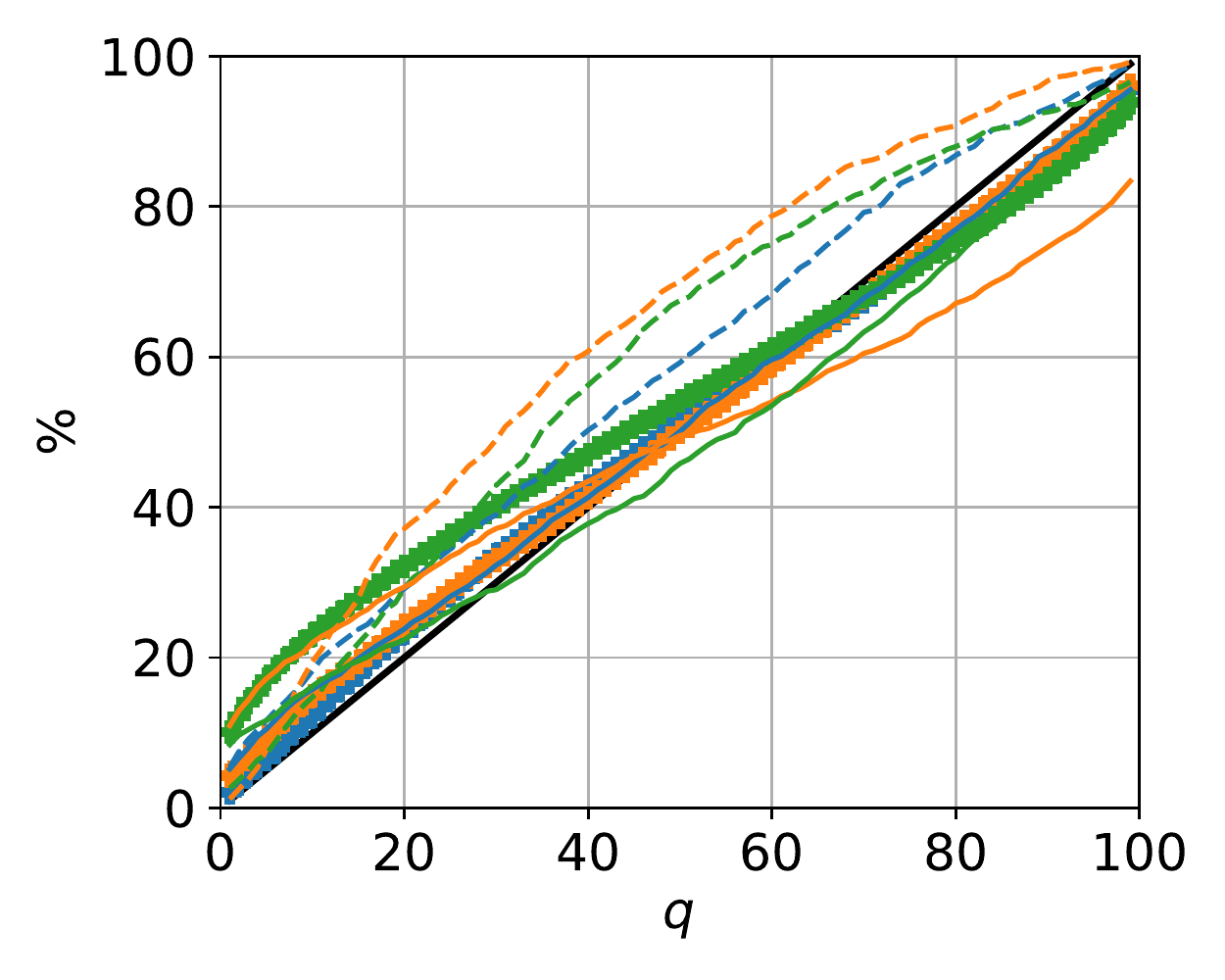}
		\caption{Reliability diagram.}
	\end{subfigure}%
	\begin{subfigure}{.33\textwidth}
		\centering
		\includegraphics[width=\linewidth]{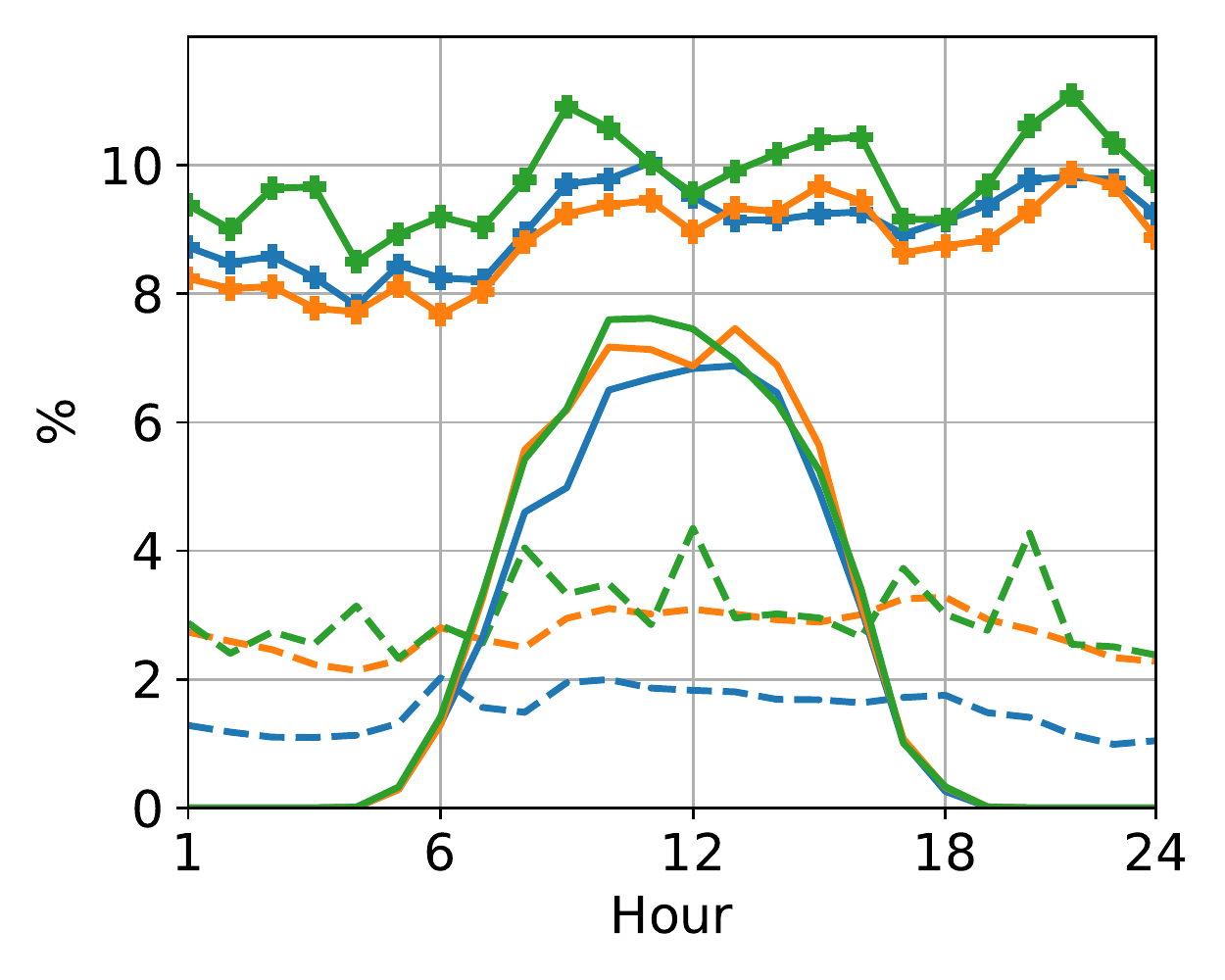}
		\caption{CRPS.}
	\end{subfigure}
	\caption{Quality standard metrics comparison on the wind (markers), PV (plain), and load (dashed) tracks. \\
	Quantile score (a): the lower and the more symmetrical, the better. Note: the quantile score has been averaged over the marginals (the 24 time periods of the day). Reliability diagram (b): the closer to the diagonal, the better. Continuous ranked probability score per marginal (c): the lower, the better. 
	NF outperforms the VAE and GAN for both the PV and load tracks and is slightly outperformed by the VAE on the wind track. Note: all models tend to have more difficulties forecasting the wind power that seems less predictable than the PV generation or the load.}
	\label{fig:ijf-quality_comparison}
\end{figure}

In addition to the generative models, a naive approach is designed (RAND). The scenarios of the learning, validation and testing sets are sampled randomly from the learning, validation, and testing sets, respectively. Intuitively, it boils down to assume that past observations are repeated. These scenarios are realistic but may not be compatible with the context.
Each model generates a set of 100 scenarios for each day of the testing set. And the metrics are computed. Figure~\ref{fig:ijf-quality_comparison} compares the QS, reliability diagram, and CRPS of the wind (markers), PV (plain), and load (dashed) tracks. Overall, for the wind track in terms of CRPS, QS, and reliability diagrams, the VAE achieves slightly better scores, followed by the NF and the GAN.
The ES and VS multivariate scores confirm this trend with 54.82 and 18.87 for the VAE \textit{vs} 56.71 and 18.54 for the NF, respectively.

\begin{figure}[tb]
\centering
	\begin{subfigure}{.33\textwidth}
		\centering
    \includegraphics[width=\linewidth]{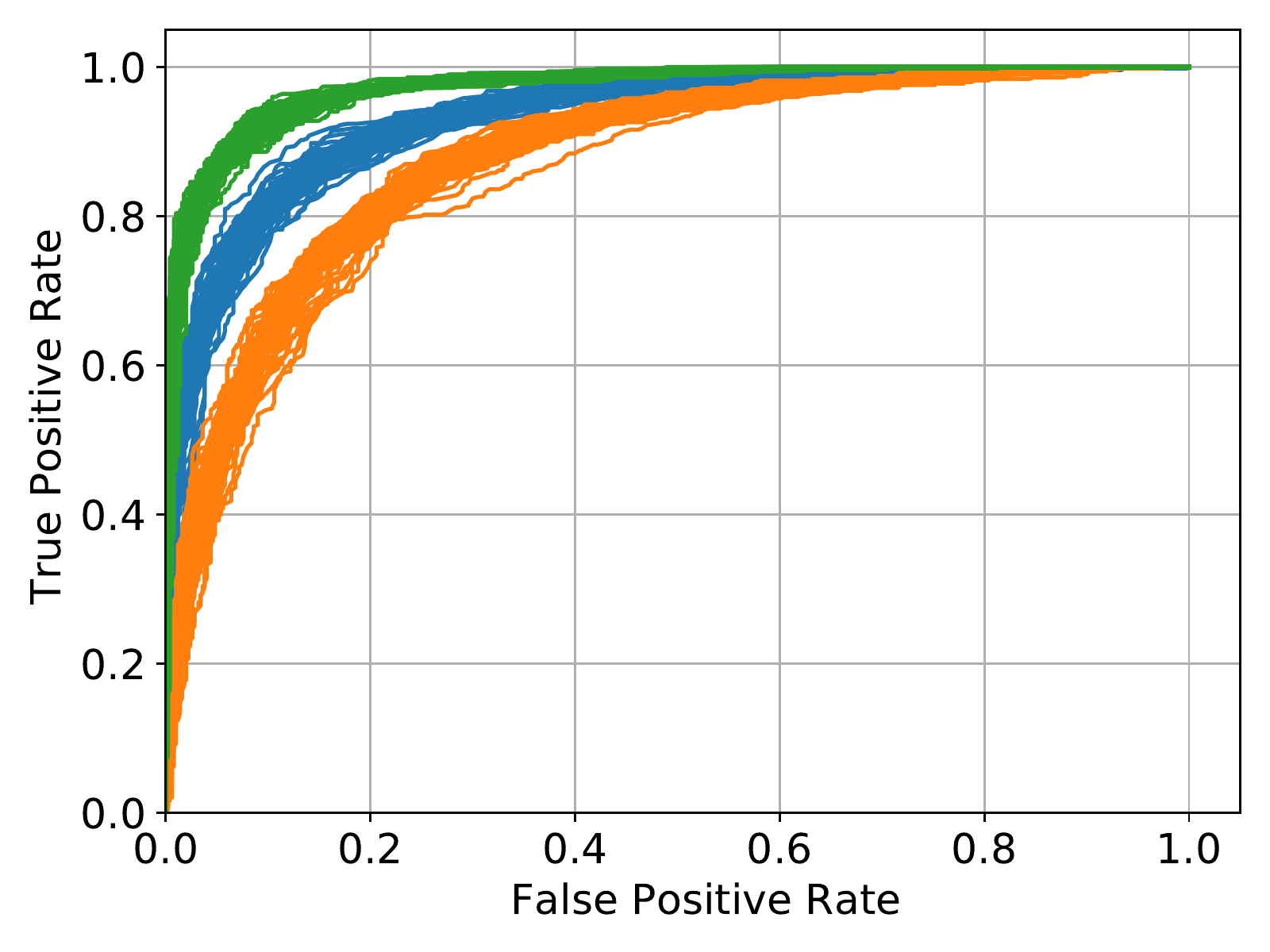}
		\caption{ROC Wind.}
	\label{fig:AE-ROC_wind}
	\end{subfigure}%
	\begin{subfigure}{.33\textwidth}
		\centering
		\includegraphics[width=\linewidth]{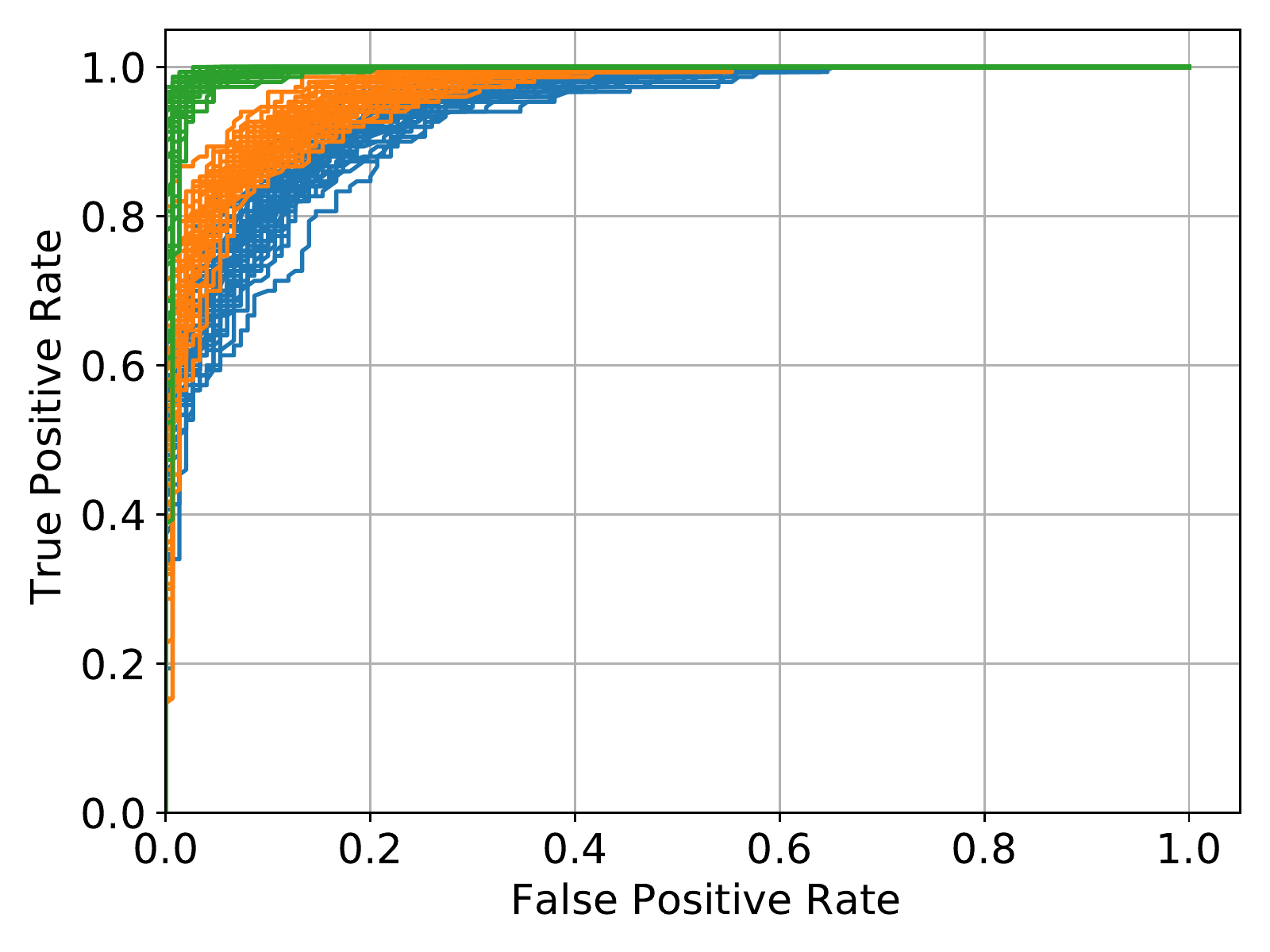}
		\caption{ROC PV.}
	\end{subfigure}%
		\begin{subfigure}{.33\textwidth}
		\centering
		\includegraphics[width=\linewidth]{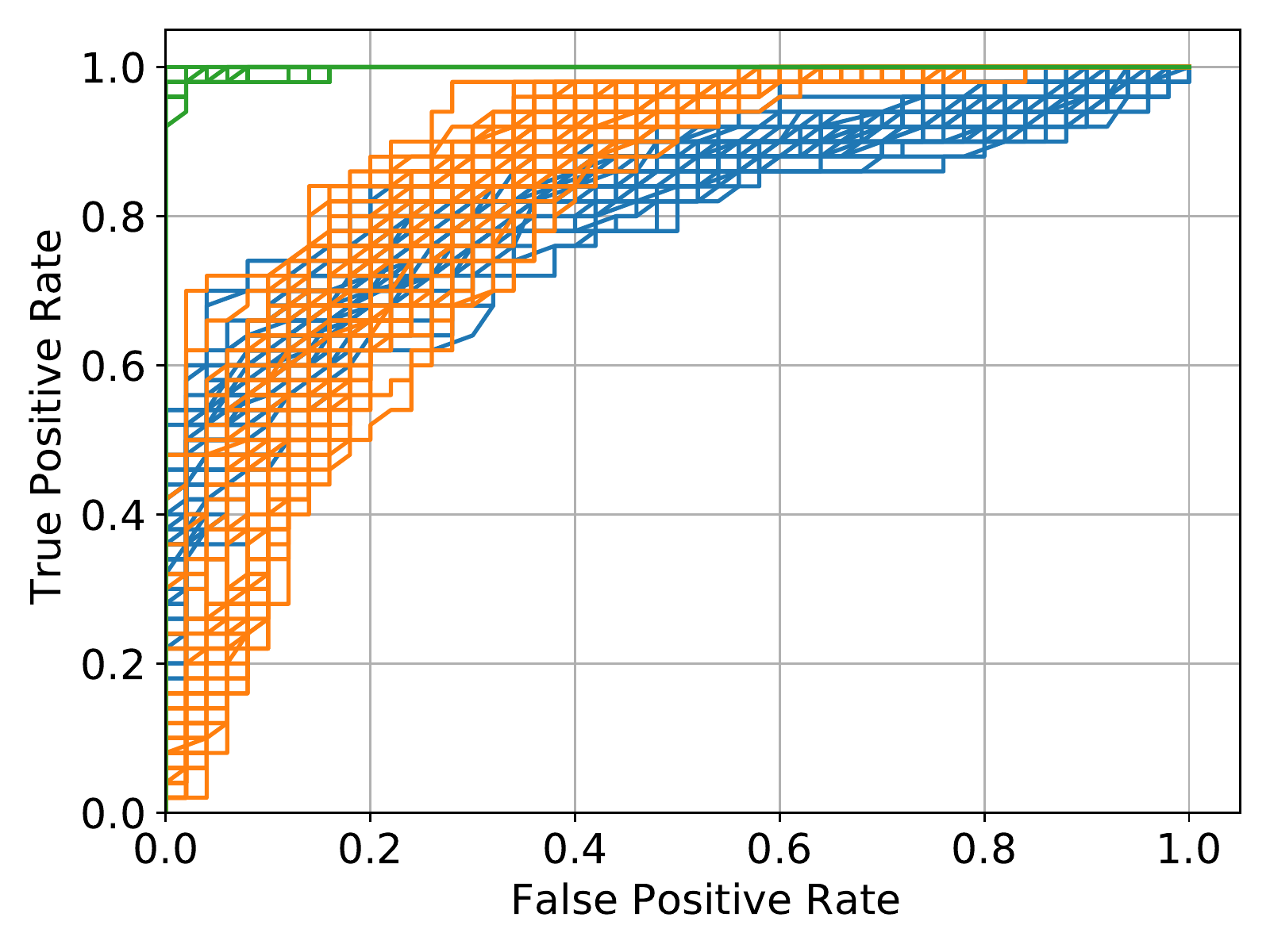}
		\caption{ROC load.}
	\end{subfigure}
\caption{Wind, PV, and load tracks classifier-based metric. \\
Wind: the VAE (orange) is the best to mislead the classifier, followed by the NF (blue) and GAN (green). PV and load: the NF (blue) is the best to fake the classifier, followed by the VAE (orange) and the GAN (green). Note: there are 50 ROC curves depicted for each model, each corresponding to a scenario generated used as input of the classifier. It allows taking into account the variability of the scenarios to avoid having results dependent on a particular scenario.
}	
\label{fig:AE-ROC-wind-pv-load}
\end{figure}

Figure~\ref{fig:AE-DM-test-wind-pv-load} provides the results of the DM tests for these metrics. The heat map indicates the range of the $p$-values. The closer they are to zero, dark green, the more significant the difference between the scores of two models for a given metric. The statistical threshold is five \%, but the scale color is capped at ten \% for a better exposition of the relevant results.
For instance, when considering the DM test for the RAND CRPS, all the columns of the RAND row are in dark green, indicating that the RAND scenarios are always significantly outperformed by the other models.
These DM tests confirm that the VAE outperforms the NF for the wind track considering these metrics. Then, the NF is only outperformed by the VAE and the GAN by both the VAE and NF.
%
These results are consistent with the classifier-based metric depicted in Figure~\ref{fig:AE-ROC_wind}, where the VAE is the best to mislead the classifier followed by the NF, and GAN. 

\begin{figure}[tb]
	\centering
	\begin{subfigure}{.5\textwidth}
		\centering
		\includegraphics[width=\linewidth]{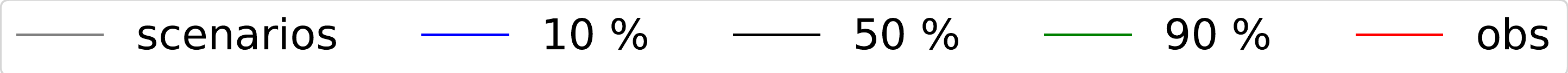}
	\end{subfigure}
	\begin{subfigure}{.4\textwidth}
		\centering
		\includegraphics[width=\linewidth]{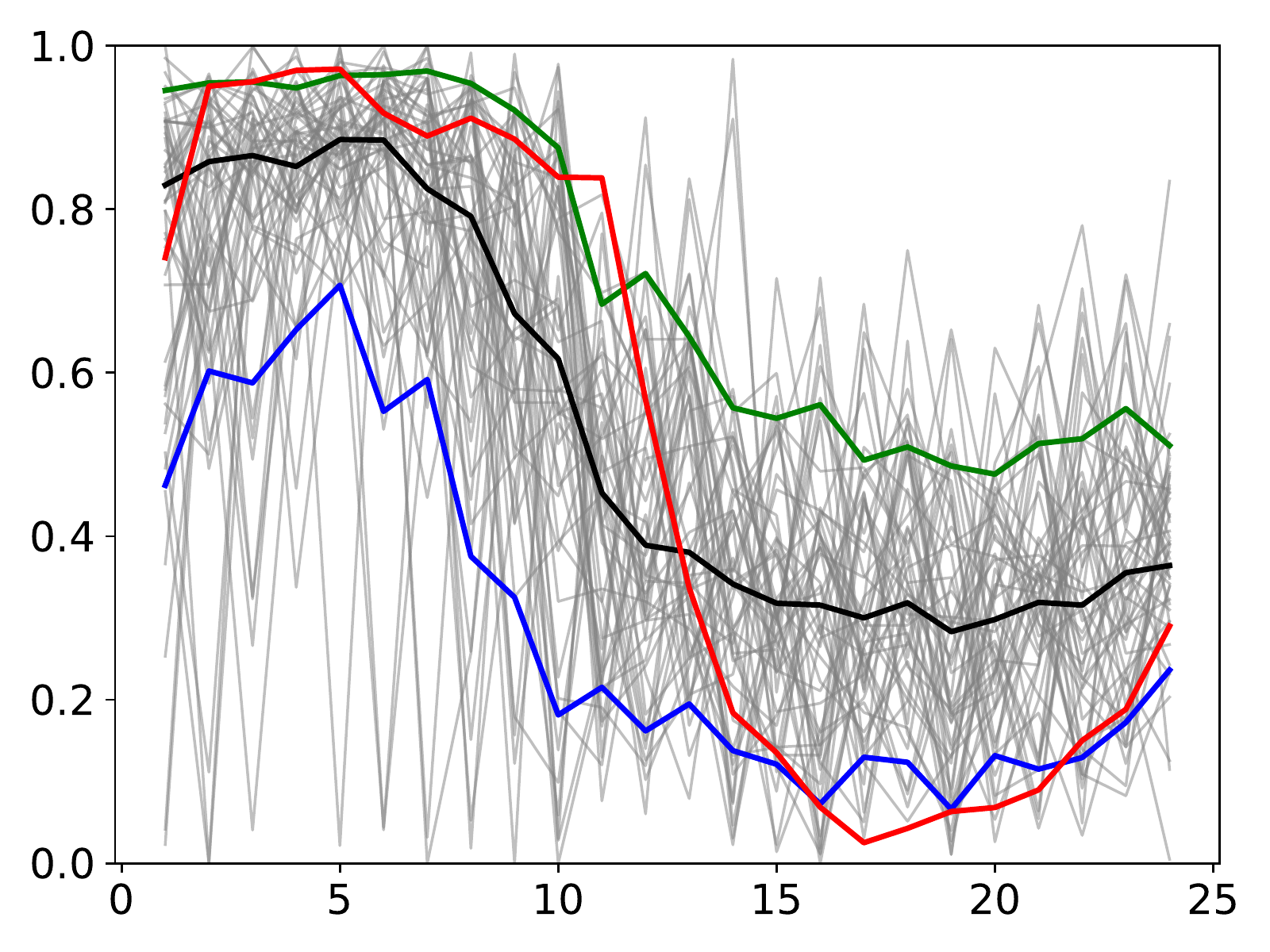}
		\caption{NF.}
	\end{subfigure}%
	\begin{subfigure}{.4\textwidth}
		\centering
		\includegraphics[width=\linewidth]{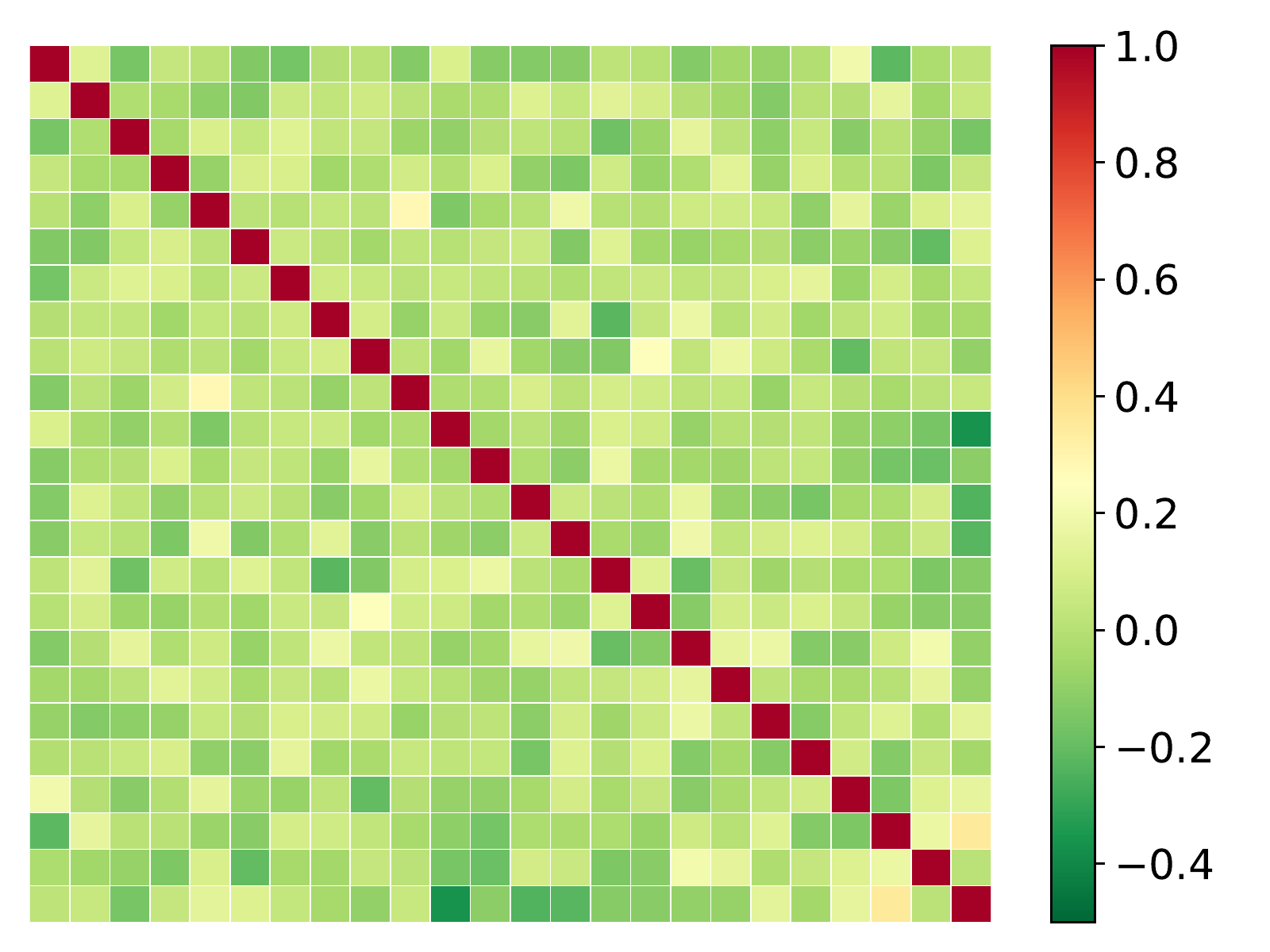}
		\caption{NF.}
	\end{subfigure}
	\begin{subfigure}{.4\textwidth}
		\centering
		\includegraphics[width=\linewidth]{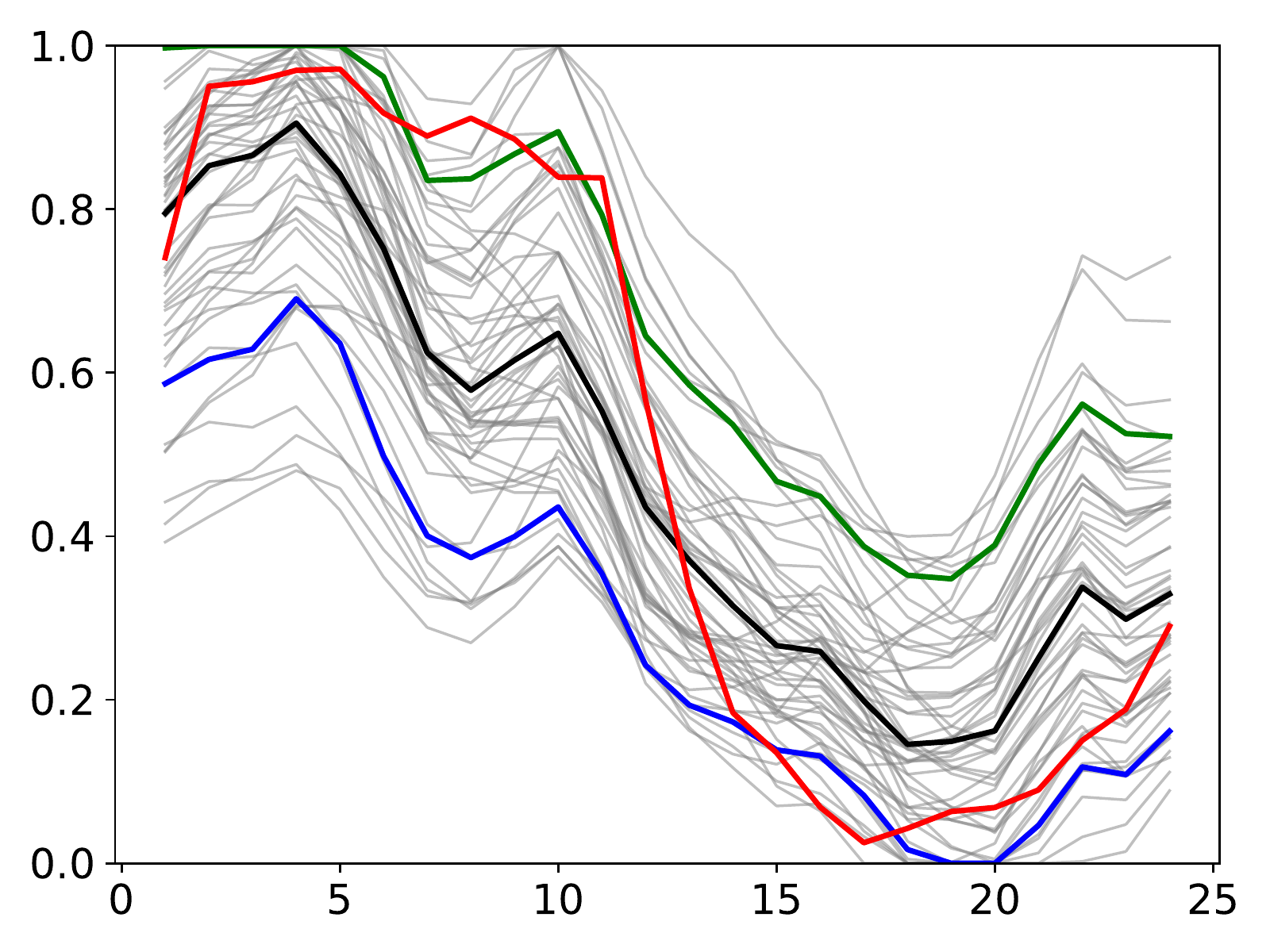}
		\caption{GAN.}
	\end{subfigure}%
	\begin{subfigure}{.4\textwidth}
		\centering
		\includegraphics[width=\linewidth]{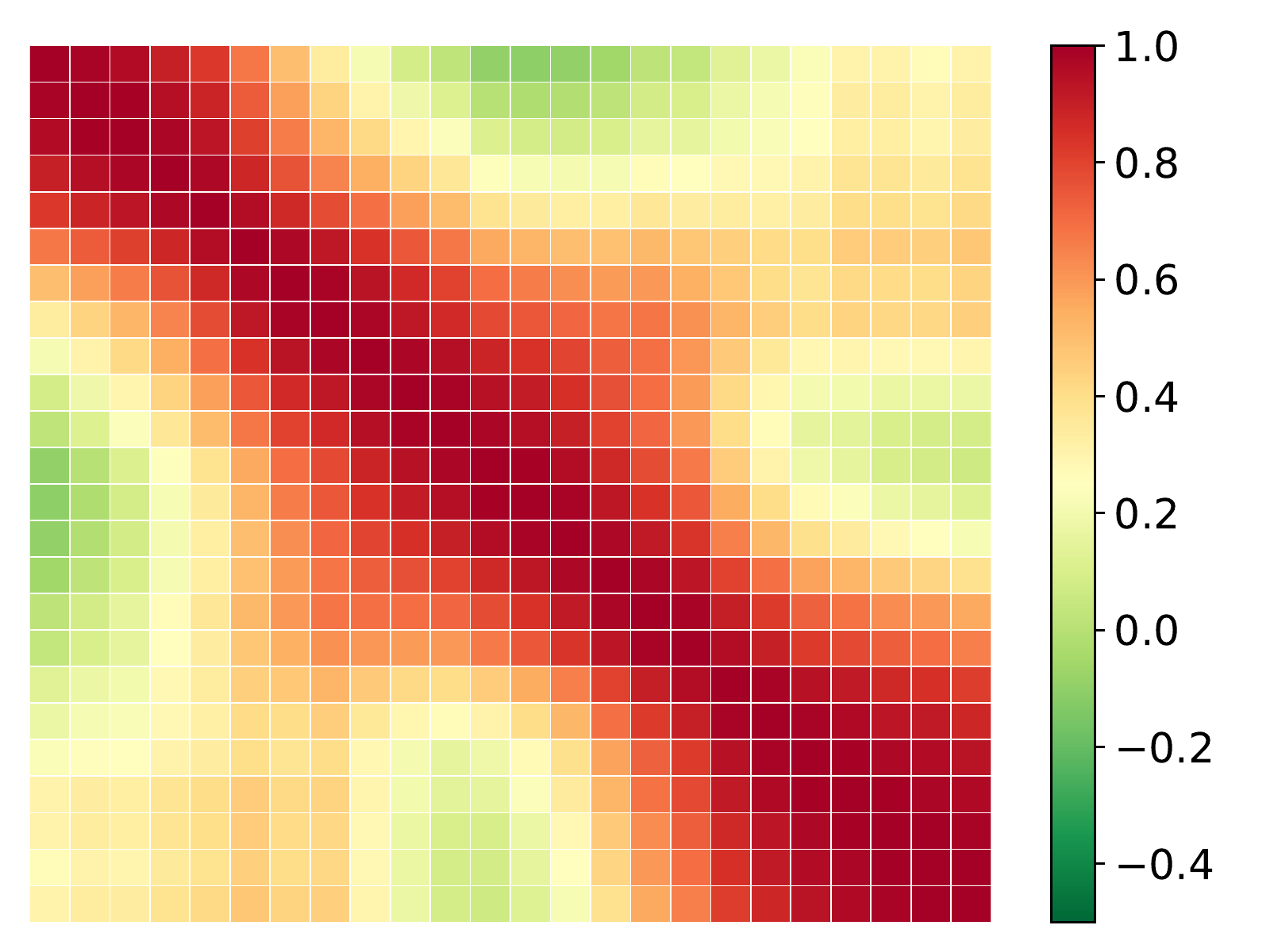}
		\caption{GAN.}
	\end{subfigure}
	\begin{subfigure}{.4\textwidth}
		\centering
		\includegraphics[width=\linewidth]{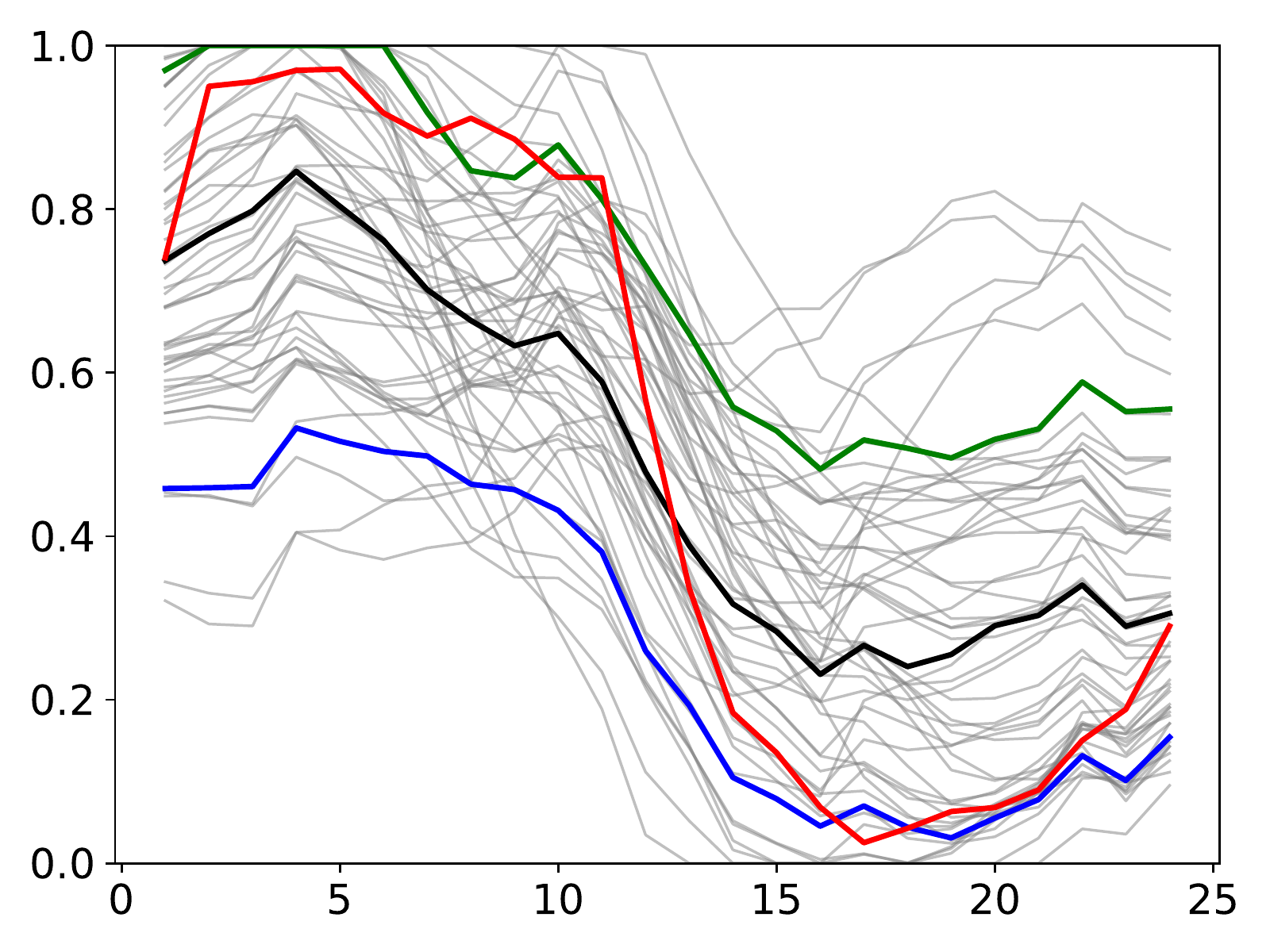}
		\caption{VAE.}
	\end{subfigure}%
	\begin{subfigure}{.4\textwidth}
		\centering
		\includegraphics[width=\linewidth]{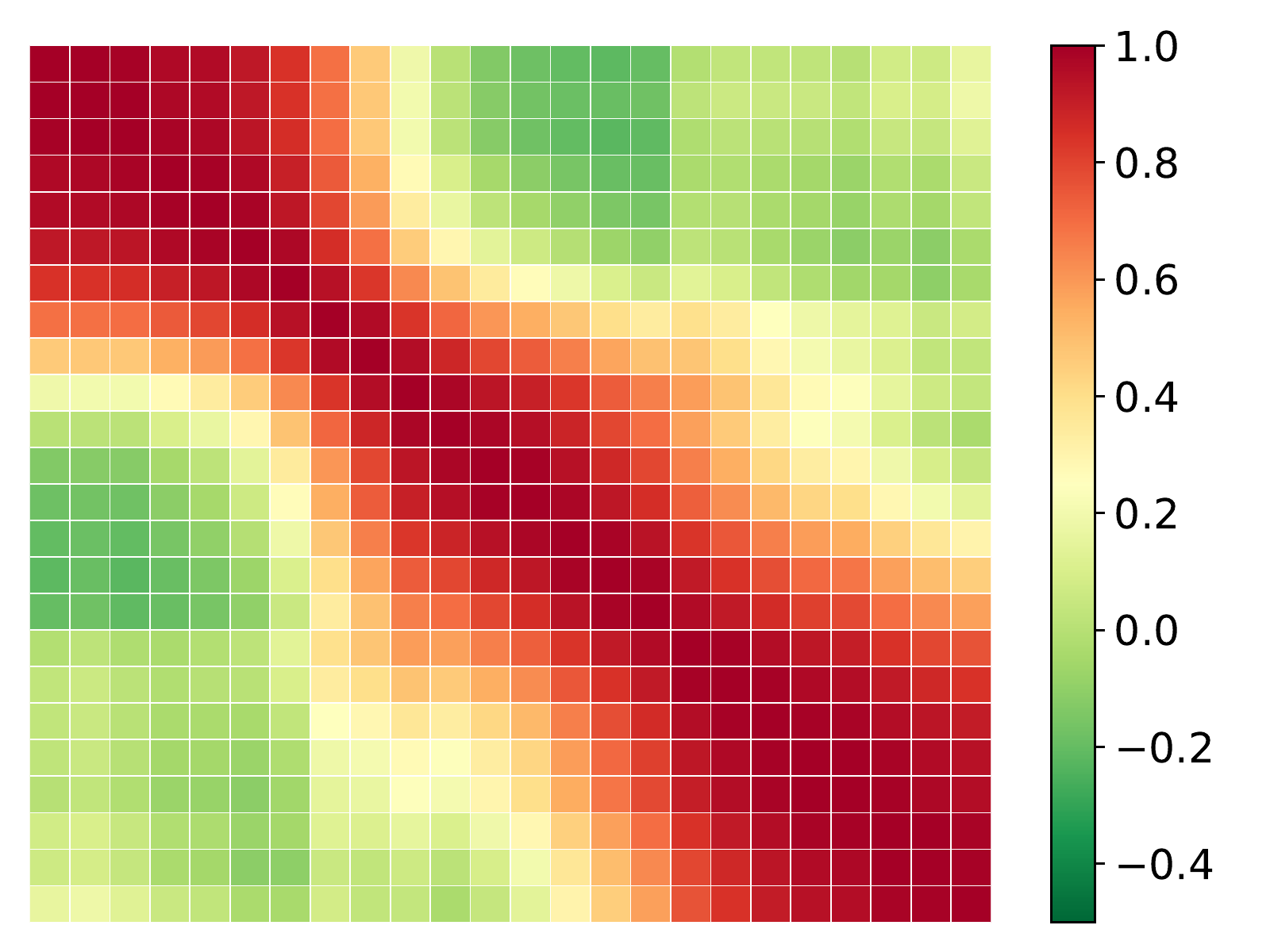}
		\caption{VAE.}
	\end{subfigure}
	\caption{Wind power scenarios shape comparison and analysis. \\
	Left part (a) NF, (c) GAN, and (e) VAE: 50 wind power scenarios (grey) of a randomly selected day of the testing set along with the ten \% (blue), 50 \% (black), and 90 \% (green) quantiles, and the observations (red). Right part (b) NF, (d) GAN, and (f) VAE: the corresponding Pearson time correlation matrices of these scenarios with the periods as rows and columns. The NF tends to exhibit no time correlation between scenarios. In contrast, the VAE and GAN tend to be partially time-correlated over a few periods.}
	\label{fig:AE-wind_scenarios}
\end{figure}

The left part of Figure~\ref{fig:AE-wind_scenarios} provides 50 scenarios, (a) NF, (c) GAN and (e) VAE, generated for a given day selected randomly from the testing set. Notice how the shape of the NF's scenarios differs significantly from the GAN and VAE as they tend to be more variable with no identifiable trend. In contrast, the VAE and GAN scenarios seem to differ mainly in nominal power but have similar shapes. This behavior is even more pronounced for the GAN, where the scenarios rarely crossed over the periods. For instance, there is a gap in generation around periods 17 and 18 where all the GAN's scenarios follow this trend.
These observations are confirmed by computing the corresponding time correlation matrices, depicted by the right part of Figure~\ref{fig:AE-wind_scenarios} demonstrating there is no correlation between NF's scenarios. On the contrary, the VAE and GAN correlation matrices tend to be similar with a time correlation of the scenarios over a few periods, with more correlated periods when considering the GAN. This difference in the scenario's shape is striking and not necessarily captured by metrics such as the CRPS, QS, or even the classifier-based metric and is also observed on the PV and load tracks, as explained in the next paragraph.
%
%
%

\subsection{All tracks}\label{sec:all_tracks_res}

\begin{figure}[htb]
	\centering
		\begin{subfigure}{.33\textwidth}
		\centering
		\includegraphics[width=\linewidth]{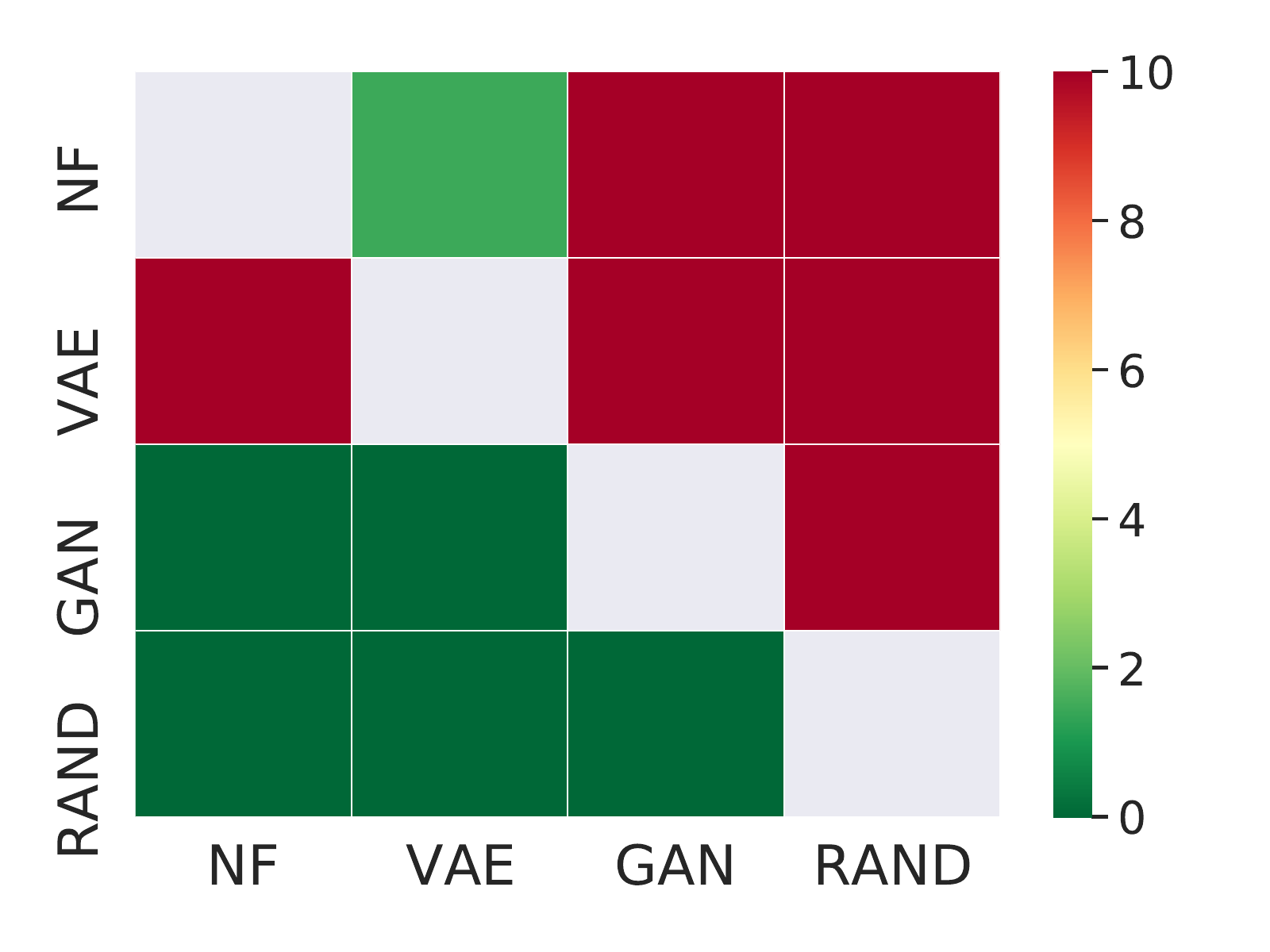}
		\caption{Wind CRPS DM.}
	\end{subfigure}%
	\begin{subfigure}{.33\textwidth}
		\centering
		\includegraphics[width=\linewidth]{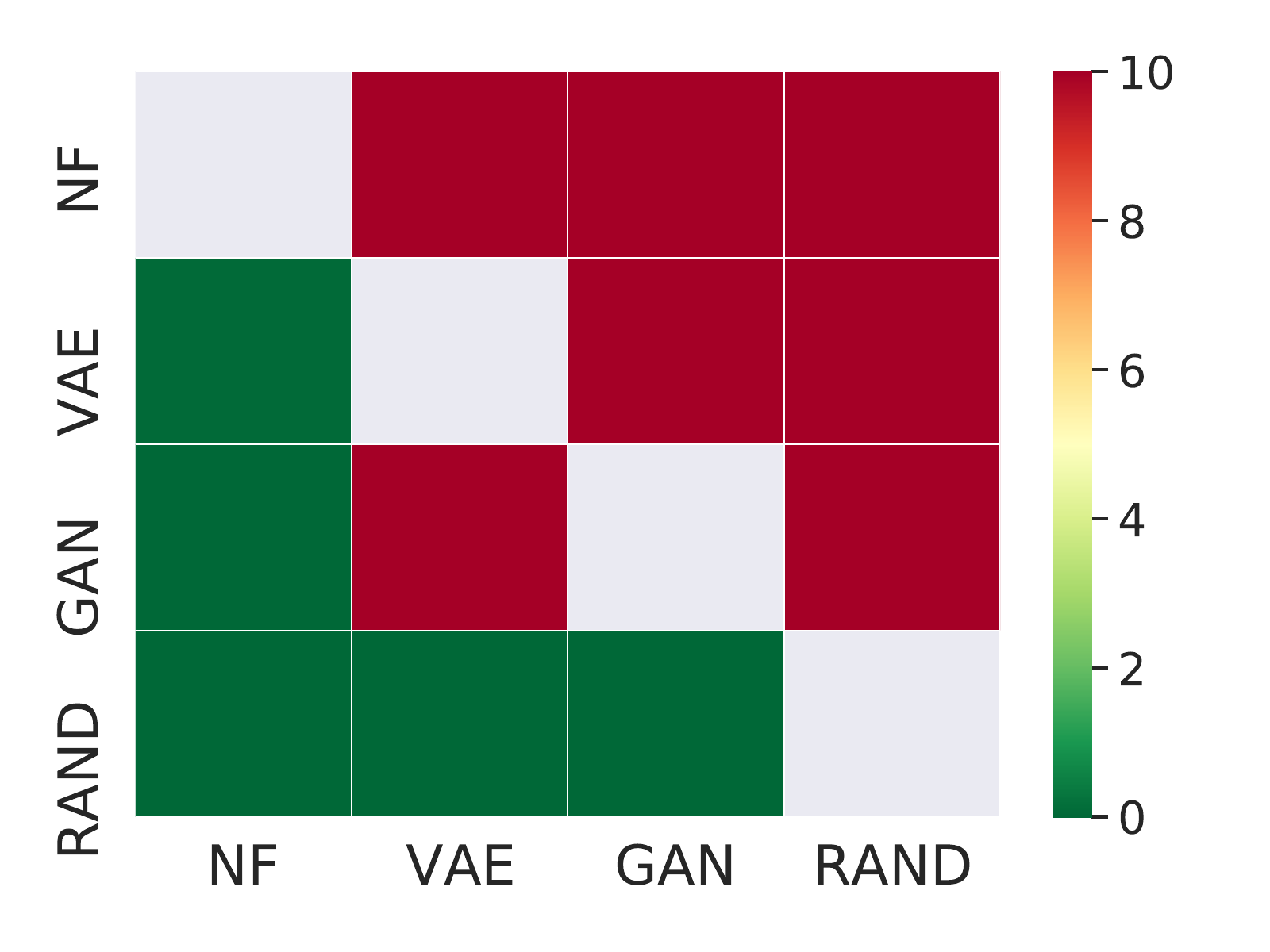}
		\caption{PV CRPS DM.}
	\end{subfigure}%
	\begin{subfigure}{.33\textwidth}
		\centering
		\includegraphics[width=\linewidth]{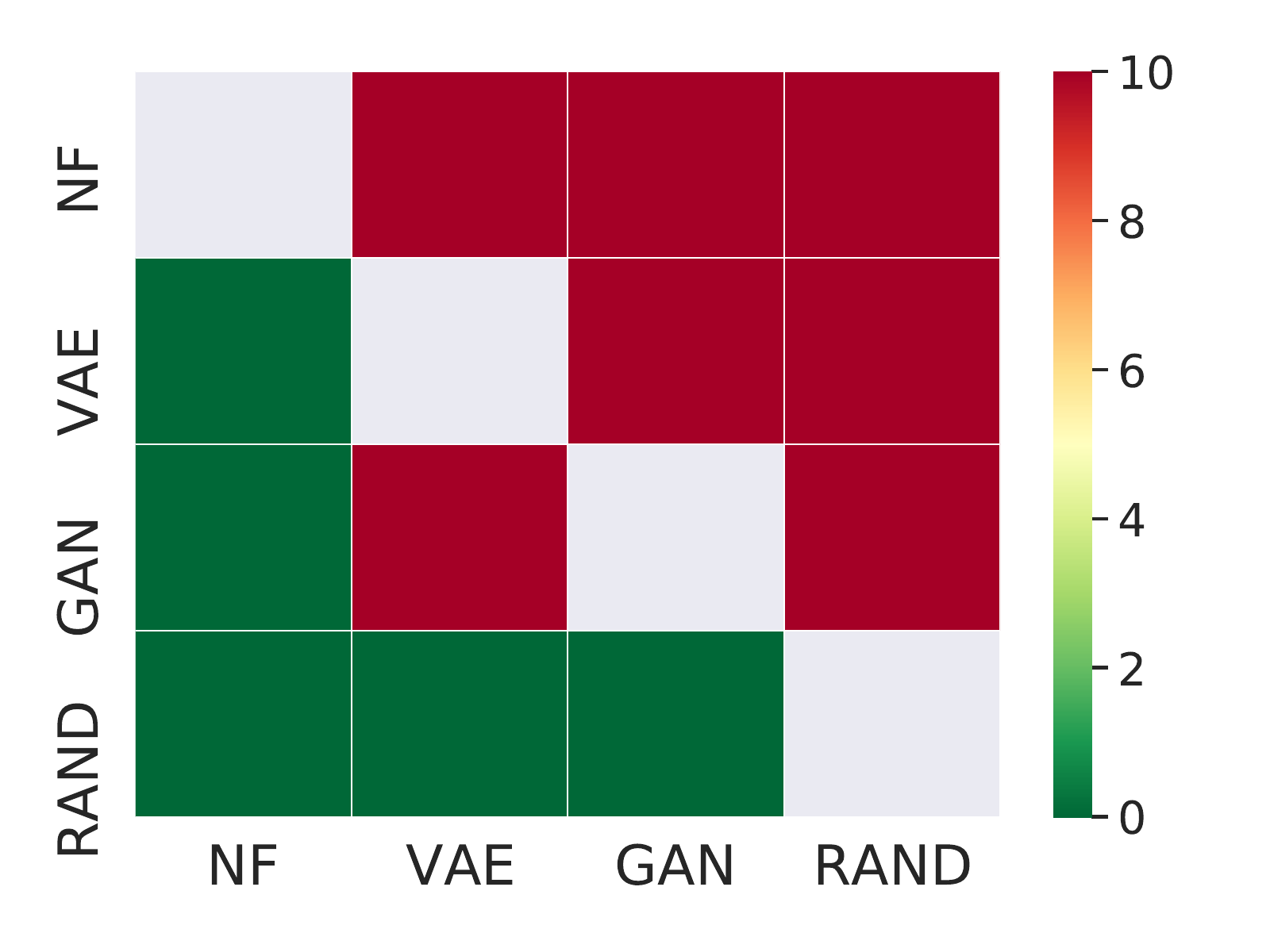}
		\caption{Load CRPS DM.}
	\end{subfigure}
	\begin{subfigure}{.33\textwidth}
		\centering
		\includegraphics[width=\linewidth]{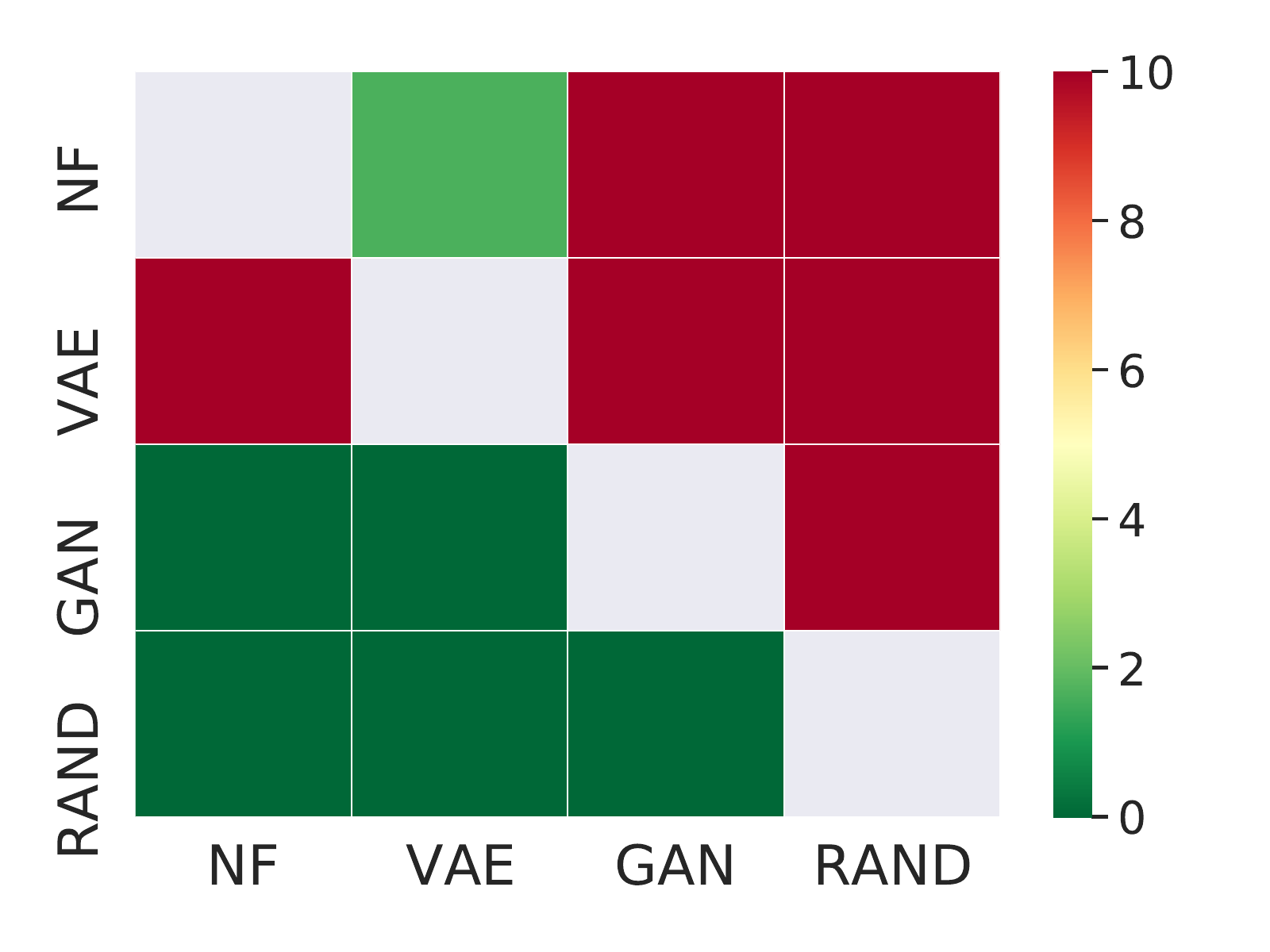}
		\caption{Wind QS DM.}
	\end{subfigure}%
	\begin{subfigure}{.33\textwidth}
		\centering
		\includegraphics[width=\linewidth]{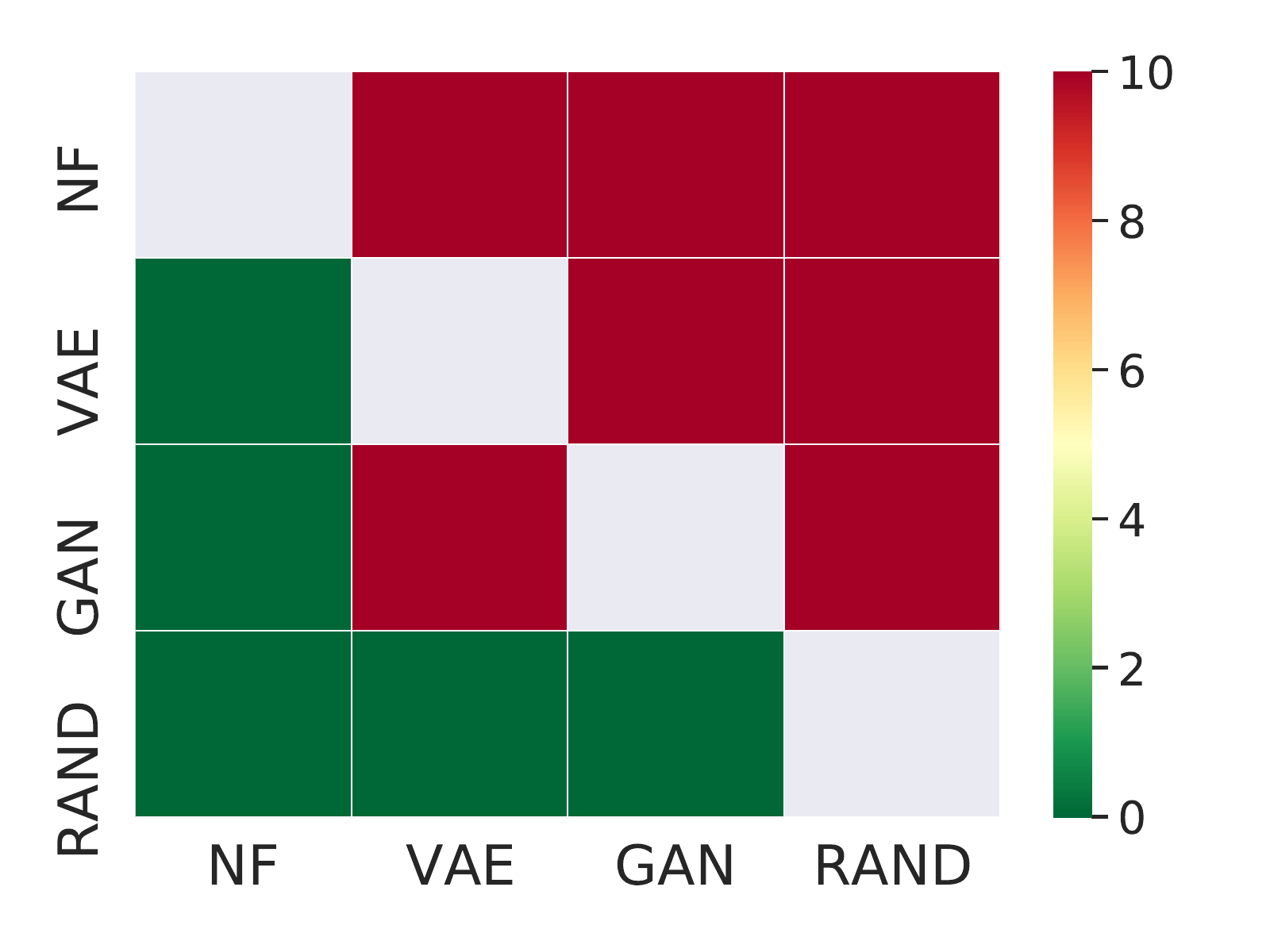}
		\caption{PV QS DM.}
	\end{subfigure}%
	\begin{subfigure}{.33\textwidth}
		\centering
		\includegraphics[width=\linewidth]{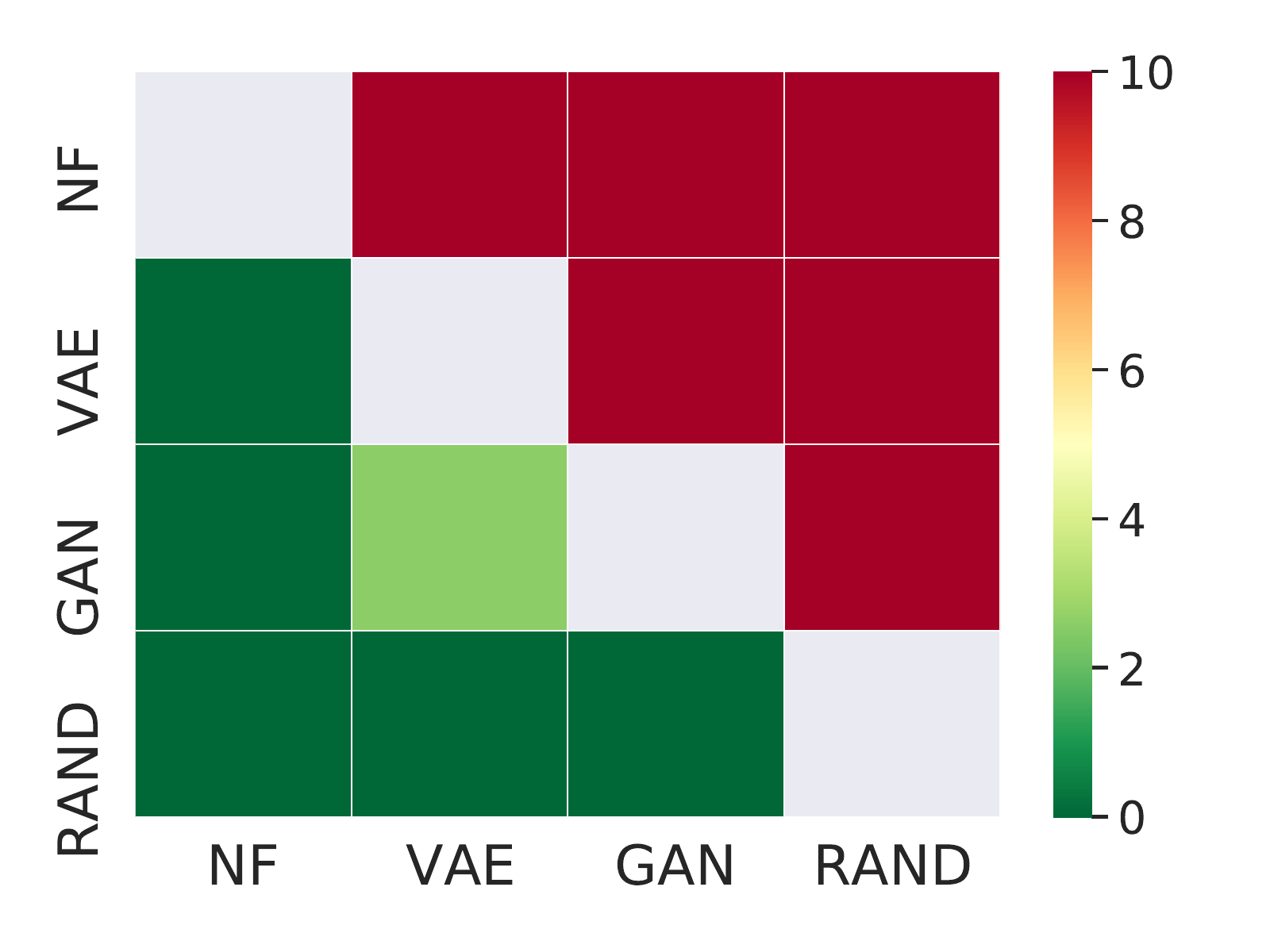}
		\caption{Load QS DM.}
	\end{subfigure}
	\begin{subfigure}{.33\textwidth}
		\centering
		\includegraphics[width=\linewidth]{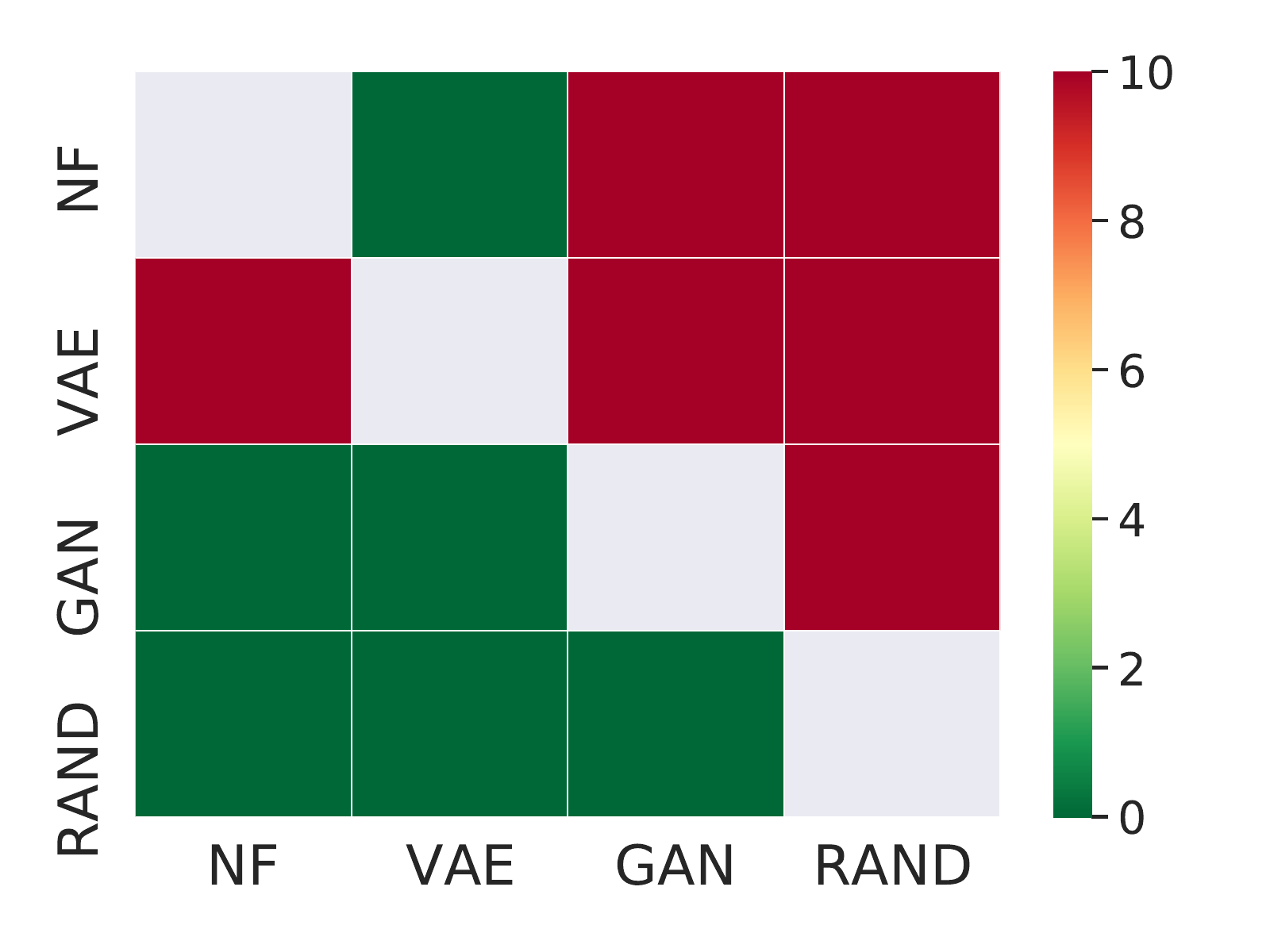}
		\caption{Wind ES DM.}
	\end{subfigure}%
	\begin{subfigure}{.33\textwidth}
		\centering
		\includegraphics[width=\linewidth]{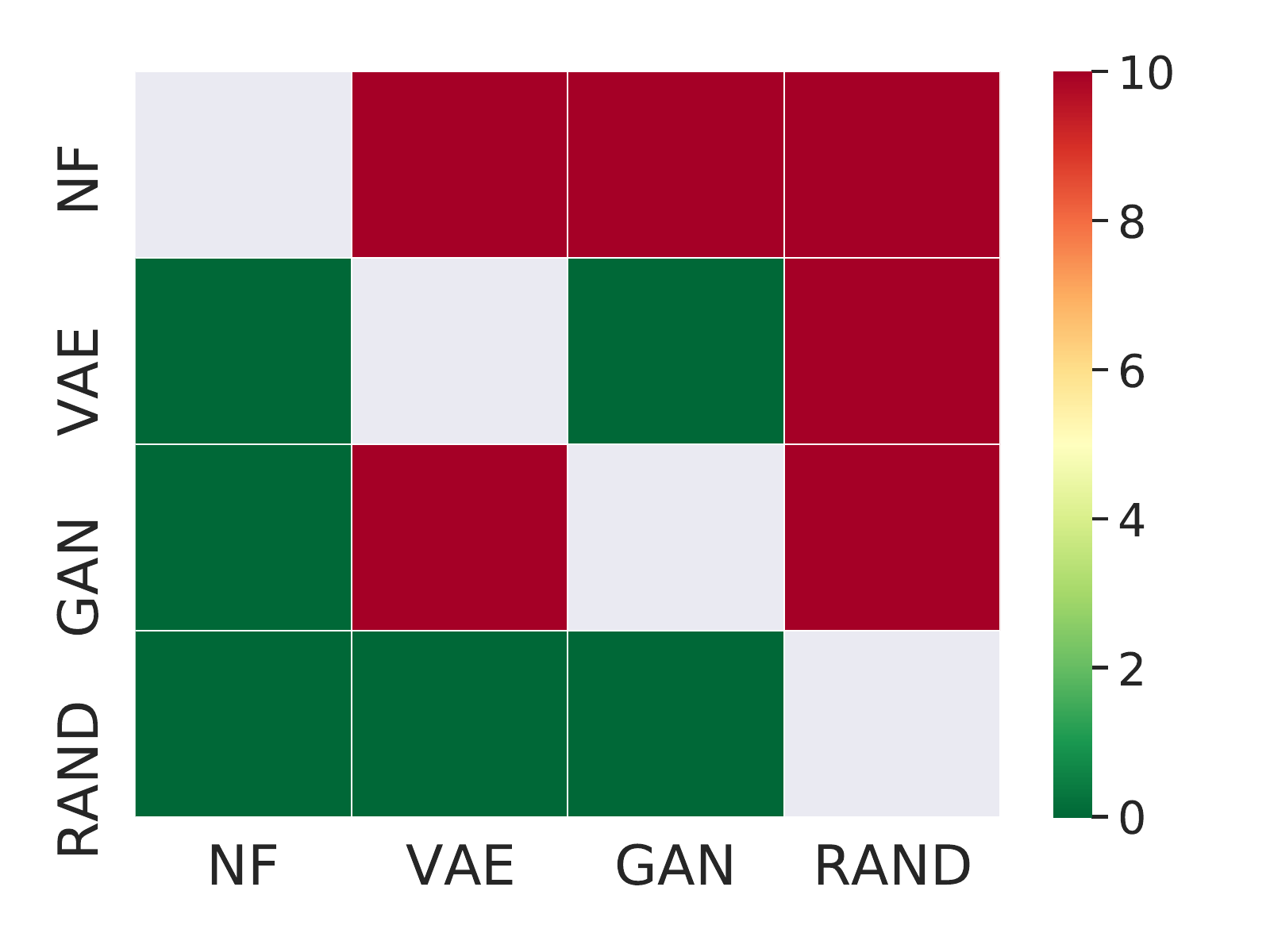}
		\caption{PV ES DM.}
	\end{subfigure}%
	\begin{subfigure}{.33\textwidth}
		\centering
		\includegraphics[width=\linewidth]{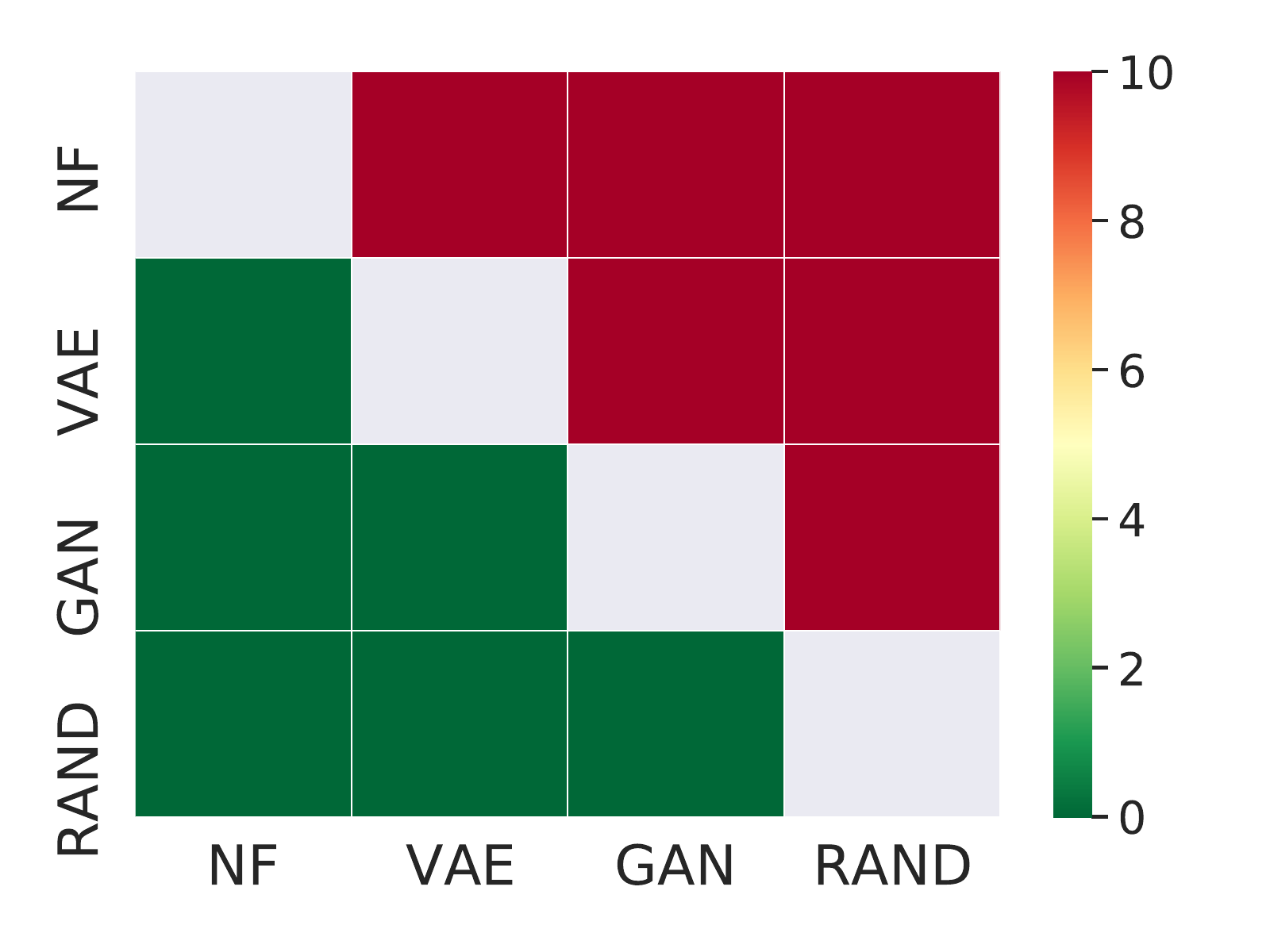}
		\caption{Load ES DM.}
	\end{subfigure}
	\begin{subfigure}{.33\textwidth}
		\centering
		\includegraphics[width=\linewidth]{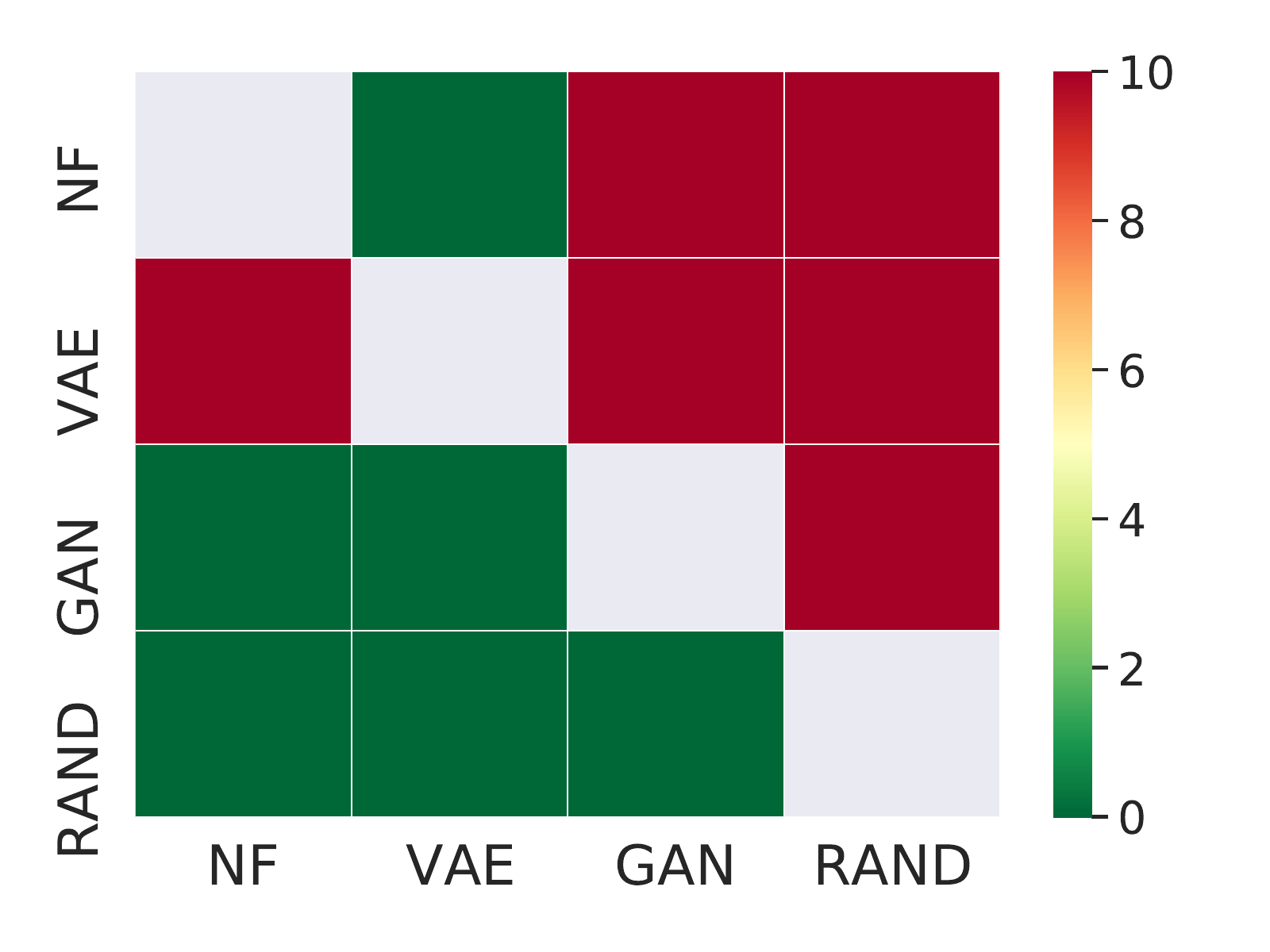}
		\caption{Wind VS DM.}
	\end{subfigure}%
	\begin{subfigure}{.33\textwidth}
		\centering
		\includegraphics[width=\linewidth]{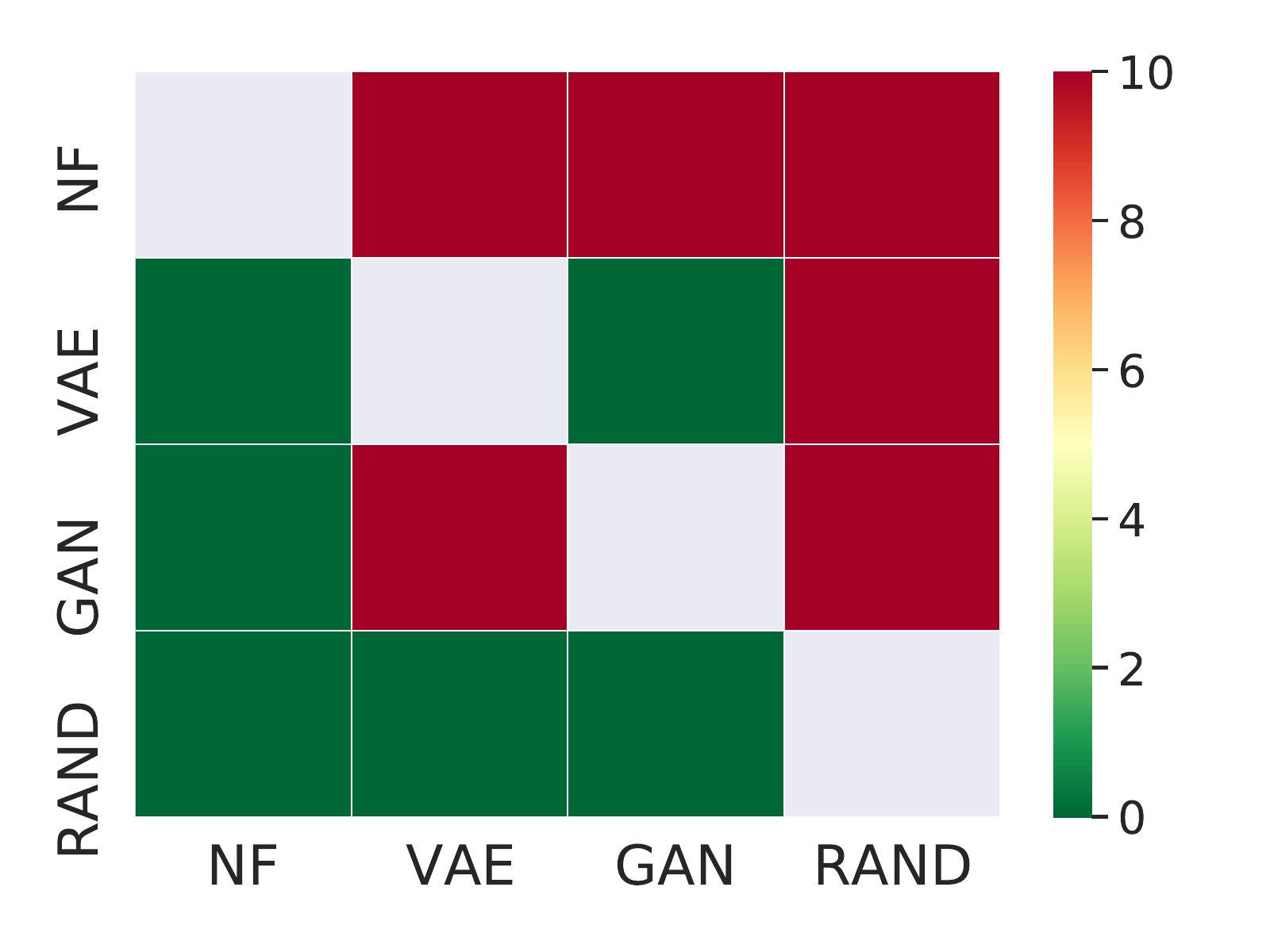}
		\caption{PV VS DM.}
	\end{subfigure}%
	\begin{subfigure}{.33\textwidth}
		\centering
		\includegraphics[width=\linewidth]{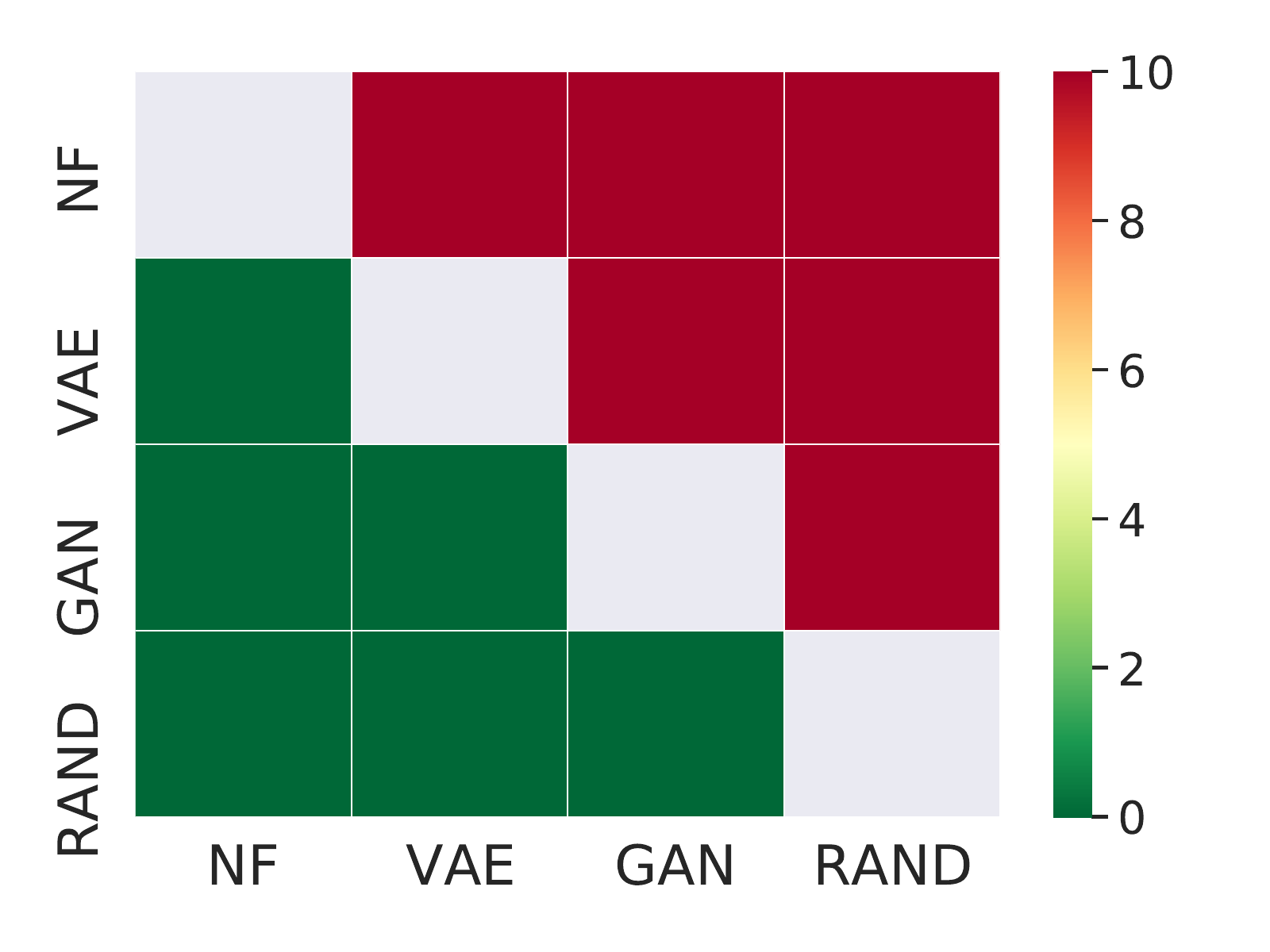}
		\caption{Load VS DM.}
	\end{subfigure}
\caption{Wind, PV and load tracks Diebold-Mariano tests. \\
Wind track: the DM tests of the CRPS, QS, ES, and VS confirm that the VAE outperforms the NF for these metrics. 
PV and load tracks: the NF outperforms the VAE and GAN for all metrics. Note: the GAN outperforms the VAE for both the ES and VS for the PV track. However, the VAE outperforms the GAN on this dataset for both the CRPS and QS.
%
}
\label{fig:AE-DM-test-wind-pv-load}
\end{figure}

Table~\ref{tab:AE-quality_average_scores} provides the averaged quality scores. The CRPS is averaged over the 24 time periods $\overline{\text{CRPS}}$. The QS over the 99 percentiles $\overline{\text{QS}}$. The MAE-r is the mean absolute error between the reliability curve and the diagonal, and $\overline{\text{AUC}}$ is the mean of the 50 AUC.
\begin{table}[htbp]
\renewcommand{\arraystretch}{1.25}
\begin{center}
\begin{tabular}{m{1cm}lrrrr}
			\hline  \hline
& &  NF &  VAE & GAN & RAND \\ \hline
\multirow{6}{*}{Wind}
& $\overline{\text{CRPS}}$  & 9.07  & \textbf{8.80} & 9.79 & 16.92 \\ 
& $\overline{\text{QS}}$    & 4.58  & \textbf{4.45} & 4.95 & 8.55 \\ 
& MAE-r &  2.83    &  \textbf{2.67} &  6.82  & 1.01  \\ 
& $\overline{\text{AUC}}$  & 0.935   &  \textbf{0.877} &  0.972 & 0.918 \\
& ES & 56.71 & \textbf{54.82} & 60.52 & 96.15 \\
& VS & 18.54 & \textbf{17.87} & 19.87 & 23.21 \\  \hline 
\multirow{6}{*}{PV}
& $\overline{\text{CRPS}}$  & \textbf{2.35}  & 2.60 & 2.61 & 4.92 \\ 
& $\overline{\text{QS}}$    & \textbf{1.19}  & 1.31 & 1.32 & 2.48 \\ 
& MAE-r &  \textbf{2.66}    & 9.04  &  4.94  & 3.94  \\ 
& $\overline{\text{AUC}}$  & \textbf{0.950}  & 0.969 &  0.997 & 0.947 \\
& ES & \textbf{23.08} & 24.65 & 24.15 & 41.53 \\
& VS & \textbf{4.68} & 5.02 & 4.88 & 13.40 \\  \hline 
\multirow{6}{*}{Load}
& $\overline{\text{CRPS}}$  & \textbf{1.51}  & 2.74 & 3.01 & 6.74 \\ 
& $\overline{\text{QS}}$    & \textbf{0.76}  & 1.39 & 1.52 & 3.40 \\ 
& MAE-r &  \textbf{7.70}    & 13.97  &  9.99  & 0.88  \\ 
& $\overline{\text{AUC}}$  & \textbf{0.823}  & 0.847 &  0.999 & 0.944 \\
& ES & \textbf{9.17} & 15.11 & 17.96 & 38.08 \\
& VS & \textbf{1.63} & 1.66 & 3.81 & 7.28 \\  \hline  \hline
\end{tabular}
\caption{Averaged quality scores per dataset. \\
The best performing deep learning generative model for each track is written in bold. 
The CRPS, QS, MAE-r, and ES are expressed in \%. Overall, for both the PV and load tracks, the NF outperforms the VAE and GAN and slightly outperforms the VAE on the wind track.
}
\label{tab:AE-quality_average_scores}
\end{center}
\end{table}
%
Overall, for the PV and load tracks in CRPS, QS, reliability diagrams, AUC, ES, and VS, the NF outperforms the VAE and GAN and is slightly outperformed by the VAE on the wind track. On the load track, the VAE outperforms the GAN. However, the VAE and GAN achieved similar results on the PV track, and the GAN performed better in terms of ES and VS.
These results are confirmed by the DM tests depicted in Figure~\ref{fig:AE-DM-test-wind-pv-load}.
%
The classifier-based metric results for both the load and PV tracks, provided by Figure~\ref{fig:AE-ROC-wind-pv-load}, confirm this trend where the NF is the best to trick the classifier followed by the VAE and GAN. 

\begin{figure}[tb]
\centering
	\begin{subfigure}{.5\textwidth}
		\centering
		\includegraphics[width=\linewidth]{Part2/Figs/scenarios/generative_models/legend_scenarios}
	\end{subfigure}
	\begin{subfigure}{0.25\textwidth}
		\centering
		\includegraphics[width=\linewidth]{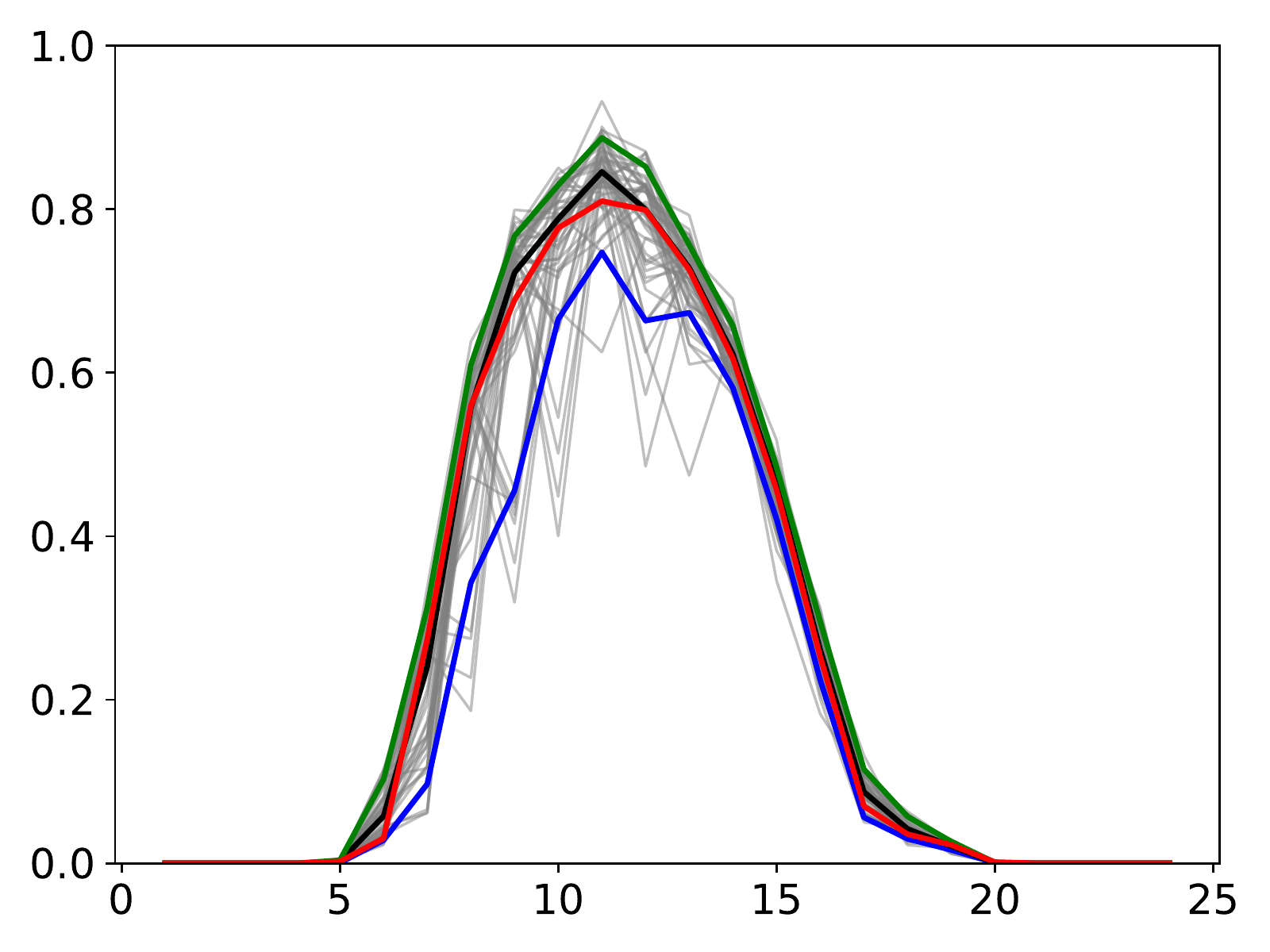}
		\caption{NF PV.}
	\end{subfigure}%
	\begin{subfigure}{0.25\textwidth}
		\centering
		\includegraphics[width=\linewidth]{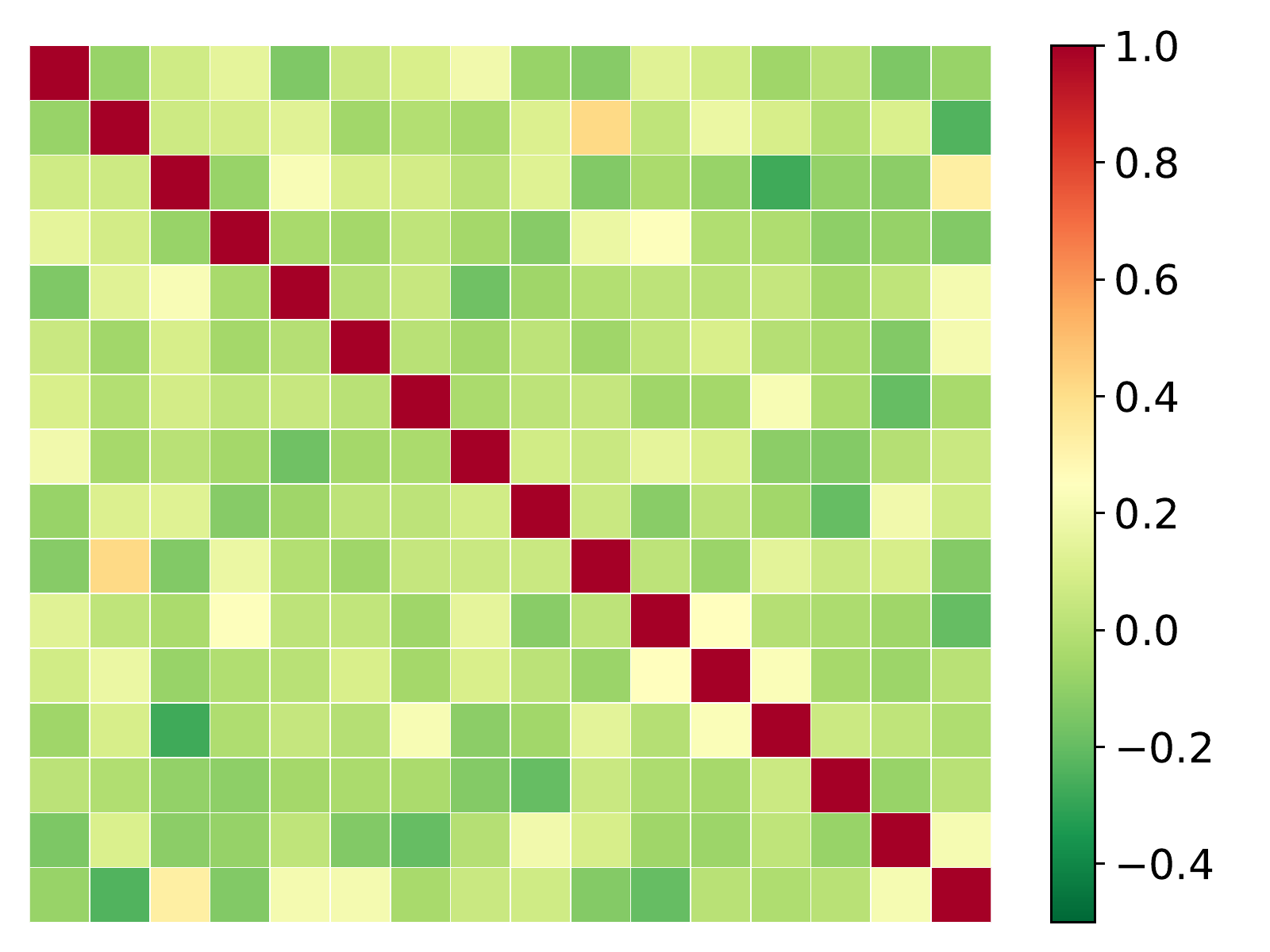}
		\caption{NF PV matrix.}
	\end{subfigure}%
	\begin{subfigure}{0.25\textwidth}
		\centering
		\includegraphics[width=\linewidth]{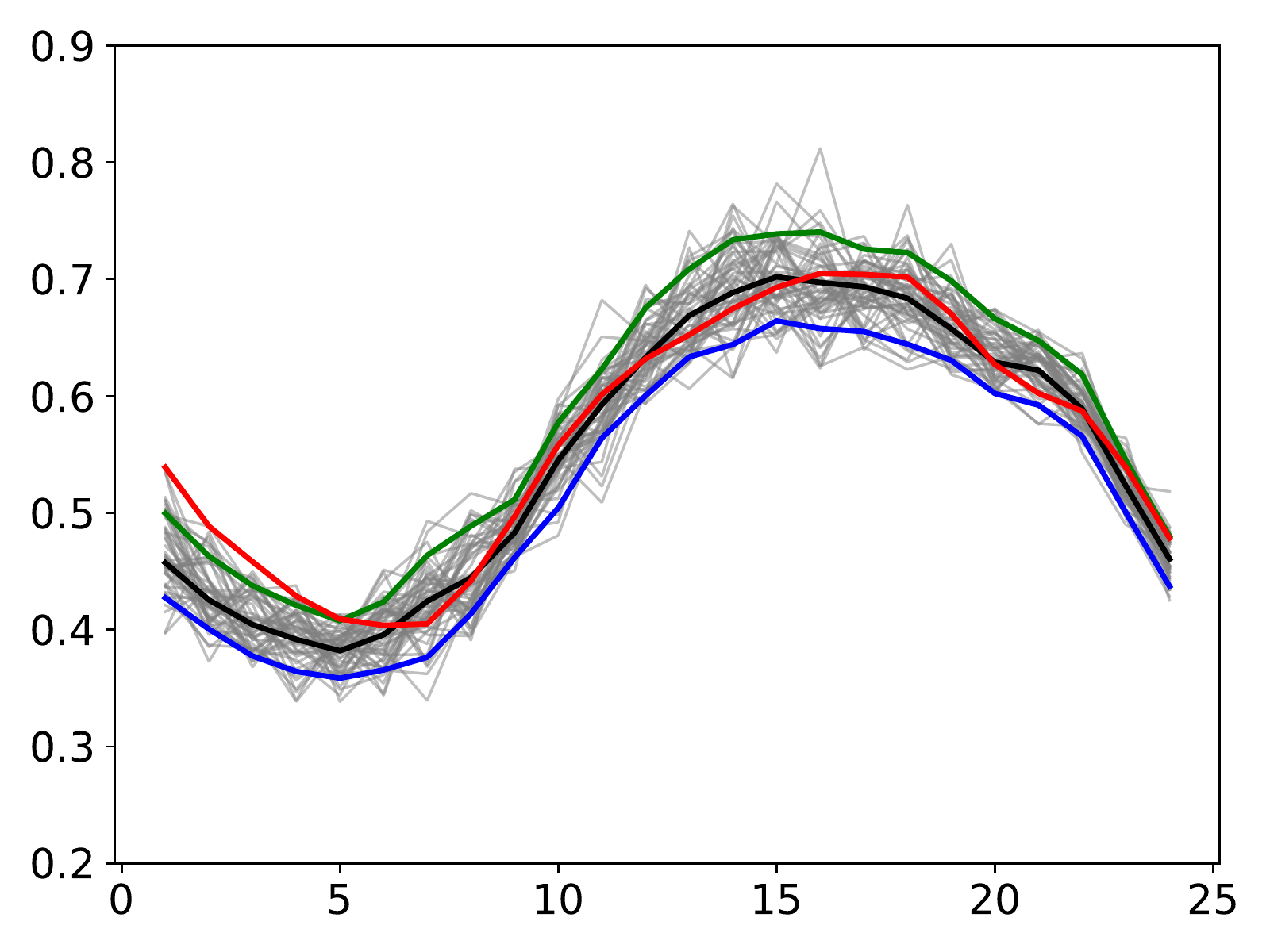}
		\caption{NF load.}
	\end{subfigure}%
	\begin{subfigure}{0.25\textwidth}
		\centering
		\includegraphics[width=\linewidth]{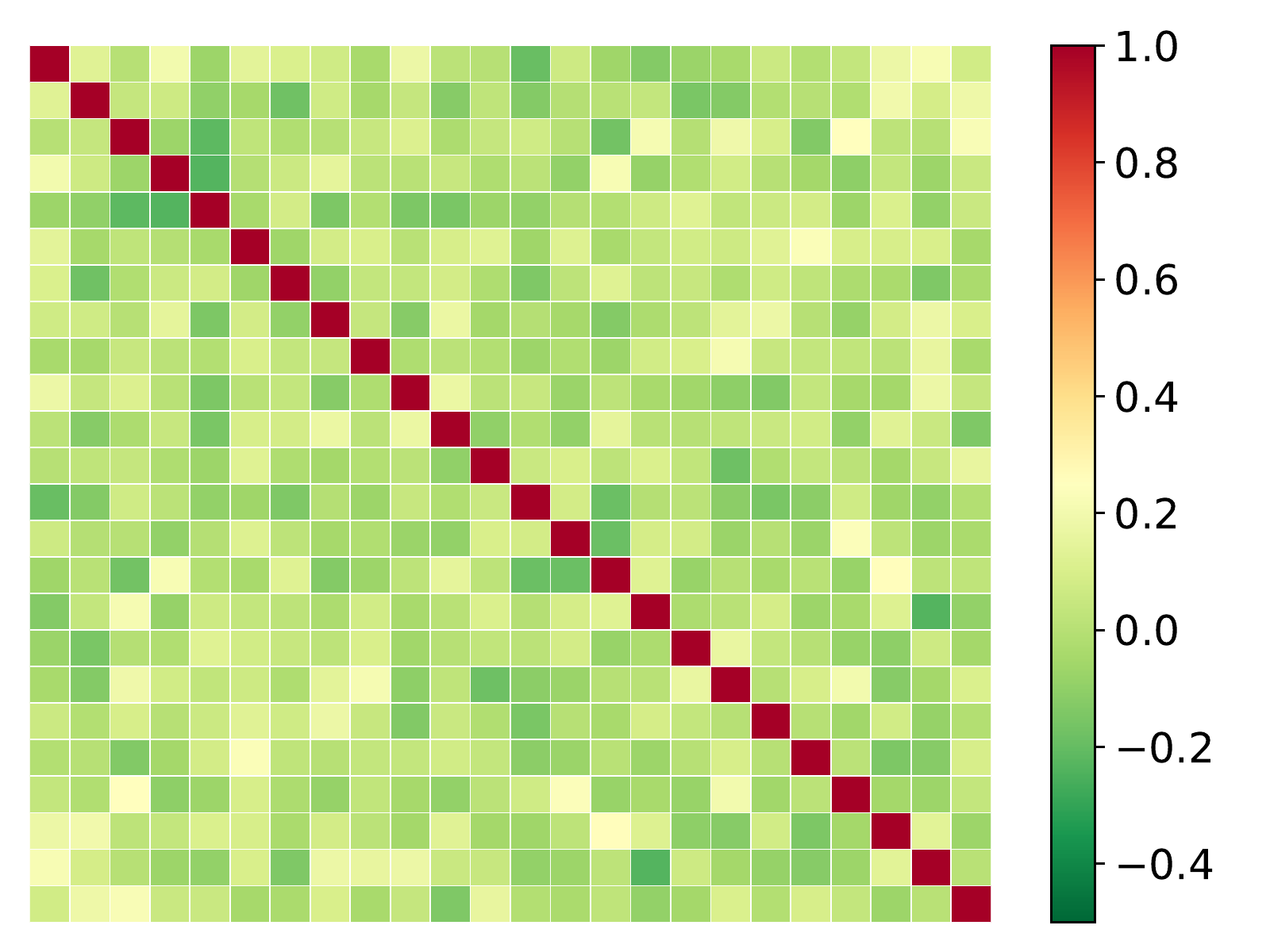}
		\caption{NF load matrix.}
	\end{subfigure}
	\begin{subfigure}{0.25\textwidth}
		\centering
		\includegraphics[width=\linewidth]{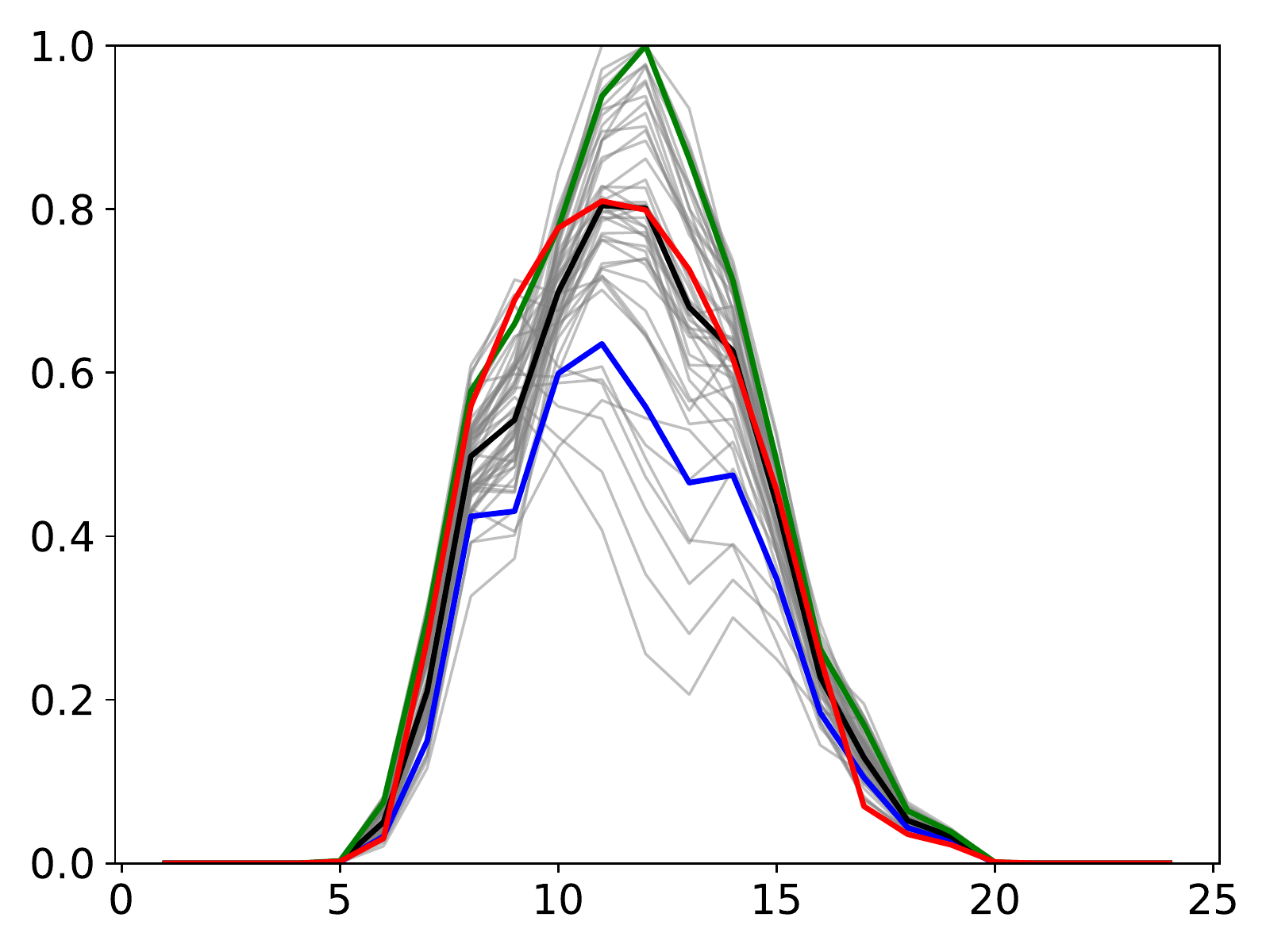}
		\caption{GAN.}
	\end{subfigure}%
	\begin{subfigure}{0.25\textwidth}
		\centering
		\includegraphics[width=\linewidth]{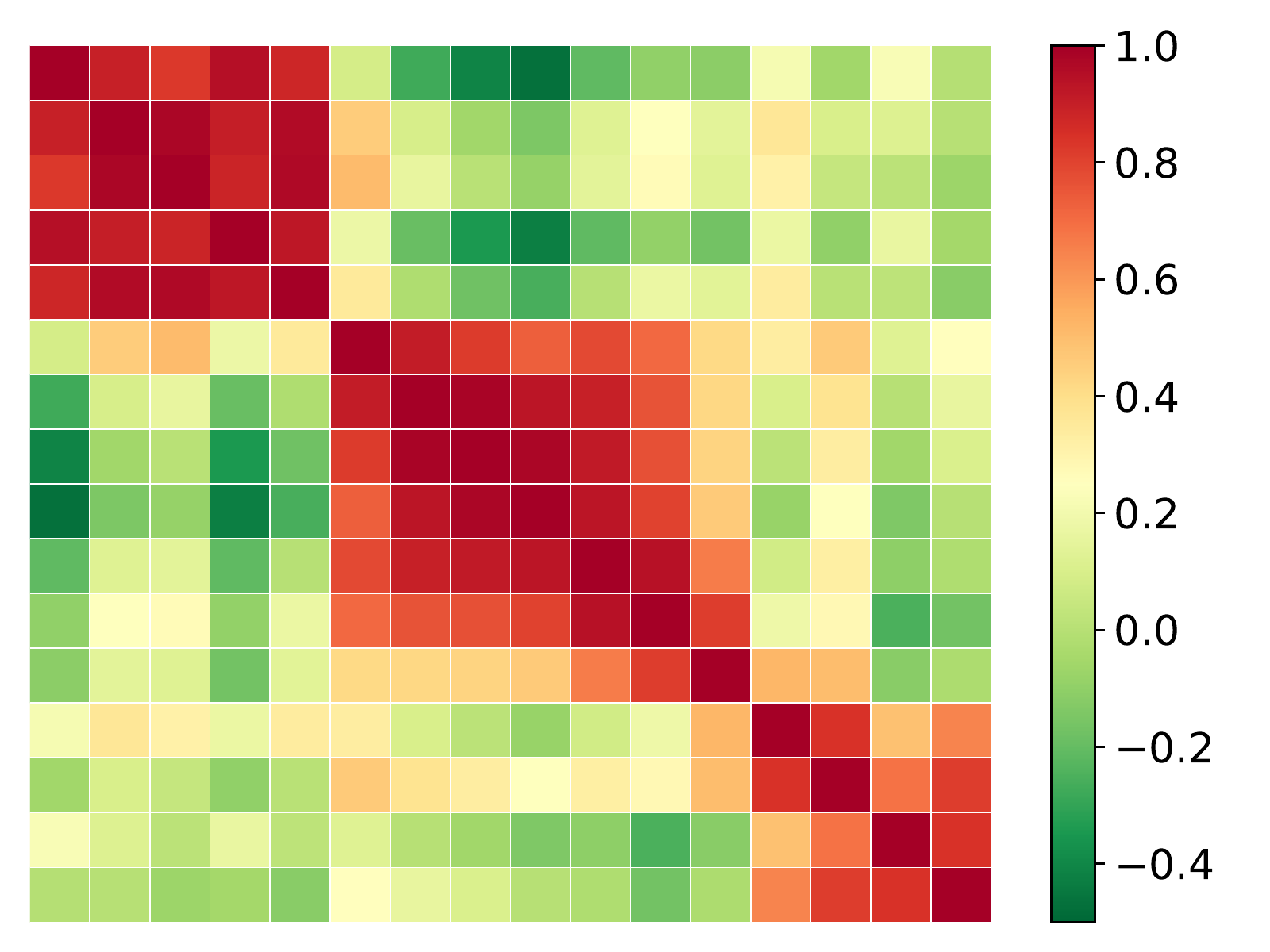}
		\caption{GAN.}
	\end{subfigure}%
	\begin{subfigure}{0.25\textwidth}
		\centering
		\includegraphics[width=\linewidth]{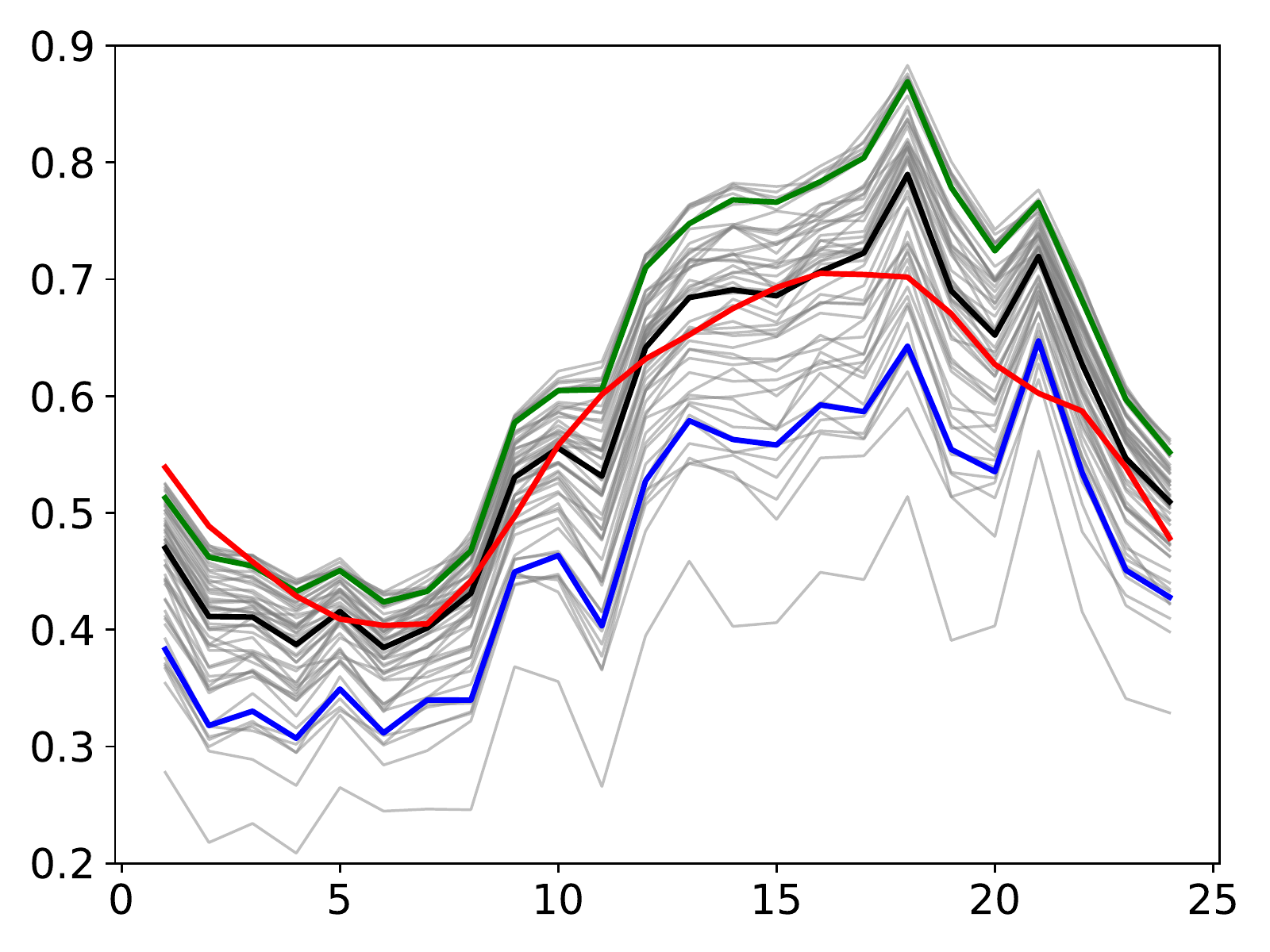}
		\caption{GAN.}
	\end{subfigure}%
	\begin{subfigure}{0.25\textwidth}
		\centering
		\includegraphics[width=\linewidth]{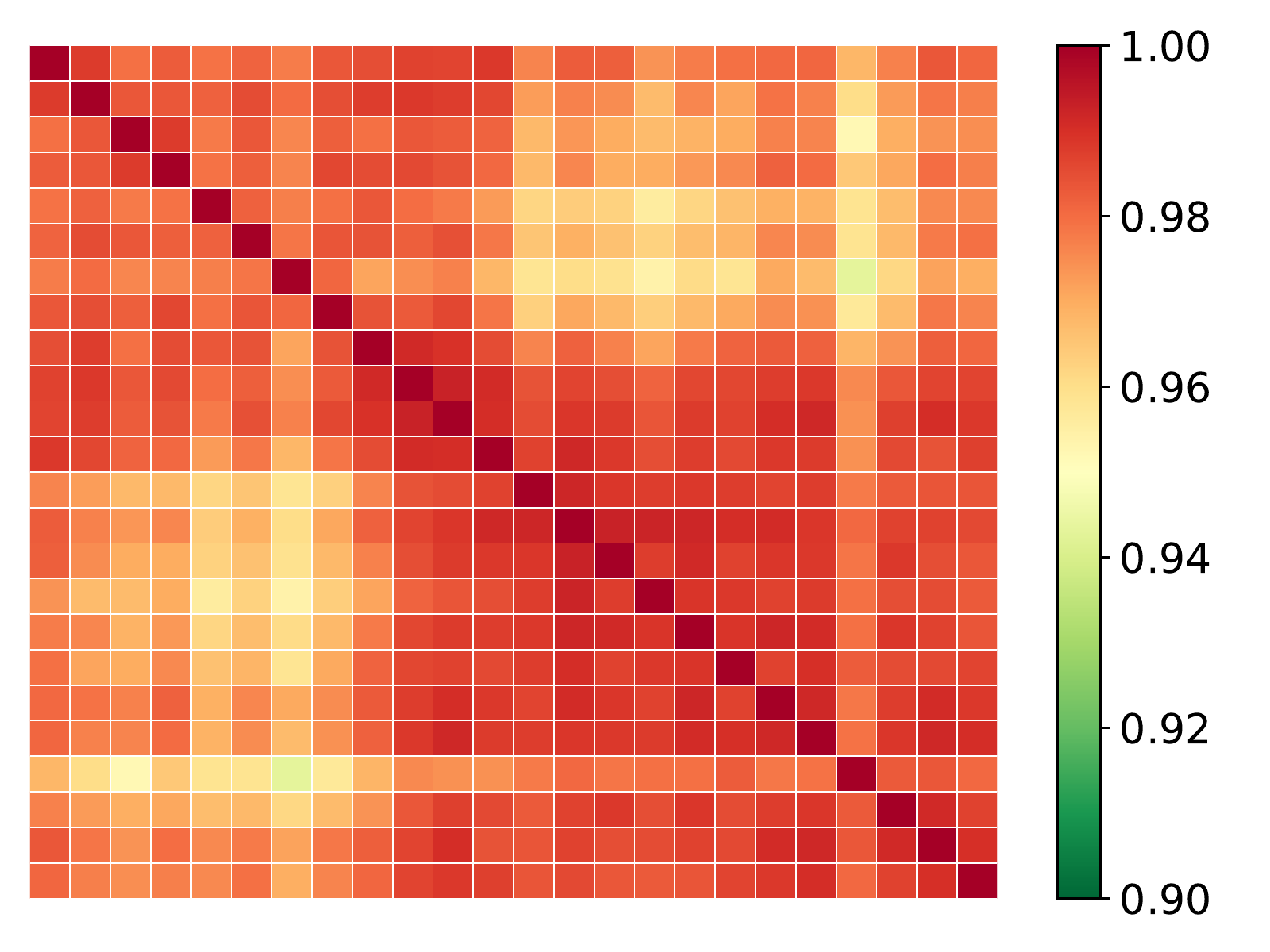}
		\caption{GAN.}
	\end{subfigure}
	\begin{subfigure}{0.25\textwidth}
		\centering
		\includegraphics[width=\linewidth]{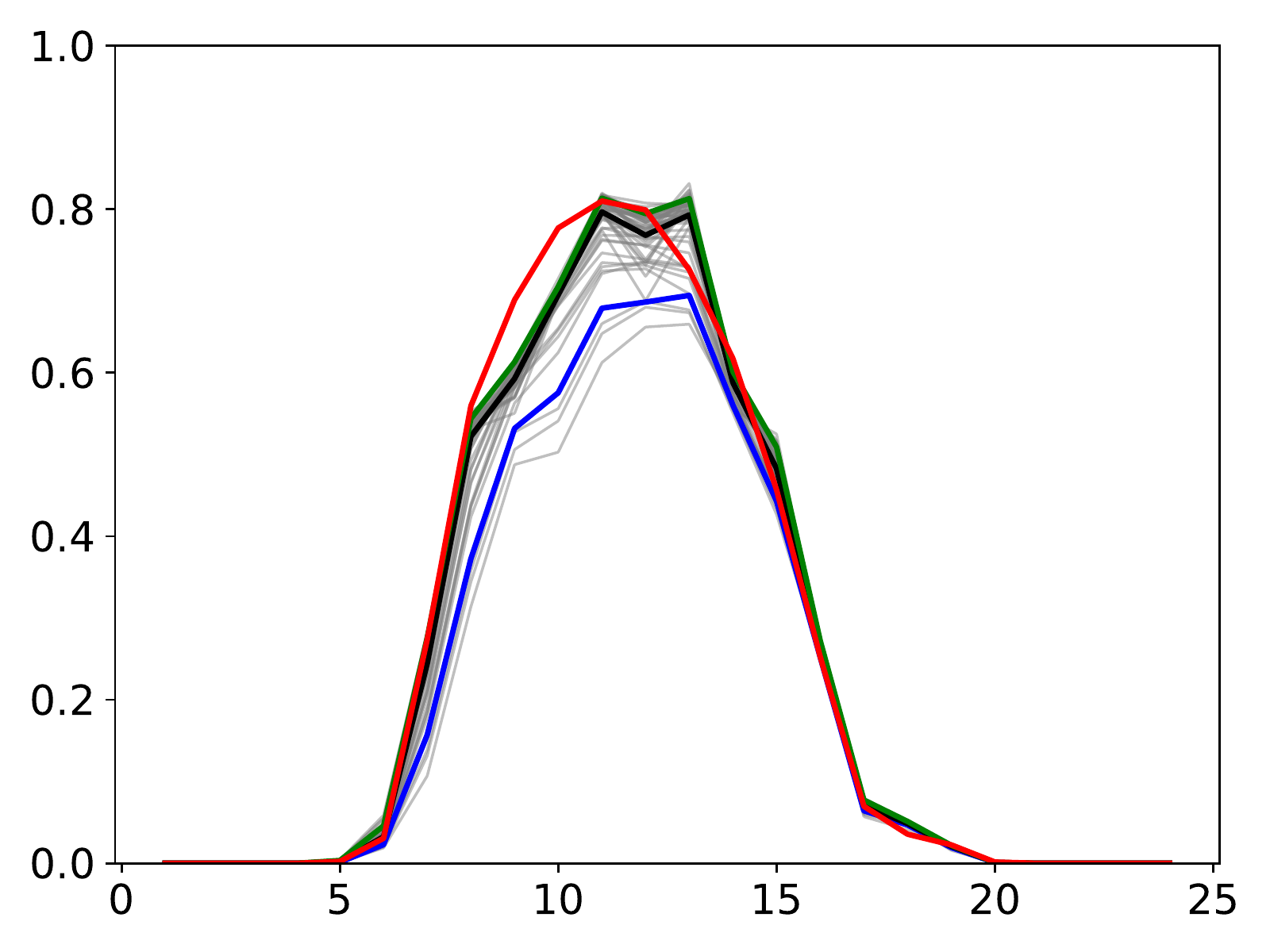}
		\caption{VAE.}
	\end{subfigure}%
	\begin{subfigure}{0.25\textwidth}
		\centering
		\includegraphics[width=\linewidth]{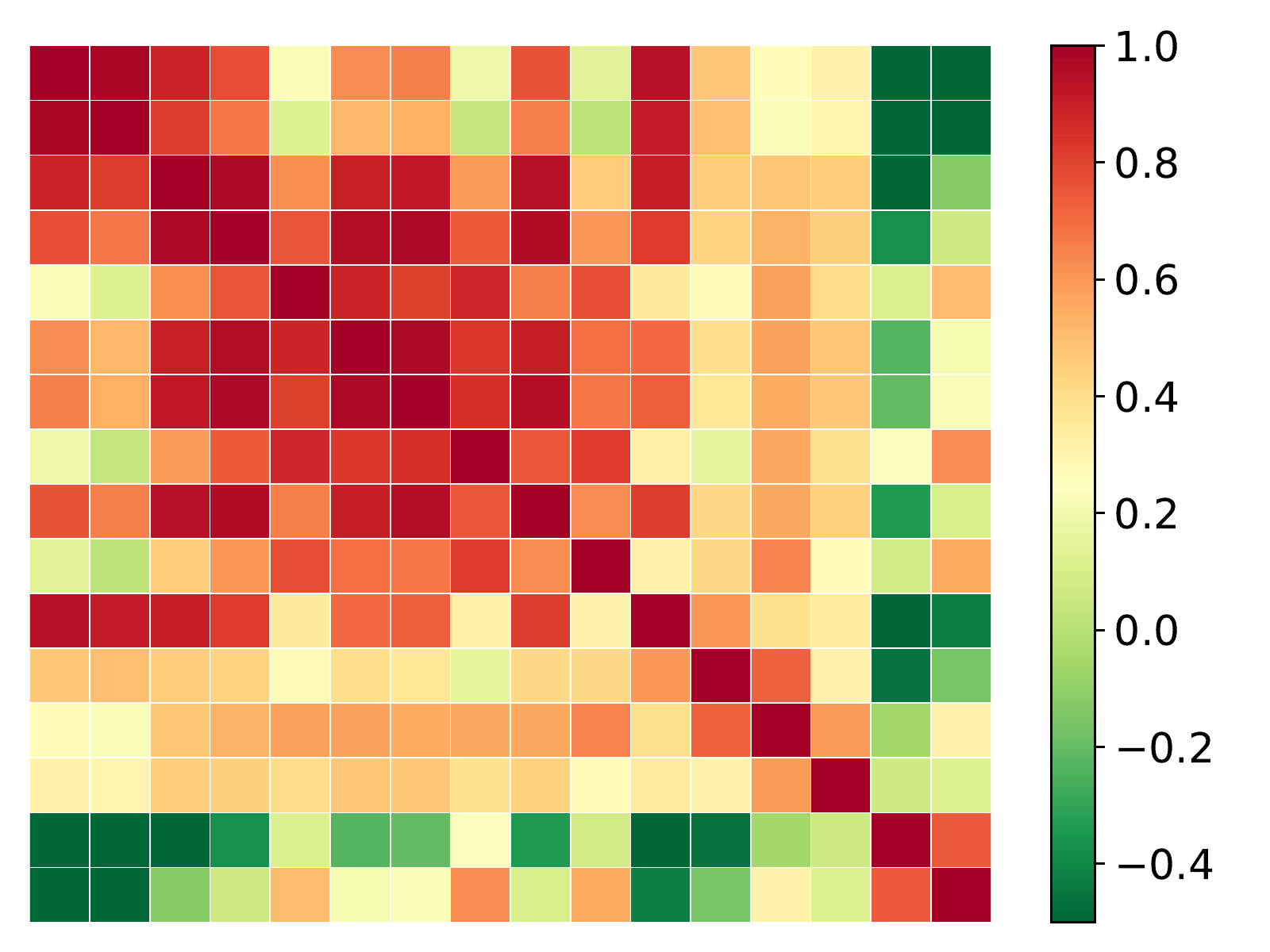}
		\caption{VAE.}
	\end{subfigure}%
	\begin{subfigure}{0.25\textwidth}
		\centering
		\includegraphics[width=\linewidth]{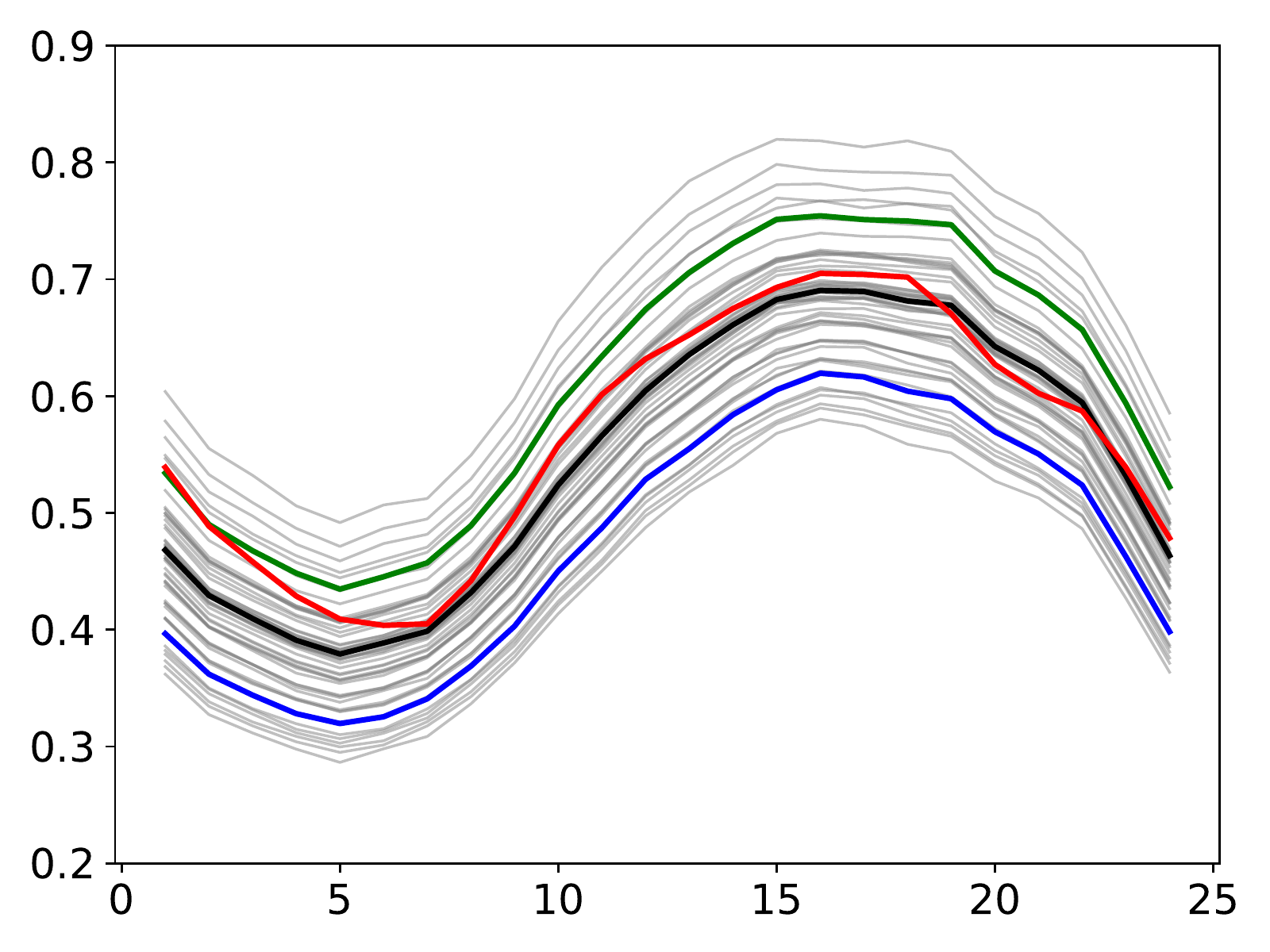}
		\caption{VAE.}
	\end{subfigure}%
	\begin{subfigure}{0.25\textwidth}
		\centering
		\includegraphics[width=\linewidth]{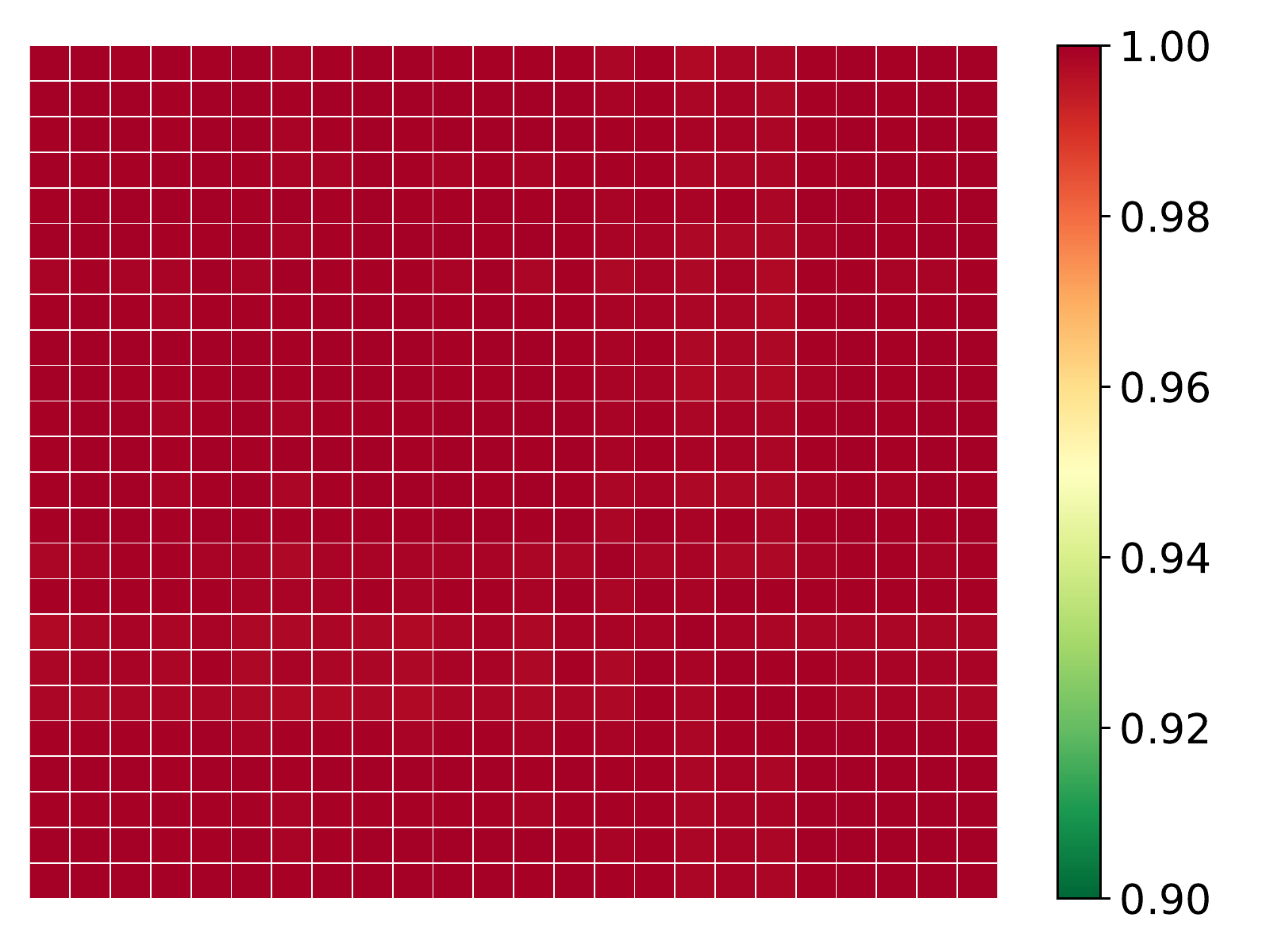}
		\caption{VAE.}
	\end{subfigure}
	\caption{PV and load scenarios shape comparison and analysis. \\
    For each model, there are 50 PV and load scenarios (grey) of a randomly selected day of the testing set along with the 10 \% (blue), 50 \% (black), and 90 \% (green) quantiles, and the observations (red). The corresponding Pearson time correlation matrices of these scenarios with the periods as rows and columns are displayed. 
    Like wind power scenarios, NF tends to exhibit no time correlation between PV and load tracks scenarios. In contrast, the VAE and GAN tend to be partially time-correlated over a few periods for the PV track. Furthermore, the VAE and GAN tend to be highly time-correlated for the load track. 
	}
	\label{fig:pv-load-scenarios}
\end{figure}

\begin{figure}[tb]
	\centering
	\begin{subfigure}{.33\textwidth}
		\centering
		\includegraphics[width=\linewidth]{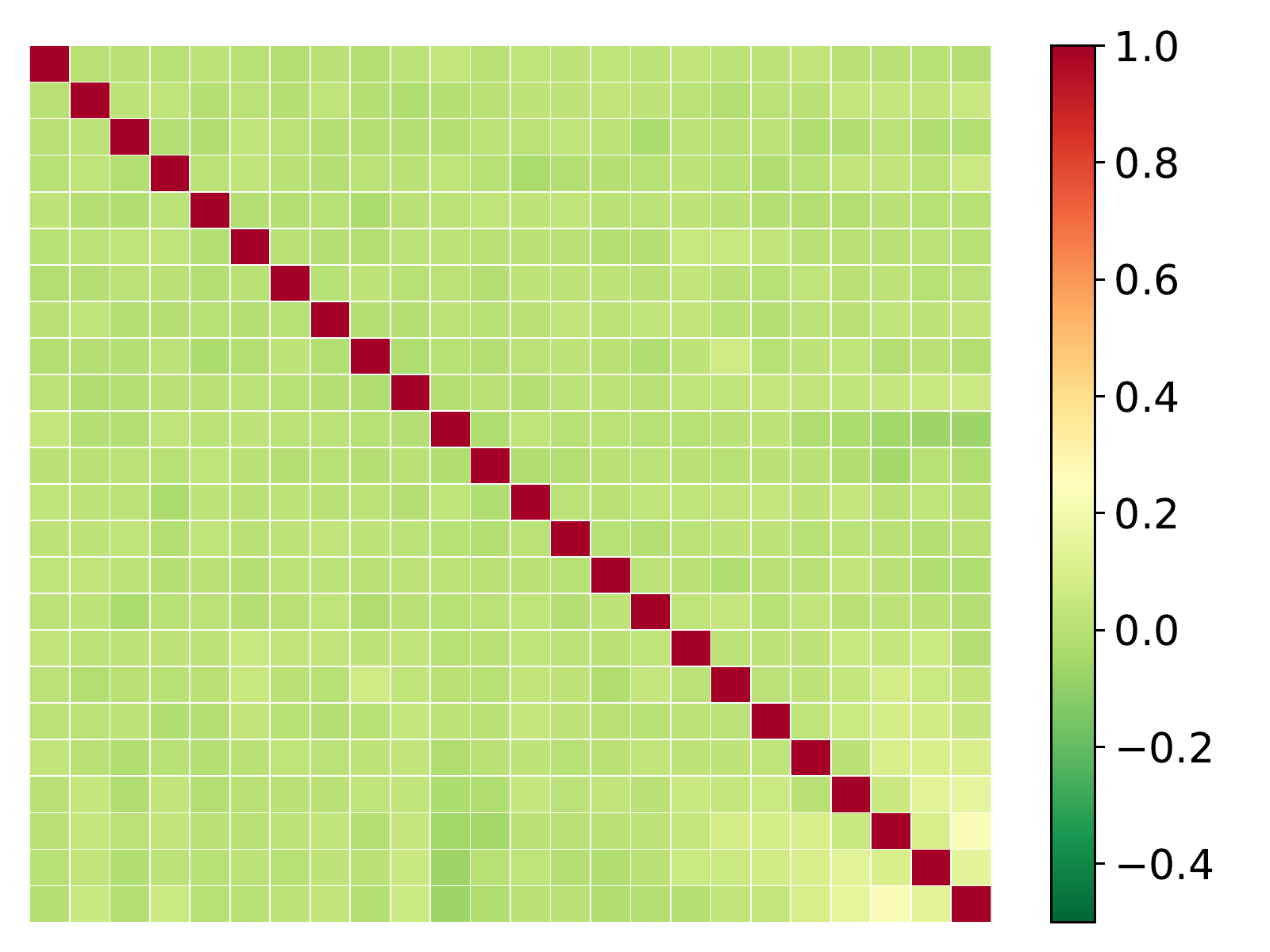}
		\caption{Wind-NF.}
	\end{subfigure}%
	\begin{subfigure}{.33\textwidth}
		\centering
		\includegraphics[width=\linewidth]{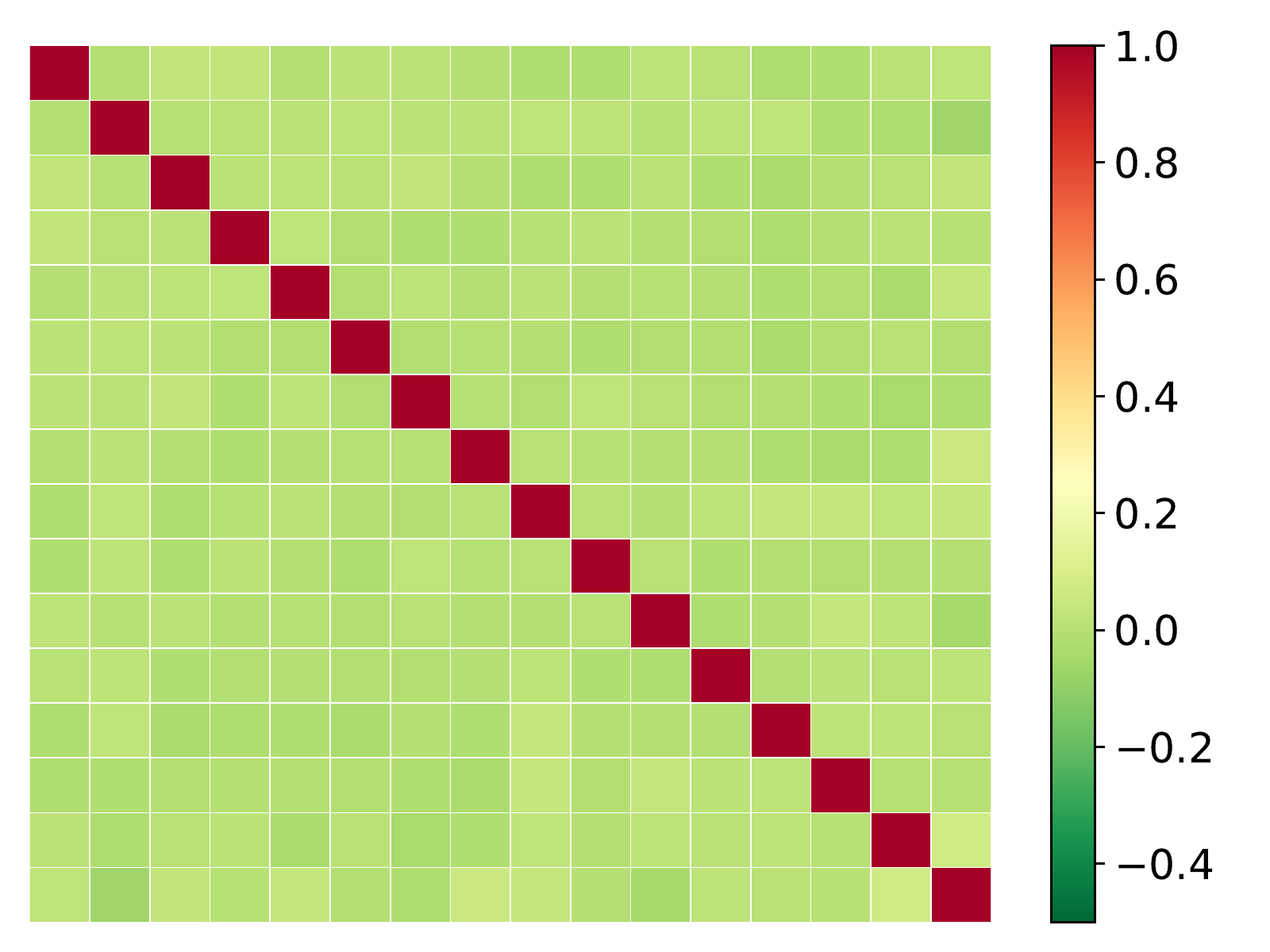}
		\caption{PV-NF.}
	\end{subfigure}%
	\begin{subfigure}{.33\textwidth}
		\centering
		\includegraphics[width=\linewidth]{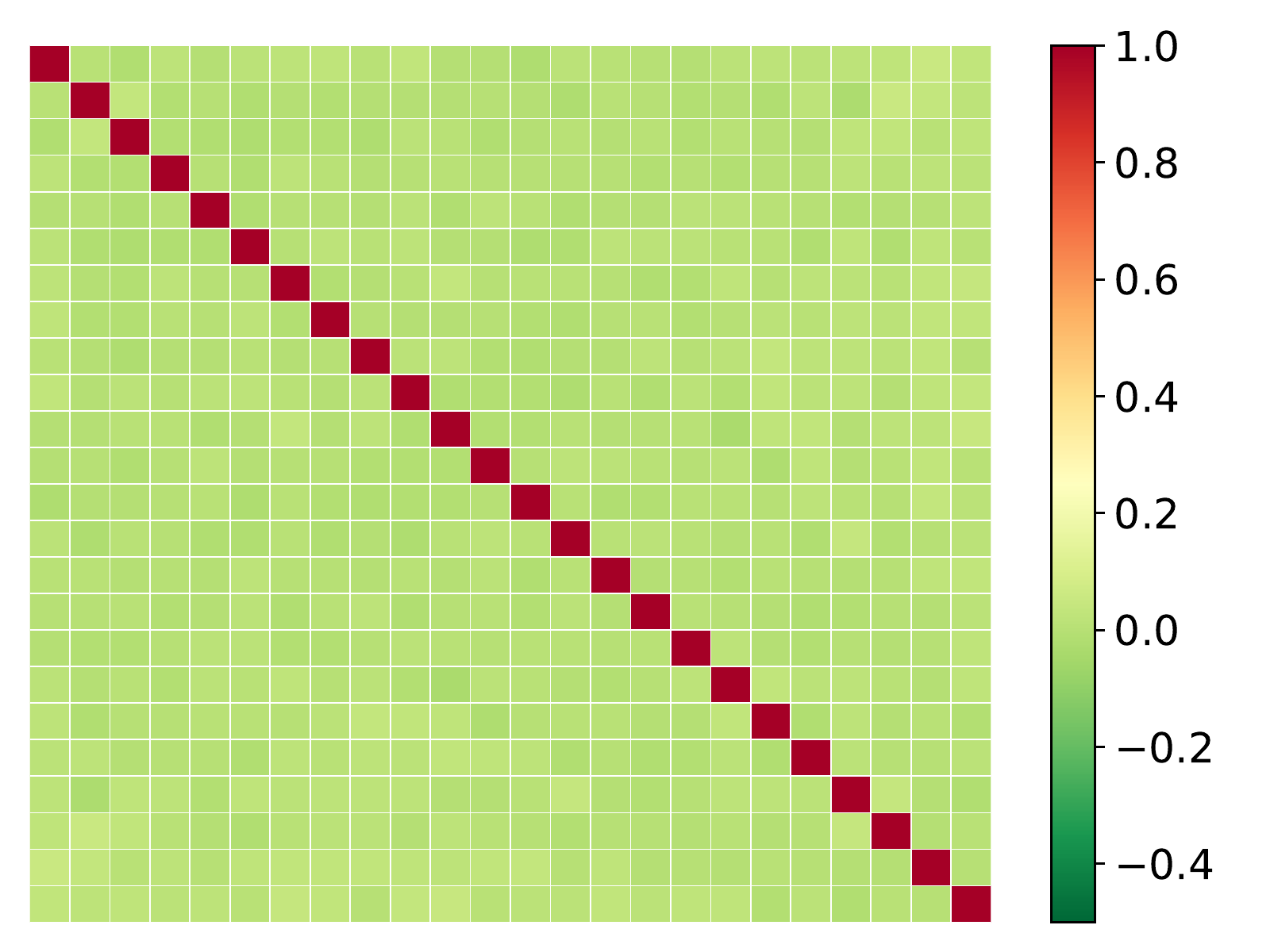}
		\caption{NF-load.}
	\end{subfigure}
	\begin{subfigure}{.33\textwidth}
		\centering
		\includegraphics[width=\linewidth]{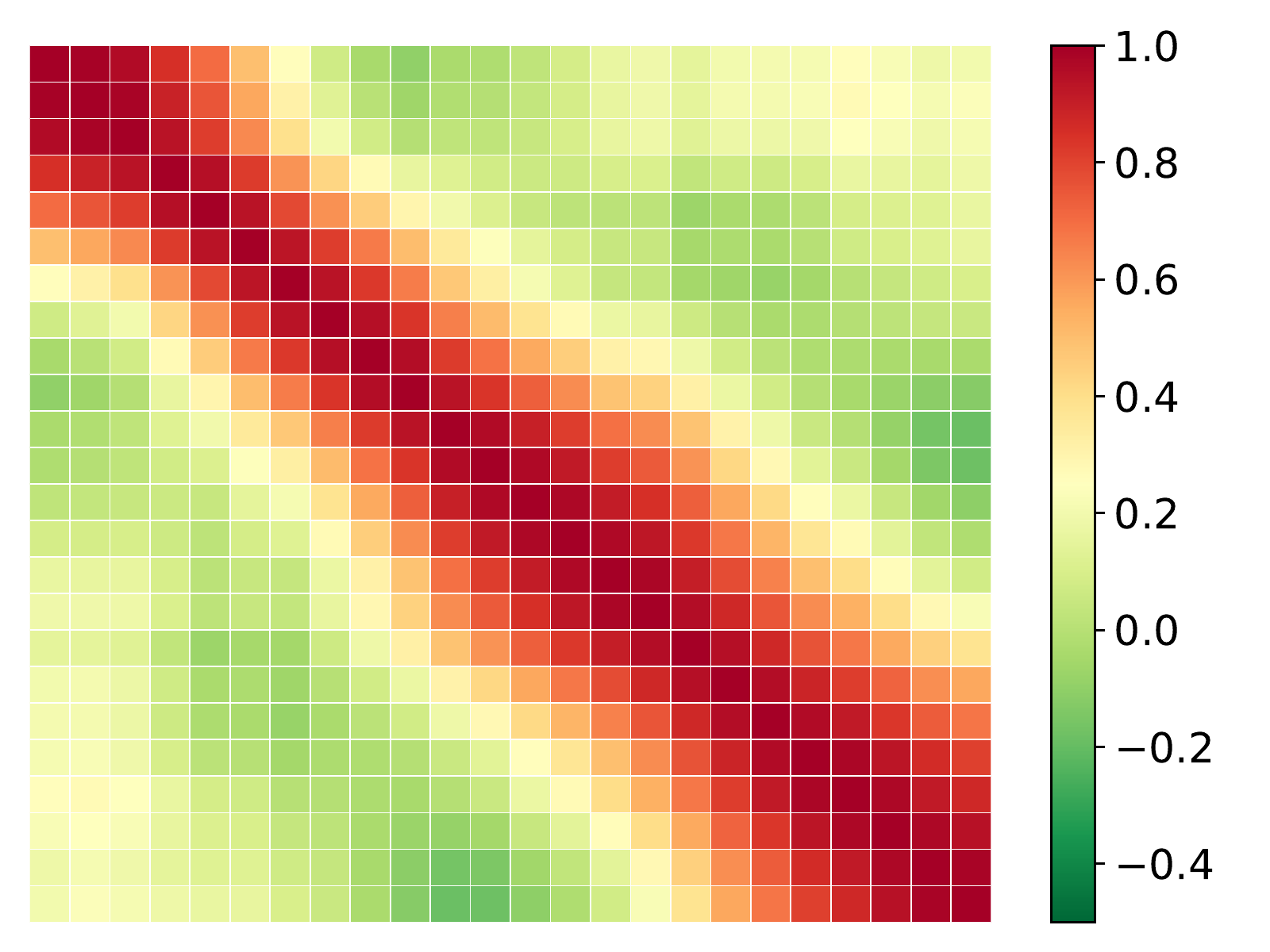}
		\caption{Wind-GAN.}
	\end{subfigure}%
	\begin{subfigure}{.33\textwidth}
		\centering
		\includegraphics[width=\linewidth]{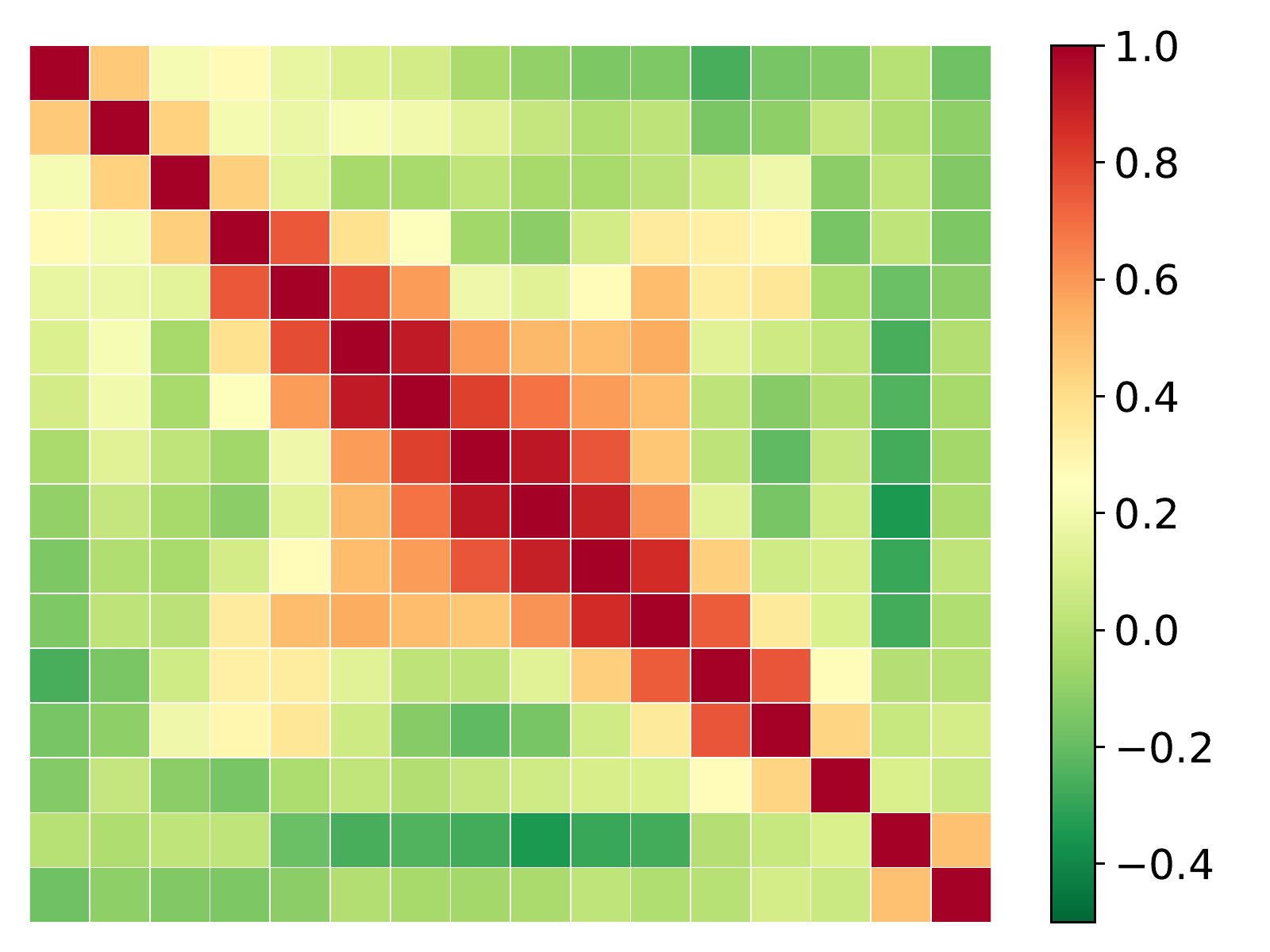}
		\caption{PV-GAN.}
	\end{subfigure}%
	\begin{subfigure}{.33\textwidth}
		\centering
		\includegraphics[width=\linewidth]{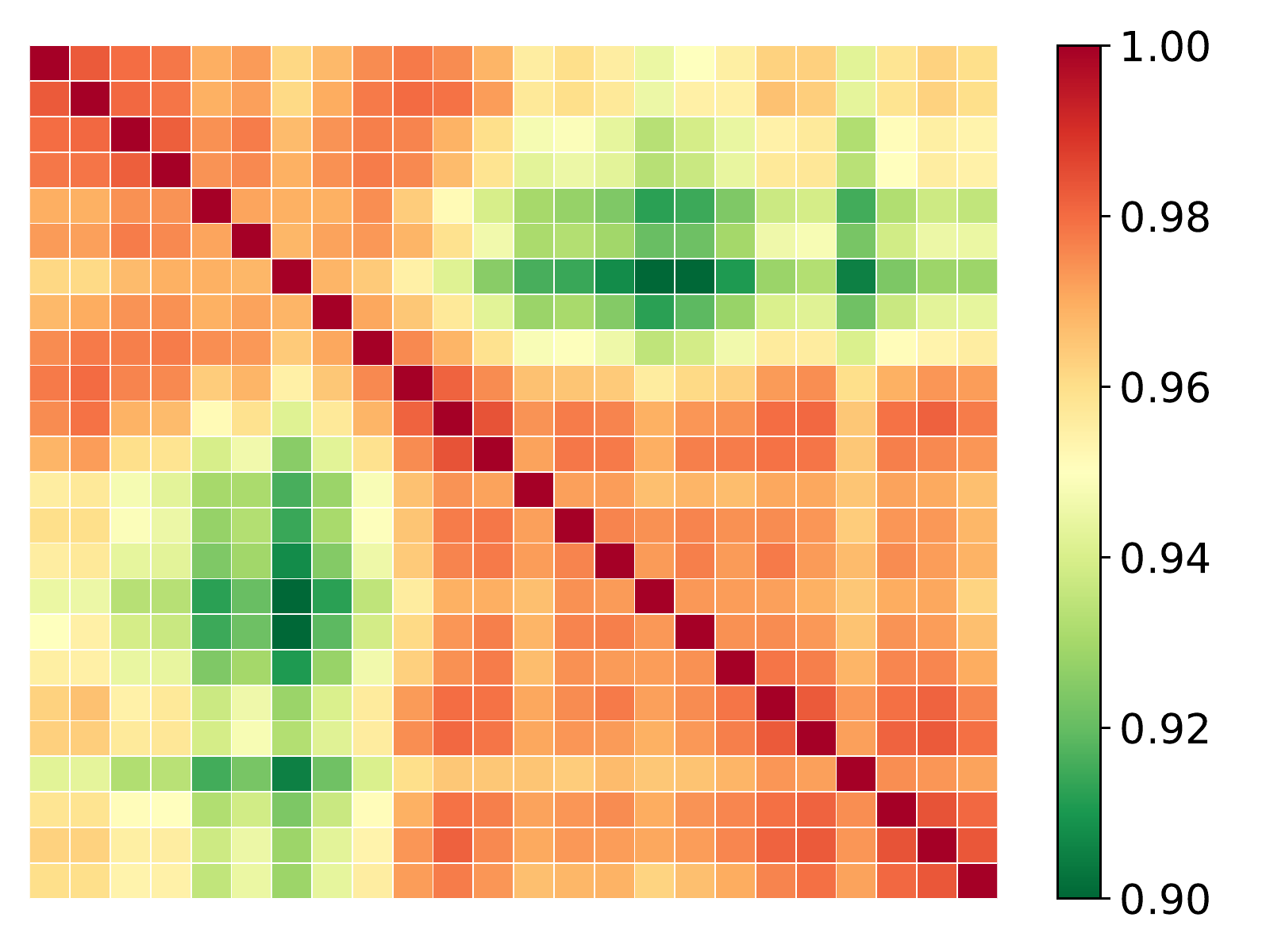}
		\caption{GAN-load.}
	\end{subfigure}
	\begin{subfigure}{.33\textwidth}
		\centering
		\includegraphics[width=\linewidth]{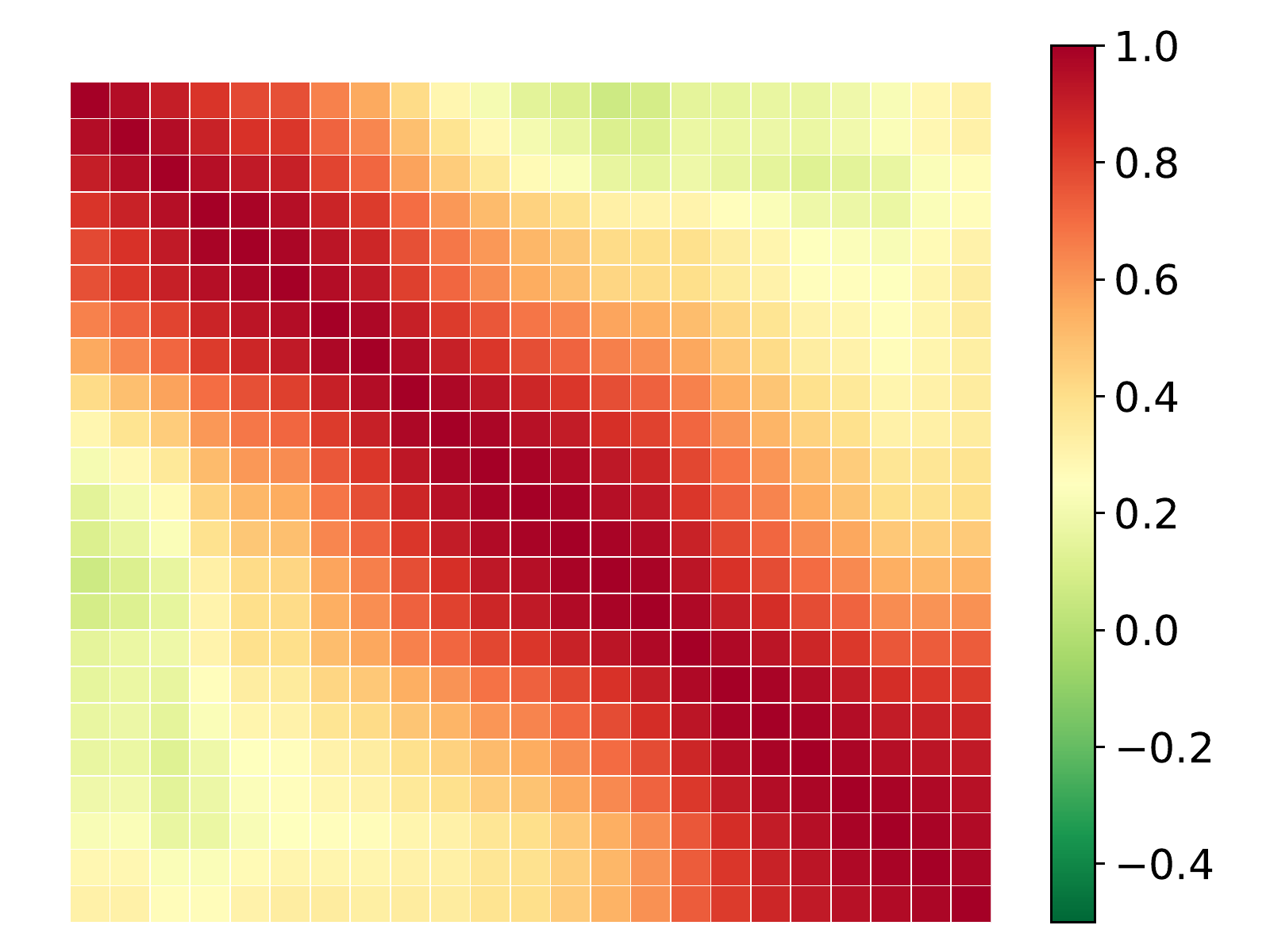}
		\caption{Wind-VAE.}
	\end{subfigure}%
	\begin{subfigure}{.33\textwidth}
		\centering
		\includegraphics[width=\linewidth]{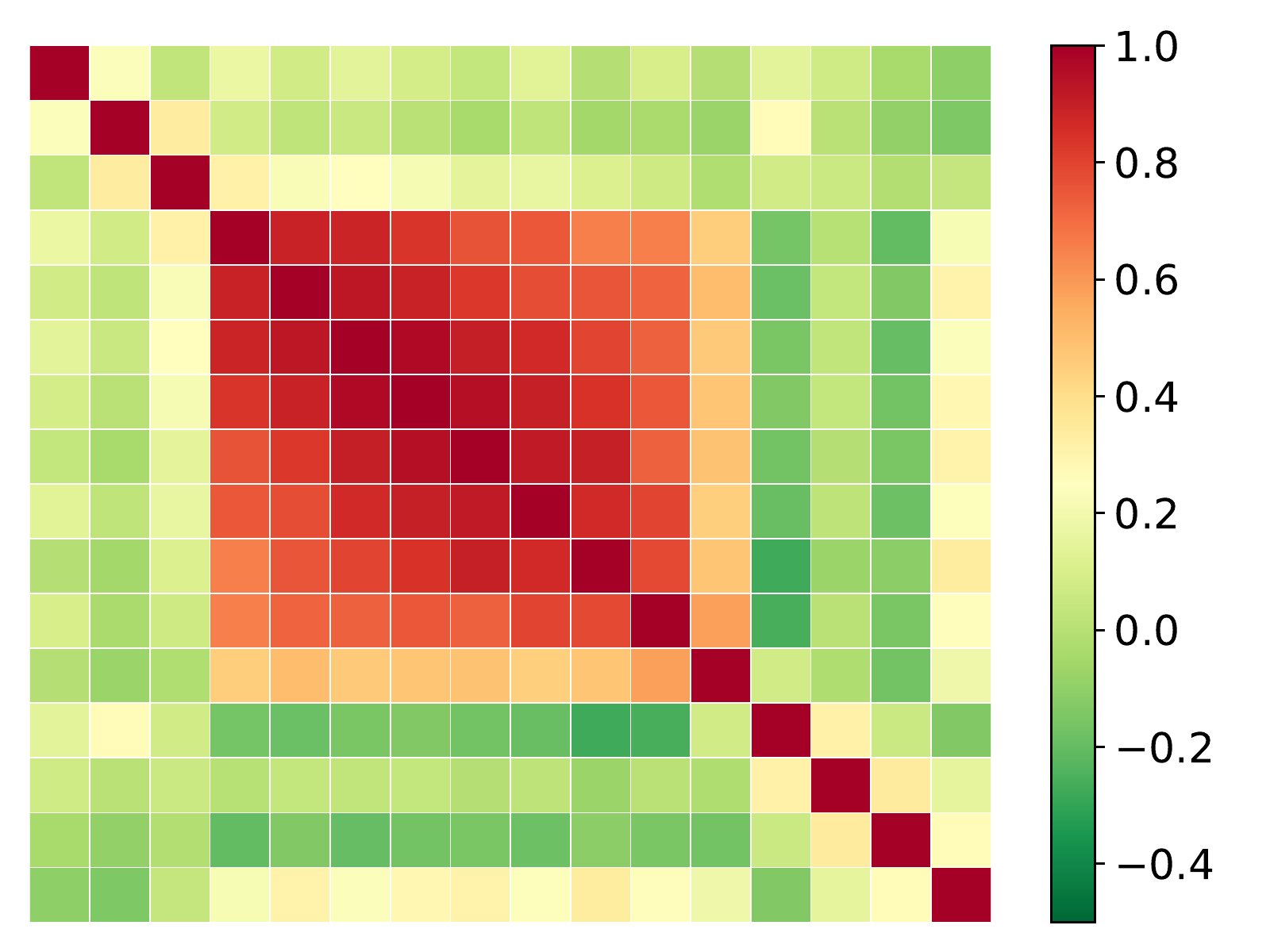}
		\caption{PV-VAE.}
	\end{subfigure}%
	\begin{subfigure}{.33\textwidth}
		\centering
		\includegraphics[width=\linewidth]{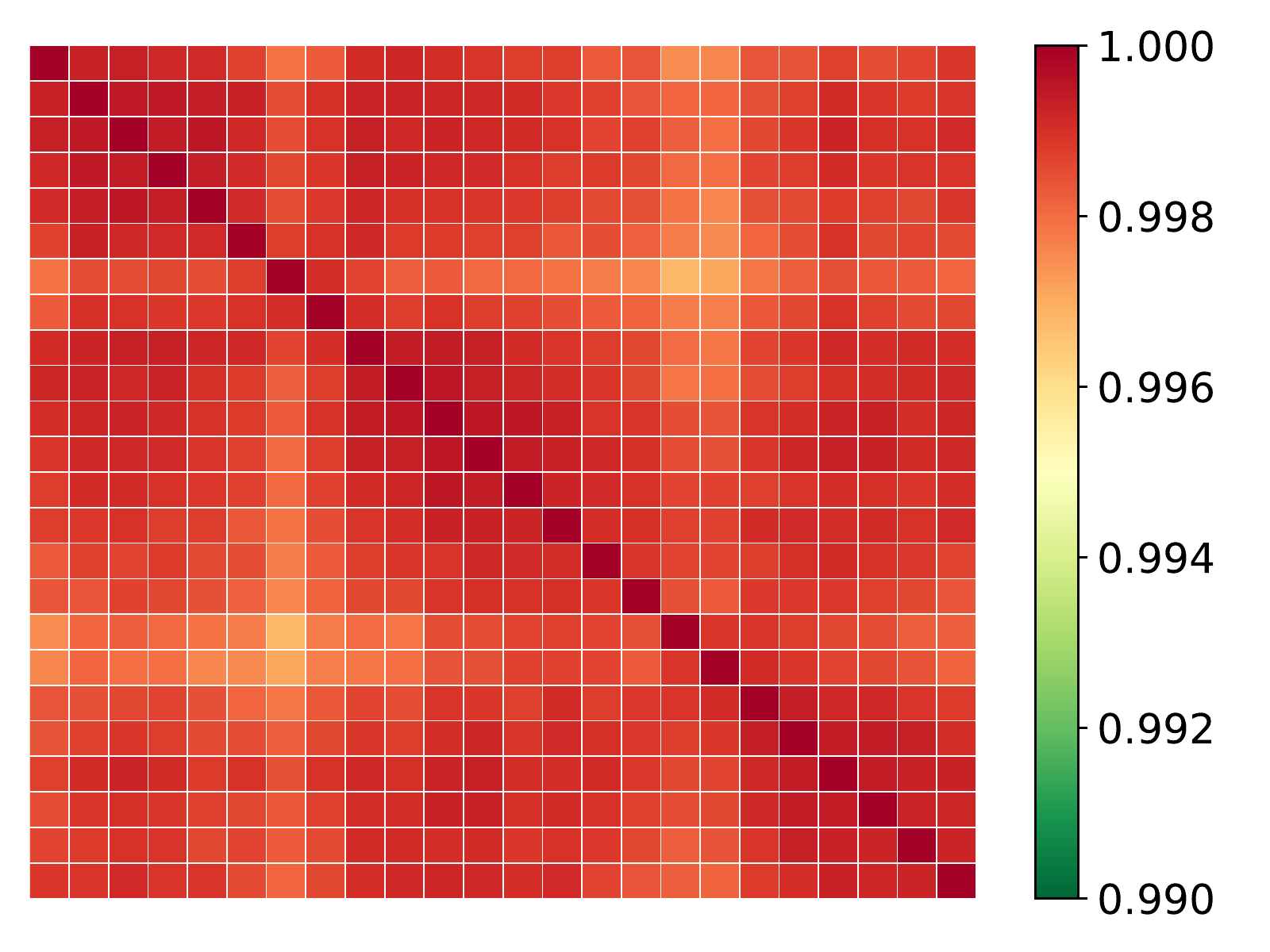}
		\caption{VAE-load.}
	\end{subfigure}
	\caption{Average of the correlation matrices over the testing set for the three datasets. \\
    Left: wind power; center: PV; right:load. The trend in terms of time correlation is observed on each day of the testing set for all tracks. The NF scenarios are not correlated. In contrast, the VAE and GAN scenarios tend to be time-correlated over a few periods. In particular, the VAE generates highly time-correlated scenarios for the load dataset.
	}
	\label{fig:average-correlation}
\end{figure}

Similar to the wind track, the shape of the scenarios differs significantly between the NF and the other models for both the load and PV tracks as indicated by Figure \ref{fig:pv-load-scenarios}. Note: the load track scenarios are highly correlated for both the VAE and GAN. Finally, Figure \ref{fig:average-correlation} provides the average of the correlation matrices over all days of the testing set for each dataset. The trend depicted above is confirmed.
This difference between the NF and the other generative model may be explicated by the design of the methods. The NF explicitly learns the probability density function (PDF) of the multi-dimensional random variable considered. Thus, the NF scenarios are generated according to the learned PDF producing multiple shapes of scenarios. In contrast, the generator of the GAN is trained to fool the discriminator, and it may find a shape particularly efficient leading to a set of similar scenarios. Concerning the VAE, it is less obvious. However, by design, the decoder is trained to generate scenarios from the latent space assumed to follow a Gaussian distribution that may lead to less variability.
%
%
%
%

\clearpage

\section{Conclusions and perspectives}\label{sec:ijf-conclusion}

This Chapter proposed a fair and thorough comparison of NFs with the state-of-the-art deep learning generative models, GANs and VAEs, both in quality and value. The numerical experiments employ the open data of the Global Energy Forecasting Competition 2014. The generative models use the conditional information to compute improved weather-based PV, wind power, and load scenarios. This Chapter demonstrated that NFs can challenge GANs and VAEs as they are, overall, more accurate both in terms of quality and value (see Chapter \ref{chap:energy-retailer}). In addition, they can be used effectively by non-expert deep learning practitioners. 
In addition, NFs have several advantages over more traditional deep learning approaches that should motivate their introduction into power system applications:
\begin{enumerate}
	\item NFs directly learn the stochastic multivariate distribution of the underlying process by maximizing the likelihood. Therefore, in contrast to VAEs and GANs, NFs provide access to the exact likelihood of the model's parameters, hence offering a sound and direct way to optimize the network parameters \citep{wehenkel2020graphical}. It may open a new range of advanced applications benefiting from this advantage. For instance, to transfer scenarios from one location to another based on the knowledge of the probability density function. A second application is the importance sampling for stochastic optimization based on a scenario approach. Indeed, NFs provide the likelihood of each generated scenario, making it possible to filter relevant scenarios in stochastic optimization.
	\item In our opinion, NFs are easier to use by non-expert deep learning practitioners once the libraries are available, as they are more reliable and robust in terms of hyper-parameters selection. GANs and VAEs are particularly sensitive to the latent space dimension, the structure of the neural networks, the learning rate, \textit{etc}. GANs convergence, by design, is unstable, and for a given set of hyper-parameters, the scenario's quality may differ completely. In contrast, it was easier to retrieve relevant NFs hyper-parameters by manually testing a few sets of values, satisfying training convergence, and quality results.  
\end{enumerate}
Nevertheless, their usage as a base component of the machine learning toolbox is still limited compared to GANs or VAEs.

\section{Appendix: Table \ref{tab:AE-contributions} justifications}\label{annex:AE-table1}

\citet{wang2020modeling} use a Wasserstein GAN with gradient penalty to model both the uncertainties and the variations of the load. First, point forecasting is conducted, and the corresponding residuals are derived. Then, the GAN generates residual scenarios conditional on the day type, temperatures, and historical loads. 
The GAN model is compared with the same version without gradient penalty and two quantile regression models: random forest and gradient boosting regression tree. 
The quality evaluation is conducted on open load datasets from the Independent System Operator-New England\footnote{\url{https://www.iso-ne.com/}} with five metrics: (1) the continuous ranked probability score; (2) the quantile score; (3) the Winkler score; (4) reliability diagrams; (5) Q-Q plots. Note: the forecast value is not assessed. \\

\citet{qi2020optimal} propose a concentrating solar power (CSP) configuration method to determine the CSP capacity in multi-energy power systems. The configuration model considers the uncertainty by scenario analysis. A $\beta$ VAE generates the scenarios. It is an improved version of the original VAE
However, it does not consider weather forecasts, and the model is trained only by using historical observations.
The quality evaluation is conducted on two wind farms and six PV plants using three metrics. (1) The leave-one-out accuracy of the 1-nearest neighbor classifier. (2) The comparison of the frequency distributions of the actual data and the generated scenarios. (3) Comparing the spatial and temporal correlations of the actual data and the scenarios by computing Pearson correlation coefficients. 
The value is assessed by considering the case study of the CSP configuration model, where the $\beta$ VAE is used to generate PV, wind power, and load scenarios. However, the VAE is not compared to another generative model for both the quality and value evaluations. 
Note: the dataset does not seem to be in open-access. 
Finally, the value evaluation case study is not trivial due to the mathematical formulation that requires a certain level of knowledge of the system. Thus, the replicability criterion is partially satisfied. \\

\citet{ge2020modeling} compared NFs to VAEs and GANs for the generation of daily load profiles. The models do not take into account weather forecasts but only historical observations. However, an example is given to illustrate the principle of generating conditional daily load profiles by using three groups: light load, medium load, and heavy load.
The quality evaluation uses five indicators. Four to assess the temporal correlation: (1) probability density function; (2) autocorrelation function;  (3) load duration curve; (4) a wave rate is defined to evaluate the volatility of the daily load profile. Furthermore, one additional for the spatial correlation: (5) Pearson correlation coefficient is used to measure the spatial correlation among multiple daily load profiles.
The simulations use the open-access London smart meter and Spanish transmission service operator datasets of Kaggle.
The forecast value is not assessed.

\section{Appendix: background}\label{annex:ijf-background}
\subsection{NFs}\label{annex:ijf-nf}

\subsubsection*{NF computation}

\noindent Evaluating the likelihood of a distribution modeled by a normalizing flow requires computing (\ref{eq:change_formula}), \textit{i.e.}, the normalizing direction, as well as its log-determinant. Increasing the number of sub-flows by $K$ of the transformation results in only $\mathcal{O}(K)$ growth in the computational complexity as the log-determinant of $ J_{f_\theta} $ can be expressed as
\begin{subequations}
\label{eq:jacobian_composite}	
\begin{align}
\log |\det J_{f_\theta}(\mathbf{x}) | & = \log \bigg| \prod_{k=1}^K \det J_{f_{k,\theta}}(\mathbf{x}) \bigg|, \\
                             & = \sum_{k=1}^K \log| \det J_{f_{k,\theta}}(\mathbf{x})|.
\end{align}
\end{subequations}
However, with no further assumption on $f_\theta$, the computational complexity of the log-determinant is $\mathcal{O}(K \cdot T^3)$, which can be intractable for large $T$. Therefore, the efficiency of these operations is essential during training, where the likelihood is repeatedly computed. There are many possible implementations of NFs detailed by \citet{papamakarios2019normalizing,kobyzev2020normalizing} to address this issue. 

\subsubsection*{Autoregressive flow}

\noindent The Jacobian of the autoregressive transformation $f_\theta$ defined by (\ref{eq:f_autoregressive}) is lower triangular, and its log-absolute-determinant is
\begin{subequations}
\label{eq:jacobian_autoregressive}	
\begin{align}
	\log | \det J_{f_\theta}(\mathbf{x})  |& = \log  \prod_{i=1}^T \bigg|\dfrac{\partial f^i}{\partial x_i} (x_i; h^i) \bigg|, 	\\
	& = \sum_{i=1}^T \log \bigg|\dfrac{\partial f^i}{\partial x_i} (x_i; h^i) \bigg|,
\end{align}
\end{subequations}
that is calculated in $\mathcal{O}(T)$ instead of $\mathcal{O}(T^3)$. 

\subsubsection*{Affine autoregressive flow}

\noindent A simple choice of transformer is the class of affine functions
\begin{align}
\label{eq:f_affin_autoregressive}	
	f^i(x_i; h^i) & = \alpha_i x_i + \beta_i ,
\end{align}
where $f^i(\cdot; h^i ) : \mathbb{R} \rightarrow \mathbb{R}$ is parameterized by $h^i = \{ \alpha_i,  \beta_i\}$, $\alpha_i$ controls the scale, and $\beta_i$ controls the location of the transformation. Invertibility is guaranteed if $\alpha_i \ne 0$, and this can be easily achieved by \textit{e.g.} taking $\alpha_i = \exp{(\tilde{\alpha}_i)}$, where $\tilde{\alpha}_i$ is an unconstrained parameter in which case $h^i = \{ \tilde{\alpha}_i,  \beta_i\}$. The derivative of the transformer with respect to $x_i$ is equal to $\alpha_i$. Hence the log-absolute-determinant of the Jacobian becomes
\begin{align}
\label{eq:jacobian_affine_autoregressive}	
	\log | \det J_{f_\theta}(\mathbf{x})  |& = \sum_{i=1}^T \log |\alpha_i | =  \sum_{i=1}^T \tilde{\alpha}_i.
\end{align}
Affine autoregressive flows are simple and computation efficient. However, they are limited in expressiveness requiring many stacked flows to represent complex distributions. It is unknown whether affine autoregressive flows with multiple layers are universal approximators \citep{papamakarios2019normalizing}.

\subsection{VAEs}\label{annex:ijf-vae}

\begin{figure}[tb]
	\centering
	\includegraphics[width=0.5\linewidth]{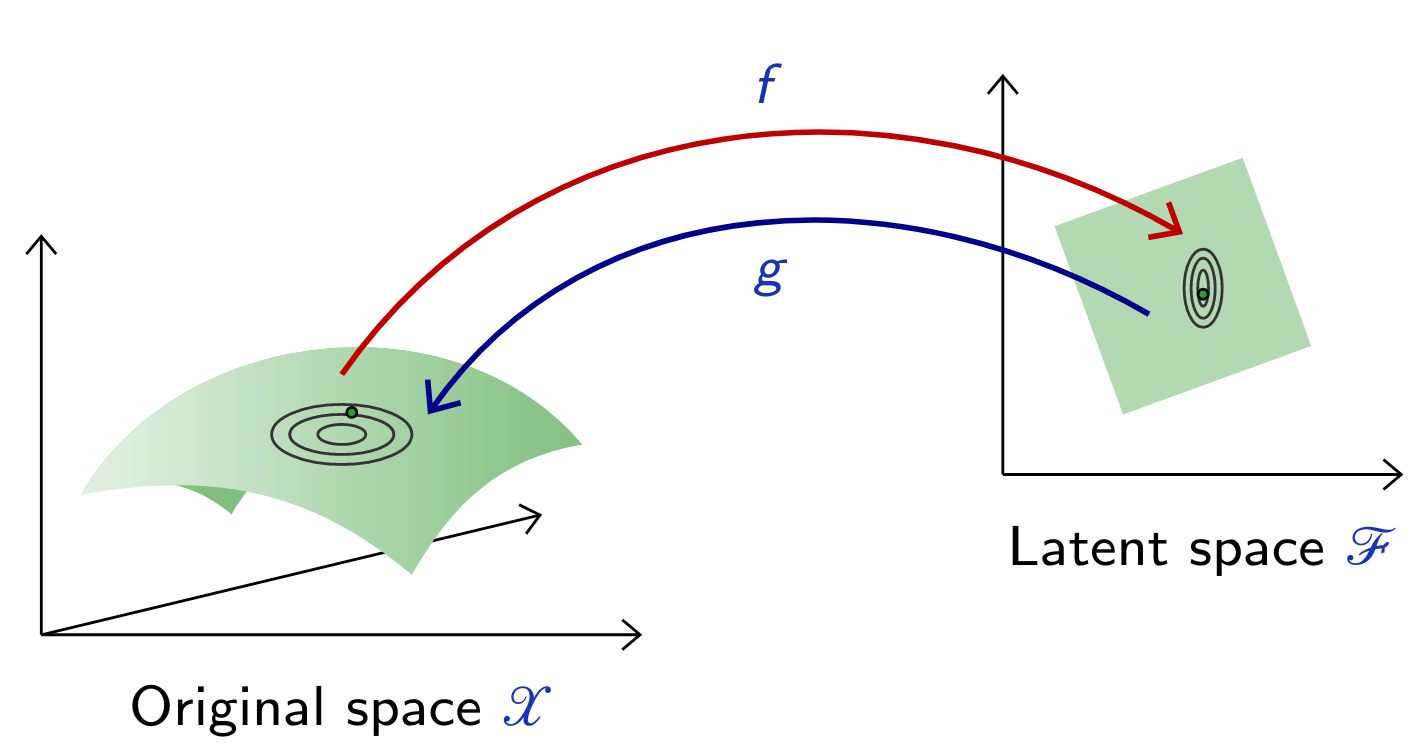}
	\caption{VAE process. Credits: Francois Fleuret, EE559 Deep Learning, EPFL \url{https://fleuret.org/dlc/}}
	\label{fig:scenarios-vae-scheme}
\end{figure}

Figure \ref{fig:scenarios-vae-scheme} illustrates the VAE process with $f$ the encoder and $g$ the decoder.

\subsubsection*{Gradients computation}
\noindent By using (\ref{eq:variational_lower_bound}) $\mathcal{L}_{\theta, \varphi}$ is decomposed in two parts
\begin{align}
\label{eq:lower_bound}	
\mathcal{L}_{\theta, \varphi}(\mathbf{x},\mathbf{c}) = & \underset{q_\varphi(\mathbf{z}|\mathbf{x},\mathbf{c})}{\mathbb{E}} [\log p_\theta(\mathbf{x}|\mathbf{z},\mathbf{c}) ] -\text{KL}[ q_\varphi(\mathbf{z}|\mathbf{x},\mathbf{c}) || p(\mathbf{z})] .
\end{align}
$\nabla_\theta  \mathcal{L}_{\theta, \varphi}$ is estimated with the usual Monte Carlo gradient estimator. However, the estimation of $\nabla_\varphi  \mathcal{L}_{\theta, \varphi}$ requires the reparameterization trick proposed by \citet{kingma2013auto}, where the random variable $\mathbf{z}$ is re-expressed as a deterministic variable 
\begin{align}
	\label{eq:reparameterization_trick1}	
	\mathbf{z} = & g_\varphi(\epsilon, \mathbf{x}),
\end{align}
with $\epsilon$ an auxiliary variable with independent marginal $p_\epsilon$, and $g_\varphi(\cdot)$ some vector-valued function parameterized by $\varphi$.
%
Then, the first right hand side of (\ref{eq:lower_bound}) becomes
\begin{align}
\label{eq:RHS_2}	
\underset{q_\varphi(\mathbf{z}|\mathbf{x},\mathbf{c})}{\mathbb{E}} [\log p_\theta(\mathbf{x}|\mathbf{z},\mathbf{c})]  = &  \underset{p(\epsilon)}{\mathbb{E}} [\log p_\theta(\mathbf{x}|g_\varphi(\epsilon, \mathbf{x}),\mathbf{c})].
\end{align}
$\nabla_\varphi  \mathcal{L}_{\theta, \varphi}$ is now estimated with Monte Carlo integration.

\subsubsection*{Conditional VAE implemented}

\noindent Following \citet{kingma2013auto}, we implemented Gaussian multi-layer perceptrons (MLPs) for both the encoder $\text{NN}_\varphi$ and decoder $\text{NN}_\theta$. In this case, $p(\mathbf{z})$ is a centered isotropic multivariate Gaussian, $p_\theta(\mathbf{x}|\mathbf{z}, \mathbf{c})$ and $q_\varphi(\mathbf{x}|\mathbf{z},\mathbf{c})$ are both multivariate Gaussian with a diagonal covariance and parameters $\boldsymbol{\mu}_\theta,\boldsymbol{\sigma}_\theta$ and $\boldsymbol{\mu}_\varphi, \boldsymbol{\sigma}_\varphi$, respectively. Note that there is no restriction on the encoder and decoder architectures, and they could as well be arbitrarily complex convolutional networks. Under these assumptions, the conditional VAE implemented is
\begin{subequations}
	\label{eq:VAE_NN_assumptions}	
	\begin{align}
	p(\mathbf{z}) & = \mathcal{N} (\mathbf{z}; \mathbf{0}, \mathbf{I}), \\
	p_\theta(\mathbf{x}|\mathbf{z},\mathbf{c}) & = \mathcal{N}(\mathbf{x};\boldsymbol{\mu}_\theta,\boldsymbol{\sigma}_\theta^2 \mathbf{I}), \\
	q_\varphi(\mathbf{z}|\mathbf{x},\mathbf{c}) & = \mathcal{N}(\mathbf{z};\boldsymbol{\mu}_\varphi,\boldsymbol{\sigma}_\varphi^2\mathbf{I} ), \\
	\boldsymbol{\mu}_\theta, \log \boldsymbol{\sigma}_\theta^2 & =\text{NN}_\theta (\mathbf{x},\mathbf{c}), \\
	\boldsymbol{\mu}_\varphi, \log \boldsymbol{\sigma}_\varphi^2 & =\text{NN}_\varphi (\mathbf{z},\mathbf{c}).
	\end{align}
\end{subequations}
%
Then, by using the valid reparameterization trick proposed by \citet{kingma2013auto}
\begin{subequations}
	\label{eq:reparameterization_trick2}	
	\begin{align}
	\boldsymbol{\epsilon} \sim & \mathcal{N}(\mathbf{0},\mathbf{I}) ,\\
	\mathbf{z} : = & \boldsymbol{\mu}_\varphi + \boldsymbol{\sigma}_\varphi \boldsymbol{\epsilon},
	\end{align}
\end{subequations}
$\mathcal{L}_{\theta, \varphi}$ is computed and differentiated without estimation using the expressions
\begin{subequations}
\label{eq:lower_bound_exact}	
\begin{align}
\text{KL} [ q_\varphi(\mathbf{z}|\mathbf{x},\mathbf{c}) || p(\mathbf{z})] = & -\frac{1}{2} \sum_{j=1}^d (1 + \log \boldsymbol{\sigma}_{\varphi,j}^2 - \boldsymbol{\mu}_{\varphi,j}^2 -\boldsymbol{\sigma}_{\varphi,j}^2) , \\
\underset{p(\boldsymbol{\epsilon})}{\mathbb{E}} [\log p_\theta(\mathbf{x}|\mathbf{z},\mathbf{c})] \approx & -\frac{1}{2} \left \Vert \frac{\mathbf{x} - \boldsymbol{\mu}_\theta}{\boldsymbol{\sigma}_\theta} \right \Vert^2 ,
\end{align}
\end{subequations}
with $d$ the dimensionality of $\mathbf{z}$.

\subsection{GANs}\label{annex:ijf-gan}

\subsubsection*{GAN}

\noindent The original GAN value function $V(\phi, \theta)$ proposed by \citet{goodfellow2014generative} is
\begin{subequations}
\label{eq:GAN_value_function}	
\begin{align}
V(\phi, \theta) = &   \underbrace{\underset{\mathbf{x}}{\mathbb{E}}  [\log d_\phi (\mathbf{x}|\mathbf{c})]  + \underset{\mathbf{\hat{x}}}{\mathbb{E}} [\log(1-d_\phi(\mathbf{\hat{x}}|\mathbf{c}))]}_{:=- L_d} , \\
L_g  := &  -\underset{\mathbf{\hat{x}}}{\mathbb{E}} [\log(1-d_\phi(\mathbf{\hat{x}}|\mathbf{c}))],
\end{align}
\end{subequations}
where $L_d$ is the cross-entropy, and $L_g$ the probability the discriminator wrongly classifies the samples.

\subsubsection*{WGAN}

\noindent 
The divergences which GANs typically minimize are responsible for their training instabilities for reasons investigated theoretically by \citet{arjovsky2017towards}. 
\citet{arjovsky2017wasserstein} proposed instead using the \textit{Earth mover} distance, also known as the Wasserstein-1 distance
\begin{align}
\label{eq:Wasserstein_distance}	
W_1(p,q) = & \inf_{\gamma \in \Pi(p,q)} \mathbb{E}_{(x,y)\sim \gamma}  [\left \Vert x - y \right \Vert ],
\end{align}
where $\Pi(p,q)$ denotes the set of all joint distributions $\gamma(x,y)$ whose marginals are
respectively $p$ and $q$, $\gamma(x,y)$ indicates how much mass must be transported from $x$ to $y$ in order to transform the distribution $p$ into $q$, $\left \Vert \cdot \right \Vert$ is the L1 norm, and $\left \Vert x - y \right \Vert$ represents the cost of moving a unit of mass from $x$ to $y$. 
%
However, the infimum in (\ref{eq:Wasserstein_distance}) is intractable. Therefore, \citet{arjovsky2017wasserstein} used the Kantorovich-Rubinstein duality \citep{villani2008optimal} to propose the Wasserstein GAN (WGAN) by solving the min-max problem
\begin{align}
\label{eq:GAN_wasserstein}	
 \theta^\star = \arg \min_\theta \max_{\phi \in \mathcal{W}} & \ \underset{\mathbf{x}}{\mathbb{E}}  [ d_\phi (\mathbf{x}|\mathbf{c})]  - \underset{\mathbf{\hat{x}}}{\mathbb{E}} [ d_\phi (\mathbf{\hat{x}}|\mathbf{c}) ],
\end{align}
where $\mathcal{W} = \{\phi : \Vert d_\phi(\cdot) \Vert_L \leq 1 \}$ is the 1-Lipschitz space, and the classifier $d_\phi(\cdot) : \mathbb{R}^T \times \mathbb{R}^{|\mathbf{c}|} \rightarrow [0, 1]$ is replaced by a critic function $d_\phi(\cdot) : \mathbb{R}^T \times \mathbb{R}^{|\mathbf{c}|}  \rightarrow \mathbb{R}$.
%
%
%
%
However, the weight clipping used to enforce $d_\phi$ 1-Lipschitzness can lead sometimes the WGAN to generate only poor samples or failure to converge \citep{gulrajani2017improved}. Therefore, we implemented the WGAN-GP to tackle this issue. 

\clearpage

\chapter{Part \ref{part:forecasting} conclusions}\label{chap:forecasting-conclusions}

\epi{If you're lonely when you're alone, you're in bad company.}{Jean-Paul Sartre }

Part \ref{part:forecasting} presents the forecasting tools and metrics required to produce and evaluate reliable point and probabilistic forecasts used as input of decision-making models. The forecasts take various forms: point forecasts, quantiles, prediction intervals, density forecasts, and scenarios. An example of forecast quality evaluation is given with standard deep learning models such as recurrent neural networks that compute PV and electrical consumption point forecasts on a real case study. 
Then, several case studies are used to assess the forecast quality of probabilistic forecasts.
\begin{itemize}
    \item Deep learning models such as the encoder-decoder architecture and recurrent neural networks compute PV quantile forecasts. These models are trained by quantile regression. The forecast quality is evaluated on a real case study composed of the PV generation of the parking rooftops of the Li\`ege University. The quantile regression models are compared to quantiles derived from deep learning generative models. In terms of forecast quality, the latter models outperform, in this case, the quantile regression models. The forecast value is assessed in Part \ref{part:optimization} where the quantile forecasts are used as input of robust optimization planners in the form of prediction intervals. 
    \item A density forecast-based approach produces probabilistic forecasts of imbalance prices, focusing on the Belgian case. The two-step probabilistic approach computes the net regulation volume state transition probabilities. Then, it infers the imbalance price. A numerical comparison of this approach to standard forecasting techniques is performed on the Belgium case. The proposed model outperforms other approaches on probabilistic error metrics but is less accurate at predicting the imbalance prices. Deep learning models could improve this probabilistic approach to avoid simplifying assumptions and adding input features to describe the market situation better.
    \item Finally, a recent class of deep learning generative models, the normalizing flows, is investigated. A fair and thorough comparison is conducted with the state-of-the-art deep learning generative models, generative adversarial networks and variational autoencoders. The experiments employ the open data of the Global Energy Forecasting Competition 2014. The models use the conditional information to compute improved weather-based PV, wind power, and load scenarios. The results demonstrate that normalizing flows can challenge generative adversarial networks and variational autoencoders. Indeed, they are overall more accurate both in terms of quality and value (see Chapter \ref{chap:energy-retailer}). Furthermore, they can be used effectively by non-expert deep learning practitioners. 
\end{itemize}

\part{Planning and control}\label{part:optimization}

\begin{infobox}{Overview}
Part \ref{part:optimization} presents the decision-making tools under uncertainty required to address the day-ahead planning of a microgrid. It investigates several approaches: deterministic planning using linear programming based on point forecasts, stochastic programming using scenarios, and robust optimization using quantile forecasts. Several case studies are considered: a grid-connected microgrid using a value function approach to propagate the information from planning to real-time optimization, a grid-connected microgrid in the capacity firming framework, and an energy retailer on the day-ahead market. The case studies use the forecasting techniques studied in Part \ref{part:optimization}. 
\end{infobox}
\epi{The straight line, a respectable optical illusion which ruins many a man.}{Victor Hugo}

\begin{figure}[htbp]
	\centering
	\includegraphics[width=1\linewidth]{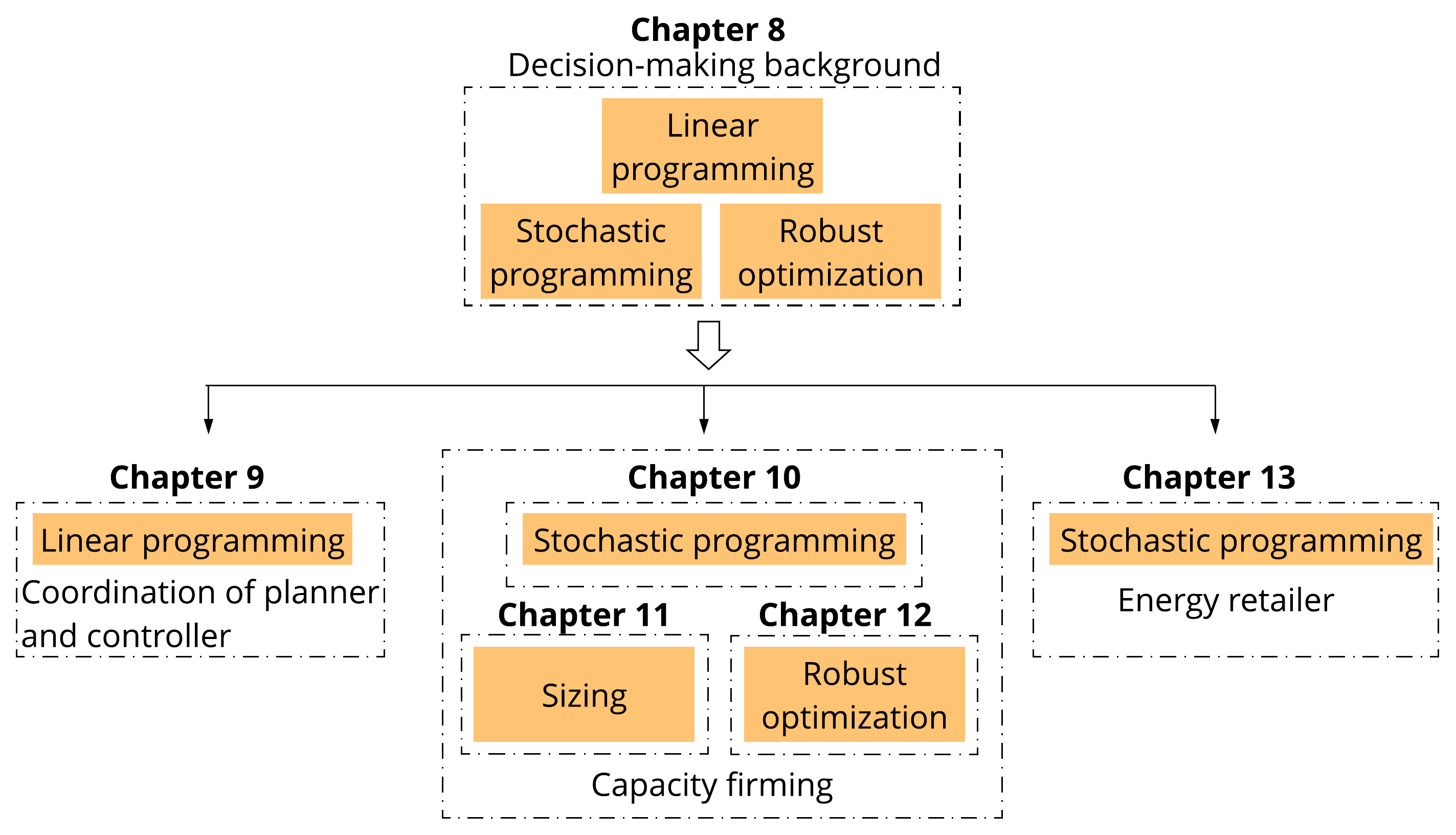}
	\caption{Part \ref{part:optimization} skeleton.}
	\label{fig:planning_part}
\end{figure}

Figure \ref{fig:planning_part} illustrates the organization of the second part. It addresses the energy management of a grid-connected microgrid by using the forecasting techniques studied in Part \ref{part:forecasting}.
Chapter \ref{chap:optimization-general_background} provides the optimization basics. Chapter \ref{chap:coordination-planner-controller} proposes a value function-based approach as a way to propagate information from operational planning to real-time optimization.
Chapters \ref{chap:capacity-firming-stochastic}, \ref{chap:capacity-firming-sizing}, and \ref{chap:capacity-firming-robust} propose stochastic, sizing, and robust approaches to address the energy management of a grid-connected renewable generation plant coupled with a battery energy storage device in the capacity firming market, respectively. The capacity firming framework has been designed for isolated markets, such as the Overseas France islands.
Chapter \ref{chap:energy-retailer} is the extension of Chapter \ref{chap:scenarios-forecasting}. It investigates the forecast value assessment of the deep generative models by considering an energy retailer portfolio on the day-ahead market.
Finally, Chapter \ref{chap:optimization-conclusions} draws the conclusions of Part \ref{part:optimization}.

\chapter{Decision-making background}\label{chap:optimization-general_background}

\begin{infobox}{Overview}
This Chapter introduces some basics of linear programming and optimization methodologies to address uncertainty in decision-making used to formulate the problems considered in Part \ref{part:optimization}. \\
The interested reader is referred to more general textbooks for further information \citep{morales2013integrating,bertsimas1997introduction,birge2011introduction} and the lectures of the courses "Renewables in Electricity Markets"\footnote{\url{http://pierrepinson.com/index.php/teaching/}} and "Advanced Optimization and Game Theory for Energy Systems"\footnote{\url{https://www.jalalkazempour.com/teaching}} given by professor Pierre Pinson and associate professor Jalal Kazempour, respectively, at the Technical University of Denmark.
\end{infobox}
\epi{Nothing is more difficult, and therefore more precious, than to be able to decide.}{Napoleon Bonaparte}
\begin{figure}[htbp]
	\centering
	\includegraphics[width=1\linewidth]{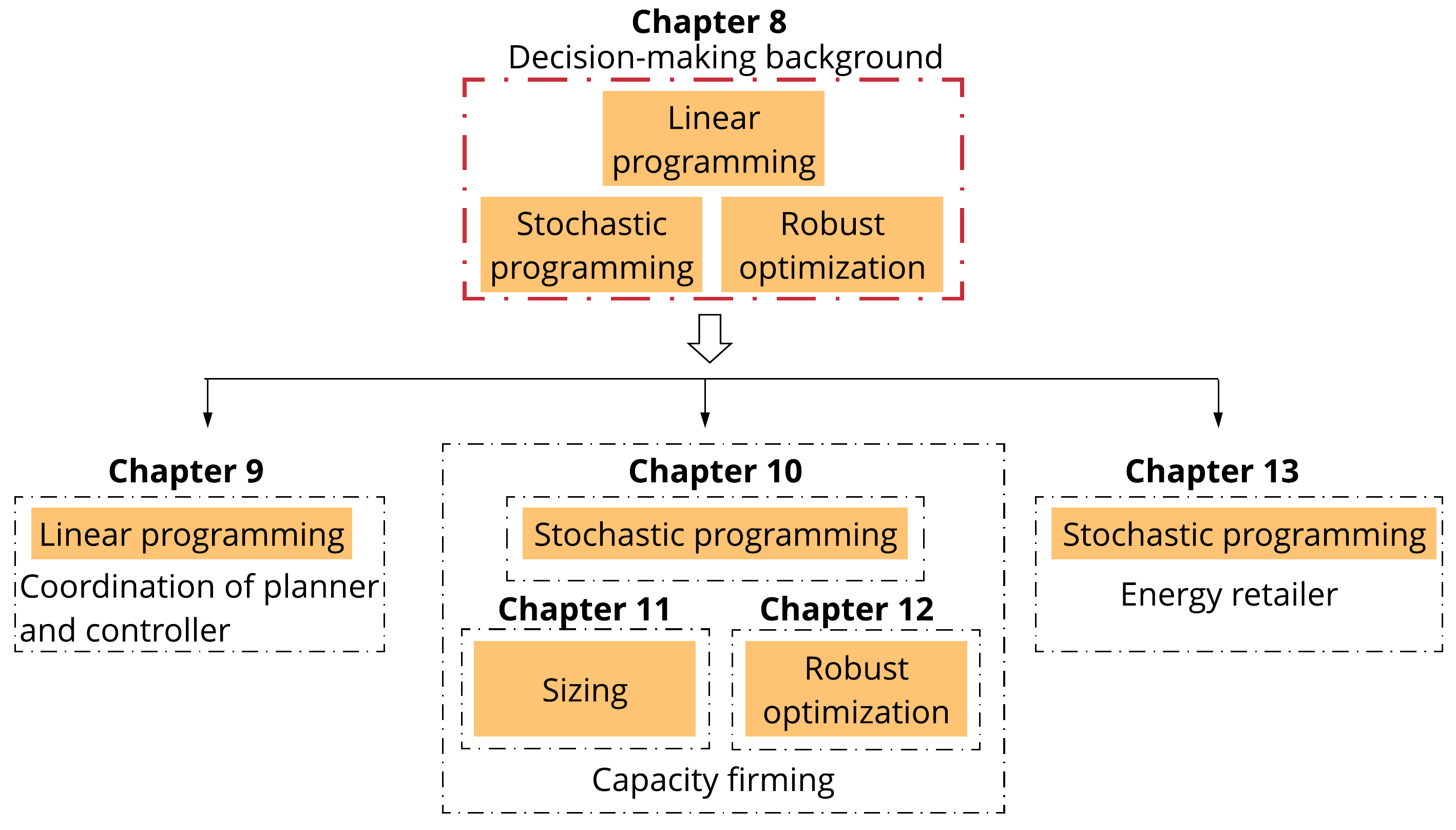}
	\caption{Chapter \ref{chap:optimization-general_background} position in Part \ref{part:optimization}.}
\end{figure}
\clearpage

Section \ref{sec:optimization-background-lp} introduces the mathematical formulation of a linear optimization problem. Sections \ref{sec:optimization-background-stochastic} and \ref{sec:optimization-background-ro} present the mathematical formulations of a linear optimization problem when considering uncertainty in the parameters of the problem. First, by considering a stochastic programming approach. Second, by using a robust methodology.

\section{Linear programming}\label{sec:optimization-background-lp}

\subsection{Formulation of a linear programming problem}

The most straightforward instance of an optimization problem is a linear programming problem.
\begin{definition}[Linear programming]
Linear programming (LP) minimizes a linear cost function subject to linear equality and inequality constraints.
\end{definition}
In a LP problem, we are given a cost vector $\mathbf{c} \in \mathbb{R}^n$ and we seek to minimize a linear cost function $\mathbf{c}^\intercal \mathbf{x}$ over all $n$-dimensional vectors $\mathbf{x} \in \mathbb{R}^n$ subject to a set of linear equality and inequality constraints. Suppose that there is a total of $m$ such constraints and let $\mathbf{b}\in \mathbb{R}^m$ and $\mathbf{A}$ be the $m \times n$ matrix. The constraints can be expressed compactly in the form $\mathbf{A} \mathbf{x} \geq \mathbf{b}$. 
\begin{definition}[LP in standard form] \citep[Chapter 1]{bertsimas1997introduction}
A linear programming problem of the form
\begin{align}\label{eq:optimization-background-LP}
    \min_{\mathbf{x}} & \ \mathbf{c}^\intercal \mathbf{x} \\
    \text{s.t.} & \ \mathbf{A} \mathbf{x} = \mathbf{b}, \notag \\
    & \ \mathbf{x} \geq \mathbf{0}, \notag
\end{align}
is said to be in the \textit{standard form}.
\end{definition}
LP problems model a wide variety of real-world problems, such as day-ahead planning of a microgrid. 

\subsection{Duality in linear programming}

Duality theory deals with the relation between a LP problem, called the \textit{primal}, and another LP problem, called the \textit{dual}. It uncovers the deeper structure of a LP problem and is a powerful tool that has various applications. The duality theory is motivated by the Lagrange multiplier method, where a price variable is associated with each constraint. It explores prices under which the presence or absence of the constraints does not affect the optimal cost. Then, the correct prices can be found by solving a new LP problem, called the original's dual \citep{bertsimas1997introduction}.
%

\begin{definition}[Dual of a LP problem]\citep[Chapter 4]{bertsimas1997introduction}
Given a primal problem in the standard form (\ref{eq:optimization-background-LP}), its dual is defined as
\begin{align}
    \max_{\mathbf{p}} & \ \mathbf{p}^\intercal \mathbf{b} \\
    \text{s.t.} & \ \mathbf{p}^\intercal \mathbf{A} \leq \mathbf{c}^\intercal, \notag \\
    & \ \mathbf{x} \geq \mathbf{0}, \notag
\end{align}
with $\mathbf{p}$ the price vector of the same dimension as $\mathbf{b}$.
\end{definition}
Note that the dual of the dual problem is the primal problem. The \textit{strong duality} theorem is the central result of linear programming duality.
\begin{quote}
'If a linear programming problem has an optimal solution so does its dual, and the respective optimal costs are equal.' \citep[Chapter 4, Theorem 4.4]{bertsimas1997introduction}
\end{quote}

\section{Stochastic optimization}\label{sec:optimization-background-stochastic}

Most decision-making problems are subject to uncertainty due to the intrinsic stochasticity of natural events conditioning our choices, \textit{e.g.}, the weather, or, more generally, to the inaccurate knowledge of input information. Therefore, decision-makers are interested in methods and tools that provide solutions less sensitive to environmental influences or inaccurate data while simultaneously reducing cost, increasing profit, or improving reliability \citep[Appendix C]{morales2013integrating}.


Let consider a LP in the standard form (\ref{eq:optimization-background-LP}). If parameters $\mathbf{A}$, $\mathbf{b}$, and $\mathbf{c}$ are perfectly known, solution algorithms for linear optimization problems, \textit{e.g.}, the simplex method, can be used to find the best value of the decision variable vector $\mathbf{x}$.

However, suppose some of these parameters are contingent on the realization $\xi(\omega)$ of a particular random vector $\xi$. In that case, determining the optimal solution to the problem (\ref{eq:optimization-background-LP}) may become further challenging. 
First, the issue of how to guarantee the feasibility of decision vector $\mathbf{x}$ becomes remarkably more involved when optimizing under uncertainty because $\mathbf{A}$ and $\mathbf{b}$ are not completely known in advance. 
Second, the issue of how to guarantee the optimally of decision vector $\mathbf{x}$ is at stake because $\mathbf{c}$ is not completely known in advance.
Finally, the LP problem (\ref{eq:optimization-background-LP}) needs to be recast so that solution algorithms for linear programming problems can be used to obtain the optimal value of decision vector $\mathbf{x}$ taking into account the uncertainty of the parameters. 

\textit{Stochastic programming} provides the concepts and tools required to deal with the implications of having uncertain data in an optimization problem for decision making to address the three issues previously stated. It assumes that an accurate probabilistic description of the random variable is assumed available, under the form of the probability distributions or densities.
In the following, we consider stochastic programming with \textit{recourse} where the set of decisions is divided into two groups: (1) decisions have to be taken before the realization of uncertain parameters. These decisions are known as \textit{first-stage}, $\mathbf{x}$, or here-and-now decisions and do not depend on the realization of the random parameters; (2) decisions can be taken after the realization of uncertain parameters. These decisions are called \textit{second-stage}, $\mathbf{y}(\mathbf{x}, \omega)$, or recourse decisions and are dependent on each plausible value of the random parameters.
Note: the term recourse points to the fact that second-stage decisions enable the decision-maker to adapt to the actual outcomes of the random events.

\begin{definition}[Two-stage program with fixed recourse]\citep[Chapter 2]{birge2011introduction}
A classical two-stage stochastic linear program with fixed recourse is 
\begin{align}\label{eq:optimization-two-stage-stochastic1}
    \min_{\mathbf{x}} & \ \mathbf{c}^\intercal \mathbf{x} + \mathop{\mathbb{E}}_\xi[Q(\mathbf{x}, \xi(\omega))] \\
    \text{s.t.} & \ \mathbf{A} \mathbf{x} = \mathbf{b},  \notag \\
    & \ \mathbf{x} \geq \mathbf{0}, \notag
\end{align}
with 
\begin{align}\label{eq:optimization-two-stage-stochastic2}
    Q(\mathbf{x}, \xi(\omega)) = \min_{\mathbf{y}} & \ \mathbf{q}(\omega)^\intercal \mathbf{y} \\
     \text{s.t.} & \ \mathbf{W} \mathbf{y} = \mathbf{h}(\omega) - T(\omega)\mathbf{x}, \quad \forallw \notag \\
    & \ \mathbf{y} \geq \mathbf{0}, \quad  \forallw \notag
\end{align}
the second-stage value function.
\end{definition}
Note: the uncertainty involved in problem (\ref{eq:optimization-two-stage-stochastic1}) and (\ref{eq:optimization-two-stage-stochastic2}) is assumed to be properly represented by means of a finite set $\Omega$ of scenarios $\omega$, with a probability $\alpha_\omega$ such that $\sum_{\omega \in \omega}\alpha_\omega = 1$.
\begin{definition}[Deterministic equivalent problem]\citep[Chapter 2]{birge2011introduction}
The \textit{deterministic equivalent} problem of the stochastic programming problem (\ref{eq:optimization-two-stage-stochastic1})-(\ref{eq:optimization-two-stage-stochastic2}) is 
\begin{align}
    \min_{\mathbf{x}, \mathbf{y}} & \ \mathbf{c}^\intercal \mathbf{x} + \sum_{\omega \in \omega}\alpha_\omega  \mathbf{q}(\omega)^\intercal \mathbf{y}\\
    \text{s.t.} & \ \mathbf{A} \mathbf{x} = \mathbf{b}, \notag \\
     & \ \mathbf{W} \mathbf{y} = \mathbf{h}(\omega) - \mathbf{T}(\omega)\mathbf{x}, \quad \forallw \notag \\
    & \ \mathbf{x} \geq \mathbf{0}, \notag \\
    & \ \mathbf{y} \geq \mathbf{0}, \quad  \forallw \notag
\end{align}
and can be directly processed by off-the-shelf optimization software for linear programs.
\end{definition}

\section{Robust optimization}\label{sec:optimization-background-ro}

Robust optimization considers optimization problems with uncertain parameters not modeled using probability distributions but with uncertainty sets. A robust optimization strategy investigates a solution to an optimization problem that is feasible for any realization of the uncertain parameters within the uncertainty set and optimal for the worst-case realization of these uncertain parameters \citep[Appendix D]{morales2013integrating}.

\begin{definition}[Two-stage robust optimization with fixed recourse]
The general form of two-stage robust optimization formulation is
\begin{align}\label{eq:optimization-ro}
    \min_{\mathbf{x}} & \ \mathbf{c}^\intercal \mathbf{x} + \max_{\mathbf{u} \in \mathcal{U}} \min_{\mathbf{y} \in \Omega(\mathbf{x}, \mathbf{u})} \mathbf{q}^\intercal \mathbf{y} \\
    \text{s.t.} & \ \mathbf{A} \mathbf{x} = \mathbf{b}, \notag \\
    & \ \mathbf{x} \geq \mathbf{0}, \notag 
\end{align}
with $\mathcal{U}$ the uncertainty set and
\begin{align}
    \Omega(\mathbf{x}, \mathbf{u}) & = \{\mathbf{y} \geq \mathbf{0} :  \mathbf{W} \mathbf{y} = \mathbf{h} - \mathbf{T}\mathbf{x} -\mathbf{M}\mathbf{u} \}.
\end{align}
\end{definition}

The objective of the problem (\ref{eq:optimization-ro}) is to make the best decisions represented by variable vector $\mathbf{x}$ for the worst realization of parameters in vector $\mathbf{u}$ and considering the recourse decisions described by variable vector $\mathbf{y}$.
If the right-hand-side problem, the $\mathbf{y}$-problem, is convex, it can be replaced by its dual and merged with the middle $\mathbf{u}$-problem rendering it a conventional single-level maximization problem. Overall, the resulting problem is a min-max problem that in some cases can be solved using decomposition such as the Benders-dual cutting plane method \citep{bertsimas2012adaptive} or column-and-constraint generation algorithm \citep{zeng2013solving}. Both the Benders-dual method and the column-and-constraint generation procedure are implemented in a master sub-problem framework. 

\subsection{Benders-dual cutting plane algorithm}

The key idea of the Benders-dual cutting plane (BD) algorithm is to gradually construct the value function of the first-stage decisions using dual solutions of the second-stage decision problems.

Consider the case where the second-stage decision problem is a linear programming (LP) problem in $\mathbf{x}$. We first take the \textit{relatively complete recourse} assumption that this LP is feasible for any given $\mathbf{x}$ and $\mathbf{u}$. Let $\mathbf{p}$ be its dual variables. Then, we obtain its dual problem, which is a maximization problem and can be merged with the maximization over $\mathbf{u}$. As a result, we have the following dispatch problem, which yields the sub-problem (SP) in the BD algorithm.
\begin{definition}[BD sub-problem]
\begin{align}\label{eq:optimization-ro-bd-SP}
    R(\mathbf{x}) = \max_{\mathbf{u}, \mathbf{p}} & \ (\mathbf{h} - \mathbf{T}\mathbf{x} -\mathbf{M}\mathbf{u})^\intercal \mathbf{p} \\
    \text{s.t.} & \ \mathbf{W}^\intercal \mathbf{p} \geq \mathbf{q}, \notag \\
    & \ \mathbf{u} \in \mathcal{U}. \notag 
\end{align}
Note: that the resulting problem in (\ref{eq:optimization-ro-bd-SP}) is a bilinear optimization problem. Several approaches have been developed to address this issue and depend on the case study.
\end{definition}

Suppose, we have an oracle that can solve (\ref{eq:optimization-ro-bd-SP}) for a given first stage variable $\mathbf{x}$. Then, the optimal solution is $(\mathbf{p}^\star, \mathbf{u}^\star)$, and a cutting plane in the form of 
\begin{align}\label{eq:optimization-ro-bd-cutting-plane}
\theta \geq (\mathbf{h} - \mathbf{T}\mathbf{x} -\mathbf{M}\mathbf{u}^\star)^\intercal \mathbf{p}^\star
\end{align}
can be generated, and included in the master problem (MP).
\begin{definition}[BD master problem]
At iteration $k$ of the BD master problem is
\begin{align}\label{eq:optimization-ro-bd-MP}
    \min_{\mathbf{x},  \theta} & \ \mathbf{c}^\intercal \mathbf{x} + \theta  \\
    \text{s.t.} & \ \mathbf{A} \mathbf{x} = \mathbf{b}, \notag \\
    & \ \theta \geq (\mathbf{h} - \mathbf{T}\mathbf{x} -\mathbf{M}\mathbf{u}_l^\star)^\intercal \mathbf{p}_l^\star \quad \forall l \leq k\\
     & \ \mathbf{x} \geq \mathbf{0}, \quad \theta \in \mathbb{R} \notag ,
\end{align}
which can compute an optimal first stage solution $\mathbf{x}_k^\star$.
\end{definition}

The MP and SP provide the lower and upper bounds with $\mathbf{c}^\intercal \mathbf{x}_k^\star + \theta_k^\star$ and $R(\mathbf{x}_{k-1}^\star)$, respectively,  at iteration $k$.
When considering the relatively complete recourse assumption, \citep{zeng2013solving,bertsimas2012adaptive,bertsimas2018scalable} demonstrated the BD algorithm converges to the optimal solution of the two-stage robust optimization problem. The lower and upper bounds converge in a finite number of steps by iteratively introducing cutting planes (\ref{eq:optimization-ro-bd-cutting-plane}) and computing the MP (\ref{eq:optimization-ro-bd-MP}). 
Note: if the SP is a MILP, this result is not straightforward. It is the case in the capacity firming problem studied in Chapter \ref{chap:capacity-firming-robust} where this issue is addressed.

\subsection{Column and constraints generation algorithm}

The column and constraints generation (CCG) algorithm does not create constraints using dual solutions of the second-stage decision problem. It dynamically generates constraints with recourse decision variables in the primal space for an identified scenario.
Let assume the uncertainty set U is a finite discrete set with $\mathcal{U} = \{\mathbf{u}_1, \ldots, \mathbf{u}_N \} $ and $ \{\mathbf{y}_1, \ldots, \mathbf{y}_N \}$ are the corresponding recourse decision variables. Then, the 2-stage RO (\ref{eq:optimization-ro}) can be reformulated as follows
\begin{align}\label{eq:optimization-ro-ccg}
    \min_{\mathbf{x},  \theta} & \ \mathbf{c}^\intercal \mathbf{x} + \theta  \\
    \text{s.t.} & \ \mathbf{A} \mathbf{x} = \mathbf{b}, \\
    & \ \theta \geq  \mathbf{q}^\intercal \mathbf{y}^l \quad \forall l \leq N \label{eq:optimization-ro-ccg-cst1} \\ 
    & \ \mathbf{W} \mathbf{y}^l = \mathbf{h} - \mathbf{T}\mathbf{x} -\mathbf{M}\mathbf{u}^l \quad \forall l \leq N\\
     & \ \mathbf{x} \geq \mathbf{0}, \quad \mathbf{y}^l \geq \mathbf{0} \ \forall l \leq N, \quad \theta \in \mathbb{R}.
\end{align}
Thus, it reduces to solve an equivalent, probably large-scale, MILP, which is very close to a 2-stage stochastic programming model if the probability distribution over $\mathcal{U}$ is known. When the uncertainty set is large or is a polyhedron, enumerating all the possible uncertain scenarios in $\mathcal{U}$ is not feasible. 
Constraints (\ref{eq:optimization-ro-ccg-cst1}) indicate that not all scenarios, and their corresponding variables and constraints, are necessary for defining the optimal value $\mathbf{x}^\star$ of the 2-stage RO. It is likely that a few relevant scenarios, a small subset of the uncertainty set, play a significant role in the formulation. Therefore, the key idea of the CCG procedure is to generate recourse decision variables for the significant scenarios. 

\begin{definition}[CCG sub-problem]
Similar to the BD method, the CCG algorithm uses a master sub-problem framework. Let assume an oracle can solve the dispatch problem for a given first-stage $\mathbf{x}$
\begin{align}\label{eq:optimization-ro-ccg-SP}
Q(\mathbf{x}) = \max_{\mathbf{u} \in \mathcal{U}} \min_{\mathbf{y} \in \Omega(\mathbf{x}, \mathbf{u})} \mathbf{q}^\intercal \mathbf{y}.
\end{align}
Then, the optimal solution $\mathbf{u}^\star$ can be derived to build constraints and variables in the master problem.
\end{definition}

\begin{definition}[CCG master problem]
At iteration $k$ of the CCG master problem is
\begin{align}\label{eq:optimization-ro-ccg-MP}
    \min_{\mathbf{x},  \theta} & \ \mathbf{c}^\intercal \mathbf{x} + \theta  \\
    \text{s.t.} & \ \mathbf{A} \mathbf{x} = \mathbf{b}, \\
    & \ \theta \geq  \mathbf{q}^\intercal \mathbf{y}^l \quad \forall l \leq k  \\ 
    & \ \mathbf{W} \mathbf{y}^l = \mathbf{h} - \mathbf{T}\mathbf{x} -\mathbf{M}\mathbf{u}^{\star, l} \quad \forall l \leq k\\
     & \ \mathbf{x} \geq \mathbf{0}, \quad \mathbf{y}^l \geq \mathbf{0} \ \forall l \leq k, \quad \theta \in \mathbb{R},
\end{align}
which can compute an optimal first stage solution $\mathbf{x}_k^\star$.
\end{definition}

Similar to BD, the MP and SP provide the lower and upper bounds with $\mathbf{c}^\intercal \mathbf{x}_k^\star + \theta_k^\star$ and $Q(\mathbf{x}_{k-1}^\star)$, respectively, at iteration $k$.

\section{Conclusions}\label{sec:optimization-background-conclusions}

This Chapter introduces the basics of linear programming and approaches to handle uncertainty in the parameters with the stochastic programming and robust approach.
Depending on the application, each approach has its pros and cons. When considering stochastic programming, the number of scenarios needed to describe the most plausible outcomes of the uncertain parameters may be huge, leading to large-scale optimization problems that may become difficult to solve or intractable. In this case, a robust approach provides an alternative and compact manner to describe uncertain parameters. However, the robust counterpart optimization problem may be challenging to solve and requires a decomposition technique such as a Benders-dual cutting plane or column and constraints generation algorithm that is not trivial to implement numerically.

\chapter{Coordination of the planner and controller}\label{chap:coordination-planner-controller}

\begin{infobox}{Overview}
This Chapter presents a two-layer approach with a value function to propagate information from operational planning to real-time optimization. The value function-based approach shares some similarities with the coordination scheme proposed in \citet{kumar2018stochastic}, which is based on stochastic dual dynamic programming. This study brings new contributions:
\begin{enumerate}
	\item The approach is tested by accounting for forecasting errors and high-resolution data monitored on-site corresponding to a "real-life" case.
	\item The value function approach allows to deal with indeterminacy issues. When there are several optimal solutions to the upper-level problem, this is accounted for in the lower level part, and a bias term can be added to favor one type of behavior over another, \textit{e.g.}, charge early. 
	\item This methodology is fully compatible with the energy markets as it can deal with imbalance, reserve, and dynamic selling/purchasing prices.
\end{enumerate}
This study reports results on an industrial microgrid capable of on/off-grid operation. Generation and consumption forecasts are based on weather forecasts obtained with the MAR model~\citep{fettweis2017reconstructions}. 

\textbf{\textcolor{RoyalBlue}{References:}} This chapter  is an adapted version of the following publication: \\[2mm]\bibentry{dumas2021coordination}.
\end{infobox}
\epi{When making a decision of minor importance, I have always found it advantageous to consider all the pros and cons. In vital matters, however, such as the choice of a mate or a profession, the decision should come from the unconscious, from somewhere within ourselves. In the important decisions of personal life, we should be governed, I think, by the deep inner needs of our nature.}{Sigmund Freud}
\begin{figure}[htbp]
	\centering
	\includegraphics[width=1\linewidth]{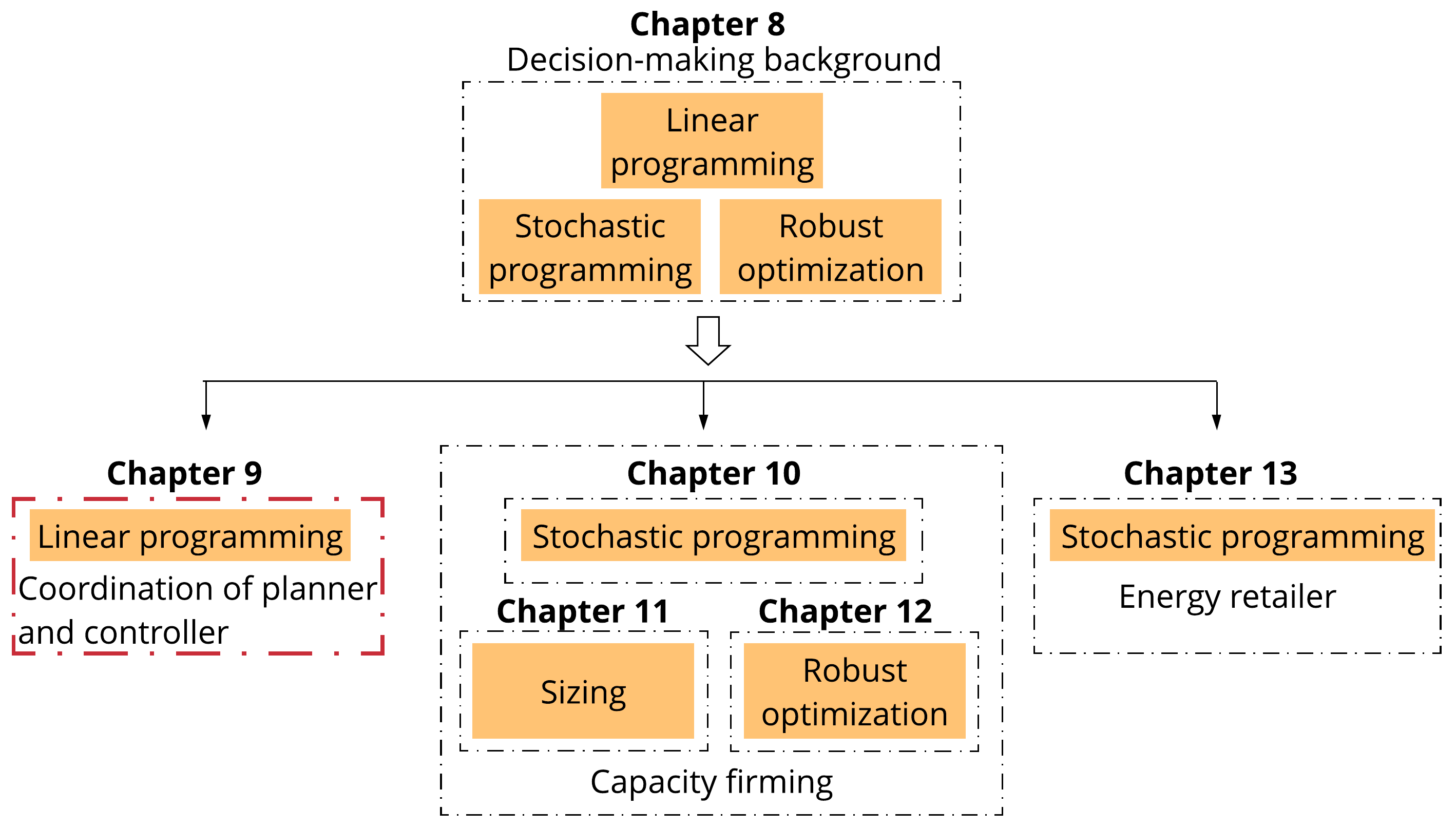}
	\caption{Chapter \ref{chap:coordination-planner-controller} position in Part \ref{part:optimization}.}
\end{figure}
\clearpage

The hierarchical microgrid control levels divide a global microgrid control problem in time and space \citep{palizban2014microgrids}. Control levels range from distributed device level controllers that run at a high frequency to centralized controllers optimizing market integration that run much less frequently. For computation time reasons, centralized controllers are often subdivided into two levels. Operational planning controllers that optimize decisions over a time horizon of one or several days but with a market period resolution, \textit{e.g.}, 15 minutes. Real-time optimization controllers that deal with actions within the current market period. The coordination of these two levels is paramount to achieving the safest and most profitable operational management of microgrids.
Microgrid control and management can be achieved in several ways. Control techniques and the principles of energy-storage systems are summarized in \citet{palizban2014microgrids}. A classification of microgrid control strategies into primary, secondary, and tertiary levels is done in \citet{olivares2014trends}. The two-level approach has been intensively studied. 
A double-layer coordinated control approach, consisting of the scheduling layer and the dispatch layer, is adopted in \citet{jiang2013energy}. The schedule layer provides an economical operation scheme, including state and power of controllable units based on the look-ahead multi-step optimization. In contrast, the dispatch layer follows the scheduling layer by considering power flow and voltage limits.  
A two-stage dispatch strategy for grid-connected systems is discussed in \citet{wu2014hierarchical}, where the first stage deals with the day-ahead schedule, optimizing capital and operational cost. At the same time, the lower level handles the rescheduling of the units for few hours ahead with a time resolution of 15 min. 
A two-stage control strategy for a PV BESS-ICE (Internal Combustion Engine) microgrid is implemented in \citet{sachs2016two}. 
Discrete Dynamic Programming is used in the first layer, while the second layer problem is posed as a Boundary Value Problem. 
An approach with a high-level deterministic optimizer running at a slow timescale, 15 min, coupled to a low-level stochastic controller running at a higher frequency, 1 min, is studied in \citet{cominesi2017two}.
A two-layer predictive energy management system for microgrids with hybrid energy storage systems consisting of batteries and supercapacitors is considered in \citet{ju2017two}. This approach incorporates the degradation costs of the hybrid energy storage systems.
A practical Energy Management System for isolated microgrid which considers the operational constraints of Distributed Energy Resources, active-reactive power balance, unbalanced system configuration, and loading, and voltage-dependent loads is studied in \citet{solanki2018practical}. 
A two-layer mixed-integer linear programming predictive control strategy was implemented and tested in simulation and experimentally in \citet{Polimeni2019}. Finally, \citet{moretti2019assessing} implemented a two-layer predictive management strategy for an off-grid hybrid microgrid featuring controllable and non-controllable generation units and a storage system. 

It is organized as follows. Section \ref{sec:optimization-pscc-notation} summarizes the notation. Section \ref{sec:optimization-pscc-pb_statement} formulates the problem in an abstract manner. Section \ref{sec:optimization-pscc-proposed_method} introduces the novel two-level value function-based approach and the assumptions made. Section \ref{sec:optimization-pscc-test-description} describes the numerical tests. Section \ref{sec:optimization-pscc-numerical results} reports the results. Conclusions are drawn in Section \ref{sec:optimization-pscc-conclusion}. The methodology used for forecasting is reported in Section \ref{sec:point-forecasting} of Chapter \ref{chap:forecast_evaluation}.

\section{Notation}\label{sec:optimization-pscc-notation}

\newcommand{\devices}[1]{\ensuremath{\mathcal{D}^\text{#1}}}
\newcommand{\sheddableDevices}{\ensuremath{\devices{she}}}
\newcommand{\nonflexibleDevices}{\ensuremath{\devices{nfl}}}
\newcommand{\steerableDevices}{\ensuremath{\devices{ste}}}
\newcommand{\nonsteerableDevices}{\ensuremath{\devices{nst}}}
\newcommand{\price}[1]{\ensuremath{\pi^{\text{#1}}}}
\newcommand{\curtailmentPrice}{\price{nst}}
\newcommand{\sheddingPrice}{\price{she}}
\newcommand{\steerablePrice}{\price{ste}}
\newcommand{\gridBuyPrice}{\price{i}}
\newcommand{\gridSalePrice}{\price{e}}
\newcommand{\curtail} {\ensuremath{a^\text{nst}}}
\newcommand{\steer}{\ensuremath{a^\text{ste}}}
\newcommand{\nonSteerable}{\ensuremath{P^\text{nst}}}
\newcommand{\steerable}{\ensuremath{P^\text{ste}}}
\newcommand{\shed} {\ensuremath{a^\text{she}}}
\newcommand{\nonFlexible}{\ensuremath{C^\text{nfl}}}
\newcommand{\sheddable}{\ensuremath{C^\text{she}}}
\newcommand{\BSSs}{\ensuremath{\mathcal{D}^{sto}}}
\newcommand{\SOC}{\ensuremath{s}}
\newcommand{\maxcharge}{\ensuremath{\overline{S}}}
\newcommand{\mincharge}{\ensuremath{\underline{S}}}
\newcommand{\chargerate}{\ensuremath{\overline{P}}}
\newcommand{\dischargerate}{\ensuremath{\underline{P}}}
\newcommand{\retentionRate}{\ensuremath{\eta^{\text{retention}}}}
\newcommand{\chargeEff}{\ensuremath{\eta^\text{cha}}}
\newcommand{\dischargeEff}{\ensuremath{\eta^\text{dis}}}
\newcommand{\initialCharge}{\ensuremath{S^\text{init}}}
\newcommand{\charge}{\ensuremath{a^\text{cha}}}
\newcommand{\discharge}{\ensuremath{a^\text{dis}}}
\newcommand{\finalCharge}{\ensuremath{S^{\text{end}}}}
\newcommand{\BSSsFee}{\ensuremath{\gamma^\text{sto}}}
\newcommand{\TotalCharge}{\ensuremath{S^{\text{tot}}}}
\newcommand{\MaxTotalCharge}{\ensuremath{\overline{S^{\text{tot}}}}}
\newcommand{\exportGrid}{\ensuremath{e}^\text{gri}}
\newcommand{\importGrid}{\ensuremath{i}^\text{gri}}
\newcommand{\foralltOP}{\ensuremath{\forall t' \in \mathcal{T}^m_a(t)}}
\newcommand{\foralltRTO}{\ensuremath{\forall t \in \mathcal{T}_i(t)}}
\newcommand{\forallb}{\ensuremath{\forall d \in \BSSs}}
\newcommand{\reserve}[1]{\ensuremath{r^\text{#1}}}
\newcommand{\reserveBSSInc}{\ensuremath{r^{s+}_{d,t'}}}
\newcommand{\reserveBSSDec}{\ensuremath{r^{s-}_{d,t'}}}
\newcommand{\maxExportToGrid}{\ensuremath{E}^{\text{cap}}}
\newcommand{\maxImportFromGrid}{\ensuremath{I}^{\text{cap}}}

\subsection*{Sets and indices}
\begin{supertabular}{l p{0.8\columnwidth}}
	Name & Description \\
	\hline
	$d$ & Device index. \\
	$t$, $t'$ & RTO and OP time periods indexes.  \\
    $\tau(t)$ & Beginning of the next market period at time $t$. \\
    $\mathcal{T}_i(t)$ &  Set of RTO time periods $= \{t, t+\Delta t, ..., t+T_i\}$. \\
    $\mathcal{T}^m_a(t)$ & Set of OP time periods $= \{\tau(t), \tau(t)+\Delta \tau,..., \tau(t+T_a)\}$. \\
    $T_a$, $T_l$ & Time durations, with $T_a \ll T_l$ \\
    $\mathcal{D}^k$ & Set of non-flexible loads ($k = \text{nfl}$), sheddable loads ($k = \text{she}$), steerable generators ($k = \text{ste}$),  non-steerable generators ($k = \text{nst}$), storage devices  ($k = \text{sto}$). \\
\end{supertabular}

\subsection*{Parameters}
\begin{supertabular}{l p{0.7\columnwidth} l}
	Name & Description & Unit \\
	\hline
	$\Delta t$ & Time delta between $t$ and the market period. & minutes\\
	$\Delta \tau$ & Market period. & minutes\\
    $\chargeEff$, $\dischargeEff$ & Charge and discharge efficiencies. & \% \\
	$\chargerate$, $\dischargerate$ & Maximum charging and discharging powers. & kW \\
	$\nonFlexible_{d,t}$ & Non-flexible power consumption. & kW \\
	$\sheddable_{d,t}$ & Flexible power consumption. & kW \\
	$\initialCharge_{d,t}$ & Initial state of charge of battery~$d$. & kWh \\
	$p_h$ & Maximum peak over the last twelve months. & kW\\
	$\price{p}$ & Yearly peak power cost. & \euro /kW\\
	$\price{s}_{OP}$ & Unitary revenue for providing reserve. & \euro /kW\\
	$\price{s}_{RTO}$ & Unitary RTO symmetric reserve penalty. & \euro /kW\\
	$\pi^k_{d,t}$ & Cost of load shedding ($k = \text{she}$), generating energy ($k = \text{ste}$),  curtailing generation ($k = \text{nst}$). & \euro /kWh\\
	$\BSSsFee_{d,t}$ & Fee to use the battery~$d$. & \euro /kWh\\
	$\gridSalePrice_t$, $\gridBuyPrice_t$ & Energy prices of export and import.  & \euro /kWh\\
	$\gridBuyPrice_d$, $\gridBuyPrice_n$ & Energy prices of day and night  imports. & \euro /kWh\\
	$\maxImportFromGrid$, $\maxExportToGrid$ & Maximum import and export limits. & kW \\
    $\mbox{PV}_p$, $\mbox{C}_p$ & PV and consumption capacities. & kW\\
	$\mbox{S}_p$ & Storage capacity. & kWh\\
	$\maxcharge$, $\mincharge$ & Maximum and minimum battery capacities. &  kWh\\
\end{supertabular}

\subsection*{\textbf{Forecasted or computed variables}}

\begin{supertabular}{l p{0.7\columnwidth} l}
	Name & Description & Unit \\
	\hline
	$a_t$ & Action at $t$. & - \\
	$a^m_t$ & Purely market related actions. & -\\
	$a^d_t$ & Set-points to the devices of the microgrid & -\\
	$a^k_{d,t} $ & Fraction of load shed ($k = \text{she}$), generation activated ($k = \text{ste}$),  generation curtailed ($k = \text{nst}$) ($ [0,1]$). & -\\
	$\charge_{d,t} $, $\discharge_{d,t} $ & Fraction of the maximum charging and discharging powers used for battery~$d$ ($ [0,1]$). & -\\
	$\exportGrid_t$, $\importGrid_t$ & Energy export and import. & kWh\\
	$\delta p_{t'}$ & OP peak difference between peak at~$t'$ and $p_h$. & kW \\
	$\delta p_{\tau(t-\Delta \tau),\tau(t)}$ & RTO peak difference between peak at~$\tau(t)$ and $p_h$. & kW \\
	$s^{TSO}_{t}$ & TSO symmetric reserve signal ($0;1$). & -\\
	$\reserve{sym}$ & Symmetric reserve. & kW\\
	$\Delta \reserve{sym}$ & Reserve difference between OP and RTO. & kW \\
	$\reserveBSSInc$, $\reserveBSSDec$ & Upward and downward reserves of power available and provided by storage device~$d$. & kW\\
	$\SOC_{d,t}$ & State of charge of battery~$d$. &  kWh\\
	$s_t$ & Microgrid state at time $t$. & -\\
	$s^m_t$ & Information related to the current market position. & -\\
	$s^d_t$ & State of the devices. & -\\
	$v_t$ & The cost-to-go function. & k\euro\\
	$\hat{\omega}$ & Forecast of a random vector $\omega$. & -\\
	$\overline{\mbox{X}}$ & Average of a variable $X$. & kW\\
	$\mbox{X}_{max}$, $\mbox{X}_{min}$ & Maximum and minimum of $X$. & kW\\
	$\mbox{X}_{std}$ & Standard deviation of $X$. & kW\\
	$c_E$, $c_p$, $c_t$ & Energy, peak and total costs. & k\euro\\
	$\Delta_p$ & Peak increment. & kW\\
	$\mbox{I}_{tot}$, $\mbox{E}_{tot}$ & Total import and export. & MWh\\
\end{supertabular}

\section{Problem statement}\label{sec:optimization-pscc-pb_statement}

A global microgrid control problem can be defined, for a given microgrid design and configuration, as operating a microgrid safely and in an economically efficient manner, by harvesting as much renewable energy as possible, operating the grid efficiently, optimizing the service to the demand side, and optimizing other side goals. We refine this definition below and start by making a few assumptions.

\subsection{Assumptions}
In this study, the control optimizes economic criteria, which are only related to active power. All devices areconnected to the same electrical bus, which can be connected or disconnected from a public grid permanently or dynamically. 
Device-level controllers offer an interface to communicate their operating point and constraints, \textit{e.g.}, maximum charge power as a function of the current state, and implement control decisions to reach power set-points. Fast load-frequency control, islanding management, as well as reactive power control are not in scope. The microgrid is a price taker in energy and reserve markets.

\subsection{Formulation}
Abstractly, a microgrid optimization problem can be formulated as follows
\begin{subequations}
	\begin{align}
	\min_{\mathbf{a}} \quad & \sum_{t\in \mathcal{T}_l} c(a_t, s_t, \omega_t) \\
	\text{s.t.} \ \forall t \in \mathcal{T}_l, \ &  s_{t+\Delta t} = f(a_t, s_t, \omega_t, \Delta t), \\
	& s_t \in \mathcal{S}_t .
	\end{align}
\end{subequations}
A controller has to return a set of actions $a_t = (a^m_t, a^d_t)$ at any time $t$ over the life of the microgrid ($\mathcal{T}_l$). Actions should be taken as frequently as possible to cope with the economic impact of the variability of the demand and generation sides, but not too often to let transients vanish, \textit{e.g.}, every few seconds. The time delta between action $a_t$ and the next action taken is denoted by $\Delta t$, and is not necessarily constant.  Some of these actions are purely market-related $a^m_t$, while other actions are communicated as set-points to the devices of the microgrid $a^d_t$. The state $s_t= (s^m_t, s^d_t)$ of the microgrid at time $t$ is thus also made of two parts. (1) $s^d_t$ represents the state of the devices, such as a storage system or a flexible load. (2) $s^m_t$ gathers information related to the current market position, such as the nominated net position of the microgrid over the next market periods. The cost function $c$ gathers all the economic criteria considered. The transition function $f$ describes the physical and net position evolution of the system. 
At time instants $t \in \{\Delta \tau, 2 \Delta \tau, \ldots\}$, with $\Delta \tau$ the market period, some costs are incurred based on the value of some state variables, which are then reset for the following market period. 
This problem is challenging to solve since the system's evolution is uncertain, actions have long-term consequences, and are both discrete and continuous. Furthermore, although functions $f$ and $c$ are assumed time-invariant, they are generally non-convex and parameterized with stochastic variables $\omega_t$. 

\section{Proposed method}\label{sec:optimization-pscc-proposed_method}
\begin{figure}[tb]
	\centering
	\includegraphics[width=0.7\linewidth]{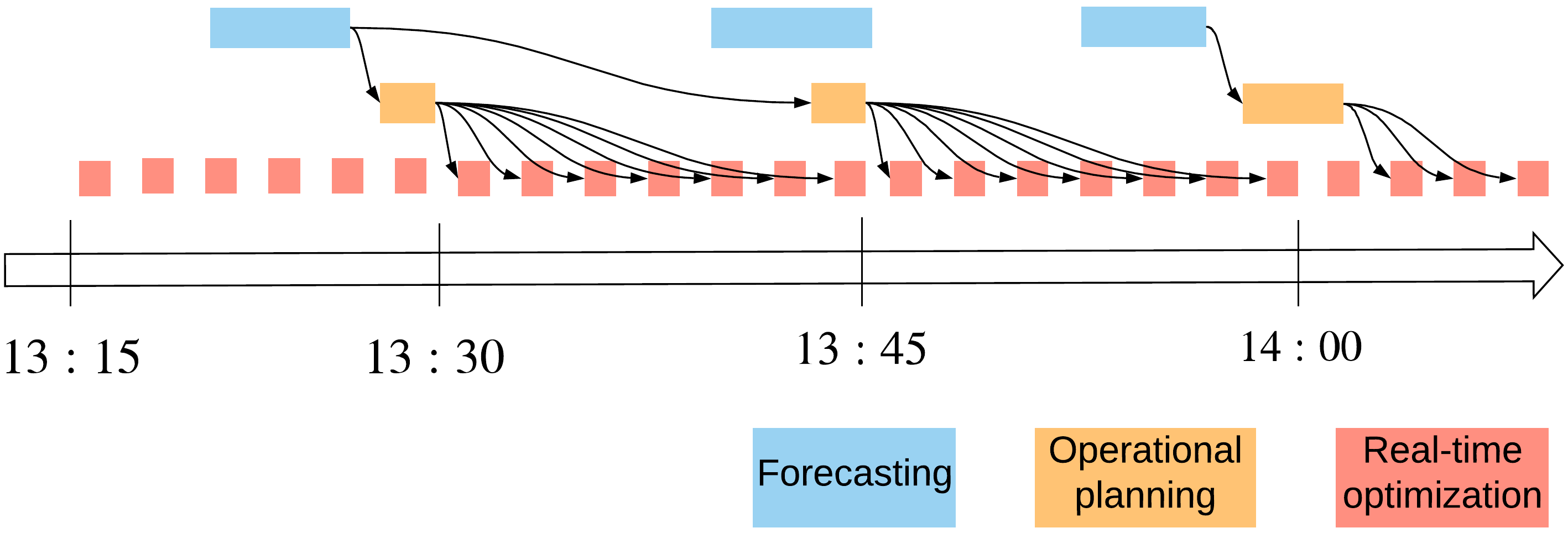}
	\caption{Hierarchical control procedure illustration.}
	\label{fig:pscc-emsschedulingpaperversion}
\end{figure}

In practice, solving the microgrid optimization problem above amounts, at every time $t$, to forecasting the stochastic variables $\boldsymbol{\omega}_{\mathcal{T}_l(t)} $, then solving the problem\footnote{Which is here expressed as a deterministic problem for simplicity but should be treated as a stochastic problem in practice.}
\begin{subequations}
	\begin{align}
	\mathbf{a}^\star_{\mathcal{T}_l(t)}  = \arg\min \ & \sum_{t'\in \mathcal{T}_l(t)} c(a_{t'}, s_{t'}, \hat{\omega}_{t'}) \\
	\text{s.t.} \ \forall t' \in \mathcal{T}_l(t), \ & s_{t'+\Delta t'} = f(a_{t'}, s_{t'}, \hat{\omega}_{t'}, \Delta t'),\\
	& s_{t'} \in S_t' ,
	\end{align}
\end{subequations}
and applying $a^\star_t$ (potentially changing $a^{m, \star}_t$ at some specific moments only).
Forecasts are valid only for a relatively near future, and optimizing over a long time would be incompatible with the real-time operation. Thus, this problem is approximated by cropping the lookahead horizon to $\mathcal{T}_a(t) \subset \mathcal{T}_l(t)$.
However, market decisions must be refreshed much less frequently than set-points. We thus propose to further decompose the problem in an operational planning problem (OPP) for $\mathcal{T}^m_a(t)$ that computes market decisions
\begin{subequations}
	\label{eq:OPP}
	\begin{align}
	\mathbf{a}^{m, \star}_{\mathcal{T}^m_a(t)}  = \arg\min \ & \sum_{t'\in \mathcal{T}^m_a(t)} c^m(a^m_{t'}, s_{t'}, \hat{\omega}_{t'}) \\
	\text{s.t.} \ \forall t' \in \mathcal{T}^m_a(t),  \ & s_{t'+\Delta \tau} = f^m(a^m_{t'}, s_{t'}, \hat{\omega}_{t'}, \Delta \tau) \\
	& s_{t'} \in S_{t'} ,
	\end{align}
\end{subequations}
and a real-time problem (RTP) that computes set-points for time $t$
\begin{subequations}
	\label{eq:RTP}
	\begin{align}
	a^{d, \star}_{t}  = \arg\min \ & c^d(a^d_{t}, s_{t}, \hat{\omega}_{t}) + v_{\tau(t)}(s_{\tau(t)}) \\
	\text{s.t.} \ & s_{\tau(t)} = f^d(a^d_{t}, s_{t}, \hat{\omega_{t}}, \tau(t) - t)\\
	& s_{\tau(t)} \in S_{\tau(t)} ,
	\end{align}
\end{subequations}
with $c(a_{t}, s_{t}, w_{t}) = c^m(a^m_{t}, s_{t}, w_{t}) + c^d(a^d_{t}, s_{t}, \omega_{t})$. 
The function $v_t$ is the cost-to-go as a function of the system's state at the end of the ongoing market period. It regularizes decisions of RTP to account for the longer-term evolution of the system.
We detail hereunder how we obtain $v_t$.  An overview of the approach is depicted in Figure~\ref{fig:pscc-emsschedulingpaperversion}.

\subsection{Computing the cost-to-go function}
The function $v_t$ represents the optimal value of (\ref{eq:OPP}) as a function of the initial state $s_{\tau(t)}$ of this problem. If we make the assumption that (\ref{eq:OPP}) is modeled as a linear program, the function $v_{\tau(t)}$ is thus convex and piecewise linear. Every evaluation of (\ref{eq:OPP}) with the additional constraint\footnote{The $\perp \mu$ notation means that $\mu$ is the dual variable of the constraint.}
\begin{align}
s_{\tau(t)} = s' \perp \mu
\end{align}
yields the value $v_{\tau(t)}(s')$ and a supporting inequality (a cut)
\begin{align}
v_{\tau(t)}(s) \geq v_{\tau(t)}(s') + \mu^T  s.
\end{align}
The algorithm to approximate $v_{\tau(t)}(s')$ works as follows:
\begin{enumerate}
	\item estimate the domain of $v_{\tau(t)}$, \textit{i.e.}, the range of states reachable at time $\tau(t)$ and the most probable state that will be reached $s^\star_{\tau(t)}$;
	\item evaluate $v_{\tau(t)}(s^\star_{\tau(t)})$ and the associated $\mu^\star$;
	\item repeat step 2 for other state values until all regions of $v_{\tau(t)}$ are explored. 
\end{enumerate}
Note: that if the state is of dimension one and  (\ref{eq:OPP}) is a linear program, simplex basis validity information can be used to determine for which part of the domain of $v_{\tau(t)}$ the current cut is tight, else a methodology such as proposed in \citet{bemporad2003greedy} can be used.

\subsection{OPP formulation}

The OPP objective function implemented for the case study is
\begin{align}	
	J_{\mathcal{T}^m_a(t)}^{OP} = \ & \sum_{t'\in \mathcal{T}^m_a(t)} \bigg ( C_{t'}^{OP} + D_{t'}^{OP} \bigg )
\end{align}
with Operational Planner (OP) the name of this planer. $\mathcal{T}^m_a(t)$ is composed of 96 values with $\Delta \tau = 15 $ minutes and $T_a = 24$ hours. $C_{t'}^{OP}$ models the immediate costs and $D_{t'}^{OP}$ the delayed costs at  $t'$.
$C_{t'}^{OP}$ takes into account different revenues and costs related to energy flows: the costs of shed demand, steered and non steered generation, the revenues from selling energy to the grid, the costs of purchasing energy from the grid and the costs for using storage
\begin{align}\label{eq:OP_immediate_cost}
	C_{t'}^{OP} = \ & \sum_{t'\in \mathcal{T}^m_a(t)} \bigg (  \sum_{d\in \sheddableDevices} \Delta \tau  \sheddingPrice_{d,t'}  \sheddable_{d,t'}  \shed_{d,t'}  + \sum_{d\in \steerableDevices} \Delta \tau  \steerablePrice_{d,t'}   \steerable_{d,t'}  \steer_{d,t'} 
	+ \sum_{d\in \nonsteerableDevices} \Delta \tau  \curtailmentPrice_{d,t'}  \nonSteerable_{d,t'}  \curtail_{d,t'} \notag\\
	&+ \sum_{d\in \BSSs} \Delta \tau  \BSSsFee_{d}  [\chargerate_{d}  \chargeEff_{d}  \charge_{d,t'} + \frac{\dischargerate_d}{\dischargeEff_d}  \discharge_{d,t'} ]  - \gridSalePrice_{t'}  \exportGrid_{t'} + \gridBuyPrice_{t'}  \importGrid_{t'} \bigg ) .
	\end{align}
$D_{t'}^{OP}$  is composed of the peak cost and symmetric reserve revenue
\begin{align}	\label{eq:OP_delayed_cost}
	D_{t'}^{OP} = \ & \price{p}  \delta p_{t'} - \price{s}_{OP}  \reserve{sym}_{t'} , 
\end{align}
$\delta p_{t'}$ is the peak difference between the previous maximum historic peak $p_h$ and the current peak within the market period $t'$. $\reserve{sym}_{t'}$ is the symmetric reserve provided to the grid within the current market period $t'$.

\subsection{OP constraints}
The first set of constraints defines bounds on state and action variables, $\foralltOP$
\begin{subequations}
	\label{eq:OP_action_constraint_set}	
	\begin{align}
	&a^k_{d,t'} \leq 1     &&\forall d \in \mathcal{D}^k, \forall k \in \{\text{ste}, \text{she}, \text{nst}\}   \\
	&\charge_{d,t'} \leq 1    &&\forallb       \\
	&\discharge_{d,t'} \leq 1 &&\forallb    \\
	&\mincharge_d \leq \SOC_{d,t'} \leq \maxcharge_d &&\forallb .
	\end{align}
\end{subequations}
The energy flows are constrained, $\foralltOP$, by
\begin{subequations}
	\label{eq:OP_energy_flows_constraint_set}	
	\begin{align}
	&(\exportGrid_{t'} - \importGrid_{t'})/\Delta \tau
	-  \sum_{d\in \nonsteerableDevices} (1-\curtail)  \nonSteerable_{d,t'} + \sum_{d\in \steerableDevices} \steer_{d,t'}  \steerable_{d,t'} \notag\\
	& + \sum_{d\in\nonflexibleDevices} \nonFlexible_{d,t'}
	+ \sum_{d\in \sheddableDevices}(1 - \shed_{d,t'})  \sheddable_{d,t'} + \sum_{d\in\BSSs} \left(\chargerate_d  \charge_{d,t'} - \dischargerate_d  \discharge_{d,t'} \right)= 0 \\
	&(\exportGrid_{t'} - \importGrid_{t'})/ \Delta \tau  \leq \maxExportToGrid_{t'} \\
	&(\importGrid_{t'} - \exportGrid_{t'})/ \Delta \tau \leq \maxImportFromGrid_{t'} .
	\end{align}
\end{subequations}
The dynamics of the state of charge are, $\forallb$
\begin{subequations}
	\label{eq:OP_soc_dynamic}
	\begin{align}
	&s_{d,1} - \Delta \tau (\chargerate_{d}  \chargeEff_{d}  \charge_{d,1} - \frac{\dischargerate_{d}}{\dischargeEff_d}  \discharge_{d,1} ) = \initialCharge_{d}\\
	&s_{d,t'} - s_{d, t'-\Delta \tau} - \Delta \tau (\chargerate_{d}  \chargeEff_{d}  \charge_{d,t'} - \frac{\dischargerate_{d}}{\dischargeEff_d}  \discharge_{d,t'} )=0, \quad \foralltOP  \\
	&s_{d,\tau(t+T_a)} = \finalCharge_{d} .
	\end{align}
\end{subequations}
The set of constraints related to the peak power $\foralltOP$
\begin{subequations}
	\label{eq:OP_peak_constraints}
	\begin{align}
	(\importGrid_{t'} - \exportGrid_{t'})/ \Delta \tau &\leq p_{t'}  \\
	- \delta p_{t'} &\leq 0 \\
	- \delta p_{t'} &\leq - (p_{t'} - p_h) .
	\end{align}
\end{subequations}
The last constraints define symmetric reserve $\foralltOP$
\begin{subequations}
	\label{eq:OP_reserve_constraints}
	\begin{align}
	&\reserveBSSInc \leq \dfrac{\left( \SOC_{d,t'} - \mincharge_{d}\right)\dischargeEff_d}{\Delta \tau} \quad  \forallb  \\
	&\reserveBSSInc \leq \dischargerate_{d} (1-\discharge_{d,t'}) \quad  \forallb \quad   \\
	&\reserveBSSDec \leq \dfrac{\left( \maxcharge_{d} - \SOC_{d,t'} \right)}{\chargeEff_d \Delta \tau} \quad \forallb  \\
	&\reserveBSSDec \leq \chargerate_{d}(1-\charge_{d,t'}) \quad \forallb  \\
	&\reserve{sym, OP} \leq  \sum_{d\in\BSSs} \reserveBSSInc + \sum_{d\in \steerableDevices}\steerable_{d,t'} (1 - \steer_{d,t'}) + \sum_{d\in \nonsteerableDevices} \nonSteerable_{d,t'} (1-\curtail)  \\
	&\hspace{4em} +\sum_{d\in \sheddableDevices} \sheddable_{d,t'} (1-\shed_{d,t'})   \notag\\
	&\reserve{sym, OP} \leq \sum_{d\in\BSSs} \reserveBSSDec + \sum_{d\in \steerableDevices} \steerable_{d,t'} \steer_{d,t'}  +\sum_{d\in \nonsteerableDevices} \nonSteerable_{d,t'} \curtail +\sum_{d\in \sheddableDevices} \sheddable_{d,t'} \shed_{d,t'} .
	\end{align}
\end{subequations}

\subsection{RTP formulation}

The RTP objective function implemented for the case study is
\begin{align}\label{eq:RTO_objective_function}
	J_{t}^{RTO} = \ & C_{t}^{RTO} + D_{t}^{RTO} + v_{\tau(t)}(s_{\tau(t)})
\end{align}
with Real-Time Optimizer (RTO) the name of this controller. $C_{t}^{RTO}$ models the immediate costs, $D_{t}^{RTO}$ the delayed costs and $v_{\tau(t)}(s_{\tau(t)})$ the cost-to-go function of the state of the system at time $t$ within a current market period. 
$C_{t}^{RTO}$ is the same as $C_{t'}^{OP}$ by replacing $t'$ by $t$, $\Delta \tau$ by $\Delta t$ and considering only one period of time $t$.
$D_{t}^{RTO}$ is composed of the peak cost and symmetric reserve penalty costs
\begin{align}\label{eq:RTO_delayed_cost}
	D_{t}^{RTO} = \ & \price{p}  \delta p_{\tau(t-\Delta \tau),\tau(t)} + s^{TSO}_{t} \price{s}_{RTO}  \Delta \reserve{sym} ,
\end{align}
$\delta p_{\tau(t-\Delta \tau),\tau(t)}$ is the peak difference between the previous maximum historic peak $p_h$ and the current peak within the market period computed by RTO. The difference with OP relies on its computation as at $t$ the market period is not finished. Thus, the peak within this market period is made of two parts. (1) The peak from the beginning of the market period to $t$. (2) The peak from the actions taken from $t$ to the end of the market period. $\Delta \reserve{sym}$ is the difference between the symmetric reserve computed by OP and the current reserve within the market period computed by RTO. $s^{TSO}_{t}$ is the reserve activation signal to activate the tertiary symmetric reserve. It is set by the TSO, 0 if activated, else 1. The activation occurs at the beginning of the following market period.

\subsection{RTO constraints}
The set of constraints that defines the bounds on state and action variables and the energy flows are the same as the OP (\ref{eq:OP_action_constraint_set}) and (\ref{eq:OP_energy_flows_constraint_set}) by replacing $t'$ by $t$, $\Delta \tau$ by $\Delta t$ and considering only one period of time $t$.
The following constraint describes the dynamics of the state of charge $ \forallb$ and $ \foralltRTO$
\begin{align}
&s_{d,\tau(t)} - \Delta t (\chargerate_{d}  \chargeEff_{d}  \charge_{d,t} - \frac{\dischargerate_{d}}{\dischargeEff_d}  \discharge_{d,t} ) = \initialCharge_{d, t} .
\end{align}
The set of constraints related to the peak power $\foralltRTO$
\begin{subequations}
	\label{eq:RTO_peak_constraints}
	\begin{align}
	(\importGrid_t - \exportGrid_t)/ \Delta t &\leq p_{t,\tau(t)}  \\
	- \delta p_{\tau(t)} &\leq 0   \\
	- \delta p_{\tau(t)} &\leq - (p_{\tau(t-\Delta \tau),\tau(t)} - p_h)   \\
	p_{\tau(t-\Delta \tau),\tau(t)} &= \beta  p_{\tau(t-\Delta \tau),t} + (1-\beta )  p_{t,\tau(t)}  
	\end{align}
\end{subequations}
with $\beta = 1 - \Delta t / \Delta \tau$.
The last set of constraints defining the symmetric reserve are the same as the OP (\ref{eq:OP_reserve_constraints}) by replacing $t'$ by $t$, $\reserve{sym, OP}$ by $\reserve{sym, RTO}$ and adding $\foralltRTO$
\begin{subequations}
	\label{eq:RTO_reserve_constraints}
	\begin{align}
	& - \Delta \reserve{sym} \leq 0  \\
	& - \Delta \reserve{sym} \leq - (\reserve{sym, OP} - \reserve{sym, RTO}) .
	\end{align} 
\end{subequations}

\section{Test description}\label{sec:optimization-pscc-test-description}

Our case study is the MiRIS microgrid located at the John Cockerill Group's international headquarters in Seraing, Belgium\footnote{\url{https://johncockerill.com/fr/energy/stockage-denergie/}}. It is composed of PV, several energy storage devices, and a non-sheddable load. The load and PV data we use come from on-site monitoring. All data, including the weather forecasts, are available on the Kaggle platform\footnote{\url{https://www.kaggle.com/jonathandumas/liege-microgrid-open-data}}. 
The case study consists of comparing $\mbox{RTO-OP}$ to a Rule-Based Controller (RBC) for three configurations of the installed PV capacity, cf. Table~\ref{tab:simulation_parameters}. The RBC prioritizes the use of PV production for the supply of electrical demand. If the microgrid is facing a long position, it charges the battery. Furthermore, if this one is fully charged, it exports to the main grid. If the microgrid is facing a short position, it prioritizes using the battery to supply the demand. Moreover, if this one is fully discharged, it imports from the main grid. This controller does not take into account any future information, \textit{e.g.}, PV, consumption forecasts, energy prices, or market information such as the peak of the symmetric reserve. Case 3 results from a sizing study that defined the optimal device sizes given the PV and consumption data. The sizing methodology used is described in \citet{dakir2019sizing}.

\begin{figure}[tb]
	\centering
	\includegraphics[width=0.7\linewidth]{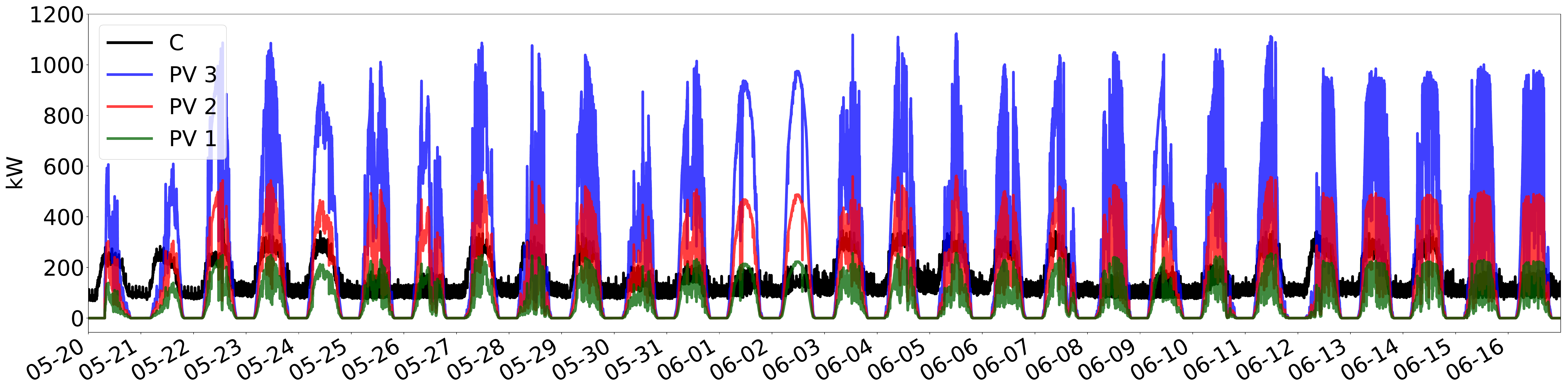} \\
	\includegraphics[width=0.7\linewidth]{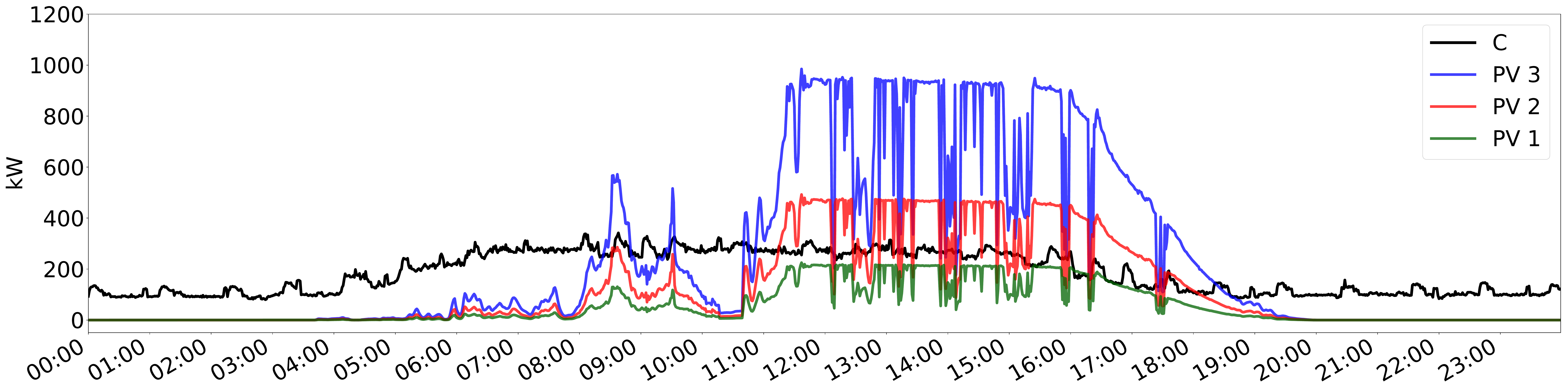}
	\captionsetup{justification=centering}
	\caption{Top: PV \& consumption simulation data. Bottom: zoom on $\pscctwelveofjune$.}
	\label{fig:energy_data}
\end{figure}

Figure~\ref{fig:energy_data} shows the PV \& consumption data over the simulation period: from $\pscccasetwostart$ to $\pscccasetwoend$. The selling price $\gridSalePrice$ is constant, and the purchasing price is composed of a day $\gridBuyPrice_d$ and night prices $\gridBuyPrice_n$. Day prices apply from 5 a.m. to 8 p.m. (UTC) during the weekdays, and night prices apply from 8 p.m. to 5 a.m. during weekdays and the entire weekend. The peak mechanism is taken into account with a constant peak price $\price{p}$ and an initial maximum historic peak $p_h$. Storage systems are initially fully charged.
The PV and consumption data have a 1-second resolution, meaning the RTO could compute its optimization problem each five to ten seconds in operational mode. CPLEX 12.9 is used to solve all the optimization problems on an Intel Core i7-8700 3.20 GHz-based computer with 12 threads and 32 GB of RAM. The average computation time per optimization problem composed of the OP and RTO is a few seconds. However, to maintain a reasonable simulation time, RTO is called every minute. The dataset is composed of 28 days with an average computation time of two hours to solve 1440 optimization problems per day, with a one-minute resolution, leading to two days for the entire dataset. The OP computes a quarterly planning corresponding to the Belgian market period. The computation time of the RTO on a regular computer is around a few seconds and the OP around twenty seconds. In total, the simulation computation time is up to a few hours. The OP computes quarterly planning based on PV and consumption twenty-four ahead forecasts. The weather-based forecast methodology is described in detail in Section \ref{sec:point-forecasting} of Chapter \ref{chap:forecast_evaluation}. Two "classic" deterministic techniques are implemented, a Recurrent Neural Network (RNN) with the Keras Python library \citep{chollet2015keras} and a Gradient Boosting Regression (GBR) with the Scikit-learn Python library \citep{scikit-learn}. These models use as input the weather forecasts provided by the Laboratory of Climatology of the Li\`ege University, based on the MAR regional climate model \citep{fettweis2017reconstructions}. It is an atmosphere model designed for meteorological and climatic research, used for a wide range of applications, from km-scale process studies to continental-scale multi-decade simulations.
To estimate the impact of the PV and consumption forecast errors on the controllers, the simulation is performed with the OP having access to the PV and consumption future values ($\mbox{RTO-OP}^{\star}$).
Then, the simulation is performed with the symmetric reserve mechanisms to cope with the forecast errors. A constant symmetric reserve price $\price{s}_{OP}$ for the OP and a penalty reserve $\price{s}_{RTO}$ for the RTO are set to 20 (\euro / kW).
\begin{table}[tb]
	\begin{center}
		\renewcommand\arraystretch{1.25}
		\begin{tabular}{lrrrrr}
			\hline  \hline
			Case & $\mbox{PV}_p$ &  $\overline{\mbox{PV}}$ & $\mbox{PV}_{max}$ & $\mbox{PV}_{min}$ & $\mbox{PV}_{std}$ \\ \hline 
			1	& 400  &  61 & 256  & 0 & 72\\ \hline 
			2	& 875  &  133 & 561 & 0 & 157 \\ \hline 
			3	& 1750 & 267 & 1122 & 0 & 314\\ \hline  \hline
			Case & $\mbox{C}_p$ &  $\overline{\mbox{C}}$ & $\mbox{C}_{max}$ & $\mbox{C}_{min}$ & $\mbox{C}_{std}$ \\ \hline 
			1 - 3	& 1000 & 153 & 390 & 68 & 72 \\ \hline  \hline
			Case & $\mbox{S}_p$ &  $\maxcharge$, $\mincharge$ & $\dischargerate$, $\chargerate$ & $\chargeEff$, $\dischargeEff$ & $\initialCharge$ \\ \hline 
			1 - 3 & 1350 & 1350, 0 & 1350, 1350 & 0.95, 0.95 & 100 \\ \hline  \hline
			Case & $p_h$, $\price{p}$ &  $\maxImportFromGrid$ & $\maxExportToGrid$ & $\gridBuyPrice_d$, $\gridBuyPrice_n$ & $\gridSalePrice$ \\ \hline 
			1 - 3	& 150, 40 & 1500 & 1500 & 0.2, 0.12 & 0.035 \\ \hline \hline 
		\end{tabular}
		\caption{Case studies parameters and data statistics.}
		\label{tab:simulation_parameters}
	\end{center}
\end{table}

\section{Numerical results}\label{sec:optimization-pscc-numerical results}

\subsection{No symmetric reserve}
Table \ref{tab:results_wht_reserve} provides the simulation results without taking into account the symmetric reserve. The smaller the PV installed capacity, the higher the peak and energy costs. The $\mbox{RTO-OP}^{\star}$ provides the minimal peak cost, whereas the RBC provides the minimal energy cost on all cases. However, $\mbox{RTO-OP}^{\star}$ achieves the minimal total cost, composed of the energy and peak costs.
\begin{table}[tb]
	\begin{center}
		\renewcommand\arraystretch{1.25}
		\begin{tabular}{lrrrrrr}
			\hline  \hline 
			Case 1 & $c_E$ & $c_p$ & $c_t$ & $\Delta_p$ & $\mbox{I}_{tot}$ & $\mbox{E}_{tot}$ \\ \hline  
			RBC	                     & 10.13 & 6.68  & 16.81 & 167 & 61 &  0    \\ \hline 
			$\mbox{RTO-OP}^{RNN}$	 & 10.37 & 3.62  & 13.99 & 91  & 64 & 1    \\ \hline 
			$\mbox{RTO-OP}^{GBR}$	 & 10.25 & 5.27  & 15.53 & 132 & 63 & 1    \\ \hline 
			$\mbox{RTO-OP}^{\star}$	 & 10.24 & 0.99  & 11.23 & 25  & 64 & 1    \\ \hline  \hline 
			Case 2 & $c_E$ & $c_p$ & $c_t$ & $\Delta_p$ & $\mbox{I}_{tot}$ & $\mbox{E}_{tot}$ \\ \hline  
			RBC	                     & 3.19  & 4.85  & 8.04 & 121 & 22 &  7  \\ \hline 
			$\mbox{RTO-OP}^{RNN}$	 & 4.78  & 2.87  & 7.65 & 72  & 31 & 15   \\ \hline 
			$\mbox{RTO-OP}^{GBR}$	 & 4.30  & 4.90  & 9.2  & 123 & 28 & 13 \\ \hline 
			$\mbox{RTO-OP}^{\star}$	 & 4.06  & 0     & 4.06 & 0   & 26 & 10\\ \hline \hline 
			Case 3 & $c_E$ & $c_p$ & $c_t$ & $\Delta_p$ & $\mbox{I}_{tot}$ & $\mbox{E}_{tot}$ \\ \hline  
			RBC	                     & -2.13 & 4.12  & 1.99 & 105   & 3 & 77  \\ \hline 
			$\mbox{RTO-OP}^{RNN}$	 & -1.66 & 4.12  & 2.46 & 105   & 7 & 80   \\ \hline 
			$\mbox{RTO-OP}^{GBR}$	 & -1.67 & 4.23  & 2.56 & 106   & 7 & 81 \\ \hline 
			$\mbox{RTO-OP}^{\star}$	 & -1.90 & 0     & 0 & 0     & 5 & 79 \\ \hline  \hline 
		\end{tabular}
				\caption{Results without symmetric reserve.}
		\label{tab:results_wht_reserve}
	\end{center}
\end{table}
This simulation illustrates the impact of the forecasts on the $\mbox{RTO-OP}$ behavior. The RNN forecaster provides the best results, but the $\mbox{RTO-OP}^{RNN}$ is still a long way to manage the peak as $\mbox{RTO-OP}^{\star}$ due to the forecasting errors. The peak cost strongly penalizes the benefits as it applies to the entire year ahead once it has been reached. 

%
\begin{figure}[tb]
	\centering
	\includegraphics[width=0.7\linewidth]{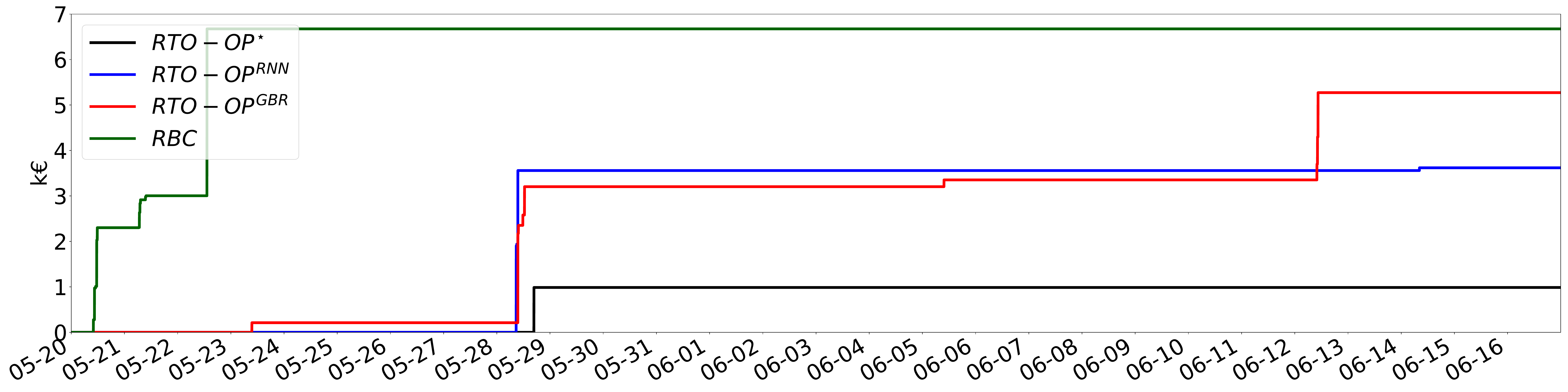} \\[1mm] 
	\includegraphics[width=0.7\linewidth]{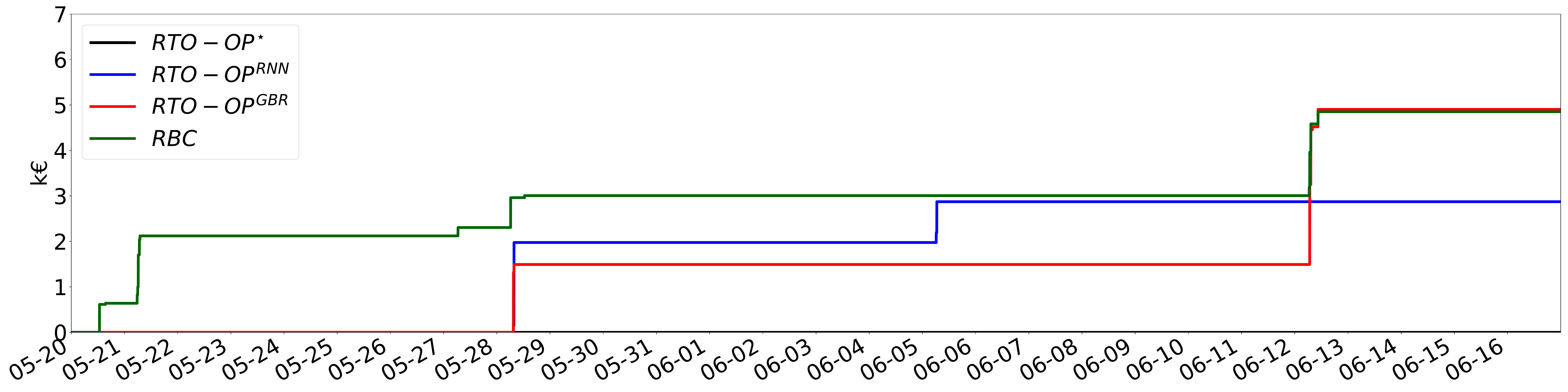} \\[1mm] 
	\includegraphics[width=0.7\linewidth]{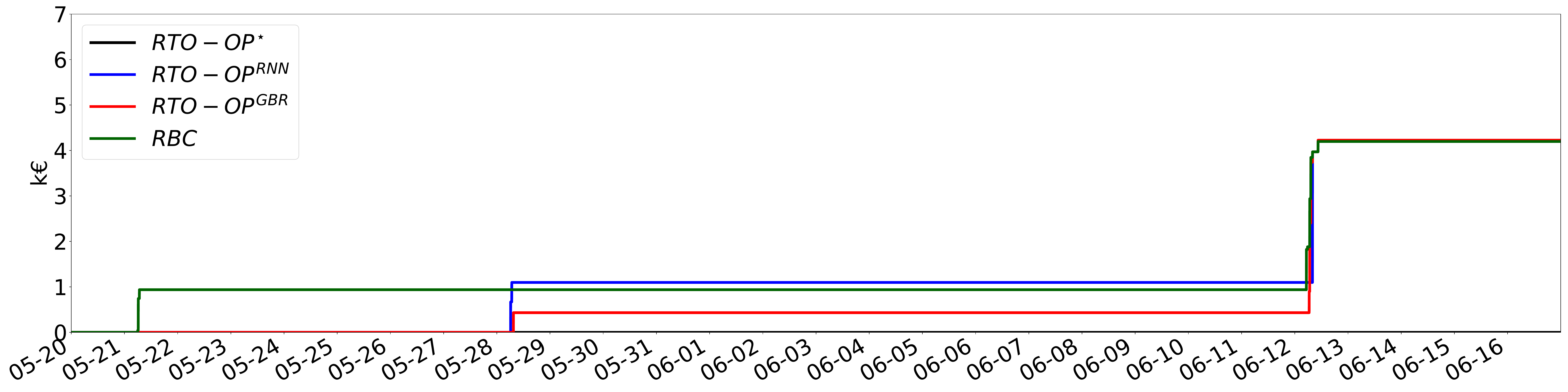} 
	\captionsetup{justification=centering}
	\caption{Case 1 (top), 2 (middle), 3 (bottom) cumulative peak costs.}
	\label{fig:case321_peak_cost}
\end{figure}
\begin{figure}[tb]
	\centering
	\includegraphics[width=0.7\linewidth]{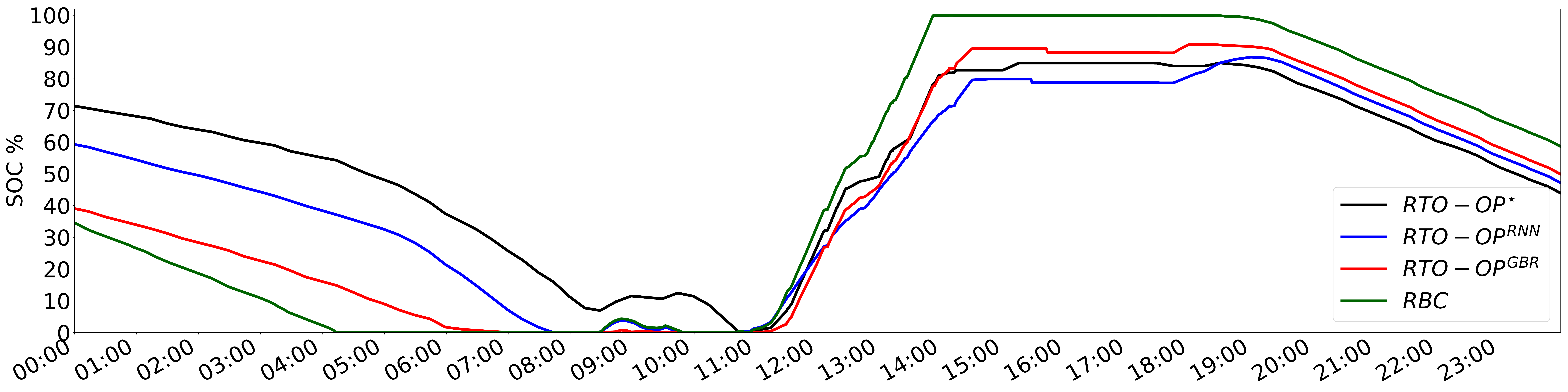}  \\[1mm] 
	\includegraphics[width=0.75\linewidth]{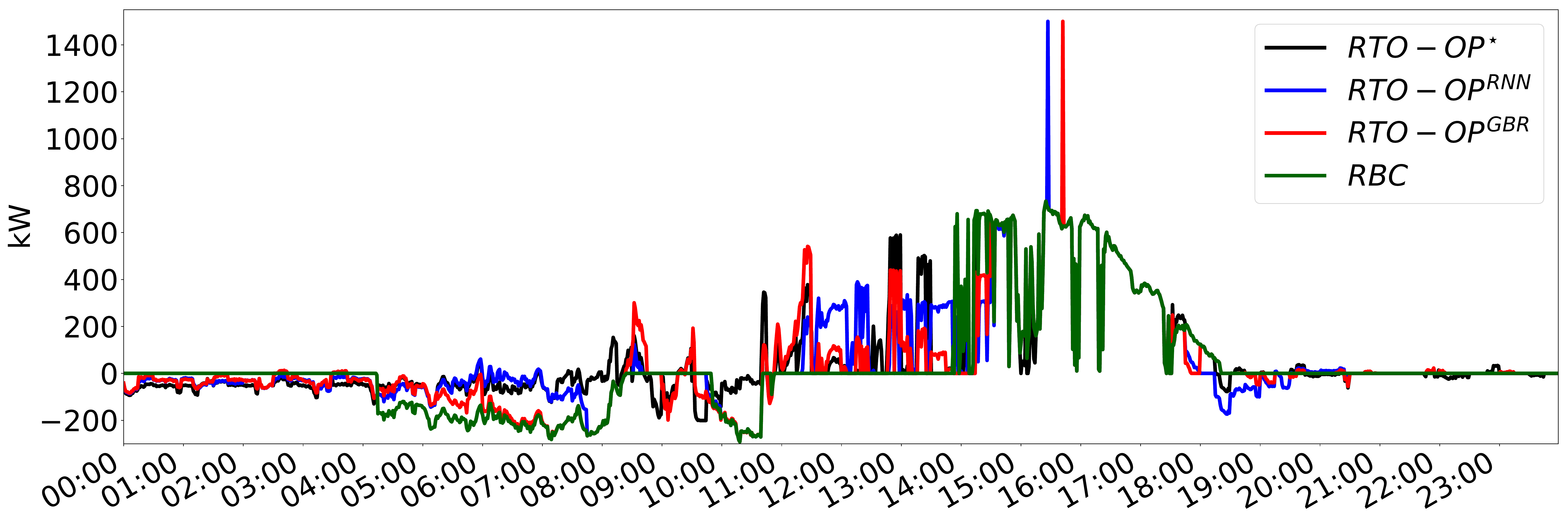} 
	\captionsetup{justification=centering}
	\caption{Case 3 SOC (top) and net export power (bottom) on $\pscctwelveofjune$.}
	\label{fig:case3_soc_net_1206}
\end{figure}

In case 3, all the controllers except $\mbox{RTO-OP}^{\star}$ reached the maximum peak on $\pscctwelveofjune$ around 10:30 a.m. as shown on Figure~\ref{fig:case321_peak_cost}. Figure~\ref{fig:energy_data} shows a sudden drop in the PV production around 10 a.m. that is not accurately forecasted by the RNN and GBR forecasters as shown in Figure~\ref{fig:case3_pv_forecast_12062019}. This prediction leads to the non-accurate planning of OP. Thus, the RTO cannot anticipate this drop and has to import energy to balance the microgrid at the last minute. Figure~\ref{fig:case3_soc_net_1206} shows the controllers behavior on $\pscctwelveofjune$ where the peak is reached.
In case 2, all controllers reached the same peak as in case 3 except $\mbox{RTO-OP}^{RNN}$ that reached a smaller one on $\psccfiveofjune$. The forecast's accuracy explains this behavior as in case 3.
Finally, in case 1, each controller reached a different peak. The smallest one is achieved by the $\mbox{RTO-OP}^{\star}$, followed by the $\mbox{RTO-OP}^{RNN}$.
These cases show that the $\mbox{RTO-OP}$ controller optimizes PV-storage usage and thus requires less installed PV capacity for a given demand level. This result was expected as the peak management is not achieved by the RBC and becomes critical when the PV production is smaller than the consumption. This simulation also demonstrates the forecast accuracy impact on the $\mbox{RTO-OP}$ behavior. 

\subsection{Results with symmetric reserve}

\begin{figure}[tb]
	\centering
	\includegraphics[width=0.7\linewidth]{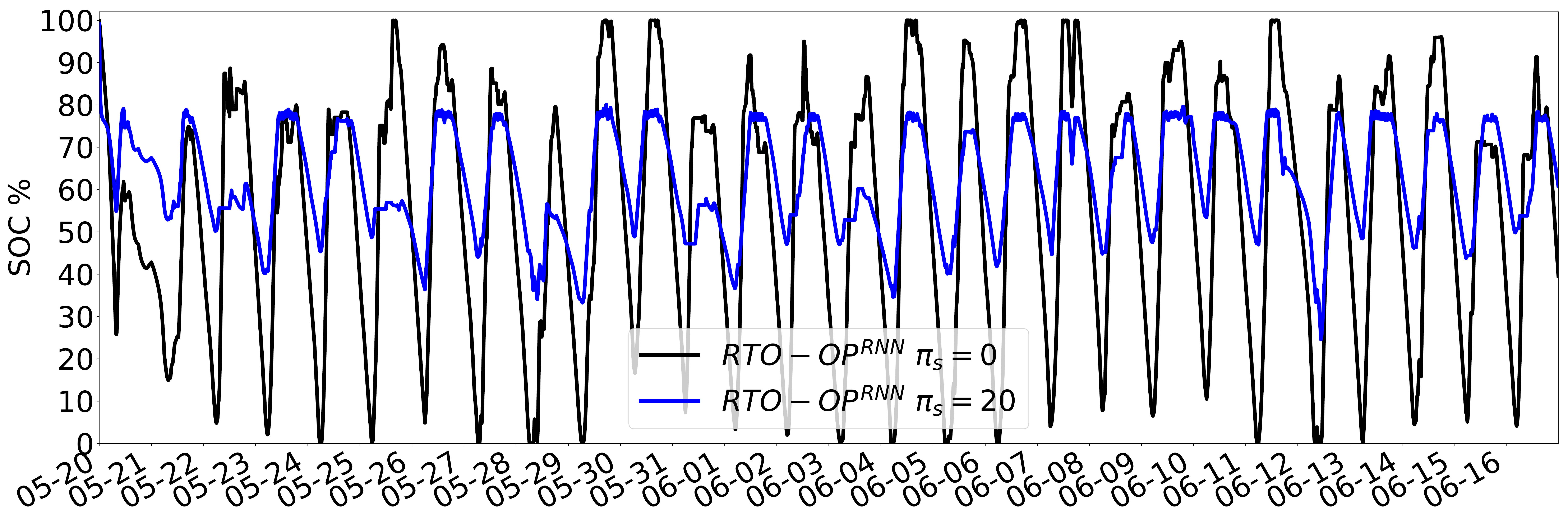} 
	\caption{Case 3 SOC comparison for $\mbox{RTO-OP}^{RNN}$ with and without symmetric reserve.}
	\label{fig:case3_soc_r0_vs_20}
\end{figure}

\begin{figure}[tb]
	\centering
	\includegraphics[width=0.7\linewidth]{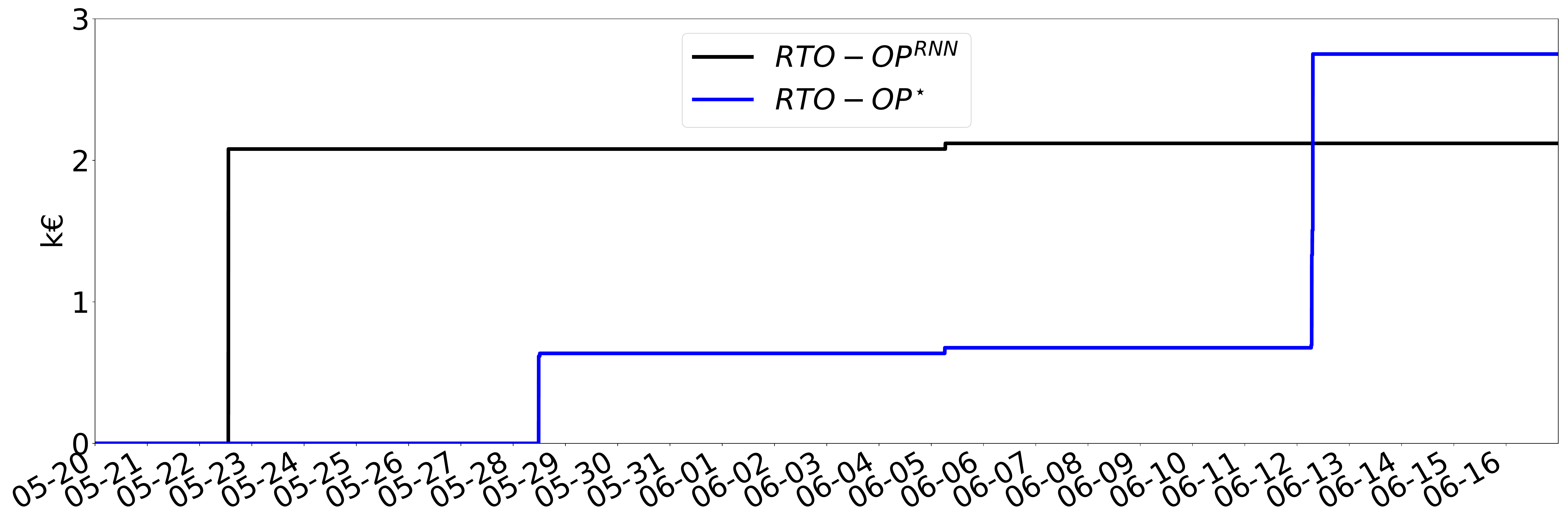}  \\[1mm] 
	\includegraphics[width=0.7\linewidth]{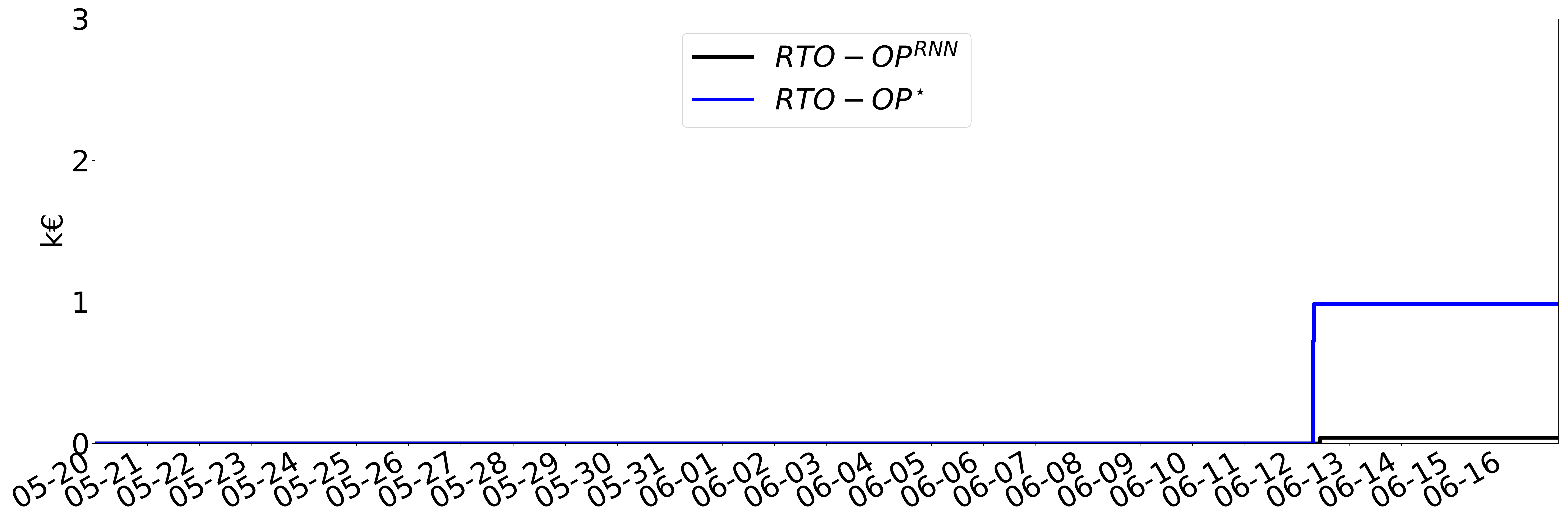} 
	\captionsetup{justification=centering}
	\caption{Case 1 (top) and 2 (bottom) cumulative peak costs.}
	\label{fig:case21_with_reserve_peak_costs}
\end{figure}

\begin{figure}[tb]
	\centering
	\includegraphics[width=0.7\linewidth]{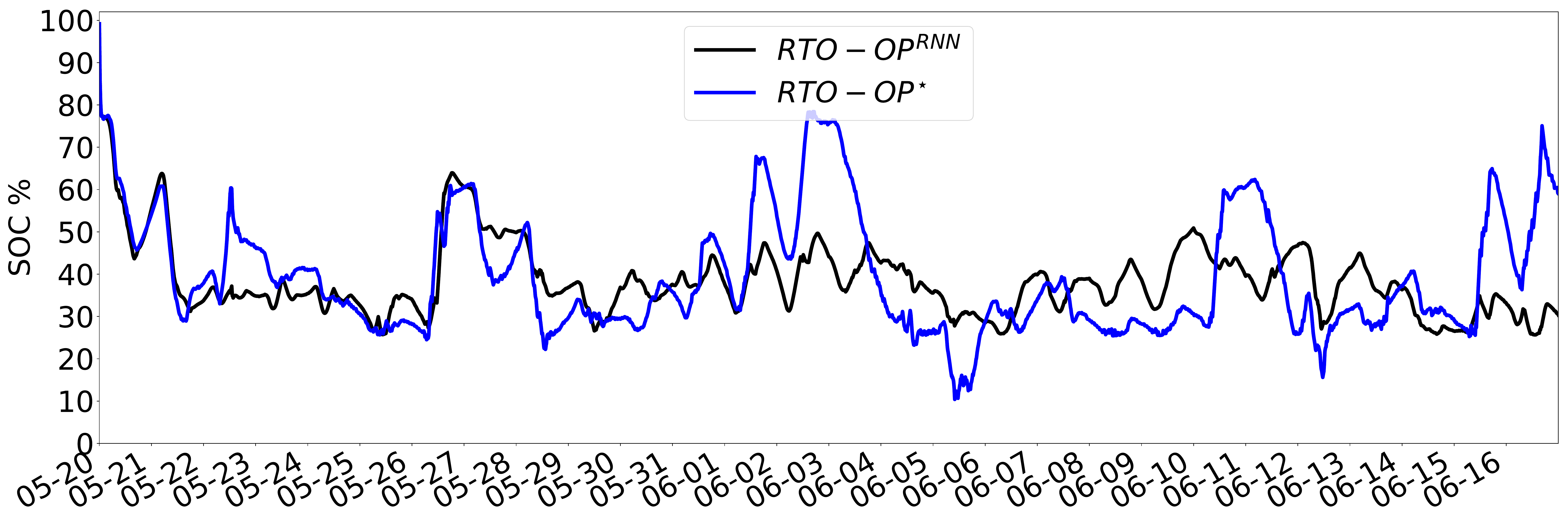} \\[1mm]
	\includegraphics[width=0.7\linewidth]{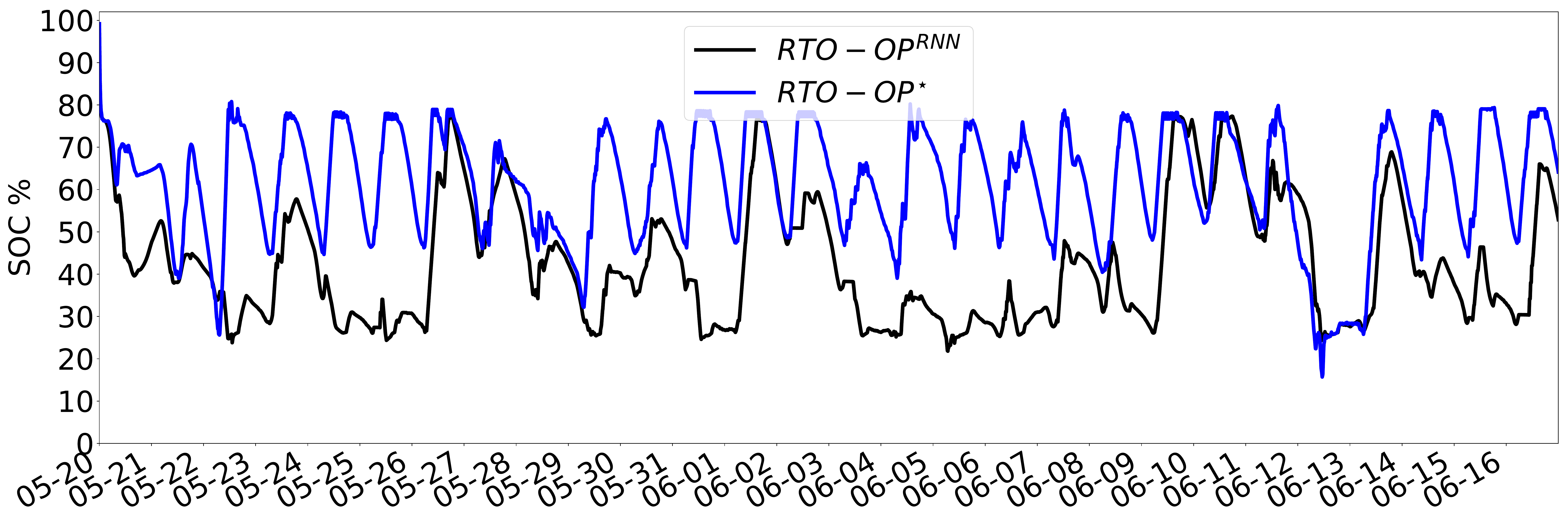}  
	\captionsetup{justification=centering}
	\caption{Case 1 (top) and 2 (bottom) SOC.}
	\label{fig:case21_soc_with_reserve}
\end{figure}

Table \ref{tab:results_w_reserve} provides the simulation results by taking into account the symmetric reserve. Figure \ref{fig:case3_soc_r0_vs_20} depicts on case 3 the behavior differences between $\mbox{RTO-OP}^{RNN}$ without and with symmetric reserve. Figures \ref{fig:case21_with_reserve_peak_costs} and \ref{fig:case21_soc_with_reserve} show the SOC and peaks costs evolution of case 2 \& 1.
The controller tends to maintain a storage level that allows $\mbox{RTO-OP}^{RNN}$ to better cope with forecast errors. Indeed for case 3, there is no more peak reached by $\mbox{RTO-OP}^{RNN}$, only 1 kW for case 2, and it has been almost divided by two for case 1. However, this behavior tends to increase the energy cost if the PV production is large compared to the consumption, such as in case 3. Indeed, the controller will tend to store more energy in the battery instead of exporting it. 
$\mbox{RTO-OP}^{\star}$ did not perform better with the symmetric reserve. The symmetric reserve competes with the peak management, and the $\mbox{RTO-OP}^{\star}$ tends not to discharge the battery entirely even if it is required to avoid a peak. In case 2, the peak is reached on $\pscctwelveofjune$ around 08:00. The controller could have avoided it by totally discharging the battery but did not maintain the reserve level. It is the same behavior in case 1, where the peak could have been limited if all the battery was discharged. There is an economic trade-off to manage the peak and the reserve simultaneously. It depends on the valorization or not on the market of the symmetric reserve. The reserve can also be valorized internally to cope with non or complex forecastable events. Such as a sudden drop in export or import limits due to loss of equipment or grid congestion.
\begin{table}[tb]
	\begin{center}
		\renewcommand\arraystretch{1.25}
		\begin{tabular}{l|r|r|r|r|r|r}
			\hline  \hline 
			Case 1 & $c_E$ & $c_p$ & $c_t$ & $\Delta_p$ & $\mbox{I}_{tot}$ & $\mbox{E}_{tot}$ \\ \hline  
			$\mbox{RTO-OP}^{RNN}$	 & 10.50 & 2.12  & 12.62 & 53  & 65 &  3    \\ \hline 
			$\mbox{RTO-OP}^{\star}$	 & 10.47 & 2.75  & 13.22 & 69  & 65 & 2 \\ \hline  \hline 
			Case 2 & $c_E$ & $c_p$ & $c_t$ & $\Delta_p$ & $\mbox{I}_{tot}$ & $\mbox{E}_{tot}$ \\ \hline  
			$\mbox{RTO-OP}^{RNN}$	 & 5.33  & 0.04 & 5.37 & 1  & 41 & 27   \\ \hline 
			$\mbox{RTO-OP}^{\star}$	 & 4.78  & 0.99 &  5.77 & 25     & 35 & 20\\ \hline \hline 
			Case 3 & $c_E$ & $c_p$ & $c_t$ & $\Delta_p$ & $\mbox{I}_{tot}$ & $\mbox{E}_{tot}$ \\ \hline  
			$\mbox{RTO-OP}^{RNN}$	 & -0.04 & 0 & -0.04 & 0   & 24   & 99   \\ \hline 
			$\mbox{RTO-OP}^{\star}$	 & -0.15 & 0 & -0.15 & 0   & 23.2 & 98 \\ \hline  \hline 
		\end{tabular}
		\caption{Results with symmetric reserve.}
		\label{tab:results_w_reserve}
	\end{center}
\end{table}


\section{Conclusions}\label{sec:optimization-pscc-conclusion}

A two-level value function-based approach is proposed as a solution method for a multi-resolution microgrid optimization problem. The value function computed by the operational planner based on PV and consumption forecasts allows coping with the forecasting uncertainties. The real-time controller solves an entire optimization problem, including the future information propagated by the value function.
This approach is tested on the MiRIS microgrid case study with PV and consumption data monitored on-site. The results demonstrate the efficiency of this method to manage the peak in comparison with a Rule-Based Controller. This test case is completely reproducible as all the data used are open, PV, consumption monitored, and forecasted, including the weather forecasts.
The proposed method can be extended in three ways. First, a stochastic formulation of the operational planning problem to cope with probabilistic forecasts. Second, adding the balancing market mechanisms. Finally, considering a community composed of several entities inside the microgrid.

\chapter{Capacity firming using a stochastic approach}\label{chap:capacity-firming-stochastic}

\begin{infobox}{Overview}
This Chapter proposes a stochastic approach to address the energy management of a grid-connected renewable generation plant coupled with a battery energy storage device in the capacity firming market. 
Both deterministic and stochastic approaches result in optimization problems formulated as quadratic problems with linear constraints. The considered case study is a real microgrid with PV production monitored on-site.

\textbf{\textcolor{RoyalBlue}{References:}} This chapter is an adapted version of the following publication: \\[2mm]\bibentry{dumas2020stochastic}. 
\end{infobox}
\epi{We are our choices.}{Jean-Paul Sartre }
\begin{figure}[htbp]
	\centering
	\includegraphics[width=1\linewidth]{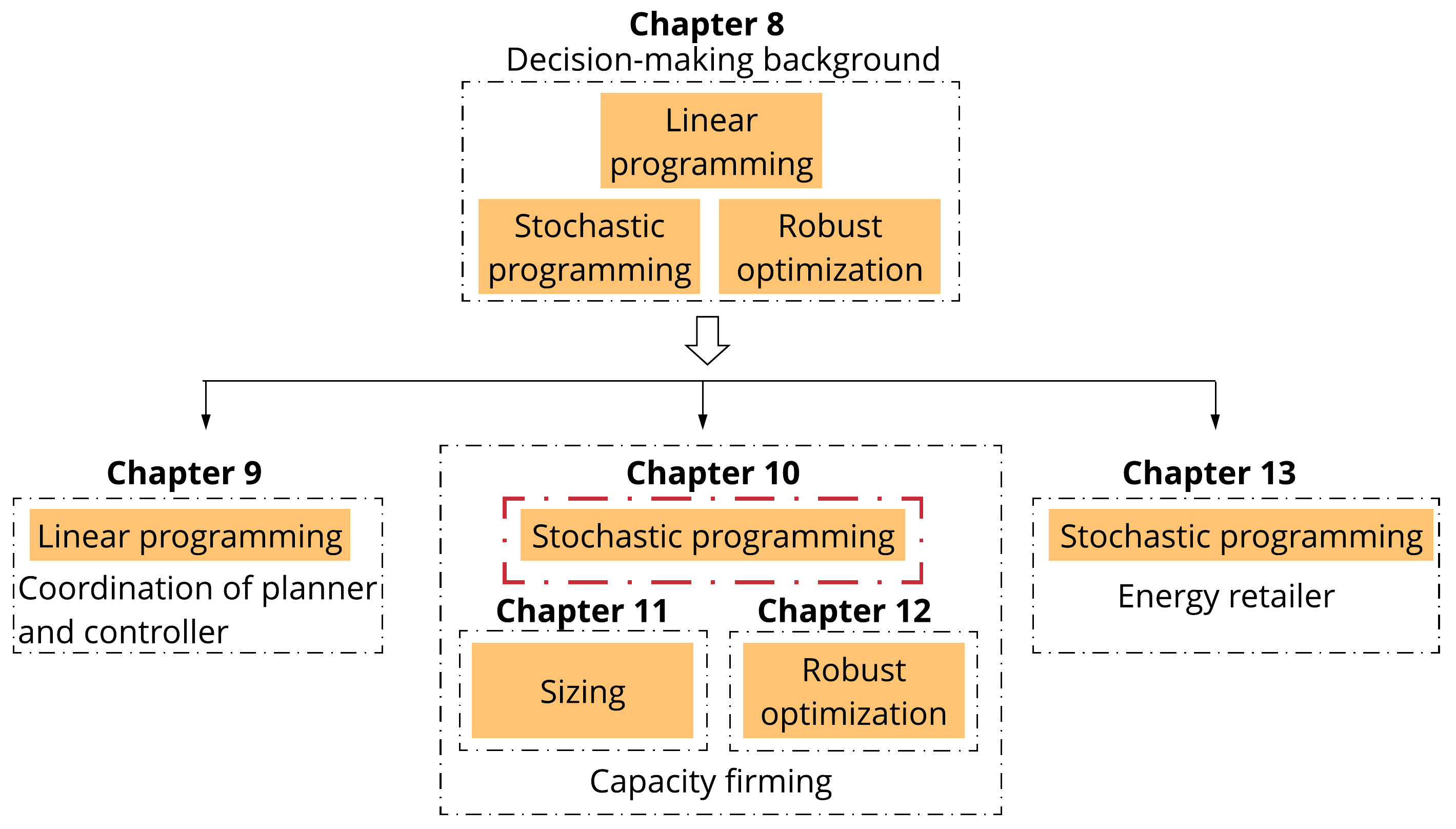}
	\caption{Chapter \ref{chap:capacity-firming-stochastic} position in Part \ref{part:optimization}.}
\end{figure}
\clearpage

The capacity firming framework is mainly designed for isolated markets, such as the Overseas France islands. For instance, the French Energy Regulatory Commission (CRE) publishes capacity firming tenders and specifications. The system considered is a grid-connected renewable energy power plant, \textit{e.g.}, photovoltaic or wind-based, with a battery energy storage system (BESS) for firming the renewable generation. At the tendering stage, offers are selected on the electricity selling price. Then, the successful tenderer builds its plant and sells the electricity exported to the grid at the contracted selling price, but according to a well-defined daily engagement and penalization scheme specified in the tender specifications. The electricity injected in or withdrawn from the grid must be nominated the day-ahead, and engagements must satisfy ramping power constraints. The remuneration is calculated a posteriori by multiplying the realized exports by the contracted selling price minus a penalty. The deviations of the realized exports from the engagements are penalized through a function specified in the tender. A peak option can be activated in the contract for a significant selling price increase during a short period defined a priori. Therefore, the BESS must shift the renewable generation during peak hours to maximize revenue and manage renewable energy uncertainty.

The problem of modeling a two-phase engagement/control with an approach dealing with uncertainty in the context of the CRE capacity framework is still an open issue. This framework has received less attention in the literature than more traditional energy markets such as day-ahead and intraday markets of European countries.

The optimal day-ahead bidding strategies of a plant composed of only a production device have been addressed in, \textit{e.g.},  \citet{pinson2007trading,bitar2012bringing,giannitrapani2014bidding,giannitrapani2015bidding}. The optimal offer turns out to be a suitable percentile of the PV/wind power cumulative distribution function. Under the assumption of time-invariant power generation statistics, the cumulative distribution functions can be estimated from historical data of the power generated by the plant. This assumption is not always justified, especially for PV power generation. In \citet{giannitrapani2015bidding}, the authors investigate two approaches to properly take into account the effects of seasonal variation and non-stationary nature of PV power generation in the estimation of PV power statistics. However, incorporating energy storage in the framework is still an open problem, and the literature provides several approaches and methodologies to this end.
An optimal power management mechanism for a grid-connected PV system with storage is implemented in \cite{riffonneau2011optimal} using Dynamic Programming (DP) and is compared with simple ruled-based management. The sizing and control of an energy storage system to mitigate wind power uncertainty is addressed by \citet{haessig2014dimensionnement,haessig2013aging,haessig2015energy} using stochastic dynamic programming (SDP). The framework is similar to the CRE PV capacity firming tender with a wind farm operator committed on a day-ahead basis to a production engagement. Finally, three distinct optimization strategies, mixed-integer quadratic programming, simulation-based genetic algorithm, and expert-based heuristic, are empirically compared by \citet{n2019optimal} in the CRE framework. 

This study addresses the energy management of a grid-connected PV plant and BESS. This topic is studied within the capacity firming specifications of the CRE, in line with the tender AO-CRE-ZNI 2019 published on $\CREspecifications$, using the MiRIS microgrid case study.
The capacity firming problem can be decomposed into two steps. The first step consists of computing the day-ahead nominations. The second step consists of computing the renominations and the set-points in real-time to minimize the energy and ramp power deviations from nominations. This study focuses on the first step and proposes both a stochastic and a deterministic formulation. The main goal of this study is to validate the stochastic approach by using an ideal predictor providing unbiased PV scenarios. Thus, the BESS efficiencies are perfect, and the degradation is not taken into account for the sake of simplicity. Different levels of prediction accuracy are evaluated. Then, the results are compared with those of the deterministic formulation, assuming perfect forecasts returned by an oracle. 
Both deterministic and stochastic approaches result in optimization problems formulated as quadratic problems with linear constraints. The considered case study is a real microgrid with PV production monitored on-site.

The study is organized as follows. Section \ref{sec:optimization-pmaps-notation} provides the notation that is also used for Chapters \ref{chap:capacity-firming-sizing} and \ref{chap:capacity-firming-robust}. Section \ref{sec:optimization-pmaps-capacity-firming-process} describes the capacity firming framework. Section \ref{sec:optimization-pmaps-problem formulation} proposes the deterministic and stochastic formulations of the nomination process. Section \ref{sec:optimization-pmaps-case_study} introduces the MiRIS microgrid case study and presents the results.
Conclusions are drawn in Section \ref{sec:optimization-pmaps-conclusions}. Appendix \ref{appendix:optimization-pmaps-scenario} describes the methodology to generate the set of unbiased PV scenarios.\\

\section{Notation}\label{sec:optimization-pmaps-notation}

\newcommand{\PVcapacity} {\ensuremath{P_c}} 
\newcommand{\PVgenerationTrue} {\ensuremath{y^\text{pv,m}}} 
\newcommand{\PVgeneration} {\ensuremath{y^\text{pv}}} 
\newcommand{\PVforecast} {\ensuremath{\hat{y}^\text{pv}}} 
\newcommand{\ijfWgeneration} {\ensuremath{y^\text{w}}} 
\newcommand{\ijfWforecast} {\ensuremath{\hat{y}^\text{w}}} %
\newcommand{\ijfLoad} {\ensuremath{y^\text{l}}} 
\newcommand{\ijfLoadforecast} {\ensuremath{\hat{y}^\text{l}}} 
\newcommand{\Discharge} {\ensuremath{y^\text{dis}}} 
\newcommand{\Charge} {\ensuremath{y^\text{cha}}} 
\newcommand{\Soc} {\ensuremath{y^s}} 
\newcommand{\ijfSoc} {\ensuremath{s}} 
\newcommand{\SocMin} {\ensuremath{s^\text{min}}} 
\newcommand{\SocMax} {\ensuremath{s^\text{max}}} 
\newcommand{\SocIni} {\ensuremath{s^\text{ini}}} 
\newcommand{\SocEnd} {\ensuremath{s^\text{end}}} 
\newcommand{\DischargeMax} {\ensuremath{y^\text{dis}_\text{max}}} 
\newcommand{\ChargeMax} {\ensuremath{y^\text{cha}_\text{max}}} 
\newcommand{\yMin} {\ensuremath{y^\text{min}}} 
\newcommand{\yMax} {\ensuremath{y^\text{max}}} 
\newcommand{\xMin} {\ensuremath{x^\text{min}}} 
\newcommand{\xMax} {\ensuremath{x^\text{max}}} 
\newcommand{\BESSbinary} {\ensuremath{y^b}} 

\subsection{Sets and indices}
\begin{supertabular}{l p{0.8\columnwidth}}
	Name & Description \\
	\hline
	$t$ & Time period index. \\
	$\omega$ & PV scenario index. \\
	$q$ & PV quantile index. \\
	$T$ & Number of time periods per day. \\
	$\mathcal{T}$ & Set of time periods, $\mathcal{T}= \{1,2, \ldots, T\}$. \\
	$\mathcal{D}$ & $ \cup_{i=1}^{i=D} \mathcal{T}_i$ set of $D$ days of market periods. \\
	$\Omega$ & PV scenarios set, $\Omega= \{1,2, \ldots, \#\Omega\}$. \\
	$\#\Omega$ & Number of PV scenarios in $\Omega$. \\
	$\mathcal{U}$ & PV uncertainty set for robust optimization. \\
\end{supertabular}

\subsection{Parameters}
\begin{supertabular}{l p{0.7\columnwidth} l}
	Name & Description & Unit \\
	\hline
	$\PVcapacity$ & PV installed capacity. & kWp \\
    $\maxcharge$ & BESS installed capacity. & kWh \\
	$y^m_t$ & Measured power at the grid connection point. & kW\\
	$\PVgenerationTrue_t$ & Measured PV generation. & kW \\
	$\PVforecast_t$ & PV point forecast. & kW \\
	$\PVforecast_{t,\omega}$ & PV scenario $\omega$. & kW \\
	$\hat{y}^{\text{pv}, (q)}_t$ & PV quantile forecast $q$. & kW \\
	$\alpha_\omega$ & Probability of PV scenario $\omega$ &  -\\
	$\DischargeMax$, $\ChargeMax$ & BESS maximum (dis)charging power. & kW \\
	$\dischargeEff$, $\chargeEff$ & BESS (dis)charging efficiency. & - \\
	$\SocMin$, $\SocMax$ & BESS minimum/maximum capacity with $\SocMax \leq \maxcharge$. & kWh\\
	$\SocIni$, $\SocEnd$ & BESS initial/final state of charge. & kWh  \\
	$\xMin_t$, $\xMax_t$  & Minimum/maximum engagement. & kW \\
	$p P_c$ &  Engagement tolerance, $ 0 \leq p \leq 1$, kW.  \\
	$\Delta X$  & Ramping-up and down limits for the engagement. & kW \\
	$\yMin_t$, $\yMax_t$  & Minimum/maximum power at the grid connection point. & kW \\
	$\pi_t$ & Contracted selling price. & \euro/kWh \\
	$\pi^e$ & Slack price. & $\frac{\text{\euro}}{\text{kWh}^2}$ \\
	$\pi_{\maxcharge}$ & BESS CAPEX price. & \euro/kWh \\
	$\Delta t$ & Time period duration. & minutes \\
	$c$ & Penalty function. & \euro \\
	$u^\text{min}_t$, $u^\text{max}_t$ & Minimum/maximum of the PV uncertainty interval. & kW \\
	$\Gamma$ & Uncertainty budget for robust optimization. & - \\
	$d_q$, $d_\Gamma$ & Uncertainty and budget depths for robust optimization. &\% \\
	$M_t^-$, $M_t^+$ & Big-M’s values for robust optimization. & -  \\
\end{supertabular}

\subsection{Variables}
\noindent For the sake of clarity the subscript $\omega$ is omitted.

\begin{supertabular}{l l p{0.7\columnwidth} l}
	Name & Range & Description & Unit \\
	\hline
	$x_t$ & $[\xMin_t, \xMax_t]$ & Engagement. & kW\\
	$y_t$ & $[\yMin_t, \yMax_t]$ & Net power at the grid connection point. & kW\\
	$\PVgeneration_t $ & $ [0,\PVforecast_t]$ & PV generation. & kW \\
	$\Charge_t $    & $ [0, \ChargeMax]$ & BESS charging power.  & kW \\
	$\Discharge_t $ & $ [0, \DischargeMax]$ & BESS discharging power. & kW \\
    $\Soc_t$ & $  [\SocMin, \SocMax]$ & BESS state of charge. & kWh\\
    $d^-_t$, $d^+_t$  & $ \mathbb{R}_+$ & Short/long deviation.  & kW\\
	$\BESSbinary_t$ & $ \{0, 1\}$ & BESS binary variable. & - \\
	$z_{t}$ & $ \{0, 1\}$ & PV uncertainty set binary variable for robust optimization. & - \\
	$\alpha_t$ & $ [M_t^-,M_t^+]$ & Variables to linearize $z_t \phi^{\PVgeneration}_t$ for robust optimization. & -  \\
\end{supertabular}

\subsection*{Dual variables, and corresponding constraints}
\noindent Dual variables of constraints are indicated with brackets $[ \cdot ]$.

\begin{supertabular}{l l p{0.7\columnwidth} }
	Name & Range & Description \\
	\hline
	$\phi^\text{cha}_t $, $\phi^\text{dis}_t $ & $ \mathbb{R}^-$ & Maximum storage (dis)charging.  \\
	$\phi^{\SocMin}_t $, $\phi^{\SocMax}_t $ & $ \mathbb{R}^-$ & Minimum/maximum storage capacity.   \\
	$\phi^y_t$ & $ \mathbb{R}$ & Net power balance. \\
	$\phi^{\yMin}_t$, $\phi^{\yMax}_t$ & $ \mathbb{R}^-$ & Minimum/maximum net power. \\
	$\phi^{\SocIni}_t $, $\phi^{\SocEnd}_t $ & $ \mathbb{R}^-$ & Initial/final state of charge.   \\
	$\phi^{\SOC}_t$ & $ \mathbb{R}^-$ & BESS dynamics. \\
	$\phi^{d^-}_t$, $\phi^{d^+}_t$ & $ \mathbb{R}^-$ & Under/overproduction.  \\
	$\phi^{\PVgeneration}_t$ & $ \mathbb{R}^-$ & Renewable generation.  \\
\end{supertabular}

\section{The Capacity Firming Framework}\label{sec:optimization-pmaps-capacity-firming-process}

The capacity firming framework can be decomposed into a day-ahead engagement process, Section \ref{sec:engagement_process}, and a real-time control process, Section \ref{sec:control_process}. Each day is discretized in $T$ periods of duration $\Delta t$. In the sequel, the period duration is the same for day-ahead engagement and the real-time control, $t$ is used as a period index, and $\mathcal{T}$ is the set of periods in a day. 

\subsection{Day-ahead engagement}\label{sec:engagement_process}

\begin{figure}[tb]
	\centering
	\includegraphics[width=0.7\linewidth]{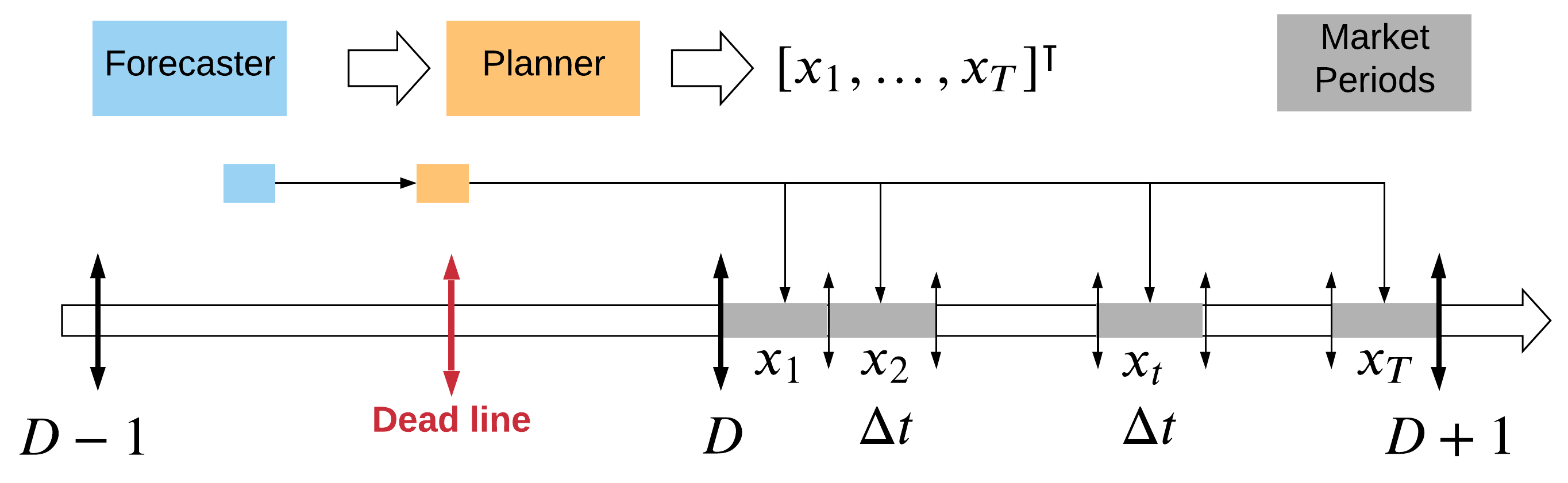}
	\caption{Day-ahead nomination process.}
	\label{fig:optimization-pmaps-dayahead-process}
\end{figure}

Each day, the operator of the renewable generation plant is asked to provide the generation profile to be followed the next day to the grid operator, based on renewable generation forecasts. More formally, a planner computes on a day-ahead basis, before a deadline, a vector of engagements composed of $T$ values $[ x_1, \ldots,x_T]^\intercal$. Figure \ref{fig:optimization-pmaps-dayahead-process} illustrates the day-ahead nomination process. The grid operator accepts the engagements if they satisfy the constraints
\begin{subequations}\label{eq:pmaps-engagement-csts}
	\begin{align}
	|x_t-x_{t-1}| & \leq  \Delta X , \ \forallt \setminus \{1\}	\\
	- x_t &  \leq -  \xMin_t, \ \forallt  \\
	x_t &  \leq \xMax_t, \ \forallt ,
	\end{align}
\end{subequations}
with $\Delta X$ a ramping power constraint, a fraction of the total installed capacity $\PVcapacity$ determined at the tendering stage and imposed by the grid operator. 

\subsection{Real-time control}\label{sec:control_process}

Then, in real-time, a receding-horizon controller computes at each period the generation level and the charge or discharge set-points from $t$ to $T$, based on forecasts of renewable generation and the engagements. Only the set-points of the first period are applied to the system.
The remuneration is calculated ex-post based on the realized power $y^m_t$ at the grid coupling point. For a given control period, the net remuneration $r_t$ of the plant is the gross revenue $ \Delta t \pi_t  y^m_t$ minus a penalty $c(x_t, y^m_t)$, with $\pi_t$ the contracted selling price set at the tendering stage
\begin{align}\label{eq:pmaps-penalty}
r_t = \Delta t \pi_t  y^m_t  - c(x_t, y^m_t),  \ \forallt.
\end{align}
The penalty function $c$ depends on the specifications of the tender. For the sake of simplicity for the rest of this Chapter, $c$ is assumed to be symmetric, convex, and quadratic piecewise-linear
\begin{align}
c(x_t, y^m_t)  & = \pi^e (\Delta t)^2\bigg(  \max \big(0, |x_t- y^m_t|- p \PVcapacity \big) \bigg)^2
\end{align} 
with $ 0\leq p \leq 1$, and $\pi^e$ is a slack price (\euro/$\text{kWh}^2$).

\section{Problem formulation}\label{sec:optimization-pmaps-problem formulation}

The problem statement follows the abstract formulation defined in Chapter \ref{chap:coordination-planner-controller} where a global microgrid control problem can be defined as operating a microgrid safely and in an economically efficient manner. In the capacity firming context, a two-stage approach is considered with a planner and a controller. A quadratic formulation with linear constraints models the CRE non-convex penalty. In this study, the planner and controller optimize economic criteria, which are only related to active power. The ancillary or grid services are not in the scope of the capacity firming specifications. The BESS degradation is not taken into account. The planner and controller horizons are cropped to twenty-four hours. 

Deterministic (D) and stochastic (S) formulations of the day-ahead nomination problem are compared. The deterministic formulation is used as a reference to validate the stochastic approach by considering perfect knowledge of the future (D$^\star$). In this Chapter, both approaches consider only exports to the grid\footnote{The imports from the grid are allowed only under specific conditions into the contract.}. The optimization variables and the parameters are defined in Section~\ref{sec:optimization-pmaps-notation}.

\subsection{Deterministic approach}
The objective function $J_D $ to minimize is the opposite of the net revenue
\begin{align}\label{eq:pmaps-D-obj}
J_D (x_t, y_t) & \ =  \sum_{t \in \mathcal{T}} - \pi_t \Delta t  y_t + \pi^e (\Delta t)^2 [(d^-_t)^2 + (d^+_t)^2] .
\end{align}
The deterministic formulation is the following Mixed-Integer Quadratic Program (MIQP)
\begin{align}\label{eq:pmaps-D-formulation}
\min_{x_t \in \mathcal{X}, y_t \in \Omega(x_t, \PVforecast_t)} & \ J_D (x_t, y_t)   \notag \\
\mathcal{X} & =   \big \{ x_t : (\ref{eq:pmaps-engagement-D-csts}) \big \} \\
\Omega(x_t, \PVforecast_t) & =  \big \{ y_t : (\ref{eq:pmaps-D-BESS-csts})- (\ref{eq:pmaps-D-PV-csts})  \big \}  , \notag
\end{align}
where $\mathcal{X}$ and $\Omega(x_t, \PVforecast_t)$ are the sets of feasible engagements $x_t$ and dispatch solutions $y_t$ for a fixed engagement $x_t$ and renewable generation point forecast $\PVforecast_t$. The optimization variables of (\ref{eq:pmaps-D-formulation}) are the engagement variables $x_t$, the dispatch variables $y_t$ (the net power at the grid connection point), $\Discharge_t$ (BESS discharging power), $\Charge_t$ (BESS charging power), $\SOC_t$ (BESS state of charge), $\BESSbinary_t$ (BESS binary variables), $\PVgeneration_t$ (renewable generation), and $d^-_t, d^+_t$ (deviation variables) (cf. the notation Section~\ref{sec:optimization-pmaps-notation}).
From (\ref{eq:pmaps-engagement-csts}), the engagement constraints are
\begin{subequations}
	\label{eq:pmaps-engagement-D-csts}	
	\begin{align}
	x_t-x_{t-1} & \leq  \Delta X, \ \forallt \setminus \{1\}  \\
	x_{t-1}-x_t & \leq   \Delta X, \ \forallt \setminus \{1\}\\
	- x_t &  \leq -  \xMin_t, \ \forallt  \\
	x_t &  \leq \xMax_t, \ \forallt.
	\end{align}
\end{subequations}
The ramping constraint on $x_1$ is deactivated to decouple consecutive days of simulation. In reality, the updated value of the last engagement of the previous day would be taken to satisfy the constraint.
The set of constraints that bound $\Charge_t$, $\Discharge_t$, and $y_t^s$ variables are $\forallt$ 
\begin{subequations}\label{eq:pmaps-D-BESS-csts}	
	\begin{align}
	\Charge_t  & \leq  \BESSbinary_t \ChargeMax  & [\phi^\text{cha}_t] \label{eq:ieee-MILP_BESS_charge_discharge_cst1}	 \\
	\Discharge_t  & \leq (1- \BESSbinary_t) \DischargeMax   & [\phi^\text{dis}_t]\label{eq:ieee-MILP_BESS_charge_discharge_cst2}	\\
	- \SOC_t & \leq -\SocMin & [\phi^{\SocMin}_t]	 \\
	\SOC_t & \leq  \SocMax , & [\phi^{\SocMax}_t] 
	\end{align}
\end{subequations}
where $\BESSbinary_t$ are binary variables that prevent the simultaneous charge and discharge of the BESS. The power balance equation and the constraints on the net power at the grid connection point are $\forallt$ 
\begin{subequations}\label{eq:pmaps-D-balance-csts}	
	\begin{align}
	y_t  &  = \PVgeneration_t + (\Discharge_t - \Charge_t )	& [\phi^y_t] \\
	- y_t&  \leq - \yMin  &  [\phi^{\yMin}_t]  \\
	y_t &  \leq \yMax .  & [\phi^{\yMax}_t] 
	\end{align}
\end{subequations}
The dynamics of the BESS state of charge are 
\begin{subequations}\label{eq:pmaps-D-soc-dyn-csts}
	\begin{align}
	\SOC_1 - \SocIni  & =  \Delta t ( \chargeEff  \Charge_1 - \frac{\Discharge_1}{\dischargeEff}   ) & [\phi^{\SocIni}] \\
	\SOC_t - \SOC_{t-1}   & = \Delta t (  \chargeEff \Charge_t - \frac{\Discharge_t}{\dischargeEff}  ) , \quad \forallt \setminus \{1\} & [\phi^{\SOC}_t] \\
	 \SOC_T & = \SocEnd = \SocIni, & [\phi^{\SocEnd}]
	\end{align}
\end{subequations}
where the parameters $\SocEnd$ and $\SocIni$ are introduced to decouple consecutive days of simulation. In reality, $\SocIni$ would be the updated value of the last measured state of charge of the previous day.
The variables $d^-_t, d^+_t $ are defined $\forallt$ to model the penalty
\begin{subequations} \label{eq:pmaps-D-deviation-csts}	
	\begin{align}
	- d^-_t & \leq \big( y_t -(x_t - p \PVcapacity) \big)  & [\phi^{d^-}_t] \\
	- d^+_t & \leq \big( (x_t + p \PVcapacity) - y_t \big), & [\phi^{d^+}_t]
	\end{align}
\end{subequations}
with $ 0\leq p \leq 1$.
Finally, the PV generation is bounded $\forallt$ by the point forecast $ \PVforecast_t$ 
\begin{align}\label{eq:pmaps-D-PV-csts}	
\PVgeneration_t & \leq  \PVforecast_t . & [\phi^{\PVgeneration}_t]
\end{align}

\subsection{Deterministic approach with perfect forecasts}

With perfect forecasts, (\ref{eq:pmaps-D-formulation}) becomes
\begin{align}\label{eq:pmaps-D-perfect-obj}
\min_{x_t \in \mathcal{X}, y_t \in \Omega(x_t, \PVforecast_t=\PVgenerationTrue_t)}& \ J_{D^\star} (x_t, y_t ),
\end{align}
with $\PVforecast_t=\PVgenerationTrue_t$ $\forallt$ in (\ref{eq:pmaps-D-PV-csts}).

\subsection{Stochastic approach}

In the stochastic formulation, the objective is given by 
\begin{align}\label{eq:pmaps-S-obj-1}	
J_S &  = \mathop{\mathbb{E}} \bigg[ \sum_{t\in \mathcal{T}}- \pi_t \Delta t  y_t +  c(x_t, y_t) \bigg] 
\end{align}
where the expectation is taken with respect to $\PVforecast_t$. Using a scenario-based approach, (\ref{eq:pmaps-S-obj-1}) is approximated by
\begin{align}\label{eq:pmaps-S-obj-2}	
J_S &  \approx  \sum_{\omega \in \Omega} \alpha_\omega  \sum_{t\in \mathcal{T}} \bigg[ - \pi_t \Delta t  y_{t,\omega} +  \pi^e (\Delta t)^2 \big ( (d_{t,\omega}^-)^2 +  (d_{t,\omega}^+)^2 \big)   \bigg],
\end{align}
with $\alpha_\omega$ the probability of scenario $\omega\in \Omega$, and $\sum_{\omega \in \Omega} \alpha_\omega = 1$. Then, the problem to solve becomes
\begin{align}\label{eq:pmaps-S-obj-3}
\min_{x_t \in \mathcal{X}, y_{t,\omega} \in \Omega(x_t, \PVforecast_{t,\omega})}& \ J_S (x_t, y_{t,\omega} ).
\end{align}
All the optimization variables but $x_t$ are now defined $\forall \omega \in \Omega$.

\subsection{Evaluation methodology}

The second step of capacity firming, \textit{i.e.}, computing the set-points in real-time, is required to assess the quality of the nomination process. However, since this study focuses on the computation of day-ahead engagements, we simulate the second step with an ideal real-time controller\footnote{Using a real-time controller with intraday forecasts is required to assess the planner-controller. However, this study focus only on the nomination step.} once the engagements are fixed. The methodology to assess the engagements consists of solving 
\begin{align}\label{eq:pmaps-eval_objective}	
J^\text{eval} &  =  \sum_{t\in \mathcal{T}}  - \pi_t  x_t  + c(x_t, y_t)
\end{align}
s.t (\ref{eq:pmaps-D-BESS-csts})-(\ref{eq:pmaps-D-PV-csts}) with $\PVforecast_t = \PVgenerationTrue_t$ in (\ref{eq:pmaps-D-PV-csts}) and given engagements $x_t$ previously computed by the planner S. The optimization variables of (\ref{eq:pmaps-eval_objective}) are $y_t$, $d_t^-$, $d_t^+$, $\PVgeneration_t$, $\Charge_t$, $\Discharge_t$, and $\SOC_t$. The optimal value of $J^\text{eval}_S$ is compared with the optimal value of $J_{D^\star}$ in (\ref{eq:pmaps-D-perfect-obj}).

\section{MiRIS microgrid case study}\label{sec:optimization-pmaps-case_study}

\begin{figure}[tb]
	\centering
	\includegraphics[width=0.6\linewidth]{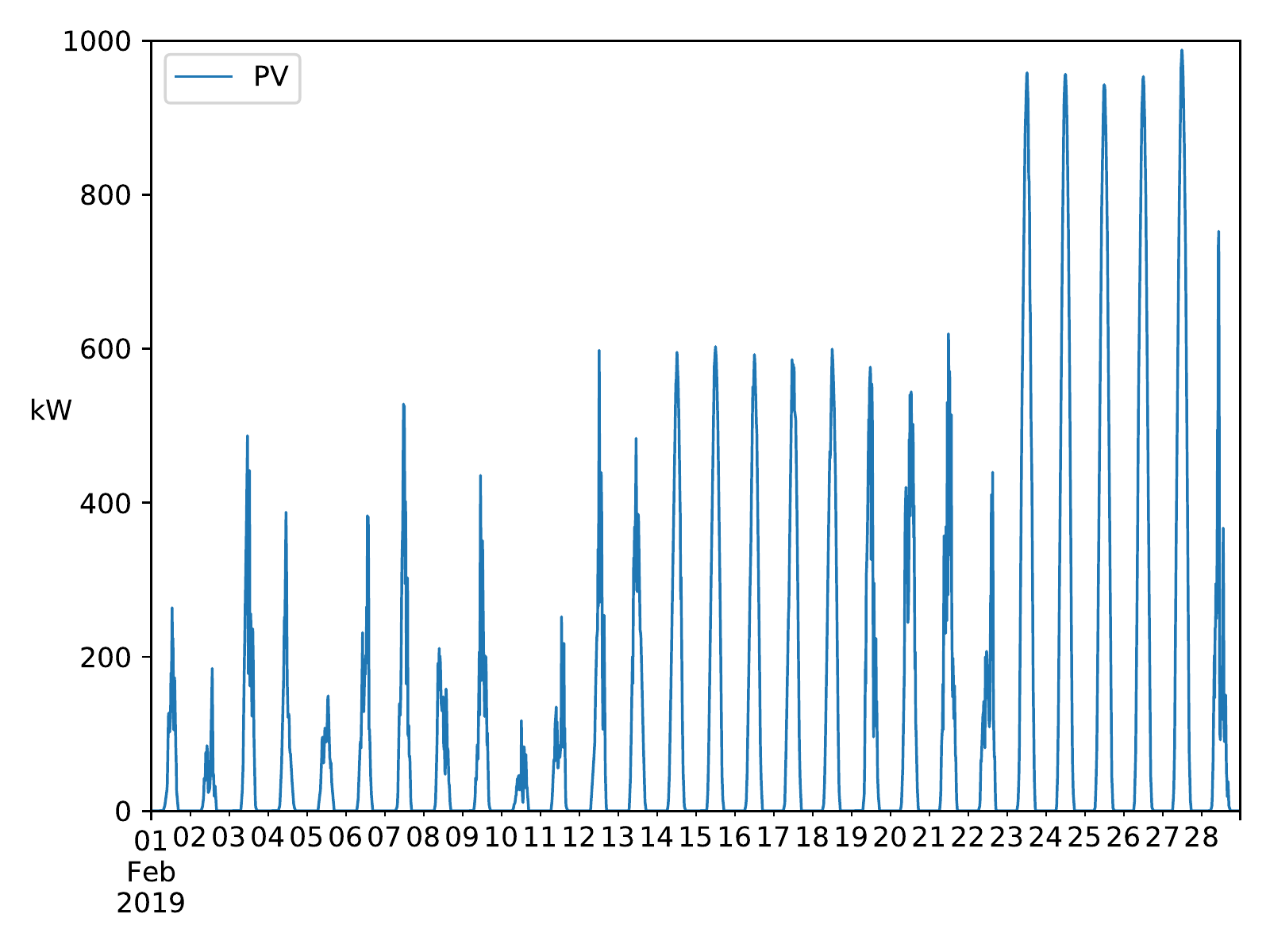}
	\caption{MiRIS February 2019 PV production.}
	\label{fig:pmaps-pv_production}
\end{figure}

The MiRIS\footnote{\url{https://johncockerill.com/fr/energy/stockage-denergie/}} microgrid case study, located at the John Cockerill Group’s international headquarters in Seraing, Belgium, is composed of a PV production plant, a BESS, and a load. For the need of this study, only historical data of PV generation are required. The BESS capacity $\maxcharge$ is 1000 kWh, $\SocMax=\maxcharge$, and the total PV installed capacity $\PVcapacity$ is 2000 kWp. The market period duration $\Delta t$ is 15 minutes. The simulation dataset $\mathcal{D}$ is the month of February 2019. Figure~\ref{fig:pmaps-pv_production} illustrates the MiRIS PV production and Table~\ref{tab:pmaps-dataset_statistics} provides some key statistics.
Table~\ref{tab:pmaps-indicators} defines the indicators used in this section.
\begin{table}[!htb]
\renewcommand{\arraystretch}{1.25}
	\centering
	\begin{tabular}{rrrrrr} \hline \hline
		$\PVcapacity$	& $ \overline{\PVgenerationTrue}$ & $\PVgenerationTrue_\text{std}$ & $\PVgenerationTrue_{\max} $ & $\PVgenerationTrue_{\%,\max} $ & $\PVgenerationTrue_\text{tot}$ \\ \hline
		2000 & 104.6  & 202.4 & 988.1 & 49.4 & 70.3   \\ \hline \hline
	\end{tabular}
	\caption{MiRIS February 2019 dataset statistics.}\label{tab:pmaps-dataset_statistics}
\end{table}
\begin{table}
\renewcommand{\arraystretch}{1.25}
	\begin{tabularx}{\linewidth}{l X l}
		\hline \hline
		Name & Description & Unit \\
		\hline
		$\overline{\PVgenerationTrue}$ & Averaged PV generation. & kW\\
		$\PVgenerationTrue_\text{std}$ & PV generation standard deviation. & kW \\
		$\PVgenerationTrue_{\max} $ & Maximum PV generation. & kW \\
		$\PVgenerationTrue_\text{tot}$ & Total PV energy produced. & MWh \\
		$\PVgenerationTrue_{\%,\max} $ & $\PVgenerationTrue_{\max} $ divided by the total installed PV capacity $\PVcapacity$. &\% \\
		$\overline{t}_\text{CPU}$ & Averaged computation time per optimization problem. & s\\
		$[X]^D $ & Total of a variable $X_t$: $ \sum_{t\in \mathcal{D}} X_t$. & $X$ unit\\
		$[\PVgeneration]^D$ & Total PV generation. & MWh\\
		$\PVgeneration_\%  $ & PV generation ratio: $\frac{[\PVgeneration]^D }{[\PVgenerationTrue]^D}$. &\% \\
		$\Charge_\% $ & BESS charge ratio: $\frac{[\Charge]^D }{[\PVgeneration]^D}$. &\% \\
		${\ChargeMax}_\% $ & Percentage of days of the dataset where the BESS achieved its maximum storage capacity. &\% \\
		$y_\% $ & Export ratio: $\frac{[y]^D }{[x]^D}$. &\% \\
		$R_{\max}$ & Maximum achievable revenue: $\pi  [\PVgenerationTrue]^D$. & k\euro \\
		$R^e$ & Gross revenue: $\pi [y]^D$. & k\euro \\
		$r^e $ & Maximum achievable revenue ratio: $ \frac{R^e}{R_{\max}}$. &\% \\
		$C^e $ & Quadratic penalty: $[c]^D$. & k\euro \\
		$R^{n,e}$ & Net revenue with quadratic penalty: $R^e - C^e$. & k\euro \\
		\hline \hline
	\end{tabularx}
		\caption{Comparison indicators.}\label{tab:pmaps-indicators}
\end{table}
It is of paramount importance to notice that the results of this case study are only valid for this dataset and cannot be extrapolated over an entire year without caution.
CPLEX\footnote{\url{https://www.ibm.com/products/ilog-cplex-optimization-studio}} 12.9 is used to solve all the optimization problems, on an Intel Core i7-8700 3.20 GHz based computer with 12 threads and 32 GB of RAM. Tables~\ref{tab:pmaps-case_study_parameters} and~\ref{tab:pmaps-battery_parameters} provide the case study and BESS parameters.
\begin{table}[tb]
\renewcommand{\arraystretch}{1.25}
	\centering
	\begin{tabular}{rrrrrrr} \hline  \hline
		$\pi$ & $\pi^e$& $\Delta t$ & $\Delta X$ & $\xMax$&  $\yMax$ & $p$\\ \hline
		0.045  & 0.0045  & 15 & 10  & 2000 & 2000 & 5\% \\ \hline   \hline
	\end{tabular}
	\caption{Case study parameters.}\label{tab:pmaps-case_study_parameters}
\end{table}
\begin{table}[tb]
\renewcommand{\arraystretch}{1.25}
	\centering
	\begin{tabular}{rrrrrrrr} \hline  \hline
		$\SocMax$ 	& $\SocMin$  & $\ChargeMax$&  $\DischargeMax$  & $\chargeEff$ & $\dischargeEff$ & $\SocIni$ & $\SocEnd$  \\ \hline
		1000 & 0 & 1000 & 1000 & 1 & 1 & 0 & 0 \\ \hline  \hline
	\end{tabular}
	\caption{BESS parameters.}\label{tab:pmaps-battery_parameters}
\end{table}

\subsection{Results for unbiased PV scenarios with fixed variance}

\begin{figure}[tb]
	\centering
	\begin{subfigure}{.4\textwidth}
	\centering
	\includegraphics[width=\linewidth]{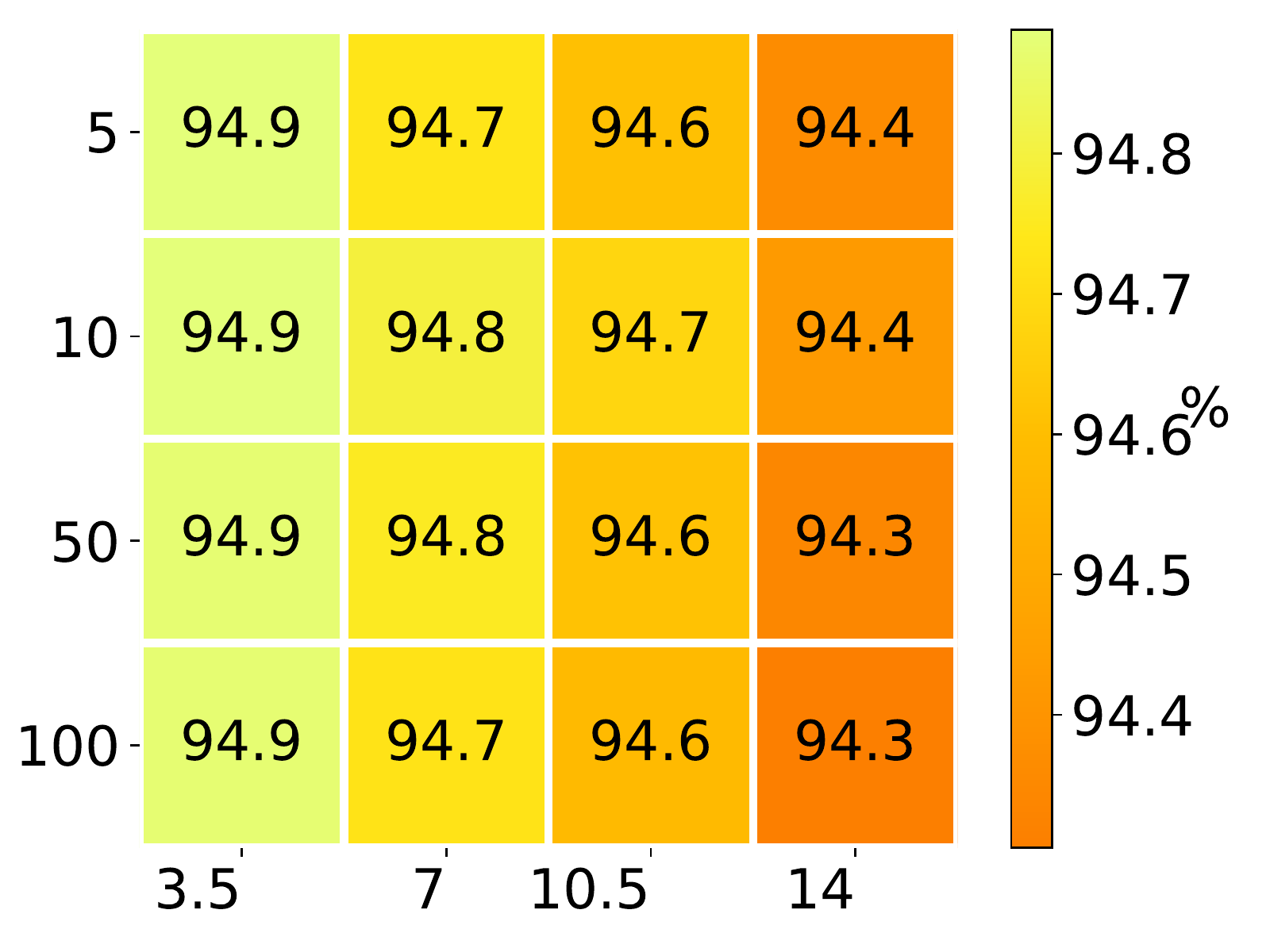}
	\caption{$\PVgeneration_\%$.}
	\end{subfigure}%
	\begin{subfigure}{.4\textwidth}
	\centering
	\includegraphics[width=\linewidth]{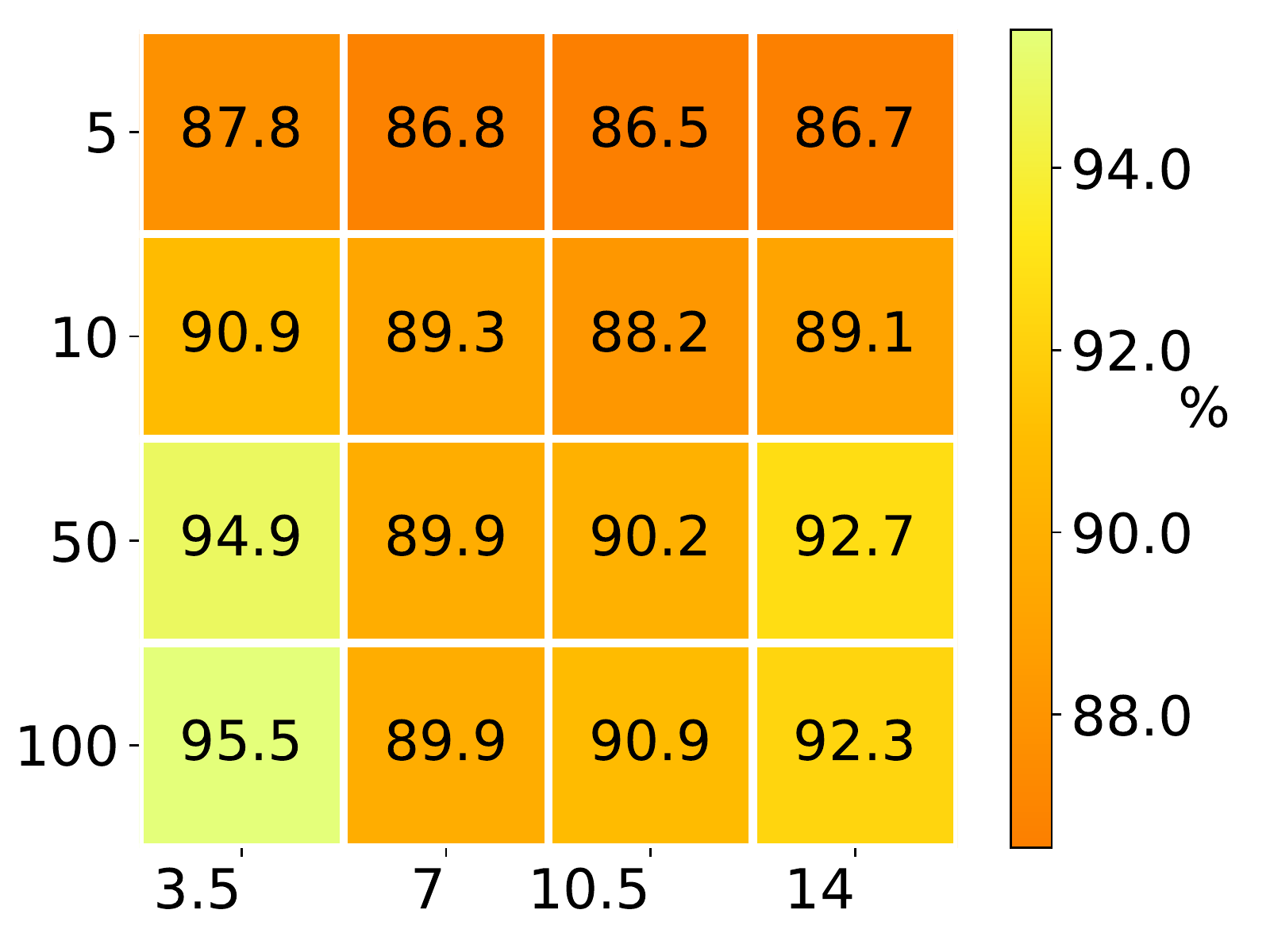}
	\caption{$y_\%$.}
	\end{subfigure}
	\caption{Planner S ratio indicator with $\#\Omega=5, 10, 50, 100$, 	$\sigma = 3.5, 7, 10.5, 14\%$.}
	\label{fig:pmaps-ratio_indicators_S}
\end{figure}

A set of unbiased PV scenarios is generated for several values of the standard deviation $\sigma$ of the prediction error. Table~\ref{tab:pmaps-scenario_generation_parameters} shows the considered values of $\sigma$, expressed as a fraction of the actual PV generation. Moreover, Table~\ref{tab:pmaps-scenario_generation_parameters} reports the cardinality of the generated scenario sets.
Table~\ref{tab:pmaps-time_computation} compares the average computation time per optimization problem between planners S and D$^\star$. Note, The optimization problem of planner S with $\#\Omega = 1$ has the same number of variables and constraints as the planner D$^\star$. The computation time is compatible with a day-ahead process even with 100 scenarios, as it takes on average 7 seconds to compute the nominations for the day-ahead. 
Table~\ref{tab:pmaps-ratio_indicators_D} and Figure~\ref{fig:pmaps-ratio_indicators_S} provide the results of the ratio indicators, respectively, for the planners D$^\star$ and S.
\begin{table}[tb]
\renewcommand{\arraystretch}{1.25}
	\centering
	\begin{tabular}{rrrrr} \hline  \hline
		$\sigma$ & 3.5\% & 7\% & 10.5\% & 14\%\\ \hline
		$\#\Omega$ & 5& 10 & 50 & 100 \\ \hline  \hline
	\end{tabular}
		\caption{Scenario generation parameters.}\label{tab:pmaps-scenario_generation_parameters}
\end{table}
\begin{table}[tb]
\renewcommand{\arraystretch}{1.25}
	\centering
	\begin{tabular}{lrrrrr} \hline  \hline
		$\#\Omega$ & 1 & 5& 10 & 50 & 100 \\ \hline
		\# variables & 769& 3 457 & 6 817  & 33 697& 67 297\\ \hline
		\# constraints& 1 248& 5 092 & 9 897  & 48 337& 96 387\\ \hline
		$\overline{t}_\text{CPU}$ S& - & 0.3 & 0.8  & 3& 7\\ \hline
		$\overline{t}_\text{CPU}$ D$^\star$&  0.1& - & -  & -& -\\ \hline \hline
	\end{tabular}
		\caption{Averaged computation times.}\label{tab:pmaps-time_computation}
\end{table}
\begin{table}[tb]
\renewcommand{\arraystretch}{1.25}
	\centering
	\begin{tabular}{rrrrr} \hline \hline
		$[\PVgeneration]^D$ & $\PVgeneration_\% $& $\Charge_\%  $ & ${\ChargeMax}_\%$ & $y_\% $	\\ \hline
		66.7 & 94.9 &  29.6  & 17.9 & 76.2 \\ \hline \hline
	\end{tabular}
		\caption{Planner D$^\star$ ratio indicators.}
	\label{tab:pmaps-ratio_indicators_D}
\end{table}

For all indicators, the results of both planners are almost equal with the smaller value of $\sigma$ and the highest value of $\#\Omega$, as expected. On average, the curtailment of PV generation equals $5\%$. The maximum $y_\% $ value is achieved with $\sigma = 3.5\%$ because the nominations are more conservative when the variance increases, leading to a smaller ratio. On average, 30\% (27\%) of the production, for planner D$^\star$ (S), is stored in the BESS over the entire dataset. $\maxcharge_\%$ is equal to 17.9\% (17.9\%\footnote{The value is the same for the $\#\Omega$ and $\sigma$ values considered.}) for the planner D$^\star$ (S), meaning the BESS reached its maximum storage level 5 days out of the 28 days of the dataset. In fact, during sunny days, the BESS is fully charged. A larger BESS capacity should decrease the curtailment and improve the gross revenue. Note: it is a winter month where the maximum generated PV power reached only half of the installed PV capacity. During a summer month, the maximum production should reach at least 80\% of the total installed capacity on sunny days. Thus, with a storage capacity of 1 MWh, the curtailment is expected to be higher during sunny summer days.

\begin{figure}[tb]
	\centering
	\begin{subfigure}{.4\textwidth}
	\centering
	\includegraphics[width=\linewidth]{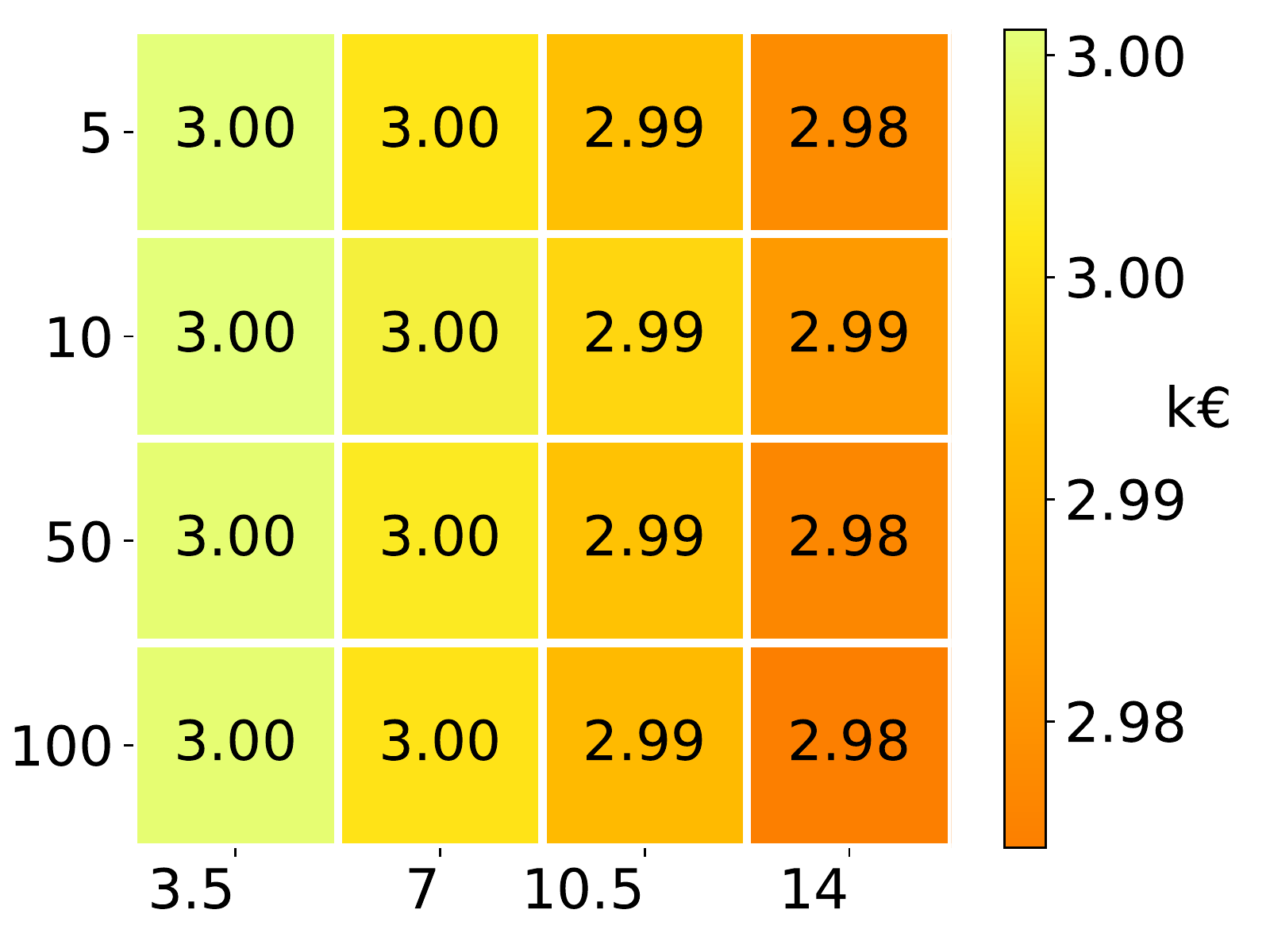}
	\caption{$R^e$.}
	\end{subfigure}%
	\begin{subfigure}{.4\textwidth}
	\centering
	\includegraphics[width=\linewidth]{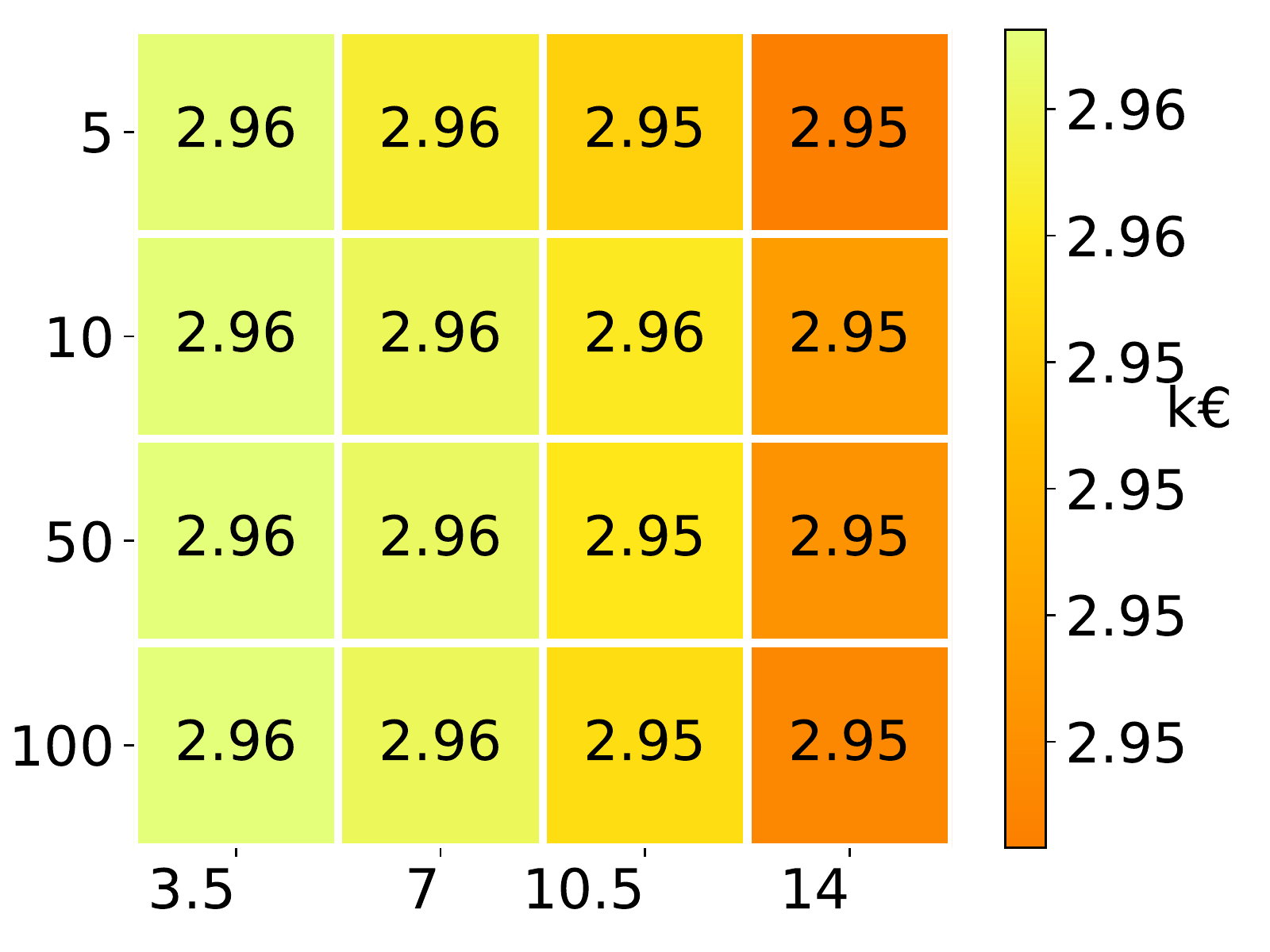}
	\caption{$R^{n,e}=-J^\text{eval}_S$.}
	\end{subfigure}
	\caption{Planner S revenue indicators with $\#\Omega=5, 10, 50, 100$, $\sigma = 3.5, 7, 10.5, 14\%$.}
	\label{fig:pmaps-rev_ind_S_bat_1000}
\end{figure}

Table~\ref{tab:pmaps-rev_ind_perfect} and Figure~\ref{fig:pmaps-rev_ind_S_bat_1000} provide the results of the revenue indicators for the planners D$^\star$ and S, respectively. It should be noted that in this case, $R^{n,e}=-J^\text{eval}_S$.
\begin{table}[tb]
\renewcommand{\arraystretch}{1.25}
	\centering
	\begin{tabular}{rrrrr} \hline \hline
	$R^e$& $r^e$ & $C^e$ & $R^{n,e}$ & $J^\text{eval}_{D^\star}$	\\ \hline
		3.0 & 94.9 & 0.04  & 2.96 & -2.96  \\ \hline \hline
	\end{tabular}
	\caption{Planner D$^\star$ revenue indicators.}
	\label{tab:pmaps-rev_ind_perfect}
\end{table}
The smallest value of the objective function is achieved by the planner D$^\star$. It is followed closely by the planner S, even for the highest value of $\sigma$. This result demonstrates the validity of the approach when exploiting an unbiased stochastic predictor. 

In terms of net revenue, both planners achieved 93.7\% of $R_{\max} \approx 3.16$ k\euro, which results in a loss of 6.3\%. Most of this loss is due to the curtailment of PV generation. For both planners, the net revenue increases with the generation.

For sunny days, the difference between the nominations and the exports is higher than the tolerance just before the production occurs, between 5 and 8 am. Indeed, the planner tends to maximize the revenue by maximizing the exports. However, the ramping power constraints (\ref{eq:pmaps-engagement-D-csts}) impose a maximum difference between two consecutive nominations. To maximize the net revenue over the entire day, the planner computes nominations that are not achievable at the beginning of the day to maximize the exports during the day. This results in a penalty between 5 and 8 am. 

\subsection{BESS capacity sensitivity analysis}

\begin{figure}[!htb]
	\centering
	\begin{subfigure}{.4\textwidth}
		\centering
		\includegraphics[width=\linewidth]{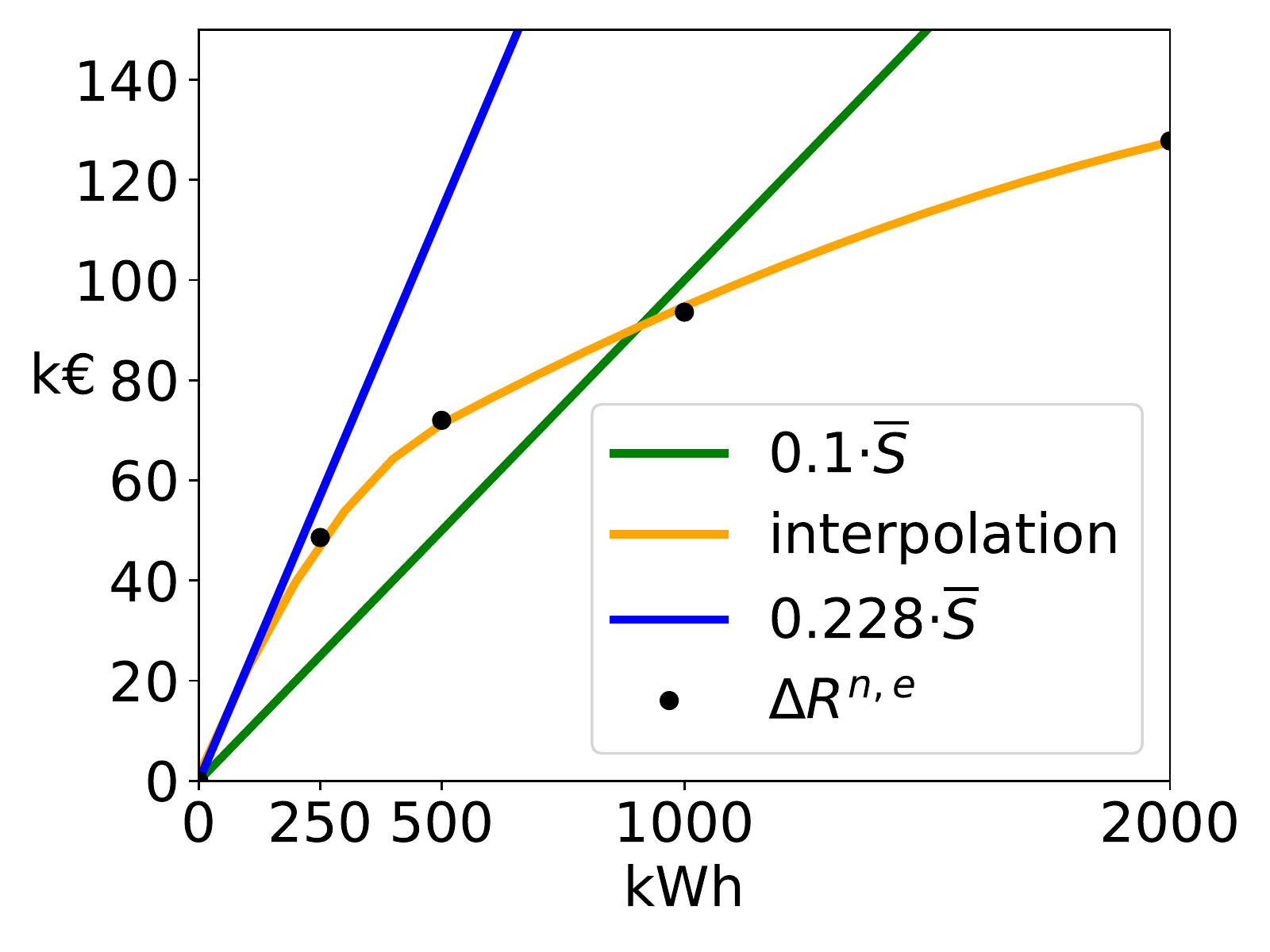}
		\caption{Variation of the net revenue.}
		\label{fig:pmaps-delta_marginal_gain}
	\end{subfigure}%
	\begin{subfigure}{.4\textwidth}
		\centering
		\includegraphics[width=\linewidth]{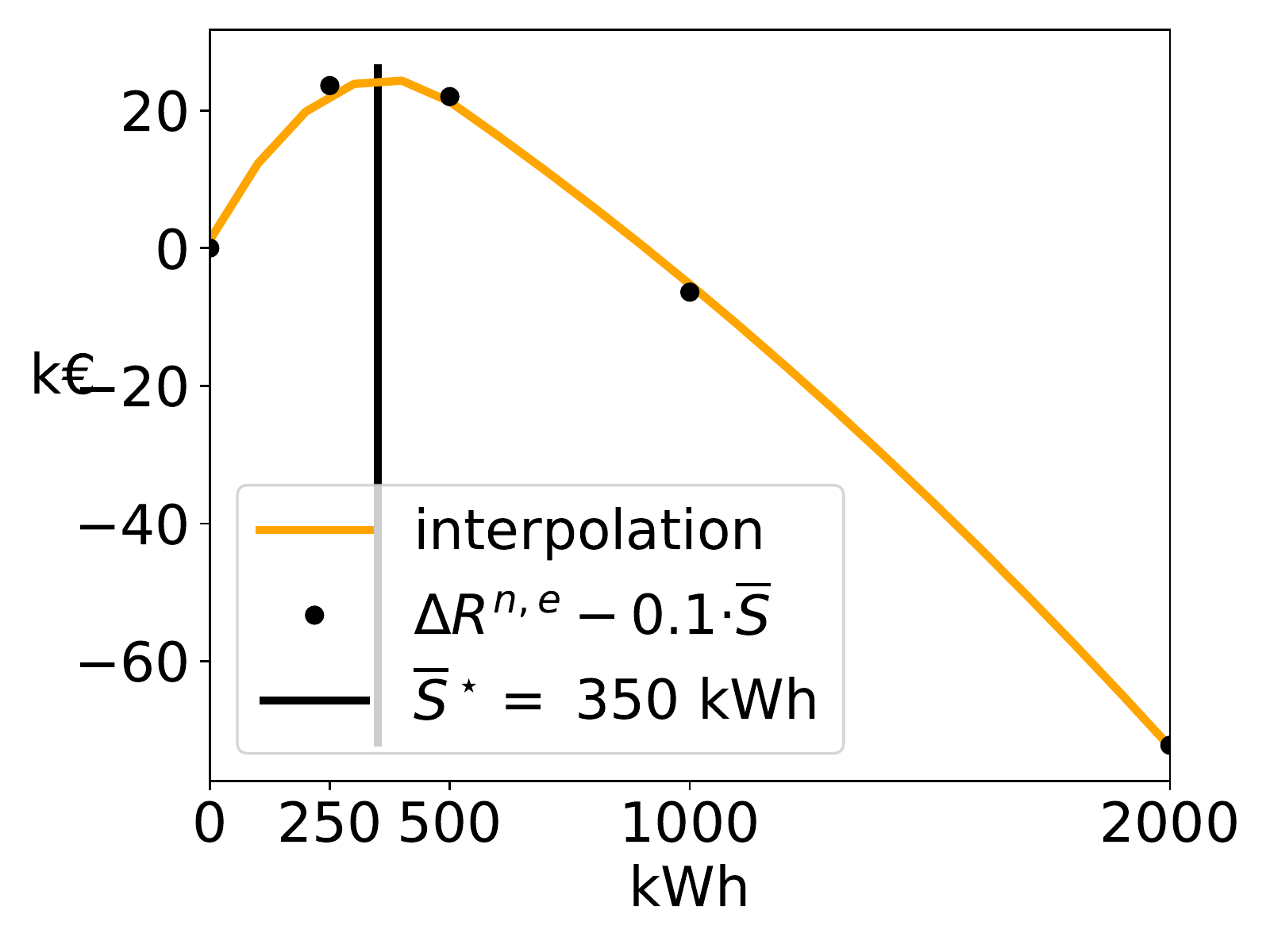}
		\caption{Optimal BESS.}
		\label{fig:pmaps-optimal_battery_size}
	\end{subfigure}
	\caption{Optimal BESS for a given CAPEX price.}
\end{figure}

The goal is to conduct a sensitivity analysis on the BESS capacity $\maxcharge$ to determine its marginal value and the optimal BESS size ${\maxcharge}^\star$ for a given CAPEX $\pi_{\maxcharge}$. The efficiencies are still assumed to be unitary. $\SocIni$ and $\SocEnd$ are set to 0 kWh. Table~\ref{tab:pmaps-bat_param_sensitivity} provides the other BESS parameters for the five cases.
The scenarios are generated using $\#\Omega=100$, and $\sigma = 3.5, 7, 10.5, 14\%$. 
\begin{table}[tb]
\renewcommand{\arraystretch}{1.25}
	\centering
	\begin{tabular}{crrrr} \hline \hline
		Case & $\maxcharge$ [kWh] 	& $\SocMin$ [kWh] & $\ChargeMax$ [kW]&  $\DischargeMax$ [kW]\\ \hline
		1	& 2000 & 0 & 2000 & 2000  \\
		2	& 1000 & 0 & 1000 & 1000  \\
		3	& 500 & 0 & 500 & 500 \\
		4	& 250 & 0 & 250 & 250 \\
		5	& 0 & 0 & 0 & 0 \\ \hline \hline
	\end{tabular}
	\caption{BESS parameters.}\label{tab:pmaps-bat_param_sensitivity}
\end{table}
A new indicator, expressed in k\euro, is defined to quantify the gain provided by the BESS over fifteen years
\begin{align}\label{eq:pmaps-price_indicator}
& \Delta R^{n,e}_i = 15 \times 12\times (R^{n,e}_i-R^{n,e}_5 ) \quad \forall i \in \{1, 2, 3, 4\}.
\end{align}
It is a lower bound of the total gain as it relies on the results of a winter month. A summer month should provide higher revenue. Table~\ref{tab:pmaps-bess_sens_cap_ratio_indic} provide the planner D$^\star$ indicators. The results demonstrate the interest in using a BESS to optimize the bidding. The larger the BESS is, the lower the curtailment is. Thus, the net revenue increases with the BESS capacity. The maximum achievable revenue is reached with a storage capacity of 2 MWh. However, the larger the BESS is, the smaller $ \Delta R^{n,e}$ increases. It means the marginal benefit decreases with the increase of BESS capacity. A trade-off should be found between the BESS capacity and its CAPEX. Figure~\ref{fig:pmaps-delta_marginal_gain} provides $ \Delta R^{n,e}$ and its quadratic interpolation in comparison with two BESS prices $\pi_{\maxcharge} = $ 0.1 and 0.228 k\euro/kWh. The value of the derivative $ [\frac{d \Delta R^{n,e}}{d \maxcharge}]_{\maxcharge=0}$ provides the maximum CAPEX that provides a profitable BESS. Then, the optimal storage capacity ${\maxcharge}^\star$ for a given CAPEX is provided solving $ [\frac{d \Delta R^{n,e}}{d \maxcharge}]_{\maxcharge} = \pi_{\maxcharge}$. For instance, with a CAPEX of 0.1 k\euro/kWh, ${\maxcharge}^\star$ is approximately 350 kWh. Figure~\ref{fig:pmaps-optimal_battery_size} provides the values of $  \Delta R^{n,e} - \pi_{\maxcharge} \maxcharge$ with a quadratic interpolation.

\begin{figure}[!htb]
	\centering
	\begin{subfigure}{.4\textwidth}
		\centering
		\includegraphics[width=\linewidth]{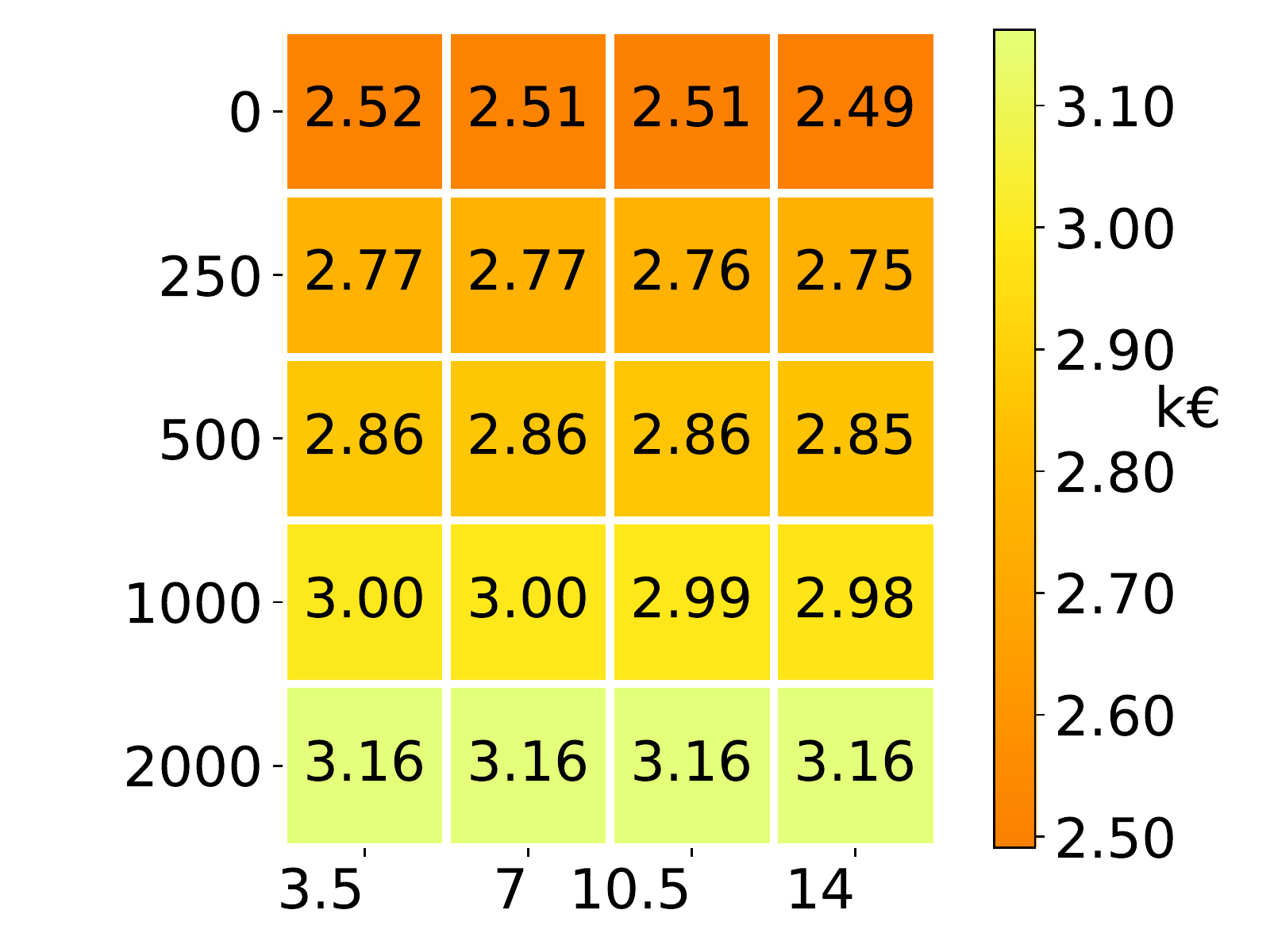}
		\caption{$R^e$.}
	\end{subfigure}%
	\begin{subfigure}{.4\textwidth}
		\centering
		\includegraphics[width=\linewidth]{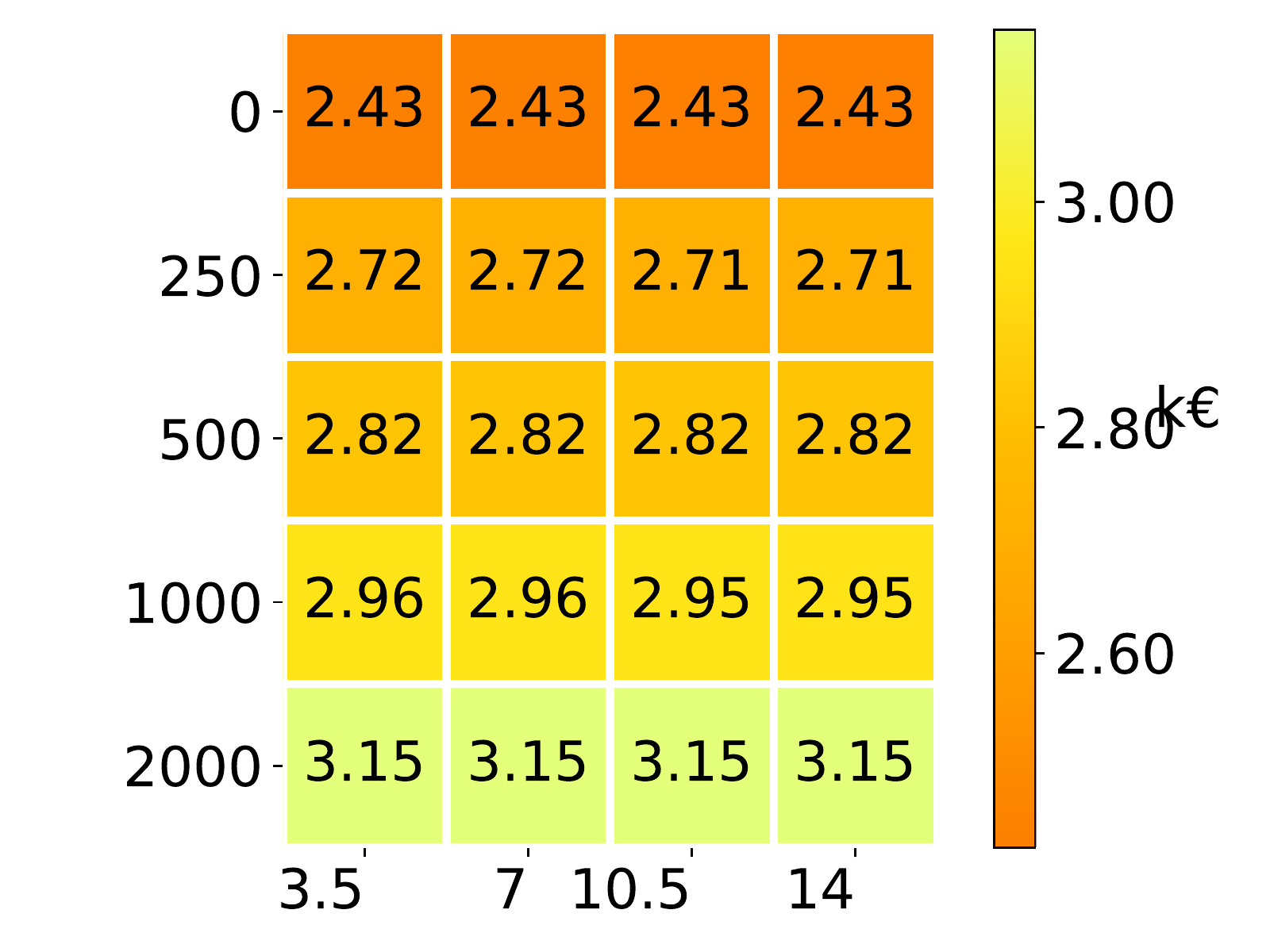}
		\caption{$R^{n,e}=-J^\text{eval}_S$.}
	\end{subfigure}
	\caption{Planner S revenue indicators BESS capacity sensitivity analysis with $\#\Omega=100$, $\sigma = 3.5, 7, 10.5, 14\%$.}
	\label{fig:pmaps-rev_indic_S_bess_sensitivity}
\end{figure}

Figure~\ref{fig:pmaps-rev_indic_S_bess_sensitivity} provides the planner S revenue indicators. The results are still almost identical for all indicators for the smallest value of $\sigma$ and very close with the highest one, as expected.   
\begin{table}[tb]
\renewcommand{\arraystretch}{1.25}
	\centering
	\begin{tabular}{crrrrr} \hline \hline
		Case   & $[\PVgeneration]^D$ & 	$\PVgeneration_\% $& $\Charge_\%  $ & ${\ChargeMax}_\%$ & $y_\% $	\\ \hline
		1 &   70.3 & 	100  & 45.6 & 17.9 & 77.7 \\ 
		2 &   66.7 &	94.9 & 29.6 & 17.9 & 76.2 \\ 
		3 &   63.6 & 	90.5 & 17.3 & 39.3 & 76.4 \\ 
		4 &   61.7 & 	87.7 & 11.1 & 46.4 & 75.2 \\ 
		5 &   56.0 & 	79.6 & -    & -    & 70.4 \\ \hline
		Case   & $R^e$& $C^e$  & $R^{n,e}$ & $J^\text{eval}_{D^\star}$	& $\Delta R^{n,e}$\\ \hline
		1 &  3.16 & 0.01  & 3.15 & -3.15 & 128\\
		2 &  3.0  & 0.04  & 2.96 & -2.96 & 94\\
		3 &  2.86 & 0.04  & 2.84 & -2.84 & 72\\
		4 &  2.77 & 0.06  & 2.71 & -2.71 & 49\\
		5 &  2.52 & 0.08  & 2.44 & -2.44 & 0 \\\hline \hline
	\end{tabular}
	\caption{Planner D$^\star$ ratio and revenue indicators BESS capacity sensitivity analysis.}
	\label{tab:pmaps-bess_sens_cap_ratio_indic}
\end{table}
%
%
%
%

\clearpage

\section{Conclusions and perspectives}\label{sec:optimization-pmaps-conclusions}

This Chapter addresses the energy management of a grid-connected PV plant coupled with a BESS within the capacity firming framework. The method is composed of two steps: computing the day-ahead nominations, then computing the renominations and the set-points in real-time to minimize the energy and ramp power deviations from nominations. This study investigates the first step by comparing a stochastic and a deterministic formulation. The main goal is to validate the stochastic approach by using an ideal predictor providing unbiased PV scenarios. 

The results of the stochastic planner are comparable with those of the deterministic planner, even when the prediction error variance is non-negligible. Finally, the BESS capacity sensitivity analysis results demonstrate the advantage of using a BESS to optimize the bidding day-ahead strategy. However, a trade-off must be found between the marginal gain provided by the BESS and its investment and operational costs.

Several extensions of this work are under investigation.
The first is to assess the planner's behavior better using a full year of data. Then, the next challenge is to use a more realistic methodology to generate PV generation scenarios. Several scenario generation approaches could be investigated, based on a point forecast model such as the PVUSA model \citep{dows1995pvusa,bianchini2013model,bianchini2020estimation}, combined with Gaussian copula \cite{papaefthymiou2008using,pinson2012evaluating,golestaneh2016generation}. Alternatively, using the deep generative models introduced in Chapter \ref{chap:scenarios-forecasting}.
Another challenge is to consider the non-convex penalty function specified by the CRE into the objective.
Finally, the last challenge is to investigate the second step of the capacity firming problem, for instance, by adapting the approach implemented in \citet{dumas2021coordination}. 

\section{Appendix: PV scenario generation}\label{appendix:optimization-pmaps-scenario}

This Annex describes the methodology to generate the set of unbiased PV scenarios. The goal is to define an ideal unbiased predictor with a fixed variance over all lead times. In this section, let $t$ be the current time index, $k$ be the lead time of the prediction, $K$ be the maximum lead time of the prediction, $y_{t+k}$ be the true value of the signal $y$ at time $t + k$, and $\hat{y}_{t+k|t}$ be the value of $y_{t+k}$ predicted at time $t$. The forecasts are computed at 4 pm (nominations deadline) for the day-ahead. With a market period duration of fifteen minutes, $K$ is equal to 128. The PV forecasts are needed for lead times from $k=33$ (00:00 to 00:15 am) to $k=K=128$ (11:45 to 12:00 pm). Then, $\hat{y}_{t+k|t}$ and $y_{t+k}$ are assumed to be related by
\begin{align}\label{eq:prediction_error_definition}	
\hat{y}_{t+k|t}  & =  y_{t+k} (1+\epsilon_k).
\end{align}
The error term $\epsilon_k$ is generated by the moving-average model \citep[Chapter 3]{box2015time}
\begin{subequations}
	\begin{align}\label{eq:epsilon_definition}	
	\epsilon_1 & =  \eta_1 \\ 
	\epsilon_k & =  \eta_k + \sum_{i=1}^{k-1} \alpha_i \eta_{k-i} \quad \forall k \in \{2, ..., K\}
	\end{align}
\end{subequations}
with $\{ \alpha_i \}_{i=1}^{K-1} $ scalar coefficients, $\{\eta_k \}_{k=1}^K $ independent and identically distributed sequences of random variables from a normal distribution $\mathcal{N}(0, \sigma)$. Thus, the variance of the error term is
\begin{subequations}
	\begin{align}\label{eq:epsilon_var_definition}	
	& \mathrm{Var} (\epsilon_1)  =  \sigma^2 \\ 
	& \mathrm{Var} (\epsilon_k)  =  \big( 1 + \sum_{i=1}^{k-1} \alpha_i^2 \big) \sigma^2 \quad \forall k \in \{2, ..., K\}.
	\end{align}
\end{subequations}
It is possible to simulate with this model an increase of the prediction error variance with the lead time $k$ by choosing
\begin{align}\label{eq:alpha_definition}	
\alpha_i & =  p^i \quad \forall i \in \{1, ..., K-1\}.
\end{align}
(\ref{eq:epsilon_var_definition}) becomes, $\forall k \in \{1, ..., K\}$
\begin{align}\label{eq:epsilon_definition_2}	
\mathrm{Var} (\epsilon_k) &  = \sigma^2 A_{\epsilon_k} 
\end{align}
with $A_{\epsilon_k}$ defined $\forall k \in \{1, ..., K\}$ by
\begin{align}\label{eq:Ak_definition}	
A_{\epsilon_k} &  =  \sum_{i=0}^{k-1} (p^2)^i = \frac{1-(p^2)^{k}}{1-p^2}.
\end{align}
Then, with $0 \leq p < 1 $, it is possible to make the prediction error variance independent of the lead time as it increases. Indeed:
\begin{align}\label{eq:Ak_limit}	
\lim_{k \to\infty}  A_{\epsilon_k} & = A_\infty = \frac{1}{1-p^2}.
\end{align}
For instance, with $p=0.9$ and $K=128$, for $k \geq 33$, $A_{\epsilon_k}\approx A_\infty$ that is approximately 5.26. Thus, $\forall k \in \{33, ..., K\}$
\begin{align}\label{eq:epsilon_variance_value}	
& \mathrm{Var} (\epsilon_k)  \approx 5.26  \sigma^2.
\end{align}
Finally, the $\sigma$ value to set a maximum $\epsilon_\text{max}$ with a high probability of 0.997, corresponding to a three standard deviation confidence interval from a normal distribution, is found by imposing $\epsilon_\text{max} = 3 \sqrt{\mathrm{Var} (\epsilon_K)}$:
\begin{align}\label{eq:epsilon_max_value}	
\sigma & \approx \frac{\epsilon_\text{max}}{3\sqrt{A_\infty}},
\end{align}
with $\epsilon_\text{max} = 0.25, 0.50, 0.75, 1.0$, $\sigma = 3.5, 7, 10.5, 14\%$.


\chapter{Capacity firming sizing}\label{chap:capacity-firming-sizing}

\begin{infobox}{Overview}
This Chapter proposes an approach to size a grid-connected renewable generation plant coupled with a battery energy storage device in the capacity firming market.
The main novelties in the CRE capacity framework context are three-fold.
\begin{enumerate}
	\item First, a MIQP formulation is proposed to address the planning stage of the two-phase engagement control that is compatible with a scenario approach to approximate the Mixed-Integer Non-Linear Programming problem generated by the CRE non-convex penalty function. It is compared to the deterministic formulation using perfect knowledge of the future and PV point forecasts on empirical data from the PV production monitored on-site at the Li\`ege University (ULi\`ege), Belgium.
	\item Second, a transparent and easily reproducible Gaussian copula methodology is implemented to generate PV scenarios based on the parametric PVUSA model using a regional climate model.
	\item Finally, the sizing of the system is addressed by a grid search to approximate the optimal sizing for a given selling price using both the deterministic and stochastic approaches. 
\end{enumerate}

\textbf{\textcolor{RoyalBlue}{References:}} This chapter  is an adapted version of the following publication: \\[2mm]\bibentry{dumas2020probabilistic}.
\end{infobox}
\epi{Each player must accept the cards life deals him or her: but once they are in hand, he or she alone must decide how to play the cards in order to win the game.}{Voltaire}
\begin{figure}[htbp]
	\centering
	\includegraphics[width=1\linewidth]{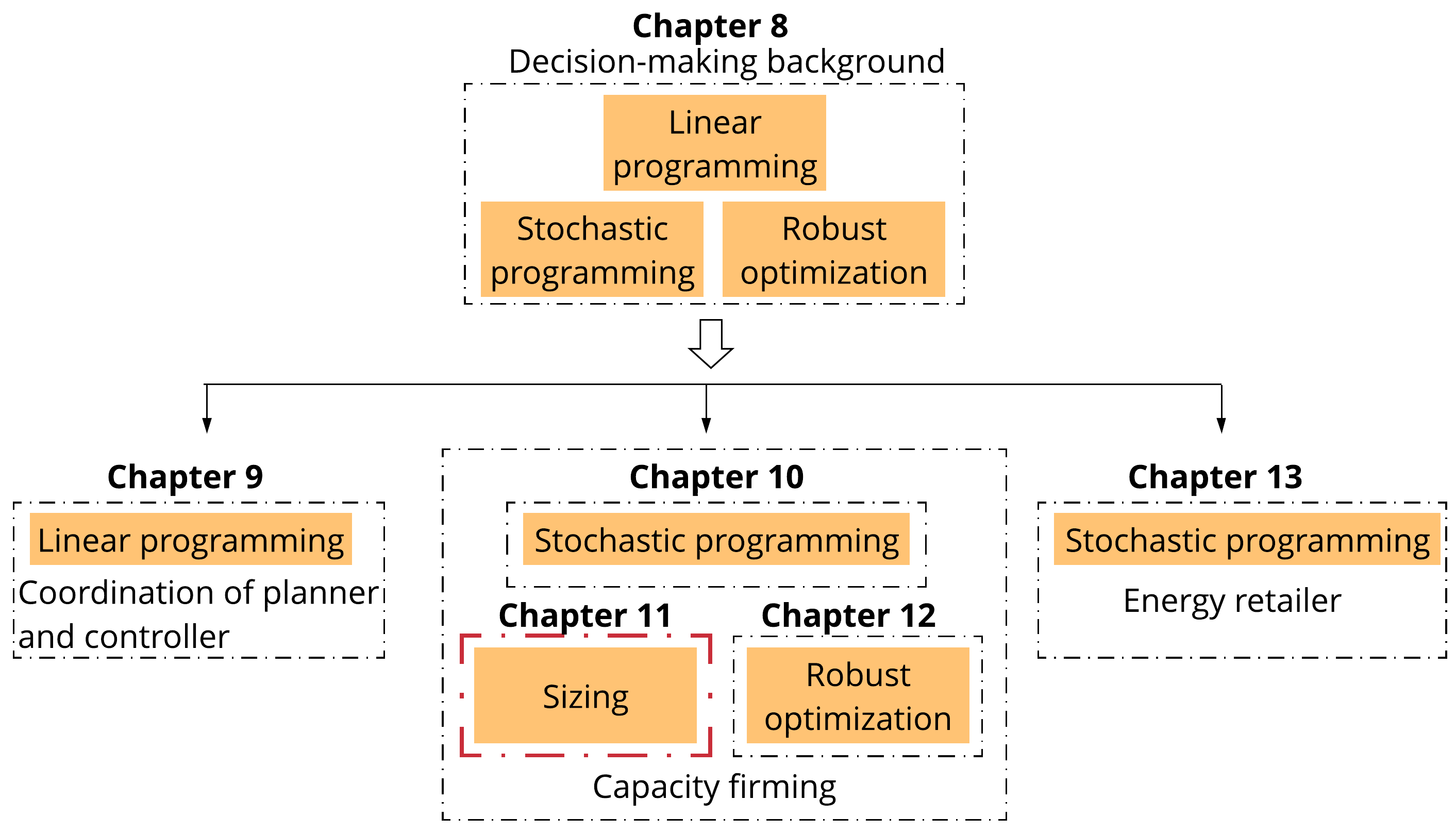}
	\caption{Chapter \ref{chap:capacity-firming-sizing} position in Part \ref{part:optimization}.}
\end{figure}
\clearpage

The sizing has only been investigated by \citet{haessig2014dimensionnement} in a similar context but with a wind farm. The main thread motivating the contribution of this Chapter is to extend Chapter \ref{chap:capacity-firming-stochastic} to address the sizing of a grid-connected PV plant and a BESS subject to grid constraints in this context. 
However, the sizing problem is difficult due to this two-phase engagement control with a day-ahead nomination and an intraday control to minimize deviations from the planning. A single sizing optimization problem would result in a non-linear bilevel optimization problem with an upper level, the sizing part, and a lower level, the two-phase engagement control. Thus, a grid search is conducted to approximate the optimal sizing for a given selling price using both the deterministic and stochastic approaches. 
In the two-phase engagement control, the PV uncertainty is considered at the planning stage by considering a stochastic approach that uses PV scenarios generated by a Gaussian copula methodology. The planner determines the engagement profile on a day-ahead basis given a selling price profile, the PV scenarios, and the system's current state, including the battery state of charge. Then, the engagement plan is sent to the grid operator and to the controller that computes every 15 minutes the production, \textit{i.e.}, injection or withdrawal, and set-points, \textit{e.g.}, BESS charge or discharge, until the end of the day. The optimization problems are formulated as Mixed-Integer Quadratic Problems (MIQP) with linear constraints. 

The remainder of this Chapter is organized as follows. Section \ref{sec:optimization-powertech-formulation} defines the problem statement. Section \ref{sec:optimization-powertech-forecasting} provides the PVUSA parametric point forecast model, and the Gaussian Copula approach to generate PV scenarios. Section \ref{sec:optimization-powertech-sizing} investigates the system sizing using both the deterministic and stochastic MIQP approaches. Finally, Section \ref{sec:optimization-powertech-conclusions} summarizes the main findings and highlights ideas for further work. Note, the capacity firming process is described in Section \ref{sec:optimization-pmaps-capacity-firming-process} of Chapter \ref{chap:capacity-firming-stochastic}.

\section{Problem statement}\label{sec:optimization-powertech-formulation}

In this Chapter, the problem formulation is almost strictly identical to Chapter \ref{chap:capacity-firming-stochastic}. The only difference lies in the definition (\ref{eq:pmaps-penalty}) of the penalty $c$. The penalty defined in the CRE specifications of the tender AO-CRE-ZNI 2019 published on $\CREspecifications$ is adopted to conduct the sizing. The optimization variables and the parameters are defined in Section~\ref{sec:optimization-pmaps-notation}.

\subsection{Stochastic approach}

A stochastic planner with a MIQP formulation and linear constraints is implemented using a scenario-based approach. The planner computes on a day-ahead basis the engagement plan $x_t, \ \forallt$ to be sent to the grid. The problem formulation is given by (\ref{eq:pmaps-S-obj-3}), where only short deviations are considered with the penalty defined in the CRE specifications of the tender AO-CRE-ZNI 2019 published on $\CREspecifications$. In compact form, the optimization problem is
\begin{subequations}
\label{eq:powertech-S-formulation}	
\begin{align}
\min  & \sum_{\omega \in \Omega} \alpha_\omega  \sum_{t\in \mathcal{T}} \bigg[ - \Delta t \pi_t  y_{t,\omega} +  \frac{\Delta t \pi_t }{\PVcapacity} d^-_{t, \omega}  \big( d^-_{t, \omega}  + 4 p\PVcapacity\big)   \bigg] \quad \text{s.t.:}  \\
x_t-x_{t-1} & \leq  \Delta X_t,  \forallt \setminus \{1\} \label{eq:S_engagement_cst1}\\
x_{t-1}-x_t & \leq   \Delta X_t,  \forallt \setminus \{1\} \\
x_t &  \leq \xMax_t,  \forallt \\
- x_t &  \leq -  \xMin_t,  \forallt \label{eq:S_engagement_cst2}\\
- d^-_{t, \omega} & \leq   y_{t, \omega} -(x_t - p\PVcapacity)   , \forallt \label{eq:S_underprod_cst}	\\
y_{t, \omega} &  \leq (x_t + p\PVcapacity )  , \forallt \label{eq:S_overprod_cst}	 \\
\PVgeneration_{t,\omega} & \leq \PVforecast_{t,\omega}, \forallt \label{eq:S_action_cst1}		\\
\Charge_{t,\omega} & \leq \BESSbinary_{t,\omega} \ChargeMax , \forallt \\
\Discharge_{t,\omega} &\leq (1-\BESSbinary_{t,\omega})\DischargeMax  , \forallt  \\
- \SOC_{t,\omega} & \leq -\SocMin , \forallt\\
\SOC_{t,\omega} & \leq \SocMax , \forallt \label{eq:S_action_cst2}	\\
y_{t,\omega} &  = \PVgeneration_{t,\omega} + \Discharge_{t,\omega} - \Charge_{t,\omega}  , \forallt \label{eq:S_energy_flows}	\\
y_{t,\omega} &  \leq \yMax_t , \forallt \label{eq:S_production1}\\
- y_{t,\omega}&  \leq - \yMin_t , \forallt \label{eq:S_production2}\\
\SOC_{1,\omega} - \SocIni & =  \Delta t (  \chargeEff  \Charge_{1,\omega} - \frac{\Discharge_{1,\omega}}{\dischargeEff}   )  \label{eq:S_soc_dynamic1}\\
\SOC_{t,\omega} - \SOC_{t-1, \omega} & =  \Delta t (  \chargeEff  \Charge_{t,\omega} - \frac{\Discharge_{t,\omega}}{\dischargeEff}   ),  \forallt \setminus \{1\}\\
\SOC_{T,\omega} & = \SocEnd = \SocIni. \label{eq:S_soc_dynamic2}
\end{align}
\end{subequations}
The optimization variables are $x_t$ (engagement at the coupling point), $y_{t,\omega}$ (net power at the coupling point), $\forallw$, $d^-_{t,\omega}$ (underproduction), $\PVgeneration_{t,\omega}$ (PV generation), $\BESSbinary_{t,\omega}$ (BESS binary variable), $\Charge_{t,\omega}$ (BESS charging power), $\Discharge_{t,\omega}$ (BESS discharging power), and $\SOC_{t,\omega}$ (BESS state of charge) (cf. the notation Section~\ref{sec:optimization-pmaps-notation}).
The engagement constraints are (\ref{eq:S_engagement_cst1})-(\ref{eq:S_engagement_cst2}), where the ramping constraint on $x_1$ is deactivated to decouple consecutive days of simulation. 
The CRE non-convex piecewise quadratic penalty function is modeled by the constraints (\ref{eq:S_underprod_cst}) $\forallw$, that defines the variables $d^-_{t, \omega}  \in \mathbb{R}_+$ to model the quadratic penalty for underproduction, and (\ref{eq:S_overprod_cst}) $\forallw$ forbidding overproduction that is non-optimal as curtailment is allowed \cite{n2019optimal}.
The set of constraints that bound $\PVgeneration_{t,\omega}$, $\Charge_{t,\omega}$, $\Discharge_{t,\omega}$, and $\SOC_{t,\omega}$ variables are (\ref{eq:S_action_cst1})-(\ref{eq:S_action_cst2}) $\forallw$ where $\PVforecast_{t,\omega}$ are PV scenarios, and $\BESSbinary_{t,\omega}$ are binary variables that prevent the BESS from charging and discharging simultaneously. The power balance equation and the production constraints are (\ref{eq:S_energy_flows}) and (\ref{eq:S_production1})-(\ref{eq:S_production2}) $\forallw$. 
The dynamics of the BESS state of charge is provided by constraints (\ref{eq:S_soc_dynamic1})-(\ref{eq:S_soc_dynamic2}) $\forallw$. Note, the parameters $\SocEnd$ and $\SocIni$ are introduced to decouple consecutive days of simulation.

\subsection{Deterministic approach}
The deterministic (D) formulation of the planner is a specific case of the stochastic formulation by considering only one scenario where $\PVgeneration_{t,\omega}$ become $\PVforecast_t$, PV point forecasts. The deterministic formulation with perfect forecasts ($D^\star$) is D with $\PVforecast_t = \PVgenerationTrue_t$ $\forallt$. For both the deterministic planners D and $D^\star$, the optimization variables are $x_t$, $y_t$, $d^-_t$, $\PVgeneration_t$, $b_t$, $\Charge_t$, $\Discharge_t$, and $\SOC_t$.

\subsection{Oracle controller}
The oracle controller is an ideal real-time controller that assumes perfect knowledge of the future by using as inputs the engagement profile to compute the set-points, maximize the revenues and minimize the deviations from the engagements. The oracle controller is $D^\star$ where the engagements are parameters.

\section{Forecasting methodology}\label{sec:optimization-powertech-forecasting}

The Gaussian copula approach has been widely used to generate wind and PV scenarios in power systems \citep{pinson2009probabilistic,golestaneh2016generation}. However, to the best of our knowledge, there is almost no guidance available on which copula family can describe correlated variations in PV generation \citep{golestaneh2016generation}, hence the Gaussian copula family is selected instead of copulas like Archimedean or Elliptical.

\subsection{Gaussian copula-based PV scenarios}

In this section, let $t$ be the current time index, $k$ be the lead time of the prediction, $Z = \{Z_1, ..., Z_T \}$ a multivariate random variable, $F_{Z_k}(\cdot)$, $k = 1, ..., T $ the marginal cumulative distribution functions, and $R_Z$ the correlation matrix. The goal is to generate samples of $Z$. The Gaussian copula methodology consists of generating a trajectory $u =[u_1, \ldots, u_T]$ from the multivariate Normal distribution $\mathcal{N}(0,R_Z)$. Then, to transform each entry $u_k$ through the standard normal cumulative distribution function $\phi(\cdot)$: $\tilde{u}_k = \phi(u_k)$, $k = 1, ..., T $, and finally, to apply to each entry $\tilde{u}_k$ the inverse marginal cumulative distribution function of $Z_k$: $z_k = F_{Z_k}^{-1}(\tilde{u}_k)$, $k = 1, ..., T $.
In our case, $Z$ is defined as the error between the PV measurement $x$ and the PV point forecast $\PVforecast$
\begin{align}\label{eq:Z_definition}
Z_k = \PVgenerationTrue_{t+k} - \PVforecast_{t+k|t} \quad k = 1, ..., T .
\end{align}
$\hat{F}_{Z_k}(\cdot)$ and $\hat{R}_Z$ are estimated from the data, and following the methodology described above, a PV scenario $i$ at time $t$ is generated for time $t+k$
\begin{align}\label{eq:PV_scenario_generation}
\hat{y}^{\text{pv}, (i)}_{t+k|t} = \PVforecast_{t+k|t} + z_k^i \quad k = 1, ..., T .
\end{align}
The PV point forecasts $\PVforecast_{t+k|t}$ are computed by using the PVUSA model presented in the following Section.

\subsection{PV point forecast parametric model}

A PV plant can be modeled using the well-known PVUSA parametric model \citep{dows1995pvusa}, which expresses the instantaneous generated power as a function of irradiance and air temperature
\begin{align}\label{eq:PVUSA}	
\PVgenerationTrue_t &  = aI_t + bI^2_t + cI_tT_t ,
\end{align}
where $\PVgenerationTrue_t$, $I_t$ and $T_t$ are the generated power, irradiance and air temperature at time $t$, respectively, and $a > 0$, $b < 0$, $c < 0$ are the PVUSA model parameters. These parameters are estimated following the algorithm of \citet{bianchini2013model} that efficiently exploits only the power generation measurements and the theoretical clear-sky irradiance and is characterized by lightweight computational effort. The same implementation of the algorithm is used with a sliding window of 12 hours. The parameters reached the steady-state values in 50 days on the Uli\`ege case study, described in the following Section, with
\begin{align}\label{eq:theta_uliege_values}	
[\tilde{a}, \tilde{b}, \tilde{c}]^\intercal &  = [0.573, -7.68 \cdot 10^{-5}, -1.86 \cdot 10^{-3}]^\intercal .
\end{align}
%
The weather hindcasts from the MAR regional climate model \citep{fettweis2017reconstructions}, provided by the Laboratory of Climatology of the Li\`ege University, are used as inputs of the parametric PVUSA model. The ERA5\footnote{The fifth-generation European Centre for Medium-Range Weather Forecasts atmospheric reanalysis of the global climate.} reanalysis database forces the MAR regional climate to produce weather hindcasts. Finally, the PV point forecasts are computed 
\begin{align}	
\PVforecast_{t+k|t} &  = \tilde{a}\hat{I}_{t+k|t} + \tilde{b}\hat{I}^2_{t+k|t} + \tilde{c}\hat{I}_{t+k|t} \hat{T}_{t+k|t} , \quad  \forall k = k_1, \ldots, k_T,
\end{align}%
and use as inputs to generate PV scenarios following the Gaussian copula approach.

\subsection{PV scenarios}

\begin{figure}[!htb]
	\centering
	\begin{subfigure}{0.4\textwidth}
		\centering
		\includegraphics[width=\linewidth]{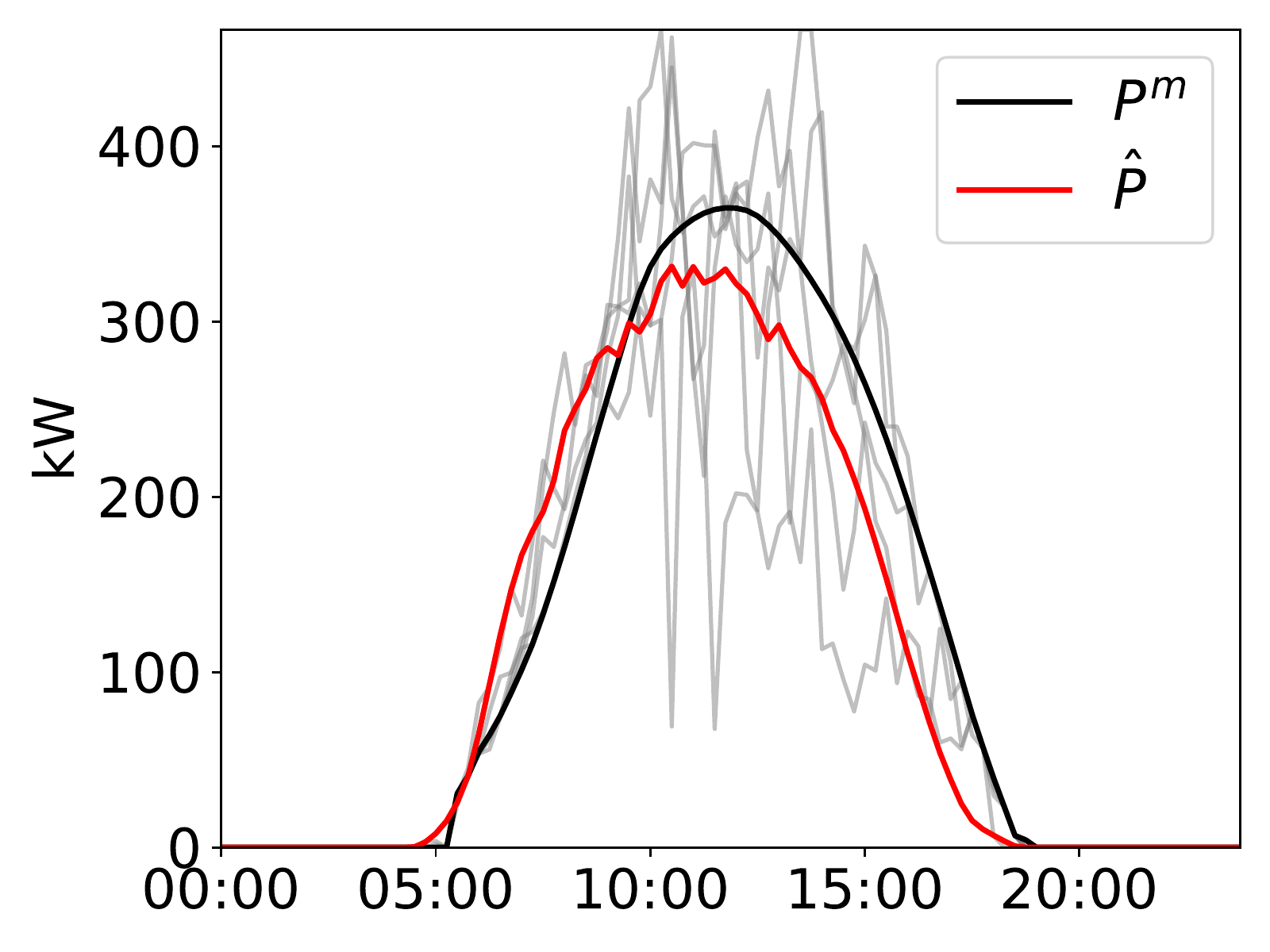}
		\caption{$\scenariocopuladateone$.}
	\end{subfigure}%
	\begin{subfigure}{0.4\textwidth}
		\centering
		\includegraphics[width=\linewidth]{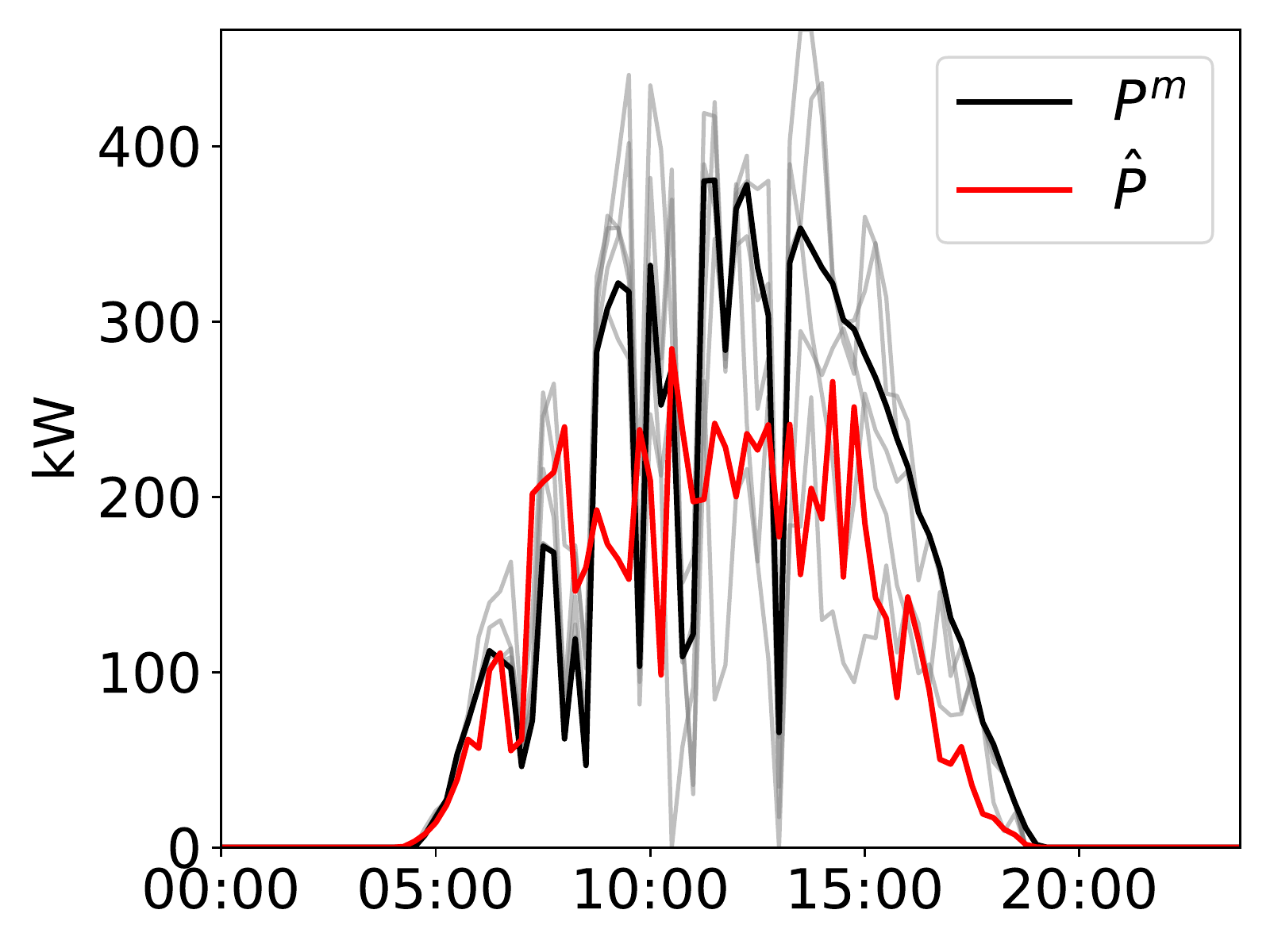}
		\caption{$\scenariocopuladatetwo$.}
	\end{subfigure}
	\begin{subfigure}{0.4\textwidth}
		\centering
		\includegraphics[width=\linewidth]{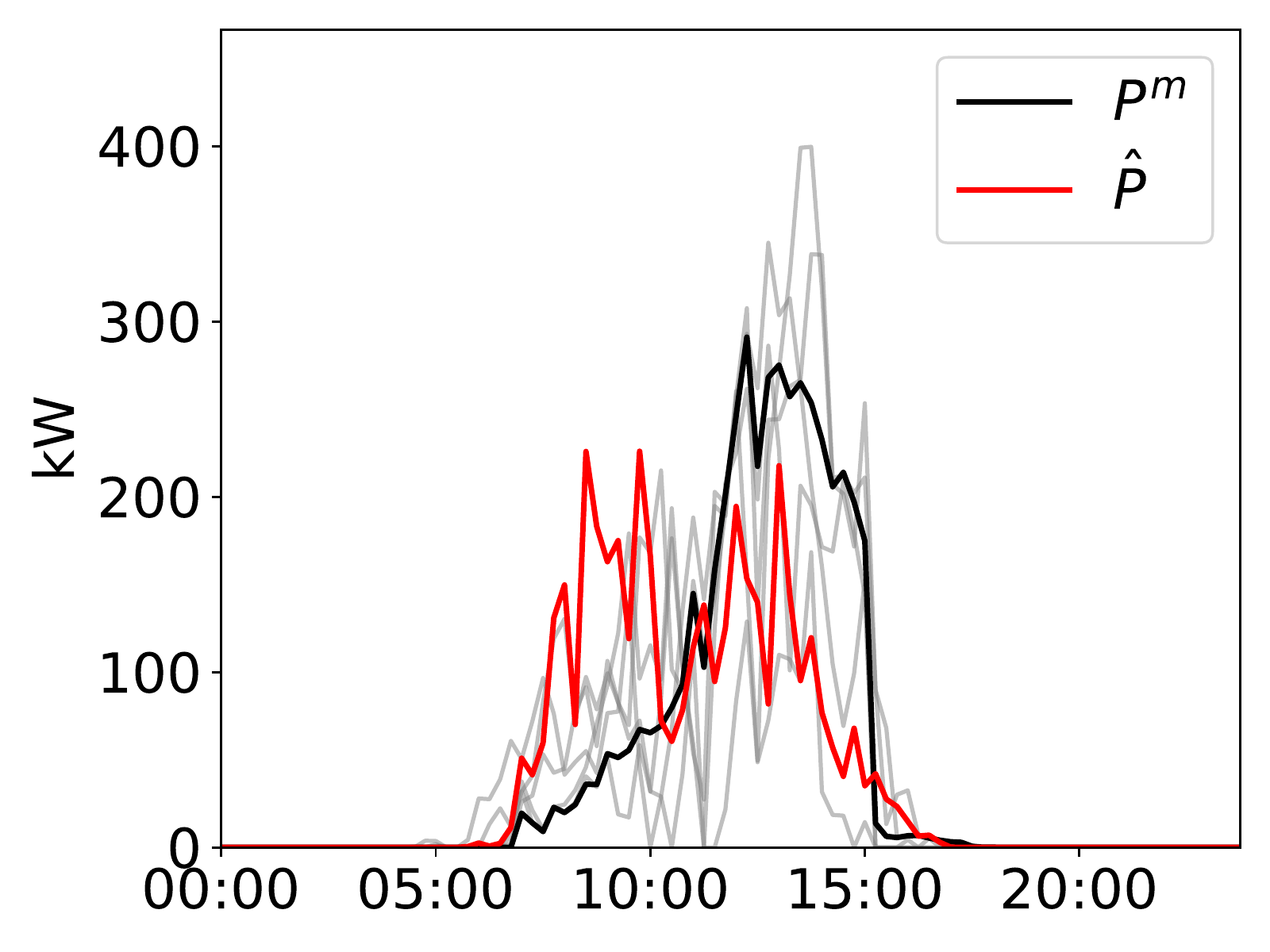}
		\caption{$\scenariocopuladatethree$.}
	\end{subfigure}%
	\begin{subfigure}{0.4\textwidth}
		\centering
		\includegraphics[width=\linewidth]{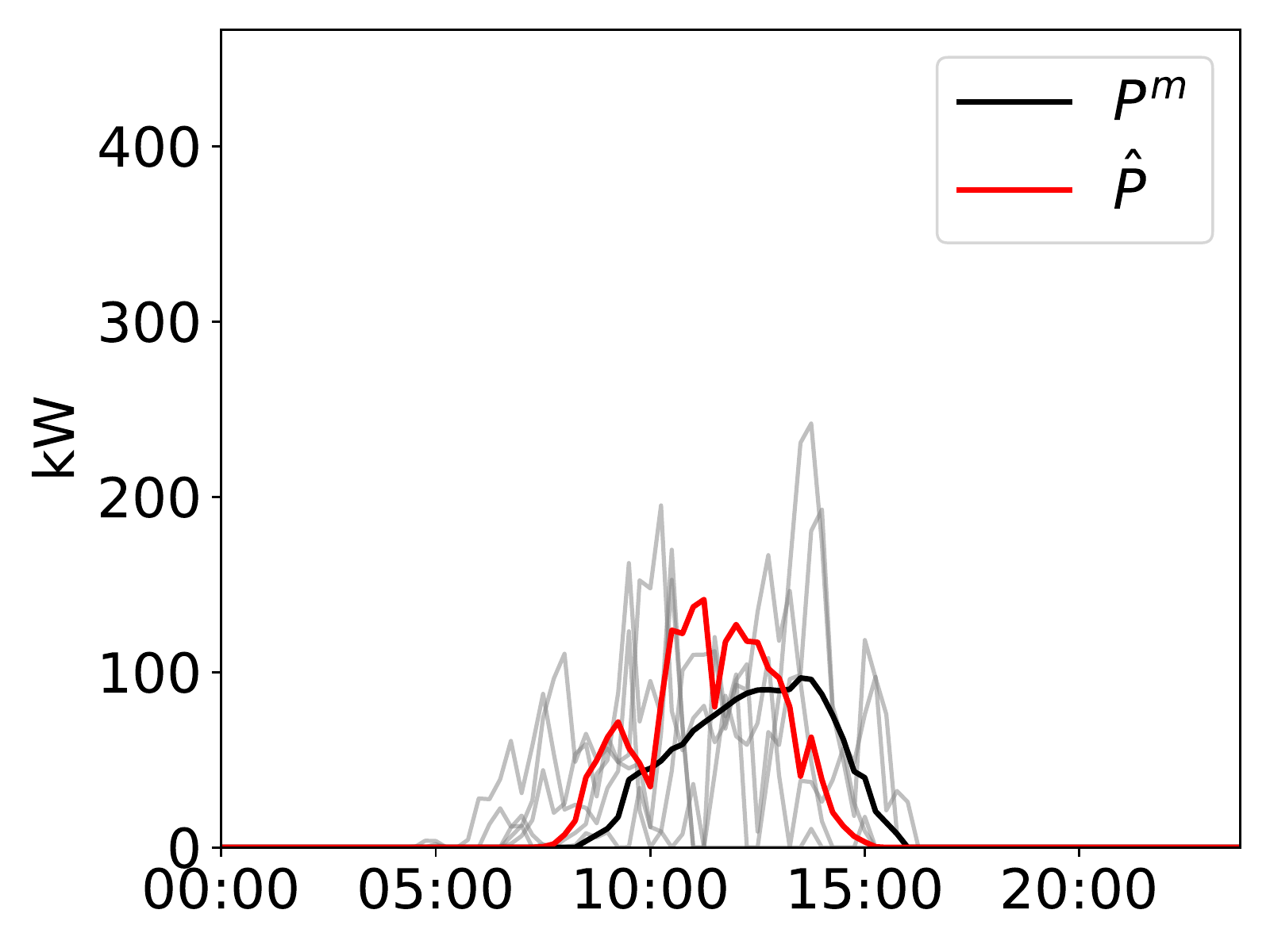}
		\caption{$\scenariocopuladatefour$.}
	\end{subfigure}
	\caption{5 PV scenarios (gray), PVUSA point forecats (red), and PV measurements (black).}
	\label{fig:uliege_pv_scenario}
\end{figure}

The Uli\`ege case study is composed of a PV generation plant with an installed capacity of 466.4 kWp. The simulation dataset is composed of August 2019 to December 2019, 151 days in total, with a total production of 141.3 MWh. The Uli\`ege PV generation is monitored on a minute basis and is resampled to 15 minutes.
The set of PV scenarios is generated using the Gaussian copula approach based on the PVUSA point forecasts. Figure~\ref{fig:uliege_pv_scenario} illustrates a set of 5 PV scenarios on four days of the dataset. 
The PV scenarios are used as input of a stochastic optimization problem.

\section{Sizing study}\label{sec:optimization-powertech-sizing}

The goal is to determine the optimal sizing of the BESS and PV for a given selling price and its related net revenue over the lifetime project to bid at the tendering stage optimally. 

\subsection{Problem statement and assumptions}

\begin{figure}[!htb]
	\centering
	\includegraphics[width=0.7\linewidth]{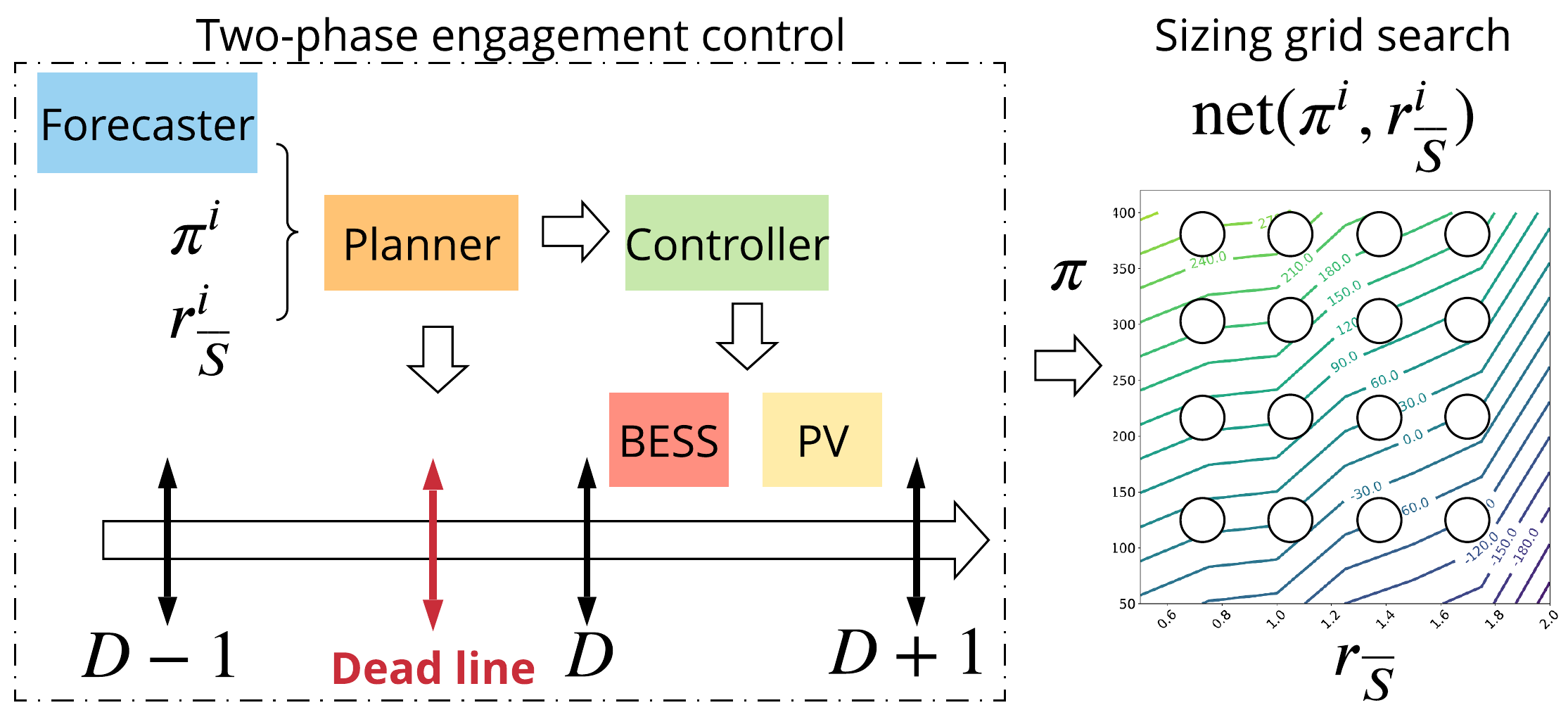}
	\caption{Sizing approach.}
	\label{fig:powertech-2021-approach}
\end{figure}

The ratio $r_{\maxcharge} = \frac{\maxcharge}{\PVcapacity}$, BESS maximum capacity over the PV installed capacity, is introduced to model several BESS and PV system configurations. The total exports, imports, deviation costs, number of charging and discharging cycles for several $r_{\maxcharge}$ and selling prices are computed. Based on the BESS and PV CapEx (Capital Expenditure) I and OpEx (Operational Expenditure) O\&M, the LCOE (Levelized Cost Of Energy) is calculated
\begin{align}\label{eq:powertech-lcoe_used}
\text{LCOE} & = \frac{\text{CRF}\cdot \text{I} + \text{O\&M} + \text{W} + \text{C}}{\text{E}},
\end{align}
with CRF the Capital Recovery Factor (or Annuity Factor), and E, W, C, the annual export, withdrawal, and deviation costs, respectively. Then, the net revenue over the lifetime project is defined as the annual gross revenue R divided by the annual export to the grid minus the LCOE
\begin{align}\label{eq:powertech-net}
\text{net}(\pi, r_{\maxcharge}) &: = \frac{R}{E} - \text{LCOE} .
\end{align}
The higher the net revenue, the more profitable the system. Section \ref{sec:powertech2021-LCOE} details the LCOE definition and the assumptions to establish (\ref{eq:powertech-lcoe_used}) and (\ref{eq:powertech-net}). Finally, the optimal sizing for a given selling price is provided by
\begin{align}
r_{\maxcharge}^\star (\pi)& = \arg \max_{r_{\maxcharge} \in \mathcal{R}_{\maxcharge}} \text{net}(\pi, r_{\maxcharge}) ,
\end{align}
with $\mathcal{R}_{\maxcharge}$ the sizing space, and the sizing approach is depicted in Figure (\ref{fig:powertech-2021-approach}). 

\subsection{Levelized cost of energy (LCOE)}\label{sec:powertech2021-LCOE}

The LCOE (\euro / MWh) is "the unique break-even cost price where discounted revenues, price times quantities, are equal to the discounted net expenses"\footnote{Annex II Metrics \& Methodology of the AR5 IPCC report.}. 
It means the LCOE is the lower bound of the selling price to be profitable over the lifetime of the project and is defined as
\begin{align}
\sum_{t=0}^{n} \frac{\text{E}_t \cdot \text{LCOE}}{(1+i)^t}  & =  \sum_{t=0}^{n} \frac{\text{Expenses}_t}{(1+i)^t},
\end{align}
with $i$ (\%) the discount rate, $n$ (years) the lifetime of the project, and $\text{E}_t$ (MWh) the annual energy at year $t$. The lifetime expenses comprise investment costs I that are the upfront investment (or CapEx) (\euro /kW for PV and \euro /kWh for BESS), operation and maintenance cost (or OpEx) \text{O\&M} (\euro /kW for PV and \euro /kWh for BESS), the annual cost of the energy withdrawn from the grid W (\euro), and the annual deviation cost penalty C (\euro)
\begin{align}\label{eq:powertech-lcoe_def2}
\text{LCOE} & =  \frac{\sum_{t=0}^{n} \frac{\text{I}_t + \text{O\&M}_t + \text{W}_t + \text{C}_t}{(1+i)^t}}{\sum_{t=0}^{n} \frac{\text{E}_t}{(1+i)^t}}.
\end{align}
We assume the annual cost of the energy $\text{E}_t$ ($\text{W}_t$) exported (withdrawn), the annual OpEx $\text{O\&M}_t$, and the annual deviation cost $\text{C}_t$ are constant during the lifetime of the project: E, W, C and $\text{O\&M} \ \forall t > 0$. The investment costs I are the sum of all capital expenditures needed to make the investment fully operational discounted to $t=0$. Thus, $\text{I}_t = 0 \ \forall t > 0$ and $\text{I}_0 = \text{I}$. Finally, the system is operational at year $t=1$: $\text{E}_0 = 0$, $\text{O\&M}_0 = 0$, $\text{W}_0 = 0$,  and $\text{C}_0 = 0$. (\ref{eq:powertech-lcoe_def2}) becomes
\begin{align}
\text{LCOE} & =  \frac{\text{I} + (\text{O\&M} + \text{W} + \text{C}) \sum_{t=1}^{n} \frac{1}{(1+i)^t}}{\text{E}\sum_{t=1}^{n} \frac{1}{(1+i)^t}},
\end{align}
that is re-written
\begin{align}
\text{LCOE} & = \frac{\text{CRF}\cdot \text{I} + \text{O\&M} + \text{W} + \text{C}}{\text{E}},
\end{align}
with $\text{CRF} = (\sum_{t=1}^{n} \frac{1}{(1+i)^t})^{-1} = \frac{i}{1-(1+i)^{-n}}$.
The gross revenue is the energy exported to the grid that is remunerated at the selling price. The annual gross revenue $R$ is assumed to be constant over the project lifetime. Then, the LCOE can be compared to R divided by the annual export $E$ to assess the financial viability of the project by calculating (\ref{eq:powertech-net}).

\subsection{Case study description}

The ULi\`ege case study comprises a PV generation plant with an installed capacity $\PVcapacity$ of 466.4 kW. The period from August 2019 to December 2019, 151 days in total, composes the dataset. The PV generation is monitored on a minute basis and resampled to 15 minutes.

The simulation parameters of the planners and the oracle controller are identical. The planning and controlling periods are $\Delta t = 15$ minutes. The peak hours are between 7 pm and 9 pm (UTC+0). 
The specifications of the tender AO-CRE-ZNI 2019 published on $\CREspecifications$ define the ramping power constraint on the engagements $\Delta X = 7.5\% \PVcapacity$ ($15\% \PVcapacity$) during off-peak (peak) hours. The lower bound on the engagement is $\xMin_t = -5\% \PVcapacity$  ($\xMin_t  = 20\% \PVcapacity$) during off-peak (peak) hours. The lower bound on the production is $\yMin_t = -5\% \PVcapacity$ ($\yMin_t  = 15\% \PVcapacity$) during off-peak (peak) hours. The upper bounds on the engagement and production are $\xMax_t= \yMax_t =  \PVcapacity$. Finally, the engagement deadband is $ 5\% $ of $\PVcapacity$. 

The Python Pyomo\footnote{\url{www.http://www.pyomo.org/}} 5.6.7 library is used to implement the algorithms in Python 3.7. IBM ILOG CPLEX Optimization Studio\footnote{\url{https://www.ibm.com/products/ilog-cplex-optimization-studio}} 12.9 is used to solve all the mixed-integer quadratic optimization problems. Numerical experiments are performed on an Intel Core i7-8700 3.20 GHz based computer with 12 threads and 32 GB of RAM running on Ubuntu\footnote{\url{https://ubuntu.com}} 18.04 LTS. 
The average computation time per optimization problem of the S planner with $\#\Omega = 20$ scenarios is 3 (s) for an optimization problem with 15 000 variables and 22 000 constraints.

\subsection{Sizing parameters}

The BESS and PV CAPEX are 300 \euro /kWh and 700 \euro /kW, the BESS and PV OPEX are 1\% of the CAPEX, the project lifetime is 20 years, and the weighted average cost of capital is 5\%. The BESS lifetime in terms of complete charging and discharging cycles is 3 000.
The BESS is assumed to be capable of fully charging or discharging in one hour $
\ChargeMax= \DischargeMax = \SocMax / 1$, with charging and discharging efficiencies $\chargeEff = \dischargeEff = 0.95$. Each simulation day is independent with a discharged battery at the first and last period to its minimum capacity $\SocIni = \SocEnd = 10\% \SocMax$. The BESS minimum ($\SocMin$) and maximum ($\SocMax$) capacities are 10\% and 90\% of the total BESS storage capacity $\maxcharge$. 
The sizing space is a grid composed of 56 values with $\mathcal{R}_{\maxcharge} =  \{0.5, 0.75, 1, 1.25, 1.5, 1.75, 2\}$ and $\mathcal{P} =  \{50, 100, 150, 200, 250, 300, 350, 400\}$.

\subsection{Sizing results}

\begin{figure}[!htb]
	\centering
	\begin{subfigure}{0.4\textwidth}
		\centering
		\includegraphics[width=\linewidth]{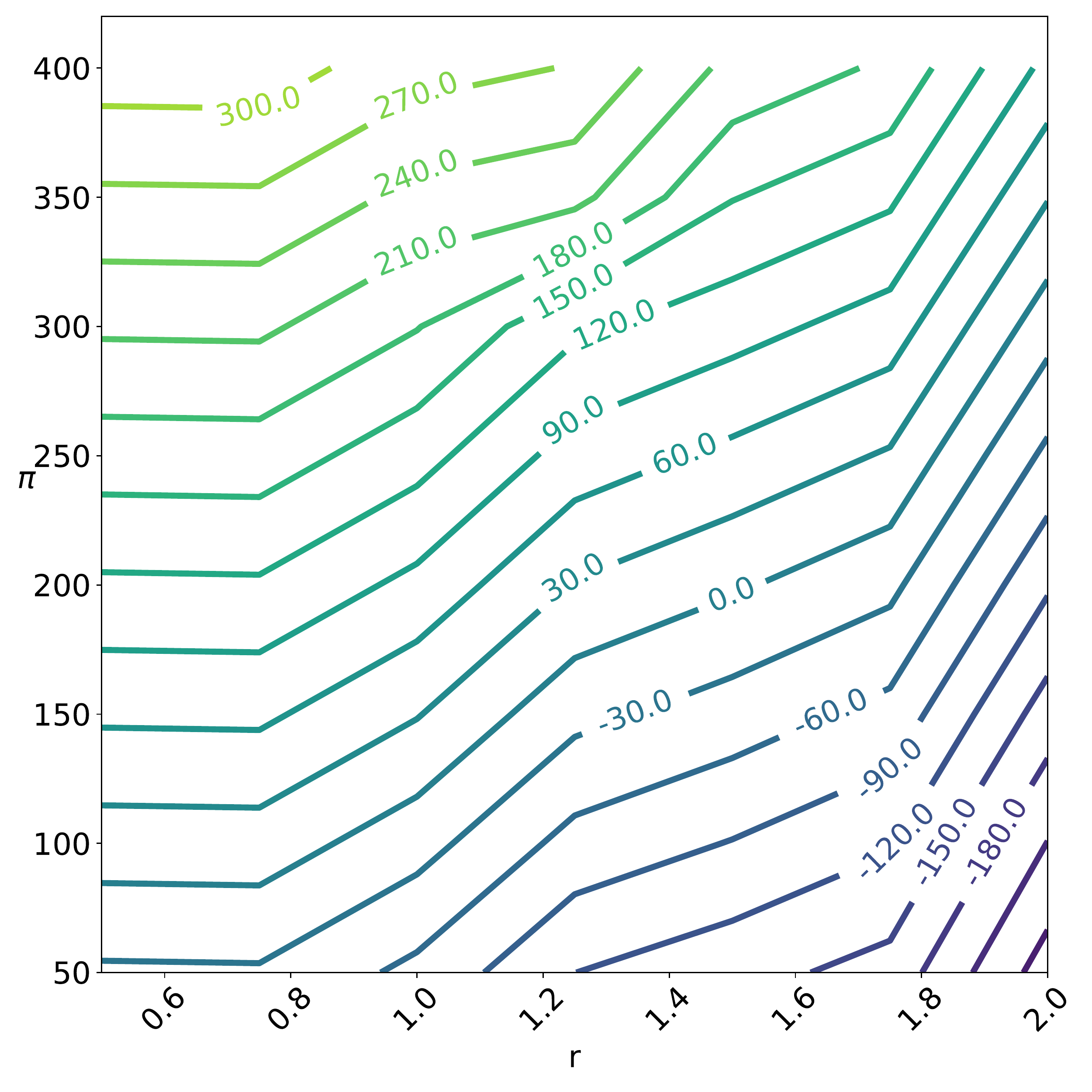}
		\caption{$D^\star$-oracle.}
	\end{subfigure}%
	\begin{subfigure}{0.4\textwidth}
		\centering
		\includegraphics[width=\linewidth]{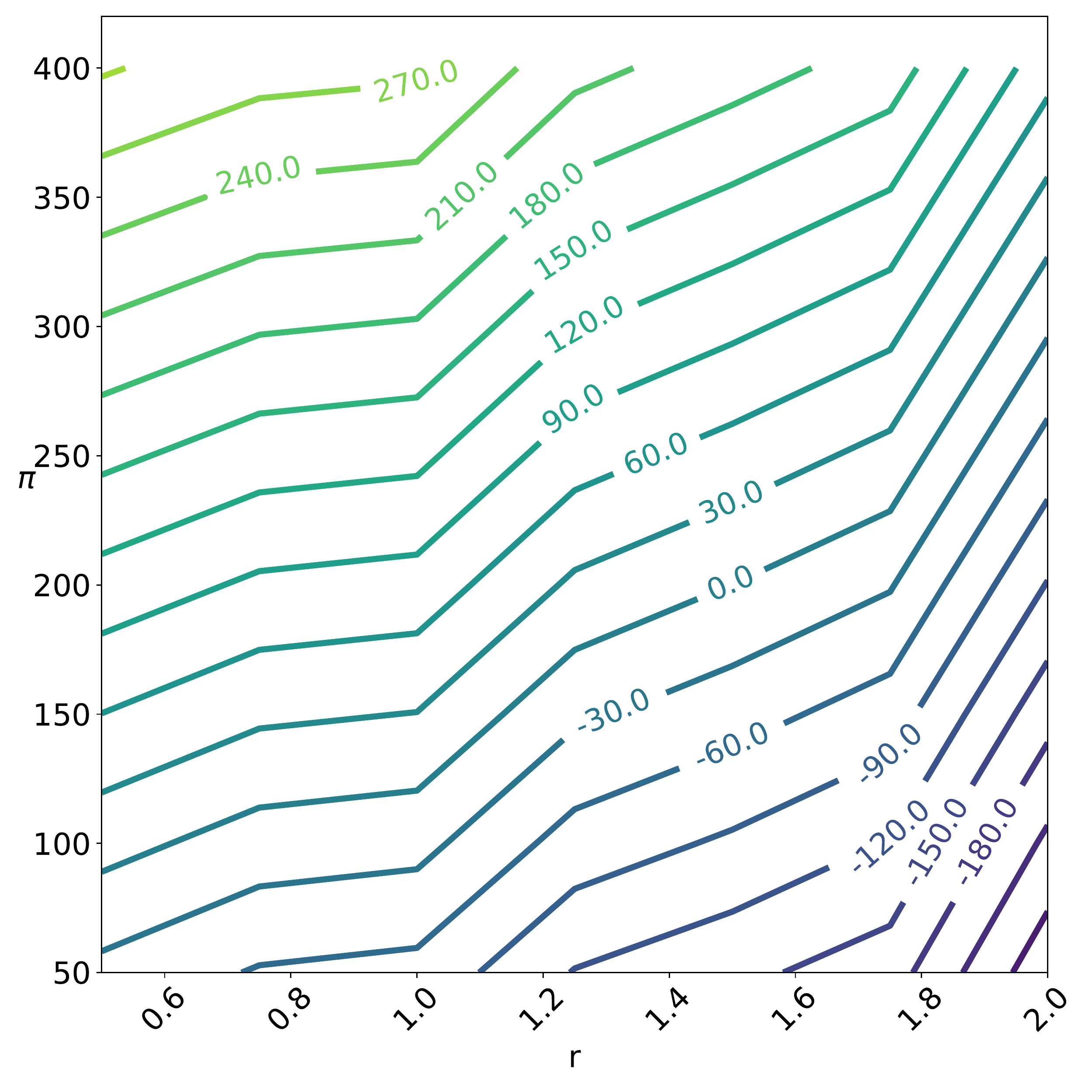}
		\caption{$D$-oracle.}
	\end{subfigure}
	\begin{subfigure}{0.4\textwidth}
		\centering
		\includegraphics[width=\linewidth]{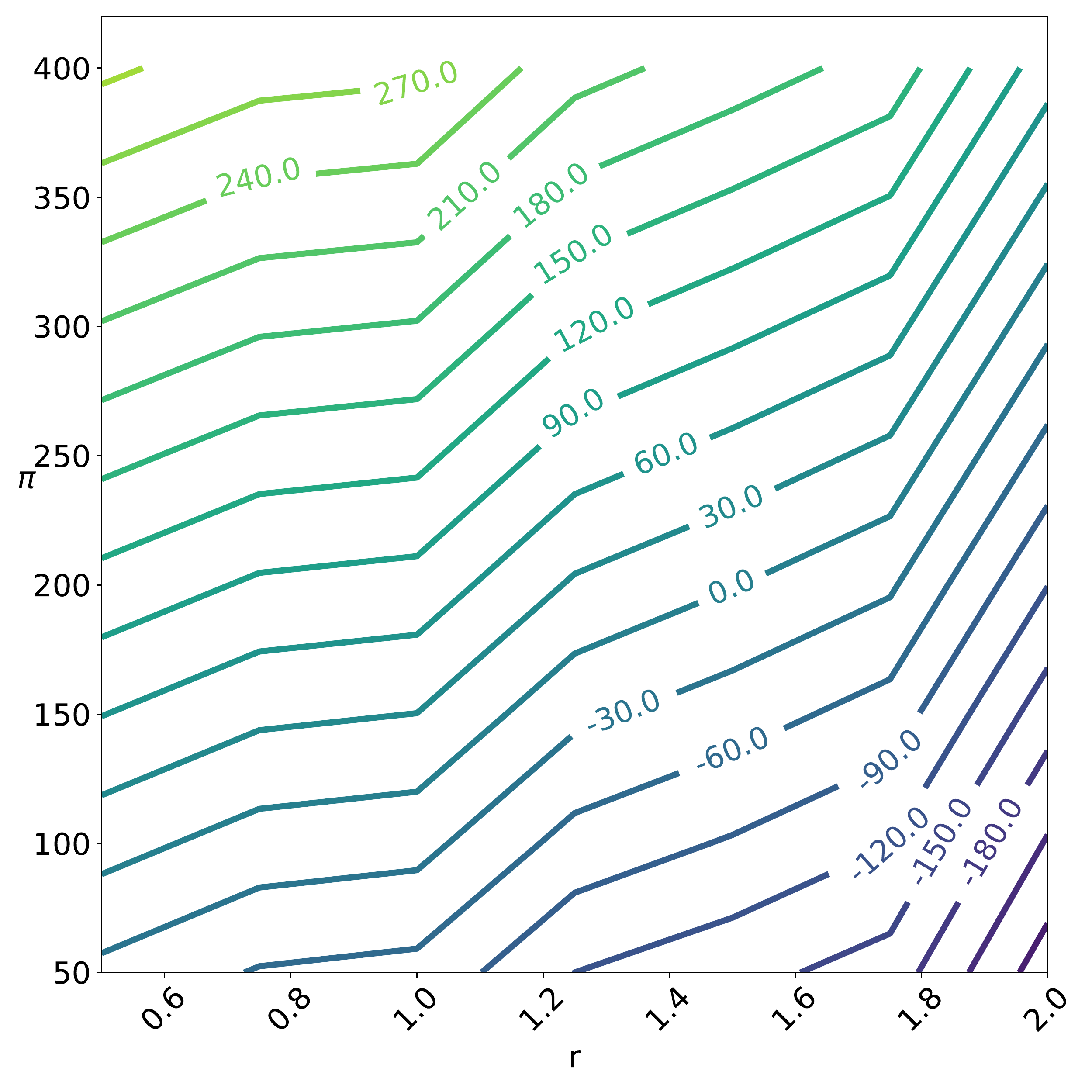}
		\caption{$S^{20}$-oracle.}
	\end{subfigure}%
	\begin{subfigure}{0.4\textwidth}
		\centering
		\includegraphics[width=\linewidth]{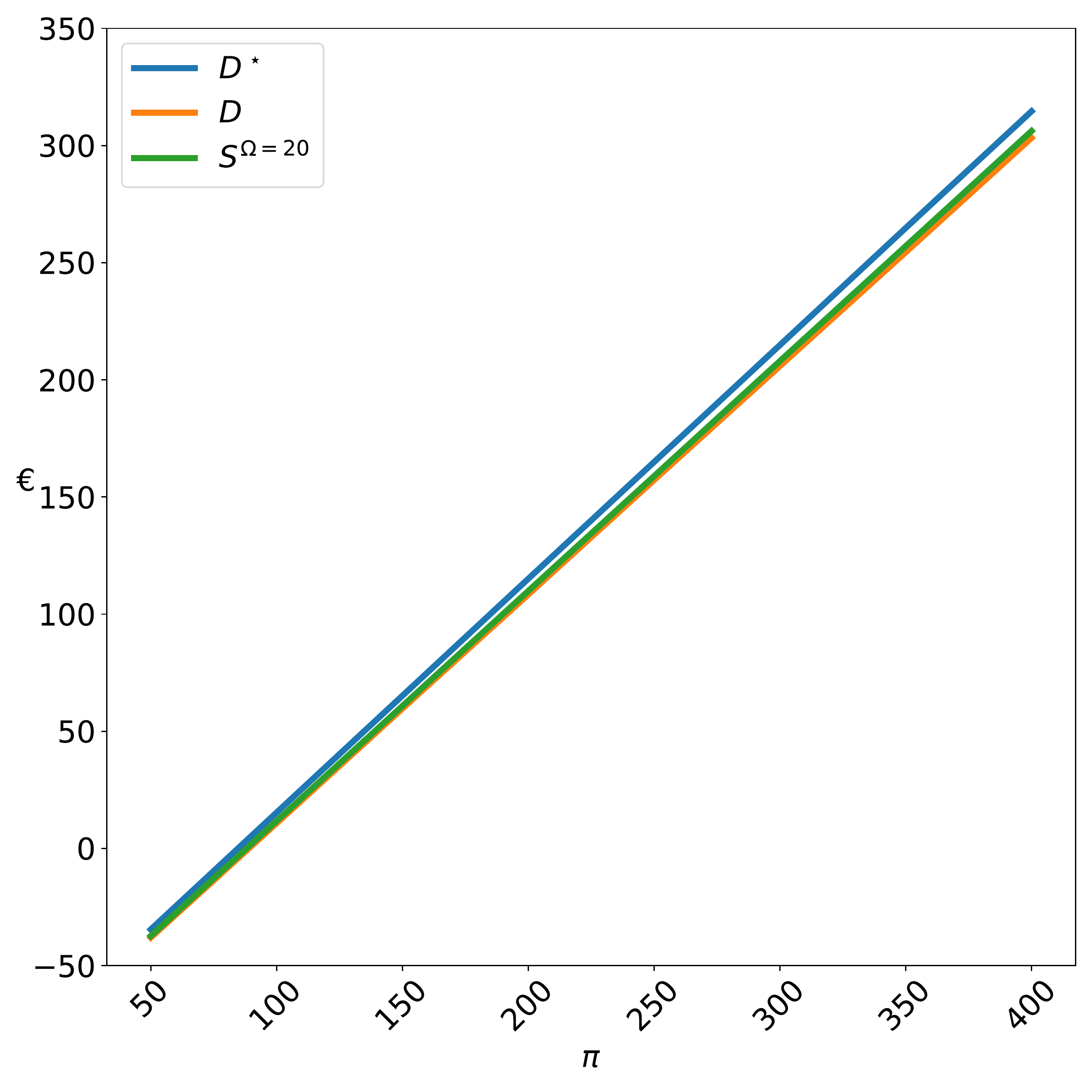}
		\caption{$ \text{net}(\pi, r_{\maxcharge}^\star = 0.5)$.}
	\end{subfigure}
	\caption{Sizing results: $\text{net}(\pi, r_{\maxcharge})$ (\euro / MWh).}
	\label{fig:powertech2021-sizing_QP_oracle_net}
\end{figure}

The $D^\star$, $D$, and $S^{\Omega = 20}$ planners are used with the oracle controller over the simulation dataset. E, W, C, and the number of complete charging and discharging cycles are calculated by extrapolating the 151 days to one year.
Figure~\ref{fig:powertech2021-sizing_QP_oracle_net} provides the grid search sizing results for the three planners. For a given selling price, the net is maximal when $r_{\maxcharge} = $ 0.5. When $r_{\maxcharge} $ increases, the BESS is more and more used to withdraw and export during peak hours. It leads to an increase in the number of charging/discharging cycles that implies an increase of the number of BESS required during the project lifetime and consequently an increase of the BESS CAPEX. As the BESS CAPEX mainly drives the LCOE, $\frac{R}{E}$ is not capable of compensating the LCOE increase resulting in a net decrease. The number of charging and discharging cycles is approximately the same for both the $D$ and $S$ planners, independently of the selling price, and rises from 4 700 with $r_{\maxcharge} = $ 0.5 to 13 000 with $r_{\maxcharge} = $ 2. 

The differences between $D^\star$, $D$, and $S$ planners are minor. $D^\star$ tends to overestimate the net by underestimating the LCOE (underestimating the deviation, BESS CAPEX, and withdrawal costs) and overestimating $\frac{R}{E}$. 
However, the minimal selling price to be profitable with $r_{\maxcharge} = 0.5$, is approximately 80 \euro / MWh for all planners as shown by Figure~\ref{fig:powertech2021-sizing_QP_oracle_net}. Then, the higher the selling price, the higher the net. In the CRE specifications, the best tender is mainly selected based on the selling price criterion. A trade-off should be reached between the net and the selling price to be selected.

This sizing study seems to indicate that it is not very sensitive to the control policy, \textit{i.e}, deterministic with perfect knowledge, deterministic with point forecasts, and stochastic with scenarios.
However, it may be dangerous not considering the uncertainty at the sizing stage and could lead to overestimating the system performance and underestimating the sizing.
Indeed, the two-phase engagement control net revenues are similar between planners explaining the minor differences in terms of sizing. Two main limitations could explain this result. First, large deviations (15-20\%) from the engagement plan at the control step occur rarely. Indeed, the oracle controller may compensate for inadequate planning and limits the deviations. A more realistic controller with point forecasts should be considered. Second, when such deviations occur, they are usually within the tolerance where there is no penalty. Furthermore, when the deviations are outside, the penalty is relatively small in comparison with the gross revenue. A sensitivity analysis of the numerical settings of the CRE specifications should be performed.

\section{Conclusions and perspectives}\label{sec:optimization-powertech-conclusions}

The key idea of this Chapter is to propose a methodology to size the PV and BESS in the context of the capacity firming framework. Indeed, the two-phase engagement control cannot easily be modeled as a single sizing optimization problem. Such an approach would result in a non-linear bilevel optimization problem challenging to solve with an upper level, the sizing part, and a lower level, the two-phase engagement control. 
The two-phase engagement control is decomposed into two steps: computing the day-ahead engagements, then recomputing the set-points in real-time to minimize the deviations from the engagements. The CRE non-convex penalty function is modeled by a threshold-quadratic penalty that is compatible with a scenario approach. The stochastic formulation using a scenario approach is compared to the deterministic formulation. The PV scenarios are generated using a Gaussian copula methodology and PV point forecasts computed with the PVUSA model. 
%
The minimal selling price to be profitable, on this dataset, in the context of the capacity firming framework is approximately 80 \euro / MWh with a BESS having a maximal capacity, fully charged or discharged in one hour, of half the total PV installed power.
The sizing study indicates that it is not very sensitive to the control policy. The differences are minor between the deterministic with perfect knowledge (or point forecasts) and stochastic with scenarios strategies. However, further investigations are required to implement a more realistic controller that uses intraday point forecasts and conduct a sensitivity analysis on the simulation parameters. 

Several extensions are under investigation. 
\begin{itemize}
    \item A PV generation methodology that is less dependent on the PV point forecasts and considers the PV power's error dependency should be implemented. 
    \item PV scenarios clustering and reduction techniques could be considered to select relevant PV scenarios and improve the stochastic planner results. 
    \item A sizing formulation as a single optimization problem with the PV and BESS capacities as variables. It allows to compute the optimum directly and avoid doing a grid search. However, this formulation is not trivial due to the specific two-phase engagement control of the capacity firming framework.
    \item Finally, the BESS aging process could be modeled in the sizing study. A dataset with at least a full year of data should be considered to consider the PV seasonality fully. 
\end{itemize}


\chapter{Capacity firming using a robust approach}\label{chap:capacity-firming-robust}

\begin{infobox}{Overview}
This Chapter proposes a robust approach to address the energy management of a grid-connected renewable generation plant coupled with a battery energy storage device in the capacity firming market. It is an extension of Chapter \ref{chap:capacity-firming-stochastic} by considering another optimization formulation to handle the PV uncertainty.
The main contributions are two-fold:
\begin{enumerate}
	\item The core contribution is applying the robust optimization framework to the capacity firming market in a tractable manner thanks to a Benders decomposition of the optimization problem and a warm start of the algorithm. In addition, a dynamic risk-averse parameters selection taking advantage of the quantile forecast distribution is proposed.
	\item The secondary contribution is the use of the Normalizing Flows, which is a new advanced forecasting technique, to provide the uncertainty estimation in the form of PV quantiles for the robust planner. To the best of our knowledge, it is the first study to use NFs in a power system application.
\end{enumerate}

\textbf{\textcolor{RoyalBlue}{References:}} This chapter  is an adapted version of the following publication: \\[2mm]\bibentry{dumas2021probabilistic}. 
\end{infobox}
\epi{You have to start with the truth. The truth is the only way that we can get anywhere. Because any decision-making that is based upon lies or ignorance can't lead to a good conclusion.}{Julian Assange}
\begin{figure}[htbp]
	\centering
	\includegraphics[width=1\linewidth]{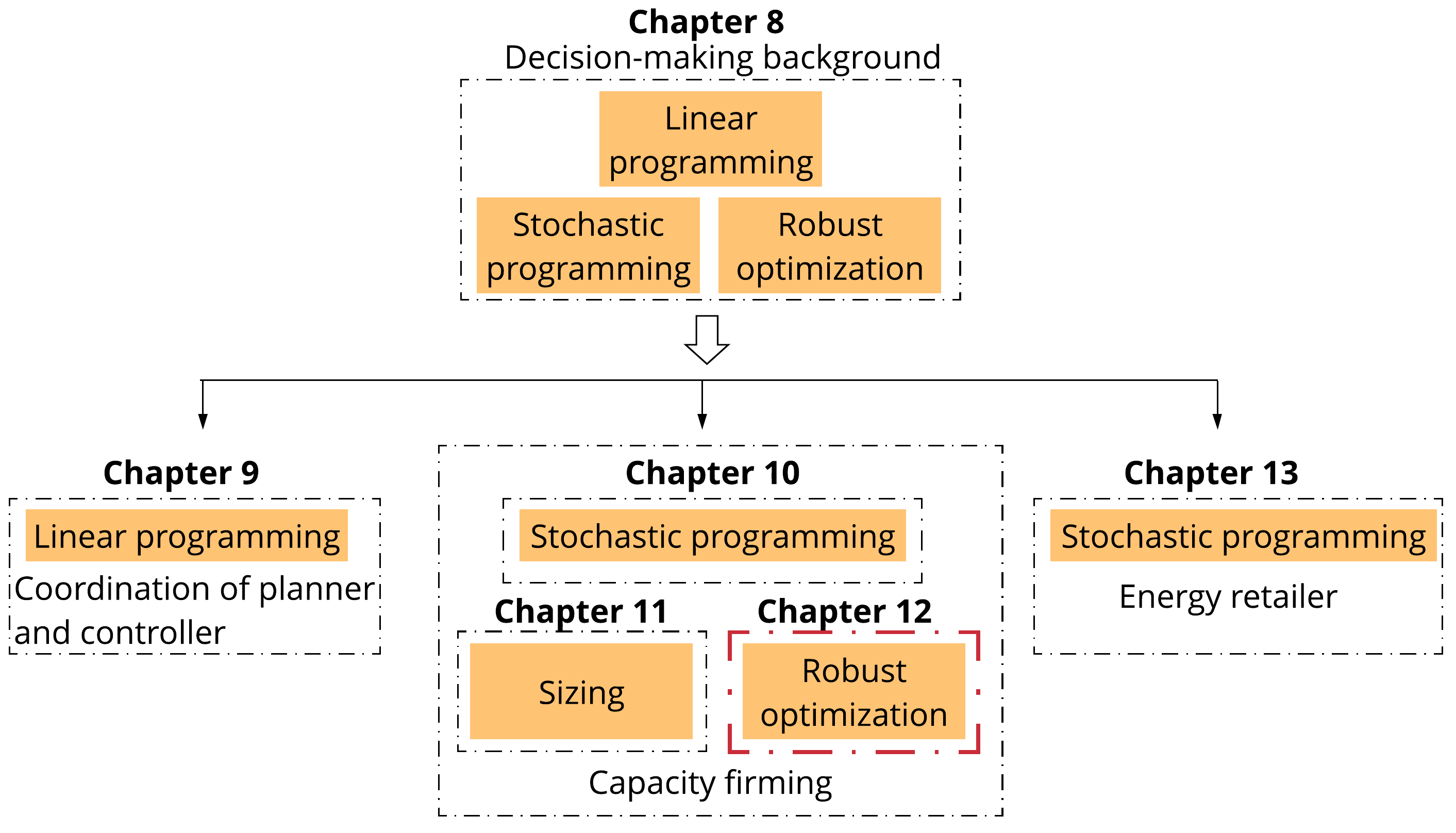}
	\caption{Chapter \ref{chap:capacity-firming-robust} position in Part \ref{part:optimization}.}
\end{figure}
\clearpage

There are several approaches to deal with renewable energy uncertainty. One way is to consider a two-stage stochastic programming approach \citep{birge2011introduction}. It has already been applied to the capacity firming framework \citep{dumas2020probabilistic,n2020controle,haessig2014dimensionnement,parisio2016stochastic}. 
The generation uncertainty is captured by a set of scenarios modeling possible realizations of the power output. However, this approach has three drawbacks. First, the problem size and computational requirement increase with the number of scenarios, and a large number of scenarios are often required to ensure the good quality of the solution. Second, the accuracy of the algorithm is sensitive to the scenario generation technique. Finally, it may be challenging to identify an accurate probability distribution of the uncertainty.
Another option is to consider robust optimization (RO) \cite{ben2009robust,bertsimas2011theory}, applied to unit commitment by \cite{bertsimas2012adaptive,jiang2011robust}, and in the capacity firming setting \cite{n2020controle}. RO accounts for the worst generation trajectory to hedge the power output uncertainty, where the uncertainty model is deterministic and set-based. Indeed, the RO approach puts the random problem parameters in a predetermined uncertainty set containing the worst-case scenario.
It has two main advantages \cite{bertsimas2012adaptive}: (1) it only requires moderate information about the underlying uncertainty, such as the mean and the range of the uncertain data; (2) it constructs an optimal solution that immunizes against all realizations of the uncertain data within a deterministic uncertainty set. Therefore, RO is consistent with the risk-averse fashion way to operate power systems.
However, the RO version of a tractable optimization problem may not itself be tractable, and some care must be taken in choosing the uncertainty set to ensure that tractability is preserved.

Traditionally, a two-stage RO model is implemented for the unit commitment problem in the presence of uncertainty. However, it is challenging to compute and often NP-hard. 
Two classes of cutting plane strategies have been developed to overcome the computational burden. The Benders-dual cutting plane (BD) algorithms are the most used and seek to derive exact solutions in the line of Benders’ decomposition \cite{benders1962partitioning} method. They decompose the overall problem into a master problem involving the first-stage commitment decisions at the outer level and a sub-problem associated with the second-stage dispatch actions at the inner level. Then, they gradually construct the value function of the first-stage decisions using dual solutions of the second-stage decision problems \cite{bertsimas2012adaptive,jiang2011robust}.
In contrast, the column-and-constraint generation (CCG) procedure, introduced by \cite{zhao2012robust,zeng2013solving} does not create constraints using dual solutions of the second-stage decision problem. Instead, it dynamically generates constraints with recourse decision variables in the primal space for an identified scenario. The generated variables and constraints in the CCG procedure are similar to those in the deterministic equivalent of a two-stage stochastic programming model.
The BD and CCG algorithms have not been compared in the capacity firming framework to the best of our knowledge.

\begin{figure}[tb]
	\centering
	\includegraphics[width=0.7\linewidth]{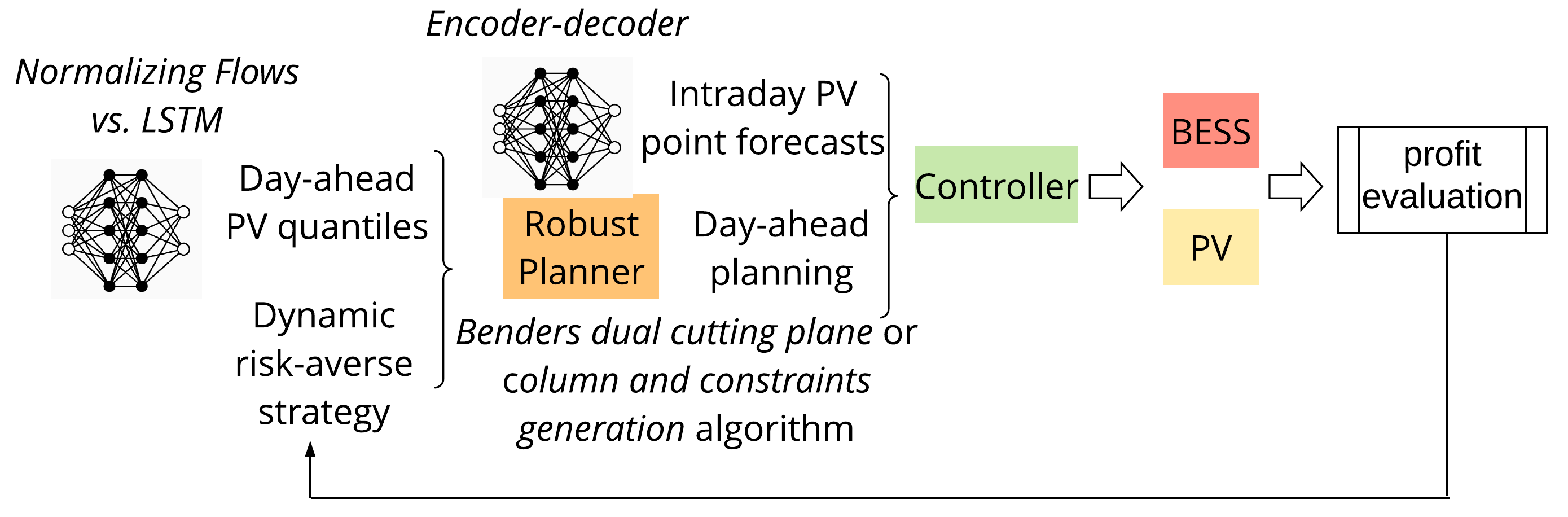}
	\caption{Forecast-driven robust optimization strategy.}
	\label{fig:process}
\end{figure}

This Chapter proposes a reliable and computationally tractable probabilistic forecast-driven robust optimization strategy. It can use either a BD or CGG algorithm in the capacity firming framework, depicted in Figure \ref{fig:process}. 
Our work goes several steps further than \cite{n2020controle}. The main contributions of this Chapter are three-fold:
\begin{enumerate}
%
\item The core contribution is applying the robust optimization framework to the capacity firming market in a tractable manner by using a Benders decomposition. The non-linear robust optimization problem is solved both using the Benders-dual cutting plane and the column-and-constraint generation algorithms. To the best of our knowledge, it is the first time that a comparison of these algorithms is performed in the capacity firming framework.
In addition, the convergence of the BD algorithm is improved with a warm-start procedure. It consists of building an initial set of cuts based on renewable generation trajectories assumed to be close to the worst-case scenario.
The results of both the CCG and BD two-stage RO planners are compared to the deterministic planner using perfect knowledge of the future, the nominal point forecasts, \textit{i.e.}, the baseline to outperform, and the quantiles (a conservative approach). The case study is the photovoltaic (PV) generation monitored on-site at the University of Li\`ege (ULi\`ege), Belgium.
%
\item Second, a dynamic risk-averse parameters selection taking advantage of the quantile forecast distribution is investigated and compared to a strategy with fixed risk-averse parameters.
%
\item Finally, the \textit{normalizing flows} (NFs) is implemented. A new class of probabilistic generative models has gained increasing interest from the deep learning community in recent years. NFs are used to compute day-ahead quantiles of renewable generation for the robust planner. Then, an encoder-decoder architecture forecasting model \cite{bottieau2019very} computes the intraday point forecasts for the controller. To the best of our knowledge, it is the first study to use NFs in a power system application.
\end{enumerate}

In addition to these contributions, this study also provides open access to the Python code\footnote{\url{https://github.com/jonathandumas/capacity-firming-ro}} to help the community to reproduce the experiments.
The rest of this Chapter is organized as follows. Section~\ref{sec:optimization-ieee-formulation} provides the mathematical formulations of the robust and deterministic planners. Section~\ref{sec:optimization-ieee-ro_benders} develops the Benders-dual cutting plane algorithm. The case study and computational results are presented in Section~\ref{sec:optimization-ieee-case_study}. Finally, Section~\ref{sec:optimization-ieee-conclusions} concludes this study and draws some perspectives of future works. Note, the capacity firming process is described in Section \ref{sec:optimization-pmaps-capacity-firming-process} of Chapter \ref{chap:capacity-firming-stochastic}, and Section \ref{sec:forecasting-quantile-comparison} in Chapter \ref{chap:quantile-forecasting} introduces the forecasting techniques and proposes a quality evaluation.

\section{Problem formulation}\label{sec:optimization-ieee-formulation}

For the sake of simplicity in this Chapter, the penalty $c$ defined in (\ref{eq:pmaps-penalty}) is assumed to be symmetric, convex, and piecewise-linear
\begin{align}
c(x_t, y^m_t)  & = \pi\Delta t \beta \bigg(  \max \big(0, |x_t- y^m_t|- p \PVcapacity \big) \bigg),
\end{align} 
with $\beta$ a penalty factor. Note, in this Chapter, both the import and export are considered in contrast to Chapter \ref{chap:capacity-firming-stochastic}.
A two-stage robust optimization formulation is built to deal with the engagement for the uncertain renewable generation that is modeled with an uncertainty set. The deterministic and robust formulations of the planner are presented in Sections \ref{sec:optimization-ieee-det_planner} and \ref{sec:optimization-ro_planner}. The robust optimization problem with recourse has the general form of a min-max-min optimization problem. The uncertainty set is defined by quantiles forecasts and a budget of uncertainty $\Gamma$. Section \ref{sec:optimization-ieee-ro_planner_transformed} uses the dual of the inner problem to formulate a min-max optimization problem. Finally, Section \ref{sec:optimization-ieee-controller} presents the formulation of the controller. The optimization variables and the parameters are defined in Section~\ref{sec:optimization-pmaps-notation}.

\subsection{Deterministic planner formulation}\label{sec:optimization-ieee-det_planner}

The deterministic formulation of the planner is provided in Section \ref{sec:optimization-pmaps-problem formulation} of Chapter \ref{chap:capacity-firming-stochastic}. The only difference lies in the definition of the penalty $c$ that is assumed to be symmetric, convex, and piecewise-linear instead of a quadratic penalty in (\ref{eq:pmaps-D-formulation}). Therefore, in this Chapter, the objective function $J $ of the deterministic planner to minimize is 
\begin{align}\label{eq:ieee-objective}
J \big(x_t, y_t \big) & \ =  \sum_{t\in \mathcal{T}}\pi_t \Delta t  [-y_t + \beta (d_t^- + d_t^+)] .
\end{align}
The deterministic formulation is the following Mixed-Integer Linear Program (MILP)
\begin{align}\label{eq:ieee-D-formulation}
\min_{x_t \in \mathcal{X}, y_t \in \Omega(x_t, \PVforecast_t)} & \ J (x_t, y_t)  \notag \\
\mathcal{X} & =  \big \{ x_t : (\ref{eq:pmaps-engagement-D-csts}) \big \}  \\
\Omega(x_t, \PVforecast_t)&  = \big \{ y_t : (\ref{eq:pmaps-D-BESS-csts})- (\ref{eq:pmaps-D-PV-csts})  \notag \big \} \notag 
\end{align}
where $\mathcal{X}$ and $\Omega(x_t, \PVforecast_t)$ are the sets of feasible engagements $x_t$ and dispatch solutions $y_t$, respectively. The optimization variables of (\ref{eq:ieee-D-formulation}) are the engagement variables $x_t$, the dispatch variables $y_t$ (the net power at the grid connection point), $\Discharge_t$ (BESS discharging power), $\Charge_t$ (BESS charging power), $\SOC_t$ (BESS state of charge), $\BESSbinary_t$ (BESS binary variables), $\PVgeneration_t$ (renewable generation), and $d^-_t, d^+_t$ (deviation variables) (cf. the notation Section~\ref{sec:optimization-pmaps-notation}).

\subsection{Robust planner formulation}\label{sec:optimization-ro_planner}

The uncertain renewable generation $\PVforecast_t$ of (\ref{eq:pmaps-D-PV-csts}) is assumed to be within an interval $\mathcal{U} = [u^\text{min}_t, u^\text{max}_t]$ that can be obtained based on the historical data or an interval forecast composed of quantiles. In the following $\PVforecast_t$ is replaced by $u_t$ in (\ref{eq:pmaps-D-PV-csts}).
The proposed two-stage robust formulation of the capacity firming problem consists of minimizing the objective function over the worst renewable generation trajectory
\begin{align}\label{eq:RO_min_max}	
\max_{u_t \in \mathcal{U}} \bigg[\min_{x_t \in \mathcal{X}, \ y_t \in \Omega(x_t, u_t)}  &  \  J \big(x_t, y_t \big) \bigg],
\end{align}
that is equivalent to
\begin{align}\label{eq:ieee-RO_min_max_min}	
\min_{x_t \in \mathcal{X}} \bigg[ \max_{u_t \in \mathcal{U}} \min_{y_t \in \Omega(x_t, u_t)} & \ J \big(x_t, y_t \big) \bigg] .
\end{align}
The worst-case dispatch cost has a max-min form, where $$\min_{y_t \in \Omega(x_t, u_t)}  J \big(x_t, y_t \big)$$ determines the economic dispatch cost for a fixed engagement and a renewable generation trajectory, which is then maximized over the uncertainty set $\mathcal{U}$.

In the capacity firming framework, where curtailment is allowed, the uncertainty interval consists only in downward deviations $[u^\text{min}, \hat{y}^{\text{pv}, (0.5)}_t]$, with $\hat{y}^{\text{pv}, (0.5)}_t$ the PV 50\% quantile forecast. 
\begin{demonstration}\label{demonstration}
Let consider $\mathcal{U}_1 =  [ u^\text{min}_t, \hat{y}^{\text{pv}, (0.5)}_t]$, $\mathcal{U}_2 = [ \hat{y}^{\text{pv}, (0.5)}_t, u^\text{max}]$, $u_t^1 \in \mathcal{U}_1$, $ u_t^2 \in \mathcal{U}_2$, $y_t^1 \in  \Omega_1 =  \Omega(x_t, u_t^1 \in \mathcal{U}_1) $, and $y_t^2 \in  \Omega_2 =  \Omega(x_t, u_t^2 \in \mathcal{U}_2) $.
Then, for a given engagement $x_t$

\begin{align}
\max_{u_t \in \mathcal{U}} \min_{y_t \in \Omega(x_t, u_t)}  J \big(x_t, y_t \big)  & = \max \bigg[  \max_{u_t^1 \in \mathcal{U}_1} \min_{y_t^1 \in \Omega_1}  J \big(x_t, y_t^1 \big),  \max_{u_t^2 \in \mathcal{U}_2} \min_{y_t^2 \in \Omega_2}  J \big(x_t, y_t^2 \big) \bigg].
\end{align}
The only difference between $\Omega_1$ and $\Omega_2$ is provided by (\ref{eq:pmaps-D-PV-csts}) where $\PVgeneration_{t,1} \leq u_t^1$ and $\PVgeneration_{t,2} \leq u_t^2$.
By definition of the uncertainty sets $u_t^1 \leq u_t^2$ $\forallt$, and it is straightforward that $\PVgeneration_{t,1} \leq u_t^2$. In addition, the variables $y_t^1$ satisfy all the other constraints of $\Omega_2$ $\forallt$.
Therefore, $y_t^1 \in \Omega_2$ $\forallt$, and $\Omega_1 \subseteq \Omega_2$. Thus, for a given engagement $x_t$, $\forall u_1 \in \mathcal{U}_1$, and $\forall u_2 \in \mathcal{U}_2$ 
\begin{align}
\min_{y_t^2 \in \Omega_2} J (x_t, y_t^2 ) =J_2^\star \leq \min_{y_t^1 \in \Omega_1} J (x_t, y_t^1 ) =J_1^\star.
\end{align}
Finally, $\max [J_2^\star, J_1^\star] = J_1^\star$. It means the worst-case is in $\Omega_1$ that corresponds to $\mathcal{U}_1$.
\end{demonstration}
In addition, the worst generation trajectories, in robust unit commitment problems, are achieved when the uncertain renewable generation $u_t$ reaches the lower or upper bounds of the uncertainty set \citep[Proposition 2]{zhao2012robust}. Thus, the uncertainty set at $t$ is composed of two values and $u_t \in \{u^\text{min}; \hat{p}^{(0.5)}_t \}$.

Following \cite{bertsimas2012adaptive,jiang2011robust}, to adjust the degree of conservatism, a budget of uncertainty $\Gamma$ taking integer values between 0 and 95 is employed to restrict the number of periods that allow $u_t$ to be far away from its nominal value, \textit{i.e.}, deviations are substantial.
Therefore, the uncertainty set of renewable generation $\mathcal{U}$ is defined as follows
\begin{align}\label{eq:G_set}	
\mathcal{U} =  \bigg \{u_t \in \mathbb{R}^{T} :  \sum_{t\in \mathcal{T}}  z_t  \leq \Gamma , \ z_t \in \{0;1 \}, u_t  = \hat{y}^{\text{pv}, (0.5)}_t - z_t u^\text{min}_t \ \forallt \bigg \},
\end{align}
where $u^\text{min}_t = \hat{y}^{\text{pv}, (0.5)}_t -\hat{y}^{\text{pv}, (q)}_t$, with $0 \leq q \leq 0.5$. When $\Gamma = 0$, the uncertainty set $\mathcal{U} =\{ \hat{y}^{\text{pv}, (0.5)}_t \} $ is a singleton, corresponding to the nominal deterministic case. As $\Gamma$ increases the size of $\mathcal{U}$ enlarges. This means that a larger total deviation from the expected renewable generation is considered, so that the resulting robust solutions are more conservative and the system is protected against a higher degree of uncertainty. When $\Gamma = T$, $\mathcal{U}$ spans the entire hypercube defined by the intervals for each $u_t$.

\subsection{Second-stage planner transformation}\label{sec:optimization-ieee-ro_planner_transformed}

The robust formulation (\ref{eq:ieee-RO_min_max_min}) consists of solving a min-max-min problem, which cannot be solved directly by a commercial software such as CPLEX or GUROBI. A scenario-based approach, \textit{e.g.}, enumerating all possible outcomes of $u_t$ that could lead to the worst-case scenario for the problem, results in at least $2^\Gamma$ possible trajectories\footnote{There are $n =\sum_{k=0}^{\Gamma}  \binom{96}{k} $ possible trajectories where $n$ is within the interval $[2^\Gamma, 2^{96}]$ as $(1+1)^\Gamma = \sum_{k=0}^{\Gamma}  \binom{\Gamma}{k}$ and $(1+1)^{96} = \sum_{k=0}^{96}  \binom{96}{k}$ by using the binomial formula.}. Thus, to deal with the huge size of the problem a Benders type decomposition algorithm is implemented.

Constraints (\ref{eq:ieee-MILP_BESS_charge_discharge_cst1})-(\ref{eq:ieee-MILP_BESS_charge_discharge_cst2}) make the dispatch problem a MILP, for which a dual formulation cannot be derived. In view of this, following \cite{jiang2011robust}, the constraints (\ref{eq:ieee-MILP_BESS_charge_discharge_cst1})-(\ref{eq:ieee-MILP_BESS_charge_discharge_cst2}) are relaxed (the convergence of the relaxed dispatch problem is discussed in Section \ref{sec:ieee-convergence-checking}).
%
Then, by applying standard tools of duality theory in linear programming, the constraints and the objective function of the dual of the dispatch problem are derived.
The dual of the feasible set $\Omega(x_t, u_t)$, with (\ref{eq:ieee-MILP_BESS_charge_discharge_cst1})-(\ref{eq:ieee-MILP_BESS_charge_discharge_cst2}) relaxed, provides the dual variables $\phi_t$ and the following objective 
\begin{align}
	G \big(x_t, u_t, \phi_t \big) = & \sum_{t\in \mathcal{T}} \bigg[ \phi^\text{cha}_t \ChargeMax + \phi^\text{dis}_t \DischargeMax  - \phi^{\SocMin}_t \SocMin  + \phi^{\SocMax}_t \SocMax- \phi^{\yMin}_t \yMin + \phi^{\yMax}_t  \yMax  \notag\\
	& + \phi^{\SocIni}  \SocIni +   \phi^{\SocEnd} \SocEnd  - \phi^{d^-}_t (x_t - p P_c)   + \phi^{d^+}_t (x_t + p P_c)  + \phi^{\PVgeneration}_t u_t \bigg]  . 
\end{align}
Then, the dual of the dispatch problem $\min_{y_t \in \Omega(x_t, u_t)}  J \big(x_t, y_t \big)$ is
\begin{align}\label{eq:ieee-dispatch_dual}
\max_{\boldsymbol{\phi_t \in \Phi}} & \ G \big(x_t, u_t, \phi_t \big)  \\
\Phi & = \big \{ (\ref{eq:ieee-dispatch_dual_cst1}) - (\ref{eq:ieee-dispatch_dual_cst11}) \big\}, \notag
\end{align}
with the set of constraints $\Phi$ defined by
\begin{subequations}\label{eq:ieee-dispatch_dual_cst}
	\begin{align}	
\phi^y_t - \phi^{\yMin}_t +  \phi^{\yMax}_t - \phi^{d^-}_t +  \phi^{d^+}_t = & - \pi_t \Delta t, & \forallt  \quad &  [y_t] \label{eq:ieee-dispatch_dual_cst1} \\
- \phi^{d^-}_t \leq & \beta \pi_t \Delta t,  &  \forallt \quad & [d^-_t] \label{eq:ieee-dispatch_dual_cst2}  \\
- \phi^{d^+}_t  \leq & \beta \pi_t \Delta t,  &  \forallt \quad & [d^+_t] \label{eq:ieee-dispatch_dual_cst3}  \\
\phi^\text{dis}_1 - \phi^{y}_1 + \phi^{\SocIni} \frac{\Delta t}{\dischargeEff} \leq & 0   \label{eq:ieee-dispatch_dual_cst4}  & \quad & [\Discharge_1] \\
\phi^\text{dis}_t - \phi^{y}_t + \phi^{\SOC}_t \frac{\Delta t}{\dischargeEff} \leq & 0,  & \forallt \setminus \{1\}   \label{eq:ieee-dispatch_dual_cst5} \quad & [\Discharge_t] \\
\phi^\text{cha}_1 + \phi^{y}_1 - \phi^{\SocIni} \chargeEff \Delta t   \leq & 0 &  \label{eq:ieee-dispatch_dual_cst6} \quad & [\Charge_1] \\
\phi^\text{cha}_t + \phi^{y}_t - \phi^{\SOC}_t \chargeEff \Delta t \leq & 0 , & \forallt \setminus \{1\}   \label{eq:ieee-dispatch_dual_cst7}  \quad & [\Charge_t] \\	
- \phi^{\SocMin}_1 + \phi^{\SocMax}_1 + \phi^{\SocIni} - \phi^{\SOC}_2 \leq & 0 & \label{eq:ieee-dispatch_dual_cst8} \quad & [\SOC_1] \\
- \phi^{\SocMin}_t + \phi^{\SocMax}_t + \phi^{\SOC}_{t-1} - \phi^{\SOC}_t \leq & 0, &  \forallt \setminus \{1, 2, T\}  \label{eq:ieee-dispatch_dual_cst9} \quad & [\SOC_t]\\
 - \phi^{\SocMin}_{T} + \phi^{\SocMax}_{T} + \phi^{\SocEnd}  + \phi^{\SOC}_T\leq &  0 & \label{eq:ieee-dispatch_dual_cst10} \quad & [\SOC_{T}] \\
- \phi^{y}_t + \phi^{\PVgeneration}_t \leq & 0 , & \forallt \label{eq:ieee-dispatch_dual_cst11} \quad & [\PVgeneration_t].
	\end{align}
\end{subequations}
\begin{definition}[Sub-Problem]
As a result, the worst-case dispatch problem $ \max_{u_t \in \mathcal{U}} \big[\min_{y_t \in \Omega(x_t, u_t)}  J \big(x_t, y_t \big) \big] $ is equivalent to the following problem, which yields the sub-problem (SP) in the Benders-dual cutting plane and column-and-constraint generation algorithms
\begin{align}\label{eq:ieee-SP}
\textbf{SP}: R(x_t) &  = \max_{u_t \in \mathcal{U}, \ \phi_t \in \Phi}  \ G \big(x_t, u_t, \phi_t \big) .
\end{align}
\end{definition}
\noindent Overall, (\ref{eq:ieee-RO_min_max_min}) becomes a min-max problem
\begin{align}\label{eq:ieee-RO_min_max_final}	
\min_{x_t \in \mathcal{X}}  \bigg[ \max_{u_t \in \mathcal{U}, \ \phi_t \in \Phi} & \ G \big(x_t, u_t, \phi_t \big)\bigg]  ,
\end{align}
that can be solved using a Benders decomposition technique such as BD or CCG, between a master problem, that is linear, and a sub-problem, that is bilinear, since 
Indeed, $G$ has the following bilinear terms $\phi^{\PVgeneration}_t u_t = \phi^{\PVgeneration}_t  \hat{p}^{(0.5)}_t - \phi^{\PVgeneration}_t z_t u^\text{min}_t$. 
It is possible to linearize the products of the binary and continuous variables $z_t \phi^{\PVgeneration}_t$ of $G$ by using a standard integer algebra trick \citep{savelli2018new} with the following constraints $\forallt$
\begin{subequations}
	\begin{align}		
	- M_t^- z_t& \leq \alpha_t \leq M_t^+ z_t\\
	- M_t^-(1-z_t) & \leq \phi^{\PVgeneration}_t-\alpha_t \leq M_t^+(1-z_t) ,
	\end{align}
\end{subequations}
where  $M_t^\pm$ are the big-M's values of $\phi^{\PVgeneration}_t$ and $\alpha_t$ is an auxiliary continuous variable.
The definition of the uncertainty set (\ref{eq:G_set}) with binary variables, based on \citep[Proposition 2]{zhao2012robust}, is essential to linearize $G$.

\subsection{Controller formulation}\label{sec:optimization-ieee-controller}

last measured values, and renewable generation intraday point forecasts. It computes at each period $t$ the set-points from $t$ to the last period $T$ of the day. The formulation is the following MILP
\begin{align}\label{eq:ieee-D-controller}
\min_{y_t \in \Omega(x_t, \PVforecast_t)}& \ J \big(x_t, y_t \big)  .
\end{align}

\section{Solution methodology}\label{sec:optimization-ieee-ro_benders}

Following the methodology described by \citet{bertsimas2012adaptive,jiang2011robust}, a two-level algorithm can be used to solve the min-max (\ref{eq:ieee-RO_min_max_min}) problem with a Benders-dual cutting plane algorithm.
\begin{definition}[Master Problem]
The following master problem (MP) is solved iteratively by adding new constraints to cut off the infeasible or non-optimal solutions and is defined at iteration $j$ by
\begin{subequations}\label{eq:ieee-MP}	
\begin{align}
\textbf{MP}: \min_{x_t \in \mathcal{X}, \ \theta} &    \ \theta & \\
& \theta \geq G \big(x_t, \alpha_{t,l}, \phi_{t,l} \big),  \quad \forall l \leq j \quad & \text{optimality cuts} \label{eq:ieee-MP-opti-cut} \\
&  G \big(x_t, \tilde{\alpha}_{t,k}, \tilde{\phi}_{t,k} \big) \leq 0, \quad  \forall k \leq j , \quad & \text{feasibility cuts} \label{eq:ieee-MP-feas-cut}
\end{align}
where constraints (\ref{eq:ieee-MP-opti-cut}) represent the optimality cuts, generated by retrieving the optimal values $\alpha_{t,l},  \phi_{t,l}$ of the SP (\ref{eq:ieee-SP}), while constraints (\ref{eq:ieee-MP-feas-cut}) represent the feasibility cuts, generated by retrieving the extreme rays $\tilde{\alpha}_{t,k}, \tilde{\phi}_{t,k}$ of (\ref{eq:ieee-SP}), and $\theta$ is the optimal value of the second-stage problem.
\end{subequations}
\end{definition}
The MP can compute an optimal solution at iteration $j$ $(x_{t,j}, \theta_j)$. Note, that $R(x_{t,j})$ provides an upper bound and $\theta_{j+1}$ provides a lower bound to the optimal value of (\ref{eq:ieee-RO_min_max_min}).
Therefore, by iteratively introducing cutting planes (\ref{eq:ieee-MP-opti-cut} and (\ref{eq:ieee-MP-feas-cut}) from the SP and computing MP (\ref{eq:ieee-MP}), lower and upper bounds will converge and an optimal solution of (\ref{eq:ieee-RO_min_max_min}) can be obtained.

\subsection{Convergence warm start}

A four-dimension taxonomy of algorithmic enhancements and acceleration strategies is proposed by \cite{rahmaniani2017benders}: solution generation, solution procedure, decomposition strategy, cut generation. The solution generation is the method used to set trial values for the SP. The quality of these solutions impacts the number of iterations, as the SP uses them to generate cuts and bounds. The standard strategy is to solve the MP without modification. However, heuristics can be used as a warm-start strategy to generate an initial set of tight cuts to strengthen the MP. A simple heuristic is proposed by \cite{lin2013exact} to generate feasible solutions and a set of good initial cuts. Computational evidence demonstrated the efficiency of this approach in terms of solution quality and time. 
Therefore, we designed the following warm-start method to improve the Benders convergence by building an initial set of cuts $\{\theta_i\}_{1\leq i \leq I}$ for the master problem (\ref{eq:ieee-MP}). It consists of sampling renewable generation trajectories that are assumed to be close to the worst trajectory of $\mathcal{U}$. 
%
Let $t_1$ and $t_f$ be the time periods corresponding to the first and last non null PV 50\% quantile forecast values. If $m = t_f -(t_1 + \Gamma-1) > 0 $, $m$ trajectories are sampled. The $m^\text{th}$ sampled trajectory is built by setting the $\Gamma$ values of the PV 50\% quantile forecast to the $u_t^{min}$ lower bound for time periods $t_1 + (m-1)\leq t \leq t_1 + \Gamma -1 + (m-1)$.
An additional trajectory is built by setting the $\Gamma$ maximum values of the PV 50\% quantile forecast to $u_t^{min}$ lower bound. 
%
Then, for each sampled trajectory $u_{t,i}$, the MILP formulation (\ref{eq:ieee-D-formulation}) is used to compute the related engagement plan $x_{t,i}$. Finally, the cut $\theta_i$ is built by solving (\ref{eq:ieee-SP}) where the uncertainty set is a singleton $\mathcal{U} =\{ u_{t,i }\}$, and the engagement plan is $x_{t,i}$ to retrieve the optimal values with (\ref{eq:ieee-MP-opti-cut}) $$\theta_i = G \big(x_{t,i}, \alpha_{t,i}, \phi_{t,i} \big),  \ i = 1 \ldots I.$$

\subsection{Algorithm convergence}\label{sec:ieee-convergence-checking}

First\footnote{
The comments of this subsection apply to the BD and CCG algorithms.}, we make the \textit{relatively complete recourse} assumption that the SP is feasible for any engagement plan $x_t$ and generation trajectory $u_t$. This assumption is valid in the capacity firming framework where curtailment is allowed. If the system faces underproduction where $x_t$ is large, the generation is 0, and the BESS discharged, penalties are applied. If it encounters overproduction where $x_t$ is close to 0, the generation is large, and the BESS is charged, the excess of generation is curtailed. In both cases, there is always a feasible dispatch.
Notice, when the relatively complete recourse assumption does not hold, \cite{bertsimas2018scalable} propose an extension of the CCG algorithm.

Second, the convergence of the relaxed SP is checked at each iteration of the algorithm by ensuring there is no simultaneous charge and discharge. However, such a situation should not occur because, in case of overproduction, the excess of generation can be curtailed. Simultaneous charging and discharging could indeed be an equivalent solution to dissipate the excess energy. That solution can be avoided in practice by adding a minor penalty for using the storage system. However, we never observed simultaneous charge and discharge over the hundreds of simulations carried out.
Thus, it is not required to implement an extension of the BD or CCG algorithm that handles a linear two-stage robust optimization model with a mixed-integer recourse problem such as proposed by \cite{zhao2012exact}.

Finally, the overall convergence of the algorithm toward the optimal solution is checked. Indeed, depending on the big-M's values, the algorithm may converge by reducing the gap between the MP and SP. However, it does not ensure an optimal solution. Therefore, once the convergence between the MP and SP is reached at iteration $j=J$, the objective of the MP at $J$ is compared to the objective of the MILP formulation (\ref{eq:ieee-D-formulation}) using the worst-case generation trajectory $u_t^{\star,J}$ as parameters. If the absolute gap $|MILP^J - MP^J |$ is higher than a convergence threshold $\epsilon$, the convergence is not reached. Then, larger big-M's values are set, and the algorithm is restarted until convergence, or a stopping criterion is reached.

\subsection{Benders-dual cutting plane algorithm}

The Benders-dual cutting plane algorithm consists of solving (\ref{eq:ieee-RO_min_max_final}) without constraints [(\ref{eq:ieee-MILP_BESS_charge_discharge_cst1})-(\ref{eq:ieee-MILP_BESS_charge_discharge_cst2})] following the procedure previously described, and to obtain a day-ahead robust schedule $x_{t,J}$ at the last iteration $J$. 
\begin{definition}[Benders-dual cutting plane algorithm]
The initialization step consists of setting the initial big-M's values $M_t^- = 1$ and $M_t^+ = 0 \ \forallt$, the time limit resolution of the sub-problem (\ref{eq:ieee-SP}) to 10 s, and the threshold convergence $\epsilon $ to 0.5 \euro. Let $MP^j $, $SP^j$, be the MP and SP objective values at iteration $j$, the lower and upper bounds, respectively, and $MILP^J$ the MILP objective value using the worst renewable generation trajectory $u_t^{\star,J}$ at iteration $J$.
\begin{algorithmic}
		\STATE Initialization.
		\STATE Warm-start: build the initial set of cuts $\{\theta_i\}_{1\leq i \leq I}$.
		\WHILE{$|MILP^J - MP_2^J |>  \epsilon$ and $M_t^- <500$ }
		\STATE Initialize $j=0$, solve the MP (\ref{eq:ieee-MP}) and retrieve $x_{t,0}$.
		\WHILE{the 10 last $|MP^j - SP^j| $ are not $< \epsilon$ }
		\STATE Solve the SP (\ref{eq:ieee-SP}) with $x_{t,j}$ as parameters:
		\IF{the SP is unbounded}
		\STATE Retrieve the extreme rays $ \tilde{\alpha}_{t,k}, \tilde{\phi}_{t,k}, \ 0 \leq k \leq j$.
		\STATE Add the $k$-th feasibility cut: $ G \big(x_{t,j},\tilde{\alpha}_{t,k}, \tilde{\phi}_{t,k} \big) \leq 0$.
		\ELSE 
		\STATE Retrieve the optimal values $\alpha_{t,l}, \phi_{t,l}, \ 0 \leq l \leq j$.
		\STATE Add the $l$-th optimality cut: $ \theta \geq G \big(x_{t,j}, \alpha_{t,l}, \phi_{t,l} \big)$.
		\STATE Update the upper bound $SP^j= R(x_{t,j})$.
		\STATE SP check: no simultaneous charge and discharge.
		\ENDIF
		\STATE Solve the MP (\ref{eq:ieee-MP}): get the optimal values $\theta_j, x_{t,j}$.
		\STATE Update the lower bound $MP^j=\theta_j$ and  $j=j+1$.
		\ENDWHILE
		\STATE $j=J$: convergence between the SP and MP is reached. Check convergence with MILP: get $u_t^{\star,J}$ from $SP^J$ and compute $MILP^J$ (\ref{eq:ieee-D-formulation}).
		\IF{$|MILP^J - MP^J |>  \epsilon$}
		\IF{$M_t^- \leq 50$}
		\STATE Update big-M's values $M_t^- = 10 + M_t^- \ \forallt$. 
		\ELSE
		\STATE Update big-M's values $M_t^- = 100 + M_t^- \ \forallt$.
		\ENDIF
		\STATE Reset $j$ to 0 and restart algorithm with a new $MP$.
		\ENDIF
		\ENDWHILE
		\STATE Retrieve the final $x_{t,J}$ engagement.
	\end{algorithmic}
\end{definition}

\subsection{Column and constraints generation algorithm}

We implemented the column and constraints generation procedure proposed by \cite{zhao2012robust,zeng2013solving}.
The following master problem ($\text{MP}_2$) is solved at iteration $j$
\begin{subequations}\label{eq:ieee-ccg-MP}	
	\begin{align}
	\min_{x_t \in \mathcal{X}, \ \theta, \ \{y_t^s\}_{0 \leq s\leq j}} &    \ \theta \\
	& \theta \geq J \big(x_t, y_t^s \big),  \quad s = 0 \ldots j \label{eq:ccg-optimality-cut} \\
	&  y_t^s\in \Omega(x_t, u_t^{\star,s}), \quad s = 0 \ldots j , \label{eq:ccg-feasability-cut}
	\end{align}
\end{subequations}
where constraints (\ref{eq:ccg-optimality-cut}) and (\ref{eq:ccg-feasability-cut}) serve as optimality and feasibility, respectively. $\{y_t^s\}_{0 \leq s\leq j}$ are the new variables added to the $\text{MP}_2$, and $u_t^{\star,s}$ represent the worst PV trajectory computed by the SP at iteration $0 \leq s\leq j$.
Note: in our CCG implementation, we solve the SP with the same approach as the SP of the BD algorithm. 
\begin{definition}[Column and constraints generation algorithm]
The initialization step is identical to BD, and the CCG algorithm implemented is similar to BD. 
%
Note: there is a maximum of 50 iterations between the SP and the $\text{MP}_2$ before checking the convergence with the MILP. If the criterion is not reached, the big-M's values are increased. Indeed, at each iteration $j$ the $y_t^j$ variables are added to the $\text{MP}_2$. In our case, it represents approximately 1 000 new variables at each iteration. With 50 iterations, the $\text{MP}_2$ is a MILP with approximately 50 000 variables which begins to be hard to solve within a reasonable amount of time.
\begin{algorithmic}
		\STATE Initialization.
		\WHILE{$|MILP^J - MP_2^J |>  \epsilon$ and $M_t^- <500$ }
		\STATE Initialize $j=0$, solve the $MP_2$ (\ref{eq:ieee-ccg-MP}) and retrieve $x_{t,0}$.
		\WHILE{the two last $|MP_2^j - SP^j| $ are not $< \epsilon$ and $j < 50$ }
		\STATE Solve the SP (\ref{eq:ieee-SP}) with $x_{t,j}$ as parameters:
		\STATE Create variables $y_t^j$ in $MP_2$.
		\STATE Retrieve $u_t^{\star,j}$ from the SP.
		\STATE Add the feasibility cut to the $MP_2$: $y_t^j\in \Omega(x_t, u_t^{\star,j})$.
		\IF{the SP is bounded}
		\STATE Add the optimality cut: $ \theta \geq J \big(x_t, y_t^j \big)$.
		\STATE Update the upper bound: $SP^j= R(x_{t,j})$.
		\STATE SP check: no simultaneous charge and discharge.
		\ENDIF
		\STATE Solve the $MP_2$ (\ref{eq:ieee-ccg-MP}): get the optimal values $\theta_j, x_{t,j}$.
		\STATE Update the lower bound: $MP^j_2=\theta_j$ and $j = j+1$.
		\ENDWHILE
		\STATE $j=J$: convergence between the SP and MP is reached. Check convergence with MILP: get $u_t^{\star,J}$ from $SP^J$ and compute $MILP^J$ (\ref{eq:ieee-D-formulation}).
		\IF{$|MILP^J - MP^J_2 |>  \epsilon$}
		\IF{$M_t^- \leq 50$}
		\STATE Update big-M's values $M_t^- = 10 + M_t^- \ \forallt$. 
		\ELSE
		\STATE Update big-M's values $M_t^- = 100 + M_t^- \ \forallt$.
		\ENDIF
		\STATE Reset $j$ to 0 and restart algorithm with a new $MP_2$.
		\ENDIF
		\ENDWHILE
	\STATE Retrieve the final $x_{t,J}$ engagement.
\end{algorithmic}
\end{definition}

\section{Case Study}\label{sec:optimization-ieee-case_study}

The BD and CCG algorithms are compared on the ULi\`ege case study.
It comprises a PV generation plant with an installed capacity $P_c =$ 466.4 kWp. The PV generation is monitored on a minute basis, and the data are resampled to 15 minutes. The dataset contains 350 days from August 2019 to November 2020, missing data during March 2020.
%
The NFs approach is compared to a widely used neural architecture, referred to as Long Short-Term Memory (LSTM). In total, eight versions of the planner are considered. Four RO versions: BD-LSTM, BD-NF, CCG-LSTM, and CCG-LSTM.
Four deterministic versions: the oracle that uses perfect knowledge of the future, a benchmark that uses PV nominal point forecasts, and two versions using NFs and LSTM PV quantiles. The set of PV quantiles is $\mathcal{Q} =  \{q =10 \%, \ldots, 50 \%\}$.
The controller uses PV intraday point forecasts and the day-ahead engagements computed by the planners to compute the set-points and the profits. They are normalized by the profit obtained with the oracle planner and expressed in \%.

Section \ref{sec:optimization-ieee-numerical_settings} presents the numerical settings. Section \ref{sec:optimization-ieee-result1} provides the results of the sensitivity analysis for several risk-averse pairs $[u_t^{min}=\hat{y}^{\text{pv}, (q)}_t , \Gamma]$, with $q = 10, \ldots, 40\%$, and $\Gamma = 12, 24, 36, 48$. Section \ref{sec:optimization-ieee-result2} investigates a dynamic risk-averse parameter selection. Section \ref{sec:optimization-ieee-result3} presents the improvement in terms of computation time provided by the initial set of cuts. Finally, Section \ref{sec:optimization-ieee-result4} compares the BD and CCG algorithms.

\subsection{Numerical settings}\label{sec:optimization-ieee-numerical_settings}

The testing set is composed of thirty days randomly selected from the dataset.
The simulation parameters of the planners and the controller are identical. The planning and controlling periods duration are $\Delta t = 15$ minutes. The peak hours are set between 7 pm and 9 pm (UTC+0). 
The ramping power constraint on the engagements are $\Delta X_t = 7.5\% \PVcapacity$ ($15\% \PVcapacity$) during off-peak (peak) hours. The lower bounds on the engagement $\xMin$ and the net power $\yMin$ are set to 0 kW. The upper bound on the engagement $\xMax$ and the net power $\yMax$ are set to $\PVcapacity$. Finally, the engagement tolerance is $ p \PVcapacity = 1\% \PVcapacity$, and the penalty factor $\beta = 5$.
%
The BESS minimum $\SocMin$ and maximum $\SocMax$ capacity are 0 kWh and 466.4 kWh, respectively. It is assumed to be capable of fully charging or discharging in one hour $\DischargeMax = \ChargeMax = \SocMax / 1$ with charging and discharging efficiencies $\chargeEff = \dischargeEff = 95$\%. Each simulation day is independent with a fully discharged battery at the first and last period $\SocIni = \SocEnd = 0$ kWh. 
%
The Python Gurobi library is used to implement the algorithms in Python 3.7, and Gurobi\footnote{\url{https://www.gurobi.com/}} 9.0.2 to solve all the optimization problems. Numerical experiments are performed on an Intel Core i7-8700 3.20 GHz based computer with 12 threads and 32 GB of RAM running on Ubuntu 18.04 LTS. 

\begin{figure}[htbp]
	\centering
	\begin{subfigure}{0.4\textwidth}
		\centering
		\includegraphics[width=\linewidth]{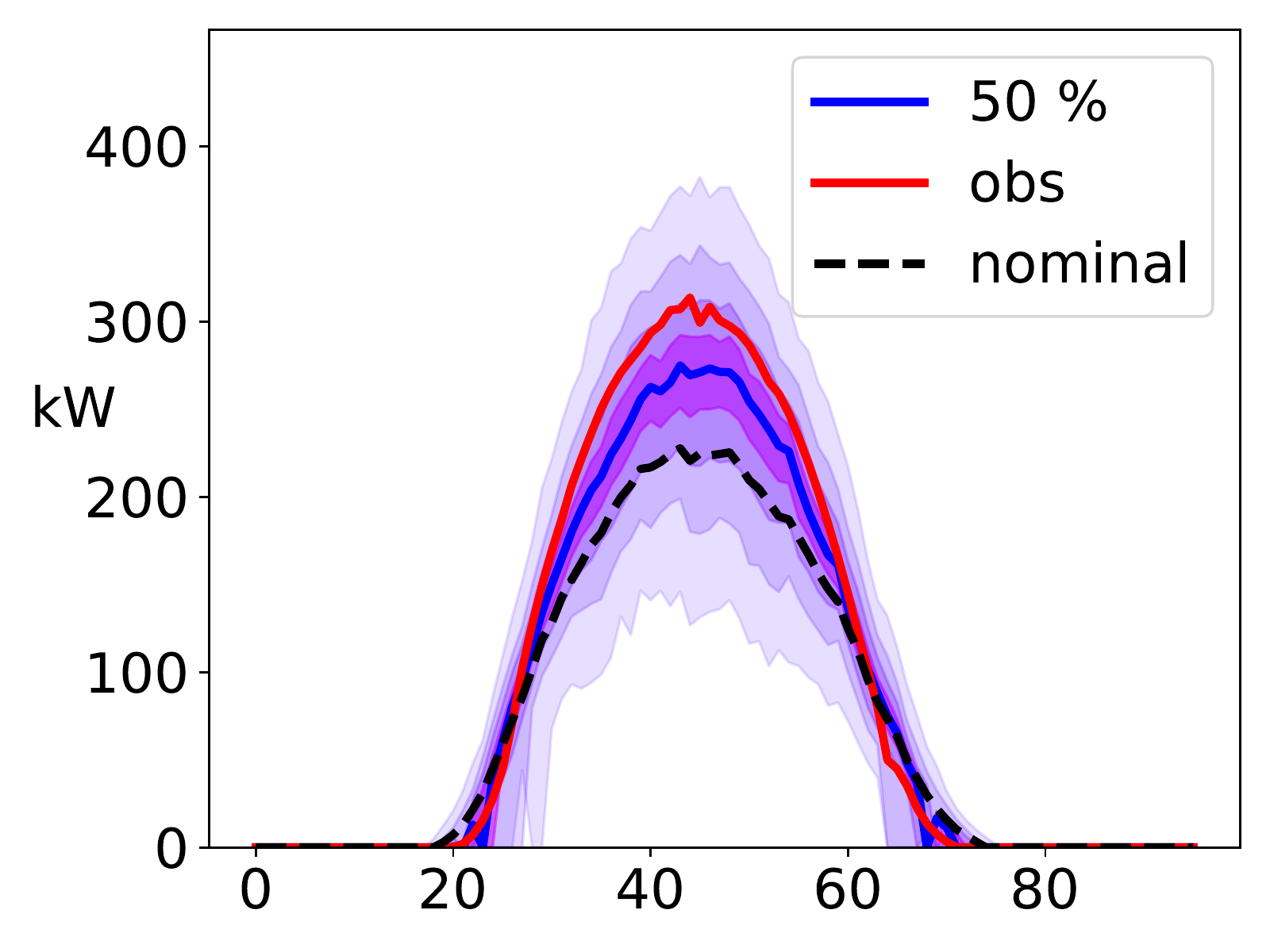}
		\caption{LSTM PV quantile forecasts.}
		\label{fig:ieee-quantile_forecast_LSTM}
	\end{subfigure}%
	\begin{subfigure}{0.4\textwidth}
		\centering
		\includegraphics[width=\linewidth]{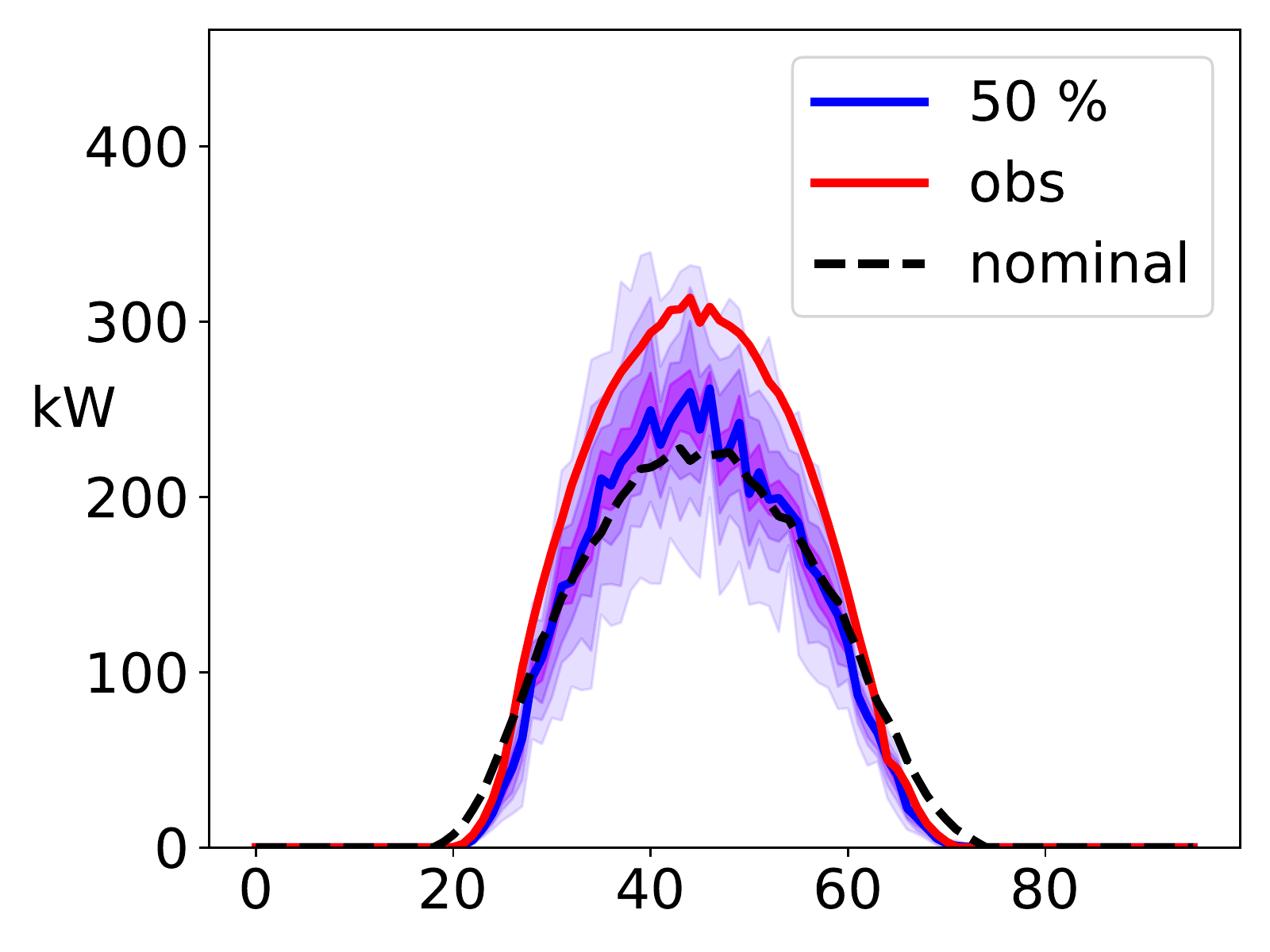}
		\caption{NFs PV quantile forecasts.}
		\label{fig:ieee-quantile_forecast_NF}
	\end{subfigure}	
	\begin{subfigure}{0.4\textwidth}
		\centering
		\includegraphics[width=\linewidth]{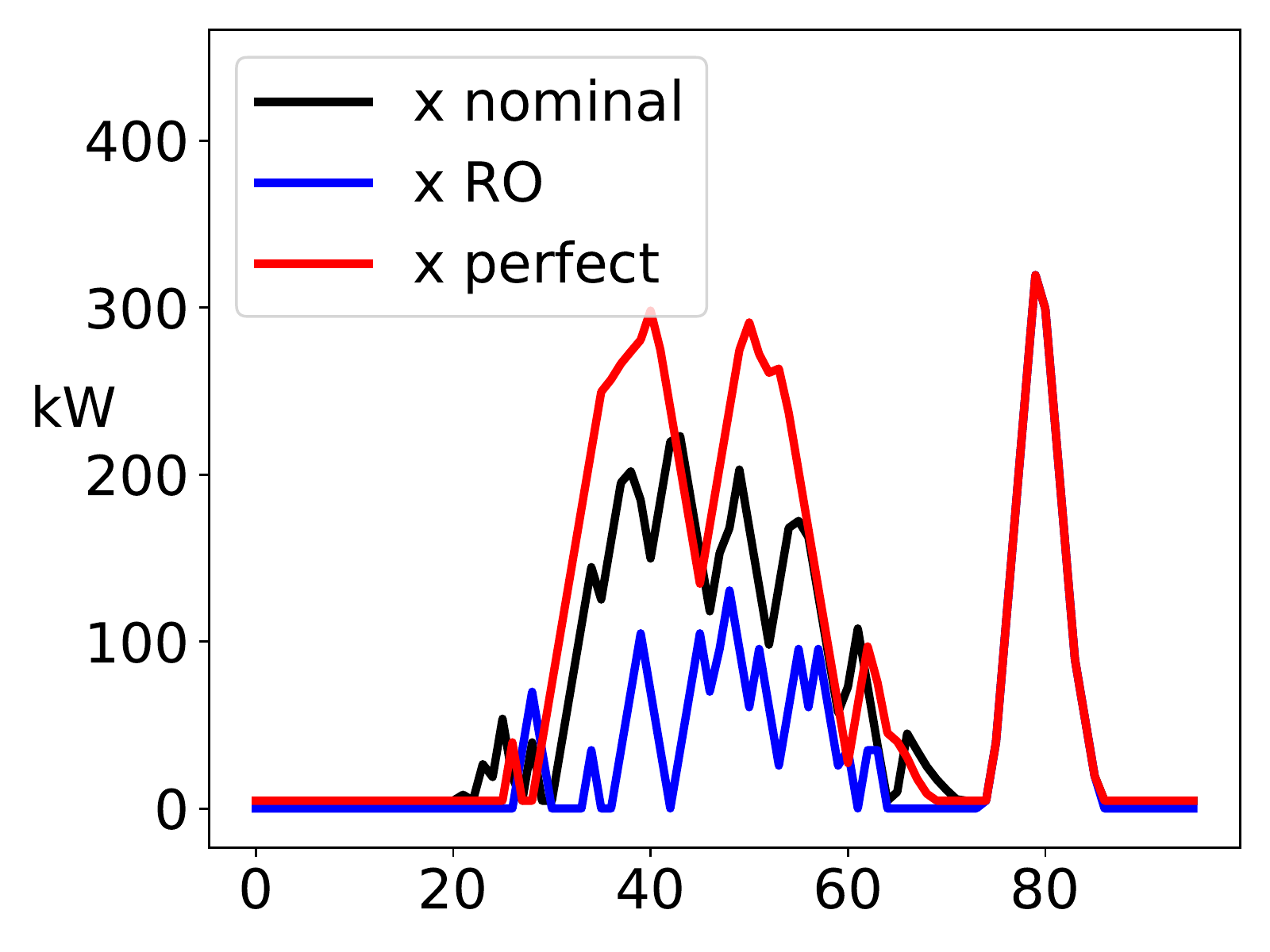}
		\caption{Engagement plan.}
		\label{fig:ieee-x}
	\end{subfigure}%
	\begin{subfigure}{0.4\textwidth}
		\centering
		\includegraphics[width=\linewidth]{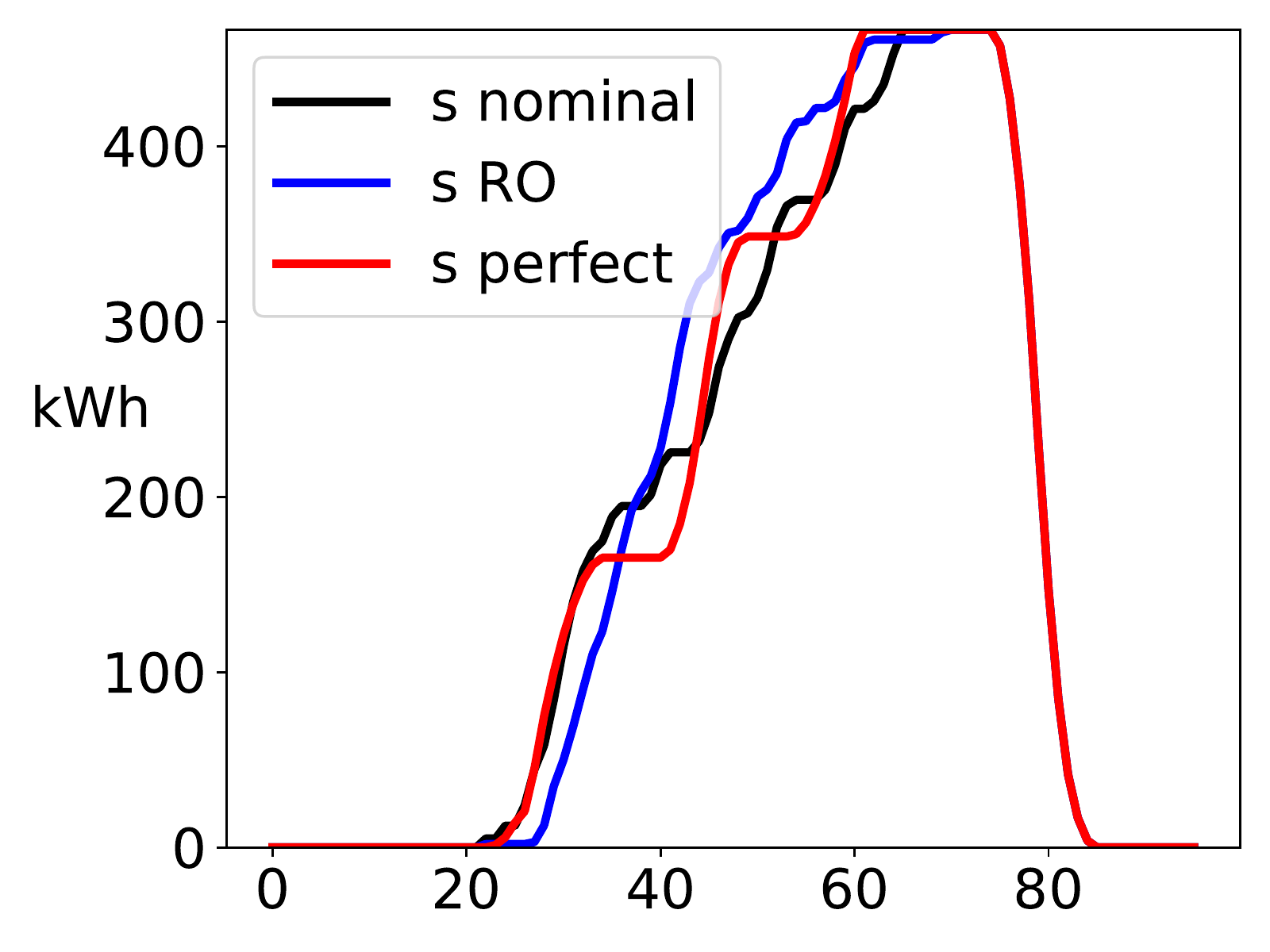}
		\caption{BESS state of charge.}
		\label{fig:ieee-s}
	\end{subfigure}
	\caption{Results illustration on $\ieeeconvdate$.}
	\label{fig:ieee-x_s_res}
\end{figure}

Figures \ref{fig:ieee-quantile_forecast_LSTM} and \ref{fig:ieee-quantile_forecast_NF} illustrate the LSTM and NFs PV quantile forecasts, observation, and nominal point forecasts on $\ieeeconvdate$. Figures \ref{fig:ieee-x} and \ref{fig:ieee-s} provide the engagement plan (x) and the BESS state of charge (s) computed with the RO planner, the deterministic planner with the nominal point forecasts, and the perfect knowledge of the future.

\subsection{Constant risk-averse parameters strategy}\label{sec:optimization-ieee-result1}

\begin{figure}[tb]
	\centering
	\begin{subfigure}{0.4\textwidth}
	\centering
	\includegraphics[width=\linewidth]{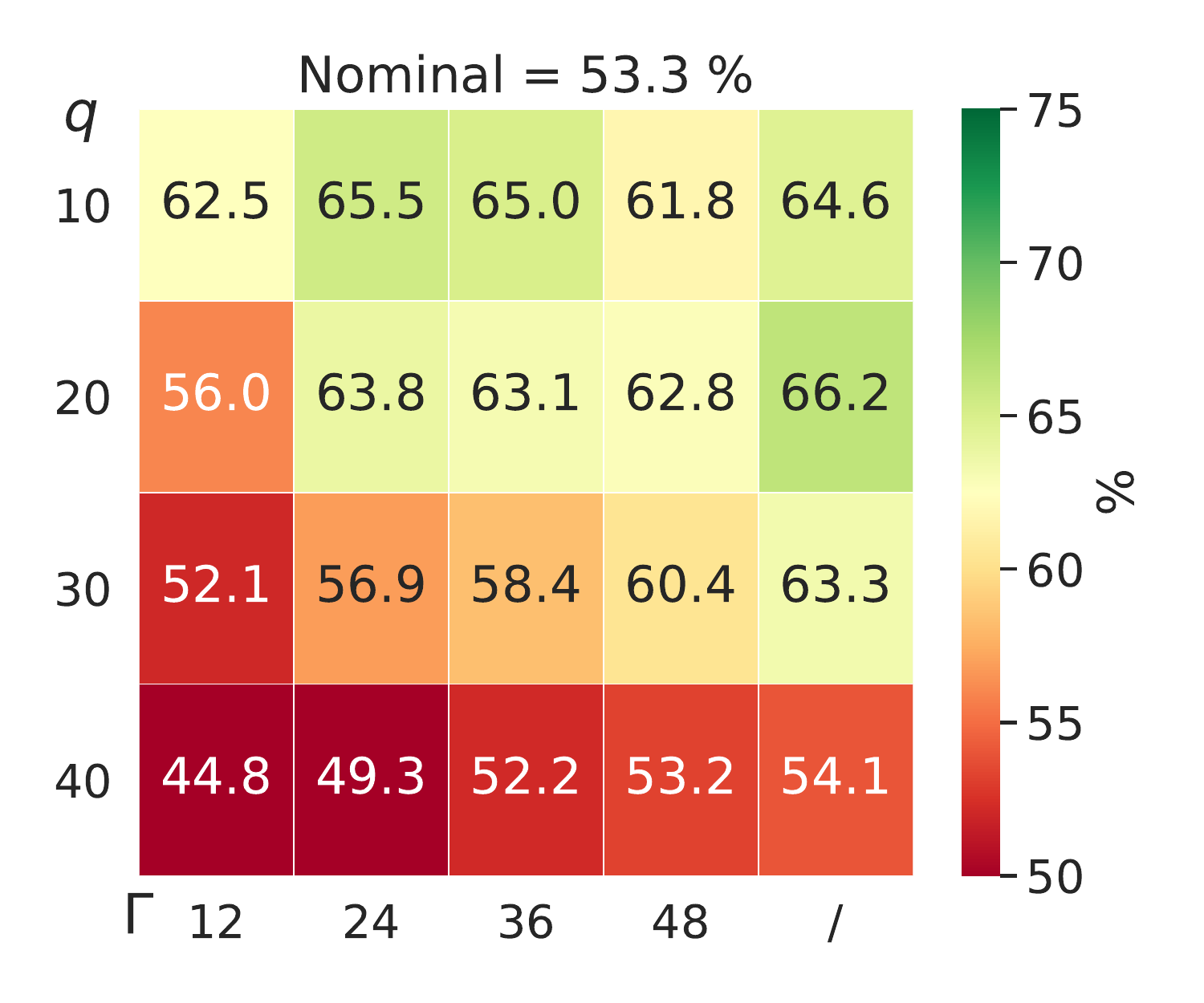}
	\caption{BD-LSTM.}
	\end{subfigure}%
	\begin{subfigure}{0.4\textwidth}
		\centering
	\includegraphics[width=\linewidth]{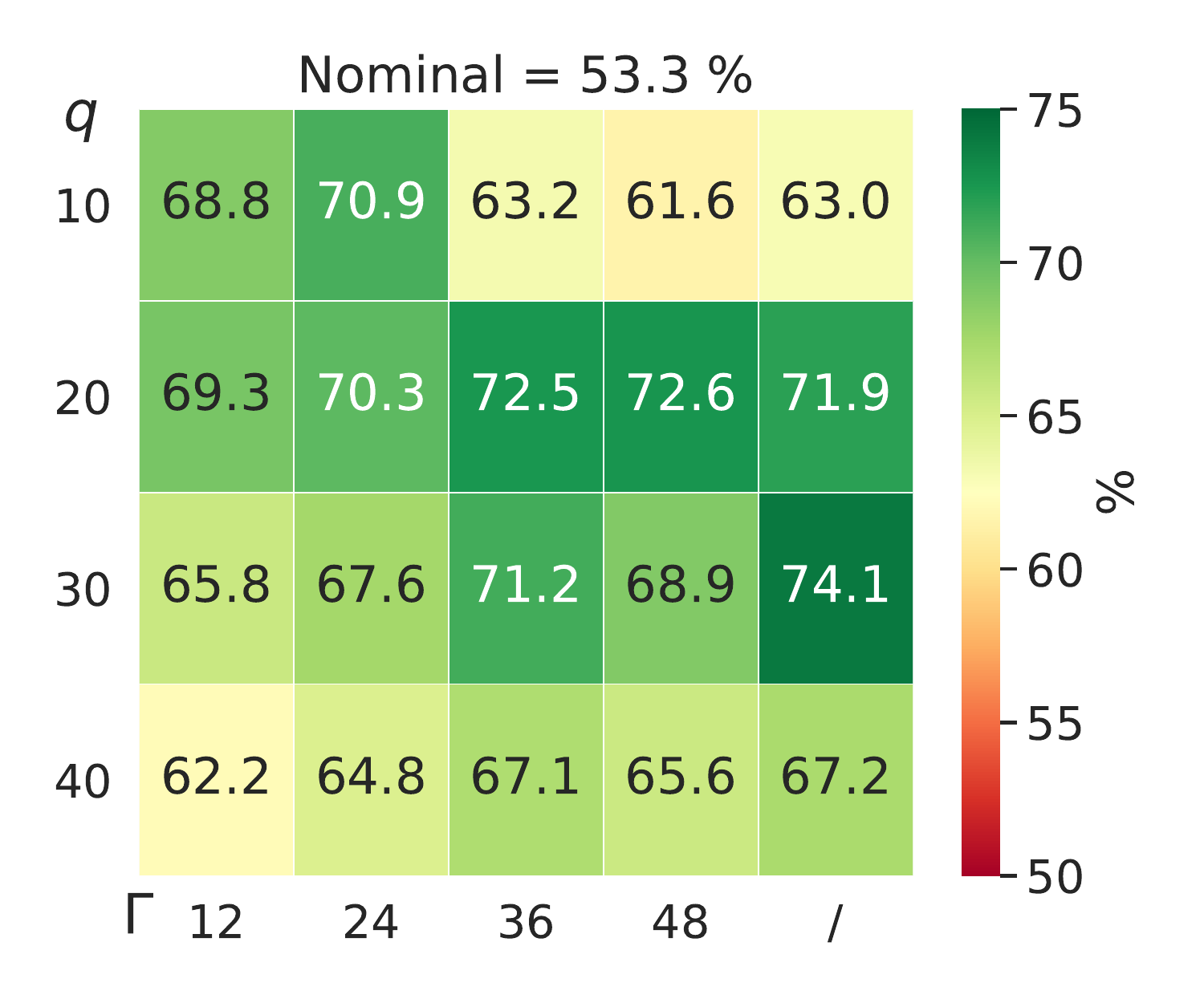}
	\caption{BD-NF.}
	\end{subfigure}
	 \begin{subfigure}{0.4\textwidth}
		\centering
	\includegraphics[width=\linewidth]{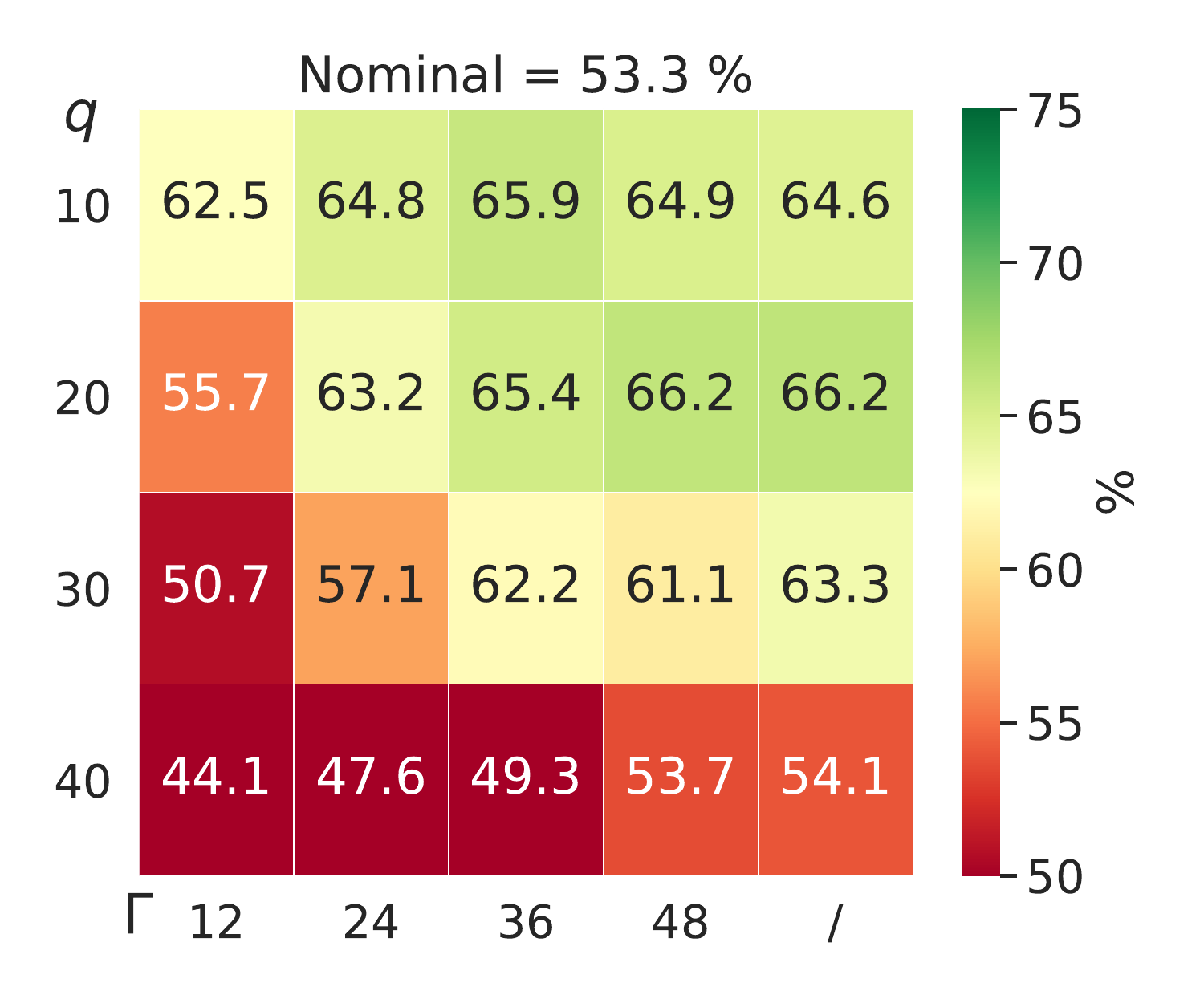}
	\caption{CCG-LSTM.}
	\end{subfigure}%
	\begin{subfigure}{0.4\textwidth}
		\centering
	\includegraphics[width=\linewidth]{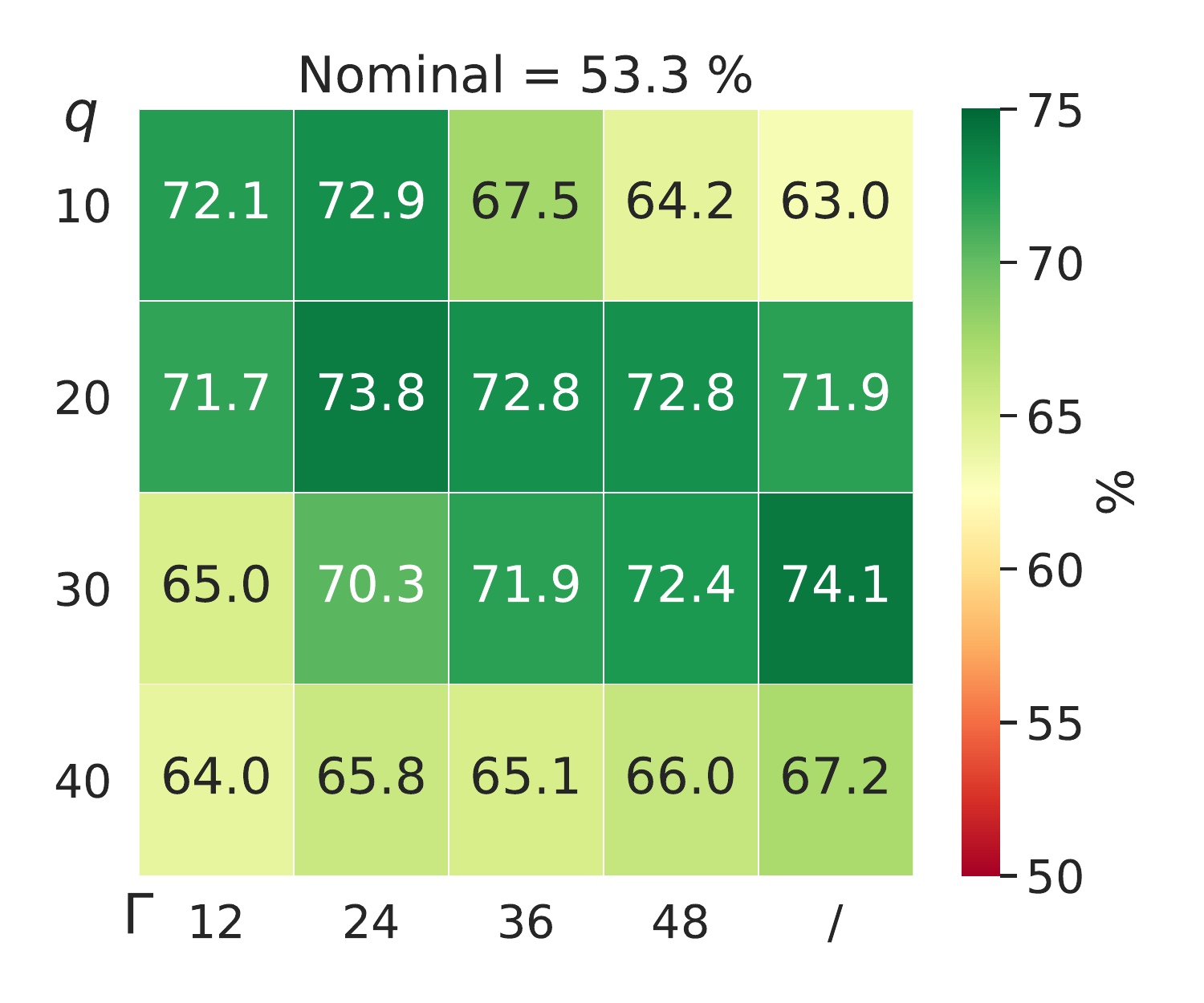}
	\caption{CCG-NF.}
	\end{subfigure}
	\caption{Results with constant risk-averse parameters. Normalized profit (\%) of the BD and CCG RO planners ($[\Gamma, q]$), deterministic ($[/,q]$) planner, and the reference that is the deterministic planner with point-forecasts (Nominal). Left part: LSTM quantiles, right part: NF quantiles.}
	\label{fig:ieee-results1}
\end{figure}

The risk-averse parameters of the RO approach $[\hat{y}^{\text{pv}, (q)}_t, \Gamma]$ are constant over the dataset. One way to identify the optimal pair is to perform a sensitivity analysis \citep{wang2015robust}. 
Figure \ref{fig:ieee-results1} provides the normalized profits of the BD-RO, CCG-RO, and deterministic planners using PV quantiles, left with LSTM and right with NFs, and nominal point forecasts. The RO and deterministic planners outperform by a large margin the baseline. The latter, the deterministic planner with nominal point forecasts, cannot deal with PV uncertainty and achieved only 53.3 \%. Then, the planners using NFs quantiles significantly outperform the planners with LSTM quantiles. 
Overall, the CCG algorithm achieved better results for almost all pairs of risk-averse parameters.
The highest profits achieved by the CCG-NF, BD-NF and NF-deterministic planners are 73.8 \%, 72.6 \% and 74.1 \%, respectively, with the risk-averse parameters $[q = 20 \%, \Gamma = 24]$, $[q = 20 \%, \Gamma = 48]$, and the quantile 30 \%.  
It should be possible to improve the RO results by tuning the risk-averse parameters $[\hat{y}^{\text{pv}, (q)}_t , \Gamma]$. However, these results emphasize the interest in considering a deterministic planner with the relevant PV quantile as point forecasts, which are easy to implement, fast to compute (a few seconds), and less prone to convergence issues than the RO approach.

\subsection{Dynamic risk-averse parameters strategy}\label{sec:optimization-ieee-result2}

In this section, the risk-averse parameters $[u_t^{min} , \Gamma]$ of the RO approach are dynamically set based on the day-ahead quantile forecasts distribution, and $u_t^{min}$ is not necessarily equal to the same quantile $\hat{y}^{\text{pv}, (q)}_t \ \forallt$.
The motivation of this strategy is to assume that the sharper the quantile forecast distribution around the median is, the more risk-averse the RO approach should be.

Two parameters are designed to this end: (1) the PV uncertainty set max depth $d_q$ to control $u_t^{min}$; (2) the budget depth $d_\Gamma$ to control $\Gamma$. $d_q$ is a percentage of the distance between the median and the 10\% quantile $d_{50-10}$, and $d_\Gamma$  is a percentage of the total installed capacity $\PVcapacity$. Then, two rules are designed to dynamically set the risk-averse parameters $[u_t^{min} , \Gamma]$ for each day of the dataset.
%
For a given day, and the set of time periods where the PV median is non null, the distances between the PV median and the PV quantiles 20, 30, and 40\% are computed: $d_{50-20}$, $d_{50-30}$, $d_{50-40}$. $u_t^{min}$ is dynamically set at each time period $t$ as follows
\begin{equation}\label{eq:min_rule}
u_t^{min} =  
\begin{cases} 
\hat{y}^{\text{pv}, (0.1)}_t   & \text{if } d^{50-20/30/40}_t  > d_q  d^{50-10}_t \\
\hat{y}^{\text{pv}, (0.2)}_t   & \text{if } d^{50-20/30}_t  > d_q  d^{50-10}_t  \\
\hat{y}^{\text{pv}, (0.3)}_t   & \text{if } d^{50-20}_t  > d_q  d^{50-10}_t  \\
\hat{y}^{\text{pv}, (0.4)}_t   & \text{otherwise}  
\end{cases} .
\end{equation}
For a given day, the budget of uncertainty $\Gamma$ is dynamically set based on the following rule
\begin{align}\label{eq:gamma_rule}	
\Gamma =   \#  \{t :  d^{50-10}_t>  d_\Gamma \PVcapacity \} .
\end{align}

\begin{figure}[tb]
	\centering
	\begin{subfigure}{0.4\textwidth}
		\centering
		\includegraphics[width=\linewidth]{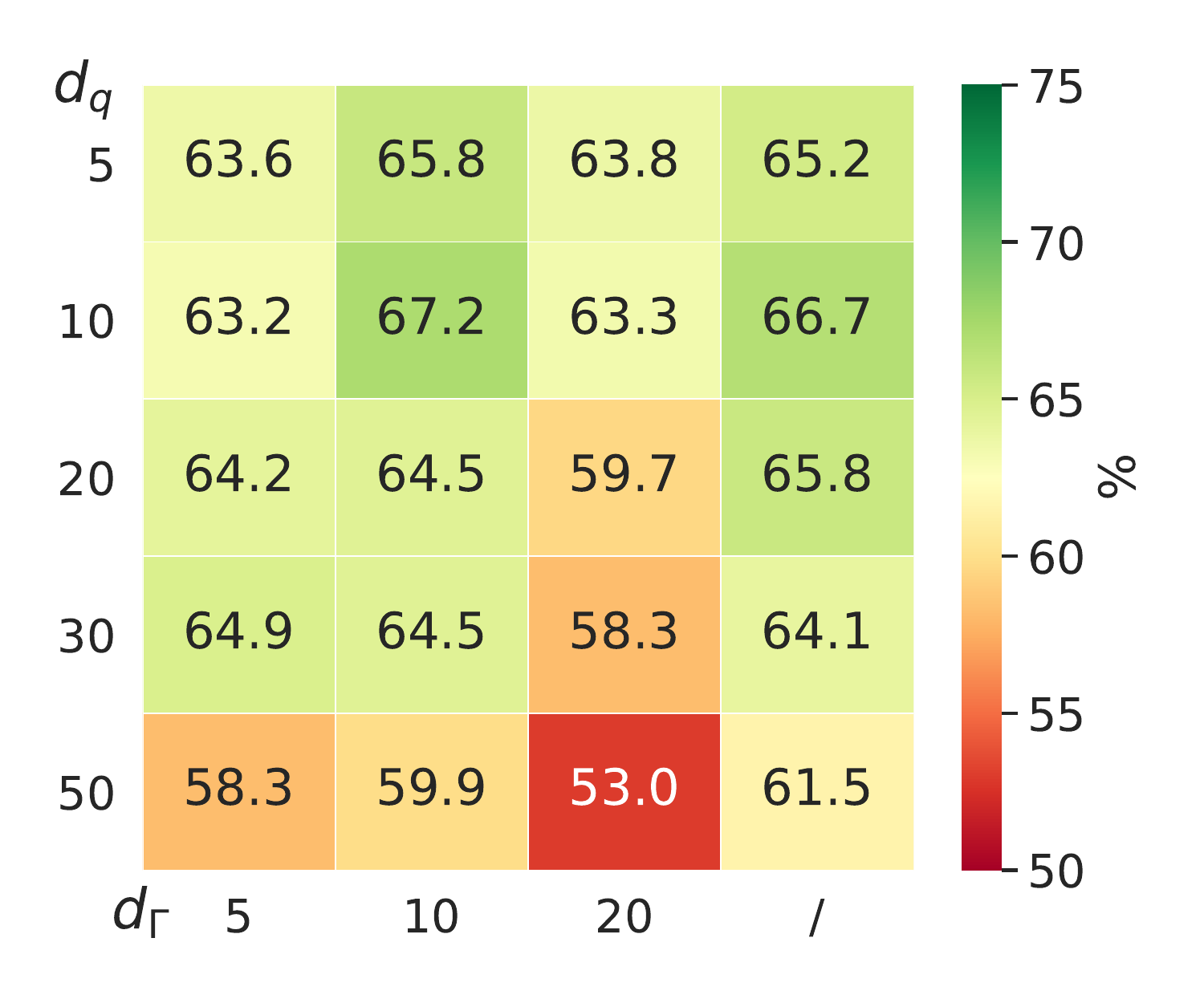}
		\caption{BD-LSTM.}
	\end{subfigure}%
	\begin{subfigure}{0.4\textwidth}
		\centering
		\includegraphics[width=\linewidth]{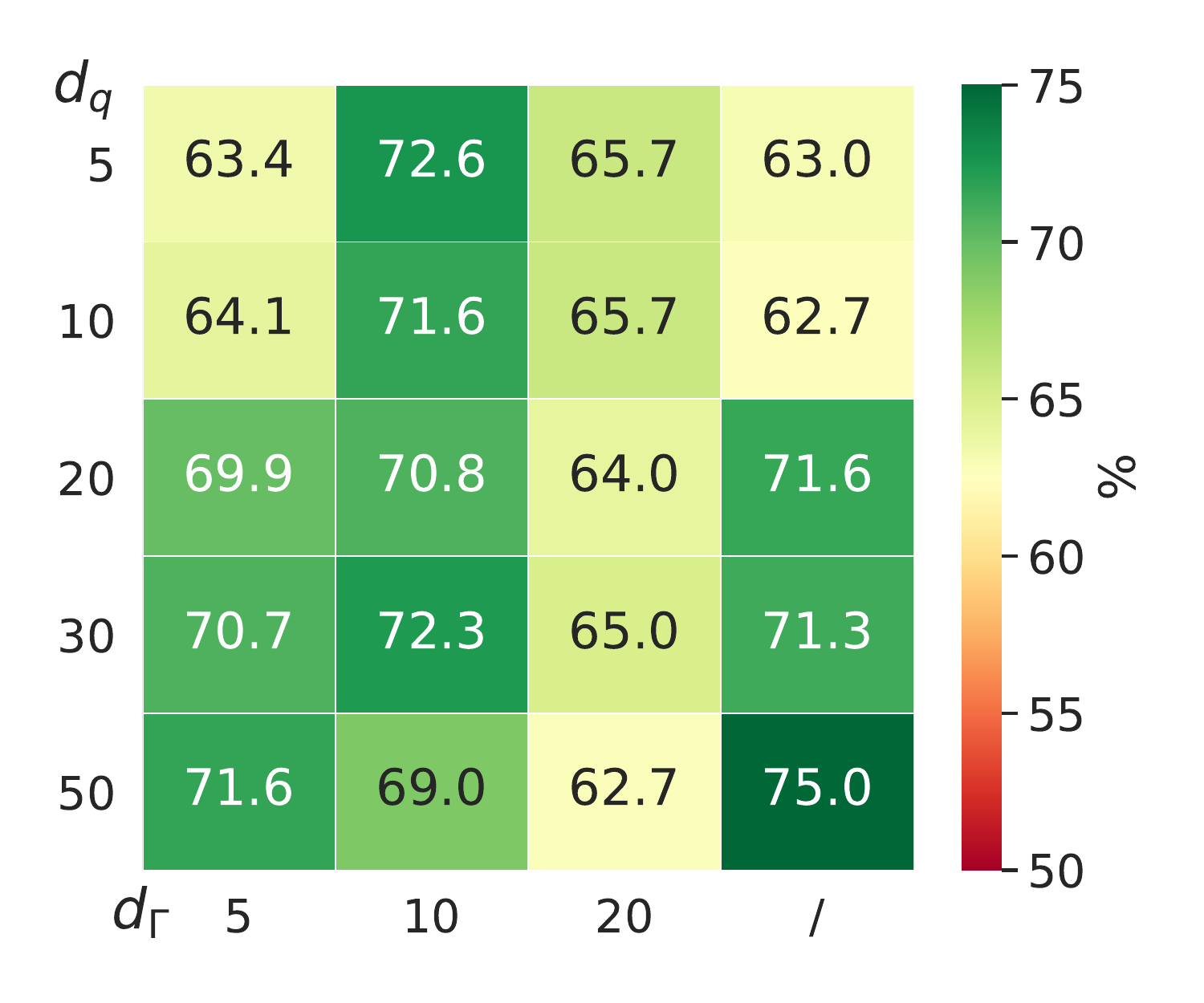}
		\caption{BD-NF.}
	\end{subfigure}
	 	\begin{subfigure}{0.4\textwidth}
		\centering
		\includegraphics[width=\linewidth]{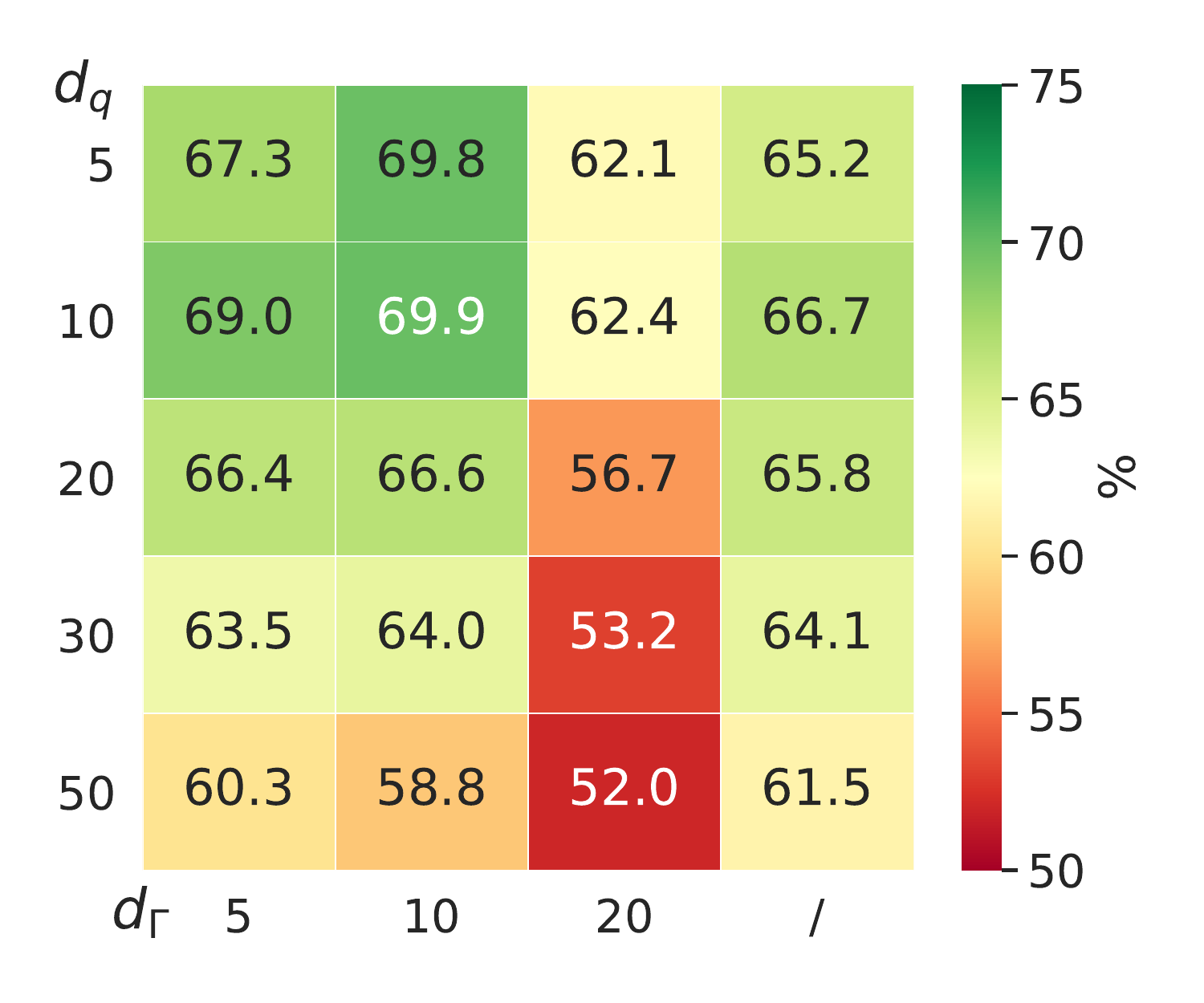}
		\caption{CCG-LSTM.}
	\end{subfigure}%
	\begin{subfigure}{0.4\textwidth}
		\centering
		\includegraphics[width=\linewidth]{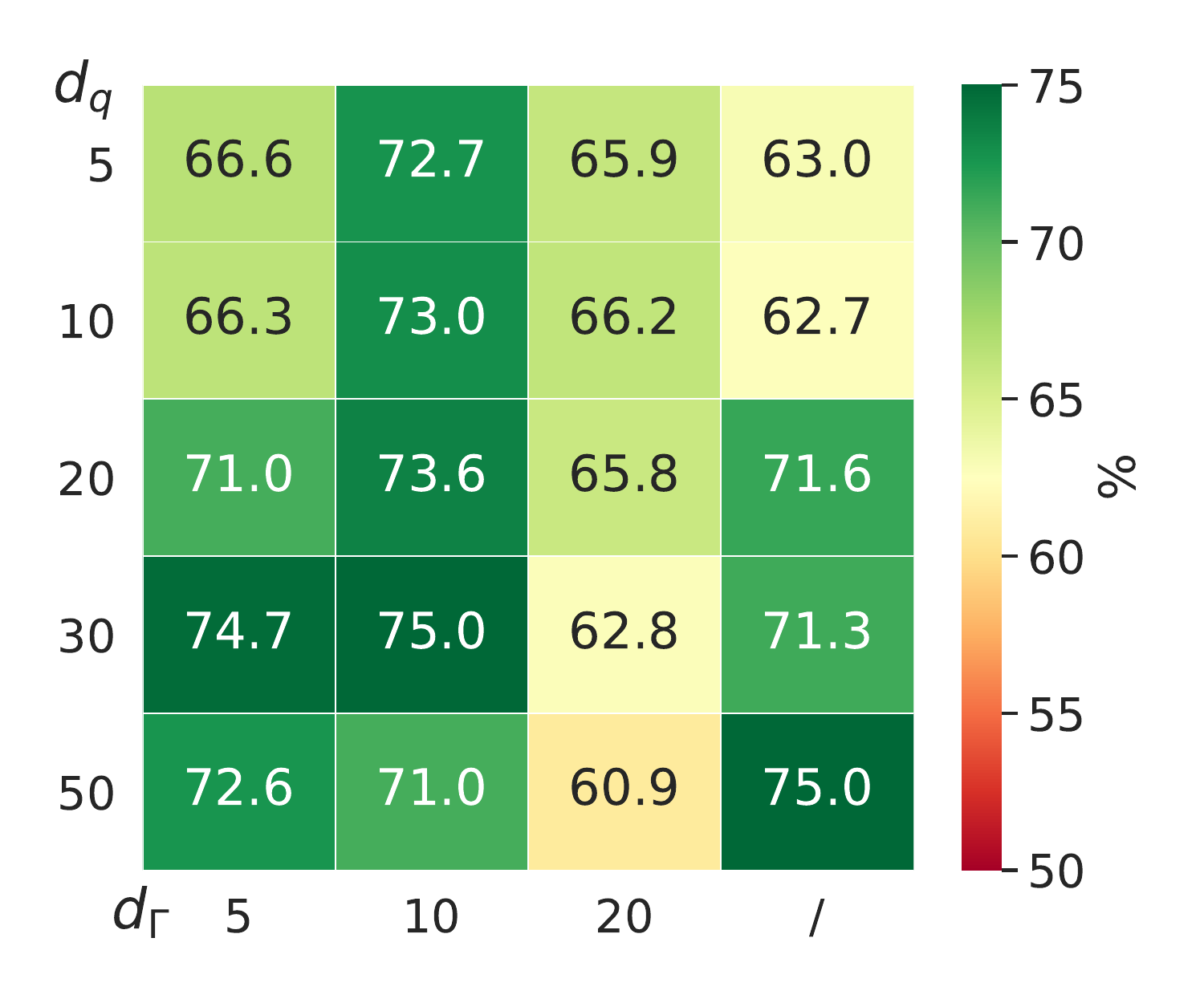}
		\caption{CCG-NF.}
	\end{subfigure}
	\caption{Results with dynamic risk-averse parameters. Normalized profit (\%) of the BD and CCG RO planners ($[d_\Gamma, d_q]$), and deterministic ($[/,d_q]$) planner. Left part: LSTM quantiles, right part: NF quantiles.}
	\label{fig:ieee-results2}
\end{figure}

Figure \ref{fig:ieee-results2} provides the normalized profits of the CCG-RO, BD-RO, and deterministic planners for several pairs $[d_\Gamma, d_q]$ using both the LSTM and NF quantiles. 
The planners achieved better results when using the NF quantiles. 
%
%
Overall, the results are improved compared to fixed risk-averse parameters for all the planners. 
The highest profits achieved by the CCG-NF, BD-NF and NF-deterministic planners are 75.0 \%, 72.6 \% and 75.0 \%, respectively, with $[d_\Gamma, d_q] = [10, 30]$, $[d_\Gamma, d_q] = [10, 5]$, and $d_q = 50 \%$.

\clearpage

\subsection{BD convergence warm start improvement}\label{sec:optimization-ieee-result3}

\begin{figure}[tb]
		\centering
	\begin{subfigure}{0.4\textwidth}
		\centering
	\includegraphics[width=\linewidth]{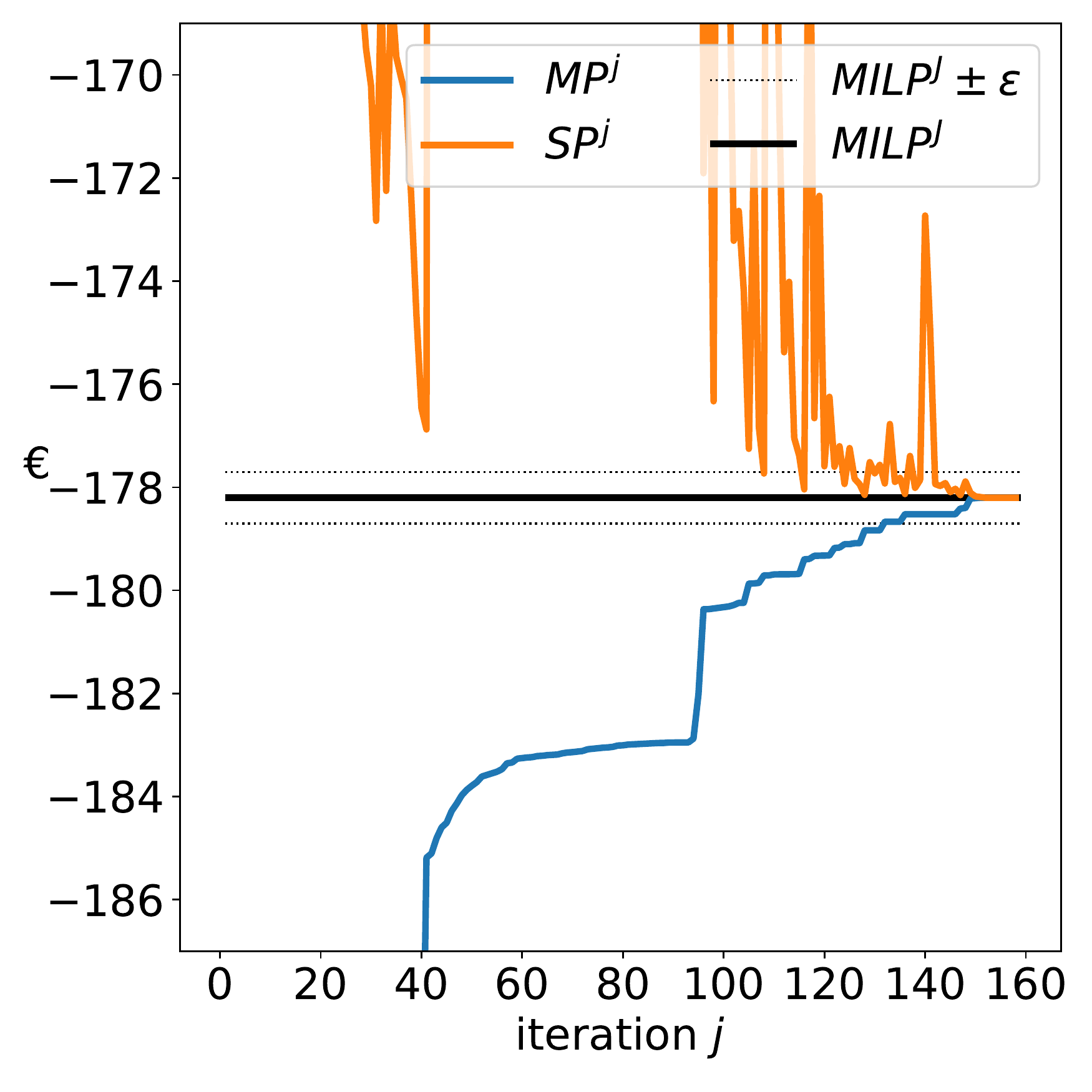}
	\end{subfigure}%
	\begin{subfigure}{0.4\textwidth}
		\centering
	\includegraphics[width=\linewidth]{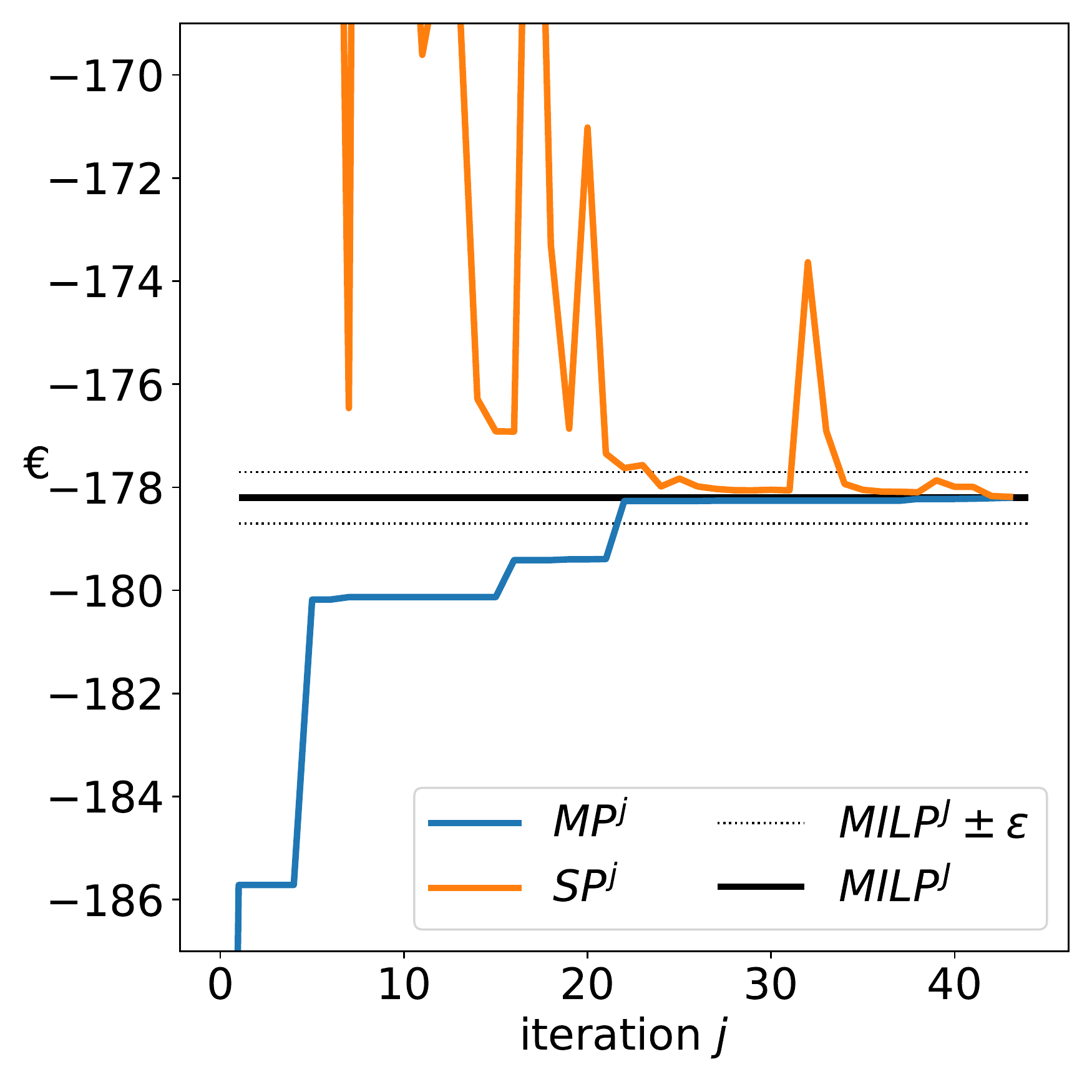}
	\end{subfigure}
	\caption{Benders convergence without (left) and with (right) an initial set of cuts on $\ieeeconvdate$.}
	\label{fig:ieee-conv_res}
\end{figure}

Overall, the warm-start procedure of the BD algorithm improves the convergence by reducing the number of iterations and allows to reach more frequently optimal solution. In addition, it reduces the number of times the big-M's values need to be increased before reaching the final convergence criterion with the MILP.
It is illustrated by considering the dynamic risk-averse parameters strategy with $[d_\Gamma, d_q] = [10,10]$.
Figure \ref{fig:ieee-conv_res} illustrates the reduction of the total number of iteration $J$ required to converge below the threshold $\epsilon$, on a specific day of the dataset. It is divided by 3.6 from 159 to 44. The computation time is divided by 4.1 from 7.4 min to 1.8 min. Table \ref{tab:ieee-computation_times} provides the computation times (min) statistics over the entire dataset with and without warm start. The averaged $t^{av}$ and total $t^{tot}$ computation times are drastically reduced when using the warm-start.
\begin{table}[tb]
\renewcommand{\arraystretch}{1.25}
	\begin{center}
		\begin{tabular}{lrrrrr}
			\hline \hline
			$\{\theta_i\}_{1\leq i \leq I}$ & $t^{av}$ & $t^{50\%}$ & $t^{min}$ & $t^{max}$ & $t^{tot}$ \\ \hline
			False & 3.5 & 2.0 & $<0.1$ & 34.1  & 105.4 \\
			True  & 2.0 & 0.7 & $<0.1$ & 30.4  & 61.3  \\ \hline \hline
		\end{tabular}
		\caption{Computation times (min) statistics.}
		\label{tab:ieee-computation_times}
	\end{center}
\end{table}

\clearpage

\subsection{BD and CCG comparison}\label{sec:optimization-ieee-result4}

Table \ref{tab:BD_vs_CCG} provides a comparison of the BD and CCG algorithms when using NF quantiles for both the static and dynamic robust optimization strategies.
Overall, the CCG algorithm converges in 5-10 iterations instead of 50-100 for BD. Therefore, the CCG computes the day-ahead planning in approximately 10 seconds, ten times faster than BD. This observation is consistent with \cite{zeng2013solving} that demonstrated the CCG algorithm converges faster than BD.
Let $n$ be the number of extreme points of the uncertainty set $\mathcal{P}$ and $m$ of the space $\Phi$ defined by constraint (\ref{eq:ieee-dispatch_dual_cst}). The BD algorithm computes an optimal solution in $O(nm)$ iterations, and the CCG procedure in $O(n)$ iterations \cite{zeng2013solving}.
Note: the BD algorithm is still competitive in an operational framework as it takes on average 1-2 minutes to compute the day-ahead planning.
However, we observed that the CCG does not always converge to an optimal solution (see Section \ref{sec:ieee-convergence-checking}), which never happened with the BD algorithm. Fortunately, these cases amount to only a few \% of the total instances. 
Overall, the CCG algorithm achieved better results than the BD for almost all the risk-averse parameters.
Finally, in our opinion, both algorithms require the same amount of knowledge to be implemented. Indeed, the only difference is the MP as the SP are solved identically. 
\begin{table}[tb]
\renewcommand{\arraystretch}{1.25}
	\begin{center}
		\begin{tabular}{lrrrr}
			\hline \hline
Algorithm & RO-type & $\overline{t}$ & $1_\%$ & $J^{max}$  \\ \hline
BD-NF   & static  & 85.2 (151.9)   & 0.0 & 72.6 \\
CCG-NF  & static  & 7.5 (6.0)      & 1.9 & 73.8  \\ \hline
BD-NF   & dynamic & 102.3 (107.3)  & 0.0 & 72.6 \\
CCG-NF  & dynamic & 9.2 (5.5)      & 4.2 & 75.0 \\ \hline\hline
\end{tabular}
\caption{BD vs CCG statistics. \\
$\overline{t}$ (s) is the averaged computation time per day with the standard deviation in the bracket. $1_{\%}$ (\%) is the \% of instances that did not terminate with optimality. $\overline{t}$ and $1_{\%}$ are computed over all days of the testing set and for all pair of constant (dynamic) risk-averse parameters $[u_t^{min} , \Gamma]$ ($[d_\Gamma, d_q]$). $J^{max}$ (\%) is the best-normalized profit achieved using the NF quantiles over all risk-averse parameters.
}
\label{tab:BD_vs_CCG}
\end{center}
\end{table}
%

\clearpage
\section{Conclusion}\label{sec:optimization-ieee-conclusions}

The core contribution of this study is to address the two-phase engagement/control problem in the context of capacity firming. 
A secondary contribution is to use a recent deep learning technique, Normalizing Flows, to compute PV quantiles. It is compared to a typical neural architecture, referred to as Long Short-Term Memory.
We developed an integrated forecast-driven strategy modeled as a min-max-min robust optimization problem with recourse that is solved using a Benders decomposition procedure.
Two main cutting plane algorithms used to address the two-stage RO unit commitment problems are compared: the Benders-dual cutting plane and the column-and-constraint generation algorithms.
The convergence is checked by ensuring a gap below a threshold between the final objective and the corresponding deterministic, objective value.
A risk-averse parameter assessment selects the optimal robust parameters and the optimal conservative quantile for the deterministic planner. Both the NF-based and LSTM-based planners outperformed the deterministic planner with nominal point PV forecasts. The NF model outperforms the LSTM in forecast value as the planner using the NF quantiles achieved higher profit than the planner with LSTM quantiles. Finally, a dynamic risk-averse parameter selection strategy is built by taking advantage of the PV quantile forecast distribution and provides further improvements. 
The CCG procedure converges ten times faster than the BD algorithm in this case study and achieves better results. However, it does not always converge to an optimal solution.

Overall, the RO approach for both the BD and CCG algorithms allows finding a trade-off between conservative and risk-seeking policies by selecting the optimal robust optimization parameters, leading to improved economic benefits compared to the baseline. Therefore, offering a probabilistic guarantee for the robust solution. However, the deterministic planner with the relevant PV quantile achieved interesting results. It emphasizes the interest to consider a well-calibrated deterministic approach. Indeed, it is easy to implement, computationally tractable for large-scale problems, and less prone to convergence issues. Note: this approach can be used in any other case study. It only requires a few months of data, renewable generation, and weather forecasts to train the forecasting models to compute reliable forecasts for the planner.


Several extensions are under investigation: (1) a stochastic formulation of the planner with improved PV scenarios based on Gaussian copula methodology or generated by a state-of-the-art deep learning technique such as Normalizing Flows, Generative Adversarial Networks or Variational AutoEncoders; (2) an improved dynamic risk-averse parameter selection strategy based on a machine learning tool capable of better-taking advantage of the PV quantiles distribution.


\chapter{Energy retailer}\label{chap:energy-retailer}

\begin{infobox}{Overview}
This Chapter investigates the forecast value assessment of the deep generative models studied in Chapter \ref{chap:scenarios-forecasting}. The reader is referred to Chapter \ref{chap:scenarios-forecasting} for the context, primary contributions, the NFs, GANs, and VAEs background, and the description of the Global Energy Forecasting Competition 2014 (GEFcom 2014) case study.

\textbf{\textcolor{RoyalBlue}{References:}} This chapter  is an adapted version of the following publication: \\[2mm]\bibentry{dumas2021nf}.
\end{infobox}
\epi{Be willing to make decisions. That's the most important quality in a good leader.}{George S. Patton}
\begin{figure}[htbp]
	\centering
	\includegraphics[width=1\linewidth]{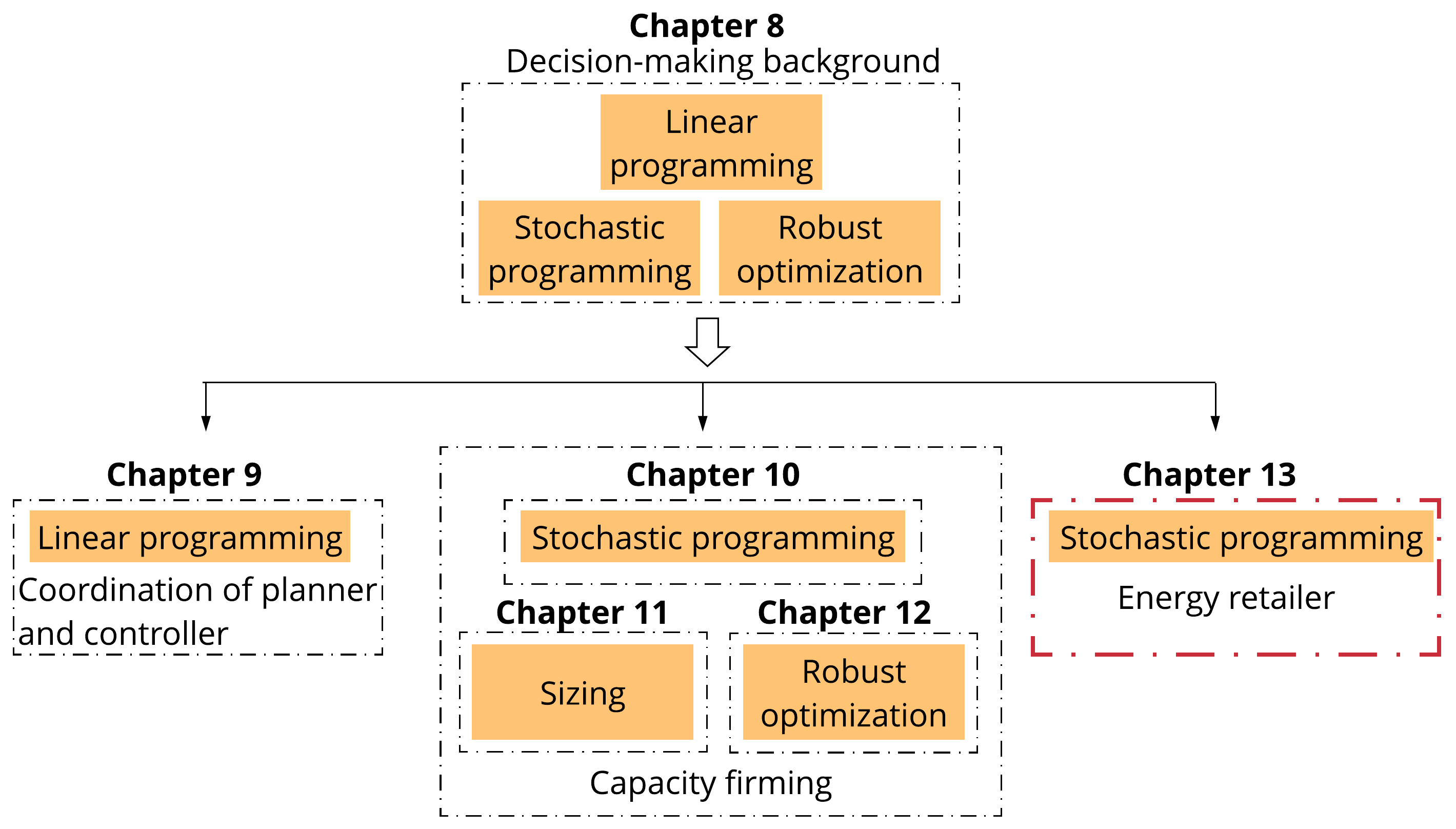}
	\caption{Chapter \ref{chap:energy-retailer} position in Part \ref{part:optimization}.}
\end{figure}
\clearpage

A model that yields lower errors in terms of forecast quality may not always point to a more effective model for forecast practitioners \citep{hong2020energy}. To this end, similarly to \citet{toubeau2018deep}, the forecast value is assessed by considering the day-ahead market scheduling of electricity aggregators, such as energy retailers or generation companies. The energy retailer aims to balance its portfolio on an hourly basis to avoid financial penalties in case of imbalance by exchanging the surplus or deficit of energy in the day-ahead electricity market. The energy retailer may have a battery energy storage system (BESS) to manage its portfolio and minimize imports from the main grid when day-ahead prices are prohibitive.

Section~\ref{sec:optimization-ijf-formulation} introduces the notations used in this Chapter. Section~\ref{sec:optimization-ijf-formulation} presents the formulation of the energy retailer case study. Section~\ref{sec:optimization-ijf-results} details empirical value results on the GEFcom 2014 dataset, and Section~\ref{sec:optimization-ijf-conclusions} summarizes the main findings and highlights ideas for further work. 

\section{Notation}\label{sec:optimization-ijf-notations}

\subsection*{Sets and indices}
\begin{supertabular}{l p{0.8\columnwidth}}
	Name & Description \\
	\hline
	$t$ & Time period index. \\
	$\omega$ & Scenario index. \\
	$T$ & Number of time periods per day. \\
	$\#\Omega$ & Number of scenarios. \\
	$\mathcal{T}$ & Set of time periods, $\mathcal{T}= \{1,2, \ldots, T\}$. \\
	$\Omega$ & Set of scenarios, $\Omega= \{1,2, \ldots, \#\Omega\}$. \\
\end{supertabular}

\subsection*{Parameters}
\begin{supertabular}{l p{0.6\columnwidth} l}
	Name & Description & Unit \\
	\hline
	$e_t^{min}$, $e_t^{max}$ & Minimum/maximum day-ahead bid. & MWh\\
	$\yMin_t$, $\yMax_t$ & Minimum/maximum retailer net position. & MWh\\
	$\DischargeMax$, $\ChargeMax$ & BESS maximum (dis)charging power.  & MW \\
	$\eta^\text{dis}$, $\eta^\text{cha}$ & BESS (dis)charging efficiency. & -\\
	$\SocMin$, $\SocMax$ & BESS minimum/maximum capacity. & MWh\\
	$\SocIni$, $\SocEnd$ & BESS initial/final state of charge.  & MWh\\
	$\pi_t$ & Day-ahead price.   & \euro/MWh \\
	$\bar{q}_t$, $\bar{\lambda}_t$ & Negative/positive imbalance price.  & \euro/MWh \\
	$\Delta t$ & Duration of a time period.  &  hour\\
\end{supertabular}

\subsection*{Variables}
\noindent For the sake of clarity the subscript $\omega$ is omitted.

\begin{supertabular}{l l p{0.5\columnwidth} l}
	Name & Range & Description & Unit \\
	\hline
	$e_t$ & $ [e_t^{min},e_t^{max}]$ & Day-ahead bid. & MWh\\
	$y_t$ & $[\yMin_t, \yMax_t]$ & Retailer net position. & MWh\\
	$\PVgeneration_t $ & $ [0, 1]$ & PV generation.  & MW\\
	$\ijfWgeneration_t $ & $ [0, 1]$ & Wind generation.   & MW \\
	$\ijfLoad_t $ & $ [0, 1]$ & Load.   & MW \\
	$\Charge_t $ & $ [0,\ChargeMax]$ & Charging power.    & MW \\
	$\Discharge_t $ & $ [0,\DischargeMax]$ & Discharging power.   & MW \\
	$\ijfSoc_t$ & $ [\SocMin, \SocMax]$ & BESS state of charge. & MWh\\
	$d^-_t$, $d^+_t$  & $ \mathbb{R}_+$ & Short/long deviation. & MWh \\
	$y_t^b$ & $ \{0, 1\}$ & BESS binary variable. & - \\
\end{supertabular}

\section{Problem formulation}\label{sec:optimization-ijf-formulation}

Let $e_t$ [MWh] be the net energy retailer position on the day-ahead market during the $t$-th hour of the day, modeled as a first stage variable. Let $y_t$ [MWh] be the realized net energy retailer position during the $t$-th hour of the day, which is modeled as a second stage variable due to the stochastic processes of the PV generation, wind generation, and load. Let $\pi_t$ [\euro / MWh] the clearing price in the spot day-ahead market for the $t$-th hour of the day, $q_t$ ex-post settlement price for negative imbalance $y_t < e_t$, and $\lambda_t$ ex-post settlement price for positive imbalance $y_t > e_t$. 
The energy retailer is assumed to be a price taker in the day-ahead market. It is motivated by the individual energy retailer capacity being negligible relative to the whole market. The forward settlement price $\pi_t$ is assumed to be fixed and known. As imbalance prices tend to exhibit volatility and are difficult to forecast, they are modeled as random variables, with expectations denoted by $\bar{q}_t = \mathop{\mathbb{E}}  [ q_t  ]$ and $\bar{\lambda}_t = \mathop{\mathbb{E}}  [ \lambda_t  ]$. They are assumed to be independent random variables from the energy retailer portfolio.

\subsection{Stochastic planner}

A stochastic planner with a linear programming formulation and linear constraints is implemented using a scenario-based approach. The planner computes the day-ahead bids $e_t$ that cannot be modified in the future when the uncertainty is resolved. The second stage corresponds to the dispatch decisions $y_{t,\omega}$ in scenario $\omega$ that aims at avoiding portfolio imbalances modeled by a cost function $f^c$. The second-stage decisions are therefore scenario-dependent and can be adjusted according to the realization of the stochastic parameters. 
The stochastic planner objective to maximize is
\begin{align}\label{eq:S_objective_1}	
J_S &  = \mathop{\mathbb{E}} \bigg[ \sum_{t\in \mathcal{T}}  \pi_t e_t +  f^c(e_t, y_{t,\omega}) \bigg] ,
\end{align}
where the expectation is taken with respect to the random variables, the PV generation, wind generation, and load. Using a scenario-based approach, (\ref{eq:S_objective_1}) is approximated by
\begin{align}\label{eq:S_objective_2}	
J_S &  \approx  \sum_{\omega \in \Omega} \alpha_\omega  \sum_{t\in \mathcal{T}} \bigg[ \pi_t e_t  +  f^c(e_t, y_{t,\omega})  \bigg],
\end{align}
with $\alpha_\omega$ the probability of scenario $\omega \in \Omega$, and $\sum_{\omega \in \Omega} \alpha_\omega = 1$. The mixed-integer linear programming (MILP) optimization problem to solve is 
\begin{subequations}
\label{eq:S_obj}	
\begin{align}
	\max_{e_t \in \mathcal{X}, y_{t, \omega} \in \mathcal{Y}(e_t)}  & \sum_{\omega \in \Omega} \alpha_\omega  \sum_{t\in \mathcal{T}} \bigg[  \pi_t e_t -  \bar{q}_t  d^-_{t, \omega}  - \bar{\lambda}_t d^+_{t, \omega} \bigg], \\
	\mathcal{X} =  & \bigg \{ e_t : e_t \in [e_t^{min},e_t^{max}] \bigg \}, \\
	\mathcal{Y}(e_t) = & \bigg \{ y_{t, \omega} : (\ref{eq:ijf-short_dev})- (\ref{eq:ijf-BESS_dyn_last_period})  \bigg \}.
\end{align}
\end{subequations}
The optimization variables are $e_t$, day-ahead bid of the net position, $\forallw$, $y_{t,\omega}$, retailer net position in scenario $\omega$, $d^-_{t, \omega}$, short deviation, $d^+_{t, \omega}$, long deviation, $\PVgeneration_{t,\omega}$, PV generation, $\ijfWgeneration_{t,\omega}$, wind generation, $\Charge_{t,\omega}$, battery energy storage system (BESS) charging power, $\Discharge_{t,\omega}$, BESS discharging power, $\ijfSoc_{t,\omega}$, BESS state of charge, and $\BESSbinary_{t,\omega}$ a binary variable to prevent from charging and discharging simultaneously. 
The imbalance penalty is modeled by the constraints (\ref{eq:ijf-short_dev})-(\ref{eq:ijf-long_dev}) $\forallw$, that define the short and long deviations variables $d^-_{t, \omega}, d^+_{t, \omega}  \in \mathbb{R}_+$. The energy balance is provided by (\ref{eq:ijf-energy_balance})  $\forallw$. The set of constraints that bound $\PVgeneration_{t,\omega}$ and $\ijfWgeneration_{t,\omega}$  variables are (\ref{eq:ijf-PV_cst})-(\ref{eq:ijf-W_cst}) $\forallw$ where $\PVforecast_{t,\omega}$ and $\ijfWforecast_{t,\omega}$ are PV and wind generation scenarios. The load is assumed to be non-flexible and is a parameter (\ref{eq:ijf-L_cst}) $\forallw$ where $\ijfLoadforecast_{t,\omega}$ are load scenarios. The BESS constraints are provided by (\ref{eq:ijf-BESS_charge})-(\ref{eq:ijf-BESS_max_soc}), and the BESS dynamics by  (\ref{eq:ijf-BESS_dyn_first_period})-(\ref{eq:ijf-BESS_dyn_last_period}) $\forallw$.
\begin{subequations}
	\label{eq:ijf-dispatch-csts}	
	\begin{align}
	- d^-_{t, \omega} & \leq  - ( e_t - y_{t,\omega} )   , \forallt \label{eq:ijf-short_dev}	\\
	- d^+_{t, \omega} & \leq -  ( y_{t,\omega} -e_t)   , \forallt  \label{eq:ijf-long_dev}	\\
	\frac{y_{t,\omega}}{\Delta t}&  =  \PVgeneration_{t,\omega} +  \ijfWgeneration_{t,\omega} -  \ijfLoad_{t,\omega}  + \Discharge_{t,\omega} - \Charge_{t,\omega}, \forallt \label{eq:ijf-energy_balance}	\\
	\PVgeneration_{t,\omega} & \leq \PVforecast_{t,\omega}, \forallt \label{eq:ijf-PV_cst}		\\
	\ijfWgeneration_{t,\omega} & \leq \ijfWforecast_{t,\omega}, \forallt \label{eq:ijf-W_cst}		\\
	\ijfLoad_{t,\omega} & = \ijfLoadforecast_{t,\omega}, \forallt \label{eq:ijf-L_cst} \\
	\Charge_{t,\omega} & \leq \BESSbinary_{t,\omega} \ChargeMax, \forallt \label{eq:ijf-BESS_charge}		\\
	\Discharge_{t,\omega} & \leq (1-\BESSbinary_{t,\omega}) \DischargeMax, \forallt \label{eq:ijf-BESS_discharge}		\\
	-\ijfSoc_{t,\omega} & \leq -s^\text{min}, \forallt \label{eq:ijf-BESS_min_soc}		\\
	\ijfSoc_{t,\omega} & \leq s^\text{max}, \forallt \label{eq:ijf-BESS_max_soc}		\\
	\frac{\ijfSoc_{1,\omega} - \SocIni}{\Delta t} & =  \eta^\text{cha} \Charge_{1,\omega} - \frac{\Discharge_{1,\omega}}{\eta^\text{dis}},  \label{eq:ijf-BESS_dyn_first_period}		\\
	\frac{\ijfSoc_{t,\omega} - \ijfSoc_{t-1,\omega}}{\Delta t}& =  \eta^\text{cha} \Charge_{t,\omega} - \frac{\Discharge_{t,\omega}}{\eta^\text{dis}}, \forallt \setminus \{1\} \label{eq:ijf-BESS_dyn_all_period}		\\
	\ijfSoc_{T,\omega}& = \SocEnd = \SocIni \label{eq:ijf-BESS_dyn_last_period} .
	\end{align}
\end{subequations}
Notice that if $\bar{\lambda}_t <0$, the surplus quantity is remunerated with a non-negative price. 
In practice, such a scenario could be avoided provided that the energy retailer has curtailment capabilities, and $(\bar{q}_t , \bar{\lambda}_t)$ are strictly positive in our case study.
%
The deterministic formulation with perfect forecasts, the oracle (O), is a specific case of the stochastic formulation by considering only one scenario where $\PVgeneration_{t,\omega}$, $\ijfWgeneration_{t,\omega}$, and $\ijfLoad_{t,\omega}$ become the actual values of PV, wind, and load $\forallt$. The optimization variables are $e_t$, $y_t$, $d_t^-$, $d_t^+$, $\PVgeneration_t$, and $\ijfWgeneration_t$, $\Charge_t$, $\Discharge_t$, $\ijfSoc_t$, and $\BESSbinary_t$.

\subsection{Dispatching}

Once the bids $e_t$ have been computed by the planner, the dispatching consists of computing the second stage variables given observations of the PV, wind power, and load. The dispatch formulation is a specific case of the stochastic formulation with $e_t$ as parameter and by considering only one scenario where  $\PVgeneration_{t,\omega}$, $\ijfWgeneration_{t,\omega}$, and $\ijfLoad_{t,\omega}$ become the actual values of PV, wind, and load $\forallt$. The optimization variables are $y_t$, $d_t^-$, $d_t^+$, $\PVgeneration_t$, and $\ijfWgeneration_t$, $\Charge_t$, $\Discharge_t$, $\ijfSoc_t$, and $\BESSbinary_t$.

\section{Value results}\label{sec:optimization-ijf-results}

The energy retailer portfolio comprises wind power, PV generation, load, and a battery energy storage device. The 50 days of the testing set are used and combined with the 30 possible PV and wind generation zones (three PV zones and ten wind farms), resulting in 1 500 independent simulated days. 
%
A two-step approach is employed to evaluate the forecast value:
\begin{itemize}
    \item First, for each generative model and the 1 500 days simulated, the two-stage stochastic planner computes the day-ahead bids of the energy retailer portfolio using the PV, wind power, and load scenarios. After solving the optimization, the day-ahead decisions are recorded.
    \item Then, a real-time dispatch is carried out using the PV, wind power, and load observations, with the day-ahead decisions as parameters.
\end{itemize}
This two-step methodology is applied to evaluate the three generative models, namely the NF, GAN, and VAE.
\begin{figure}[tb]
	\centering
    \includegraphics[width=0.5\linewidth]{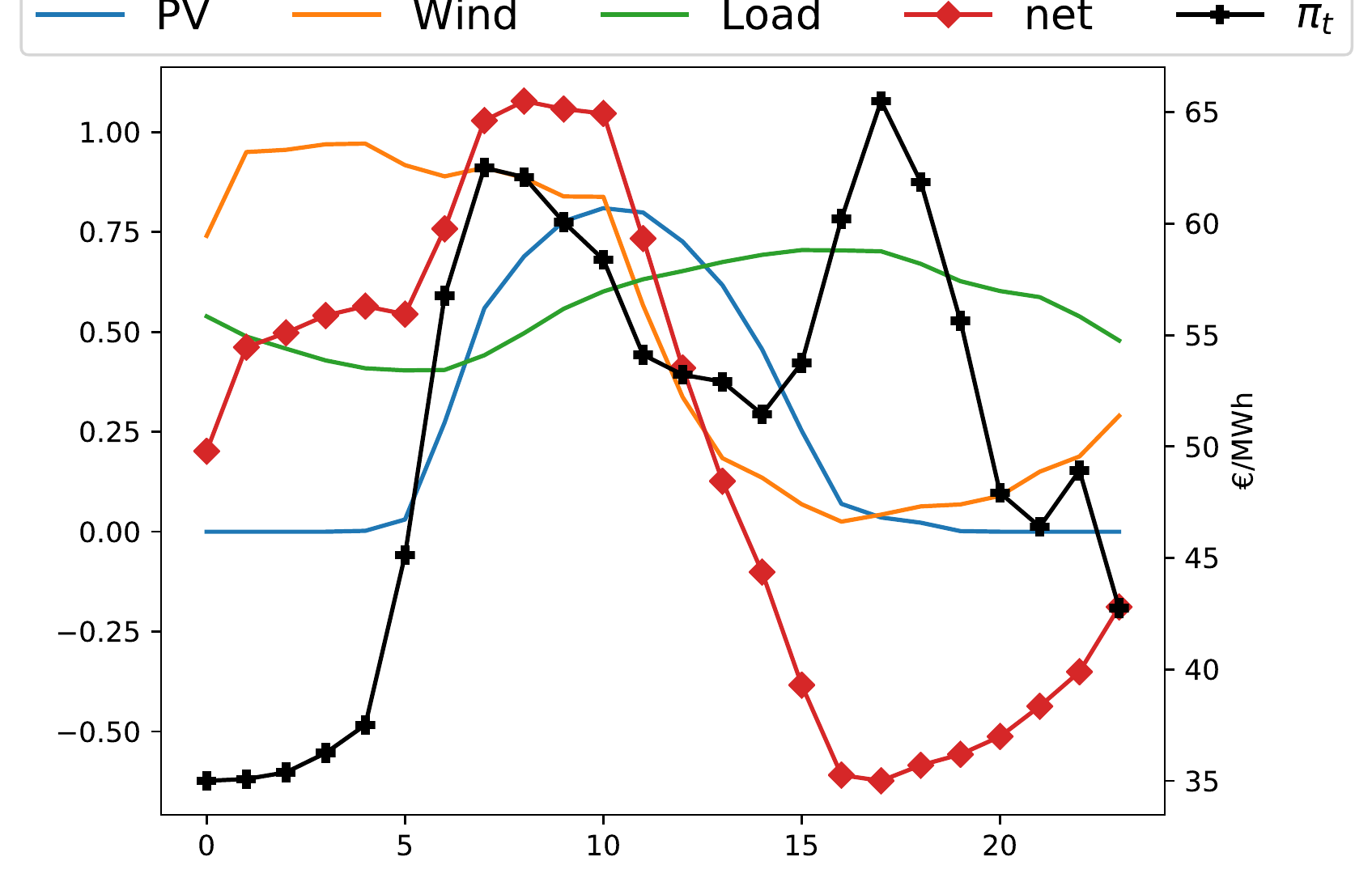}
	\caption{Energy retailer case study: illustration of the observations on a random day of the testing set. \\
	The energy retailer portfolio comprises PV generation, wind power, load, and storage device. The PV, wind power, and load scenarios from the testing set are used as inputs of the stochastic day-ahead planner to compute the optimal bids. The net is the power balance of the energy retailer portfolio. The day-ahead prices $\pi_t$ are obtained from the Belgian day-ahead market on $\ijfdaddate$.}
	\label{fig:value_dataset}
\end{figure}
%
Figure~\ref{fig:value_dataset} illustrates an arbitrary random day of the testing set with the first zone for both the PV and wind. $\pi_t$~[\euro / MWh] is the day-ahead prices on $\ijfdaddate$ of the Belgian day-ahead market used for the 1 500 days simulated. The negative $\bar{q}_t$ and positive $\bar{\lambda}_t$ imbalance prices are set to $ 2 \times \pi_t$, $\forallt$.
The retailer aims to balance the net power, the red curve in Figure~\ref{fig:value_dataset}, by importing/exporting from/to the main grid. Usually, the net is positive (negative) at noon (evening) when the PV generation is maximal (minimal), and the load is minimal (maximal). As the day-ahead spot price is often maximal during the evening load peak, the retailer seeks to save power during the day by charging the battery to decrease the import during the evening. Therefore, the more accurate the PV, wind generation, and load scenarios are, the better is the day-ahead planning.

The battery minimum $\SocMin$ and maximum $\SocMax$ capacities are 0 and 1, respectively. It is assumed to be capable of fully (dis)charging in two hours with $\DischargeMax = \ChargeMax = \SocMax / 2$, and the (dis)charging efficiencies are $\eta^\text{dis} =\eta^\text{cha} = 95$\%. Each simulation day is independent with a fully discharged battery at the first and last period of each day $\SocIni = \SocEnd = 0$. 
The 1 500 stochastic optimization problems are solved with 50 PV, wind generation, and load scenarios. The python Gurobi library is used to implement the algorithms in Python 3.7, and Gurobi\footnote{\url{https://www.gurobi.com/}} 9.0.2 is used to solve the optimization problems. Numerical experiments are performed on an Intel Core i7-8700 3.20 GHz based computer with 12 threads and 32 GB of RAM running on Ubuntu 18.04 LTS. 

The net profit, that is, the profit minus penalty, is computed for the 1 500 days of the simulation and aggregated in the first row of Table~\ref{tab:value-res}. The ranking of each model is computed for the 1 500 days. The cumulative ranking is expressed in terms of percentage in Table~\ref{tab:value-res}. NF outperformed both the GAN and VAE with a total net profit of 107 k\euro. There is still room for improvement as the oracle, which has perfect future knowledge, achieved 300 k\euro. NF, ranked first 39.0\% during the 1 500 simulation days and achieved the first and second ranks 69.6\%.
\begin{table}[htbp]
\renewcommand{\arraystretch}{1.25}
	\begin{center}
		\begin{tabular}{lrrrr}
			\hline \hline
			        &  NF & VAE & GAN \\ \hline
			Net profit (k\euro) &  \textbf{107}  &  97    &  93\\   \hline
			1 (\%)  &  \textbf{39.0} &  31.8  &  29.2\\  
			1 \& 2 (\%)  &  \textbf{69.6}  &  68.3  &  62.1\\  
			1 \& 2 \& 3 (\%)  &  100  &  100   &  100\\  \hline \hline
		\end{tabular}
\caption{Total net profit (k\euro) and cumulative ranking (\%). \\
Using the NF PV, wind power, and load scenarios, the stochastic planner achieved the highest net profit with 107 k\euro, ranked first 39.0 \%, second 30.6 \%, and third 30.4 \% over 1 500 days of simulation. The second-best model, the VAE, achieved a net profit of 97 k\euro, ranked first 31.8 \%, second 36,5 \%, and third 31.7 \%.}
\label{tab:value-res}
	\end{center}
\end{table}
Overall, in terms of forecast value, the NF outperforms the VAE and GAN. However, this case study is "simple," and stochastic optimization relies mainly on the quality of the average of the scenarios. Therefore, one may consider taking advantage of the particularities of a specific method by considering more advanced case studies. In particular, the specificity of the NFs to provide direct access to the probability density function may be of great interest in specific applications. It is left for future investigations as more advanced case studies would prevent a fair comparison between models.

\subsection{Results summary}\label{sec:discussion}

Table \ref{tab:AE-comparison} summarizes the main results of this study by comparing the VAE, GAN, and NF implemented through easily comparable star ratings. The rating for each criterion is determined using the following rules - 1 star: third rank, 2 stars: second rank, and 3 stars: first rank. Specifically, training speed is assessed based on reported total training times for each dataset: PV generation, wind power, and load; sample speed is based on reported total generating times for each dataset; quality is evaluated with the metrics considered; value is based on the case study of the day-ahead bidding of the energy retailer; the hyper-parameters search is assessed by the number of configurations tested before reaching satisfactory and stable results over the validation set; the hyper-parameters sensitivity is evaluated by the impact on the quality metric of deviations from the optimal the hyper-parameter values found during the hyper-parameter search; the implementation-friendly criterion is appraised regarding the complexity of the technique and the amount of knowledge required to implement it.
\begin{table}[htbp]
\renewcommand{\arraystretch}{1.25}
\begin{center}
\begin{tabular}{lccc}
\hline \hline
Criteria			     & VAE          &  GAN         & NF            \\ \hline
Train speed              & $\nstars[3]$ & $\nstars[2]$ & $\nstars[1]$    \\
Sample speed             & $\nstars[3]$ & $\nstars[2]$ &  $\nstars[1]$   \\
Quality                  & $\nstars[2]$ & $\nstars[1]$ & $\nstars[3]$    \\
Value                    & $\nstars[2]$ & $\nstars[1]$ & $\nstars[3]$    \\
Hp search                & $\nstars[2]$ & $\nstars[1]$ & $\nstars[3]$  \\
Hp sensibility 	         & $\nstars[2]$ & $\nstars[1]$ & $\nstars[3]$  \\
Implementation           & $\nstars[3]$ & $\nstars[2]$ & $\nstars[1]$  \\
\hline \hline
\end{tabular}
\caption{Comparison between the deep generative models. \\
The rating for each criterion is determined using the following rules - 1 star: third rank, 2 stars: second rank, and 3 stars: first rank.
Train speed: training computation time; Sample speed: scenario generation computation time; Quality: forecast quality based on the eight complementary metrics considered; Value: forecast value based on the day-ahead energy retailer case study; Hp search: assess the difficulty to identify relevant hyper-parameters; Hp sensibility: assess the sensitivity of the model to a given set of hyper-parameters (the more stars, the more robust to hyper-parameter modifications); Implementation: assess the difficulty to implement the model (the more stars, the more implementation-friendly). Note: the justifications are provided in Appendix \ref{annex:AE-table2}.
}
\label{tab:AE-comparison}
\end{center}
\end{table}

\section{Conclusions}\label{sec:optimization-ijf-conclusions}

This Chapter proposed a fair and thorough comparison of NFs with the state-of-the-art deep learning generative models, GANs and VAEs, in terms of value. The experiments are performed using the open data of the Global Energy Forecasting Competition 2014. The generative models use the conditional information to compute improved weather-based PV, wind power, and load scenarios. This study demonstrated that NFs are capable of challenging GANs and VAEs. They are overall more accurate both in terms of quality (see Chapter \ref{chap:scenarios-forecasting}) and value and can be used effectively by non-expert deep learning practitioners. 
Note, Section \ref{sec:ijf-conclusion} in Chapter \ref{chap:scenarios-forecasting} presents the NFs advantages over more traditional deep learning approaches that should motivate their introduction into power system applications.

\section{Appendix: Table \ref{tab:AE-comparison} justifications}\label{annex:AE-table2}

The VAE is the fastest to train, with a recorded computation time of 7 seconds on average per dataset. The training time of the GAN is approximately three times longer, with an average computation time of 20 seconds per dataset. Finally, the NF is the slowest, with an average training time of 4 minutes. This ranking is preserved with the VAE the fastest concerning the sample speed, followed by the GAN and NF models. The VAE and the GAN generate the samples over the testing sets, 5 000 in total, in less than a second. However, the NF considered takes a few minutes. In contrast, the affine autoregressive version of the NF is much faster to train and generate samples. Note: even a training time of a few hours is compatible with day-ahead planning applications. In addition, once the model is trained, it is not necessarily required to retrain it every day.

The quality and value assessments have already been discussed in Sections \ref{sec:ijf-numerical_results} and \ref{sec:optimization-ijf-results}. Overall, the NF outperforms both the VAE and GAN models.

Concerning the hyper-parameters search and sensibility, the NF tends to be the most straightforward model to calibrate. Compared with the VAE and GAN, we found relevant hyper-parameter values by testing only a few combinations. In addition, the NF is robust to hyper-parameter modifications. In contrast, the GAN is the most sensitive. Variations of the hyper-parameters may result in very poor scenarios both in terms of quality and shape. Even for a fixed set of hyper-parameters values, two separate training may not converge towards the same results illustrating the GAN training instabilities. The VAE is more accessible to train than the GAN but is also sensitive to hyper-parameters values. However, it is less evident than the GAN.

Finally, we discuss the implementation-friendly criterion of the models. Note: this discussion is only valid for the models implemented in this study. There exist various architectures of GANs, VAEs, and NFs with simple and complex versions.
In our opinion, the VAE is the effortless model to implement as the encoder and decoder are both simple feed-forward neural networks. The only difficulty lies in the reparameterization trick that should be carefully addressed. The GAN is a bit more difficult to deploy due to the gradient penalty to handle but is similar to the VAE with both the discriminator and the generator that are feed-forward neural networks. The NF is the most challenging model to implement from scratch because the UMNN-MAF approach requires an additional integrand network. An affine autoregressive NF is easier to implement. However, it may be less capable of modeling the stochasticity of the variable of interest. 
However, forecasting practitioners do not necessarily have to implement generative models from scratch and can use numerous existing Python libraries. 

\chapter{Part \ref{part:optimization} conclusions}\label{chap:optimization-conclusions}

\epi{Everything has been figured out, except how to live.}{Jean-Paul Sartre}

Part \ref{part:optimization} presents several approaches to address the energy management of a grid-connected microgrid. It proposes handling the renewable generation uncertainty by considering deterministic, stochastic, and robust approaches that rely on point forecasts, scenarios, and quantiles forecasts.
Several numerical experiments evaluate these approaches.
\begin{itemize}
    \item The MiRIS microgrid located at the John Cockerill Group's international headquarters in Seraing, Belgium, assesses a value function-based approach to propagate information from operational planning to real-time optimization. This methodology relies on point forecasts of PV and electrical demand to conduct the deterministic optimization. The results demonstrate the efficiency of this method to manage the peak in comparison with a Rule-Based Controller.
    \item The MiRIS case study is also employed in the capacity firming market. It allows studying a stochastic approach to address the energy management of a grid-connected renewable generation plant coupled with a battery energy storage device. The results of the stochastic planner are comparable with those of the deterministic planner, even when the prediction error variance is non-negligible. However, further investigations are required to use a more realistic methodology to generate PV scenarios and a realistic controller. 
    \item A sizing study is performed using the capacity firming framework in a different case: the ULi\`ege case study. The results indicate that the approach is not very sensitive to the control policy. The difference between the deterministic with perfect knowledge (or point forecasts) and stochastic with scenarios is minor. However, further investigations are required to implement a more realistic controller that uses intraday point forecasts and conduct a sensitivity analysis on the simulation parameters. 
    \item A robust day-ahead planner is compared to its deterministic counterpart using the ULi\`ege case study in the capacity firming framework. Overall, it allows finding a trade-off between conservative and risk-seeking policies by selecting the optimal robust optimization parameters, leading to improved economic benefits compared to the baseline.
    \item Finally, a stochastic approach deals with the optimal bidding of an energy retailer portfolio on the day-ahead market to assess the value of a recent class of deep learning generative models, the normalizing flows. The numerical experiments use the open-access data of the Global Energy Forecasting Competition 2014. The results demonstrate that normalizing flows can challenge Variational AutoEncoders and Generative Adversarial Networks. They are overall more accurate both in terms of quality (see Chapter \ref{chap:scenarios-forecasting}) and value and can be used effectively by non-expert deep learning practitioners. 
\end{itemize}
\chapter{General conclusions and perspectives}

\begin{infobox}{Overview}
This chapter concludes the thesis and identifies four potential research directions in both forecasting and planning for microgrids. (1) Forecasting techniques of the future with the development of new machine learning models that take advantage of the underlying physical process. (2) Machine learning for optimization with models that simplify optimization planning problems by learning a sub-optimal space. (3) Modelling and simulation of energy systems by applying forecasting and decomposition techniques investigated in this thesis into open-source code. (4) Machine learning and psychology where models could identify appropriate behaviors to reduce carbon footprint. Then, inform individuals, and provide constructive opportunities by modeling individual behavior.
\end{infobox}
\epi{Words are loaded pistols.}{Jean-Paul Sartre}

\section{Summary}

The IPCC report 'AR5 Climate Change 2014: Mitigation of Climate Change'\footnote{\url{https://www.ipcc.ch/report/ar5/wg3/}}, states that electrical systems are responsible for about a quarter of human-caused greenhouse gas emissions each year. According to the IPCC, the transition to a carbon-free society goes through an inevitable increase in the renewable generation's share in the energy mix and a drastic decrease in the total consumption of fossil fuels. However, a high share of renewables is challenging for power systems that have been designed and sized for dispatchable units. In addition, a major challenge is to implement these changes across all countries and contexts, as electricity systems are everywhere.
Machine learning can contribute on all fronts by informing the research, deployment, and operation of electricity system technologies. High leverage contributions in power systems include \citep{rolnick2019tackling}: accelerating the development of clean energy technologies, improving demand and clean energy forecasts, improving electricity system optimization and management, and enhancing system monitoring.
This thesis focuses on two leverages: (1) the supply and demand forecast; (2) the electricity system optimization and management.

Since variable generation and electricity demand both fluctuate, they must be forecast ahead of time to inform real-time electricity scheduling and longer-term system planning. 
Better short-term forecasts enable system operators to reduce reliance on polluting standby plants and proactively manage increasing amounts of variable sources. Better long-term forecasts help system operators and investors to decide where and when to build variable plants. Forecasts need to become more accurate, span multiple horizons in time and space, and better quantify uncertainty to support these use cases. 
Therefore, probabilistic forecasts have become a vital tool to equip decision-makers, hopefully leading to better decisions in energy applications. 

When balancing electricity systems, system operators use scheduling and dispatch to determine how much power every controllable generator should produce. This process is slow and complex, governed by NP-hard optimization problems \citep{rolnick2019tackling} such as unit commitment and optimal power flow that must be coordinated across multiple time scales, from sub-second to days ahead. 
Scheduling becomes even more complex as electricity systems include more storage, variable generators, and flexible demand. Indeed, operators manage even more system components while simultaneously solving scheduling problems more quickly to account for real-time variations in electricity production. Thus, scheduling must improve significantly, allowing operators to rely on variable sources to manage systems.

These two challenges raise the following two central research questions:
\begin{itemize}
    \item How to produce reliable probabilistic forecasts of renewable generation, consumption, and electricity prices?
    \item How to make decisions with uncertainty using probabilistic forecasts to improve scheduling?
\end{itemize}
This thesis studies the day-ahead management of a microgrid system and investigates both questions into two parts.

Part \ref{part:forecasting} investigates forecasting techniques and quality metrics required to produce and evaluate point and probabilistic forecasts. Then, Part \ref{part:optimization} use them as input of decision-making models. 
Chapters \ref{chap:forecasting-general_background} and \ref{chap:forecast_evaluation} present the forecasting basics. They introduce the different types of forecasts to characterize the behavior of stochastic variables, such as renewable generation or electrical consumption, and the assessment procedures. An example of forecast quality evaluation is provided by employing standard deep-learning models such as recurrent neural networks to compute PV and electrical consumption point forecasts.
The following Chapters of Part \ref{part:forecasting} study the point forecasts, quantile forecasts, scenarios, and density forecasts.
%
%
%
Chapter \ref{chap:quantile-forecasting} proposes to implement deep-learning models such as the encoder-decoder architecture to produce PV quantile forecasts. Then, a day-ahead robust optimization planner uses these forecasts in the form of prediction intervals in the second part of this thesis.
Chapter \ref{chap:density-forecasting} presents a density forecast-based approach to compute probabilistic forecasting of imbalance prices with a particular focus on the Belgian case.
Finally, Chapter \ref{chap:scenarios-forecasting} analyzes the scenarios of renewable generation and electrical consumption of a new class of deep learning generative models, the normalizing flows. Relevant metrics assess the forecast quality. A thorough comparison is performed with the variational autoencoders and generative adversarial networks models.

Part \ref{part:optimization} presents several approaches and methodologies based on optimization for decision-making under uncertainty. It employs the forecasts computed in the first part of the thesis.
Chapter \ref{chap:optimization-general_background} introduces the different types of optimization strategies for decision making under uncertainty.
Chapter \ref{chap:coordination-planner-controller} presents a value function-based approach as a way to propagate information from operational planning to real-time optimization in a deterministic framework.
Chapters \ref{chap:capacity-firming-stochastic}, \ref{chap:capacity-firming-sizing}, and \ref{chap:capacity-firming-robust} consider a grid-connected renewable generation plant coupled with a battery energy storage device in the capacity firming market. This framework has been designed to promote renewable power generation facilities in small non-interconnected grids. First, a stochastic optimization strategy is adopted for the day-ahead planner. Second, a sizing study of the system is conducted to evaluate the impact of the planning strategy: stochastic or deterministic. Finally, a robust day-ahead planner is considered to optimize the results regarding risk-aversion of the decision-maker. 
Finally, Chapter \ref{chap:energy-retailer} is the extension of Chapter \ref{chap:scenarios-forecasting}. It presents the forecast value of the deep learning generative models by considering the day-ahead market scheduling of an energy retailer. It is an easily reproducible case study designed to promote the normalizing flows in power system applications.

\section{Future directions}

We propose four primary research future directions. 

\subsection{Forecasting techniques of the future}

Nowadays, the renewable energy forecasting field is highly active and dynamic. Hence, new forecasting methods will likely be proposed and used in operational problems related to power system applications in the coming years.
The type of forecasts to be used as input to operational problems will depend upon the problem's nature and the formulation of the corresponding optimization problem. 
Machine learning algorithms of the future will need to incorporate domain-specific insights. For instance, the study \citep{rolnick2019tackling} defines exciting research directions. We propose in the following several research directions.
First, it is well known that the weather fundamentally drives both variable generation and electricity demand. Thus, machine learning algorithms should draw from climate modeling, weather forecasting innovations, and hybrid physics-plus-ML modeling techniques \citep{voyant2017machine,das2018forecasting}. 
Second, machine learning models could be designed to directly optimize system goals \citep{donti2019task,elmachtoub2021smart}. An example is given by \citet{donti2019task}. They use a deep neural network to produce demand forecasts that optimize electricity scheduling costs rather than forecast accuracy.
Third, better understanding the value of improved forecasts is an interesting challenge.  In this line, the benefits of specific solar forecast improvements in a region of the United States are described by \citet{martinez2016value}.
Finally, studying techniques that take advantage of the power system characteristics, such as graphical neural networks \citep{wehenkel2020graphical,donon2019graph}. They are capable of learning the power network structure and could provide successful contributions to hierarchical forecasting.
Indeed, probabilistic graphical models reduce to Bayesian networks with a pre-defined topology and a learnable density at each node \citep{wehenkel2020graphical}. From this new perspective, the graphical normalizing flow provides a promising way to inject domain knowledge into normalizing flows while preserving Bayesian networks' interpretability and the representation capacity of normalizing flows. 
In this line, a graphical model is used to detect faults in rooftop solar panels \citep{iyengar2018solarclique}. This paper proposes Solar-Clique, a data-driven approach that can flag anomalies in power generation with high accuracy.

\subsection{Machine learning for optimization}

There is a broad consensus in the power system community that the uncertain nature of renewable energy sources like wind and solar is likely to induce significant changes in the paradigms of power systems management \citep{morales2013integrating}.
The electricity market results from traditional practices, such as unit commitment or economic dispatch, designed given a generation mix mainly formed by dispatchable plants. Therefore, they are now to be reexamined so that stochastic producers can compete on equal terms. For instance, the capacity firming market is a new design conceived to promote renewable generation in isolated markets.
The type of decision-making tools to deal with these new market designs and systems, such as microgrids, depends upon the nature of the problem itself. These tools have been studied intensively, and new methods will likely be proposed and used in power system applications. 
In particular, machine learning for optimization is nowadays a hot topic. It learns partially or totally the sizing space to provide a fast and efficient sizing tool.  It also simplifies optimization planning problems by learning a sub-optimal space. This approach has already been investigated in a few applications. For instance:
\begin{itemize}
    \item Machine learning can be used to approximate or simplify existing optimization problems \citep{bertsimas2021online,zamzam2020learning}, and to find good starting points for optimization \citep{jamei2019meta}.
    \item Dynamic scheduling \citep{essl2017machine} and safe reinforcement learning can also be used to balance the electric grid in real-time.
    \item \citet{misyris2020physics} propose a framework for physics-informed neural networks in power system applications. In this line, \citet{fioretto2020predicting} present a deep learning approach to the optimal power flows. The learning model exploits the information available in the similar states of the system and a dual Lagrangian method to satisfy the physical and engineering constraints present in the optimal power flows. 
    \item \citet{donon2019graph} propose a neural network architecture that emulates the behavior of a physics solver that solves electricity differential equations to compute electricity flow in power grids. It uses proxies based on graph neural networks.
    \item \citet{tsaousoglou2021managing} consider an economic dispatch problem for a community of distributed energy resources, where energy management decisions are made online and under uncertainty. The economic dispatch problem is a multi-agent Markov Decision Process. The difficulties lie in the curse of dimensionality and in guaranteeing the satisfaction of constraints under uncertainty. A novel method that combines duality and deep learning tackles these challenges.
\end{itemize}

\subsection{Modelling and simulation of energy systems}

As already previously stated, the transition towards more sustainable fossil-free energy systems requires a high penetration of renewables, such as wind and solar. 
These new energy resources and technologies will lead to profound structural changes in energy systems, such as an increasing need for storage and drastic electrification of the heating and mobility sectors. 
Therefore, new flexible and open-source optimization modeling tools are required to capture the growing complexity of such future energy systems.
To this end, in the past few years, several open-source models for the strategic energy planning of urban and regional energy systems have been developed. A list of such models is depicted in \citep{limpens2019energyscope}.
We select and present two\footnote{In both cases, machine learning investigation for optimization may be an interesting research direction to simplify and improve the models.} of them where we think it may be relevant to implement and test the forecasting techniques and scheduling strategies developed in this thesis.

First, EnergyScope TD \citep{limpens2019energyscope} is a novel open-source model for the strategic energy planning of urban and regional energy systems. EnergyScope TD optimizes an entire energy system's investment and operating strategy, including electricity, heating, and mobility. Its hourly resolution, using typical days, makes the model suitable for integrating intermittent renewables, and its concise mathematical formulation and computational efficiency are appropriate for uncertainty applications. A new research direction could be integrating multi-criterion optimization to consider the carbon footprint, the land use, the energy return on investment, and the cost could be an exciting and challenging research direction \citep{muyldermans2021}. Solving such a complex optimization problem may require decomposition techniques such as the ones investigated in this thesis.

Second, E4CLIM \citep{tantet2019e4clim} is open-source Python software integrating flexibility needs from variable renewable energies in the development of regional energy mixes.  It aims at evaluating and
optimizing energy deployment strategies with higher shares of variable renewable energies. It also assesses the impact of new technologies and climate variability and allows conducting sensitivity studies. The
E4CLIM's potential was already illustrated at the country scale with an optimal recommissioning study of the 2015 Italian PV-wind mix. A new research direction could be to adapt and develop E4CLIM at the infra-regional scale and consider the electrical grid constraints. It may imply solving complex optimization problems requiring state-of-the-art decomposition techniques.

\subsection{Machine learning and psychology}

This last research direction goes off the beaten tracks. In my opinion, achieving sustainability goals requires as much the use of relevant technology as psychology. Therefore, one of the main challenges is not designing relevant technological tools but changing how we consume and behave in our society. In other words, technology will not save us. Nevertheless, it is the way we use technology that could help us to meet the sustainability targets. Therefore, it requires research in transdisciplinary fields to design collaborative solutions and guidances towards this goal. 
Psychology is a comprehensive tool to study and understand why there is a gap between rhetoric and actions. 
An increase in psychological research on climate change has been conducted since the last decade \citep{clayton2018psychology}. We select and provide three useful references among the vast literature on this topic to the interested reader.
Seven categories of psychological barriers or \textit{dragons of inaction} are studied by \citet{gifford2011dragons} where five strategies are proposed to help overcome them.
A discussion is proposed by \citet{nielsen2021psychology} to consider how further research can make an even more significant difference in limiting Climate Change.
Finally, \citet{abrahamse2013social} analyze the effect of social influence. Social influence refers to how our behavior is affected by what other people do or think. This study compares the effectiveness of different social influence approaches to sustainability using a meta-analysis. It also presents what remains to be investigated about the mechanisms linking social influence to behavior change. 

This topic raises several exciting research questions. (1) How can we consider psychology and human behavior when designing deep learning algorithms or decision-making tools? (2) How can we use algorithms to influence behavior towards sustainable goals? (3) How can we design new tools and propose relevant climate change solutions without leading to a rebound effect? (4) Do we need high-tech solutions to make a significant shift? Or is it more a matter of behavior and psychology? What is the optimal trade-off?

Some actions can meaningfully reduce each person’s carbon footprint and, if widely adopted, could significantly mitigate global emissions \citep{williamson2018climate}. Machine learning could help identify those behaviors, inform individuals, and provide constructive opportunities by modeling individual behavior \citep{rolnick2019tackling}.

First, by understanding how is composed our carbon footprint. Indeed, as individuals, we may lack the data and knowledge to know which decisions are most impactful regarding the carbon footprint. Therefore, machine learning can help determine an individual’s carbon footprint based on their behavior and personal and household data. For instance, a natural language processing model can identify the products we purchased from a bill or a type of transportation from email. The algorithm can also monitor household electricity and gas consumption or the food we eat. Then, making it possible to predict the associated emissions and help consumers who wish to identify which behaviors result in the highest emissions. In addition, such a machine learning model could use counterfactual reasoning to demonstrate to consumers how much their emissions would be reduced for each behavior they changed. Therefore, such information can empower people to understand how they can best help mitigate climate change through behavior change.

Second, by facilitating behavior change. Indeed, machine learning can model and cluster individuals based on their climate knowledge, preferences, demographics, and consumption characteristics \citep{gabe2016householders}. Thus, it predicts who will be most responsive to new technologies and sustainable behavior change. For instance, such techniques have improved the enrollment rate of customers in an energy savings program by 2-3x \citep{albert2016predictive}. It can also make the effects of climate change more tangible for consumers and help motivate those who wish to act. In this line, \citet{konstantakopoulos2019deep} has proposed gamification through deep learning to allow further individuals to explore their energy usage.


\begin{spacing}{0.9}


\bibliographystyle{plainnat} 
\cleardoublepage
\bibliography{References/references} 



\end{spacing}


\begin{appendices} 

\end{appendices}

\printthesisindex 

\end{document}